\documentclass[12pt]{article}

\usepackage{fancyhdr}                 
\usepackage[paperheight=11in,paperwidth=8.5in,outer=1in,inner=1in,bottom=1in,top=1in,includeheadfoot]{geometry}  
\usepackage{rotating,axodraw4j,pstricks,color,float,subfig}
\usepackage{amsmath}
\usepackage{amsthm}
\usepackage{amssymb}
\usepackage{accents}
\usepackage{bbm}
\usepackage{empheq}
\usepackage{mathrsfs}
\usepackage[colorlinks,linkcolor=black,pagebackref,bookmarksnumbered=true,breaklinks=true]{hyperref}
\usepackage{cancel,ulem,amsfonts,graphicx,latexsym,feyn,wick,dsfont}

\newcount\n
\newcommand{\sektion}[2]{\stepcounter{section}
\renewcommand{\thesection}{#1}\n=\count0 \newpage
\ifnum\count0=\n \ifnum\count0>1\ \newpage \fi\fi\section{#2}
\gdef\sectionname{#1. #2}
}

\newcommand\sectionname{}
\numberwithin{equation}{section}

\newcommand{\descriptionone}{\addcontentsline{toc}{subsection}{Units; Translational invariance, Rotational invariance and Lorentz invariance for a single free particle.}}

\newcommand{\descriptiontwo}{\addcontentsline{toc}{subsection}{Position operator; Violation of Causality; Pair Production; Fock Space; Occupation number representation; SHO; Oscillator-like formalism for Fock space.}}

\newcommand{\descriptionthree}{\addcontentsline{toc}{subsection}{Causality, observables and quantum fields; Constructing the quantum field from Fock space operators axiomatically; Translational invariance; Lorentz invariance and Relativistic normalization; Field constructed satisfies Klein-Gordon equation.}}

\newcommand{\descriptionfour}{\addcontentsline{toc}{subsection}{Constructing Fock space from the quantum field axiomatically; The Method of the Missing Box, Classical Particle Mechanics, Quantum Particle Mechanics, Classical Field Theory, Quantum Field Theory; Quantum field from free scalar theory.}}

\newcommand{\descriptionfive}{\addcontentsline{toc}{subsection}{Hamiltonian recovered in free scalar theory up to infinite constant; Normal ordering; Symmetries and conservation laws, Noether's theorem; Noether's theorem in field theory, conserved currents; ambiguity in currents; Energy-momentum tensor.}}

\newcommand{\descriptionsix}{\addcontentsline{toc}{subsection}{Lorentz transformations; Angular momentum conservation; Internal symmetries; $SO(2)$ internal symmetry; Charged field; $SO(n)$ internal symmetry.}}

\newcommand{\descriptionseven}{\addcontentsline{toc}{subsection}{Lorentz transformation properties of conserved quantities; Discrete symmetries; $\phi\rightarrow-\phi$; Charge conjugation; Parity; Ambiguity of choice of parity; Time reversal; Unitary and anti-unitary operators, angular momentum; Dilatations.}}

\newcommand{\descriptioneight}{\addcontentsline{toc}{subsection}{Scattering theory overview; Low budget scattering theory; Turning on and off function; Schr\"odinger picture, Heisenberg picture and interaction picture; Evolution operator; Time ordered product; Three models; Wick's theorem.}}

\newcommand{\descriptionnine}{\addcontentsline{toc}{subsection}{Diagrammatic perturbation theory in Model 3; Vertex in model 1; Connected diagrams; Thm: $\sum$ all Wick diagrams = $:e^{\sum\text{connected Wick diagrams}}:$; Model 1 solved; Model 2 begun.}}

\newcommand{\descriptionten}{\addcontentsline{toc}{subsection}{Model 2 finished; Vacuum energy c.t.; $S$ matrix is 1; Ground state energy; Yukawa potential; Ground state wave function; Model 3 and Mass renormalization; Renormalized perturbation theory.}}

\newcommand{\descriptioneleven}{\addcontentsline{toc}{subsection}{Feynman diagrams in Model 3; Feynman rules in model 3; A catalog of all Feynman diagrams in model 3 to $\mathcal{O}(g^2)$; Scattering amplitude at $\mathcal{O}(g^2)$; Direct and exchange Yukawa potentials.}}

\newcommand{\descriptiontwelve}{\addcontentsline{toc}{subsection}{``Nucleon"-anti``nucleon" scattering at $\mathcal{O}(g^2)$; Energy eigenstate pole; Meson-``nucleon" scattering; ``Nucleon"-anti``nucleon" annihilation; Assembling the amplitudes for various processes into one amplitude; Mandelstam variables; Mandelstam-Kibble plot; crossing symmetry; CPT; Phase space and the $S$ matrix; \\ $\frac{\text{Differential tran. prob.}}{\text{unit time}}$.}}

\newcommand{\descriptionthirteen}{\addcontentsline{toc}{subsection}{Applications of $\frac{\text{Differential tran. prob.}}{\text{unit time}}$; Decay; Cross sections, flux; Final state phase space simplified for two bodies; $\frac{d\sigma}{d\Omega}$; Optical theorem; Final state phase space for three bodies; Feynman diagrams with external lines off the mass shell; they could be an internal part of a larger diagram.}}

\newcommand{\descriptionfourteen}{\addcontentsline{toc}{subsection}{Fourier transform of the new blob; A second interpretation of Feynman diagrams with lines off the mass shell; they are the coefficients of $\rho^n$ in $\langle 0|S|0\rangle_\rho$; A third interpretation of the blob; they are the Fourier transform of the VEV of a string of Heisenberg fields; Reformulation of scattering theory; $S$ matrix elements without the turning on and off function; LSZ formula stated.}}

\newcommand{\descriptionfifteen}{\addcontentsline{toc}{subsection}{LSZ formula proved; A second look at Model 3 and its renormalization; Renormalization conditions.}}

\newcommand{\descriptionsixteen}{\addcontentsline{toc}{subsection}{Perturbative determination of a c.t.; Problems with derivative couplings; Rephrasing renormalization conditions in terms of Green's functions; Lehmann-Kallen spectral representation for the propagator; Rephrasing renormalization conditions in terms of 1PI functions.}}

\newcommand{\descriptionseventeen}{\addcontentsline{toc}{subsection}{Perturbative determination of c.t.; Corrections to external lines in the computation of $S$ matrix elements; One loop correction to meson self energy; Feynman's trick for combining 2 denominators; Shift to make denominator $SO(3,1)$ invariant; Wick notation to make denominator $O(4)$ invariant; Integral tables for convergent combinations; Self-energy at one loop studied; Combining lots of denominators; The shift in the general case to reduce any multi-loop integral to an integral over Feynman parameters.}}

\newcommand{\descriptioneighteen}{\addcontentsline{toc}{subsection}{Rephrasing coupling constant renormalization in terms of a 1PI function; Experimental significance of the definition; Renormalization versus the infinities; Renormalizable Lagrangians; Unstable particles, Decay products.}}

\newcommand{\descriptionnineteen}{\addcontentsline{toc}{subsection}{Unstable particles, lifetime, method of stationary phase; Where it begins again; Lorentz transformation laws of fields; Equivalent representations; Reducible reps; The finite dimensional inequivalent irreducible representations of SO(3); Unitarity; Complex conjugation; Direct product; Projection operators and reducibility.}}

\newcommand{\descriptiontwenty}{\addcontentsline{toc}{subsection}{Parametrizing the Lorentz group; Commutation relations for the generators; decomposition into two sets obeying $SO(3)$ commutation relations; The catalog; Complex conjugation properties; Tensor product properties; Restriction to $SO(3)$; The vector; Rank 2 tensors; Spinors.}}

\newcommand{\descriptiontwentyone}{\addcontentsline{toc}{subsection}{Lagrangian made of two component spinors; Solution of Weyl equations of motion; Weyl particles; Dirac Lagrangian; Four-component spinors; Weyl basis, Dirac basis; Plane wave solutions of the Dirac equation.}}

\newcommand{\descriptiontwentytwo}{\addcontentsline{toc}{subsection}{Plane wave solutions of the Dirac equation; Pauli's theorem; Dirac adjoint; Pauli-Feynman notation; Parity; Bilinears; Orthogonality; Completeness; Summary.}}

\newcommand{\descriptiontwentythree}{\addcontentsline{toc}{subsection}{Canonical quantization of the Dirac Lagrangian.}}

\newcommand{\descriptiontwentyfour}{\addcontentsline{toc}{subsection}{Perturbation theory for spinors; Time ordered product; Wick's theorem; Calculation of the contraction (propagator); Wick diagrams; Feynman diagrams; Matrix multiplication; Spin averages and spin sums.}}

\newcommand{\descriptiontwentyfive}{\addcontentsline{toc}{subsection}{Parity for spinors; Fermion and antifermion have opposite intrinsic parity; Charge conjugation; Majorana basis; Charge conjugation properties of fermion bilinears; Decay of ortho and para positronium; $U_PU_C=U_CU_P(-1)^{N_F}$; PT.}}

\newcommand{\descriptiontwentysix}{\addcontentsline{toc}{subsection}{Effect of PT on states; Proof of CPT theorem in perturbation theory; Renormalization of spinor theories; Propagator.}}

\newcommand{\descriptiontwentyseven}{\addcontentsline{toc}{subsection}{1PI part of propagator; Spectral representation for propagator;$\sum^\prime(\not{\!p})$ to $\mathcal{O}(g^2)$ in meson-nucleon theory; Coupling constant renormalization; Is renormalization sufficient to eliminate $\infty$'s.}}

\newcommand{\descriptiontwentyeight}{\addcontentsline{toc}{subsection}{Regularization; Regulator fields; Dim'l regularization; Minimal subtraction; BPHZ
renormalizability; Renormalization and symmetry; Renormalization of composite operators.}}

\newcommand{\nm}{^{\nu\mu}}
\newcommand{\mnd}{_{\mu\nu}}

\newcommand{\ab}{^{\alpha\beta}}

\newcommand{\mn}{^{\mu\nu}}
\newcommand{\mnls}{^{\mu\nu\lambda\sigma}}
\newcommand{\pp}{^{(+)}}
\newcommand{\mm}{^{(-)}}
\newcommand{\one}{^{-1}}

\newcommand{\ve}{\vec{e}}
\newcommand{\vf}{\vec{f}}
\newcommand{\vj}{\vec{J}}
\newcommand{\vl}{\vec{L}}
\newcommand{\vm}{\vec{M}}
\newcommand{\vp}{\vec{p}} 
\newcommand{\vs}{\vec{\sigma}} 
\newcommand{\vn}{\vec{\nabla}} 
\newcommand{\va}{\vec{\alpha}} 
\newcommand{\vx}{\vec{x}} 
\newcommand{\vy}{\vec{y}} 
\newcommand{\vpp}{\vec{p}\,'}
\newcommand{\vz}{\vec{0}}

\newcommand{\eijk}{\epsilon_{ijk}}
\newcommand{\id}{\text{Id}}

\newcommand{\pmu}{\partial_\mu}
\newcommand{\pmuu}{\partial^\mu}
\newcommand{\pnu}{\partial_\nu}

\newcommand{\plu}{\partial_\lambda}

\newcommand{\psu}{\partial_\sigma}

\newcommand{\po}{\partial_0} 

\newcommand{\SO}{\text{SO}}

\newcommand{\up}{u_+} 
\newcommand{\upd}{u_+^\dagger} 
\newcommand{\um}{u_-} 
\newcommand{\umd}{u_-^\dagger} 
\newcommand{\uvp}{u_{\vec{p}}} 
\newcommand{\vvp}{v_{\vec{p}}} 
\newcommand{\bvp}{b_{\vec{p}}}
\newcommand{\cvp}{c_{\vec{p}}}

\newcommand{\bvpd}{b_{\vec{p}}^\dagger}
\newcommand{\cvpd}{c_{\vec{p}}^\dagger}
\newcommand{\uvpr}{u_{\vec{p}}^{(r)}}
\newcommand{\vvpr}{v_{\vec{p}}^{(r)}}
\newcommand{\bvpr}{b_{\vec{p}}^{(r)}}
\newcommand{\cvpr}{c_{\vec{p}}^{(r)}}
\newcommand{\uvps}{u_{\vec{p}}^{(s)}}
\newcommand{\vvps}{v_{\vec{p}}^{(s)}}

\newcommand{\uvprd}{u_{\vec{p}}^{(r)\dagger}}
\newcommand{\vvprd}{v_{\vec{p}}^{(r)\dagger}}
\newcommand{\bvprd}{b_{\vec{p}}^{(r)\dagger}}
\newcommand{\cvprd}{c_{\vec{p}}^{(r)\dagger}}

\newcommand{\uvpp}{u_{\vec{p}\,'}}
\newcommand{\vvpp}{v_{\vec{p}\,'}}
\newcommand{\bvpp}{b_{\vec{p}\,'}}
\newcommand{\cvpp}{c_{\vec{p}\,'}}

\newcommand{\bvppd}{b_{\vec{p}\,'}^\dagger}
\newcommand{\cvppd}{c_{\vec{p}\,'}^\dagger}

\newcommand{\uvo}{u_{\vec{0}}}
\newcommand{\vvo}{v_{\vec{0}}}

\newcommand{\sta}{|0\rangle}
\newcommand{\stai}{\langle 0|}

\newcommand{\aonep}{A_1^{(+)}}
\newcommand{\aonem}{A_1^{(-)}}
\newcommand{\aone}{A_1}
\newcommand{\atwop}{A_2^{(+)}}
\newcommand{\atwom}{A_2^{(-)}}
\newcommand{\atwo}{A_2}

\newcommand{\eipx}{e^{ip\cdot x}}
\newcommand{\eipxm}{e^{-ip\cdot x}}

\newcommand{\eipxym}{e^{-ip\cdot (x-y)}}
\newcommand{\eipxy}{e^{ip\cdot (x-y)}}

\newcommand{\psib}{{\overline{\psi}}}
\newcommand{\psipb}{{\overline{\psi'}}}
\newcommand{\psid}{\psi^\dagger} 
\newcommand{\phih}{\phi_H}
\newcommand{\phii}{\phi_I}

\newcommand{\intp}{\frac{d^3 p}{(2\pi)^{3/2} \sqrt{2E_{\vec{p}}}}}
\newcommand{\intpp}{\frac{d^3 p'}{(2\pi )^{3/2} \sqrt{2E_{\vec{p}\,'}}}}
\newcommand{\intps}{\frac{d^3 p}{(2\pi)^3 2E_{\vec{p}}}}

\newcommand{\deltat}{\delta^{(3)}}

\newcommand{\dtp}{d^3 \vec{p} \;}
\newcommand{\tptep}{ \frac{1}{(2\pi)^{3/2}} \; \frac{1}{\sqrt{2E_{\vec{p}}}}}
\newcommand{\evp}{E_{\vec{p}}}

\newcommand{\ipme}{\frac{i}{p^2-m^2 + i \epsilon}}
\newcommand{\ipmpme}{\frac{i (\cancel{p}+m)}{p^2 -m^2 + i \epsilon}}

\newcommand{\notd}{\cancel{\partial}}
\newcommand{\notp}{\cancel{p}}
\newcommand{\notq}{\cancel{q}}

\newcommand{\ie}{i \epsilon}

\newcommand{\mh}{\mathcal{H}}
\newcommand{\ml}{\mathcal{L}} 

\newcommand{\bpm}{\begin{pmatrix}} 
\newcommand{\epm}{\end{pmatrix}} 

\newcommand{\thirteensuminta}{\quad\;\int\!\!\!\!\!\!\!\!\!\!\!\!\!\!\!\sum}
\newcommand{\thirteensumintb}{\;\,\,\int\!\!\!\!\!\!\!\!\!\sum}

\newcommand{\fourteensuminta}{\quad\;\int\!\!\!\!\!\!\!\!\!\!\!\!\!\!\!\sum}
\newcommand{\fourteensumintb}{\;\,\,\int\!\!\!\!\!\!\!\!\!\sum}

\newcommand{\sixteensumintaaa}{\quad\;\int\!\!\!\!\!\!\!\!\!\!\!\!\!\!\!\!\!\!\!\!\!\!\!\!\!\!\sum}
\newcommand{\sixteensumintaa}{\quad\;\int\!\!\!\!\!\!\!\!\!\!\!\!\!\!\!\!\!\sum}
\newcommand{\sixteensuminta}{\quad\;\int\!\!\!\!\!\!\!\!\!\!\!\!\!\!\!\sum}
\newcommand{\sixteensumintb}{\;\,\,\int\!\!\!\!\!\!\!\!\!\sum}

\newcommand{\eighteensumintaa}{\quad\;\int\!\!\!\!\!\!\!\!\!\!\!\!\!\!\!\!\!\sum}
\newcommand{\eighteensumintb}{\;\,\,\int\!\!\!\!\!\!\!\!\!\sum}
\newcommand{\eighteensumintc}{\;\,\int\!\!\!\!\!\!\!\sum}

\newcommand{\suminta}{\quad\;\int\!\!\!\!\!\!\!\!\!\!\!\!\!\!\!\sum}
\newcommand{\sumintb}{\;\,\,\int\!\!\!\!\!\!\!\!\!\sum}

\begin{document}{

\title{Notes from Sidney Coleman's Physics 253a}
\author{Sidney Coleman\thanks{Edited and typeset by Yuan-Sen Ting and Bryan Gin-ge Chen from scans of the handwritten notes of Brian Hill. Numerous additional edits by Richard Sohn. Please contact Yuan-Sen Ting (yuan-sen.ting@cfa.harvard.edu), Bryan Gin-ge Chen (bryangingechen@gmail.com) and Richard Sohn (sonar@mobiusweb.com). If you have any questions or comments.}}
\date{Harvard, Fall 1986\thanks{\LaTeX{} version of February 21, 2013}}
\maketitle

\sektion{0}{Preface}
It's unexpected and heart-warming to be asked by Bryan Chen and Yuan-Sen Ting to write something about these notes, 25 years after taking them. I was the teaching assistant for Sidney's quantum field theory course for three years. In the first year, I sat in, because frankly, I hadn't learned quantum field theory well enough the first time that I took it.

When I have the good fortune to hear a really good lecturer, I often re-copy my notes, preferably the evening on the day that I took them.

Once in a while, students would miss a class, and then ask me if they could look at my notes. At some point, the requests started happening enough that it was suggested that a copy be put on reserve in the Harvard physics library. From there, copies of the notes just kept spreading.

Sidney once expressed disappointment about the spread of the notes. For one thing, I even wrote down some of his anecdotes and jokes, and that made it less fun for him to re-tell them. For another, he wrote {\em Aspects of Symmetry} which shared a lot of material with what he taught in Physics 253b. He may have had in mind that he would write a field theory book as a companion volume.

Of course, he never did write a field theory book, or you'd be reading that, and he never tried to rein the copies in. Now that he is gone, we are lucky that his clarity lives on.

Thanks to Bryan Chen and Yuan-Sen Ting for creating this lovely \LaTeX{} version. Sometimes transcription can seem tedious, but I hope it was as valuable for them as the first re-copying was for me, and that for you -- fellow student of quantum field theory -- the existence of these notes is similarly valuable.

--Brian Hill, \href{http://www.lingerhere.org}{www.lingerhere.org}, March 10, 2011

\subsubsection*{Editors' notes}
The great field theorist Sidney Coleman for many years taught the course Physics 253 at Harvard on Quantum Field Theory.  The notes you are reading were typeset from a scanned version of handwritten lecture notes by Brian Hill from the Fall 1986 of the first half of the course: Physics 253a.  The Harvard Physics Department has made films of the lectures from the 1975-1976 version of the course available  on their website as well.

The typesetting for lectures 1-11 was done by Bryan Gin-ge Chen and for lectures 12-28 by Yuan-Sen Ting, who also recreated most of the figures. We have attempted to stay as faithful as possible to the scanned notes, aside from correcting some obvious errors in the notes and changing some of the in-text references.  

We thank Richard Sohn, Leonard Gamberg, Avraham Gal for pointing out typographical errors in the previous versions. The list of typographical errors can be found in Yuan-Sen Ting's homepage \href{https://www.cfa.harvard.edu/~yuanting/}{www.cfa.harvard.edu/$\sim$yuanting}. 

\newpage

\tableofcontents

\vspace{1in}

\noindent \textbf{Class:} Why not use Feynman's lecture notes?\\
\noindent \textbf{Gell-Mann:} Because Feynman uses a different method than we do.\\
\noindent \textbf{Class:} What is Feynman's method?\\
\noindent \textbf{Gell-Mann:} You write down the problem.  Then you look at it and you think. \\ Then you write down the answer.

 \sektion{1}{September 23}
\descriptionone
In NRQM, rotational invariance simplifies scattering problems. Why does the addition of relativity, the addition of L.I., complicate quantum mechanics?

The addition of relativity is necessary at energies $E\ge mc^2$. At these energies
\[p+p\rightarrow p+p+\pi^0\]

\noindent 
is possible. At slightly higher energies
\[p+p\rightarrow p+p+p+\overline{p}\]

\noindent 
can occur. The exact solution of a high energy scattering problem necessarily involves many particle processes.

You might think that for a given $E$, only a finite number, even a small number, of processes actually contribute, but you already know from NRQM that that isn't true.
\begin{align*} 
H&\rightarrow H+\delta V & \delta E_0&=\langle 0 |\delta V|0\rangle + \sum_n\frac{|\langle 0 |\delta V|n\rangle |^2}{E_0-E_n}+\cdots
\end{align*}

Intermediate states of all energies contribute, suppressed by energy denominators.

For calculations of high accuracy effects at low energy, relativistic effects of order $(v/c)^2$ can be included. Intermediate states with extra particles will contribute corrections of order 
$\frac{E}{mc^2}\begin{smallmatrix} & \text{\tiny Typical} \\ & \text{\tiny energies} \\ \leftarrow & \text{\tiny in problem} \\ \\ \leftarrow & \text{\tiny Typical} \\ & \text{\tiny energy} \\ & \text{\tiny denominator} \end{smallmatrix} \sim \frac{mv^2}{mc^2} = \left(\frac{v} {c}\right)^2$
. As a general conclusion: the corrections of relativistic kinematics and the corrections from multiparticle intermediate states are comparable; the addition of relativity forces you to consider many-body problems. We can't even solve the zero-body problem. (It is a phenomenal fluke that relativistic kinematic corrections for the Hydrogen atom work. If the Dirac equation is used, without considering multi-particle intermediate states, corrections of $\mathcal{O}\left(\frac{v}{c}\right)$ can be obtained. This is a \underline{fluke} caused by some unusually low electrodynamic matrix elements.)

We will see that you cannot have a consistent relativistic picture without pair production.

\subsection*{\sc{Units}}
\begin{align*}
& \hbar=c=1 & [m]=[E]=[T^{-1}]=[L^{-1}]\\
& \begin{matrix}
\text{Because we're doing}\\
\text{relativistic $(c)$}\\
\text{quantum mechanics } (\hbar)
\end{matrix} & \begin{matrix}
\text{Sometimes } 1=-1=2\pi\\
\text{and } \frac{1}{2\pi}=\text{\sout{1}=``one-bar''}\\
\\
\text{(1 fermi)}^{-1}\approx 197\text{ MeV}
\end{matrix}
\end{align*}

We say things like the inverse Compton wavelength of the proton is ``1 GeV".

\subsection*{\sc{Lorentz Invariance}}
Every Lorentz transformation is the product of an element of the connected Lorentz group, $SO(3,1)$, and 1, $P$ (reflects all three space components), $T$ (time reversal), or $PT$. By Lorentz invariance we mean $SO(3,1)$.

Metric convention: $+ - - -$.

\section*{\sc{Theory of a single free spinless particle of mass $\mu$}}
The components of momentum form a complete set of commuting variables.
\begin{align*}
\text{Momentum operator}\rightarrow & \underset{ \substack{\text{\tiny State of a spinless} \\ \text{\tiny particle is completely} \\ \text{\tiny specified by its momentum.} }}{ \vec{P}|\underbrace{\vec{k}} \rangle} = \vec{k}|\vec{k}\rangle \\
\text{Normalization } &\langle\vec{k}|\vec{k}\,'\rangle=\delta^{(3)} (\vec{k}-\vec{k}\,')
\end{align*}

The statement that this is a complete set of states, that there are no others, is
\begin{align*}
1 &= \int d^3k|\vec{k}\rangle\langle\vec{k}| \\
|\psi\rangle &= \int d^3k\psi(\vec{k})|\vec{k}\rangle & \psi(\vec{k})\equiv \langle \vec{k}|\psi\rangle 
\end{align*}

(If we were doing NRQM, we'd finish describing the theory by giving the Hamiltonian, and thus the time evolution: $H|\vec{k}\rangle =
\frac{|\vec{k}|^2}{2\mu}|\vec{k}\rangle$.)

We take $H|\vec{k}\rangle= \sqrt{|\vec{k}|^2+\mu^2}|\vec{k}\rangle \equiv \omega_{\vec{k}}|\vec{k}\rangle$

That's it, the theory of a single free spinless particle, made relativistic.

How do we know this theory is L.I.? Just because it contains one relativistic formula, it is not necessarily relativistic. The theory is not manifestly L.I..

The theory is manifestly rotationally and translationally invariant. Let's be more precise about this.

\subsection*{\underline{Translational Invariance}}
Given a four-vector, $a$, specifying a translation (active), there should be a linear operator, $U(a)$, satisfying:
\begin{align}\label{eq:01-trans1}
U(a)U(a)^\dagger&=1, \text{ to preserve probability amplitudes}\\
U(0)&= 1 \\ \label{eq:01-trans3}
U(a)U(b)&=U(a+b)
\end{align}

The $U$ satisfying these is $U(a)=e^{iP\cdot a}$ where $P=(H,\vec{P})$.

(This lecture is in pedagogical, not logical order. The logical order would be to state:
\begin{enumerate} 
\item That we want to set up a translationally invariant theory of a spinless particle. The theory would contain unitary translation operators $U(\vec{a})$.
\item Define $P^i=-i\frac{\partial U}{\partial a^i}\big|_{\vec{a}=0}$, (by \eqref{eq:01-trans3} $[P_i,P_j]=0$, by \eqref{eq:01-trans1} $\vec{P}=\vec{P}^\dagger$).
\item Declare $P^i$ to be a complete set and classify the states by momentum.
\item Define $H=\sqrt{\vec{P}^2+\mu^2}$, thus giving the time evolution.)
\end{enumerate}

\subsection*{More translational invariance}
\begin{align*}
U(a)|0\rangle &=|a\rangle & U(a)=e^{iP\cdot a}
\end{align*}

\noindent 
where $|0\rangle$ here means state centered at zero and $|a\rangle$ means state centered at $a$.

\begin{align*}
O(x+a)&=U(a)O(x)U(a)^\dagger\\
\langle a|O(x+a)|a\rangle&=\langle 0|O(x)|0\rangle
\end{align*}

\subsection*{Non-relativistic reduction}
\begin{align*}
U(\vec{a})&=e^{-i\vec{P}\cdot\vec{a}}\\
e^{-i\vec{P}\cdot\vec{a}}|\vec{q}\rangle &= |\vec{q}+\vec{a}\rangle
\end{align*}

\subsection*{Something hard to digest, but correct:}
\begin{align*}
\widehat{\vec{q}}e^{-i\vec{P}\cdot\vec{a}}|\vec{q}\rangle &=(\vec{q}+\vec{a})|\vec{q}+\vec{a}\rangle\\
e^{i\vec{P}\cdot\vec{a}}\widehat{\vec{q}}e^{-i\vec{P}\cdot\vec{a}}|\vec{q}\rangle &= (\vec{q}+\vec{a})|\vec{q}\rangle\\
\Rightarrow e^{i\vec{P}\cdot\vec{a}}\widehat{\vec{q}}e^{-i\vec{P}\cdot\vec{a}}&= \widehat{\vec{q}}+\vec{a}\\
\text{Looks like the opposite } &\text{of}\\
e^{-i\vec{P}\cdot\vec{a}}O(\vec{x})e^{i\vec{P}\cdot\vec{a}}&=O(\vec{x}+\vec{a})
\end{align*}

The $\widehat{\vec{q}}$ operator is not an operator localized at $\vec{q}$. No reason for these (last two equations) to look alike.

\subsection*{\underline{Rotational Invariance}}
Given an $R\in SO(3)$, there should be a $U(R)$ satisfying
\begin{align}
\label{eq:01-rot1} U(R)U(R)^\dagger&=1\\
\label{eq:01-rot2} U(1)&=1\\
\label{eq:01-rot3} U(R_1)U(R_2)&=U(R_1R_2)
\end{align}

Furthermore denote $|\psi'\rangle=U(R)|\psi\rangle$
\[ \langle \psi'|\vec{P}|\psi'\rangle = R\langle\psi|\vec{P}|\psi\rangle \]
\begin{align}\label{eq:01-rot4}
\text{ for any $|\psi\rangle$ i.e.~} U(R)^\dagger\vec{P}U(R)&=R\vec{P} \\ \label{eq:01-rot5}
\text{ and } U(R)^\dagger HU(R) &=H
\end{align}

A $U(R)$ satisfying all these properties is given by \[U(R)|\vec{k}\rangle =|R\vec{k}\rangle \]

That \eqref{eq:01-rot2} and \eqref{eq:01-rot3} are satisfied is trivial.

Proof that \eqref{eq:01-rot1} is satisfied
\begin{align*}
U(R)U(R)^\dagger &= U(R)\int d^3k|\vec{k}\rangle\langle\vec{k}|U(R)^\dagger \\
&=\int d^3k|R\vec{k}\rangle\langle R\vec{k}| & k'=Rk, d^3k'=d^3k\\
&= \int d^3k'|\vec{k'}\rangle\langle \vec{k'}| =1
\end{align*}

Proof that \eqref{eq:01-rot4} is satisfied
\begin{align*}
U(R)^\dagger\vec{P}U(R) &\underset{\text{by \eqref{eq:01-rot1}}}{=} U(R)^{-1}\vec{P}(U(R)^{-1})^\dagger & \downarrow\text{by \eqref{eq:01-rot2} and \eqref{eq:01-rot3}}\\
&=U(R^{-1})\vec{P}U(R^{-1})^\dagger \\
&=U(R^{-1})\vec{P}\int d^3k|\vec{k}\rangle\langle\vec{k}| U(R^{-1})^\dagger \\
&=U(R^{-1})\int d^3k\vec{k}|\vec{k}\rangle\langle\vec{k}| U(R^{-1})^\dagger \\
&=\int d^3k\vec{k}|R^{-1}\vec{k}\rangle\langle R^{-1}\vec{k}| & k=Rk', d^3k=d^3k'\\
&=\int d^3k'R\vec{k}\,'|\vec{k}\,'\rangle\langle \vec{k}\,'|\\
&= RP
\end{align*}

You supply proof of \eqref{eq:01-rot5}.

This is the template for studying L.I.

Suppose a silly physicist took 
\begin{align*}
|\vec{k}\rangle_s &= \sqrt{1+k_z^2}|\vec{k}\rangle \\
_s\langle\vec{k}|\vec{k}\,'\rangle_s &= (1+k_z^2)\delta^{(3)}(\vec{k}-\vec{k}\,')\\
1 &= \int d^3k\frac{1}{1+k_z^2}|\vec{k}\rangle_s\ _s\langle\vec{k}|
\end{align*}

If he took $U_s(R)|\vec{k}\rangle_s = |R\vec{r}\rangle_s$ his proofs of \eqref{eq:01-rot1}, \eqref{eq:01-rot4}, \eqref{eq:01-rot5} would break down because
\begin{align*}
\frac{d^3k}{1+k_z^2}&\neq \frac{d^3k'}{1+k_z'^2} & \text{i.e.~$\frac{d^3k}{1+k_z^2}$ is not a rotationally invariant measure!}
\end{align*}

\subsection*{\underline{Lorentz Invariance}}
$\langle \vec{k}|\vec{k}\,'\rangle=\delta^{(3)}(\vec{k}-\vec{k}\,')$ is a silly normalization for Lorentz invariance. 

$d^3k$ is not a Lorentz invariant measure.

We want a Lorentz invariant measure on the hyperboloid $k^2=\mu^2, k^0>0$.

$d^4k$ is a Lorentz invariant measure.

Restrict it to the hyperboloid by multiplying it by a Lorentz invariant $d^4k\delta(k^2-\mu^2)\Theta(k^0)$.

This yields the measure on the hyperboloid\footnote{Think of the $\delta$ function as a function of $k^0$ and use the general formula
$\delta(f(k^0))=\sum_{\substack{\text{zeroes}\\ \text{of $f$, $K_i$} }}\frac{\delta(k^0-K_i)}{|f'(K_i)|}$.}
$\frac{d^3k}{2\omega_{\vec{k}}}$, $\omega_{\vec{k}}=\sqrt{\vec{k}^2 +\mu^2}$, $k=(\omega_{\vec{k}},\vec{k})$.

So we take $\underset{\substack{\text{relativistically }\\ \text{normalized}}}{|k\rangle} =\underset{\substack{\text{So factors of $2\pi$} \\ \text{ come out right in the} \\ \text{ Feynman rules a few}\\\text{ months from now}}}{\underbrace{\sqrt{(2\pi)^3}} \sqrt{2\omega_{\vec{k}}}|\vec{k} \rangle}$.
\begin{center}
\includegraphics[width=8cm]{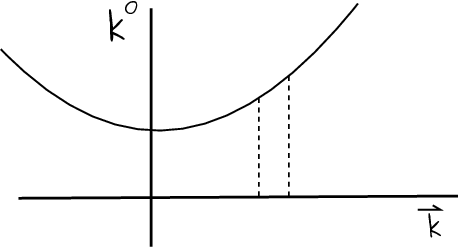}
\end{center}

(Using $d^3x\delta(\vec{x}^2-R^2)$ can get $\frac{R}{2}\sin\theta d\theta d\phi$)

Looks like factor multiplying $d^3k$ ought to get larger as $|\vec{k}|$ gets large. This is an illusion, caused by graphing on Euclidean paper. This is the same illusion as in the twin paradox. The moving twin's path looks longer, but in fact, its proper time is shorter.

Now the demonstration of Lorentz invariance: Given any Lorentz transformation $\Lambda$ define
\[U(\Lambda)|k\rangle = |\Lambda k\rangle\]

$U(\Lambda)$ satisfies
\begin{align}\label{eq:01-lor1}
U(\Lambda)U(\Lambda)^\dagger &=1\\ \label{eq:01-lor2}
U(1)&=1\\ \label{eq:01-lor3}
U(\Lambda_1)U(\Lambda_2)&=U(\Lambda_1\Lambda_2)\\ \label{eq:01-lor4}
U^\dagger(\Lambda)PU(\Lambda)&=\Lambda P
\end{align}

The proofs of these are exactly like the proofs of rotational invariance, using 
\begin{align*}
1&=\int \frac{d^3k}{(2\pi)^32\omega_{\vec{k}}}|k\rangle\langle k| & \frac{d^3k}{(2\pi)^32\omega_{\vec{k}}}&=\frac{d^3k'}{(2\pi)^3 2\omega_{\vec{k}\,'}}
\end{align*}

We have a fairly complete theory except we still don't know where anything is.

We need a position operator, satisfying
\begin{align}\label{eq:01-pos1}
\vec{X}&=\vec{X}^\dagger\\ \label{eq:01-pos2}
R\vec{X}&=U(R)^\dagger\vec{X}U(R)\\ \label{eq:01-pos3}
e^{i\vec{P}\cdot\vec{a}}\vec{X}e^{-i\vec{P}\cdot\vec{a}} &= \vec{X}+\vec{a}
\end{align}

Take $\frac{\partial}{\partial a_i}$ of \eqref{eq:01-pos3} to get $i[P_i,X_j]=\delta_{ij}$.

\subsection*{Determination of $\vec{X}$ in position space}
\begin{align*}
\psi(\vec{k})&\equiv\langle\vec{k}|\psi\rangle \qquad\qquad \langle\vec{k}|\vec{P}|\psi\rangle =\vec{k}\psi(\vec{k}) \\
\langle \vec{k}|\vec{X}|\psi\rangle &=\underset{\substack{\text{to satisfy}\\ \text{inhomogeneous part of} \\ \text{commutation relation}}}{i\underbrace{\frac{\partial}{\partial\vec{k}}\psi}(\vec{k})}+\underset{\substack{\text{an arbitrary vector} \\ \text{commuting with the $\vec{P}$'s,} \\ \text{a complete set, can be}\\ \text{written this way}}}{ \underbrace{\vec{k}F(|\vec{k}|^2)}\psi(\vec{k})}
\end{align*}

The extra arbitrary vector can be eliminated by redefining the phases of the states. In the momentum state basis let
\begin{align*}
|\vec{k}\rangle\rightarrow|\vec{k}\rangle_N&=e^{iG(|\vec{k}|^2)}|\vec{k}\rangle & \begin{matrix}\text{(This is a unitary transformation, call it
$U$.}\\ \text{In effect, $U^\dagger\vec{X}U$ is our new position operator.)}\end{matrix} \\ \text{Here }\vec{\nabla}G(|\vec{k}|^2)&=\vec{k}F(|\vec{k}|^2)
\end{align*}

The only formula this affects in all that we have done so far is the expression for $\langle\vec{k}|\vec{X}|\psi\rangle$. With the redefined states, it is
\begin{align*}
_N\langle\vec{k}|\vec{X}|\psi\rangle_n &=\underset{e^{-iG(|\vec{k}|^2)}\delta^{(3)}(\vec{k}-\vec{k}\,')}{\int d^3k' \underbrace{_N \langle \vec{k}|\vec{k}\,'\rangle}}\ \ \underset{\left[i\frac{\partial}{\partial\vec{k}\,'}+\vec{k}\,'F(|\vec{k}\,'|^2)\right]e^{iG(|\vec{k}\,'|^2)} \psi(\vec{k}\,')}{\underbrace{\langle\vec{k'}|\vec{X}|\psi\rangle_N}\ \ \ \ \ \ \ \ \ \ \ \ \ \ \ \ \ \ \ \ \ \ \ \ } \\
&=\int d^3k'e^{-iG(|\vec{k}|^2)}\delta^{(3)}(\vec{k}-\vec{k}\,')\left\{\left[ \cancel{-\frac{\partial}{\partial\vec{k}\,'}G(|\vec{k}\,'|^2)} +\cancel{\vec{k}\,' F(|\vec{k}\,'|^2)}\right]e^{iG}\psi+e^{iG} \frac{\partial} {\partial\vec{k}\,'}\psi\right\}\\
&=i\frac{\partial}{\partial\vec{k}}\psi(\vec{k})
\end{align*}

Up to an unimportant phase definition, we have shown that the obvious definition for $\vec{X}$ is the unique definition, and we have done it without using L.I.~or the explicit form of $H$.
}{
 \sektion{2}{September 25}
\descriptiontwo
It is possible to measure a particle's position in our theory, $\vec{x}$ is an observable, and this leads to a conflict with causality.

Introduce position eigenstates 
\[\langle\vec{k}|\vec{x}\rangle = \frac{1}{(2\pi)^{3/2}}e^{-i\vec{k}\cdot\vec{x}}\]

We will evaluate $\langle\vec{x}|e^{-iHt}|\vec{x}=0\rangle$ and see that our particle has a nonzero amplitude to be found outside its forward light cone
\footnote{By translational invariance and superposition we could easily get the evolution of any initial configuration from this calculation.}.
\begin{align*} 
\langle\vec{x}|e^{-iHt}|\vec{x}=0\rangle &= \int d^3k\langle \vec{x}|\vec{k}\rangle \langle\vec{k}|e^{-iHt}|\vec{x}=0\rangle & H|\vec{k}\rangle&=\omega_{\vec{k}}|\vec{k}\rangle \\
&= \int d^3k\frac{1}{(2\pi)^3}e^{i\vec{k}\cdot\vec{x}} e^{-i\omega_{\vec{k}}t} & \omega_{\vec{k}} &=\sqrt{\vec{k}^2+\mu^2}\\
&= \int_0^\infty \frac{k^2dk}{(2\pi)^3} \underbrace{\int_0^\pi \sin\theta d\theta}_{\int_{-1}^1d(\cos\theta)} \underbrace{\int_0^{2\pi} d\phi}_{2\pi} e^{ikr\cos\theta} e^{-i\omega_{k}t}& r&=|\vec{x}|, k=|\vec{k}|\\
&=\frac{1}{(2\pi)^2}\frac{1}{ir}\int_0^\infty k dk(e^{ikr}-e^{-ikr}) e^{-i\omega_{k}t}& \omega_{k}
&=\sqrt{k^2 +\mu^2} \\
&=\frac{-i}{(2\pi)^2r}\int_{-\infty}^\infty k dk e^{ikr} e^{-i\omega_{k}t}
\end{align*}

These steps can be applied to the F.T.~of any function of $|\vec{k}|$.
\begin{center}
\includegraphics[width=8 cm]{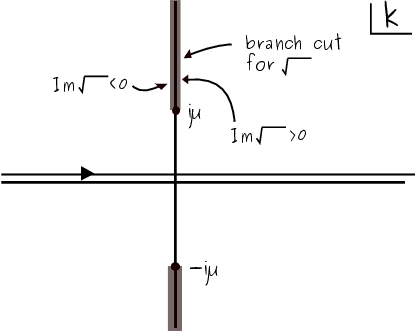}
\end{center}

$e^{-i\sqrt{k^2+\mu^2}t}$ is a growing exponential as you go up the right side of the upper branch cut, and a decreasing exponential on the left side. Given $r>0$ and $r>t$ the product $e^{ikr}e^{-i\omega_{k}t}$ decreases exponentially as you go up the branch cut.

Given $r>0$ and $r>t$ deform the contour to:
\begin{center}
\includegraphics[width=7 cm]{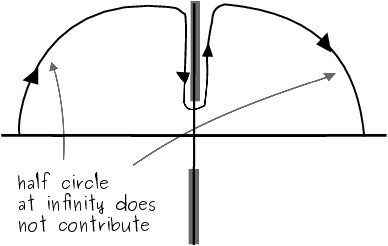}
\end{center}

The integral becomes
\begin{align*}
\frac{-i}{(2\pi)^2r}\int_\mu^\infty \overbrace{(-z)}^{k=iz}d(iz)e^{-zr}[e^{\sqrt{z^2-\mu^2}t}- e^{-\sqrt{z^2-\mu^2}t}]
&= -\frac{e^{-\mu r}}{2\pi^2r}\int_\mu^\infty z dz e^{-(z-\mu)r} \sinh(\sqrt{z^2-\mu^2}t)
\end{align*}

The integrand is positive definite, the integral is nonzero.

We can measure a particle's position in this theory. We can trap it in a box of arbitrarily small size, and we can release it and detect it outside of its forward light cone. The particle can travel faster than light and thus it can move backwards in time, with all the associated paradoxes.

Admittedly, the chance that the particle is found outside the forward light cone falls off exponentially as you get further from the light cone, and that makes it extremely unlikely that I could go back and convince my mother to have an abortion, but if it is at all possible, it is still an unacceptable contradiction.

In practice, how does this affect atomic physics? Not at all, because we never tried to localize particles to spaces of order their Compton wavelength when doing atomic physics. We say that the electron is in the TV picture tube and there is not much chance that it is actually out in the room with you.

In principle, the ability to localize a single particle is a disaster, how does nature get out of it?
\begin{center}
\includegraphics[width=2.5 cm]{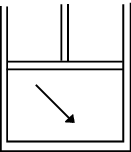}\\
Particle trapped in container with reflecting walls
\end{center}

If the particle is localized to a space with dimensions on the order of $L$ the uncertainty in the particle's momentum is $\sim\frac{1}{L}$. In the relativistic regime this tells us that the uncertainty in the particle's energy is $\sim\frac{1}{L}$. As $L$ gets less than $\frac{1}{\mu}$ states with more than one particle are energetically accessible. If the box contained a photon and the walls were mirrors the photon would pick up energy as it reflected off the descending mirror, it could turn into two photons as it reflected.
\begin{center}
\includegraphics[width=3 cm]{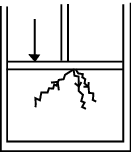}
\end{center}

If we try to localize a particle in a box with dimensions smaller or on the order of a Compton wavelength it is unknown whether what we have in the box is 1 particle, 3 particles, 27 particles or 0 particles.

Relativistic causality is inconsistent with a single particle theory. The real world evades the conflict through pair production. This strongly suggests that the next thing we should do is develop a multi-particle theory.

\section*{Any number of one type of free spinless mesons}
\footnotesize
The space we construct is called Fock space. This formalism is also used in thermodynamics with the grand canonical ensemble. Particle number instead of being fixed, fluctuates around a value determined by the chemical potential.

\normalsize
Basis for single particle states $|\vec{k}\rangle$
\begin{align*}
\langle \vec{k}\,'|\vec{k}\rangle &=\delta^{(3)}(\vec{k}-\vec{k}\,') & H|\vec{k}\rangle &= \omega_{\vec{k}}|\vec{k}\rangle & \vec{P}|\vec{k}\rangle &=\vec{k}|\vec{k}\rangle 
\end{align*}

This is the same as last time, except now this is just part of the basis.

Two particle states $|\vec{k}_1,\vec{k}_2\rangle\!\!\!\!\!\!\!\!\!\!\!\!\underbrace{=}_{\substack{\text{indistinguishability,}\\ \text{ Bose statistics}}}\!\!\!\!\!\!\!\!\!\!\!\!|\vec{k}_2,\vec{k}_1\rangle$
\begin{align*}
\langle \vec{k}_1,\vec{k}_2|\vec{k}_1',\vec{k}_2'\rangle &= \delta^{(3)}(\vec{k}_1-\vec{k}_1')\delta^{(3)}(\vec{k}_2-\vec{k}_2') +\delta^{(3)}(\vec{k}_1-\vec{k}_2')\delta^{(3)}(\vec{k}_2-\vec{k}_1') \\
H|\vec{k}_1,\vec{k}_2\rangle &= (\omega_{\vec{k}_1} +\omega_{\vec{k}_2})|\vec{k}_1,\vec{k}_2\rangle \\
\vec{P}|\vec{k}_1,\vec{k}_2\rangle &=(\vec{k}_1+\vec{k}_2)|\vec{k}_1,\vec{k}_2\rangle \\
&\text{etc.}
\end{align*}

Also need a no particle state $|0\rangle$
\begin{align*}
\langle 0\underbrace{|0\rangle} &\underbrace{=1}& H|0\rangle &= 0 & \vec{P}|0\rangle &=0\\
\text{\scriptsize{not part }} & \text{\scriptsize{of a continuum}}
\end{align*}

The vacuum is unique, it must satisfy $U(\Lambda)|0\rangle=|0\rangle$. All observers agree that the state with no particles is the state with no particles.

Completeness relation
\begin{align*}
1=|0\rangle\langle0|+\int d^3k|\vec{k}\rangle\langle\vec{k}|+&\underbrace{\frac{1}{2!}}\int d^3k_1 d^3k_2|k_1k_2\rangle\langle k_1,k_2|\\
&\substack{\text{to avoid double counting. Alternatively,}\\
\text{just check that this works on }|\vec{k},\vec{k}\,'\rangle }
\end{align*}

Now we could proceed by setting up equations for wave functions. To specify a state, a wave function contains a number, a function of three variables, a function of six variables, etc. Interactions involving a change in particle number will connect a function of six variables to a function of nine variables. This would be a mess.

We need a better description. As a pedagogical device, we will work in a periodic cubical box of side $L$ for a while. Since a translation by $L$ does nothing, the momenta must be restricted to allowed values
\begin{align*}
\vec{k}&=\left(\frac{2\pi n_x}{L},\frac{2\pi n_y}{L},\frac{2\pi n_z}{L}\right) \text{ satisfying } \begin{matrix}
\vec{k}\cdot(0,0,L)=2n_z\pi\\
\vec{k}\cdot(0,L,0)=2n_y\pi\\
\vec{k}\cdot(L,0,0)=2n_x\pi
\end{matrix}
\end{align*}

Dirac deltas become Kronecker deltas and integrals become sums
\begin{align*}
\delta^{(3)}(\vec{k}-\vec{k}\,')&\rightarrow\delta_{\vec{k}\vec{k}\,'} & \int d^3k &\rightarrow \sum_{\vec{k}} \\
\langle \vec{k}|\vec{k}\,'\rangle &= \delta_{\vec{k}\vec{k}\,'} & \langle \vec{k}_1,\vec{k}_2|\vec{k}_1',\vec{k}_2'\rangle &= \delta_{\vec{k}_1\vec{k}_1'}\delta_{\vec{k}_2\vec{k}_2'} +\delta_{\vec{k}_1\vec{k}_2'}\delta_{\vec{k}_2\vec{k}_1'}
\end{align*}

\subsection*{Occupation number representation }
Each basis state corresponds to a single function 
\begin{align*}
|\vec{k}_1\rangle&\leftrightarrow n(\vec{k})=\delta_{\vec{k}\vec{k}_1}\\
|\vec{k}_1,\vec{k}_2\rangle&\leftrightarrow n(\vec{k})=\delta_{\vec{k}\vec{k}_1}+\delta_{\vec{k}\vec{k}_2} \\
|0\rangle&\leftrightarrow n(\vec{k})=0 
\end{align*}

Given a function $n(\vec{k})$ in the occupation number description, we write the state 
\begin{align*}
& |\underbrace{n(\cdot)}\rangle & \text{Inner product } \langle n(\cdot)|n'(\cdot)\rangle = \prod_{\vec{k}}n(\vec{k})!
\delta_{n(\vec{k})n'(\vec{k})}\\
& \substack{\text{\scriptsize{no argument, to emphasize}} \\
\text{\scriptsize{that the state depends on the }} \\
\text{\scriptsize{whole function $n$, not just }} \\
\text{\scriptsize{its value for one specific }} \vec{k}. \\ }
\end{align*}

Define an occupation number operator 
\begin{align*}
N(\vec{k})|n(\cdot)\rangle &= n(\vec{k})|n(\cdot)\rangle & H &= \sum_{\vec{k}} \omega_{\vec{k}}N(\vec{k}) & \vec{P}=\sum_{\vec{k}}\vec{k} N(\vec{k})
\end{align*}

This is a better formalism, but it could still use improvement. It would be nice to have an operator formalism that did not have any wave functions at all.

Note that $H$ for our system has the form it would have if the system we were dealing with was actually a bunch of harmonic oscillators. The two systems are completely different. In ours the $N(\vec{k})$ tells how many particles are present with a given momentum. In a system of oscillators, $N(\vec{k})$ gives the excitation level of the oscillator labelled by $\vec{k}$.

\section*{Review of the simple harmonic oscillator}
No physics course is complete without a lecture on the simple harmonic oscillator. We will review the oscillator using the operator formalism. We will then exploit the formal similarity to Fock space to get an operator formulation of our multi-particle theory.
\begin{align*}
H&=\frac{1}{2}\omega[p^2+q^2-1] & [p,q]=-i
\end{align*}

If $[p,A]=[q,A]=0$ then $A= \!\!\!\!\!\underbrace{\lambda}_{\scriptsize \mbox{c-number}} \!\!\!\!\!\!\!\!\! \overbrace{I}^{\scriptsize \mbox{Identity}}$.

This is all we need to get the spectrum.

Define raising and lowering operators
\begin{align*}
a&\equiv\frac{q+ip}{\sqrt{2}} & a^\dagger &=\frac{q-ip}{\sqrt{2}} & H&=\omega a^\dagger a \\
[H,a^\dagger]&=\omega a^\dagger & [H,a]&=-\omega a & [a,a^\dagger]&=1
\end{align*}
\begin{align*}
Ha^\dagger|E\rangle &=a^\dagger H|E\rangle + \omega a^\dagger|E\rangle \\
&= (E+\omega)a^\dagger |E\rangle &\text{\scriptsize{$a^\dagger$ is the raising operator}}\\
Ha|E\rangle &= (E-\omega)a |E\rangle 
\end{align*}
\begin{center}
\includegraphics[width=8 cm]{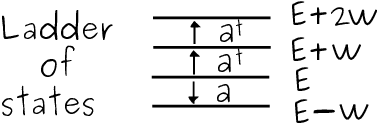}
\end{center}

Because, $\langle \psi|H|\psi\rangle = \omega\langle\psi|a^\dagger a|\psi\rangle =\omega||a|\psi\rangle||^2\geq 0$ the ladder of states must stop going down or else $E$ becomes negative. The only way this can happen is if $a|E_0\rangle=0$. Then $H|E_0\rangle=0$.

The lowest state of the ladder, having $E_0=0$ is denoted $|0\rangle$. The higher states are made by
\begin{align*}
(a^\dagger)^n|0\rangle&\propto |n\rangle & H|n\rangle&=n\omega|n\rangle \\
\text{Get normalizations right }a^\dagger|n\rangle &= c_n|n+1\rangle \\
|c_n|^2&=\langle n|aa^\dagger|n\rangle =n+1\Rightarrow &c_n&=\sqrt{n+1}\\
a|n\rangle&=d_n|n-1\rangle & |d_n|^2&=n & d_n=\sqrt{n}
\end{align*}

Now we use $[p,A]=[q,A]=0\Rightarrow A=\lambda I$ to show that this ladder built from a state with $E_0=0$ is in fact the whole space. We do this by considering the projector, $\mathcal{P}$, onto the states in the ladder. Since $a$ and $a^\dagger$ keep you within the ladder
\begin{align*}
[a,\mathcal{P}]=[a^\dagger,\mathcal{P}]=0&\Rightarrow[p,\mathcal{P}]= [q,\mathcal{P}]=0 \\
&\Rightarrow \mathcal{P}=\lambda I
\end{align*}

The projector onto the ladder is proportional to the identity. There is nothing besides the states we have found.

Now we will apply this to Fock space.

Define creation and annihilation operators for each momentum, $a_{\vec{k}}$, $a_{\vec{k}\,'}^\dagger$, satisfying
\begin{align*}
[a_{\vec{k}}, a_{\vec{k}\,'}^\dagger]&=\delta_{\vec{k}\vec{k}\,'} & [a_{\vec{k}},a_{\vec{k}\,'}]&=0 & [a_{\vec{k}}^\dagger,a_{\vec{k}\,'}^\dagger]&=0
\end{align*}

Hilbert space built by acting on $|0\rangle$ with strings of creation operators.
\begin{align*}
|\vec{k}\rangle&= a_{\vec{k}}^\dagger|0\rangle & a_{\vec{k}}|0\rangle &=0 \\
a_{\vec{k}_1}^\dagger a_{\vec{k}_2}^\dagger a_{\vec{k}_3}^\dagger |0\rangle &= |\vec{k}_1,\vec{k}_2,\vec{k}_3\rangle \\
H&=\sum_{\vec{k}}\omega_{\vec{k}}a_{\vec{k}}^\dagger a_{\vec{k}} & \vec{P}&=\sum_{\vec{k}}\vec{k}a_{\vec{k}}^\dagger a_{\vec{k}} 
\end{align*}
\begin{align*}
\text{If } &\forall \vec{k}\;\; 0=[a_{\vec{k}},A]= [a_{\vec{k}}^\dagger,A] \Longrightarrow\underbrace{ A=\lambda I}_{\substack{\text{This tells us there}\\ \text{ are no other degrees}\\ \text{ of freedom.}}}
\end{align*}

We've laid out a compact formalism for Fock space. Let's drop the box normalization and see if it is working.
\begin{align*}
[a_{\vec{k}},a_{\vec{k}\,'}^\dagger]&=\delta^{(3)}(\vec{k}-\vec{k}\,') & [a_{\vec{k}},a_{\vec{k}\,'}]&=0=[a_{\vec{k}}^\dagger, a_{\vec{k}\,'}^\dagger]\\
H&= \int d^3k\omega_{\vec{k}}a_{\vec{k}}^\dagger a_{\vec{k}} & \vec{P}&= \int d^3k \vec{k}a_{\vec{k}}^\dagger a_{\vec{k}}
\end{align*}

Check energy and normalization of one particle states.
\begin{align*}
\langle \vec{k}\,'|\vec{k}\rangle &=\langle 0|a_{\vec{k}\,'}a_{\vec{k}}^\dagger |0\rangle =\langle 0|[a_{\vec{k}\,'},a_{\vec{k}}^\dagger]|0\rangle \\
&=\delta^{(3)}(\vec{k}-\vec{k}\,')\langle0|0\rangle =\delta^{(3)}(\vec{k}-\vec{k}\,')\\
[H,a_{\vec{k}}^\dagger] &=\int d^3k'\omega_{\vec{k'}} [a_{\vec{k}\,'}^\dagger a_{\vec{k}\,'}, a^\dagger_{\vec{k}}]=\omega_{\vec{k}}a^\dagger_{\vec{k}} \Rightarrow \\
H|\vec{k}\rangle &= H a_{\vec{k}}^\dagger |0\rangle =[H,a_{\vec{k}}^\dagger]|0\rangle = \omega_{\vec{k}}|\vec{k}\rangle \\
[\vec{P},a_{\vec{k}}^\dagger] &=\vec{k}a_{\vec{k}}^\dagger \Rightarrow \vec{P}|\vec{k}\rangle=\vec{k}|\vec{k}\rangle
\end{align*}

Check normalization of two-particle states $|\vec{k}_1,\vec{k}_2\rangle = a_{\vec{k}_1}^\dagger a_{\vec{k}_2}^\dagger |0\rangle$. Using commutation relations check
\begin{align*}
\langle\vec{k}_1',\vec{k}_2'|\vec{k}_1,\vec{k}_2\rangle &=\delta^{(3)}(\vec{k}_1-\vec{k}_1')\delta^{(3)}(\vec{k}_2-\vec{k}_2') +\delta^{(3)}(\vec{k}_1-\vec{k}_2')\delta^{(3)}(\vec{k}_2-\vec{k}_1')
\end{align*}

\section*{Mathematical Footnote:}
We've been calling the $a_{\vec{k}}$ and $a_{\vec{k}}^\dagger$ operators. An operator takes any normalizable vector in Hilbert space to another normalizable vector: $A|\psi\rangle$ is normalizable whenever $|\psi\rangle$ is normalizable. Even $x$ in 1-d QM is not an operator. $\int dx|f(x)|^2<\infty \nRightarrow \int dx|f(x)|^2x<\infty$. $x$ is an unbounded operator. An unbounded operator has $A|\psi\rangle$ normalizable for a dense set\footnote{Any$|\psi\rangle$ is the limit of a sequence in the dense set.} of $|\psi\rangle$. $a_{\vec{k}}$ and $a_{\vec{k}}^\dagger$ are even more awful than unbounded operators. An extra meson in a plane wave state added to any state is enough to make it nonnormalizable. $a_{\vec{k}}$ is an operator valued distribution. You only get something a mathematician would be happy with after integration. $\int d^3k f(\vec{k})a_{\vec{k}}$ is acceptable with a ``sufficiently smooth'' function.
}{
 \sektion{3}{September 30}
\descriptionthree
In ordinary QM, \underline{any} hermitian operator is observable. This can't be true in relativistic quantum mechanics. Imagine two experiments that are at space-like separation. If $x_1\in R_1$ and $x_2\in R_2$ then $(x_1-x_2)^2<0$.
\begin{center}
\includegraphics[width=.8\textwidth]{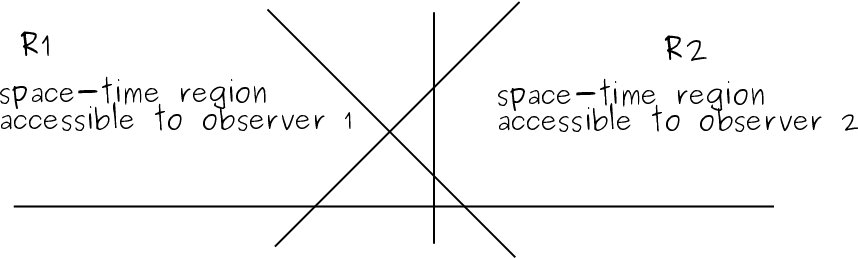}
\end{center}

Suppose observer 2 has an electron in his lab and she measures $\sigma_k$. If observer 1 can measure $\sigma_y$ of that electron it will foul up observer 2's experiment. Half the time when she remeasures $\sigma_x$ it will have been flipped. This tells her that observer 1 has made a measurement. This is faster than light communication, an impossibility. It is a little hard to mathematically state the obvious experimental fact that I can't measure the spin of an electron in the Andromeda galaxy. We don't have any way of localizing particles, any position operator, yet. We can make a mathematical statement in terms of observables:

If $O_1$ is an observable that can be measured in $R_1$ and $O_2$ is an observable that can be measured in $R_2$ and $R_1$ and $R_2$ are space-like separated then they commute:
\[ [O_1,O_2]=0\]

Observables are attached to space time points. A given observer cannot measure all observables, only the ones associated with his or her region of space-time.

It is not possible, even in principle, for everyone to measure everything. Out of the hordes of observables, only a restricted set can be measured at a space-time region. Localization of measurements is going to substitute for localization of particles.

The attachment of observables to space-time points has no analog in NRQM, nor in the classical theory of a single particle, relativistic or nonrelativistic, but it does have an analog in classical field theory. In electromagnetism, there are six observables at each point: $E_x(x)=E_x(\vec{x},t)$, $E_y(x)$, $E_z(x)$, $B_x(x)$, $B_y(x)$, $B_z(x)$. We can't design an apparatus here that measures the $E_x$ field now in the Andromeda galaxy.

In classical field theory, these observables are numbers. In quantum mechanics, observables are given by operators. The fields will become quantum fields, an operator for each spacetime point. We can see in another way that the electric field is going to have to become a quantum field: How would you measure an electric field? You mount a charged ball, pith ball, to some springs and see how much the springs stretch. The location of the ball is given by an equation like:
\[q\ddot{x} = E_x(x)\longleftarrow\begin{smallmatrix}
\text{might be $\int d^4xf(x)E_x(x)$ where $f(x)$}\\
\text{gives some suitable average over the pith ball.}
\end{smallmatrix}\]

The amount the ball moves is related to the $\vec{E}$ field, and if the world is quantum mechanical, $x$ must be an operator, and so $E_x$ must be an operator.

We don't have a proof, but what is strongly suggested is that QM and relativistic causality force us to introduce quantum fields. In fact, relativistic QM is practically synonymous with quantum field theory.

We will try to build our observables from a complete commuting set of quantum fields. 
\[\underset{a=1,\dots,N}{\phi^a(x)}\;\;\;\text{ operator valued functions of space-time}.\]

Observables in a region $R$ will be built out of $\phi^a(x)$ with $x\in R$. Observables in space-like separated regions will be guaranteed to commute if
\begin{enumerate}
\item $[\phi^a(x),\phi^b(y)]=0$ whenever $(x-y)^2<0$. \\
\null\\
We are going to construct our fields out of the creation and annihilation ops. These five conditions will determine them:
\item $\phi^a(x)=\phi^{a\dagger}(x)$ hermitian, observable\\
\null\\
and that they have proper translation and Lorentz transformation properties
\item $e^{-iP\cdot a}\phi^a(x)e^{iP\cdot a}=\phi^a(x-a)$
\item $U(\Lambda)^\dagger\phi^a(x)U(\Lambda)=\underset{\substack{\text{If this were not a scalar}\\ \text{field, there would be extra} \\ \text{factors here reflecting a} \\ \text{change of basis; as well as} \\ \text{a change of argument}}}{\underbrace{}\phi^a(\Lambda^{-1}x)}$\\
\null\\
and finally, a simplifying assumption, that the fields are a linear combination of $a_{\vec{k}}$ and $a_{\vec{k}}^\dagger$ (if that doesn't work, we'll try quadratic functions.)
\item $\phi^a(x)=\int d^3k[F^a_k(x)a_{\vec{k}}+G^a_k(x)a_{\vec{k}}^\dagger]$
\end{enumerate}

We can think of our unitary transformations in two ways; as transformations on the states $|\psi\rangle\rightarrow U|\psi\rangle$, or as transformations on the operators $A\rightarrow U^\dagger AU$. NOT BOTH!

What's embodied in assumption (3):

Given $U(\vec{a})$ the unitary operator of space translation by $\vec{a}$ ($U(\vec{a})=e^{-i\vec{P}\cdot\vec{a}}$) the translation of a state $|\psi\rangle$ is a state $|\psi'\rangle = U(\vec{a})|\psi\rangle$. Suppose the value of some observable, like charge density is
\[ f(\vec{x})=\langle\psi|\rho(\vec{x})|\psi\rangle\]
\begin{center}
\includegraphics[width=7cm]{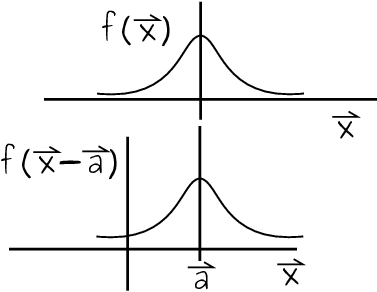}
\end{center}

\noindent 
then it should be that 
\[ \langle\psi'|\rho(\vec{x})|\psi'\rangle = f(\vec{x}-\vec{a}).\]

Rewrite the first equation with $\vec{x}\rightarrow \vec{x}-\vec{a}$
\[f(\vec{x}-\vec{a})=\langle \psi|\rho(\vec{x}-\vec{a})|\psi\rangle \]
\begin{align*}
\text{Equate }\langle\psi|\rho(\vec{x}-\vec{a})|\psi\rangle &=\langle \psi'|\rho(\vec{x})|\psi'\rangle \\
&=\langle\psi|e^{i\vec{P}\cdot\vec{a}}\rho(\vec{x})e^{-i\vec{P}\cdot \vec{a}}|\psi\rangle
\end{align*}

A hermitian operator is determined by its expectation values
\[ e^{i\vec{P}\cdot\vec{a}}\rho(\vec{x})e^{-i\vec{P}\cdot\vec{a}}= \rho(\vec{x}-\vec{a}) \]

(3) is just the full relativistic form of this equation. The equation with $a=(t,\vec{0})$ is just the time evolution for Heisenberg fields.

For the exact same reason as $x-a$ appears in the RHS of (3), $\Lambda^{-1}x$ appears in the RHS of (4). (4) gives the Lorentz transformation properties of a \uwave{scalar} field. This is not much of an assumption. We can get fields transforming as vectors or tensors by taking derivatives of the $\phi^a$. Out of vector or tensor fields, we could make scalars. 

In order to apply condition (4), it is nice to have the discussion phrased in terms of relativistically normalized creation and annihilation operators. 

Recall the relativistically normalized one particle states
\[ |k\rangle = (2\pi)^{3/2}\sqrt{2\omega_{\vec{k}}}|\vec{k}\rangle \]

Introduce $\alpha^\dagger(k)=(2\pi)^{3/2}\sqrt{2\omega_{\vec{k}}} a^\dagger_{\vec{k}}$, $\alpha^\dagger(k)|0\rangle=|k\rangle$.

Multiparticle states are made by
\[ \alpha^\dagger(k_1)\cdots\alpha^\dagger(k_n)|0\rangle=|k_1,\dots, k_n\rangle \]

The Lorentz transformation properties of the states are 
\begin{align*}
U(\Lambda)|0\rangle &=0 \\
U(\Lambda)|k_1,\dots,k_n\rangle &=|\Lambda k_1,\dots,\Lambda k_n\rangle 
\end{align*}

\noindent 
and $U(a)=e^{iP\cdot a}$ is found from 
\begin{align*}
P^\mu|0\rangle&=0\\
P^\mu|k_1,\dots,k_n\rangle &=(k_1+\cdots+k_n)^\mu|k_1,\dots,k_n\rangle 
\end{align*}

In Eqs.~(\ref{eq:01-lor1})-(\ref{eq:01-lor4}) and Eqs.~(\ref{eq:01-trans1})-(\ref{eq:01-trans3}) of the Sept.~23 lecture we set up the criteria that $U(\Lambda)$ and $U(a)$ must satisfy. At the time our Hilbert space consisted only of the one particle part of the whole Fock space we have now. You should check that the criteria are satisfied in Fock space.

We can determine Lorentz transformation and translation properties of the $\alpha^\dagger(k)$. Consider,
\begin{align*}
U(\Lambda)\alpha^\dagger(k)U(\Lambda)^\dagger|k_1,\dots,k_n\rangle &= U(\Lambda)\alpha^\dagger(k)|\Lambda^{-1}k_1,\dots,\Lambda^{-1}k_n \rangle \\
&=U(\Lambda)|k,\Lambda^{-1}k_1,\dots,\Lambda^{-1}k_n \rangle\\
& =|\Lambda k,k_1\dots,k_n\rangle
\end{align*}

That is $U(\Lambda)\alpha^\dagger(k)U(\Lambda)^\dagger|k_1,\dots,k_n\rangle =\alpha^\dagger(\Lambda k)|k_1,\dots,k_n\rangle$

$|k_1,\dots,k_n\rangle$ is an arbitrary state in our complete basis so we have determined its action completely
\[ U(\Lambda)\alpha^\dagger(k)U(\Lambda)^\dagger=\alpha^\dagger (\Lambda k)\]

Similarly, or by taking the adjoint of this equation
\[ U(\Lambda)\alpha(k)U(\Lambda)^\dagger=\alpha(\Lambda k)\]

An analogous derivation shows that
\begin{align*}
e^{iP\cdot x}\alpha^\dagger(k)e^{-iP\cdot x}&= e^{ik\cdot x} \alpha^\dagger(k)\\
e^{iP\cdot x}\alpha(k)e^{-iP\cdot x}&= e^{-ik\cdot x} \alpha(k)
\end{align*}

Now to construct the field $\phi$ (if there is more than one we'll label them when we've found them) satisfying all 5 conditions. First we'll satisfy condition (5) except we'll write the linear combination of $a_{\vec{k}}$ and $a_{\vec{k}}^\dagger$ in terms of our new $\alpha(\vec{k})$ and $\alpha^\dagger(\vec{k})$
\[ \phi(x)=\int\!\!\!\!\!\!\overbrace{\frac{d^3k}{(2\pi)^32\omega_{\vec{k}}}}^{\substack{\text{It would be stupid}\\ \text{not to use the} \\ \text{L.I.~measure}}}\!\!\!\!\!\![f_k(x)\alpha(k) +g_k(x)\alpha^\dagger(k)] \]

By (3), $\phi(x)=e^{iP\cdot x}\phi(0)e^{-iP\cdot x}$, that is,
\begin{align*}
\phi(x)&=\int \frac{d^3k}{(2\pi)^32\omega_{\vec{k}}}[f_k(0)e^{iP\cdot x}\alpha(k)e^{-iP\cdot x} +g_k(0)e^{iP\cdot x}\alpha^\dagger(k) e^{-iP\cdot x}]\\
&=\int \frac{d^3k}{(2\pi)^32\omega_{\vec{k}}}[f_k(0)e^{-ik\cdot x} \alpha(k) +g_k(0)e^{ik\cdot x}\alpha^\dagger(k) ]
\end{align*}

We have found the $x$ dependence of $f_k(x)$ and $g_k(x)$ now we will use (4) to get their $k$ dependence. A special case of (4) is
\begin{align*}
\phi(0)&=U(\Lambda)\phi(0)U(\Lambda)^\dagger \\
\int\frac{d^3k}{(2\pi)^32\omega_{\vec{k}}}[f_k(0)\alpha(k) +g_k(0) \alpha^\dagger(k) ]&=\\
\int\frac{d^3k}{(2\pi)^32\omega_{\vec{k}}}&[f_k(0) \underbrace{U(\Lambda)\alpha(k)U(\Lambda)^\dagger}_{\alpha(\Lambda k)} +g_k(0)\underbrace{U(\Lambda)\alpha^\dagger(k)U(\Lambda)^\dagger}_{\alpha^\dagger(\Lambda k)} ] \\
&\substack{\text{change variables}\\ \text{measure is unchanged}}\;\;\; k\rightarrow \Lambda^{-1}k\\
&=\int\frac{d^3k}{(2\pi)^32\omega_{\vec{k}}}[f_{\Lambda^{-1}k}(0)\alpha(k) +g_{\Lambda^{-1}k}(0)\alpha^\dagger(k) ]
\end{align*}

The coefficients of $\alpha(k)$ and $\alpha^\dagger(k)$ must be unchanged $\Rightarrow f_k(0)=f_{\Lambda^{-1}k}(0)$ and $g_k(0)=g_{\Lambda^{-1}k}(0)$.

$k$ ranges all over the mass hyperboloid ($k^0>0$ sheet), but a Lorentz transformation can turn any of these $k$'s into any other. So $f_k(0)$ and $g_k(0)$ are constants, independent of $k$.
\[\phi(x)=\int \frac{d^3k}{(2\pi)^32\omega_{\vec{k}}}[fe^{-ik\cdot x} \alpha(k) +ge^{ik\cdot x}\alpha^\dagger(k) ]\]

We have two linearly independent solutions of conditions (3), (4) and (5), the coefficients of the complex constants $f$ and $g$. We'll name them. (Switching back to our old creation and annihilation ops.)
\begin{align*}
\phi^+(x)&=\int \frac{d^3k}{(2\pi)^{3/2}\sqrt{2\omega_{\vec{k}}}} a_{\vec{k}}e^{-ik\cdot x} & \phi^-(x)&=\int \frac{d^3k}{(2\pi)^{3/2}\sqrt{2\omega_{\vec{k}}}} a^\dagger_{\vec{k}}e^{ik\cdot x}\\ \text{Note }\phi^-(x)&=\phi^+(x)^\dagger & &\substack{\text{$\pm$ convention is bananas, but it}\\ \text{was est'd by Heisenberg and Pauli 50 years ago.} }
\end{align*}

Now we'll apply hermiticity. Two independent combinations satisfying (2) are
\[\phi(x)=\phi^+(x)+\phi^-(x)\;\;\text{ and }\;\; \phi(x)=\frac{1}{i}[\phi^+(x)-\phi^-(x)]\]

These are two independent cases of the most general choice satisfying (2):
\[\phi(x) = e^{i\theta}\phi^+(x)+e^{-i\theta}\phi^-(x)\]

Now to satisfy (1). There are three possible outcomes of trying to satisfy (1).

\newcounter{threeLcount}
\begin{list}{Possibility \Alph{threeLcount}: }
{\usecounter{threeLcount}
\setlength{\rightmargin}{\leftmargin}}
\item Both of the above combinations are OK. We have two fields $\phi^1$ and $\phi^2$ commuting with themselves \uline{and each other} at spacelike separation. In this possibility $\phi^+(x)$ and $\phi^-(x)$ commute with each other at spacelike separation.
\item Only one combination is acceptable. It is of the form $\phi=e^{i\theta}\phi^++e^{-i\theta}\phi^-$. While $\theta$ may be arbitrary, only one $\theta$ is acceptable.
\item The program crashes, and we could weaken (5) or think harder.
\end{list}

Let's first calculate some commutators. Using their expansions in terms of the $a_{\vec{k}}$ and $a_{\vec{k}}^\dagger$ and the commutation relations for $a_{\vec{k}}$ and $a_{\vec{k}}^\dagger$ we find
\begin{align*}
[\phi^+(x),\phi^+(y)]&=0=[\phi^-(x),\phi^-(y)]\\
\text{and }\;\;[\phi^+(x),\phi^-(y)]&=\int \frac{d^3k}{(2\pi)^{3} 2\omega_{\vec{k}}}e^{-ik\cdot(x-y)}\equiv \underbrace{\Delta_+(x-y,\mu^2)}_{\text{or just }\Delta_+(x-y)}\\
\text{also }\;\;[\phi^-(x),\phi^+(y)]&=-\Delta_+(x-y,\mu^2)\\
\text{$\Delta_+$ is manifestly Lorentz invariant }\;\;\Delta_+(\Lambda x)&=\Delta_+(x).
\end{align*}

Possibility A runs only if $\Delta_+(x-y)=0$ for $(x-y)^2<0$. We have encountered a similar integral when we were looking at the evolution of one particle position eigenstates. In fact,
\[i\partial_0\Delta_+(x-y)= \frac{1}{2} \int\frac{d^3k}{(2\pi)^3}e^{-ik\cdot(x-y)}\]

\noindent 
is the very integral we studied, and we found that it did \uwave{not} vanish when $(x-y)^2<0$.

Possibility A is DEAD.

On to possibility B. Take
$\phi(x)=e^{i\theta}\phi^+(x)+e^{-i\theta}\phi^-(x)$ and calculate
\[ [\phi(x),\phi(y)]=\Delta_+(x-y)-\Delta_+(y-x)\;\;\substack{\text{$\theta$ dependence} \\ \text{drops out}}\]

Does this vanish when $(x-y)^2<0$? Yes, and we can see this without any calculations. A space-like vector has the property that it can be turned into minus itself by a (connected) Lorentz transformation. This and the fact that $\Delta_+$ is Lorentz invariant, tells us that $[\phi(x),\phi(y)]=0$ when $x-y$ is space-like. We can choose $\theta$ arbitrarily, but we can't choose more than one $\theta$. We'll choose $\theta=0$. Any phase could be absorbed into the $a_{\vec{k}}$ and $a_{\vec{k}}^\dagger$.

Possibility B is ALIVE, and we don't have to go on to C.

We have our free scalar field of mass $\mu$
\[\phi(x)=\phi^+(x)+\phi^-(x)=\int \frac{d^3k}{(2\pi)^{3/2}\sqrt{2\omega_{\vec{k}}}}[ a_{\vec{k}}e^{-ik\cdot x}+ a_{\vec{k}}^\dagger e^{ik\cdot x} ] \]

Our field satisfies an equation (show using $k^2=\mu^2$)
\begin{align*}
(\square+\mu^2)\phi(x)&=0 & \square&=\partial^\mu\partial_\mu 
\end{align*}

This is the Heisenberg equation of motion for the field. It is called the Klein-Gordon equation. If we had quantized the electromagnetic field it would have satisfied Maxwell's equations.

Actually, Schr\"odinger first wrote down the Klein-Gordon equation. He got it at the same time as he got the Schr\"odinger equation:
\[i\partial_0\psi=-\frac{1}{2\mu}\nabla^2\psi \]

This equation is obtained by starting with $E=\frac{\vec{p}\,^2}{2m}$ noting that $E=\omega$ (when $\hbar=1$) and $p^i=k^i$ and for plane waves $\omega=i\partial_0$ and $k^i=\frac{1}{i}\partial_i$.

Schr\"odinger was no dummy, he knew about relativity, so he also obtained $(\square+\mu^2)\phi(x)=0$ from $p^2=\mu^2$.

He immediately saw that something was wrong with the equation though. The equation has both positive and negative energy solutions. For a free particle the energies of its possible states are unbounded below$!$ This is a disgusting relativistic generalization of a single particle wave equation, but with 50 years hindsight we see that this is no problem for a field that can create and destroy particles.
}{
 \sektion{4}{October 2}
\descriptionfour
We have constructed the quantum field. It is the object observables are built from. More than that though, we can reconstruct the entire theory from the quantum field. The structure we built in the last three lectures is rigid. We can make the top story the foundation. Suppose we started with a quantum field satisfying (as our $\phi$ does),
\begin{enumerate}
\item $\phi(x)=\phi^\dagger(x)$, hermiticity
\item $(\square+\mu^2)\phi(x)=0$, K.-G.~equation
\item $[\phi(x),\phi(y)]=\Delta_+(x-y)-\Delta_+(y-x)=\int \frac{d^3k}{(2\pi)^{3}2\omega_{\vec{k}}}[e^{-ik\cdot (x-y)}- e^{ik\cdot (x-y)}]$
\item $\left.\begin{array}{l} U(\Lambda)^\dagger\phi(x)U(\Lambda)=\phi(\Lambda^{-1}x)\\
\ U(a)^\dagger\phi(x)U(a)=\phi(x-a) \end{array}\right\}\phi$ is a scalar field
\item $\phi(x)$ is a complete set of operators, i.e.~if $\forall x\;\;[A,\phi(x)]=0\Rightarrow A=\lambda I$.
\end{enumerate}

\noindent 
then from these properties, we could reconstruct the creation and annihilation operators and the whole theory. Conversely, all these properties follow from the expression for $\phi(x)$ in terms of the creation and annihilation operators. The two beginning points are logically equivalent.

\section*{Defining $a_{\vec{k}}$ and $a_{\vec{k}}^\dagger$ and recovering their properties}
Property (2) of $\phi$, that $\phi$ is a solution of the Klein-Gordon equation, tells us that 
\begin{align*}
\phi(x)&=\int d^3k[ \alpha_{\vec{k}}e^{-ik\cdot x}+ \beta_{\vec{k}} e^{ik\cdot x} ] & (k^0=\omega_{\vec{k}}=\sqrt{\vec{k}^2+\mu^2})
\end{align*}

This is because \uline{any} solution of the Klein-Gordon equation can be expanded in a complete set of solutions of the K.-G.~eqn., and the plane wave solutions are a complete set. Because $\phi$ is an operator, the coefficients in the expansion, $\alpha_{\vec{k}}$ and $\beta_{\vec{k}}$, are operators. Because of property (1), $\alpha_{\vec{k}}=\beta_{\vec{k}}^\dagger$. Just to please my little heart, let's define
\begin{align*}
& & &\substack{\text{These funny factors will make the} \\ \text{commutation relations for the $a_{\vec{k}}$ and $a_{\vec{k}}^\dagger$ } \\ \text{come out nice}}\\
\alpha_{\vec{k}}&=\frac{a_{\vec{k}}}{(2\pi)^{3/2}\sqrt{2\omega_{\vec{k}}}} & \beta_{\vec{k}}&=\alpha_{\vec{k}}^\dagger=\frac{ a_{\vec{k}}^\dagger} {(2\pi)^{3/2}\sqrt{2 \omega_{\vec{k}}}} 
\end{align*}

Then the expression is
\[\phi(x)=\int \frac{d^3k}{(2\pi)^{3/2}\sqrt{ 2\omega_{\vec{k}}}}[ a_{\vec{k}}e^{-ik\cdot x}+ a_{\vec{k}}^\dagger e^{ik\cdot x} ] \]

This defines implicitly the $a_{\vec{k}}$ and $a_{\vec{k}}^\dagger$. To find their commutation relations we'll first have to solve for them. Note that
\[\phi(\vec{x},0)=\int \frac{d^3k}{(2\pi)^{3/2}\sqrt{2\omega_{\vec{k}}}}[ a_{\vec{k}}e^{i\vec{k}\cdot \vec{x}}+ a_{\vec{k}}^\dagger e^{-i\vec{k}\cdot \vec{x}} ] \]

While
\[\dot{\phi}(x)=\int \frac{d^3k}{(2\pi)^{3/2}\sqrt{2\omega_{\vec{k}}}}[ a_{\vec{k}}(-i\omega_{\vec{k}})e^{-ik\cdot x}+ a_{\vec{k}}^\dagger (i\omega_{\vec{k}})e^{ik\cdot x} ] \]

And 
\[\dot{\phi}(\vec{x},0)=\int \frac{d^3k}{(2\pi)^{3/2}}\sqrt{\frac{\omega_{\vec{k}}}{2}}[ - i a_{\vec{k}}e^{i\vec{k}\cdot \vec{x}}+ ia_{\vec{k}}^\dagger e^{-i\vec{k}\cdot \vec{x}} ] \]

The coefficient of $\frac{1}{(2\pi)^{3/2}}e^{i\vec{k}\cdot\vec{x}}$ in the expansion for $\phi(\vec{x},0)$ is $\frac{1}{\sqrt{2\omega_{\vec{k}}}}(a_{\vec{k}}+a_{-\vec{k}}^\dagger)$ therefore $\frac{1}{\sqrt{2\omega_{\vec{k}}}}(a_{\vec{k}}+ a_{-\vec{k}}^\dagger)=\int\frac{d^3x}{(2\pi)^{3/2}}\phi(\vec{x},0) e^{-i\vec{k}\cdot\vec{x}}$

The coefficient of $\frac{1}{(2\pi)^{3/2}}e^{i\vec{k}\cdot\vec{x}}$ in the expansion for $\dot{\phi}(\vec{x},0)$ is $\sqrt{\frac{\omega_{\vec{k}}}{2}}(-ia_{\vec{k}}+ia_{-\vec{k}}^\dagger)$ therefore $\sqrt{\frac{\omega_{\vec{k}}}{2}}(-ia_{\vec{k}}+ ia_{-\vec{k}}^\dagger)=\int\frac{d^3x}{(2\pi)^{3/2}}
\dot{\phi}(\vec{x},0) e^{-i\vec{k}\cdot\vec{x}}$

That's just the inverse Fourier transformation applied.

Now I can use these two expressions to solve for $a_{\vec{k}}$. Take $\sqrt{2\omega_{\vec{k}}}$ times the expression for $\frac{1}{\sqrt{2\omega_{\vec{k}}}}(a_{\vec{k}}+a_{-\vec{k}}^\dagger)$ and add $\sqrt{\frac{2}{\omega_{\vec{k}}}} i$ times the expression for $\sqrt{\frac{\omega_{\vec{k}}}{2}}(-ia_{\vec{k}}+ia_{-\vec{k}}^\dagger)$ and divide by 2 to get 
\[a_{\vec{k}}=\frac{1}{2}\left[ \sqrt{2 \omega_{\vec{k}}} \int\frac{d^3x}{(2\pi)^{3/2}}\left(\phi(\vec{x},0)+\frac{i}{\omega_{\vec{k}}}\dot{\phi}(\vec{x},0)\right)e^{-i\vec{k}\cdot\vec{x}}\right] \]

Take the hermitian conjugate to get an expression for $a_{\vec{k}}^\dagger$.

Having solved for the $a_{\vec{k}}$ and $a_{\vec{k}}^\dagger$, we can now \uline{find} their commutation relations. Write out the double integral for $[a_{\vec{k}}, a_{\vec{k}\,'}^\dagger]$ and use property (3') below (which is weaker than property (3)). You should get $\delta^{(3)}(\vec{k}-\vec{k}\,')$.

From property (4), you can derive, what $U(a)^\dagger a_{\vec{k}}U(a)$ and $U(\Lambda)^\dagger a_{\vec{k}} U(\Lambda)$\footnote{It is easier to find the action of Lorentz transformations on $a(k)\equiv (2\pi)^{3/2}\sqrt{2\omega_{\vec{k}}}a_{\vec{k}}$.} are.

You can also derive and state the analog of property (5).

The properties of the field $\phi(x)$ are actually a little overcomplete. We can weaken one of them, (3), without losing anything. Since $\phi$ obeys the K.-G.~equation, (2),
\[ (\square_x+\mu^2)[\phi(x),\phi(y)]=0\]

We see that the commutator obeys the K.-G.~equation for any given $y$. For a given $y$, we only need to give the commutator and the time derivative on one initial time surface, and the K.-G.~equation determines its evolution off the surface. So we will weaken (3) to 
\begin{list}{3'. }
{\setlength{\rightmargin}{\leftmargin}}
\item $\begin{array}{l}
[\phi(\vec{x},t),\phi(\vec{y},t)]=0 \\[0pt]
[\partial_0\phi(\vec{x},t),\phi(\vec{y},t)]=-i\delta^{(3)}(\vec{x}-\vec{y})
\end{array}$
\end{list}

We can easily check that this is the right specialization of property (3) by doing the integrals, which are easy with $x^0=y^0$. (For a given $y=(\vec{y},t)$ we have chosen our initial surface to be $x^0=t$. The equations in (3') are called equal time commutation relations.)
\begin{align*}
[\phi(\vec{x},t),\phi(\vec{y},t)] &=\int \frac{d^3k}{(2\pi)^{3}2\omega_{\vec{k}}}\underset{\substack{\text{Change variables $\vec{k}\rightarrow -\vec{k}$}\\ \text{in second term}}}{[e^{i\vec{k}\cdot (\vec{x}-\vec{y})}- e^{-i\vec{k}\cdot
(\vec{x}-\vec{y})}]}\\
&=0\\
[\partial_0\phi(\vec{x},t),\phi(\vec{y},t)] &=\int \frac{d^3k}{(2\pi)^{3}2\omega_{\vec{k}}}\underset{\substack{\text{Again $\vec{k}\rightarrow -\vec{k}$ in second term}}}{(-i\omega_{\vec{k}})[e^{i\vec{k}\cdot (\vec{x}-\vec{y})}+e^{-i\vec{k}\cdot (\vec{x}-\vec{y})}]}\\
&=\int \frac{d^3k} {(2\pi)^{3}\cancel{2\omega_{\vec{k}}}}(-i\cancel{\omega_{\vec{k}}})\cancel{2}e^{i\vec{k}\cdot(\vec{x}-\vec{y})}\\
&=-i\delta^{(3)}(\vec{x}-\vec{y})\ \ \left(\substack{\text{Reminds us a little}\\ \text{of a bad dream in}\\\text{which $[p_a,q^b]=-i\delta^b_a$} }\right)
\end{align*}

In the remainder of this lecture we are going to develop a completely different approach to quantum field theory. We will obtain a field satisfying properties (1), (2), (3'), (4) and (5), and then we'll stop. We have just shown that given a field satisfying these properties we can recover everything we did in the first three lectures. The new approach is

\section*{The method of the missing box}
\begin{center}
\includegraphics[width=.8\textwidth]{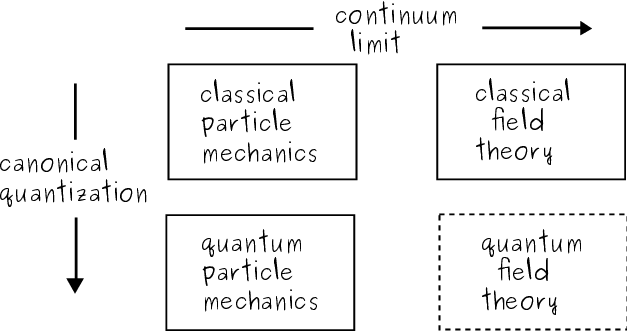}\\
Anyone looking at this diagram can see there is a missing box
\end{center}

Taking the continuum limit essentially is just letting the number of coordinates, degrees of freedom go to infinity. This is usually done in a cavalier way. It doesn't matter a whole lot if you have a discrete infinity or a continuous infinity. With a continuous infinity you can Fourier transform the coordinates to obtain a discrete set of coordinates.

Canonical quantization is a turn the crank way of getting quantum mechanics, beginning with a classical Hamiltonian.

We'll combine these two standard operations to get the missing box, but first, a lightning review of the essential principles of the three boxes we have.

\subsection*{Classical Particle Mechanics}
We start with a Lagrangian, a function of the generalized coordinates, and their time derivatives
\begin{align*}
&L(q^1,\dots,q^N,\dot{q}^1,\dots,\dot{q}^N,t) & \underbrace{q^a(t)}_{\mbox{\footnotesize real number functions of time}} \!\!\!\!\!\!\!\!\!\!\!\!\!\!\!\!\!\!\!\!\!\!\!\!, a\in1,\dots,N
\end{align*}

\noindent 
e.g.~$L=\frac{1}{2}m\dot{q}^2-V(q)$

We define the action $S=\int_{t_1}^{t_2}dtL$ and apply Hamilton's principle to get the equations of motion. That is, we vary $S$ by arbitrarily varying $\delta q^a(t)$ except at the endpoints 
\[\delta q^a(t_1)=\delta q^a(t_2)=0\]

\noindent 
and demand 
\[0=\delta S\]
\begin{align*}
\delta S&=\int_{t_1}^{t_2}dt\underbrace{\left[\sum_a\left(\frac{\partial L}{\partial q^a}\delta q^a+\frac{\partial L}{\partial\dot{q}^a}\delta \dot{q}^a\right)\right]}_{\text{parts integration}}\\
&=\int_{t_1}^{t_2}dt\sum_a\underset{\ \ \ \ \ \ \ \ \ \ p_a\equiv\frac{\partial L} {\partial \dot{q}^a}} {\left(\frac{\partial L}{\partial q^a}-\frac{d}{dt}p_a\right)}\delta q^a+\underbrace{\cancel{p_a \delta q^a\vert_{t_1}^{t_2}}}_{\substack{\text{vanishes because of}\\ \text{endpoint restriction}}}
\end{align*}

Since the variation $\delta q^a(t)$ is arbitrary 
\begin{align*}
\frac{\partial L}{\partial q^a}&=\dot{p}_a,\ \text{ for all $a$}& &\text{e.g.~}p=m\dot{q},\dot{p}=-\frac{dV}{dq}
\end{align*}

These are the Euler-Lagrange equations.

Around 1920(!) the Hamiltonian formulation of classical particle mechanics was discovered.
\[ \text{Define }\ H(p_1,\dots,p_N,q_1,\dots,q_N,t)=\sum_ap_a\dot{q}^a-L\]

$H$ must be written in terms of the $p$'s and $q$'s only, not the $\dot{q}$'s. (This is not always possible. The new variables must also be independent, so it is possible to vary them independently. Examples where the $p$'s are not complete and independent are electromagnetism and a particle on a sphere in $\mathbb{R}^3$. The Lagrangian for the latter system may be taken as
\[ L=\frac{1}{2}m\dot{\vec{r}}\,^2+\lambda(\vec{r}\,^2-a^2)-V(\vec{r}) \]

The equation of motion for the variable $\lambda$ enforces the constraint. In the passage to the Hamiltonian formulation $p_\lambda=0$;\footnote{How can I vary that?!?} $p_\lambda$ is not an independent function. In this system, one way to get to the Hamiltonian description is to first eliminate $\lambda$ and one more coordinate, taking perhaps $\theta$, $\phi$, polar coordinates for the system, rewriting the Lagrangian, and then trying again.)

Vary the coordinates and momenta
\[dH=\sum_a\biggl(\underset{\substack{\text{Fortunately we didn't}\\
\text{have to expand out $d\dot{q}^a$}}}{dp_a\dot{q}^a+\cancel{p_a d\dot{q}^a}}-\underset{\substack{\text{$\dot{p}_a$ by the E-L equations}}}{\underbrace{\frac{\partial L}{\partial q^a}}dq^a-\cancel{\frac{\partial L}{\partial \dot{q}^a}d\dot{q}^a}} \biggr)\]

Read off Hamilton's equations
\begin{align*}
\frac{\partial H}{\partial p_a}&=\dot{q}^a &\frac{\partial H}{\partial q^a}&=-\dot{p}_a & \begin{array}{l} \text{Notice: }\frac{\partial L}{\partial t}=0\Rightarrow\frac{dH}{dt}=0\\ \substack{ \text{\scriptsize in that case $H$ is called the energy (which}\\ \text{\scriptsize is the name reserved for the conserved quantity}\\ \text{\scriptsize resulting from time translation invariance)}}
\end{array}
\end{align*}

\subsection*{Quantum Particle Mechanics}
We replace the classical variables $p_a$, $q^a$ by operator valued functions of time satisfying
\begin{align*}
[q^a(t),q^b(t)]&=0=[p_a(t),p_b(t)]\\
\text{and }[p_a(t),q^b(t)]&=-i\delta_a^b & \boxed{\hbar=1}
\end{align*}

The $p$'s and $q$'s are hermitian observables. They are assumed to be complete. They determine the Hilbert space. For example 1-d particle mechanics, the space of square integrable functions. A basis set can be the eigenstates of $q$. $p$ or rather $e^{ipx}$ tells you how to relate the phases of the various eigenstates and indeed even that the range of $q$ is $\mathbb{R}$ (not [-1,1] or anything else).

The quantum Hamiltonian, which determines the dynamics, is just the classical Hamiltonian except it is now a function of the operator
$p$'s and $q$'s. For any $A$
\[\frac{dA}{dt}=i[H,A]+\frac{\partial A}{\partial t} \longleftarrow \substack{\text{if there is explicit dependence on} \\ \text{time other than that implicit in the} \\ \text{time dependence of the $p$'s and $q$'s}}\]

$H$ is the generator of infinitesimal time translations just as in classical mechanics.

$H$ suffers from ordering ambiguities. Because $p$ and $q$ don't commute, it is not clear, and it may matter, whether you write $p^2q$, $qp^2$ or $pqp$. Sometimes this ambiguity can be cleared up using other criteria. For example, in central force problems, we quantize in Cartesian coordinates and then transform to central coordinates.

\subsection*{Heisenberg equations of motion}
\begin{align*}
\begin{split}
\frac{dq^a}{dt}&=i[H,q^a]\\
&=i\left(-i\frac{\partial H}{\partial p_a}\right)
\end{split}\Downarrow\substack{\text{This step depends \uline{only} on the commutation}\\ \text{relations for the $p$'s and $q$'s. Fairly easy to see}\\ \text{for a polynomial in the $p$'s and $q$'s.}}\\
&=\frac{\partial H}{\partial p_a} \\
\text{Similarly, }\frac{dp_a}{dt}&=-\frac{\partial H}{\partial q^a}
\end{align*}

This is the motivation for the commutation relations. It is a way of putting the correspondence principle into the theory. The quantum equations resemble the classical equations, at least up to the ordering ambiguities, and any variation due to ordering ambiguities is down by a factor of $\hbar$. We wouldn't expect any general procedure for turning classical theories into quantum theories, motivated only by the correspondence principle, to be able to fix those ambiguities. We can be a little more precise about how the classical equations are actually recovered. In general, the Heisenberg equations of motion for an arbitrary operator $A$ relate one polynomial in $p$, $q$, $\dot{p}$ and $\dot{q}$ to another. We can take the expectation value of this equation to obtain (a quantum mechanical average of) an equation between observables. In the classical limit, when fluctuations are small, expectations of products can be replaced by products of expectations, $\langle p^n\rangle\rightarrow\langle p\rangle^n$, $\langle pq\rangle\rightarrow\langle p\rangle\langle q\rangle$, and this turns our equation among polynomials of quantum operators into an equation among classical variables.

The other two boxes are actually going to go quite quickly, because if you are just a little cavalier, the continuum limit is little more than some new notation.

\subsection*{Classical Field Theory}
We have an infinite set of generalized coordinates, real number functions of time, $\phi^a(\vec{x},t)$ labelled by a discrete index, $a$, and a continuous index, $\vec{x}$.
\begin{align*}
\phi^a(\vec{x},t)&\leftrightarrow q^a(t)\\
t&\leftrightarrow t\\
a&\leftrightarrow a,\vec{x}
\end{align*}

It is sometimes a handy mnemonic to think of $t$, $\vec{x}$, as a generalization of $t$, but that is not the right way to think about it. For example we are used to giving initial value data at fixed $t$ in CPM. In CFT we don't give initial value data at fixed $t$ and $\vec{x}$, that is obviously incomplete. We give initial value data for fixed $t$ and \uline{all} $\vec{x}$.

With cowboy boldness, everywhere in CPM we see a sum on $a$, we'll just replace it with a sum on $a$ and an integral over $\vec{x}$, and everywhere we see $\delta^a_b$ we'll replace it with $\delta^a_b\delta^{(3)}(\vec{x}-\vec{y})$. The Dirac delta function has the exact same properties in integrals that the Kronecker delta has in sums.

The next thing to do is write down Lagrangians. Because a CPM Lagrangian can contain products of $q^a$ with different $a$, our Lagrangian could contain products of $\phi(\vec{x},t)$ with different $\vec{x}$
\[\int d^3xd^3yd^3zf_{abc}(\vec{x},\vec{y},\vec{z})\phi^a(\vec{x},t) \phi^b(\vec{y},t)\phi^c(\vec{z},t)\leftrightarrow \sum_{a,b,c}f_{abc}q^a(t)q^b(t)q^c(t) \]

But note that a CPM Lagrangian does not contain products at different times.

With an eye to Lorentz invariance, and noticing that Lagrangians are local in time, we will specialize to Lagrangians that are local in space. Also, you know that when taking continuum limits, differences of neighboring variables become spatial derivatives and a combination like 
\[\frac{1}{a^2}[\rho((n-1)a,t)+\rho((n+1)a,t)-2\rho(na,t)]\]

\noindent 
would become $\frac{\partial^2\rho}{\partial x^2}$ in the continuum. But again with an eye to Lorentz invariance, because only first derivatives with respect to time appear in the Lagrangian, we will only consider first derivatives with respect to the $x^i$. So $L$ has the form
\begin{align*}
L(t)&=\int d^3x\mathcal{L}(\phi^a(x),\partial_\mu\phi^a(x),x)\\
\text{And the action }S&=\int_{t_1}^{t_2}dtL(t)=\int d^4x\mathcal{L}
\end{align*}

The Euler-Lagrange equations which come from varying $S$ will be Lorentz covariant if $\mathcal{L}$ is a Lorentz scalar. At three points we have used Lorentz invariance to cut down on the possible forms for $L$. These have been specializations, not generalizations. Now we'll apply Hamilton's principle
\begin{align*}
0&=\delta S\;\;\;\left(\substack{\text{under arbitrary variations $\delta\phi$}\\ \text{satisfying $\delta\phi^a(\vec{x},t_1)=\delta\phi^a(\vec{x},t_2)=0$} }\right)\\
\begin{split}
&=\sum_a\int d^4x\biggl(\frac{\partial\mathcal{L}}{\partial \phi^a(\vec{x},t)}\delta \phi^a(\vec{x},t)+\underbrace{\frac{\partial \mathcal{L}} {\partial\partial_\mu \phi^a}}_{\equiv \pi^\mu_a}\underbrace{\delta\partial_\mu\phi^a}_{\partial_\mu\delta\phi^a}\biggr)\\
&=\sum_a\int d^4x\left[\frac{\partial\mathcal{L}}{\partial\phi^a}-\partial_\mu\pi^\mu_a\right]\delta\phi^a
\end{split}\Downarrow\substack{\text{Do parts integration. As usual, the} \\ \text{restrictions on $\delta\phi^a$ make the surface} \\ \text{terms at $t_1$ and $t_2$ drop out} } \\
\Rightarrow\frac{\partial\mathcal{L}}{\partial\phi_a}&=\partial_\mu \pi^\mu_a \;\;\;\text{(for all $\vec{x}$ and
$a$)}\;\;\;\substack{\text{Euler-Lagrange}\\ \text{equations}}
\end{align*}

$\pi^\mu_a$ should not be thought of as a four-vector generalization of $p_a$. The correspondence is
\[ \pi^0_a(\vec{x},t)\leftrightarrow p_a(t) \]

In fact $\pi^0_a$ is often just written $\pi_a$.

We should say something about the surface terms at spatial infinity which we ignored when we did our parts integration. One can say ``we are only considering field configurations which fall off sufficiently rapidly at spatial infinity that we can ignore surface terms.'' Alternatively we could work in a box with periodic boundary conditions. Anyway, we'll just be slothful.

\subsection*{A simple example of a possible $\mathcal{L}$}
Most general $\mathcal{L}$ satisfying (3) conditions
\begin{enumerate}
\item Build $\mathcal{L}$ out of one real scalar field, $\phi$ \\
$\phi=\phi^*$ (not $\phi=\phi^\dagger$, we're not doing QFT yet)
\item $\mathcal{L}$ is a Lorentz scalar
\item $\mathcal{L}$ is quadratic in $\phi$ and $\partial_\mu\phi$
\end{enumerate}

(1) and (3) are for simplicity. A motivation for (3) is that a quadratic action yields linear equations of motion, the easiest ones to solve. Most general $\mathcal{L}$ is 
\[ \mathcal{L}=\frac{1}{2}a[\partial_\mu\phi\partial^\mu\phi+b\phi^2] \]

One of these constants is superfluous; we are always free to rescale $\phi$, $\phi\rightarrow\frac{\phi}{\sqrt{|a|}}$. $\mathcal{L}$ becomes 
\[ \mathcal{L}=\pm\frac{1}{2}[\partial_\mu\phi\partial^\mu\phi +b\phi^2]\]

The Euler-Lagrange equations for our example are
\begin{align*}
\pi^\mu&\equiv\frac{\partial\mathcal{L}}{\partial\partial_\mu\phi}=\pm \partial^\mu\phi\;\;\text{ and}\\
\partial_\mu\pi^\mu-\frac{\partial\mathcal{L}}{\partial\phi}&=0 \text{ or }\pm(\partial^\mu\partial_\mu\phi-b\phi)=0
\end{align*}

In general the Hamiltonian which was $\sum_ap_a\dot{q}^a-L$ in CPM is 
\begin{align*}
H&=\sum_a\int d^3x(\pi^0_a\partial_0\phi^a-\mathcal{L})=\int d^3x\mathcal{H}\\
\mathcal{H}&=\sum_a\pi^0_a\partial_0\phi^a-\mathcal{L}\text{ is the Hamiltonian density}
\end{align*}

In our example
\[H = \pm\int d^3x[\frac{1}{2}(\pi^0)^2+\frac{1}{2}(\vec{\nabla} \phi)^2-b\phi^2 ] \]

Since each of these terms separately can be made arbitrarily large, if the energy is to be bounded below, they each better have a positive coefficient.$\pm$ better be $+$ and $b$ better be
\[b=-\mu^2\;\;\;\text{ definition of }\mu\geq0\]

The E-L equation is now $\partial_\mu\partial^\mu\phi+\mu^2\phi=0$

Gosh, this is looking familiar now.

We are ready to fill in the last box. We are ready to canonically quantize classical field theory. Since we are not worrying about the passage to an infinite number of variables, this will be little more than a notational change in the canonical quantization of CPM. It would not even have required a notational change if Newton hadn't chosen two ways of writing $S$ for sum, $\int$ and $\sum$.

\section*{Quantum Field Theory}
We replace the classical variables $\phi^a$, $\pi^0_a$ by quantum operators satisfying
\begin{align*}
[\phi^a(\vec{x},t),\phi^b(\vec{y},t)] &=0=[\pi^0_a(\vec{x},t), \pi^0_b(\vec{y},t)] \\
\text{and }[\pi^0_a(\vec{x},t),\phi^b(\vec{y},t)] &=-i\delta_a^b \delta^{(3)}(\vec{x}-\vec{y})
\end{align*}

The Dirac delta for the continuous index $\vec{x}$ is just the continuum generalization of the Kronecker delta. $H=\int d^3x\mathcal{H}(\pi^0_a,\phi^a,x)$ determines the dynamics as usual. The commutation relations are set up to reproduce the Heisenberg equations of motion. Since this is just a change of notation there is no need to redo any proofs, but let's check that things are working in our example anyway.
\begin{align*}
H&=\int d^3x\frac{1}{2}[(\pi^0)^2+(\vec{\nabla}\phi)^2+\mu^2\phi^2]\\
\partial_0\phi(\vec{y},t)&=i[H,\phi(\vec{y},t)]=i(-i)\int d^3x\pi^0(\vec{x},t)\delta^{(3)}(\vec{x}-\vec{y}) \\
&=\pi^0(\vec{y},t)
\end{align*}

Similarly, $\partial_0\pi^0(\vec{y},t)=\vec{\nabla}^2\phi-\mu^2\phi$. You need the commutator $[\pi^0(\vec{y},t),\vec{\nabla}\phi(\vec{x},t)]$ which is obtained by taking the gradient of the equal time commutation relation (ETCR).

We have reproduced our quantum field satisfying (1), (2), (3'), (4) and (5). This was accomplished in one lecture rather than three because this is a mechanical method without physical insight. The physical interpretation comes at the end instead of at the beginning. In our first method, the constructive method, we shook each object we introduced to make sure it made sense, and we finally obtained a local observable, the quantum field. We can't put interactions in this method though, because in the very first steps we had to know the whole spectrum of the theory, and about the only theory we know the exact spectrum for is the free theory. In our magical canonical quantization method it's easy to put in interactions: just let $\mathcal{L}\rightarrow\mathcal{L}-\lambda\phi^4$. At the first order in perturbation theory the part of $\lambda\phi^4$ that has two creation and two annihilation operators produces two-into-two scattering$!$ At second order we'll get two-into-four and two-into-six\footnote{pair production!} scattering. Looks easy\dots if there weren't any booby traps\dots but boy are there going to be booby traps.
}{
 \sektion{5}{October 7}
\descriptionfive
The simplest example obtained from applying canonical quantization to a free scalar field led to the same theory we got by constructing a local observable painstakingly in the theory we had before.

We'll double check that we get the same Hamiltonian we had in that theory. Evaluate
\begin{align*}
H&=\frac{1}{2}\int d^3x[\pi^2+(\nabla\phi)^2+\mu^2\phi^2] &\pi&=\dot{\phi}
\end{align*}

Write $\phi$ in terms of its Fourier expansion
\[\phi(x)=\int \frac{d^3k}{(2\pi)^{3/2}\sqrt{2\omega_{\vec{k}}}}[ a_{\vec{k}}e^{-ik\cdot x}+ a_{\vec{k}}^\dagger e^{ik\cdot x} ] \]

Substitution would lead to a triple integral, but the $x$ integration is easily done yielding a $\delta$ function, which does one of the $k$ integrals\footnote{$\int d^3x \; e^{i\vec{k}\cdot\vec{x}}e^{i\vec{k}\,'\cdot\vec{x}}=(2\pi)^3\delta^{(3)}(\vec{k}+\vec{k}\,')\Rightarrow \vec{k}\,'=-\vec{k}$}. 
\begin{align*}
H=\frac{1}{2} \int \frac{d^3k}{2\omega_{\vec{k}}}\biggl\{&a_{\vec{k}}\!\!\!\!\!\!\underbrace{a_{-\vec{k}}}_{\substack{\vec{k}\,'=-\vec{k}\text{ see}\\ \text{footnote 1}}}\!\!\!\!\!\! e^{-2i\omega_{\vec{k}}t} \underset{\substack{\text{That what multiplies $e^{\pm 2i\omega_{\vec{k}}t}$ is 0 is good.}\\ \text{$H$ should not be time-dependent.}}}{\underbrace{\cancel{(-\omega_{\vec{k}}^2+|\vec{k}|^2+\mu^2)}} +a_{\vec{k}}a_{\vec{k}}^\dagger(\omega_{\vec{k}}^2+|\vec{k}|^2+\mu^2)}\\
&+a_{\vec{k}}^\dagger a_{\vec{k}}(\omega_{\vec{k}}^2+|\vec{k}|^2+\mu^2)+a_{\vec{k}}^\dagger a_{-\vec{k}}^\dagger e^{2i\omega_{\vec{k}}t} \overbrace{\cancel{(-\omega_{\vec{k}}^2+|\vec{k}|^2+\mu^2)}}\biggr\}
\end{align*}

$H$ had four types of terms: ones creating particles with momentum $\vec{k}$ and $-\vec{k}$, ones destroying particles with momentum $\vec{k}$ and $-\vec{k}$, ones creating a particle with momentum $\vec{k}$, then destroying one with momentum $\vec{k}$, and ones destroying a particle with momentum $\vec{k}$, then creating one with momentum $\vec{k}$. These all conserve momentum.
\[H=\frac{1}{2}\int d^3k \;\omega_{\vec{k}}(a_{\vec{k}} a_{\vec{k}}^\dagger+a_{\vec{k}}a_{\vec{k}}^\dagger)\]

Almost, but not quite what we had before.

Using $[a_{\vec{k}},a_{\vec{k}\,'}^\dagger]=\delta^{(3)}(\vec{k}-\vec{k}\,')$ gives $H=\int d^3k\omega_{\vec{k}}\left(a_{\vec{k}}^\dagger a_{\vec{k}}+\frac{1}{2}\delta^{(3)}(0)\right)$. $\delta^{(3)}(0)!$ Can get some idea of the meaning of this by putting the system in a box. $H$ becomes
\[\frac{1}{2}\sum_{\vec{k}}\omega_{\vec{k}}(a_{\vec{k}} a_{\vec{k}}^\dagger+a^\dagger_{\vec{k}}a_{\vec{k}})=\sum_{\vec{k}} \omega_{\vec{k}}\left( a_{\vec{k}}^\dagger a_{\vec{k}}+\frac{1}{2}\right)\]

In a box we see that this is just a zero point energy. (It is still infinite though.) Just as the spectrum of the Hamiltonian $H=\frac{1}{2}(p^2+\omega^2q^2)$ starts at $\frac{1}{2}\omega$, not zero, the spectrum of our Hamiltonian starts at $\frac{1}{2}\sum_{\vec{k}}\omega_{\vec{k}}$, not zero. That this constant is infinite even in a box is because even in a box the field has an infinite number of degrees of freedom. There can be excitations of modes with arbitrarily short wavelengths. The zero point energy is ultraviolet divergent. In general, infinite systems, like an infinite crystal, also have ``infrared divergences.'' If the energy density of an infinite crystal is changed by an infinite amount, the total energy is changed by an infinite amount. Our Hamiltonian for the system in infinite space, would still have an infinity ($\delta^{(3)}(0)$) even if the momentum integral were cut
off, say by restricting it to $|\vec{k}|<K$. That part of the infinity is eliminated by putting the system in a box, eliminating long wavelengths. Our system in infinite space has a zero-point energy that is also infrared divergent.

This is no big deal for two reasons.

\newcounter{05count1}
\begin{list}{(\Alph{05count1}) }
{\usecounter{05count1}
\setlength{\rightmargin}{\leftmargin}}
\item You can't measure absolute energies, only energy differences, so it's stupid to ask what the zero point energy is. This even occurs in introductory physics. We usually put interaction energies to be zero when particles are infinitely separated, but for some potentials you can't do that, and you have to choose your zero another place.
\item This is just an ordering ambiguity. Just as the quantum Hamiltonian for the harmonic oscillator could be chosen as
\[H=\frac{1}{2}(p+i\omega q)(p-i\omega q)\]
\noindent 
we could reorder our Hamiltonian (see below).
\end{list}

In general relativity the absolute value of the energy density does matter. Einstein's equations
\[ R_{\mu\nu}-\frac{1}{2}g_{\mu\nu}R=-8\pi GT_{\mu\nu}\]

\noindent couple directly to the energy density $T_{00}$. Indeed, introducing a change in the vacuum energy density, in a covariant way,
\[ T_{\mu\nu}\rightarrow T_{\mu\nu}-\lambda g_{\mu\nu}\]

\noindent is just a way of changing the cosmological constant, a term introduced by Einstein and repudiated by him 10 years later. No astronomer has ever observed a nonzero cosmological constant.\footnote{[BGC note: As of 1998, measurements on supernovae have indicated that $\Lambda>0$.]} Our theory is eventually going to be applied to the strong interactions, maybe even to some grand unified theory. Strong interactions have energies typically of 1 GeV and a characteristic length of a fermi, $10^{-13}$ cm. With a cosmic energy density of $10^{39}$ GeV/cm$^3$, the universe would be about 1 km long according to Einstein's equations. You couldn't even get to MIT without coming back where you started. We won't talk about why the cosmological constant is zero in this course. They don't explain it in any course given at Harvard because nobody knows why it is zero.

\section*{Reordering the Hamiltonian:}
Given a set of free fields (possibly with different masses) $\phi_1(x_1)$, $\dots$, $\phi_n(x_n)$, define the normal ordered product
\[:\phi_1(x_1)\cdots\phi_n(x_n):\]

\noindent 
as the usual product except rearranged so that all the annihilation operators are on the right and \uline{a fortiori} all creation operators are on the left. No further specification is needed since all annihilation operators commute with one another as do all creation operators. This operation is only defined for free fields satisfying the K.-G.~equation, which tells us we can write the field in terms of time independent creation and annihilation operators.

Redefine $H$ to be $:H:$. This gets rid of the normal ordering constant.\footnote{ What's wrong? 
\begin{align*}
a^\dagger a&=aa^\dagger-1\;\;\;\text{ Normal order both sides}\\
:a^\dagger a:&=:a^\dagger a:-1\\
0&=-1
\end{align*}

Answer. We don't ``normal order'' equations. Normal ordering is not derived from the ordinary product any more than the cross product is derived from the scalar product. Normal ordering cannot accurately be used as a verb.}

This is the first infinity encountered in this course. We'll encounter more ferocious ones. We ran into it because we asked a dumb question, a physically uninteresting question, about an unobservable quantity. Later on we'll have to think harder about what we've done wrong to get rid of troublesome infinities.

If we wanted to get as quickly as possible to applications of QFT, we'd develop scattering theory and perturbation theory next. But first we are going to get some more exact results from field theory.

\section*{Symmetries and Conservation Laws}
We will study the relationship between symmetries (or invariances) and conservation laws in the classical Lagrangian framework. Our derivations will only use classical equations of motion. Hopefully everything will go through all right when Poisson brackets become commutators, since the commutation relations are set up to reproduce the classical equations of motion. In any given case you can check to see if ordering ambiguities or extra terms in commutators screw up the calculations.

Given some general Lagrangian $L(q^a,\dot{q}^a,t)$ and a transformation of the generalized coordinates
\begin{align*}
q^a(t)&\rightarrow q^a(t,\lambda) & q^a(t,0)&=q^a(t)
\end{align*}

\noindent 
we can define
\[ Dq^a\equiv\left.\frac{\partial q^a}{\partial\lambda} \right|_{\lambda=0} \]

This is useful because the little transformations will be important.

Some examples:
\begin{enumerate}
\item Space translation of point particles described by $n$ vectors $\vec{r}\,^a$, $a=1,\dots,n$. The Lagrangian we'll use is
\[ \sum_{a=1}^n\frac{m_a}{2}|\dot{\vec{r}}\,^a|^2-\sum_{ab}V_{ab}(|\vec{r}\,^a-\vec{r}\,^b|) \]

The transformation is $\vec{r}\,^a\rightarrow\vec{r}\,^a+\vec{e}\lambda$, all particles in the system moved by $\vec{e}\lambda$
\[D\vec{r}\,^a=\vec{e}\]

\item Time translations for a general system. Given some evolution $q^a(t)$ of the system, the transformed system is a time $\lambda$ ahead $q^a(t)\rightarrow q^a(t+\lambda)$
\[Dq^a=\frac{\partial q^a}{\partial t}\]

\item Rotations in the system of example (1)
\begin{align*}
\vec{r}\,^a&\rightarrow\!\!\!\!\!\overbrace{R}^{\substack{\text{Rotation}\\ \text{matrix}}}\!\!\!\!\!(\underset{\substack{\text{angle, axis}}}{\underbrace{\lambda}\underbrace{ \vec{e}}})\vec{r}\,^a & D\vec{r}\,^a&=\vec{e}\times\vec{r}\,^a 
\end{align*}

(If $\vec{e}=\widehat{z}$, $Dx^a=y^a$, $Dy^a=-x^a$, $Dz^a=0$.)
\end{enumerate}

Most transformations are not symmetries (invariances). 
\begin{list}{Definition: }
\item A transformation is a symmetry iff $DL=\frac{dF}{dt}$ for some $F(q^a,\dot{q}^a,t)$. This equality must hold for arbitrary $q^a(t)$, not necessarily satisfying the equations of motion.
\end{list}

Why is this a good definition? Hamilton's principle is
\[ 0=\delta S=\int_{t^1}^{t^2}dt \; DL = \int_{t^1}^{t^2} dt \frac{dF}{dt} =F(t_2)-F(t_1) \]

Thus a symmetry transformation won't affect the equations of motion (we only consider transformations that vanish at the endpoints).

Back to our examples:
\begin{enumerate}
\item $DL=0$, $F=0$

\item If $L$ has no explicit time dependence, all of its time
dependence comes from its dependence on $q^a$ and $\dot{q}^a$,
$\frac{\partial L}{\partial t}=0$, then $DL=\frac{dL}{dt}$, $F=L$

\item $F=0$
\end{enumerate}

Theorem (E.~Noether {\scriptsize say `Nota'}): For every symmetry there is a conserved quantity.

The proof comes from considering two expressions for $DL$.
\begin{align*}
DL&=\sum_a\left(\frac{\partial L}{\partial q^a}Dq^a+p_aD\dot{q}^a\right) & \substack{\text{used }p_a\equiv\frac{\partial L}{\partial\dot{q}^a}}\\
&=\sum_a(\dot{p}_aDq^a+p_aD\dot{q}^a) & \substack{\text{used E-L equations}}\\
&=\frac{d}{dt}\sum_ap_aDq^a & \substack{\text{by equality of mixed partials} \\ D\dot{q}^a=\frac{d}{dt}Dq^a }
\end{align*}

By assumption $DL=\frac{dF}{dt}$. Subtracting these two expressions for $DL$ we see that the quantity
\[Q=\sum_ap_aDq^a-F\;\;\;\;\text{ satisfies }\;\;\;\frac{dQ}{dt}=0\]

(There is no guarantee that $Q\neq0$, or that for each independent symmetry we'll get another independent $Q$, in fact the construction fails to produce a $Q$ for gauge symmetries.)

We can write down the conserved quantities in our examples
\begin{enumerate}
\item ($p_a=m_a\dot{\vec{r}}\,^a$), $D\vec{r}\,^a=\vec{e}$, $F=0$.
\[ Q=\sum_am_a\vec{e}\cdot\dot{\vec{r}}\,^a=\vec{e}\cdot\sum_am_a \dot{\vec{r}}\,^a \]

For each of three independent $\vec{e}$'s we get a conservation law. The momentum 
\[\vec{P}=\sum_am_a\dot{\vec{r}}\,^a\;\;\;\text{ is conserved.}\;\;\;\left(\frac{d\vec{P}}{dt}=0\right).\]

Whenever we get conserved quantities from spatial translation invariance, whether or not the system looks anything like a collection of point particles, we'll call the conserved quantities the momentum.

\item $Dq^a=\frac{\partial q^a}{\partial t}$, $F=L$. Note: $Q$ is identical to $H$.
\[ Q=\sum_ap_a\dot{q}^a-L\;\;\;\text{ is conserved}\left(\text{when }\;\frac{\partial L}{\partial t}=0\right) \]

Whenever we get a conserved quantity from time translation invariance, we'll call the conserved quantity the energy.

\item $D\vec{r}\,^a=\vec{e}\times\vec{r}\,^a$, $F=0$.
\begin{align*}
Q&=\sum_a\vec{p}_a\cdot(\vec{e}\times\vec{r}\,^a)=\sum_a\vec{e}\cdot(\vec{r}\,^a\times\vec{p}_a)\\
&=\vec{e}\cdot\sum_a\vec{r}\,^a\times\vec{p}_a \;\;\;\text{three laws}\\ \vec{J}&= \sum_a\vec{r}\,^a\times\vec{p}_a\;\;\;\text{ is conserved }\left(\frac{d\vec{J}}{dt}=0\right)
\end{align*}

Whenever we get a conserved quantities from rotational invariance, we'll call them the angular momentum.
\end{enumerate}

There is nothing here that was not already in the Euler-Lagrange equations. What this theorem provides us with is a turn the crank method for obtaining conservation laws from a variety of theories. Before this theorem, the existence of conserved quantities, like the energy, had to be noticed from the equations of motion in each new theory. This theorem organizes conservation laws. It explains, for example, why a variety of theories, including ones with velocity dependent potentials all have a conserved Hamiltonian, or energy (example (2)).

From the conserved quantity, we can usually reconstruct the symmetry. This can be done in classical mechanics using Poisson brackets. We'll do it in quantum mechanics using commutators.\footnote{Show: $[Q^i,Q^j]=i\epsilon_{ijk}Q^k$ when the $Q^i$'s are generators of an $SO(3)$ internal symmetry. For a general internal symmetry group, the quantum $Q$'s recreate the algebra of the generators.}

Assume $Dq^b$ and $F$ depend only on the $q^a$, not the $\dot{q}^a$ so that their expression in the Hamiltonian formulation only depends on the $q^a$, not the $p_a$. Then
\begin{align*}
[Q,q^a]&=\left[\sum_bp_bDq^b-F,q^a\right]=\sum_b\underbrace{[p_b,q^a]}_{-i\delta_b^a}Dq^b\\
&=-iDq^a
\end{align*}

This assumption is not at all necessary. Usually the result holds. For example the energy generates time translations even though the assumption doesn't hold.

\section*{Symmetries and Conservation Laws in (Classical) Field Theory}
Field theory is a specialization of particle mechanics. There will be more that is true in field theory. What is this more?

Electromagnetism possesses a conserved quantity $Q$, the electric charge. The charge is the integral of the charge density $\rho$, $Q=\int d^3x\rho(\vec{x},t)$. There is also a current $\vec{\jmath}$ and there is a much stronger statement of charge conservation than $\frac{dQ}{dt}=0$. Local charge conservation says
\[ \frac{\partial\rho}{\partial t}+\vec{\nabla}\cdot\vec{\jmath}=0\]

Integrate this equation over any volume $V$ with boundary $S$ to get
\[\frac{dQ_V}{dt}=-\int_S dA \;\widehat{n}\cdot\vec{\jmath}\;\;\;\substack{\text{using Gauss's theorem}\\ \left(Q=\int_Vd^3x \rho(\vec{x},t)\right)} \]

This equation says you can see the charge change in any volume by watching the current flowing out of the volume. You can't have:
\begin{center}
\includegraphics[width=6cm]{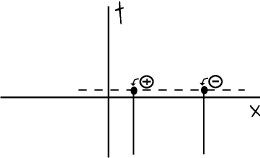}\\
simultaneously wink out of existence with nothing happening anywhere else.
\end{center}

This picture satisfies global charge conservation, but violates local charge conservation. You have to be able to account for the change in charge in any volume, and there would have to be a flow of current in between the two charges. Even if there were not a current and a local conservation law, we could invoke special relativity to show this picture is inconsistent. In another frame the charges don't disappear simultaneously, and for a moment global charge conservation is violated.

Field theory which embodies the idea of local measurements, should have local conservation laws.

Given some Lagrangian density $\mathcal{L}(\phi^a,\partial_\mu\phi^a,x)$ and a transformation of the fields 
\begin{align*}
\phi^a(x)&\rightarrow\phi^a(x,\lambda) & \phi^a(x,0)&=\phi^a(x)
\end{align*}

\noindent 
we define $D\phi^a=\left.\frac{\partial\phi^a}{\partial\lambda}\right|_{\lambda=0}$.
\begin{list}{Definition: }
\item A transformation is a symmetry iff $D\mathcal{L}=\partial_\mu F^\mu$ for some $F^\mu(\phi^a,\partial_\mu\phi^a,x)$. This equality must hold for arbitrary $\phi^a(x)$ not necessarily satisfying the equations of motion.
\end{list}

We are not using special relativity. $F^\mu$ is not necessarily a 4-vector (just some set of 4 objects).

Note that the previous definition can be obtained. To wit, 
\[DL=D\int d^3x\mathcal{L}=\int d^3x\partial_\mu F^\mu=\int d^3x\partial_0F^0=\frac{d}{dt}F\]

\noindent 
where $F=\int d^3xF^0$.

Why is this a good definition? Hamilton's principle is
\begin{align*}
0&=\int\delta S=\int d^4xD\mathcal{L}=\int d^4x\partial_\mu F^\mu \\
&=\int d^3x[F^0(\vec{x},t_2)-F^0(\vec{x},t_1)]\;\;\;\substack{\text{blithe as usual about}\\ \text{contributions from}\\ \text{spatial infinity}}
\end{align*}

Thus a symmetry transformation does not affect the equations of motion (we only consider variations that vanish at the endpoints when deriving the equations of motion.).

Theorem (maybe this is Noether's theorem?): For every symmetry there is a conserved current.

The proof comes from considering two expressions for $D\mathcal{L}$.
\begin{align*}
D\mathcal{L}&=\sum_a\left(\frac{\partial L}{\partial \phi^a}D\phi^a+\pi_a^\mu D\partial_\mu\phi^a \right) & \substack{\text{used}\pi_a^\mu\equiv\frac{\partial L}{\partial\partial_\mu\phi^a}}\\
&=\sum_a(\partial_\mu \pi_a^\mu D\phi^a+\pi_a^\mu D\partial_\mu\phi^a) & \substack{\text{used E-L equations}}\\
&=\partial_\mu\sum_a\pi_a^\mu D\phi^a & \substack{\text{by equality of mixed partials} \\ D\partial_\mu \phi^a=\partial_\mu D\phi^a }
\end{align*}

By assumption $D\mathcal{L}=\partial_\mu F^\mu$. Subtracting these two expressions for $D\mathcal{L}$ we see that the four quantities
\[J^\mu=\sum_a\pi_a^\mu D\phi^a - F^\mu \;\;\;\;\text{ satisfy}\;\;\;\partial_\mu J^\mu=0\]

\noindent which implies $\underset{\substack{\text{This equation justifies calling}\\ \text{ $J^0$ the density of stuff}\\ \text{and $J^i$ the current of stuff.}}}{\frac{dQ_V}{dt}=-\int_S dA \; \widehat{n}\cdot\vec{J}}$, $S=$boundary of $V$. The total amount of stuff, $Q$ is independent of time.

There is an ambiguity in the definition of $F^\mu$ and $J^\mu$ since $F^\mu$ can be changed by any $\chi^\mu$ satisfying $\partial_\mu\chi^\mu=0$. In the particle mechanics case, $F$ was ambiguous, but only by a time independent quantity. For arbitrary antisymmetric $A$ we can let $F^\mu\rightarrow F^\mu+\partial_\nu A^{\mu\nu}$ ($\partial_\mu\partial_\nu A^{\mu\nu}=0$). As a result of this change $J^\mu\rightarrow J^\mu-\partial_\nu A^{\mu\nu}$.
\[ Q=\int d^3x J^0\rightarrow Q-\int d^3x\partial_i A^i=Q \]

\noindent 
i.e.~$Q$ is unchanged, ignoring contributions from spatial infinity as usual.

This is a lot of freedom in the definition of $J^\mu$. For example in a theory with three fields we can let
\[ J^\mu\rightarrow J^\mu-(\phi^3)^{14}\partial_\nu(\partial^\mu\phi^1 \partial^\nu\phi^2 -\partial^\nu \phi^1\partial^\mu\phi^2) \]

This $J^\mu$ is as good as any other. There are 500 papers arguing about which energy-momentum tensor is the right one to use inside a dielectric medium. 490 of them are idiotic. It's like if someone passes you a plate of cookies and you start arguing about which copy is \#1 and which is \#2. They're all edible$!$ Sometimes one $J^\mu$ is more useful in a given calculation than another, for some reason. Instead of arguing that the most convenient $J^\mu$ is the ``right'' one, you should just be happy that you had some freedom of choice.

From cranking Hamilton's principle, we can give another derivation of the relation between symmetries and conserved quantities.
\[\delta S=\delta\int dt L=\sum_a\left.p_aDq^a\right|_{t_1}^{t_2} \]

\noindent 
for an arbitrary variation about a solution of the equations of motion. By assumption of a symmetry 
\[\delta S=\int dt DL=\int dt\left.\frac{dF}{dt}\right|_{t_1}^{t_2} \]

Subtracting we see $\sum_ap_aDq^a-F$ is time independent.

\section*{[\sc{Aside: Noether's Theorem derived at the level\footnote{i.e.~using Hamilton's principle rather than mucking around with the equations of motion} of the action}}

If without using the equations of motion, for an arbitrary function of space time $\alpha(x)$ parametrizing a transformation, to first order in $\alpha(x)$
\[\delta \mathcal{L} = \alpha(x)\partial_\mu k^\mu+\partial_\mu\alpha(x)j^\mu \]

(Equivalent to $D\mathcal{L}=\partial_\mu k^\mu$ when $\alpha$ is a constant).

Then by Hamilton's principle, for fields satisfying the equations of motion,
\begin{align*}
0&=\int d^4x\delta\mathcal{L}=\int d^4x(\alpha(x)\partial_\mu k^\mu +\partial_\mu\alpha(x)j^\mu)\\
&=\int d^4x(\alpha(x)\partial_\mu k^\mu - \alpha(x)\partial_\mu j^\mu)\;\;\;\substack{\text{Let $\alpha(x)$ vanish at $\infty$}\\ \text{to do parts integration.}}
\end{align*}

So it must be that $J^\mu=k^\mu-j^\mu$ is conserved.

\subsubsection*{\uline{Example}: Space-time translation invariance}
\begin{align*}
\delta\phi&=-\epsilon_\nu\partial^\nu\phi & \delta\partial_\tau\phi &=-\partial_\tau(\epsilon_\nu\partial^\nu\phi) \\
\delta\mathcal{L}&=\frac{\partial\mathcal{L}}{\partial\phi}\delta\phi + \frac{\partial\mathcal{L}}{\partial\partial_\tau\phi}\delta \partial_\tau\phi \\
&=-\frac{\partial\mathcal{L}}{\partial\phi}\epsilon_\nu\partial^\nu\phi -\frac{\partial\mathcal{L}}{\partial\partial_\tau\phi}\epsilon_\nu \partial^\nu\partial_\tau\phi - \frac{\partial\mathcal{L}}{\partial\partial_\tau\phi}\partial_\tau\epsilon_\nu\partial^\nu\phi &\substack{\text{$\nu$ is a parameter for the} \\ \text{type of transformation.}}\\ 
&=-\epsilon_\nu\mathcal{L}'^\nu-\pi^\mu\partial^\nu\phi\partial_\mu \epsilon_\nu &\substack{\text{$\mu$ plays the same role}}
\end{align*}

I read off $k^{\mu\nu}=-g^{\mu\nu}\mathcal{L}$, $j^{\mu\nu}=-\pi^\mu\partial^\nu\phi$ so $T^{\mu\nu}=\pi^\mu\partial^\nu\phi-g^{\mu\nu}\mathcal{L}$ is conserved.]

\section*{\uline{Noether's Theorem}: when the variation is a hermitian traceless matrix. (useful in theories with fields most easily written as matrices.)}
Suppose {\small (without using the equations of motion)} $\delta S= \int d^4x\underset{\substack{\text{hermitian, traceless}\\ \text{but otherwise arbitrary}}}{\text{Tr }\underbrace{\delta\omega}\partial_\mu \!\!\!\overbrace{F^\mu}^{\text{matrix}}}$.

Then Hamilton's principle says
\[0=\int d^4x\text{Tr }\delta\omega\partial_\mu F^\mu \]

Then $F^\mu$ is a matrix of conserved currents, except there is no conservation law associated with the trace.

If you like, write this statement as 
\[\partial_\mu\underset{\text{dimension of $F^\mu$}}{\biggl(F^\mu-\underbrace{\frac{1}{N}}\text{Id Tr }F^\mu\biggr)}=0 \]

\section*{Space-time translations in a general field theory}
We won't have to restrict ourselves to scalar fields because under space-time translations, an arbitrary field, vector, tensor, etc., transforms the same way. Just let the index $a$ also denote vector or tensor indices.

Translations are symmetries as long as $\mathcal{L}$ does not have any explicit dependence on $x$. It depends on $x$ only through its dependence on $\phi^a$ and $\partial_\mu\phi^a$, $\frac{\partial\mathcal{L}}{\partial x}=0$.
\begin{align*}
\phi^a(x)&\rightarrow \phi^a(x+\lambda e) \;\;\; \substack{\text{$e$ some fixed four vector}} \\
D\phi^a&=e^\nu\partial_\nu\phi^a & D\mathcal{L}&=e^\mu\partial_\mu \mathcal{L} = \partial_\mu(e^\mu\mathcal{L}) \\
F^\mu&= e^\mu\mathcal{L} & J^\mu&=\sum_a\pi_a^\mu e^\nu\partial_\nu\phi^a-e^\mu\mathcal{L}=e_\nu T^{\mu\nu} \\
T^{\mu\nu}&=\sum_a\pi_a^\mu \partial^\nu\phi^a- g^{\mu\nu}\mathcal{L}
\end{align*}

There are four conserved currents, four local conservation laws, one for each of the four independent directions we can point $e$, i.e.~$\partial_\mu T^{\mu\nu}=0$ since $\partial_\mu J^\mu=0$ for arbitrary $e$.

$T^{\mu0}$ is the current that is conserved as a result of time translation invariance, indeed,
\[\mathcal{H}=T^{00}=\sum_a\pi_a\dot{\phi}^a-\mathcal{L},\;\;\;H=\int d^3x \; \mathcal{H} \]

$T^{00}$ is the energy density, $T^{i0}$ is the current of energy. $T^{i\mu}$ are the three currents from spatial translation invariance.
\begin{align*}
P^i&=\int d^3x T^{0i}\\
T^{0i}&=\text{the density of the $i$th component of momentum}\\
T^{ji}&=\text{the $j$th component of the current of the $i$th component of momentum}
\end{align*}

For a scalar field theory with no derivative interactions,
$\pi_a^\mu=\partial^\mu\phi^a$ so 
\[T^{\mu\nu}=\sum_a\partial^\mu \phi^a\partial^\nu\phi^a-g^{\mu\nu} \mathcal{L}\]

Note that $T^{\mu\nu}$ is symmetric, so you don't have to remember which index is which. $T^{\mu\nu}$ can be nonsymmetric, which can lead to problems, for example gravity and other theories of higher spin. Can try to symmetrize it.
}{
 \sektion{6}{October 9}
\descriptionsix
\section*{Lorentz transformations}
Under a Lorentz transformation all vectors transform as
\[a^\mu\rightarrow\Lambda^\mu_\nu a^\nu\]

\noindent 
where $\Lambda^\mu_\nu$ specifies the Lorentz transformation. $\Lambda^\mu_\nu$ must preserve the Minkowski space inner product, that is if 
\[ b^\mu\rightarrow \Lambda^\mu_\nu b^\nu\;\;\;\text{ then }\;\;\; a_\mu b^\mu\rightarrow a_\mu b^\mu\]

This must be true for arbitrary $a$ and $b$. (The equation this condition gives is $g_{\mu\nu}\Lambda^\mu_\alpha\Lambda^\nu_\beta=g_{\alpha\beta}$.) We'll be interested in one parameter subgroups of the group of Lorentz transformations parametrized by $\lambda$. This could be rotations about some specified axis by an angle $\lambda$ or a boost in some specified direction by a rapidity $\lambda$. In any case, the Lorentz transformation is given by a family 
\[ a^\mu\rightarrow a^\mu(\lambda)=\Lambda(\lambda)^\mu_\nu a^\nu \]

Under this (active) transformation (we are not thinking of this as a passive change of coordinates) the fields $\phi^a$ transform as ($\phi^a$ is a scalar),
\[ \phi^a(x)\rightarrow \phi^a(x,\lambda)=\phi^a(\Lambda(\lambda)^{-1}x) \]

We are restricting ourselves to scalar fields. Even though we only used scalar fields in our examples of $T^{\mu\nu}$, the derivation of the conservation of $T^{\mu\nu}$ from space-time translation invariance applies to tensor, or vector, fields. With Lorentz transformations, we only consider scalars, because there are extra factors in the transformation law when the fields are tensorial. For example a vector field $A^\mu(x)\rightarrow \Lambda^\mu_\nu
A^\nu(\Lambda^{-1}x)$.

We need to get $D\phi=\left.\frac{\partial\phi}{\partial\lambda} \right|_{\lambda=0}$. We'll define
\[ D\Lambda^\mu_\nu\equiv \epsilon^\mu_\nu\;\;\;\substack{\text{defines some matrix }\epsilon^\mu_\nu}\]

From the invariance of $a^\mu b_\nu$, we'll derive a condition on $\epsilon^\mu_\nu$.
\begin{align*}
0&=D(a^\mu b_\mu)=(Da^\mu)b_\mu+a^\mu(Db_\mu)\\
\begin{split}
&=\epsilon^\mu_{\;\; \nu} a^\nu b_\mu+ a^\mu\epsilon_\mu^{\;\; \nu} b_\nu\\
&=\epsilon_{\mu\nu}a^\nu b^\mu+\epsilon_{\nu\mu}a^\nu b^\mu
\end{split}\Downarrow\substack{\text{relabel dummy indices}\\
\text{in the second term,}\\ \mu\rightarrow\nu, \nu\rightarrow\mu.}\\
&=(\epsilon_{\mu\nu}+\epsilon_{\nu\mu})a^\nu b^\mu\Rightarrow \epsilon_{\mu\nu}=-\epsilon_{\nu\mu}
\end{align*}

\noindent 
since this has to hold for arbitrary $a$ and $b$. $\mu$ and $\nu$ range from 0 to 3, so there are $\frac{4\cdot(4-1)}{2}=6$ independent $\epsilon$, which is good since we have to generate 3 rotations (about each axis) and 3 boosts (in each direction).

As a second confidence-boosting check we'll do two examples.

Take $\epsilon_{12}=-\epsilon_{21}=1$, all other components zero. 
\begin{align*}
Da^1&=\epsilon^1_2a^2=-\epsilon_{12}a^2=-a^2\\
Da^2&=\epsilon^2_1a^1=-\epsilon_{21}a^1=+a^1
\end{align*}
\begin{center}
\includegraphics[width=6cm]{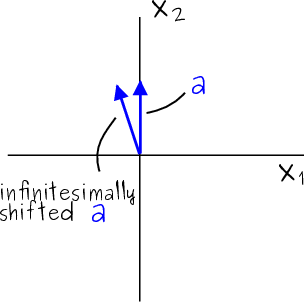}
\end{center}

This says $a^1$ gets a little negative component proportional to $a^2$ and $a^2$ gets a little component proportional to $a^1$. This is a rotation, in the standard sense about the positive $z$ axis.

Take $\epsilon_{01}=-\epsilon_{10}=+1$, all other components zero.
\begin{align*}
Da^0&=\epsilon^0_1a^1=\epsilon_{01}a^1=a^1\\
Da^1&=\epsilon^1_0a^0=-\epsilon_{10}a^0=a^0
\end{align*}

This says $x^1$, which could be the first component of the position of a particle, gets a little contribution proportional to $x^0$, the time, which is definitely what a boost in the $x^1$ direction does. In fact, $Da^0=a^1$, $Da^1=a^0$ is just the infinitesimal version of
\begin{align*}
a^0&\rightarrow \cosh\lambda a^0+\sinh\lambda a^1\\
a^1&\rightarrow \sinh\lambda a^0+\cosh\lambda a^1
\end{align*}

Without even thinking, the great index raising and lowering machine has given us all the right signs.

Now assuming $\mathcal{L}$ is a scalar, we are all set to get the 6 conserved currents.

From $\Lambda^{-1}(\lambda)\Lambda(\lambda)=1$, $0=D(\Lambda^{-1}\Lambda)$ and $D\left.\Lambda^{-1\mu}_\nu \right|_{\Lambda=1}=-\epsilon^\mu_\nu$
\begin{align*}
D\phi^a(x)&=\left.\frac{\partial}{\partial\lambda}\phi^a(\Lambda^{-1} (\lambda)^\mu_\nu x^\nu)\right|_{\lambda=0}\;\;\; \substack{\text{There are extra}\\ \text{terms in $D\phi^a$ if}\\ \text{$\phi^a$ are not scalars}}\\
&=\partial_\sigma\phi^a(x)D(\Lambda^{-1}(\lambda)^\sigma_\tau x^\tau)\\
&=\partial_\sigma\phi^a(x)(-\epsilon^\sigma_\tau)x^\tau= -\epsilon_{\sigma\tau}x^\tau\partial^\sigma\phi^a(x)\\
\end{align*}

Using the assumption that $\mathcal{L}$ is a scalar depending only on $x$ through its dependence on $\phi^a$ and $\partial_\mu\phi^a$ we have
\begin{align*}
D\mathcal{L}&=\epsilon_{\lambda\sigma}x^\lambda \partial^\sigma \mathcal{L}\\
&=\partial_\mu[\epsilon_{\lambda\sigma}x^\lambda g^{\mu\sigma} \mathcal{L}]
\end{align*}

The conserved current $J^\mu$ is
\begin{align*}
J^\mu&=\sum_a\pi^\mu_a\epsilon_{\lambda\sigma}x^\lambda\partial^\sigma \phi^a - \epsilon_{\lambda\sigma}x^\lambda g^{\mu\sigma}\mathcal{L}\\
&=\epsilon_{\lambda\sigma}\left(\sum_a\pi^\mu_ax^\lambda \partial^\sigma \phi^a - x^\lambda g^{\mu\sigma}\mathcal{L}\right)
\end{align*}

This current must be conserved for all six independent antisymmetric matrices $\epsilon_{\lambda\sigma}$, so the part of the quantity in parentheses that is antisymmetric in $\lambda$ and $\sigma$ must be conserved i.e.~$\partial_\mu M^{\mu\lambda\sigma}=0$ where
\begin{align*}
M^{\mu\lambda\sigma}&=\left(\sum_a\pi^\mu_ax^\lambda \partial^\sigma \phi^a - x^\lambda g^{\mu\sigma}\mathcal{L}\right)-(\lambda\leftrightarrow\sigma)\\
&=x^\lambda\left(\sum_a\pi^\mu_a\partial^\sigma \phi^a - g^{\mu\sigma} \mathcal{L}\right)-(\lambda\leftrightarrow\sigma)\\
&=x^\lambda T^{\mu\sigma}-x^\sigma T^{\mu\lambda}
\end{align*}

If the $\phi^a$ were not scalars, we would have additional terms, feeding in from the extra terms in $D\phi^a$. The 6 conserved charges are 
\[J^{\lambda\sigma}=\int d^3x M^{0\lambda\sigma}=\int d^3x(x^\lambda T^{0\sigma}-x^\sigma T^{0\lambda})\]

For example, $J^{12}$, the conserved quantity coming from invariance under rotations about the 3 axis, often called, $J^3$ is ($J^i=\frac{1}{2}\epsilon_{ijk}J^{jk}$),
\[ J^3=J^{12}=\int d^3x(x^1T^{02}-x^2T^{01})\]

If we had point particles with 
\[T^{0i}(\vec{x},t)=\sum_ap_a^i\delta^{(3)}(\vec{x}-\vec{r}^a(t))\]

$J^3$ would be
\[\sum_a(x^{a1}p_a^2-x^{a2}p_a^1)=\sum_a(\vec{r}^a\times\vec{p}_a)_3 \]

We have found the field theory analog of the angular momentum. The particles themselves could have some intrinsic angular momentum. Those contributions to the angular momentum are not in the $J^i$. We only have the orbital contribution. Particles that have intrinsic angular momentum, spin, will be described by fields of tensorial character, and that will be reflected in extra terms in the $J^{ij}$.

So far we have just found the continuum field theory generalization of three conserved quantities we learn about in freshman physics. But we have three other conserved quantities, the $J^{0i}$. What are they? Consider
\[ J^{0i}=\int d^3x[x^0T^{0i}-x^iT^{00}] \]

This has an explicit reference to $x^0$, the time, something we've not seen in a conservation law before, but there is nothing a priori wrong with that. We can pull the $x^0$ out of the integral over space. The conservation law is $\frac{dJ^{0i}}{dt}=0$ so we have
\begin{align*}
0&=\frac{d}{dt}J^{0i}=\frac{d}{dt}\left[ t\int d^3xT^{0i}-\int d^3x x^iT^{00} \right]\\
&=t \cancel{\frac{d}{dt}\underbrace{{\int d^3x T^{0i}}}_{p^i}}+\underbrace{\int d^3x T^{0i}}_{p^i}-\frac{d}{dt}\int d^3x x^i T^{00}
\end{align*}

Dividing through by $p^0$ gives
\[ \text{constant}=\frac{p^i}{p^0}=\frac{\frac{d}{dt}\int d^3x x^iT^{00}}{p^0}=\frac{\frac{d}{dt}(\text{``center of energy''}^i)}{\text{total energy}} \]

This says that the ``center of energy'' moves steadily. $T^{00}$ is the relativistic generalization of mass. This is the relativistic generalization of the statement that the center of mass moves steadily. You aren't used to calling this a conservation law, but it is, and in fact it is the Lorentz partner of the angular momentum conservation law.

\section*{Internal Symmetries}
There are other conservation laws, like conservation of electric charge, conservation of baryon number, and conservation of lepton number that we have not found yet. We have already found all the conservation laws that are in a general Lorentz invariant theory. These additional conservation laws will only occur in specific theories whose Lagrange densities have special properties. Conservation laws are the best guide for looking for theories that
actually describe the world, because the existence of a conservation law is a qualitative fact that greatly restricts the form of the Lagrange density. All these additional charges are scalars, and we expect the symmetries they come from to commute with Lorentz transformations. The transformations will turn fields at the same spacetime point into one another. They will not relate fields at different spacetime points. Internal symmetries are non-geometrical symmetries. Historically, the name comes from the idea that what an internal symmetry did was transform internal characteristics of a particle. For us internal just means non-geometrical. We'll study internal symmetries with two examples.

The first example is
\[\mathcal{L} = \frac{1}{2}\sum_{a=1}^2\left(\partial_\mu\phi^a \partial^\mu\phi^a-\mu^2\phi^a\phi^a\right)-g\left(\sum_a(\phi^a)^2 \right)^2\]

This is a special case of a theory of two scalar fields. Both fields have the same mass, and the potential only depends on the combination $(\phi^1)^2+(\phi^2)^2$.

This Lagrangian\footnote{We have left particle mechanics behind and I'll often use Lagrangian to mean Lagrange density.} is invariant ($D\mathcal{L}=0$) under the transformation
\begin{align*}
\begin{split}
\phi^1&\rightarrow\phi^1\cos\lambda+\phi^2\sin\lambda\\
\phi^2&\rightarrow - \phi^1\sin\lambda+\phi^2\cos\lambda
\end{split}\;\;\;\text{$SO(2)$ symmetry}
\end{align*}

The same transformation at every space time point. This is a rotation\footnote{Clockwise rotation makes the signs come out conventionally if $b$ and $c$ (which we see a little bit later) are defined like $a$ and $b$ in Itzykson and Zuber p.~121.} in the $\phi^1$, $\phi^2$ plane. The Lagrangian is invariant because it only depends on $(\phi^1)^2+(\phi^2)^2$ and that combination is unchanged by rotations.
\begin{align*}
D\phi^1&=\phi^2,\;\;\;\;\;\; D\phi^2=-\phi^1 & D\mathcal{L}&=0 &F^\mu&=0\\
J^\mu &=\pi^\mu_1D\phi^1+\pi^\mu_2 D\phi^2=(\partial^\mu\phi^1)\phi^2 -(\partial^\mu\phi^2)\phi^1\\
Q&=\int d^3xj^0=\int d^3x(\partial_0\phi^1\phi^2-\partial_0\phi^2 \phi^1)
\end{align*}

We can get some insight into this quantity by going to the case $g=0$ in which case $\phi^1$ and $\phi^2$ are both free fields and can be expanded in terms of creation and annihilation ops.
\[\phi^a(x)=\int \frac{d^3k}{(2\pi)^{3/2}\sqrt{2\omega_{\vec{k}}}} (a_{\vec{k}}^ae^{-ik\cdot x}+a_{\vec{k}}^{a\dagger}e^{ik\cdot x}) \]

Now we'll compute $Q$. Let's have faith in our formalism and assume that the terms with two creation ops or two annihilation ops go away. If they didn't $Q$ wouldn't be time independent since they are multiplied by $e^{\pm 2i\omega_{\vec{k}}t}$.
\[ Q=\int \frac{d^3k}{2\omega_{\vec{k}}} [a_{\vec{k}}^1 a_{\vec{k}}^{2\dagger}(-i\omega_{\vec{k}}-i\omega_{\vec{k}})+ 2i\omega_{\vec{k}}a_{\vec{k}}^2a_{\vec{k}}^{1\dagger}] \]

We don't have to worry about the order because $[ a_{\vec{k}}^1,a_{\vec{k}\,'}^{2\dagger}]=0$
\[Q=i\int d^3k[a_{\vec{k}}^{1\dagger}a_{\vec{k}}^2 -a_{\vec{k}}^{2\dagger}a_{\vec{k}}^1 ] \]

We are within reach of something intuitive. Define
\begin{align*}
b_{\vec{k}}&=\frac{a_{\vec{k}}^1+ia_{\vec{k}}^2}{\sqrt{2}} & b_{\vec{k}}^\dagger&=\frac{a_{\vec{k}}^{1\dagger}-ia_{\vec{k}}^{2\dagger}}{\sqrt{2}} 
\end{align*}

That's not the end of the definitions; we need the other combination to reconstruct $a_{\vec{k}}^1$, $a_{\vec{k}}^2$, $a_{\vec{k}}^{1\dagger}$, $a_{\vec{k}}^{2\dagger}$. So define,
\begin{align*}
c_{\vec{k}}&=\frac{a_{\vec{k}}^1-ia_{\vec{k}}^2}{\sqrt{2}} & c_{\vec{k}}^\dagger&=\frac{a_{\vec{k}}^{1\dagger} +ia_{\vec{k}}^{2\dagger}}{\sqrt{2}} 
\end{align*}

These linear combinations of operators are allowable\footnote{Exchanging the role of $b_{\vec{k}}$ and $c_{\vec{k}}$ is the way of making the signs come out conventionally with counterclockwise rotation.}. They are operators that create (or destroy) a superposition of states with particle 1 and particle 2. If one state is degenerate with another, and for a given $\vec{k}$, $|\vec{k},1\rangle$ is degenerate with $|\vec{k},2\rangle$, it is often convenient to work with linear combinations of these states as basis states. Because $b_{\vec{k}}^\dagger$ and $c_{\vec{k}}^\dagger$ create orthogonal states, $b_{\vec{k}}$ and $c_{\vec{k}}^\dagger$ commute with each other, as is easily checked. The useful thing about these linear combinations is that $Q$ has a simple expression.
\begin{align*}
i(a_{\vec{k}}^{1\dagger}a_{\vec{k}}^2-a_{\vec{k}}^{2\dagger} a_{\vec{k}}^1)&=i\frac{b_{\vec{k}}^\dagger+c_{\vec{k}}^\dagger}{\sqrt{2}}\cdot\frac{b_{\vec{k}}-c_{\vec{k}}}{\sqrt{2} i}+i\frac{b_{\vec{k}}^\dagger-c_{\vec{k}}^\dagger}{\sqrt{2} i }\cdot\frac{b_{\vec{k}}+c_{\vec{k}}}{\sqrt{2}}\\
&=\frac{1}{2}[2b_{\vec{k}}^\dagger b_{\vec{k}}+0b_{\vec{k}}^\dagger c_{\vec{k}}+0 c_{\vec{k}}^\dagger b_{\vec{k}}-2 c_{\vec{k}}^\dagger c_{\vec{k}}] = b_{\vec{k}}^\dagger b_{\vec{k}}-c_{\vec{k}}^\dagger c_{\vec{k}}
\end{align*}

That is, 
\[Q=\int d^3k(b_{\vec{k}}^\dagger b_{\vec{k}}-c_{\vec{k}}^\dagger c_{\vec{k}})= N_b-N_c \]

The $b$'s carry $Q$ charge +1, the $c$'s carry $Q$ charge -1. It's like particles and antiparticles; the $b$'s and $c$'s have the same mass and opposite charge. $Q=N_b-N_c$ is true as an operator equation. $b$ and $c$ type mesons are eigenstates of $Q$. 1 and 2 type mesons are not. An eigenstate of $N_a$ and $N_b$ is an eigenstate of $Q$, an eigenstate of $N_1$ and $N_2$ is not. By the way $H$ and $\vec{P}$ have familiar forms in terms of the $b$'s and $c$'s.
\begin{align*}
H&=\int d^3k \, \omega_{\vec{k}}(b_{\vec{k}}^\dagger b_{\vec{k}}+c_{\vec{k}}^\dagger c_{\vec{k}}) & \vec{P}&= \int d^3k \, \vec{k} (b_{\vec{k}}^\dagger b_{\vec{k}} +c_{\vec{k}}^\dagger c_{\vec{k}})
\end{align*}

Taking all these linear combinations suggests that we could have changed bases earlier in the calculation and simplified things.
\begin{align*}
\psi&=\frac{\phi^1+i\phi^2}{\sqrt{2}}\\
&=\int\frac{d^3k}{(2\pi)^{3/2}\sqrt{2\omega_{\vec{k}}}}[b_{\vec{k}} e^{-ik\cdot x}+c_{\vec{k}}^\dagger e^{ik\cdot x}]
\end{align*}

$\psi$ always diminishes $Q$ by 1 either by annihilating a $b$ type particle or by creating a $c$ type particle.
\[ [Q,\psi]=-\psi\;\;\;\text{ a $Q$ ``eigenfield''} \]

There is also the hermitian conjugate of $\psi$, $\psi^\dagger$,
\begin{align*}
\psi^\dagger&=\frac{\phi^1-i\phi^2}{\sqrt{2}}\\
&=\int\frac{d^3k}{(2\pi)^{3/2}\sqrt{2\omega_{\vec{k}}}}[c_{\vec{k}} e^{-ik\cdot x}+b_{\vec{k}}^\dagger e^{ik\cdot x}]
\end{align*}

($\psi^\dagger$ will be denoted $\psi^*$ in the classical limit. A quantum field $\psi^*$ will be understood to be $\psi^\dagger$.) $\psi^\dagger$ always increases $Q$ by 1. $[Q,\psi^\dagger]=+\psi^\dagger$.

$\psi$ and $\psi^\dagger$ have neat commutation relations with $Q$. $\phi^1$ and $\phi^2$ have messy commutation relations with $Q$.
\begin{align*}
[Q,\phi^1]&=\frac{1}{\sqrt{2}}[Q,\psi+\psi^\dagger]=\frac{1}{\sqrt{2}} (-\psi+\psi^\dagger)\\
\begin{split}
&=-i\phi^2\\
[Q,\phi^2]&=i\phi^1
\end{split}\;\;\;\substack{\text{In agreement with a formula which}\\ \text{is usually true $[Q,\phi^a]=-iD\phi^a$}}
\end{align*}

Under our transformation $\psi\rightarrow e^{-i\lambda}\psi$, $\psi^*\rightarrow e^{i\lambda}\psi^*$. This is called a $U(1)$ or phase transformation. It is equivalent to $SO(2)$.

{\scriptsize Digression; we'll get back to symmetries.}

As quantum fields, $\psi$ and $\psi^\dagger$ are as nice as $\phi^1$ and $\phi^2$. They obey
\begin{align*}
(\square+\mu^2)\psi&=0 & (\square+\mu^2)\psi^\dagger&=0
\end{align*}

They obey simple equal time commutation relations:
\begin{align*}
[\psi(\vec{x},t),\psi(\vec{y},t)]&=0=[\psi^\dagger(\vec{x},t),\psi^\dagger(\vec{y},t)]=[\psi(\vec{x},t),\psi^\dagger(\vec{y},t)]\\
[\psi(\vec{x},t),\dot{\psi}(\vec{y},t)]&=0=[\psi^\dagger(\vec{x},t),\dot{\psi}^\dagger(\vec{y},t)]\\
[\psi(\vec{x},t),\dot{\psi}^\dagger(\vec{y},t)]&=i\delta^{(3)}(\vec{x}-\vec{y})=[\psi^\dagger(\vec{x},t),\dot{\psi}(\vec{y},t)]
\end{align*}

These equations can be obtained by doing something completely idiotic.
Back to the classical Lagrangian which is
\[ \mathcal{L}=\partial_\mu\psi^*\partial^\mu\psi-\mu^2\psi^*\psi\;\;
\; \left(\substack{\text{no }\frac{1}{2}}\right) \]

Imagine a person who once knew a lot of quantum field theory, but has suffered brain damage, is going to canonically quantize this theory. He has forgotten that $*$ stands for complex conjugate, and he is going to treat $\psi$ and $\psi^*$ as if they were independent fields. Here he goes
\begin{align*}
\pi^\mu_\psi&=\frac{\partial\mathcal{L}}{\partial\partial_\mu\psi}=\partial^\mu\psi^* & \pi^\mu_{\psi^*}&=\partial^\mu\psi 
\end{align*}

The Euler-Lagrange equations he gets are
\[ \partial_\mu\pi^\mu_\psi=\frac{\partial\mathcal{L}}{\partial\psi}\;\;\;\text{ i.e.~}\;\;\;\square\psi^*=-\mu^2\psi^* \]

The idiot got it right. He also gets $\square\psi=-\mu^2\psi$. How about that!? Now he says ``I'm going to deduce the canonical commutation relations'':
\[i\delta^{(3)}(\vec{x}-\vec{y})=[\psi(\vec{x},t),\pi^0_\psi(\vec{y},t)]=[\psi(\vec{x},t),\partial_0\psi^*(\vec{y},t)] \]

By God, he gets that right, too. How can you work with complex fields, which can't be varied independently, treat them as if they can be and nevertheless get the right answers?

We'll demonstrate that this works for the equations of motion. It works very generally. You can demonstrate that it works for the equal time commutation relations.

Suppose I have some action $S(\psi,\psi^*)$. The equations of motion come from
\[0=\delta S=\int d^4x(A\delta\psi+A^*\delta\psi^*) \]

{\sc Naive Approach:} Treating the variations in $\delta\psi$ and $\delta\psi^*$ as independent we get 
\[ A=0 \;\;\;\text{ and }\;\;\;A^*=0 \]

{\sc Sharp Approach:} However $\delta\psi$ is not independent of $\psi^*$. We can use the fact that $\psi$ and $\psi^*$ are allowed to be complex. We can make purely real variations in $\psi$ so that $\delta \psi=\delta\psi^*$ and we can make purely imaginary variations in $\psi$ so that $\delta\psi=-\delta\psi^*$. From the real variation we deduce 
\[ A+A^*=0 \]

\noindent 
and from the imaginary variation we deduce
\[ A-A^*=0\]

\noindent 
which implies $A=A^*=0$.

For the ETCR write $\psi=\frac{\psi_r+i\psi_i}{\sqrt{2}}$, $\psi^\dagger=\frac{\psi_r-i\psi_i}{\sqrt{2}}$. Nothing tricky or slick.

Back to internal symmetries. For our second example, take
\[\mathcal{L}=\frac{1}{2}\sum_{a=1}^n(\partial_\mu\phi^a\partial^\mu \phi^a-\mu^2\phi^a\phi^a)-g\left(\sum_{a=1}^n(\phi^a)^2\right)^2\]

This is the same as our first example except we have $n$ fields instead of 2. Just as in the first example the Lagrangian was invariant under rotations mixing up $\phi^1$ and $\phi^2$, this Lagrangian is invariant under rotations mixing up $\phi^1$, $\dots$, $\phi^n$, because it only depends on $(\phi^1)^2+\cdots+(\phi^n)^2$. The notations are,
\[\phi^a\rightarrow\underset{\substack{n\times n \text{ rotation matrix}}}{\sum_b\underbrace{R^a_b}\phi^b} \]

There are $\frac{n(n-1)}{2}$ independent planes in $n$ dimensions and we can rotate in each of them, so there are $\frac{n(n-1)}{2}$ conserved currents and associated charges. This example is quite different from the first one because the various rotations don't in general commute. (They all commuted in the first one by virtue of the fact that there was only one.) We don't expect the various charges to commute. If they did they would generate symmetries that commute. Anyway, for any single rotation axis, the symmetry is just like the one we had in the first example, so we can read off the current,
\[ J_\mu^{[a,b]}=\partial_\mu\phi^a\phi^b-\partial_\mu\phi^b\phi^a \]

You can't find combinations of the fields that have simple commutation relations with all the $Q^{[a,b]}$'s. For $n=3$ you can choose the fields to have a simple commutation relation with one charge, say $Q^{[1,2]}$. This is just like isospin, which is a symmetry generated by $I_1$, $I_2$ and $I_3$. You can only choose the particle states $\pi^+$, $\pi^0$ and $\pi^-$ to be eigenstates of one of them, $I_3$.
}{
 \sektion{7}{October 14}
\descriptionseven
\section*{Lorentz transformation properties of conserved quantities}
We've worked with three currents, $J^\mu$, the current of meson number, $T^{\mu\nu}$, the current of the $\nu$th component of momentum, and $M^{\mu\nu\lambda}$, the current of the $[\nu\lambda]$ component of angular momentum. We integrated the zeroth component of each of these currents to obtain $Q$, $P^\nu$ and $J^{\nu\lambda}$ respectively. These look like tensors with one less index, but do they have the right Lorentz transformation properties?

We'll prove that $P^\nu=\int d^3xT^{0\nu}$ is indeed a Lorentz vector given that $T^{\mu\nu}$ is a two index tensor and that $T^{\mu\nu}$ is conserved, $\partial_\mu T^{\mu\nu}=0$. The generalization to currents with more indices or 1 index will be clear. We could do this proof in the classical theory or the quantum theory. We'll choose the latter because we need the practice. The assumption that $T^{\mu\nu}$ is an operator in quantum field theory that transforms as a two-index tensor is phrased mathematically as follows:

Given $U(\Lambda)$ the unitary operator that effects Lorentz transformations in the theory
\[ U(\Lambda)^\dagger T^{\mu\nu}(x)U(\Lambda)=\Lambda^\mu_\sigma \Lambda^\nu_\tau T^{\sigma\tau}(\Lambda^{-1}x) \]

Now we'll try to show that
\[ P^\nu=\int d^3x \, T^{0\nu}(\vec{x},0) \]

\noindent 
is a Lorentz vector. Introduce $n=(1,0,0,0)$, a unit vector pointing in the time direction. Then we can write $P^\nu$ in a way that makes its Lorentz transformation properties clearer.
\[ P^\nu=\int d^4x \, n_\mu T^{\mu\nu}\delta(n\cdot x) \]

Perform a Lorentz transformation on $P^\nu$
\begin{align*}
U(\Lambda)^\dagger P^\nu U(\Lambda)&=\int d^4x \, n_\mu U(\Lambda)^\dagger T^{\mu\nu}(x)U(\Lambda)\delta(n\cdot x) \\
&=\int d^4x \,  n_\mu \Lambda^\mu_\sigma\Lambda^\nu_\tau T^{\sigma\tau}(\Lambda^{-1}x)\delta(n\cdot x) 
\end{align*}

Change integration variables $x'=\Lambda^{-1}x$ and define $n'=\Lambda^{-1}n$
\begin{align*}
U(\Lambda)^\dagger P^\nu U(\Lambda)&=\int d^4x' \, \Lambda_{\mu\rho} n'^\rho \Lambda^\mu_\sigma \Lambda^\nu_\tau T^{\sigma\tau}(x') \delta(n'\cdot x') \\
&=\int d^4x' \, n'_\rho\Lambda^\nu_\tau T^{\rho\tau}(x')\delta(n'\cdot x')\;\;\;\substack{\text{Using $\Lambda_{\mu\rho}\Lambda^\mu_\sigma= g_{\rho\sigma}$}} \\
&=\Lambda^\nu_\tau \int d^4x \, n'_\rho T^{\rho\tau}(x)\delta(n'\cdot x)
\end{align*}

In the last expression we've dropped the prime on the variable of integration. The only difference between what we have and what we would like to get 
\[ U(\Lambda)^\dagger P^\nu U(\Lambda)=\Lambda^\nu_\tau P^\tau =\Lambda^\nu_\tau \int d^4x \, n_\rho T^{\rho\tau}\delta(n\cdot x) \]

\noindent 
is that $n$ has been redefined. The surface of integration is now $t'=0$, and we take the component $n'_\mu T^{\mu\nu}$ in the $t'$ direction. Our active transformation has had the exact same effect as if we had made a passive transformation changing coordinates to $x'=\Lambda^{-1}x$. It's the same old story: $\underbrace{\text{alias}}_{\text{passive}}$ (another name) versus $\underbrace{\text{alibi}}_{\text{active}}$ (another place).
\begin{center}
\includegraphics[width=8cm]{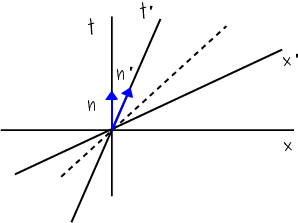}
\end{center}

You can think of this as Lorentz transforming the field or Lorentz transforming the surface.

This doesn't produce a vector $P^\nu$ for any tensor $T^{\mu\nu}$. For an arbitrary tensor, $P^\nu$ is not even independent of what time you compute at, let alone changing the tilt of the surface. We need to use that $T^{\mu\nu}$ is conserved. More or less, that the current that flows through the surface $t'=0$ is the same as the current that flows through the surface $t=0$.

Note that $n_\mu\delta(n\cdot x)=\partial_\mu\theta(n\cdot x)$, so that if I call the Lorentz transform of $P^\nu$, $P'^\nu$, what we are trying to show is 
\begin{align*}
0&=P^\nu-\Lambda^{-1\nu}_{\;\;\;\;\;\; \sigma} P'^\sigma\\
&=\int d^4x[\partial_\mu\theta(n\cdot x)-\partial_\mu\theta(n'\cdot x)] T^{\mu\nu}(x) \\
&=\int d^4x\partial_\mu [\theta(n\cdot x)-\theta(n'\cdot x)] T^{\mu\nu}(x) 
\end{align*}

This integral over all spacetime is a total divergence. In the far future or the far past $\theta(n\cdot x)=\theta(n'\cdot x)$ so there are no surface terms there, and as usual we won't worry about surface terms at spatial infinity.

\section*{{\sc Discrete Symmetries}}
A discrete symmetry is a transformation $q^a(t)\rightarrow q^{a\prime}(t)$ that leaves the Lagrangian, $L$, unchanged $L\rightarrow L$. This could be parity, this could be rotation by $\pi$ about the $z$ axis. (It does not include time reversal, wait\footnote{See Dirac \uline{Principles of QM}, pp.~103ff}.)

Since \uline{all} the properties of the theory are derived from the Lagrangian, (the canonical commutation relations, the inner product, the Hamiltonian all come from $L$) and since the Lagrangian is unchanged, we expect that there is a unitary operator effecting the transformation
\begin{align*}
U^\dagger q^a(t) U&=q^{a\prime}(t) & U^\dagger HU=H 
\end{align*}

The discrete symmetry could be an element of a continuous symmetry group. It could be rotation by 20$^\circ$. There is no conserved quantity associated with a discrete symmetry however. What is special about the continuous symmetry is that there is a parameter, and that you have a unitary operator for each value of the parameter, $\theta$, satisfying 
\[U(\theta)^\dagger HU(\theta)=H\]

You can differentiate this with respect to the parameter to find that
\begin{align*}
[I,H]&=0 & I&\equiv-i\left.\frac{dU}{d\theta}\right|_{\theta=0}\equiv -iDU 
\end{align*}

There is nothing analogous for discrete symmetries.

\section*{Examples of internal symmetries (there is not a general theory, but we'll do prototypical examples)}

\subsubsection*{Example (1). The transformation is $\phi(x)\rightarrow-\phi(x)$ at every space-time point.}

This is a symmetry for any Lagrangian with only even powers of $\phi$, in particular,
\[\mathcal{L}=\frac{1}{2}(\partial_\mu\phi)^2-\frac{1}{2}\mu^2\phi^2 -\lambda\phi^4\rightarrow\mathcal{L} \]

There should be a unitary operator effecting the transformation
\[ \phi\rightarrow U^\dagger \phi U=-\phi \]

For $\lambda=0$, we will actually be able to construct $U$, that is give its action on the creation and annihilation operators, and thus its action on the basis states.

Since $\phi$ is a linear function of $a_{\vec{k}}$ and $a_{\vec{k}}^\dagger$ it must be that 
\[ U^\dagger a_{\vec{k}}U=-a_{\vec{k}} \]

\noindent 
and the h.c.~equation 
\[ U^\dagger a_{\vec{k}}^\dagger U=-a_{\vec{k}}^\dagger \]

We'll also make a choice in the phase of $U$ by specifying
\[ U|0\rangle=|0\rangle \]

We can determine the action of $U$ on the basis states
\begin{align*}
U|k_1,\dots,k_n\rangle &= Ua_{\vec{k}_1}^\dagger a_{\vec{k}_2}^\dagger \cdots a_{\vec{k}_n}^\dagger|0\rangle \\
&= Ua_{\vec{k}_1}^\dagger U^\dagger Ua_{\vec{k}_2}^\dagger U^\dagger \cdots U a_{\vec{k}_n}^\dagger U^\dagger U|0\rangle \\
&= (-1)^na_{\vec{k}_1}^\dagger a_{\vec{k}_2}^\dagger \cdots a_{\vec{k}_n}^\dagger|0\rangle =(-1)^n |k_1,\dots,k_n\rangle 
\end{align*}

As an operator statement we see that 
\[ U=(-1)^N(=e^{i\pi N}\text{ if you prefer})\;\;\;\substack{(N=\text{meson number}\\ =\int d^3k a_{\vec{k}}^\dagger a_{\vec{k}} )}\]

Suppose we could construct this operator when $\lambda\neq0$. The existence of this unitary operator tells you that you'll never see 2 mesons scatter into 43 mesons or any odd number of mesons. Mesons are always produced in pairs. More formally, it must be that $0=\langle n|S|m\rangle$ where $|n\rangle$ is a state with $n$ mesons, $|m\rangle$ is a state with $m$ mesons and $S$ is the scattering matrix made from the Hamiltonian, whenever $n+m$ is odd. For if $n+m$ is odd
\begin{align*}
\langle n|S|m\rangle &=\langle n|U^\dagger USU^\dagger U|m\rangle = \langle n|U^\dagger\!\!\!\!\!\!\!\!\!\!\underbrace{(USU^\dagger)}_{S\text{ because
}UHU^\dagger=H}\!\!\!\!\!\!\!\!\!U|m\rangle \\
&=(-1)^{n+m}\langle n|S|m\rangle =-\langle n|S|m\rangle 
\end{align*}

\subsubsection*{Our second example of a discrete internal symmetry will turn out to be \sc Charge Conjugation.}

Recall that
\[\mathcal{L}=\sum_{a=1}^2\left(\frac{1}{2}(\partial_\mu\phi^a)^2 -\frac{1}{2}\mu^2(\phi^a)^2\right)-\lambda[(\phi^1)^2+(\phi^2)^2]^2 \]

\noindent 
had an $SO(2)$ symmetry. In fact it has an $O(2)$ symmetry, rotations and rotations with reflections in the $\phi^1 \phi^2$ plane. It is invariant under proper and improper rotations. We'll take one standard improper rotation
\begin{align*}
\phi^1&\rightarrow \phi^1 & \phi^2&\rightarrow-\phi^2
\end{align*}

Any other one can be obtained by composing this one with an element of the internal symmetry group $SO(2)$.

It is easy to write down the unitary operator in the case $\lambda=0$. Then we have two independent free scalar field theories
\[ U_C=(-1)^{N_2} \]

As mathematicians are fond of saying, we have reduced it to the previous case.

The action of $U$ is especially nice if we put it in terms of the fields $\psi$ and $\psi^\dagger$ already introduced
\begin{align*}
\psi=\frac{\phi^1+i\phi^2}{\sqrt{2}}&\rightarrow U_C^\dagger \psi U_C = \frac{\phi^1-i\phi^2}{\sqrt{2}} =\psi^\dagger \\
\psi^\dagger&\rightarrow\psi
\end{align*}

For this reason, this is sometimes called a conjugation symmetry. Because $U_C^\dagger QU_C=-Q$ it is also called charge\footnote{the charge from the $SO(2)$ symmetry} conjugation or particle-anti-particle conjugation. From the action on $\psi$ and $\psi^\dagger$ we see
\[ U_C^\dagger b_{\vec{k}} U_C=c_{\vec{k}}\;\;\;\text{ and }\;\;\; U_C^\dagger c_{\vec{k}}U_C=b_{\vec{k}} \]

\noindent 
and the equations obtained from these by hermitian conjugation. $U_C$ is unitary and hermitian:
\[U_C^2=1=U_CU_C^\dagger\Rightarrow U_C=U_C^\dagger\]

We could continue the discussion of discrete internal symmetries, but it is boring. You could write down a Lagrangian with four fields that is invariant under rotation in four-dimensional space and under permutations of any of the four fields. You could write down a theory with the icosahedral group.

\subsubsection*{\sc Parity Transformations}
Any transformation that takes 
\[ \phi^a(\vec{x},t)\rightarrow\sum_b M^a_b\phi^b(-\vec{x},t) \]

\noindent 
we'll call a parity transformation. (We have used the fact that we live in an odd number of dimensions (3) and thus that $\vec{x}\rightarrow-\vec{x}$ is an improper rotation, in our definition. If we lived in two space dimensions we could do a similar thing with only $x^2\rightarrow-x^2$.) A parity transformation transforms each fundamental observable at the point $\vec{x}$ into some linear combination of fundamental observables at the point $-\vec{x}$.

Usually parity takes $\mathcal{L}\overset{\substack{\text{The usual}\\ \text{confusing notation}}}{\overbrace{(\vec{x},t)}\rightarrow \mathcal{L}(-\vec{x},t)}$, when it is a symmetry, but all we really demand is that parity takes 
\[L(t)\rightarrow L(t)\;\;\;\text{ as usual.}\]
\newcounter{07count1}
\begin{list}{\sc Example (\arabic{07count1}) }
{\usecounter{07count1}
\setlength{\rightmargin}{\leftmargin}}
\item Parity is a symmetry of
\begin{align*}
\mathcal{L}&=\frac{1}{2}(\partial_\mu\phi)^2-\frac{\mu^2}{2}\phi^2 &
\begin{split}
P:\phi(\vec{x},t) &\rightarrow\phi(-\vec{x},t)\\
L&\rightarrow L
\end{split}\;\;\;(M=1)
\end{align*}

From
\begin{align*}
\phi(\vec{x},t)&=\int \frac{d^3k}{(2\pi)^{3/2} \sqrt{2\omega_{\vec{k}}}} [a_{\vec{k}} e^{i\vec{k}\cdot \vec{x}} e^{-i\omega_{\vec{k}}t}+ a_{\vec{k}}^{\dagger}e^{-i\vec{k}\cdot \vec{x}}e^{i\omega_{\vec{k}}t}]\\
\text{and }\;\;\;U_P^\dagger\phi(\vec{x},t)U_P&=\phi(-\vec{x},t)
\end{align*}

\noindent 
we can see that the action on the creation and annihilation operators must be
\[ U_P^\dagger\begin{Bmatrix} a_{\vec{k}}\\ a_{\vec{k}}^\dagger \end{Bmatrix}  U_p=\begin{Bmatrix} a_{-\vec{k}}\\ a_{-\vec{k}}^\dagger \end{Bmatrix}  \]

\noindent 
and on the basis states
\[ U_P|\vec{k}_1,\dots,\vec{k}_n\rangle=|-\vec{k}_1,\dots,-\vec{k}_n \rangle \]

But there is a second possibility for parity
\begin{align*}
P':\phi(\vec{x},t)&\rightarrow-\phi(-\vec{x},t) &(M=-1)\\
L&\rightarrow L
\end{align*}

\uline{Whenever} there is an internal symmetry in a theory I can multiply one definition of parity by an element of that symmetry group (discrete or continuous) and get another definition of parity. In the case at hand the unitary operator $U_{P'}$ is given by
\[ U_{P'}=(-1)^NU_P\;\;\;\text{ or }\;\;\; U_{P'}|\vec{k}_1,\dots,\vec{k}_n\rangle=(-1)^n|-\vec{k}_1,\dots, -\vec{k}_n\rangle \]

Sometimes people distinguish between a theory with invariance under $P:\phi(\vec{x},t)$ $\rightarrow\phi(-\vec{x},t)$ and a theory with invariance under $P':\phi(\vec{x},t)\rightarrow-\phi(-\vec{x},t)$, by calling the first the theory of a scalar meson and the second a theory of a pseudo scalar meson. Our theory is invariant under both; it's the old plate of cookies problem again. The theory has a set of invariances. As long as you are in agreement about the total set of invariances of the theory, you shouldn't waste time arguing about what you'll call each one. This is why the conventions on the parity of
some particles is arbitrary (relative parity). If $-\lambda\phi^4$ is added to $\mathcal{L}$ both $P$ and $P'$ are still invariances of $L$. With a $\phi^3$ interaction $P$ is a symmetry, $P'$ isn't. All the physicists in the world would agree that this is a scalar meson. One of the cookies has been poisoned.

\item This awful Lagrangian has been cooked up to illustrate a point. After this lecture you won't see anything this bad again.
\[\mathcal{L}=\sum_{a=1}^4\left[\frac{1}{2}(\partial_\mu\phi^a)^2-\frac{\mu_a^2}{2}(\phi^a)^2\right]-g\epsilon^{\mu\nu\lambda\sigma} \partial_\mu\phi^1\partial_\nu\phi^2\partial_\lambda\phi^3 \partial_\sigma\phi^4\]

All four meson masses are different. The new interaction involving the totally antisymmetric tensor in four indices is invariant under L.T.~for the same reason that $\vec{a}\cdot(\vec{b}\times\vec{c})$ is invariant under proper rotations. ($\vec{a}\cdot(\vec{b}\times \vec{c})$ is multiplied by $\det R$ under a rotation). Because $\epsilon^{\mu\nu\lambda\sigma}$ is nonzero only if one of its indices is timelike, the other three spacelike, three of the derivatives are on space coordinates. We get three minus signs under parity. An odd number of mesons are going to have to be pseudoscalar to get a net even number of minus signs. Because any $\phi^a\rightarrow-\phi^a$ is an internal symmetry of the free Lagrangian, it doesn't matter which one you choose or which three you choose to be pseudoscalar.

\item Take the perverse Lagrangian of the last example by and make it worse by adding 
\[\sum_{a=1}^4(\phi^a)^3 \]

Now there is no definition of parity that gives a symmetry. This theory violates parity.

\item Sometimes people say that because the product of two reflections is 1, the square of parity is 1. This example is cooked up to show that a theory with parity can have indeed $U_P^2\neq 1$. Indeed $U_P$ cannot be chosen to satisfy $U_P^2=1$.
\begin{multline*}
\mathcal{L}=\sum_{a=1}^4\left[\frac{1}{2}(\partial_\mu\phi^a)^2-\frac{\mu^2}{2}(\phi^a)^2\right]+\partial_\mu\psi^*\partial^\mu\psi -m^2\psi^*\psi\\
-h\sum_{a=1}^4(\phi^a)^3-g\epsilon_{\mu\nu\lambda\sigma} \partial^\mu\phi^1\partial^\nu\phi^2\partial^\lambda\phi^3 \partial^\sigma\phi^4[\psi^2+\psi^{*2}]
\end{multline*}

The transformation of the $\phi^a$'s must be $\phi^a(\vec{x},t) \rightarrow+\phi^a(-\vec{x},t)$. The only way to make the last term parity invariant is for $\psi$ to transform as 
\begin{align*}
\psi&\rightarrow\pm i\psi & \psi^*&\rightarrow\mp i\psi^*
\end{align*}

In either case $U_P^2\neq1$, $U_P^\dagger U_P^\dagger \psi U_PU_P=-\psi$. Fortunately, nothing like this occurs in nature (as far as we know). If it did, and if parity were a symmetry (or an approximate symmetry) of the world we would have a name for fields transforming like $\psi$, a ``semi-pseudo-scalar''.
\end{list}

\subsubsection*{\sc Time Reversal}
First a famous example from classical particle mechanics, a particle moving in a potential
\begin{align*}
L&=\frac{1}{2}m\dot{q}^2-V(q) & T:q(t)&\rightarrow q(-t)
\end{align*}
 
$T$ is not a discrete symmetry the way we have defined it
\[ T:L(t)\rightarrow L(-t) \;\;\;\substack{\text{in the usual confusing } \\ \text{notation where $L(t)$ refers} \\ \text{to the time dependence} \\ \text{through the coordinates}} \]

Nevertheless $T$ does take one solution of the equations of motion into another. You might still hope there is a unitary operator that does the job in the quantum theory.
\[U_T^\dagger q(t)U_T=q(-t)\]

There are two paradoxes I'll give to show this can't happen.
\begin{description}
\item{\sc 1st Paradox } Differentiate $U_T^\dagger q(t)U_T=q(-t)$ with respect to $t$. Because $p(t)\propto \dot{q}(t)$
\[ U_T^\dagger p(t)U_T=-p(t)\]

Consider $U_T^\dagger[p,q]U_T=-i$. From the two relations we have just obtained we also have
\[ U_T^\dagger[p(t),q(t)]U_T=-[p(-t),q(-t)]=-i\;\;\;\substack{\text{Particularly poignant}\\ \text{at $t=0$}} \]

Looks like we would have to give up the canonical commutation relations to implement time reversal. If that isn't enough to make you abandon the idea of a unitary time reversal operator I'll continue to the

\item{\sc 2nd Paradox } Roughly, $U_T$ should reverse time evolution, i.e.~$U_T^\dagger e^{-iHt}U_T=e^{iHt}$. I can prove this. For any operator $\mathcal{O}(t)$
\[ \mathcal{O}(t)=e^{iHt}\mathcal{O}e^{-iHt} \]

Apply $U_T^\dagger$ the unitary transformation to both sides to obtain
\begin{align*}
\mathcal{O}(-t)&=U_T^\dagger e^{+iHt}U_T\mathcal{O}U_T^\dagger e^{-iHt}U_T & \mathcal{O}&=\mathcal{O}(0) \\
\text{but }\mathcal{O}(-t)&= e^{-iHt}U_T\mathcal{O}e^{iHt} & \substack{\text{let }V\equiv U_T^\dagger e^{-iHt}U_T}
\end{align*}

I'd like to show $V=e^{iHt}$. What we have is $e^{-iHt}\mathcal{O} e^{iHt}=V^{-1}\mathcal{O}V$ which implies $Ve^{-iHt}\mathcal{O}= \mathcal{O}Ve^{-iHt}$. $Ve^{-iHt}$ commutes with any operator $\mathcal{O}$. $Ve^{-iHt}=1$. Now that I've proved
\[ U_T^\dagger e^{-iHt}U_T=e^{iHt} \]

Take $\left.\frac{d}{dt}\right|_{t=0}$ of this relation, 
\[U_T^\dagger(-\cancel{i}H)U_T=\cancel{i}H,\]

\noindent 
canceling the $i$'s, we see that $H$ is unitarily related to $-H$. The spectrum of $H$ cannot be bounded below, because (the spectrum of unitarily related operators is the same and) the spectrum of $H$ is not bounded above. AUGGH!
\end{description}

A unitary time reversal operator is an object that makes no sense whatsoever. The answer is that time reversal is implemented by an antiunitary operator. Antiunitary operators are antilinear. Dirac notation is designed to automate the handling of linear operators, so for a while we'll use some more cumbersome notation that does not automate the handling of linear operators.

$\begin{matrix}
\text{Let }& a,b &\text{denote states,}\\
& \alpha,\beta &\text{denote complex numbers and}\\
& A,B &\text{denote operators.}\\
&(a,b)&\text{is the inner product of two states.}
\end{matrix}$

A \uline{unitary operator} is an invertible operator, $U$, satisfying
\begin{align*}
(Ua,Ub)&=(a,b) \;\;\;\text{ for all }\;\;\;a,b & \substack{\text{unitarity}}
\end{align*}

This is enough of an assumption to show $U$ is linear (see related proof below). The simplest unitary operator is 1.
\begin{align*}
 U(\alpha a+\beta b)&= \alpha Ua+\beta Ub & \substack{\text{linearity}}
\end{align*}

The adjoint of a linear operator $A$ is denoted $A^\dagger$ and is the operator defined by (this definition is not consistent if $A$ is not linear)
\[(a,A^\dagger b)=(Aa,b)\;\;\;\text{ for all }\;\;\; a,b \]

I'll show that $U^\dagger=U^{-1}$ (which is sometimes given as the definition of unitarity).
\[(a,U^{-1}b)\overbrace{=}^{\substack{\text{unitarity}\\ \text{of $U$}}} (Ua,UU^{-1}b)=(Ua,b) \]

A transformation of the states, $a\rightarrow Ua$, can also be thought of as a transformation of the operators in the theory 
\[(a,Ab)\rightarrow(Ua,AUb)=(a,U^\dagger AUb)\;\;\;\text{ can alternatively be thought of as $A\rightarrow U^\dagger AU$.}\]

An antiunitary operator is an invertible operator, $\Omega$, (this is a notational gem, an upside down $U$) satisfying
\begin{align*}
(\Omega a,\Omega b)&=(b,a)\;\;\; \text{ for all } \;\;\;a,b & \substack{\text{antiunitarity}}
\end{align*}

We can immediately make a little table showing the result of taking products of unitary and antiunitary operators. (The product of a unitary operator with an antiunitary operator is an antiunitary operator, etc.) $\begin{matrix} & \begin{matrix} U \hfill & \Omega\end{matrix}\\ \begin{matrix} U \\ \Omega \end{matrix} & \begin{array}{|c|c|} \hline U & \Omega \\ \hline \Omega & U \\ \hline \end{array} \end{matrix}$

We can prove (this is the related proof referred to above) that any operator (not necessarily invertible) satisfying the antiunitarity condition is antilinear.
\begin{align*}
(\Omega a,\Omega b)&=(b,a)\Rightarrow \Omega(\alpha a+\beta
b)=\alpha^*\Omega a+\beta^*\Omega b &\substack{\text{antilinearity}}
\end{align*}

Consider $(\Omega(\alpha a+\beta b)-\alpha^*\Omega a-\beta^*\Omega b, \Omega(\alpha a+\beta b)-\alpha^*\Omega a-\beta^*\Omega b)$. (You ask
why?!) This is the inner product of $\Omega(\alpha a+\beta b)-\alpha^*\Omega a-\beta^*\Omega b$ with itself. If this is zero, the fact that the inner product is positive definite implies that $\Omega(\alpha a+\beta b)-\alpha^*\Omega a-\beta^*\Omega b=0$. The result we want$!$ Indeed, it is simply a matter of expanding this inner product out into its 9 terms, applying the antiunitarity condition to each term, and then expand the 5 terms containing $\alpha a+\beta b$ some more to show this is zero. (The analogous proof for operators satisfying the unitarity condition also only uses properties of the inner product and is even easier.)

The simplest antiunitary operator is complex conjugation, $K$. For the elements of some basis, $b_i$, $Kb_i=b_i$ and on any linear combination 
\[K(\sum_i\alpha_ib_i)=\sum_i\alpha_i^*b_i \]

For consistency $(b_i,b_j)$ must be real. This is the familiar complex conjugation of nonrelativistic quantum mechanics of position space wave functions. The basis is a complete set of real wave functions.

A useful fact (especially conceptually) is that any antiunitary operator, $\Omega$, is equal to $UK$ for some unitary $U$. Proof by construction: take $U=\Omega K$.

In a more limited sense, the transformation of the states by an antiunitary operator $\Omega$, $a\rightarrow \Omega a$, can also be thought of as a transformation of the operators in the theory. Consider the expectation value of a Hermitian operator (observable) in the state $a$. It transforms as
\begin{align*}
(a,Aa)\rightarrow(\Omega a,A\Omega a)&=(A\Omega a,\Omega a) & \substack{\text{(hermiticity)}} \\
&= (\Omega\Omega^{-1}A\Omega a,\Omega a) &\substack{\text{(invertibility)}} \\
&= (a,\Omega^{-1}A\Omega a) & \substack{\text{(antiunitarity)}}
\end{align*}

This transformation can alternatively be thought of as
\[ A\rightarrow\Omega^{-1} A\Omega \]

We don't write $\Omega^\dagger A\Omega$ because adjoint is not even defined for antilinear ops.

50 years ago [1931], Eugene Wigner proved a beautiful theorem telling us why unitary and antiunitary operators are important in QM. He showed that (up to phases) they are the only operators that preserve probabilities. It is not necessary to preserve inner products; they aren't measurable. It is the probabilities that are measurable. Look in the appendix of his book on group theory.

Given that $F(a)$ ($F:\mathcal{H}\rightarrow\mathcal{H}$) is continuous and for any $a$ and $b$
\begin{align*}
|(F(a),F(b))|^2&=|(a,b)|^2\\
\text{then }\;\;\;F(a)&=e^{i\phi(a)} \times \left\{\substack{\text{a unitary op}\\ \text{or an}\\ \text{antiunitary op}}\right. & (\phi:\mathcal{H}\rightarrow\mathbb{R})
\end{align*}

Time reversal is not a unitary operator. Time reversal is antiunitary.

It is easy to see now how the two paradoxes are avoided.
\[\Omega_T^{-1}i\Omega_T=-i\]

You can't cancel the $i$'s as we did in paradox 2.
\[\Omega_T^{-1}(-iH)\Omega_T=iH\Rightarrow \Omega_T^{-1}H\Omega_T=H \]

We can explicitly construct the time reversal operator in free field theory. The simpler thing to look at in a relativistic theory is actually $PT$. Let's find
\[\Omega_{PT}\;\;\;\text{ such that }\;\;\;\Omega_{PT}^{-1}\phi(x) \Omega_{PT}=\phi(-x) \]

The simplest candidate is just complex conjugation, in the momentum state basis. That is $\Omega_{PT}$ does nothing, \uline{absolutely nothing} to $a_{\vec{k}}$ and $a_{\vec{k}}^\dagger$
\[\Omega_{PT}^{-1}a_{\vec{k}}\Omega_{PT} = a_{\vec{k}}\;\;\;\text{ and }\;\;\;\Omega_{PT}^{-1}a_{\vec{k}}^\dagger\Omega_{PT} = a_{\vec{k}}^\dagger\]

Furthermore we'll take $\Omega_{PT}|0\rangle=|0\rangle$ and it follows that
\[\Omega_{PT}|\vec{k}_1,\dots,\vec{k}_n\rangle=|\vec{k}_1,\dots, \vec{k}_n\rangle \]

\noindent 
again nothing, they just lie there. What does this operator do to $\phi(x)$?
\[\phi(x)=\int \frac{d^3k}{(2\pi)^{3/2} \sqrt{2\omega_{\vec{k}}}} [a_{\vec{k}} e^{-ik\cdot x} + a_{\vec{k}}^{\dagger}e^{ik\cdot x}] \]

Apply $\Omega_{PT}$ to $\phi(x)$. It does nothing to $\frac{1}{(2\pi)^{3/2}\sqrt{2\omega_{\vec{k}}}}$, it does nothing to $a_{\vec{k}}$, and nothing to $a_{\vec{k}}^\dagger$.

But what about that $i$ up in the exponential. It turns that into $-i$!
\begin{align*}
\Omega_{PT}^{-1}\phi(x)\Omega_{PT}&=\int \frac{d^3k}{(2\pi)^{3/2} \sqrt{2\omega_{\vec{k}}}} [a_{\vec{k}} e^{ik\cdot x} + a_{\vec{k}}^{\dagger}e^{-ik\cdot x}] \\
&=\phi(-x)\;\;\;\text{\large !}
\end{align*}

$PT$ does nothing to momentum states. That is expected. Parity turns $\vec{k}\rightarrow-\vec{k}$ and time reversal changes it back again.
}{
 \sektion{8}{October 16}
\descriptioneight
\section*{\sc Scattering Theory}
{\sc For a wide class of quantum mechanical systems, the description of any state is simple if you go far enough into the past or far enough into the future.

We'll illustrate this by a sequence of three increasingly complicated systems.}

\begin{enumerate}
\item NRQM, two particles interacting through a repulsive force, which dies off at large separation.
\begin{align*}
H&=\frac{p_A^2}{2m_A}+\frac{p_B^2}{2m_B}+V(|\vec{r}_A-\vec{r}_B|) & V&\geq0 & V(\infty)&=0
\end{align*}

Any state of the system, a real normalizable state, not a plane wave, looks in the far past / future like two particles far apart (actually a superposition of such states). The potential always pushes the particles far apart for times in the far future, and because the potential dies off they then look like noninteracting particles
\begin{center}
\includegraphics[width=10 cm]{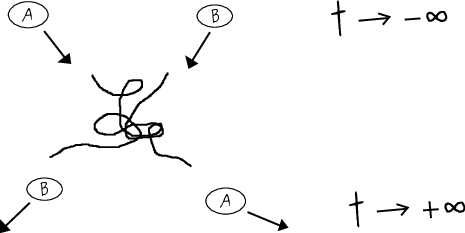}
\end{center}

There is no trace of the interaction in the far future or the far past. The states act like states with dynamics governed by a free Hamiltonian, $H_0$, which is simple, 
\[ H_0=\frac{p_A^2}{2m_A}+\frac{p_B^2}{2m_B} \]

{\sc The aim of scattering theory is to tell what (superposition of) simple state(s) in the far future a simple state in the far past evolves into.

How far in the ``far past'' you have to go depends on the initial conditions and the interaction. If the interaction is the nuclear force and we collide two neutrons at low energy which elastically scatter near $t=0$, then $t=-7$ years is far enough in the far past so that the system looks like two non interacting nucleons. If however the initial conditions are set up so that the elastic scattering occurs around $t=-1$ billion years then you might have to go to $t=-(1\text{ billion and seven})$ years to make the system look simple.

It need not be that the simple Hamiltonian in the far future is the same as the simple Hamiltonian in the far past.}

\item NRQM, Three particles $A$, $B$, $C$ which interact through attractive interactions which are strong enough to make an $AB$ bound state. Start in the far past with $C$ and the $AB$ bound state. Everything still looks like free particles governed by a free Hamiltonian but the free Hamiltonian in the far past is
\[ H_0=\frac{p_{AB}^2}{2m_{AB}}+\frac{p_C^2}{2m_C} \]

Now let scattering occur. In the far future, we can get states that look like three particles, $A$, $B$, $C$, non-interacting, governed by the free Hamiltonian 
\[ H_0=\frac{p_{A}^2}{2m_{A}}+\frac{p_B^2}{2m_B}+\frac{p_C^2}{2m_C} \]
\begin{center}
\includegraphics[width=9 cm]{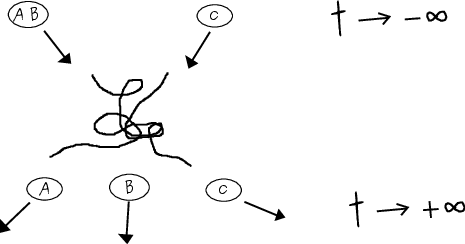}
\end{center}

If you had a sufficiently advanced QM course, you may have studied such a system: 
\[ e^+ +H\rightarrow p+e^++e^-\] 

There is no way to truncate this system's full Hamiltonian into a free part and an interacting part, for which the free part describes the evolution of the system in the far future and the far past. If you use the far future $H_0$, you don't have an $AB$ bound state.

\item (A plausible picture of) the real world.

In the real world we have loads of (stable) bound states. If the real world has a laboratory bench as a stable bound state, then I can do chalk-bench scattering, and I'll need a description of a freely flying piece of chalk and a freely flying laboratory bench. The description of states in the far past requires states with free electrons, hydrogen atoms, protons, Iron atoms, Iron nuclei, laboratory benches, chalk, and the associated free Hamiltonians.

Let's get some formalism up. (The first part of this lecture, with lots of words and few equations, is the part of a lecture that makes some people nervous and some people bored.)

Let $H$ be the actual Hamiltonian of the world and $\mathcal{H}$ be the actual Hilbert space of the world. If you go sufficiently far in the past, every state in the actual Hilbert space looks simple. Let $\mathcal{H}_0$ be the Hilbert space of simple states and let $|\psi\rangle\in\mathcal{H}_0$. Somewhere in the real world Hilbert space there is a state that looks like $|\psi\rangle$ in the far past. We'll label that state $|\psi\rangle^{\text{in}}$, $|\psi\rangle^{\text{in}}\in\mathcal{H}$. Given another state $|\phi\rangle\in\mathcal{H}_0$, there is another state in the real
world Hilbert space that looks like $|\phi\rangle$ in the far future. We'll label that state $|\phi\rangle^{\text{out}}$, $|\phi\rangle^{\text{out}}\in\mathcal{H}$. States in the complicated space are labelled by what they look like in the far past or the far future.

What we are after in scattering theory is the probability, and hence the amplitude, that a given state looking like $|\psi\rangle$ in the far past, looks like $|\phi\rangle$ in the far future. We are after
\[^{\text{out}}\langle\phi|\psi\rangle^{\text{in}}\]

The correspondence between $|\psi\rangle^{\text{in}}$ and $|\psi\rangle$ (for every state $|\psi\rangle\in\mathcal{H}_0$, there is a $|\psi\rangle^{\text{in}} \in \mathcal{H}$ that looks like $|\psi\rangle$ in the far past) and between $|\phi\rangle^{\text{in}}$ and $|\phi\rangle$ allows us to define an operator in the simple Hilbert space $\mathcal{H}_0:S$, the scattering matrix, which is defined by
\[\langle\phi|S|\psi\rangle\equiv\;^{\text{out}}\langle\phi|\psi \rangle^{\text{in}}\]
\end{enumerate}

An ideal scattering theory would have two parts
\begin{enumerate}
\item A turn the crank method of obtaining the ``descriptor'' states $|\psi\rangle$, $|\phi\rangle$, that is, generating $\mathcal{H}_0$, from the real world Hamiltonian. We also need $H_0$ which gives the evolution of the descriptor states. $H_0$ evolves the descriptor states without scattering.
\item A turn the crank method of obtaining $S$.
\end{enumerate}

90\% of the rest of this course will be devoted to calculating the matrix elements of $S$ perturbatively.

That's an ideal scattering theory. We want to get calculating so we'll start with a bargain basement, K-Mart scattering theory.

\section*{\sc Low Budget Scattering Theory}
Imagine that $H$ can be written as $H=H_0+f(t)H'$, $f(t)=0$ for large $|t|$, and $H_0$ a free Hamiltonian that evolves states simply, without scattering. Unless the interaction is with some externally specified apparatus, interesting Hamiltonians and the real world Hamiltonian are not of this form. We want a simple description of states in the far past / future. Because the interaction is off in the far past / future, the simple descriptor states are simply the states in the full theory far enough in the past / future that $f(t)=0$, $\mathcal{H}=\mathcal{H}_0$. Furthermore, the Hamiltonian that gives the evolution of the simple states, $H_0$, is just the full Hamiltonian $H$, far enough in the past / future that $f(t)=0$.

Most Hamiltonians don't have an $f(t)$ in them that goes to zero as $|t|\rightarrow\infty$. However, many Hamiltonians are of the form $H=H_0+H'$ where $H_0$ is a Hamiltonian we know the solution of. \uline{Maybe we could put an $f(t)$ into the Hamiltonian without changing scattering processes much.} We know we can't do this in system (2). No matter how far you go into the far past / future, it is the interaction that holds the stable $AB$ bound state together, and you can't shut the interaction off without the bound state falling apart, totally changing scattering processes, no matter how long you
wait to shut it off. The real world is like system (2) and we can only get a little ways studying the real world if we hack it up like this. We might get a ways studying system (1) like this. In the far past / future the particles in system (1) are far apart and noninteracting. If $f(t)\rightarrow0$ in the far past / future, it should not affect their evolution since they are not interacting then anyway.

Suppose we wanted to insert an $f(t)$ to study a theory of electrons interacting through a repulsive Coulomb force. We can see a flow developing. As a single electron goes off to $\infty$ away from all others, it still has a Coulomb field (cloud of photons) around it. If you weigh an electron, you get a contribution $\int\frac{E^2}{8\pi}d^3x$ to the mass-energy in addition to the contribution to the mass-energy at the heart of the electron. This is one and the same electric field that causes the scattering to take place, and you can't turn off scattering without turning off this cloud. Maybe, \uline{if we turn the interaction off sufficiently slowly the simple states in the real theory will turn into the states in the free theory with probability 1}. We want $f(t)$ to look like:
\begin{center}
\includegraphics[width=\textwidth]{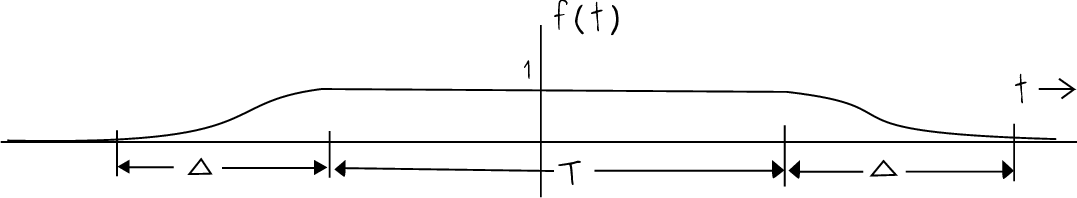}
\end{center}

$f(t)$ turns on and off adiabatically. A more precise way of stating the condition under which we hope inserting $f(t)$ into the theory won't change scattering processes much is \uline{there must be a 1-1 correspondence between the asymptotic (simple) states of the full Hamiltonian and the states of the free Hamiltonian}. That means no bound states, no confinement. We hope scattering processes won't be changed at all under this assumption in the limit
$\Delta\rightarrow\infty$, $T\rightarrow\infty$, $\frac{\Delta}{T}\rightarrow 0$. The last limit is needed so that edge effects are negligible. We want adiabatic turn on and off, but we also want the interaction to be on much longer than the amount of time we spend turning it on and off. Similar requirements must be imposed if you put a system in a spatial box, depending on what kind of quantities you want to know about. In slightly racy language, the electron without its cloud of photons is called a ``bare'' electron, and with its cloud of photons a ``dressed'' electron. The scattering process goes like this: In the far far past a bare electron moves freely along. A billion years before it is to interact it leisurely dresses itself. Then it moves along for a long time as a dressed electron, briefly interacts with another (dressed) electron and moves for a long time again, dressed. Then it leisurely undresses.

We need to develop \uline{Time dependent Perturbation Theory} for Hamiltonians of the form
\[H=H_0+H'(t)\]

$H'(t)$ may depend on time because of externally varying interactions or because of the insertion of $f(t)$. We'll do the formalism in the interaction picture developed by Dirac. This is the \uline{best} formalism for doing time-dependent perturbation theory. If you have laser light shining on an atom, and you know this formalism, it is the most efficient way of calculating what happens to the atom, although it is higher powered than the minimum formalism you need for that problem.

\subsubsection*{\uline{Schr\"odinger picture}}
We have states evolving in time according to
\[i\frac{d}{dt}|\psi(t)\underset{\substack{\text{for Schr\"odinger since we'll be}\\ \text{working in several pictures}}}{\underbrace{\rangle_S}=H(p_S,q_S,t)|\psi(t) \rangle_S} \]

The fundamental operators, $p_S$ and $q_S$ are time independent
\begin{align*}
q_S&=q_S(t)=q_S(0) & p_S&=p_S(t)=p_S(0) 
\end{align*}

The only operators that are not time independent are operators that explicitly depend on time. The time evolution operator $U(t,t')$ is given by
\[ |\psi(t)\rangle_S=U(t,t')|\psi(t')\rangle_S\]

$U$ is completely determined by the 1st order differential equation in $t$
\[ i\frac{d}{dt}U(t,t')=H(p_S,q_S,t)U(t,t') \]

\noindent 
and the initial condition $\left.U(t,t')\right|_{t=t'}=1$. Think of $t'$ as a parameter. $U$ is unitary (from the Hermiticity of $H$), i.e.~$U(t,t')^\dagger=U(t,t')^{-1}$, which expresses the conservation of probability. $U$ also obeys the composition law
\[ U(t,t')U(t',t'')=U(t,t'') \]

\noindent 
which implies 
\[U(t,t')=U(t',t)^{-1}\]

\subsubsection*{\uline{Heisenberg Picture}}
The states do not change with time
\[ |\psi (t) \rangle_H=|\psi(0)\rangle_H=|\psi(0)\rangle_S \]

If $A_S\!\!\!\!\!\!\!\!\!\!\!\!\overbrace{(t)}^{\substack{\text{possible explicit}\\ \text{time dependence}}}\!\!\!\!\!\!\!\!\!\!\!\!$ is an operator in the Schr\"odinger picture and $A_H(t)$ its counterpart in the Heisenberg picture, demand that
\begin{align*}
_S\langle \phi(t)|A_S(t)|\psi(t)\rangle_S &= _H\langle \phi(t)|A_H(t)|\psi(t)\rangle_H \\
&=_S\langle\phi(0)|A_H(t)|\psi(0)\rangle_S\\
&=_S\langle \phi(t)|U(0,t)^\dagger A_H(t) U(0,t)|\psi(t)\rangle_S \\
\therefore A_H(t)&=U(0,t)A_S(t)U(0,t)^\dagger=U(t,0)^\dagger A_S(t)U(t,0)
\end{align*}

Suppose we have some function of operators and the time in the Schr\"odinger picture, itself an operator. For example $H$ itself is a function of $p_S$, $q_S$ and $t$. To get the operator in the Heisenberg picture, all you have to do is replace $p_S$ and $q_S$ by $p_H$ and $q_H$.
\[H_H(t)=H(p_H(t),q_H(t),t)\;\;\;\substack{\text{Expand $H$ as a power} \\ \text{series and insert} \\ U(t,0)U(t,0)^\dagger\text{ all over.}} \]

\subsubsection*{\uline{Interaction Picture}}
Assume $H$ can be written as $H(p,q,t)=H_0(p,q)+H'(p,q,t)$. This defines the relation between the functions $H$, $H_0$ and $H'$, the arguments of these functions will change.

The interaction picture is intermediate between the Schr\"odinger picture and the Heisenberg picture. You make the transformation you would make to get the Schr\"odinger picture to the Heisenberg picture, but you do it using just the free part of the Hamiltonian only.
\[ |\psi(t)\rangle_I=e^{iH_0(p_S,q_S)t}|\psi(t)\rangle_S \;\;\; \substack{\text{If there were no interaction $H'$, there would}\\ \text{be no time evolution of the states. That is}\\ \text{what makes the interaction picture so useful.}} \]

If $H_0$ depended explicitly on time we would have to define a $U_0(t,0)$ which would take the place of $e^{-iH_0t}$, but we will have no occasion to be that general. If $A_S(t)$ is an operator in the Schr\"odinger picture, and $A_I(t)$ its counterpart in the interaction picture, we get
\begin{equation}\label{eq:08-interactionpicture}
A_I(t)=e^{iH_0(p_S,q_S)t} A_S(t) e^{-iH_0(p_S,q_S)t} 
\end{equation}

\noindent 
by demanding
\[_S\langle\phi(t)|A_S(t)|\psi(t)\rangle_S=_I\langle\phi(t)|A_I(t)|\psi(t)\rangle_I \]

We can find a differential equation for $|\psi(t)\rangle_I$
\begin{align*}
i\frac{d}{dt}|\psi(t)\rangle_I&=i\frac{d}{dt}\left(e^{iH_0(p_S,q_S)t}|\psi(t)\rangle_S\right)\\
&=e^{iH_0(p_S,q_S)t}[-H_0(p_S,q_S)+H(p_S,q_S,t)]|\psi(t)\rangle_S\\
&=e^{iH_0(p_S,q_S)t}[H'(p_S,q_S,t)]e^{-iH_0(p_S,q_S)t}|\psi(t)\rangle_I\\
&=H'(p_I,q_I,t)|\psi(t)\rangle_I\;\;\;\substack{\text{Expand $H'(p_S,q_S,t)$ in a}\\ \text{power series and insert}\\
e^{-iH_0t}e^{iH_0t}\text{ all over.}}\\
&\equiv H_I(t)|\psi(t)\rangle_I \;\;\;\substack{\text{As promised, if $H'$ is zero, no time evolution.}}
\end{align*}

In field theory, $H_I(t)$ will contain the free fields 
\[\phi(\vec{x},t)=e^{iH_0t}\phi_S(\vec{x})e^{-iH_0t} \]

That is why all our results about free fields are still going to be useful.

We can define $U_I(t,t')$ by
$|\psi(t)\rangle_I=U_I(t,t')|\psi(t')\rangle_I $
\begin{align*}
U_I^\dagger(t,t')&= U_I(t,t')^{-1} & U_I(t,t')U_I(t',t'')&=U_I(t,t'')\\
U_I(t,t')&=U_I(t',t)^{-1} & U_I(t,0) &= e^{iH_0(p_S,q_S)t} U(t,0)
\end{align*}

In a field theory, $\phi_I(x)$ obey free eq of motion + commutation relations.

$U_I(t,t')$ can be determined from the first order differential equation in $t$ it satisfies
\[ i\frac{d}{dt}U_I(t,t')=H_I(t)U_I(t,t')\;\;\;\text{and the initial conditions } \left.U_I(t,t')\right|_{t=t'}=1 \]

Now we'll apply interaction picture perturbation theory to scattering theory. In the interaction picture the scattering process looks like this: In the far past the interaction is not felt, both because of $f(t)$ and the fact that these particles are far apart. The states just lie there, although the $p$'s and $q$'s are changing. The time of scattering approaches and the state starts changing. After scattering, they stop scattering, like a game of musical chairs, everything freezes again.

You want to connect the simple description in the far past to the simple description in the far future. Because of $f(t)$ the simple description in the far past / future is the actual description in the far past / future. The states in the far past and future are their own descriptors (used in arrowed step).
\begin{align*}
\langle\phi|S|\psi\rangle\equiv\;^{\text{out}}\langle\phi|\psi \rangle^{\text{in}}&=\;_I\langle \phi(0)|\psi(0)\rangle _I\\
\begin{split}
&=\;_I\langle \phi(\infty) | U_I(\infty,-\infty)|\psi(-\infty)\rangle_I\\
&=\langle\phi|U_I(\infty,-\infty)|\psi\rangle
\end{split}\Downarrow\\
\therefore S&=U_I(\infty,-\infty)
\end{align*}

Our number one priority then is to evaluate $U_I(\infty,-\infty)$.
That will cause us to develop Dyson's formula and Wick's theorem.
Then we'll apply this formalism to three models.

\subsubsection*{Proof that $S=U_I(\infty,-\infty)$ in the Schr\"odinger picture.}
\begin{align*}
i\frac{d}{dt}|\psi\rangle^{\text{in}}&=H|\psi\rangle^{\text{in}} & |\psi(-\infty)\rangle&=|\psi(-\infty)\rangle^{\text{in}} \\
i\frac{d}{dt}|\psi\rangle&=H_0|\psi\rangle \\
^{\text{out}}\langle\phi|\psi\rangle^{\text{in}}&\equiv\langle\phi|S|\psi\rangle \\
^{\text{out}}\langle\phi|\psi\rangle^{\text{in}}&=\;^{\text{out}}\underset{\substack{\text{any time will do}}}{\langle\phi(t)|\psi(t)\rangle^{\text{in}}}\\
&=\;^{\text{out}}\langle\phi(\infty)|U(\infty,0)U(0,-\infty)|\psi (-\infty)\rangle^{\text{in}} \\
&=\langle\phi(\infty)|U(\infty,-\infty)|\psi(-\infty)\rangle\\
&=\langle\phi(0)|U_0(\infty,0)^{\dagger-1}U(\infty,-\infty)U_0(0,-\infty)^{-1}|\psi(0)\rangle\\
&=\langle\phi|U_I(\infty,-\infty)|\psi\rangle
\end{align*}

$[S,H_0]=0$ because $S$ turns free states of a given energy into other free states of the same energy.

\section*{Dyson's Formula}
We would like to find the solution of the equation 
\begin{align*}
i\frac{d}{dt}U_I(t,t')&=H_I(t)U_I(t,t') & U_I(t,t')\Big|_{t = t'} &=1
\end{align*}

Imagine that $[H_I(t),H_I(t')]=0$, which is not true. Then the solution would be 
\[ \cancel{U_I(t,t')=e^{-i\int_{t'}^t dt'' H_I(t'')}} \]

Let's define a new exponential so this equation is right. Given a string of operators define the time-ordered product 
\[T[A_1(t_1)\cdots A_n(t_n)] \]

\noindent 
to be the string rearranged so that later operators are to the left of earlier operators, with the operator with the latest time one the leftest. The ambiguity of what to do at equal times does not bother us when the operators commute at equal times. This is certainly the case when all the operators are the same, $H_I(t)$, evaluated at various $t$. $T$, the symbol for the time ordering operation, is not an operator in Hilbert space. Time ordering is a notation.

Now we'll show that the differential equation for $U_I(t,t')$ is satisfied by 
\begin{align*}
&Te^{-i\int_{t'}^tdt''H_I(t'')} & t>t' 
\end{align*}

Under the time ordering symbol everything commutes, so we can naively take a time derivative to get
\[i \frac{d}{dt}Te^{-i\int_{t'}^tdt''H_I(t'')} =T\left(H_I(t)e^{-i\int_{t'}^tdt''H_I(t'')} \right) \]

Now $t$ is a special time. It is the latest time, so the time ordering puts $H_I(t)$ on the leftest, and we can pull it out on the left to get
\[i \frac{d}{dt}Te^{-i\int_{t'}^tdt''H_I(t'')} =H_I(t)Te^{-i\int_{t'}^tdt''H_I(t'')}\]

This solution of the differential equation also obeys the boundary condition (any old ordering does that).

The solution of a first order differential equation with given initial value is unique. Therefore
\begin{align*} 
U_I(t,t')&=Te^{-i\int_{t'}^tdt''H_I(t'')} & t&>t' &\text{Dyson's Formula}
\end{align*} 

To illustrate what Dyson's formula means, we'll look at the second order term in the power series expansion for the exponential.
\[ \frac{(-i)^2}{2!}\int_{t'}^tdt_1\int_{t'}^tdt_2 T(H_I(t_1)H_I(t_2)) \]
\begin{center}
\includegraphics[width=8cm]{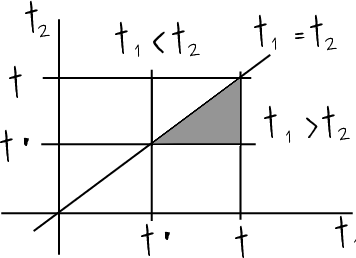}
\end{center}

Think of this as an integration over the square. If the time ordering symbol were not there, you would be integrating $H_I(t_1)H_I(t_2)$ over the whole square. Because of the time ordering symbol, you get instead twice the integral of $H_I(t_1)H_I(t_2)$ over the lower half of the square, the shaded triangle.

Our formula is only valid for $t>t'$. You can easily get the formula for $t<t'$ by taking the adjoint.

We are going to apply Dyson's formula to three model theories in order of increasing complexity.
\newcounter{08count1}
\begin{list}{\uline{\sc Model \arabic{08count1}} }
{\usecounter{08count1}
\setlength{\rightmargin}{\leftmargin}}
\item {\large $\mathcal{L}=\frac{1}{2}\partial_\mu\phi\partial^\mu\phi
-\frac{\mu^2}{2}\phi^2-g\rho(x)\phi(x) $}

$\rho(x)\rightarrow0$ as $x\rightarrow\infty$ in space or time. $\rho(x)$ is a prescribed $c$-number function of space-time, a source which will create mesons. The equation of motion is
\[(\square+\mu^2)\phi(x)=-g\rho(x) \]

Electromagnetism with an external source which generates the EM field looks like $\square A^\mu=-ej^\mu$. Except that our field is massive, has no vector index, and is a quantum field, these two theories look similar. We'll call model 1 quantum meso-dynamics. We'll be able to solve it exactly, which is not a big surprise; in momentum space it is just a bunch of independent forced harmonic oscillators.

\item {\large $\mathcal{L}=\frac{1}{2}\partial_\mu\phi\partial^\mu\phi -\frac{\mu^2}{2}\phi^2-g\rho(\vec{x})\phi(x) $}

$\rho(\vec{x})\rightarrow0$ as $|\vec{x}|\rightarrow\infty$. This is the same as model 1 except the source is static. You might think this time independent problem would be easier than model 1, but it isn't because the source does not turn off as $|t|\rightarrow\infty$. We will have to use (and thus gain experience with) our adiabatic turning on and off function, $f(t)$. This is the quantum scalar analog of electrostatics, so we'll call it ``mesostatics.''

\item {\large $\mathcal{L}=\frac{1}{2}\partial_\mu\phi\partial^\mu\phi-\frac{\mu^2}{2}\phi^2+\partial_\mu\psi^*\partial^\mu\phi-m^2\psi^*\psi-g\psi^*\psi\phi$}

The equation of motion for the $\phi$ field is
\[(\square+\mu^2)\phi=-g\psi^*\psi \]

This is beginning to look like the real thing. In real electrodynamics, the current $j^\mu$ is not prescribed, it is the current of charged particles. Here we have the charged field $\psi$ as a source for $\phi$ ($\psi^*\psi$ is like a current). The $\phi$ field in turn appears in the equation of motion for the $\psi$ field.
\[(\square+m^2)\psi=-g\psi\phi\]

This theory also looks a lot like Yukawa's theory of the interaction between mesons and nucleons, except our charged particles are spinless and we only have one meson. We'll call this ``meson-nucleon" theory. Actually, we had better not push this theory too far (we'll be doing low orders in P.T.~only). The classical Hamiltonian contains $g\psi^*\psi$ and that is not bounded below for either sign of $g$.
\end{list}

\section*{\uline{Wick's Theorem}}
When doing perturbative calculations in $g$ in any of these three models, we are going to have to evaluate time ordered products of strings of Hamiltonians between states. In model 1, $\mathcal{H}_I=g\rho(x)\phi(x)$. At fourth order in $g$ for a meson scattered by the source we would have to evaluate 
\[\langle\vec{k}\,'|T(\overbrace{\phi(x_1)\phi(x_2)\phi(x_3)\phi(x_4)}^{\substack{\text{these are free fields}}})|\vec{k}\rangle \]

The time ordered product contains 16 arrangements of creation and annihilation operators, from $a_{\vec{k}_1}a_{\vec{k}_2}a_{\vec{k}_3}a_{\vec{k}_4}$ and $a_{\vec{k}_1}a_{\vec{k}_2}a_{\vec{k}_3}a_{\vec{k}_4}^\dagger$ to $a_{\vec{k}_1}^\dagger a_{\vec{k}_2}^\dagger a_{\vec{k}_3}^\dagger a_{\vec{k}_4}^\dagger$. If we could rearrange these into normally ordered products, the only normally ordered product that could contribute would be the one with one creation
operator on the left and one annihilation operator on the right, a great simplification. In model 3 we wil have to evaluate time ordered products like
\[ T(\phi(x_1)\psi^*(x_1)\psi(x_1)\phi(x_2)\psi^*(x_2)\psi(x_2)) \]

If we had an algorithm for normal ordering the time ordered product, we would again have great simplifications when we sandwiched this between states of mesons and nucleons. Wick's theorem turns time ordered products of free fields into normal ordered products of free fields. To state Wick's theorem we'll define the contraction.
\[ \wick{1}{<1A(x)>1B(y)}\equiv T(A(x)B(y))-:A(x)B(y): \]

Suppose, without loss in generality, really, that $x^0>y^0$. Then
\[T(A(x)B(y))=A(x)B(y)=(A^{(+)}+A^{(-)})(B^{(+)}+B^{(-)})=:AB:+[A^{(+)},B^{(-)}]\] 

\noindent 
and $\wick{1}{<1A(x)>1B(y)}=[A^{(+)},B^{(-)}]$ which is a $c$-number. This is also a $c$\# when $x^0<y^0$. So whether $x^0<y^0$ or $y^0<x^0$, $\wick{1}{<1A(x)>1B(y)}$ is a $c$\# and is thus equal to its vacuum expectation value. Using its definition
\[\wick{1}{<1A(x)>1B(y)}=\langle0|\wick{1}{<1A(x)>1B(y)}|0\rangle\!\!\!\!\!\!\!\!\overset{ \substack{\text{Using the defn.}}}{ \overbrace{=}}\!\!\!\!\!\!\!\!\langle0|T(A(x)B(y))|0\rangle-\cancel{\langle0|:A(x) B(y):|0\rangle}\]

That's why the calculation of $\langle0|T(\phi(x)\phi(y))|0\rangle$ in the first problem set is going to be useful.
\[\wick{1}{<1\phi(x)>1\phi(y)}=\langle0|T(\phi(x)\phi(y))|0\rangle=\int\frac{d^4k}{(2\pi)^4}e^{\pm ik\cdot(x-y)}\frac{i}{k^2-\mu^2+i\epsilon}\]

$\lim_{\epsilon\rightarrow0^+}$ is understood. Convince yourself the $\pm$ doesn't matter. You can also see that
\[\wick{1}{<1\psi(x)>1\psi^*(y)}=\wick{1}{<1\psi^*(x)>1\psi(y)}=\int\frac{d^4k}{(2\pi)^4}e^{ ik\cdot(x-y)}\frac{i}{k^2-m^2+i\epsilon} \]

A little more obvious notation: $:A(x)\wick{1}{<1B(y)C(z)>1D(w)}:\equiv:A(x)C(z):\wick{1}{<1B(y)>1 D(w)}$ and let $\phi_1\equiv \phi^{a_1}(x_1)$, $\phi_2\equiv\phi^{a_2}(x_2)$, etc., just for this proof.

\uline{Theorem} (Gian-Carlo Wick)
\begin{align*}
T(\phi_1 \dots \phi_n)&=\overbrace{:\phi_1\cdots\phi_n:}^{\substack{\text{term with}\\ \text{no contractions}}}+:\wick{1}{<1\phi_1>1\phi_2}\cdots\phi_n:+\substack{\text{all the other}\\ \frac{n(n-1)}{2} - 1 \text{ possible}\\ \text{terms with one}\\ \text{contraction}}+:\wick{11}{<1\phi_1>1\phi_2<2\phi_3>2\phi_4} \cdots\phi_n:\\
&+\substack{\text{all the other}\\ \frac{1}{2}\frac{n(n-1)}{2}\frac{(n-2)(n-3)}{2} -1 \\ \text{possible terms with}\\ \text{two contractions}}+\substack{\vdots\\ \text{all possible}\\ \text{terms with}\\ <\frac{n-1}{2}\\ \text{contractions}\\ \vdots}
+\begin{cases}
:\wick{11}{<1\phi_1>1\phi_2<2\phi_3>2\phi_4}\cdots \wick{1}{<3\phi_{n-1}>3\phi_n}: &\text{ if $n$ is even}\\
:\wick{11}{<1\phi_1>1\phi_2<2\phi_3>2\phi_4}\cdots \wick{1}{<3\phi_{n-2}>3\phi_{n-1}\phi_n}:&\text{ if $n$ is odd}
\end{cases}
\end{align*}

You draw all possible terms with all possible contractions. That you get all that is no surprise. The remarkable and graceful thing about this theorem is that each term occurs with coefficient +1.

\uline{Proof} (By induction) Define the RHS of the expression to be $W(\phi_1\cdots\phi_n)$. We want to show $W=T$. Trivial for $n=1,2$. Choose without loss of generality $x_{10}\geq x_{20}\geq\cdots\geq x_{n0}$. Then 
\[T(\phi_1\cdots\phi_n)=\phi_1 T(\phi_2\cdots\phi_n)\!\!\!\!\overbrace{=}^{\substack{\text{induction}\\ \text{step}}}\!\!\!\!\phi_1W(\phi_2\cdots\phi_n)=\phi_1^{(-)}W+W\phi_1^{(+)} +[\phi_1^{(+)},W] \]

This expression is normal ordered. The first two terms contain all possible contractions that do not include $\phi_1$. The third term contains all possible contractions that do include $\phi_1$. Together they contain all possible contractions. Either a contraction includes $\phi_1$ or it doesn't. The right hand side is thus $W(\phi_1\cdots\phi_n)$.
}{
 \sektion{9}{October 21}
\descriptionnine
\section*{\sc Diagrammatic Perturbation Theory}
Dyson's formula applied to $S=U_I(\infty,-\infty)$ is
\[ U_I(\infty,-\infty)=Te^{-i\int dtH_I(t)} \]

(Without the use of the time ordering notation, this formula for $U_I(t,t')$ was written down by Dirac 15 years before Dyson wrote it this way, and Dyson says he should not have credit for little more than a change in notation.)

From this and Wick's theorem, which for those of you who really love combinatorics can be written
\[T(\phi_1\cdots\phi_n)=:e^{\frac{1}{2}\sum_{i,j=1}^n\wick{1}{<1\phi_i>1\phi_j}\frac{\partial}{\partial\phi_i}\frac{\partial}{\partial\phi_j}}\phi_1\cdots\phi_n: \]

We have enough work done to write down diagrammatic perturbation theory for $S=U_I(\infty,-\infty)$. The easiest way to see this is to look at a specific model and a contribution to $U_I(\infty,-\infty)$ at a specific order in $g$.

In model 3, $\mathcal{H}_I=gf(t)\psi^*\psi\phi$, $H_I=\int d^3x\mathcal{H}_I$,
\[ U_I(\infty,-\infty)=Te^{-i\int d^4x\mathcal{H}_I}=Te^{-i\int d^4xgf(t)\psi^*\psi\phi} \]

\noindent 
The contribution at second order in $g$ is
\[\frac{(-ig)^2}{2!}\int d^4x_1d^4x_2f(t_1)f(t_2)T(\psi^*\psi\phi(x_1)\psi^*\psi\phi(x_2)) \]

One of the terms in the expansion of the time ordered product into normal ordered products by Wick's theorem is
\[\frac{(-ig)^2}{2!}\int d^4x_1d^4x_2f(t_1)f(t_2):\psi^*\psi\wick{1}{<1\phi(x_1)\psi^*\psi>1\phi(x_2)}: \]

This term can contribute to a variety of physical processes. The $\psi$ field contains operators that annihilate a ``nucleon'' and operators that create an anti-``nucleon''. The $\psi^*$ field contains operators that annihilate an anti-nucleon and create a nucleon. The operator
\[:\psi^*\psi\wick{1}{<1\phi(x_1)\psi^*\psi>1\phi(x_2)}:=:\psi^*\psi(x_1)\psi^*\psi(x_2):\wick{1}{<1\phi(x_1)>1\phi(x_2)} \]

\noindent 
can contribute to $N+N\rightarrow N+N$. That is to say
\[ \langle\substack{\text{final 2}\\ \text{nucleon state}}|:\psi^*\psi(x_1)\psi^*\psi(x_2):|\substack{\text{initial 2}\\ \text{nucleon state}}\rangle\]

\noindent 
is nonzero because there are terms in the two $\psi$ fields that can annihilate the two nucleons in the initial state and terms in the 2 $\psi^*$ fields that can then create two nucleons, to give a nonzero matrix element. It can also contribute to $\overline{N} + \overline{N} \rightarrow \overline{N} + \overline{N}$ and $N + \overline{N} \rightarrow N + \overline{N}$. You can see that there is no combination of creation and annihilation operators in this operator that can contribute to
$N+N \rightarrow \overline{N} + \overline{N}$. The $\psi$ fields would have to annihilate the nucleons and the $\psi^*$ fields cannot create antinucleons. This is good because this process does not conserve the $U(1)$ symmetry charge. However it looks like our operator can contribute to vacuum$\rightarrow N + N + \overline{N} + \overline{N}$, which would be a disaster. The coefficient of that term after integrating over $x_1$ and $x_2$ had better turn out to be zero.

Another term in the expansion of the time ordered product into normal ordered products is
\[\frac{(-ig)^2}{2!}\int d^4x_1d^4x_2f(t_1)f(t_2):\psi^*\wick{1}{<1\psi\phi(x_1)>1\psi^*}\psi\phi(x_2): \]

This term can contribute to the following 2$\rightarrow$2 scattering processes: $N+\phi\rightarrow N+\phi$, $\overline{N} + \phi \rightarrow \overline{N} + \phi$, $N + \overline{N} \rightarrow 2 \phi$, $2 \phi \rightarrow N + \overline{N}$.

A single term is capable of contributing to a variety processes because a single field is capable of creating or destroying a particle.

The terms in the Wick expansion can be written down in a diagrammatic shorthand according to the following rules. At $N$th order in perturbation theory, you start by writing down $N$ interaction vertices and numbering them 1 to $N$. For model 3 at second order in perturbation theory you write down
{\large
\begin{align*}
&\substack{\text{\scriptsize This line is for the $\phi$ in}\\ \text{\scriptsize $\psi^*\psi\phi(x_1)$. It creates}\\ \text{\scriptsize or destroys a meson}}
\;\;\;\Diagram{ & & fuA \\
f \vertexlabel^1 \\
& &fdV\\
}&
\Diagram{ & & fuA \\ f\vertexlabel^{2} \\ & &fdV\\}\;\;\;\substack{\text{\scriptsize This outgoing line is for the $\psi^*$ in} \\ \text{\scriptsize $\psi^*\psi\phi(x_2)$. It can be thought of as creating}\\ \text{\scriptsize a nucleon or annihilating an antinucleon}\\ 
\\ \\
\text{\scriptsize This incoming line is for the $\psi$ in $\psi^*\psi\phi(x_2)$.} \\ \text{\scriptsize It can be thought of as annihilating}\\ \text{\scriptsize a nucleon or creating an antinucleon}\\ }
\end{align*}}

The vertex represents the factor of $f\psi^*\psi\phi$. From the fact that there are two in this diagram you know to include $\frac{(-ig)^2}{2!}\int d^4x_1d^4x_2$

Contractions are represented by connecting the lines. Any time there is a contraction, join the lines of the contracted fields. The arrows will always line up, because the contractions for which they don't are zero. An unarrowed line will never be connected to an arrowed line because that contraction is also zero.

Our first term in the expansion of the time ordered product corresponds to the diagram
{\large \begin{align*}
\Diagram{ fdV & & & fuA \\ & \vertexlabel^{1}f & f\vertexlabel^{2} \\ fuA & & & fdV }
\end{align*}}

The second term corresponds to
{\large \begin{align*}
\Diagram{\;\;\,fd & \quad\quad fu \\ fV\vertexlabel^{1} fV & fV\vertexlabel^{2} fV }
\end{align*}}

The term in the Wick expansion
\[\frac{(-ig)^2}{2!}\int d^4x_1d^4x_2 f(t_1)f(t_2):\wick{1}{<1\psi^* \psi\phi(x_1)>1\psi^*}\psi\phi(x_2): \]

\noindent 
is zero because $\wick{1}{<1\psi^*>1\psi^*}=0$, so we never write down
{\large \begin{align*}
\Diagram{\;\;\,fd & \quad\quad fu \\ fA\vertexlabel^{1} fA & fV\vertexlabel^{2} fV }
\end{align*}}

Because the arrows always line up, we can shorten
{\large \begin{align*}
\Diagram{\;\;\;fd & \quad\quad fu \\ fV\vertexlabel^{1} fV & fV\vertexlabel^{2} fV }\;\;\;\text{\normalsize to }\;\;\; \Diagram{\;\;\,fd \quad\;\;\; fu \\ fV\vertexlabel^{1} fV\vertexlabel^{2} fV }
\end{align*}}

\underline{These diagrams are in one-to-one correspondence with terms in the Wick expansion of}\\ \underline{Dyson's formula.}

We'll call them Wick diagrams. They stand for operators and the vertices are numbered. We are most of the way to Feynman diagrams which stand for matrix elements, but these aren't them yet. The vertices are numbered in Wick diagrams and 
{\large \begin{align*}
\Diagram{\;\;\,fd \quad\;\;\; fu \\ fV\vertexlabel^{2} fV\vertexlabel^{1} fV }
\;\;\;\text{\normalsize is distinct from }\;\;\; \Diagram{\;\;\,fd \quad\;\;\; fu \\ fV\vertexlabel^{1} fV\vertexlabel^{2} fV }
\end{align*}}

\noindent (there are two distinct terms in the Wick expansion) even though after integrating over $x_1$ and $x_2$ these are identical operators. In Feynman diagrams the lines will be labelled by momenta.

On the other hand 
{\large \begin{align*}
\text{\normalsize 1}\feyn{f flV fluA f }\text{\normalsize 2}\;\;\;\text{\normalsize is identical to}\;\;\;
\text{\normalsize 2}\feyn{f flV fluA f }\text{\normalsize 1}
\end{align*}}

(there is only one way of contracting all three fields at one vertex with all three at the other). Although these two have been written down to look different they aren't. Rotate the right one by 180$^\circ$ and you see they are the same.
{\large \begin{align*}
\text{\normalsize \begin{sideways}\begin{sideways}1\end{sideways}
\end{sideways}}\feyn{f flV fluA f }\text{\normalsize
\begin{sideways}\begin{sideways}2\end{sideways}\end{sideways}}
\end{align*}}

The contraction this diagram corresponds to is
\[\frac{(-ig)^2}{2!}\int d^4x_1d^4x_2f(t_1)f(t_2):\wick{213}{<1\psi^*<2\psi<3\phi(x_1)>2\psi^*>1\psi>3\phi(x_2)}: \]

In model 1 $\mathcal{H}_I=g\rho(x)\phi(x)$ (we don't have to insert a turning on and off function because the interaction goes to zero as $x\rightarrow\infty$ in any direction and in particular in the time direction in the far past / future). $\rho(x)$ is a prescribed $c$-number source, so strongly made we don't have to worry about the back reaction of the field $\phi$ on the source. The vertex in this model is
{\large \begin{align*}
\bullet\feyn{f }
\end{align*}}

That represents $\rho\phi(x)$. At $O(g)$ in $U_I$ we have $(-ig)\int d^4x_1\rho\phi(x_1)$ which is represented by $\overset{1}{\bullet}\feyn{f}$.

At $O(g^2)$ in $U_I$ we have $\overset{1}{\bullet}\!\feyn{f}\,\overset{2}{\bullet}\negmedspace\feyn{f}$ and $\overset{1}{\bullet}\feyn{f}\!\overset{2}{\bullet}$.

At $O(g^3)$ in $U_I$ we have $\overset{1}{\bullet}\feyn{f}\,\overset{2}{\bullet}\!\feyn{f}\,\overset{3}{\bullet}\!\feyn{f}$, $\overset{1}{\bullet}\feyn{f}\!
\overset{2}{\bullet}\;\overset{3}{\bullet}\!\feyn{f}$, $\overset{1}{\bullet}\feyn{f}\;\overset{2}{\bullet}\!\feyn{f}\overset{3}{\bullet}$, and $\overset{1}{\bullet}\!\feyn{f}\!\overset{3}{\bullet}\;\overset{2}{\bullet}\!\feyn{f}$.

A diagram at $O(g^4)$ is $\overset{1}{\bullet}\feyn{f}\!\!\overset{2}{\bullet}\,\overset{3}{\bullet}\!\feyn{f}\,\overset{4}{\bullet}\negmedspace\feyn{f}$.

We have been putting the normal ordering inside the integrand. Of course we could put it around the whole integral in which case we see that this $O(g^4)$ diagram corresponds to
\begin{multline*}
\frac{(-ig)^4}{4!}:\int d^4x_1d^4x_2d^4x_3d^4x_4\rho(x_1)\rho(x_2)
\rho(x_3)\rho(x_4)\wick{1}{<1\phi(x_1)>1\phi(x_2)}\phi(x_3)\phi(x_4):
=\\
\frac{(-ig)^4}{4!}:\int d^4x_1d^4x_2\wick{1}{<1\phi(x_1)>1\phi(x_2)}
\rho(x_1)\rho(x_2) \int d^4x_3\phi(x_3)\rho(x_3)\int d^4x_4\phi(x_4)
\rho(x_4):
\end{multline*}

That is, the integrands factor into products of terms corresponding to each connected\footnote{``Connected'' means (in any theory) that the diagram is in one connected piece. It doesn't mean fully contracted. $\overset{1}{\bullet}\!\!\!-\!\!\overset{2}{\bullet}\,\overset{3}{\bullet}\!\!-\!\overset{4}{\bullet}$ is a fully contracted diagram that is not connected. $\overset{1}{\bullet}\!\!-$ is not contracted, but is connected.} part of the diagram. This suggests we can sum the series and then normal order in this simple theory, because at all orders in $g$ the diagrams only contain $\bullet\!\feyn{f}\!\bullet$ and $\bullet\!\feyn{f}$ various numbers of times. We could do the sum in this theory, but instead we will prove a general theorem.
\[\sum\text{all Wick diagrams}=:e^{\sum\text{connected Wick diagrams}}: \]

In a theory with only two connected diagrams this theorem is powerful enough to solve the theory exactly in a couple of lines. It will help a lot in model 3, but since there are still an infinite number of connected diagrams in model 3, we won't solve it. This formula is also useful in condensed matter physics where you develop a perturbation theory for $\text{Tr }e^{-\beta H}$. The free energy which is the logarithm of the partition function is what is actually of interest. This theorem's analogue tells you that you don't have to calculate a huge series for $\text{Tr }e^{-\beta H}$ and then try to take its logarithm. The free energy is just the sum of the connected diagrams.

Let $D$ be a general diagram with $n(D)$ vertices. Associated with this diagram is an operator
\[ \frac{:O(D):}{n(D)!} \]

We have explicitly displayed the $n(D)!$ and we have pulled the normal ordering outside. For example for
\begin{align*}
D&= \Diagram{ fdV & & & fuA \\ & \vertexlabel^{1}f & fs\vertexlabel^{2} \\ fuA & & & fdV } & O(D)&=(-ig)^2\int d^4x_1d^4x_2 f(t_1)f(t_2)\wick{1}{<1\phi(x_1)>1\phi(x_2)}\psi^*\psi(x_1)\psi^*\psi(x_2)
\end{align*}

I will define two diagrams to be of the same ``pattern'' if they differ just by permuting the labels at the vertices, 1, 2, \dots, $n(D)$.

Since after integration over $x_1$, \dots, $x_{n(D)}$ two different diagrams of the same pattern give identical contributions to $U_I$ and since there are $n(D)!$ permutations of the numbers 1, \dots, $n(D)$, you might expect the sum over all diagrams of a given pattern to exactly cancel the $n(D)!$. This is not quite right however. For some diagrams there are permutations of the vertices that have no effect, for example
{\large\begin{align*}
\Diagram{ fd & & & fu \\ & \vertexlabel^{1}fA \vertexlabel^{2} \\ &fvA & fvV\\ & \vertexlabel_{3}fV \vertexlabel_{4} \\ fu & & & fd }
\;\;\;\text{\normalsize is not distinct from }\;\;\;
\Diagram{ fd & & & fu \\ & \vertexlabel^{4}fA \vertexlabel^{1} \\ &fvA & fvV\\ & \vertexlabel_{3}fV \vertexlabel_{2} \\ fu & & & fd }
\end{align*}}

\noindent (and there are two more cyclic permutations) but it is distinct from the diagrams with noncyclic permutations. This is in exact correspondence with the question of whether or not there is a new term in the Wick expansion from permuting $x_1$, \dots, $x_{n(D)}$.

For any pattern, there will be some symmetry number, $S(D)$, which is the number of permutations that have no effect on the diagram $D$ (and of course there is the analogous statement, that there are $S(D)$ permutations of $x_1$, \dots, $x_{n(D)}$ that do not give additional contributions in the Wick expansion). Summing over all distinct diagrams of the same pattern as $D$ yields
\[ \frac{:O(D):}{S(D)}\]

Let $D_1$, $D_2$, \dots, $D_r$, \dots be a complete set of connected diagrams, with one diagram of each pattern. A general diagram, $D$, has $n_r$ components of pattern $D_r$. Because of the factorization of the integrands and because we have explicitly pulled out the $n(D)!$
\[:O(D):=:\prod_{r=1}^\infty[O(D_r)]^{n_r}: \]

Summing over all diagrams with the same pattern as $D$ gives $\frac{:O(D):}{S(D)}$. What is $S(D)$? $S(D)$ certainly contains $\prod_r[S(D_r)]^{n_r}$. If I have 2 identical factors, I can take all the indices on one of them and exchange them with the other. If I have $n$ identical factors, there are $n!$ whole exchanges, So $S(D)$ contains $\prod_r n_r!$. The sum over all diagrams with the same pattern as $D$ gives
\[ \frac{:O(D):}{S(D)}=\frac{:\prod_{r=1}^\infty[O(D_r)]^{n_r}: }{\prod_{r=1}^{\infty}[S(D_r)^{n_r}n_r!]} \]

Now that we have done the sum over all diagrams of a given pattern, we have to sum over all patterns. Notice that there is a 1-1 correspondence between patterns and sets $\{n_r\}$. Thus summing over all patterns is the same as summing over all sets $\{n_r\}$.

So,
\begin{align*}
\sum\text{all Wick diagrams}&=\sum_{n_1=0}^\infty\sum_{n_2=0}^\infty \cdots\frac{:\prod_{r=1}^\infty[O(D_r)]^{n_r}: }{\prod_{r=1}^{\infty}[S(D_r)^{n_r}n_r!]} \\
&=:\sum_{n_1=0}^\infty\sum_{n_2=0}^\infty \cdots\prod_{r=1}^{\infty}\frac{[O(D_r)]^{n_r}}{S(D_r)^{n_r}n_r!}: \\
&=:\prod_{r=1}^{\infty}\left(\sum_{n_r=0}^\infty\frac{\left[\frac{O(D_r)}{S(D_r)}\right]^{n_r}}{n_r!}\right): \\ 
&=:\prod_{r=1}^{\infty}e^{\frac{O(D_r)}{S(D_r)}}: \\ 
&=:e^{\sum_{r=1}^\infty\frac{O(D_r)}{S(D_r)}}: \\ 
&=:e^{\sum\text{connected Wick diagrams}}: 
\end{align*}

This is a neat theorem because it expresses a fact about diagrams, pictures, algebraically.

Now we'll apply this to model 1.

\section*{\uline{Model 1 Solved}}
\begin{align*}
D_1&=\overset{1}{\bullet}\!\feyn{f},\;\;\; D_2=\overset{1}{\bullet}\!\feyn{f}\!\overset{2}{\bullet},\;\;\; S(D_2)=2\\
U_I(\infty,-\infty)&=:e^{O_1+\frac{O_2}{2}}:=:e^{\overset{1}{\bullet}\!\feyn{f}\;+\frac{\overset{1}{\bullet}\!\feyn{f}\!\overset{2}{\bullet}}{2}} \\
O_1&=-ig\int d^4x_1\rho(x_1)\phi(x_1)\\
O_2&=(-ig)^2\int d^4x_1d^4x_2\wick{1}{<1\phi(x_1)>1\phi(x_2)}\rho(x_1)\rho(x_2)=\substack{\text{some}\\
\text{number}}=\alpha+i\beta
\end{align*}

You will compute $\alpha$ in the homework. We'll get it here by a consistency argument, demanding that $U_I$ be unitary.

Let's rewrite $O_1$, using the expansion for $\phi(x)$.
\begin{align*}
O_1&=-ig\int \frac{d^3k}{(2\pi)^{3/2} \sqrt{2\omega_{\vec{k}}}}\int d^4x\rho(x)\left(e^{-ik\cdot x}a_{\vec{k}} + e^{ik\cdot x}a_{\vec{k}}^\dagger\right)\\
&=-ig\int \frac{d^3k}{(2\pi)^{3/2} \sqrt{2\omega_{\vec{k}}}} \Bigl(\underbrace{\widetilde{\rho}(-k)}_{\widetilde{\rho}(k)^*}a_{\vec{k}} +\widetilde{\rho}(k)a_{\vec{k}}^\dagger\Bigr)
\end{align*}

$\biggl($Using the Fourier transform convention (this convention will \uline{not} be adhered to, see Nov.~6)
\begin{align*}
\widetilde{f}(k)&=\int d^4x e^{ik\cdot x}f(x) & f(x)&=\int\frac{d^4k}{(2\pi)^4} e^{-ik\cdot x}\widetilde{f}(k)
\end{align*}

\noindent 
also in three space dimensions
\begin{align*}
\widetilde{f}(\vec{k})&=\int d^3x e^{-i\vec{k}\cdot \vec{x}}f(\vec{x}) & f(\vec{x})&=\int\frac{d^3\vec{k}}{(2\pi)^3} e^{i\vec{k}\cdot \vec{x}} \widetilde{f}(\vec{k})\biggr)
\end{align*}

So as not to carry around so many factors, define
\[ f(\vec{k})\equiv\frac{-ig}{(2\pi)^{3/2} \sqrt{2\omega_{\vec{k}}}} \widetilde{\rho}(\vec{k},\omega_{\vec{k}})\]

then
\[ U_I(\infty,-\infty)=e^{\frac{1}{2}(\alpha+i\beta)}e^{\int d^3kf(\vec{k})a_{\vec{k}}^\dagger}e^{-\int d^3kf(\vec{k})^* a_{\vec{k}}}\]

Now that we have solved the model we can answer the usual questions you ask about when a field is driven by an external source.

Given that you start with nothing in the far past, $|0\rangle$, what is the probability of finding $n$ mesons in the far future?

The state in the far future is
\begin{align}
U_I(\infty,-\infty)|0\rangle&=e^{\frac{1}{2}(\alpha+i\beta)}e^{\int d^3kf(\vec{k})a_{\vec{k}}^\dagger}\nonumber\\
&=e^{\frac{1}{2}(\alpha+i\beta)}\sum_{n=0}^\infty\frac{1}{n!}\int d^3k_1\cdots d^3k_n f(\vec{k}_1)\cdots f(\vec{k}_n)|\vec{k}_1,\dots,\vec{k}_n\rangle\label{eq:09-UI}
\end{align}

The probability, $P_n$, of finding $n$ mesons is thus
\begin{align*}
P_n&=\left|e^{\frac{1}{2}(\alpha+i\beta)}\frac{1}{n!}\int d^3k_1\cdots d^3k_n f(\vec{k}_1)\cdots f(\vec{k}_n)|\vec{k}_1,\dots, \vec{k}_n\rangle\right|^2\\
&=e^\alpha\frac{1}{(n!)^2}\int d^3k_1\cdots d^3k_n |f(\vec{k}_1)|^2\cdots|f(\vec{k}_n)|^2n!\\
&=e^\alpha\frac{1}{n!}\left(\int d^3k_1|f(\vec{k}_1)|^2\right)^n
\end{align*}

Now is where we demand unitarity of $U_I$ to get $\alpha$.
\begin{align*}
1&\overset{!}{=}\sum_nP_n=e^\alpha\sum_n\frac{1}{n!}\left(\int d^3 k_1|f(\vec{k}_1)|^2\right)^n=e^\alpha e^{\int d^3 k_1|f(\vec{k}_1)|^2} \\
\alpha&=-\int d^3k|f(\vec{k})|^2\\
\text{So }P_n&=e^{-|\alpha|}\frac{|\alpha|^n}{n!}\;\;\;\text{Poisson distribution.}
\end{align*}

This state, created by a classical source, is called a coherent state. Coherent states of the harmonic oscillator are 
\[|\lambda\rangle \equiv e^{\lambda a^\dagger}|0\rangle \]

They are special because they diagonalize $a$
\[ a|\lambda\rangle=ae^{\lambda a^\dagger}|0\rangle = [a,e^{\lambda a^\dagger}]|0\rangle=\lambda e^{\lambda a^\dagger}|0\rangle =\lambda|\lambda\rangle \]

$\langle\lambda|x(t)|\lambda\rangle$ and $\langle\lambda|p(t)|\lambda \rangle$ oscillate sinusoidally like the classical variables.

The coherent states we have constructed are eigenvectors of $\phi^{(+)}(x)$ with eigenvalue 
\[ \int \frac{d^3k}{(2\pi)^{3/2} \sqrt{2\omega_{\vec{k}}}} e^{-ik\cdot x} f(\vec{k}) \]

Except for the $\frac{1}{n!}$ this state's $n$ particle part is just the product of $n$ 1 particle states. It is about as uncorrelated as a state of mesons can be. If you remove a particle with $\phi^{(+)}(x)$, you get the same state back. Expectations of normal ordered products factorize. 

What is the average number of mesons created?
\begin{equation}\label{eq:09-N}
\langle N\rangle=\sum_{n=0}^\infty nP_n=\sum_{n=1}^\infty \frac{e^{-|\alpha|}|\alpha|^n}{(n-1)!}=|\alpha|\;\;\;\substack{\text{(pull out an $|\alpha|$ and}\\ \text{reindex the sum)}}
\end{equation}

What is the average energy of the final state, i.e.~the total energy of all the mesons created?
\begin{align*}
\langle H\rangle &=\sum_{n=0}^\infty\frac{e^{-|\alpha|}}{(n!)^2} \int d^3k_1\cdots d^3k_n |f(\vec{k}_1)|^2\cdots|f(\vec{k}_n)|^2 \underbrace{(\omega_{\vec{k}_1}+\cdots+\omega_{\vec{k}_n})}_{n \omega_{\vec{k}_1}}n!\\
&=\sum_{n=1}^\infty \frac{e^{-|\alpha|}}{(n-1)!}|\alpha|^{n-1}\int d^3k|f(\vec{k})|^2\omega_{\vec{k}} = \int d^3k|f(\vec{k})|^2\omega_{\vec{k}}
\end{align*}

Average momentum? 
\[\langle \vec{P}\rangle = \int d^3k|f(\vec{k})|^2\vec{k} \]

\section*{\uline{Model 2 solved} (beginning)}
Combinatorically, model 2 is identical to model 1, but physically the content is different. The interaction doesn't actually turn off in the far past / future. We put that in by hand.
\[ \mathcal{H}_I=g\phi(x)\rho(\vec{x})f(t) \quad\quad\includegraphics[width=10 cm]{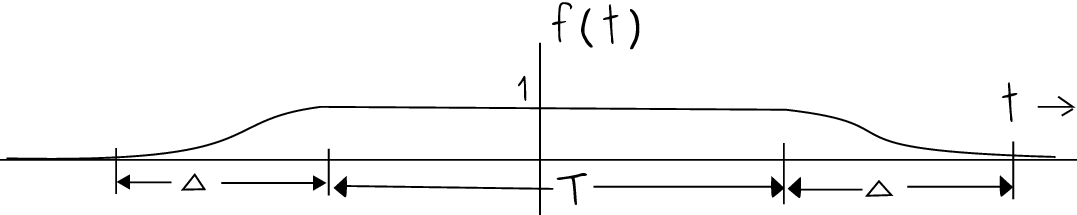}\]

Assuming the theory has a ground state, the vacuum-to-vacuum scattering matrix element ought to be easy to calculate. \uline{ex nihil nihil}. You start out with nothing you end up with nothing.

If you calculate $\langle 0|S|0\rangle$ however, you will not get one.
\begin{align*}
\text{Let }\;\;\;|0\rangle_P&=\substack{\text{ground state of the whole}\\ \text{Hamiltonian, with energy $E_0$}}\\
|0\rangle&= \substack{\text{ground state of $H_0$ as usual}}
\end{align*}

Let's look at the scattering process in the Schr\"odinger picture.

For $t<-\frac{T}{2}$ we have $|0\rangle$; At $t\approx-\frac{T}{2}$, in time $\Delta$, the interaction turns on adiabatically. The adiabatic hypothesis says that $|0\rangle$ turns into $|0\rangle_P$ with probability 1. It can pick up a phase, $e^{-i\gamma_-}$. From $t=-\frac{T}{2}$ to $t=+\frac{T}{2}$, the state evolves with the full Hamiltonian, it rotates as $e^{-iE_0t}$ and picks up a total phase $e^{-iE_0T}$. At $t\approx \frac{T}{2}$, as the interaction turns off adiabatically $|0\rangle_P$ turns back into $|0\rangle$, getting one more phase $e^{-i\gamma_+}$. The state we have for $t>\frac{T}{2}$ is $e^{-i(\gamma_-+\gamma_++E_0T}|0\rangle$.

We can transfer this to the interaction picture, to get
\[ \langle0|U_I(\infty,-\infty)|0\rangle=e^{-i(\gamma_-+\gamma_+ +E_0T)}\]

This is disgusting. A divergent phase. How will we get rid of it?

We'll change the theory. The problem is that there is a mismatch between the ground state energy of the full Hamiltonian and the ground state energy of the free Hamiltonian. Subtract the mismatch and we'll eliminate the problem.
\begin{align*}
H_I&\rightarrow \left[g\int d^3x\phi(\vec{x},t)\rho(\vec{x})-a\right] f(t) & a&=E_0
\end{align*}

It's obvious what will happen. The number, $a$, just exponentiates while the interaction is on.
\begin{align*}
\langle0|S|0\rangle&=e^{-i\left[(\gamma_++\gamma_-+E_0T)-aT(1+O\left(\frac{\Delta}{T}\right)\right]}\overset{!}{=}1\\
\text{take }\;\;\; a&=E_0+O\left(\frac{\Delta}{T}\right)
\end{align*}

This is the first example of what is called a counterterm. It counters a problem we ran into in scattering theory. It doesn't change the physics, but it fixes up the problem.

You might worry that there will be energy mismatches in the one or many particle energy states even after we get the energy mismatch in the ground states fixed up. There shouldn't be though. Because the physical states get far away from the potential at large times, we expect the energy difference between a state with one physical particle and the physical vacuum, to be the same as the energy difference between the bare particle and the bare vacuum. If the
vacuum energies are lined up, the one particle state energies should be lined up. We don't expect this to be true in model 3. The particles interact with themselves and they can never get away from that as a particle can get away from an external potential.
}{
 \sektion{10}{October 23}
\descriptionten
\section*{\uline{Model 2 solved} (conclusion)}
\[ H_I=f(t)\left[g\int d^3x\phi(\vec{x},t)\rho(\vec{x})-a\right]\]

$a$ is the vacuum energy counterterm chosen so that 
\[ \langle0|S|0\rangle=1\]

We have already argued that as $T\rightarrow\infty$, $a\rightarrow E_0$, where $E_0$ is the vacuum energy of the interacting theory (without the counterterm $a$). Finding $a$ is going to give us $E_0$. Now it is clear what the addition of a constant to the Hamiltonian does to $U_I(\infty,-\infty)$. The constant just exponentiates. Let's see this come out of our diagrammatic perturbation theory.

There are now three connected diagrams
\begin{align*}
& \underset{(1)}{\overset{1}{\bullet}\!\feyn{f}} & 
& \underset{(2)}{\overset{1}{\bullet}\!\feyn{f}\!\overset{2}{\bullet}} 
& &\underset{(3)}{\feyn{x}}
\end{align*}

Diagram (3), which has no lines coming out, is for the counterterm.
\[S=U_I(\infty,-\infty)=:e^{(1)+(2)+(3)}:=e^{(2)+(3)}:e^{(1)}: \]

(Since (2) and (3) are just numbers.) To set $\langle0|S|0\rangle=1$ is to set 
\[e^{(2)+(3)}=1\;\;\;\text{ i.e.~}\;\;\;(3)=-(2) \]

Since $a$ is an addition to the Hamiltonian, it had better be purely real. Diagram (3) is then pure imaginary, and diagram (2) in order to be cancellable had better come out pure imaginary. It didn't come out pure imaginary in Model (1), but there the source was time dependent.

Photons don't scatter off nailed down charges. Mesons don't scatter off nailed down nucleons. They only scatter off real nucleons (or off nailed down nucleons if there is some dynamical charged field in the theory).
\begin{align*}
(1)&=-ig\int f(t)\rho(x)\phi(x)d^3xdt \\
\text{and }\;\phi(x)&=\int \frac{d^3k}{(2\pi)^{3/2} \sqrt{2\omega_{\vec{k}}}}\left(e^{-ik\cdot x}a_{\vec{k}} + e^{ik\cdot x}a_{\vec{k}}^\dagger\right)\;\text{ so}\\
(1)&=-ig\int\frac{d^3k}{(2\pi)^{3/2} \sqrt{2\omega_{\vec{k}}}} \left(\widetilde{f}(\omega_{\vec{k}}\underset{\substack{ \longleftarrow\!\longrightarrow\\ \text{hermitian conjugates}}}{)a_{\vec{k}}^\dagger\widetilde{\rho} (\vec{k})+ \overbrace{\widetilde{f}(-\omega_{\vec{k}})}^{\widetilde{f} (\omega_{\vec{k}})^*}}a_{\vec{k}}\underbrace{\widetilde{\rho}(-\vec{k}) }_{\widetilde{\rho}(\vec{k})^*}\right)
\end{align*}

$f(t)$, our turning on and off function, has a Fourier transform that looks like
\begin{center}
\includegraphics[width=8 cm]{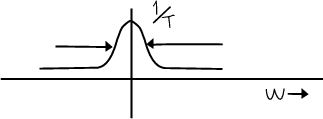}
\end{center}

As $T\rightarrow\infty$, $\widetilde{f}(\omega_{\vec{k}})$ goes to zero for every $\omega_{\vec{k}}\neq0$ and since $\omega_{\vec{k}}\geq \mu$ $\widetilde{f}(\omega_{\vec{k}})\rightarrow0$ for all $\vec{k}$. That is $(1)\rightarrow0$ as $T\rightarrow\infty$. (as long as we can set $(3)=-(2)$) We have found 
\[\boxed{S=1}\] 

This theory is a complete washout as far as scattering is concerned. While this was easy to see in the formalism we have built up, it was not easy when they were evaluating the theory in the Born approximation. Not until miraculous cancellations of all the terms at 4th order in the Born series occurred did people realize that they should try to prove $S=1$ to all orders.

Why is $S=1$? A time independent source can impart no energy. Since it can only create mesons one at a time and since $\omega=0$ is not on the mass shell, it cannot create mesons.

This result holds in the massless theory too. Since there is clearly no scattering for all $\vec{k}\neq0$, you only have to prove that for wave packets centered about $\vec{k}=0$, the failure at $\vec{k}=0$, a set of measure 0, does not screw up the wave packet.

\section*{\uline{Ground State Energy}, \uline{Ground State Wavefunction}}
In most QM courses these are discussed in a model long before scattering. You usually use time independent perturbation theory. I'll show you how to get these quantities out of the time dependent perturbation theory we have already developed.

Why is the ground state energy interesting? We have been studying the response of the meson field to a classical source. In meson-``nucleon" theory, the source will be $\psi^*\psi$. Our classical source theory is a lot like a meson-``nucleon" theory with the ``nucleons" nailed down. Take
\[\rho=``\delta"^{(3)}(\vec{x}-\vec{y}_1)+``\delta"^{(3)}(\vec{x}-\vec{y}_2)\]

The quotes are around the $\delta$ functions because we might want to smear them out a little bit. This is the charge density of two nucleons at $\vec{y}_1$ and $\vec{y}_2$. By computing the ground state energy and then by varying the positions we can find the potential between two ``nucleons".

This is the same thing we do in QM. We calculate the interaction between the two protons due to their interaction with the electron in $H_2^+$ by considering how the ground state energy of the electron varies with the separation of the protons. The protons are nailed down in that calculation, usually you say that the protons are so much heavier than the electrons and move so slowly that we can treat the response of the electron field to changes in positions of the protons as if the changes take place adiabatically. Of course in that calculation we also have a Coulomb potential between the protons. Here we are trying to get at the whole internucleon potential by saying it all comes from the interaction with the meson field. Of course the Coulomb potential in QM really comes from the interaction with the photons\dots

Now to calculate $a=E_0$ by setting $(3)=-(2)$:
\begin{align*}
(3)&=-i\int dtf(t)(-a)=iaT\left(1+O\left(\frac{\Delta}{T}\right)\right)=iE_0T \left(1+O\left(\frac{\Delta}{T}\right)\right)\\
(2)&=\frac{(-ig)^2}{2!}\int d^4x_1d^4x_2 f(t_1)\underset{\int\frac{d^4k}{(2\pi)^4}\frac{i}{k^2-\mu^2+i\epsilon} e^{ik\cdot(x_1-x_2)}}{f(t_2)\rho(\vec{x}_1) \rho (\vec{x}_2)\underbrace{\wick{1}{<1 \phi(x_1)>1 \phi(x_2)}}}\\
&=\frac{-ig^2}{2}\int \frac{d^3k}{(2\pi)^3}|\widetilde{\rho}(\vec{k})|^2 \int \frac{dk^0}{2\pi}|\widetilde{f}(\omega)|^2\frac{1}{\omega^2-\vec{k}^2 -\mu^2 +i\epsilon}
\end{align*}

Now $|\widetilde{f}(k^0)|^2$ is sharply concentrated at $k^0=0$, we can replace $\omega$ in $\frac{1}{\omega^2-\vec{k}^2-\mu^2+i\epsilon}$ by 0 (and then the $i\epsilon$ is not needed any longer). Also
\[\int_{-\infty}^\infty \underset{\substack{\text{famous theorem,}\\ \text{Parseval's theorem}}}{\frac{d\omega}{2\pi}|\widetilde{f}(\omega)|^2 \underbrace{=}}\underbrace{\int_{-\infty}^\infty dt |f(t)|^2}_{T+O(\Delta)}=T\left(1+O\left(\frac{\Delta}{T}\right)\right) \]

We could sum up these properties by saying something sloppy like
\[\lim_{T\rightarrow\infty}|\widetilde{f}(\omega)|^2=2\pi T\delta(\omega) \]

\noindent but what I have just shown is all (no more, no less) than that sloppy statement means.
\[(2)=\frac{ig^2}{2}T\left(1+O\left(\frac{\Delta}{T}\right)\right) \int\frac{d^3k}{(2\pi)^3}|\widetilde{\rho}(\vec{k})|^2\frac{1}{|\vec{k}|^2+\mu^2} \]

The moment of truth: Set $(3)=-(2)$, the $T$'s and $i$'s cancel.
\[ E_0\underset{T\rightarrow\infty}{=}\frac{-g^2}{2}\int\frac{d^3k}{(2\pi)^3}|\widetilde{\rho}(\vec{k})|^2\frac{1}{\vec{k}^2 +\mu^2} \]

The potential has come out in momentum space. To convert it to position space, 
\begin{align*}
\text{Define }\;\; V(\vec{x})&=-g^2\int\frac{d^3k}{(2\pi)^3}\frac{e^{i\vec{k}\cdot\vec{x}}}{\vec{k}^2 +\mu^2} \\
\text{Then }\;\; E_0&=\frac{1}{2}\int d^3xd^3y\rho(\vec{x})\rho(\vec{y})V(\vec{x}-\vec{y})
\end{align*}

The $\frac{1}{2}$ is the usual factor found even in electrostatics from overcounting the interaction when integrating over all space. For the two nucleon charge density, there will be four contributions. Two will be the interaction of the nucleons with themselves and two will be their interaction with each other (cancelled by the $\frac{1}{2}$).
\begin{align*}
\rho(\vec{x})&=``\delta"^{(3)}(\vec{x}-\vec{y}_1)+``\delta"^{(3)}(\vec{x}-\vec{y}_2) \\
E_0&=\underbrace{\text{something independent of $\vec{y}_1$, $\vec{y}_2$}}_{\substack{\text{If $``\delta"\rightarrow\delta$, this part $\rightarrow\infty$}\\ \text{Same problem as the self-energy}\\ \text{of a charged sphere in E.D.}}}+V(\vec{y}_1-\vec{y}_2)
\end{align*}

The usual procedure for the integration of spherically symmetric Fourier transforms, followed by a contour integration gives
\begin{align*}
V(r)&=\frac{-g^2}{4\pi r}e^{-\mu r} & r&=|\vec{y}_1-\vec{y}_2| & \text{Yukawa potential} 
\end{align*}

Looks like the Coulomb potential for $r\ll \mu^{-1}$ the Compton wavelength of the meson, and falls off rapidly for $r\gg\mu^{-1}$.

The force is attractive (between like charges) (because the particle mediating it has even integer spin) and short\footnote{$M_W$ is much larger and the weak force is thus much shorter ranged.} ranged because the mediating particle is massive. This potential has some of the essential features of the real nuclear force. Of course it doesn't include the effect of the whole family of mesons in the real world of multi-meson processes, but with $\mu=m_\pi$ it is a start.

The ground state wavefunction, is of course not a position space wavefunction (the expansion of $|\psi\rangle$ into $|x\rangle$'s), it is an expansion of $|0\rangle_P$ into the basis states $|\vec{k}_1,\dots,\vec{k}_n\rangle$ of the noninteracting theory.

To get the ground state wave function of model 2 using time dependent perturbation theory, we'll use the results of model 1.

Consider
\[\rho(\vec{x},t)=\begin{cases} \rho(\vec{x})e^{\epsilon t} & t<0,\;\; \epsilon\rightarrow0^+\\ 0 &t>0 \end{cases}\]

That is we turn it on very slowly, arbitrarily slowly so that at $t=0$ we finally have the full interaction of model 2, then we turn the interaction off abruptly.

Consider the $S$ matrix in this theory
\[\langle\vec{k}_1,\dots,\vec{k}_n|U_I(\infty,-\infty)|0\rangle=\langle\vec{k}_1,\dots,\vec{k}_n|U_I(\infty,0)U_I(0,-\infty)|0\rangle \]

Since the interaction is turned on arbitrarily slowly, $U_I(0,-\infty)$ should turn\footnote{screw the phase factor} the bare vacuum into $|0\rangle_P$. $U_I(\infty,0)$, the evolution by the free Hamiltonian alone, which is 1 in the interaction picture, does nothing on the left so 
\[\langle\vec{k}_1,\dots,\vec{k}_n|U_I(\infty,-\infty)|0\rangle= \langle\vec{k}_1,\dots,\vec{k}_n|0\rangle_P \]

\noindent 
which is what we are after.

Now we can apply the results of model 1 (Oct.~21, Eq.~(\ref{eq:09-UI}))
\begin{align*}
\langle\vec{k}_1,\dots,\vec{k}_n|0\rangle_P &=\langle\vec{k}_1,\dots,\vec{k}_n|U_I(\infty,-\infty)|0\rangle\\
&=e^{-\frac{|\alpha|}{2}}e^{\frac{i\beta}{2}}f(\vec{k}_1)\cdots f(\vec{k}_n)\\
f(\vec{k})&=\frac{1}{(2\pi)^{3/2}}\frac{1}{\sqrt{2\omega_{\vec{k}}}} (-ig)\widetilde{\rho}(\vec{k},\omega_{\vec{k}}) & |\alpha|&=\int d^3k|f(\vec{k})|^2\\
\widetilde{\rho}(k)&=\int d^4xe^{ik\cdot x}\rho(x)=\int d^3xe^{-i\vec{k} \cdot\vec{x}}\rho(\vec{x})\int_{-\infty}^0dte^{ik^0t}e^{\epsilon t}\\
&\underset{\epsilon>0}{=}\widetilde{\rho}(\vec{k})\frac{1}{ik^0+\epsilon} \xrightarrow[\epsilon\rightarrow0]{}-\frac{i}{k^0}\widetilde{\rho} (\vec{k})
\end{align*}

The probability for having $n$ mesons is
\[P_n=e^{-|\alpha|}\frac{|\alpha|^n}{n!}\]

What is $P_n$ for a point charge at the origin? That is
\[\rho(\vec{x})\rightarrow\delta^{(3)}(\vec{x})\;\;\;\substack{\text{(not $``\delta"$ smeared, the limit}\\ \text{of a real point charge)}} \]

Well, 
\[\widetilde{\rho}(\vec{k})\rightarrow1\;\;\;\text{ and} \]

\noindent 
this is bad news: at high $k$ we have a UV divergence in the integral for $|\alpha|$. 
\[|\alpha|=\int d^3k|f(\vec{k})|^2\underset{\text{high $k$}}{\sim} \int d^3k\frac{1}{\omega_{\vec{k}}^3}\underset{\text{high $k$}}{\sim} \int \frac{k^2dk}{k^3} \]

The integral is log divergent. Since $\langle N\rangle=|\alpha|$ (Oct.~21, Eq.~(\ref{eq:09-N})) we see that not only the energy of the field becomes infinite in the limit of a point nucleon, but the ground state flees Fock space.

These infinities are scary but not harmful.

Physically observable quantities like the $S$ matrix and the internucleon potential are hearteningly sensible.

Even if we don't go to the limit $``\delta"\rightarrow\delta$, but instead take the limit of massless mesons $\mu\rightarrow0$, $\langle N\rangle=|\alpha|\rightarrow\infty$, this time because of a small $k$ divergence. What about that?

Answer: So what if there are an infinite number of mesons in the ground state. An experimentalist will tell you he can only measure the existence of a meson down to some low energy, not arbitrarily low. If you say there are 1,000,000 mesons with a wavelength between $\frac{1}{2}$ light year and 1 light year, so what. It might be a problem if there were an infinite amount of energy at small $\vec{k}$, but there isn't.
\[\langle H\rangle=\int d^3k|f(\vec{k})|^2\omega_{\vec{k}}\;\;\;\substack{\text{This is manifestly positive. \uuline{Seems}}\\ \text{to contradict Yukawa pot.~result.}} \]

And the extra factor of $\omega_{\vec{k}}$ moderates the IR divergence. The integral is finite even as $\mu\rightarrow0$. Same with
\[\langle\vec{P}\rangle=\int d^3k|f(\vec{k})|^2\vec{k} \]

So it seems we have been lucky. In this simple theory, the divergences have restricted themselves to unobservable quantities. Maybe the divergences will break this quarantine in more complicated theories. In fact there is a surprisingly wide class of theories in which the divergences don't break the quarantine. They are called ``renormalizable" theories.

Next: Mass renormalization.

\section*{\uline{Model 3} and \uline{Mass Renormalization}}
The ground state energy, in perturbation theory, of this system is not necessarily zero. We will need a vacuum energy counterterm to make it zero. In this theory, that is not enough to make the 1-particle states of the Hamiltonian equal in energy to the 1-particle states of the full Hamiltonian. Indeed, the energy of a ``static nucleon" depended on its interaction with the meson field. Not only did the vacuum energy of a state with one static nucleon depend on $g$, the coupling, it depended on how smeared the nucleon was.

The change in energy of a particle due to its interaction with another field is called ``mass renormalization". The cure for this disease is also called ``mass renormalization".

Mass renormalization goes all the way back to hydrodynamics.

Suppose I have a ping pong ball, with mass equal to $1/20$ of the water it displaces.
\[m_0=\frac{1}{20}\rho V\]
\begin{center}
\includegraphics[width=2.5 cm]{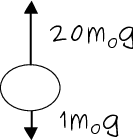}
\end{center}

Elementary hydrostatics tells you that there is an upward force on the ping pong ball equal to $g$ times the mass of the water it displaces. There is also the downward force of gravity on the ping pong ball itself. The net force on the ball is thus $19m_0g$ upward. Putting this in Newton's equation we have
\begin{align*}
m_0a&=19m_0g & a&=19g
\end{align*}

The ball accelerates upwards at $19g$.

This is \uline{nonsense}, as anyone who has ever held a ping pong ball underwater knows. The ping pong ball may accelerate up fast, but not at $19g$. The answer is not friction. You can see that the ping pong ball is not accelerating with $19g$ even when its velocity is low and friction or viscosity is negligible.

The answer is that in order to move the ping pong ball, you have to move some fluid. In order to accelerate the ping pong ball, you have to accelerate some fluid. Stokes solved the fluid motion around a sphere.
\begin{center}
\includegraphics[width=10 cm]{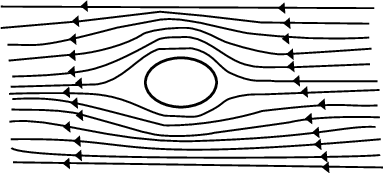}\\
Flow lines in rest frame of sphere, moving with velocity $\vec{v}$ through fluid
\end{center}

If a ball that displaces volume $V$ is moving with velocity $\vec{v}$ through the fluid, the fluid flow has a momentum, in the same direction as the ball, of $\frac{1}{2}\rho V\vec{v}$. The total momentum of the system, ball and fluid, is thus
\[\frac{1}{2}\rho V\vec{v}+m_0\vec{v}= 11m_0\vec{v}\]

\noindent 
which we set equal to the force after taking $d/dt$
\begin{align*}
\frac{dp}{dt}&=11m_0a=19m_0g=F\\
a&=\frac{19}{11}g
\end{align*}

See the derivation in Landau and Lifshitz, \uline{Fluid Mechanics}, leading up to the problem on p.36, for a more detailed understanding of the problem.

\subsection*{More motivation for mass renormalization:}
Two limits of classical field theory:

\begin{description}
\item{\uline{Point particle limit}: } Mass renormalization occurs

Example: GR

Point particle of mass $m$ creates a gravitational field which itself has energy density and creates further gravitational field.

\item{\uline{Classical field limit}: } Can read dispersion relation for low amplitude plane waves off of the quadratic part of the Lagrangian.
\end{description}

Since the quantum field theory will probably exhibit all the behavior of the worst classical limit, we better be prepared for mass renormalization.

\subsection*{Another example of renormalized perturbation theory}
You could try, in the statistical-mechanical theory of critical phenomena to calculate the critical temperature, as well as other properties of the system in terms of the microphysical parameters. However, you may be able to do computations much more easily if you trade in one of the microphysical parameters for the critical temperature.

The classical theory of the electron also suffers mass renormalization. Imagine the electron as a charged shell. The bare mass is $m_0$, the charge, $e$, the radius, $r$. There is a contribution to the measured mass of the electron other than $m_0$. There is the electrostatic energy (divided by $c^2$).
\[\underset{\text{measured, physical mass}}{\underbrace{m}=m_0+\frac{e^2} {2rc^2}} \]

Model 3 is going to suffer mass renormalization. The energy of a single meson state or a single nucleon state is going to depend on its interaction. We looked at static ``nucleons" interacting with the meson field in model 2. Recall that even for a single nucleon nailed down at $\vec{y}$
\[ \rho(\vec{x})=``\delta"^{(3)}(\vec{x}-\vec{y}) \]

\noindent 
The energy of the system depends in detail on how we smear out the $\delta$ function. In fact if we don't smear it out at all 
\[ \rho(\vec{x})=\delta^{(3)}(\vec{x}-\vec{y}) \]

\noindent 
the energy of the meson ground state $\rightarrow-\infty$. Now the energy of a one nucleon state includes the change in energy of the meson field its presence causes, and although the features of this effect may change when the coupling to the $\phi$ field goes from $\rho(\vec{x})\phi(x)$ (nailed down nucleons) to $\psi^*\psi\phi(x)$ (dynamical nucleons) there is no reason to expect it to go away.

This is going to be bad news for scattering theory. Just as the failure to match up the ground state energy for the noninteracting and full Hamiltonians in model 2 produced $T$ and $\Delta$ dependent phases in $\langle0|S|0\rangle$, the failure to match up one particle state energies in model 3 will yield $T$ and $\Delta$ dependent phases in $\langle \vec{k}|S|\vec{k}\,'\rangle$. Worse than that, it can even cause two wave packets that were arranged to collide, not to collide. I'll show how:

In our relativistic \uline{interacting} theory, for sufficiently weak coupling, we expect that there will be one nucleon states, $|\vec{p}\rangle_P$, which are eigenstates of the \uline{full} Hamiltonian, $H$:
\[H|\vec{p}\rangle_P=\sqrt{\vec{p}\,^2+m^2}|\vec{p}\rangle_P \]

There are also eigenstates of the free Hamiltonian, which because of this mass renormalization mess, can have a different mass. If I prepare a packet of these free Hamiltonian eigenstates they propagate along with group velocity
\[\vec{v}=\frac{\partial E}{\partial\vec{p}}\left(=\frac{\vec{p}}{E} \;\;\;\text{ when }\;\;\; E^2=\vec{p}\,^2+m^2\right) \]

When I turn the interaction on (slowly, so that the free Hamiltonian eigenstate (bare nucleon) turns into the dressed nucleon (full Hamiltonian eigenstate)) the group velocity changes because the mass in $E=\sqrt{\vec{p}\,^2+m^2}$ changes. I could set up a nucleon and meson to scatter, and if I turn on the interaction too early or too late, they might not even come close$!$

To fix up this problem, we are going to introduce new counterterms in our theory
\begin{multline}\label{eq:10-page13}
\mathcal{L} = \frac{1}{2}(\partial_\mu\phi)^2-\frac{\mu^2}{2}\phi^2 +\partial_\mu\psi^*\partial^\mu\psi-m^2\psi^*\psi+f(t)\Bigl[-g\psi^* \psi\phi+\underbrace{a}_{\substack{\text{vacuum energy}\\ \text{\uline{density} counterterm}}}\\
+\underbrace{\frac{b}{2}\phi^2}_{\text{meson mass counterterm}}+\underbrace{c\psi^*\psi}_{\substack{\text{``nucleon"}\\ \text{mass counterterm}}}\Bigr]
\end{multline}

$\mu$ is the \uline{measured} mass of the meson. $m$ is the \uline{measured} mass of the nucleon. 

When the interaction is off $(f(t)=0)$ this theory is a free theory with mesons of mass $\mu$ and nucleons of mass $m$.

When the interaction is turned on $(f(t)=1)$, we \uline{arrange}, by adjusting $b$ and $c$, that the one meson state has mass $\mu$ and the one nucleon state has mass $m$. This eliminates the phases in the one particle matrix elements by matching the energy of the one particle state (the vacuum energy, which in an infinite volume system may be infinite, being proportional to the volume, is adjusted to zero with the help of $a$).

To summarize, the conditions determining $a$, $b$, and $c$ are
\begin{align*}
\begin{split}
\langle0|S|0\rangle &=1\Rightarrow a \\
\overset{\substack{\text{one meson states}}}{\langle\overbrace{\vec{q}}|S|\overbrace{\vec{q}\,'}\rangle}&=\delta^{(3)}(\vec{q}-\vec{q}\,') \Rightarrow b\\
\overset{\substack{\text{one nucleon states}}}{\langle\overbrace{\vec{p}}|S|\overbrace{\vec{p}\,'}\rangle}&=\delta^{(3)}(\vec{p}-\vec{p}\,') \Rightarrow c
\end{split}\;\;\;
\substack{\text{The one meson and one nucleon states shouldn't}\\ \text{do anything; they have got nothing to scatter (only vacuum)}}
\end{align*}

This procedure should match up the energies of states of widely separated nucleons and mesons, without any additional twiddling. The energy of two widely separated mesons, even in model 3 when they are affected by self-interaction, should be the sum of their respective energies. Matching the energies of the vacuum and the one particle state matches the energy of states of widely separated mesons too. Although the particles in model 3 never become separated from their own fields, in the far past / future, they always become widely separated from each other. 

Sometimes it is useful to think about 
\[ \mu_0^2\equiv\mu^2-b \;\;\;\text{ and }\;\;\; m_0^2\equiv m^2-c \]

\noindent 
the coefficients of $\frac{1}{2}\phi^2$ and $\psi^*\psi$ in the full Lagrangian, respectively, although they have very little physical significance. What our procedure amounts to is breaking up the free and interacting parts of the Hamiltonian in a less naive way. You are always free to break up the free and interacting parts of the Hamiltonian any way you like, although you won't get anywhere unless you can solve the free Hamiltonian.

We have put $\frac{b}{2}\phi^2$ and $c\psi^*\psi$ in with the interaction $-g\psi^*\psi\phi$ because that way the mass of the meson (nucleon) is $\mu(m)$ when the interaction is off (manifestly) and the mass of the meson (nucleon) is $\mu(m)$ when the interaction is on (by our careful checks of $b$ and $c$).

This procedure gives us a BONUS.

By making $b$ and $c$ (hence $\mu_0^2$ and $m_0^2$) quantities you compute, our perturbation theory is expressed in terms of the actual physical masses, not the dumb quantities, $\mu_0$ and $m_0$.

If you treated $m_0$ and $\mu_0$ as fundamental, you would calculate all your cross sections, bound state energy levels, all quantities of interest, in terms of them, and them to make contact with reality, you would have to calculate $\mu$ and $m$, the physical masses, in terms of $m_0$ and $\mu_0$ too. Since no one is interested in your $m_0$ and $\mu_0$, to present your results, you would have to reexpress all your cross sections in terms of $\mu$ and $m$.

We have bypassed that mess by turning perturbation theory on its head. Instead of a perturbation theory for $m^2$, $\mu^2$ and all other physical quantities in terms of $m_0^2$ and $\mu_0^2$, we have a perturbation theory (for $m_0^2$ and $\mu_0^2$) and all physical quantities in terms of the observed masses, $m$ and $\mu$. 
}{
 \sektion{11}{October 28}
\descriptioneleven
\section*{\uline{{\sc Feynman Diagrams}} in Model 3}
{\small \[\langle0|(S-1)|0\rangle=\langle\vec{p}|(\underset{\substack{\text{one-nucleon}}}{S-1)|\vec{p}\,'\rangle}=\langle\vec{k}|(\underset{\substack{\text{one-meson}}}{S-1)|\vec{k}\,'\rangle} \]}
Let's look at nucleon-nucleon scattering at $O(g^2)$. That is the first order at which there is a contribution to 
\[\langle \underset{\text{two-nucleon states}}{\underbrace{p_1'p_2'}|(S-1)|\underbrace{p_1p_2}}\rangle\]

The $-1$ in $S-1$ is there because we aren't really interested in the no scattering process, $p_1=p_1'$ and $p_2=p_2'$ or $p_1=p_2'$ and $p_2=p_1'$, which comes from 1, the $O(g^0)$ term in 
\[ S=Te^{-ig\int d^4x\left(\psi^*\psi\phi-\frac{b}{2g}\phi^2-\frac{c}{g}\psi^*\psi\right)} \]

(Since the power series for $b$ and $c$ begin at order $g^1$ at the earliest (they are zero if $g=0$), it is not misleading to pull the $g$ out in front of the whole interaction and talk about the $O(g^0)$ contribution to $S$.)

There are no arrows over $p_1$, $p_2$, $p_1'$ and $p_2'$ because we are going to use the states that transform nicely under Lorentz transformations
\begin{align*}
U(\Lambda)|p_1,p_2\rangle&=|\Lambda p_1,\Lambda p_2\rangle\\
|p_1,p_2\rangle &=b^\dagger(p_1)b^\dagger(p_2)|0\rangle\\
b^\dagger(p)&=(2\pi)^{3/2}\sqrt{2\omega_{\vec{p}}} \, b^\dagger_{\vec{p}}
\end{align*}

So we don't have to worry about Bose statistics demand $p_1\neq p_2$ and $p_1'\neq p_2'$. We can recover what we've lost by building wave packets concentrated \uline{around} $p_1=p_2$.

The term in $S$ with two factors of the interaction is
\[\frac{(-ig)^2}{2!}\int d^4x_1d^4x_2 T[\psi^*\psi\phi(x_1)\psi^*\psi\phi(x_2)] \]

After all the hoopla about the turning on and off function, we are abandoning it, being sloppy: The only term in the Wick expansion of this term in $S$ that can contribute to two nucleons goes to two nucleons is
\[\frac{(-ig)^2}{2!}\int d^4x_1d^4x_2:\psi^*\psi\wick{1}{<1\phi(x_1) \psi^*\psi>1\phi(x_2)}: \]

$\wick{1}{<1\phi(x_1)>1\phi(x_2)}$ is some number. Let's look at
\begin{equation}\label{eq:11-page2}
\langle p_1'p_2'|:\psi^*(x_1)\psi(x_1)\psi^*(x_2)\psi(x_2):|p_1p_2 \rangle 
\end{equation}

The nucleon annihilation terms in $\psi(x_1)$ and $\psi(x_2)$ have to be used to annihilate the two incoming nucleons. The nucleon creation terms in $\psi^*(x_1)$ and $\psi^*(x_2)$ have to be used to create two nucleons, so as not to get zero inner product (alternatively I could say they have to be used to ``annihilate two nucleons on the left''). In equations
\[\langle p_1'p_2'|:\psi^*\psi(x_1)\psi^*\psi(x_2):|p_1p_2 \rangle=\langle p_1'p_2'|\psi^*(x_1)\psi^*(x_2)|0\rangle\langle0|\psi (x_1)\psi(x_2)|p_1p_2 \rangle \]

You can easily show that the two contributions to the second matrix element are (the c.c.~equation is also used)
\begin{equation}\label{eq:11-botpage2}
\langle0|\psi(x_1)\psi(x_2)|p_1p_2\rangle=\underset{\uparrow\text{$p_1$ absorbed at $x_1$\quad\quad\quad}}{e^{-ip_1\cdot x_1-ip_2\cdot x_2}+e^{-ip_1\cdot x_2-ip_2\cdot x_1}} 
\end{equation}

So there are four contributions to our matrix element 
\begin{align}
\langle p_1'p_2'|:\psi^*\psi(x_1)\psi^*\psi(x_2):|p_1p_2\rangle &= \bigl(e^{ip_1'\cdot x_1+ip_2'\cdot x_2}+e^{ip_1'\cdot x_2+ip_2'\cdot x_1}\bigr)\bigl(e^{-ip_1\cdot x_1-ip_2\cdot x_2}+e^{-ip_1\cdot x_2-ip_2 \cdot x_1}\bigr)\nonumber\\
&=\underset{F}{e^{ip_1'\cdot x_1+ip_2'\cdot x_2-ip_1\cdot x_1-ip_2 \cdot x_2}}+\underset{L}{e^{ip_1'\cdot x_2+ip_2'\cdot x_1-ip_1\cdot x_2-ip_2 \cdot x_1}}\nonumber\\
&+\underset{I}{e^{ip_1'\cdot x_2+ip_2'\cdot x_1 -ip_1\cdot x_1-ip_2 \cdot x_2}}+\underset{O}{e^{ip_1'\cdot x_1+ip_2' \cdot x_2-ip_1\cdot x_2-ip_2\cdot x_1}}
\end{align}

Notice that the pair of terms on the first line differ only by $x_1\leftrightarrow x_2$ and that the pair of terms on the second line only differ by $x_1\leftrightarrow x_2$. Since $x_1$ and $x_2$ are to be integrated over and since $\wick{1}{<1\phi(x_1)>1 \phi(x_2)}$ is symmetric under $x_1\leftrightarrow x_2$ these pairs give identical contributions to the matrix element. We'll just write one of the pairs, which cancels the $\frac{1}{2!}$. We have,
\begin{multline}
(-ig)^2\int d^4x_1 d^4x_2 \wick{1}{<1\phi(x_1)>1\phi(x_2)}\bigl(e^{ip_1'\cdot x_1+ip_2'\cdot x_2 -ip_1\cdot x_1-ip_2 \cdot x_2}+e^{ip_1'\cdot x_2+ip_2'\cdot x_1 -ip_1 \cdot x_1-ip_2 \cdot x_2}\bigr)\\
=(-ig)^2\int d^4x_1d^4x_2\int\frac{d^4k}{(2\pi)^4}\bigl[e^{ix_1\cdot(p_1'-p_1+k)}e^{ix_2\cdot(p_2'-p_2-k)}\\
+e^{ix_1\cdot (p_2'-p_1+k)}e^{ix_2\cdot(p_1'-p_2-k)}\bigr]\frac{i}{k^2-\mu^2+i\epsilon}
\end{multline}

I have used the expression for $\wick{1}{<1\phi(x_1)>1\phi(x_2)}$ and grouped all the exponential factors by spacetime point.

With these two integrals (there are two terms in the integrand) go two pictures
\begin{align}\label{eq:11-page4}
&\underset{(a)}{\Diagram{\momentum{fV}{p_1'} & \momentum{fV}{p_1} \\
& \momentum[ulft]{fv}{k\downarrow}\\
\momentum{fV}{p_2'} & \momentum{fV}{p_2}}}
& &\substack{N+N\rightarrow N+N\\ \text{Feynman diagrams @ $O(g^2)$}} &
&\underset{(b)}{\Diagram{\momentum{fV}{p_2'} & \momentum{fV}{p_1} \\
& \momentum[ulft]{fv}{k\downarrow}\\
\momentum{fV}{p_1'} & \momentum{fV}{p_2}}}\quad\substack{\text{Notice external lines each}\\ 
\text{have an associated momentum.}\\ \text{The vertices are not numbered.}}
\end{align} 

\noindent 
and two stories\footnote{$\begin{array}{cc} \\ \text{Conventions:}& \text{in on right}\\ &\text{out on left} \end{array}$}. The story that goes with picture (a) is this. A nucleon with momentum $p_1$ comes in and interacts. Out of the interaction point comes a nucleon with momentum $p_1'$ and a ``virtual meson.'' This ``virtual meson'' then interacts with a nucleon with momentum $p_2$ and out of the interaction point comes a nucleon with 
momentum $p_2'$. The interaction points $x_1$ and $x_2$ can occur anywhere, and so they are integrated over. Furthermore, this ``virtual meson'' can have any momentum $k$ and this is integrated over, although you can see from the factor $\frac{i}{k^2-\mu^2 +i\epsilon}$, ``Feynman's propagator'' that it likes to be on the meson mass shell, although with $k^0=\pm\sqrt{\vec{k}^2+\mu^2}$.

Fairy tales like this helped Feynman discover and think about quantum electrodynamics. In our formalism, they are little more than fairy tales, but in a formulation of quantum particle mechanics called the path integral formulation they gain some justification. The words not only match the pictures, they parallel the mathematics.

The $x_1$ and $x_2$ integrations are easy to do. We get
\begin{multline}\label{eq:11-page5}
(-ig)^2\int\frac{d^4k}{(2\pi)^4}\frac{i}{k^2-\mu^2+i\epsilon}\bigl[(2\pi)^4\delta^{(4)}(p_1'-p_1+k)(2\pi)^4\delta^{(4)}(p_2'-p_2-k)\\ +(2\pi)^4\delta^{(4)}(p_2'-p_1+k)(2\pi)^4\delta^{(4)}(p_1'-p_2-k)\bigr]
\end{multline}

Because the interaction is spacetime translationally invariant, after integrating the interaction point over all space-time we get delta functions which enforce energy-momentum conservation at every vertex.

All the features of this computation generalize to more complicated $S$ matrix element contributions. I'll give a set of rules for writing down these integral expressions for contributions to $S$ matrix elements. First, I'll explain in general why there are no combinatoric factors in model 3 to worry about, no symmetry numbers in the integral expressions.

Take a given operator in the Wick expansion, which has an associated Wick diagram
\[ D, \text{ a diagram }\longleftrightarrow \frac{:O(D):}{n(D)!} \]

Designate which of the lines leading out of the diagram annihilates each incoming particle, and which of the lines creates each outgoing particle.

This is one contribution to the $S$ matrix element.

Now consider summing over the permutation of the numbered points in the Wick diagram. While only $\frac{n(D)!}{S(D)}$ of these permutations actually correspond to different terms in the Wick expansion, in model 3, all $n(D)!$ of these permutations correspond to different contributions to the $S$ matrix element. This cancels the $\frac{1}{n(D)!}$ exactly.

There are other possible designations (in general) for the way the external lines connect to the vertices of the Wick diagram. If they differ just by a permutation of the vertices then we have already counted them (by cancelling the $\frac{1}{n(D)!}$). If they don't differ by just a permutation of the vertices then they correspond to a different Feynman diagram (the difference between (a) and (b) in Eq.~(\ref{eq:11-page4})

Only in certain theories, like Model 3, do the $n(D)!$ permutations of the vertices all make different contributions to the $S$ matrix, and in fact this is true in model 3 only when a diagram\footnote{in fact, each connected part has to have at least one external line} has at least one external line. In that case there is an unambiguous way of identifying each vertex in the diagram. Contributions to $\langle0|S|0\rangle$, which have no external lines, can have symmetry factors. The unambiguous labelling statement for diagram (a) is
\begin{align*}
&\Diagram{\momentum{fV}{p_1'} & \momentum{fV}{p_1} \\
& \momentum[ulft]{fv}{k\downarrow}\\
\momentum{fV}{p_2'} & \momentum{fV}{p_2}} & \substack{\text{The upper vertex is uniquely labelled as the one where $p_1$ is absorbed.}\\ \\ \\ \text{The lower vertex is the one connected to the upper vertex by a muon line.}}
\end{align*}

In this theory, as soon as one vertex is labelled (by an external line) they are all uniquely labelled.

\section*{Feynman Rules for Model 3}
You should convince yourself by taking some other low order terms in the Wick expansion of $S$ and looking at some simple matrix elements they contribute to that the following set of rules applied to the diagram always gives you the correct contribution to the $S$ matrix element.

For external lines $\begin{Bmatrix} \text{incoming}\\ \text{outgoing}\end{Bmatrix}$ momenta are directed $\begin{Bmatrix} \text{in}\\ \text{out}\end{Bmatrix}$.

Assign a directed momentum to every internal line.
\begin{align*}
&\text{For every} & &\text{Write}\\
& \begin{array}{c}
\text{internal meson line}\\
\Diagram{ \momentum[bot]{f}{\leftarrow k}}
\end{array}
& & \int\frac{d^4k}{(2\pi)^4}\frac{i}{k^2-\mu^2+i\epsilon}\\
& \begin{array}{c}
\text{internal nucleon line}\\
\Diagram{ \momentum[bot]{fV}{\leftarrow p}}
\end{array}
& & \int\frac{d^4p}{(2\pi)^4}\frac{i}{p^2-m^2+i\epsilon}\\
& \begin{array}{c}
\text{vertex}\\
\Diagram{ \momentum[bot]{fV}{\leftarrow p'} & \momentum[bot]{fV}{\leftarrow p} \\
& \momentum[bot]{fd}{k\nwarrow}}\end{array}
& & (-ig)(2\pi)^4\delta^{(4)}(p'-p-k)\\
& \begin{array}{cc}
\text{meson vacuum counterterm}\\
\bullet& \substack{(2\pi)^4\delta^{(4)}(0)\text{ would turn into the}\\ \text{volume of all spacetime if}\\ \text{the system were in a box. This}\\ \text{c.t.~diagram is designed to cancel}\\ \text{diagrams without external lines}\\ \text{which you will see also have a}\\ \text{factor of } \delta^{(4)}(0)\\}\end{array}
& & ia(2\pi)^4\delta^{(4)}(0)\\
& \begin{array}{cc}
\text{meson mass counterterm}\\
\Diagram{ \momentum[bot]{f}{\leftarrow k'} x \momentum[bot]{f}{\leftarrow k}}
& \substack{\text{although the meson mass}\\ \text{counterterm had a }\frac{1}{2}\text{ in}\\ \text{the Lagrangian, there is no}\\ \frac{1}{2}\text{ here because there are}\\ \text{two ways to do the interaction}\\ 
}\end{array}
& & ib(2\pi)^4\delta^{(4)}(k-k')\\
& \begin{array}{c}
\text{nucleon mass counterterm}\\
\Diagram{ \momentum[bot]{fV}{\leftarrow p'} & x &\momentum[bot]{fV}{\leftarrow p}}
\end{array}
& & ic(2\pi)^4\delta^{(4)}(p-p')
\end{align*}

A catalog of all Feynman diagrams in model 3 up to $O(g^2)$ (except those related by $C$ or $T$ will not be written down twice)

Order $g$
\newcounter{elevenLcount}
\begin{list}{(\arabic{elevenLcount})}
{\usecounter{elevenLcount}
\setlength{\rightmargin}{\leftmargin}}
\item $\Diagram{fdV \\ & f \\ fuA} = 0 $ if $\mu<2m$ by energy momentum conservation.
\item $\Diagram{f0 flSA flSu fs0 f} =0 $\\
\end{list}

Order $g^2$
\begin{list}{(\arabic{elevenLcount})}
{\usecounter{elevenLcount}
\setcounter{elevenLcount}{2}
\setlength{\rightmargin}{\leftmargin}}
\item $\begin{array}{c}
\Diagram{fdV \\ & f \\ fuA}\\\Diagram{fdV \\ & f \\ fuA} \end{array} =0^2$ if $\mu<2m$
\item $\left.\begin{array}{lc}
\text{(a)} & \Diagram{f flV fluA f}\\
\\
\\
\text{(b)} & \Diagram{flV flu f0 & f & f0 flV flu}\\
\\
\\
\text{(c)} & \begin{array}{c}\bullet\\
\substack{\uparrow\\ \text{vacuum energy c.t.~to $O(g^2)$}}\end{array}\end{array}\right\}
\substack{\text{\normalsize Because we demand there be no corrections to}\\\text{\normalsize $\langle0|S|0\rangle$ these sum to zero. This fixes the vacuum}\\ \text{\normalsize energy c.t.~to $O(g^2)$}}$
\end{list}

In 4(c) think of the c.t.~as $O(g^2)$. Its value is \underline{determined} by the fact that it has to cancel some $O(g^2)$ contributions to the vacuum-to-vacuum $S$ matrix element.
\begin{list}{(\arabic{elevenLcount})}
{\usecounter{elevenLcount}
\setcounter{elevenLcount}{4}
\setlength{\rightmargin}{\leftmargin}}
\item $\left.\begin{array}{lc}
\text{(a)} & \Diagram{fV fV fl f fV }\\
\\
\\
\text{(b)} & \Diagram{fV x fV}\\ 
& \substack{\uparrow\\ \text{nucleon 
mass c.t.~to
$O(g^2)$}}\end{array} \right\}\substack{\text{\normalsize Sum to zero because we demand that}\\\text{\normalsize there are no corrections to
$\langle\!\!\!\!\!\!\!\!\underbrace{\vec{p}|S|\vec{p}\,'}_{\substack{\text{one nucleon each}}}\!\!\!\!\!\!\!\!\rangle$}
}$ \\
\\
\\
$\begin{array}{lc}
\text{(c)} &\begin{matrix}\!\!\!\!\!\!\!\!\!\!\!\!\Diagram{f0 flV flu}\\ \\ \Diagram{fV fv fV}\end{matrix}\end{array}$ this diagram comes out zero for the same reason as (2) does.
\\
\\
\item $\left.\begin{array}{lc}
\text{(a)} & \Diagram{f f0 flA fluV f0 f }\\
\\
\\
\text{(b)} & \Diagram{f x f}\end{array}\right\}\substack{\text{\normalsize Sum to zero because we demand that}\\\text{\normalsize there are no corrections to $\langle\!\!\!\!\!\!\underbrace{\vec{k}|S|\vec{k}\,'}_{\substack{\text{one meson each}}}\!\!\!\!\!\!\rangle$}
}$ \end{list}

The remaining order $g^2$ diagrams are more interesting. They contribute to the following processes
\begin{list}{(\arabic{elevenLcount})}
{\usecounter{elevenLcount}
\setcounter{elevenLcount}{6}
\setlength{\rightmargin}{\leftmargin}}
\item $N+N\longrightarrow N+N$ (connected by $C$ to $\overline{N} + \overline{N} \longrightarrow \overline{N} + \overline{N}$)
\item $N + \overline{N} \longrightarrow N + \overline{N}$
\item $N+\phi\longrightarrow N+\phi$ (connected by $C$ to $\overline{N} + \phi \longrightarrow \overline{N} + \phi$)
\item $N + \overline{N} \longrightarrow \phi + \phi$ (connected by $T$ to $\phi + \phi \longrightarrow N + \overline{N}$)
\end{list}

Although $\phi+\phi\longrightarrow\phi+\phi$ appears in this theory, indeed it must appear, it does not do so until $O(g^4)$. The diagram is
\[ \Diagram{ fd & & & fu \\ & fA \\ &fvA & fvV\\ & fV \\ fu & & & fd } \]

We have already written down the contributions to process (7). The diagrams are 
\begin{align*}
&\underset{(a)}{\Diagram{ \momentum{fV}{p_1'} & \momentum{fV}{p_1} \\
& \momentum[ulft]{fv}{k\downarrow}\\ \momentum{fV}{p_2'} & \momentum{fV}{p_2}}} \quad \text{\normalsize and }\quad \underset{(b)}{\Diagram{\momentum{fV}{p_2'} & \momentum{fV}{p_1} \\
& \momentum[ulft]{fv}{k\downarrow}\\
\momentum{fV}{p_1'} & \momentum{fV}{p_2}}} \quad\text{\normalsize a.k.a.} \hspace{0.5cm} \includegraphics[width=2cm]{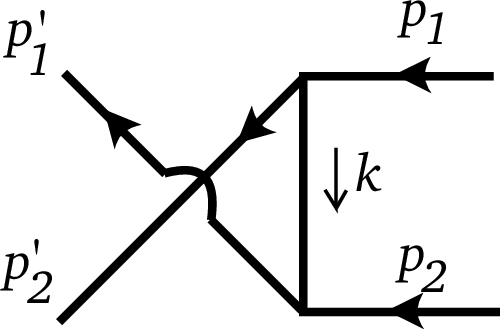}\\ 
\end{align*} 

They give (see Eq.~(\ref{eq:11-page5}))
\begin{multline*}
(-ig)^2\frac{i}{(p_1-p_1')^2-\mu^2+i\epsilon}(2\pi)^4\delta^{(4)} (p_1'+p_2'-p_1-p_2)\\
+(-ig)^2\frac{i}{(p_1-p_2')^2-\mu^2+i\epsilon}(2\pi)^4\delta^{(4)} (p_1'+p_2'-p_1-p_2)
\end{multline*}

Note that the $\frac{1}{(2\pi)^4}$ associated with $d^4k$ exactly cancels with the $(2\pi)^4$ associated with the $\delta$ functions used to do the integral. All our formulas have been arranged so that $2\pi$'s always go with $\delta$'s and $\frac{1}{2\pi}$'s always go with $\int dk$'s.

Note that you can shortcut these trivial integrations over $\delta$ functions by just assigning internal momenta so as to conserve momentum whenever an internal momentum is determined by the other momenta at a vertex.

Finally, note that performing the trivial integrals over $\delta$ functions always gives you a factor 
\begin{align*}
(2\pi)^4\delta^{(4)}(\!\!\!\!\!\!\!\underbrace{p_f}_{\substack{\text{sum of all}\\
\text{final momenta}}}\!\!\!\!\!\!\!\!-\!\!\!\!\!\!\!\!\overbrace{p_i}^{\substack{\text{sum of all}\\\text{initial momenta}}}\!\!\!\!\!\!\!\!) \quad\quad\substack{\text{\normalsize at least when the diagram}\\ \text{\normalsize is of one connected piece}}
\end{align*}

We define $a_{fi}$, the invariant Feynman amplitude by 
\[\langle f|(S-1)|i\rangle = ia_{fi}\delta^{(4)}(p_f-p_i)\]

The factor of $i$ is inserted to match the phase convention of NRQM.

For $N+N\longrightarrow N+N$
\[ia = (-ig)^2\left[\frac{i}{(p_1-p_1')^2-\mu^2+i\epsilon}+\frac{i}{(p_1-p_2')^2-\mu^2+i\epsilon}\right]\]

Let's look at this in the COM frame.
\begin{align*}
\begin{array}{lc}
p_1=(\sqrt{p^2+m^2},p\!\!\!\!\overbrace{\vec{e}}^{\text{unit vector}}\!\!\!\!)\\
p_2=(\sqrt{p^2+m^2},-p\vec{e})\\
p_3=(\sqrt{p^2+m^2},p\vec{e}\,')\\
p_4=(\sqrt{p^2+m^2},-p\vec{e}\,')\\ 
\end{array}
\end{align*}
\begin{center}
\includegraphics[width=5 cm]{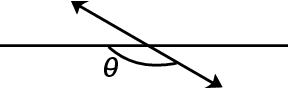}\\
$\vec{e}\cdot\vec{e}\,'=\cos\theta$, $\theta$ is the scattering angle in the COM frame.
\end{center}

$E_T=2\sqrt{p^2+m^2}$ is often used to characterize collisions. In the nonrelativistic limit, which we will be taking, $p$ is more useful.

Define the momentum transfer, $\Delta$, and the crossed momentum transfer, $\Delta_c$, by
\begin{align*}
(p_1-p_1')^2&=-\Delta^2\\
(p_1-p_2')^2&=-\Delta_c^2
\end{align*}

In our COM variables
\begin{align*}
\Delta^2&=2p^2(1-\cos\theta) & \Delta_c^2&=2p^2(1+\cos\theta)
\end{align*}

The invariant Feynman amplitude is
\[ a=g^2\left[\frac{1}{\Delta^2+\mu^2}+\frac{1}{\Delta_c^2+\mu^2} \right] \]

We have dropped the $i\epsilon$ because it is unnecessary. For physically accessible values of $\Delta^2$ and $\Delta_c^2$ the denominators are never less than $\mu^2$.

The first term is peaked (peaked sharper at higher $p$) in the forward ($\theta\approx0$) direction. The second term produces an identical peak in the backward direction. Of course when identical particles collide who is to say what is forward and what is backward. $\theta=0$ is indistinguishable from a scattering angle of $\theta=\pi$. The probability had better have come out symmetrical.

People were scattering nucleons off nucleons long before quantum field theory was around, and at low energies they could describe scattering processes adequately with NRQM. Let's try to understand our amplitude in NRQM. First we'll find the NR analog of the first term.

In the COM frame, two body scattering is simplified to the problem of scattering of a potential (classically and quantum-mechanically). P.T.~at lowest order gives 
\begin{align*}
\langle \vec{k}\,'|S-1|\vec{k}\rangle &\propto \langle \vec{k}\,'|V|\vec{k}\rangle \\
&=\int d^3r V(\vec{r})e^{-i\vec{\Delta}\cdot\vec{r}} & \text{``Born'' approximation}\\
&=\widetilde{V}(\vec{\Delta}) & \vec{\Delta}=\vec{k}\,'-\vec{k}
\end{align*}

To explain the first term in our scattering amplitude using NRQM we must have 
\[ \widetilde{V}(\vec{\Delta})\propto\frac{1}{\Delta^2+\mu^2}\Longrightarrow V(\vec{r}\,)\propto\frac{g^2e^{-\mu r}}{r} \]

Our amplitude, which is characterized by having a simple pole in a physically unobservable region, at $\Delta^2=-\mu^2$, corresponds to the Born approx.~to the Yukawa interaction$!$

The second term also has an analog in NRQM. With two identical particles, the Hamiltonian should contain an exchange potential
\begin{align*}
H&=H_0+\!\!\!\underset{\substack{\\ \text{Yukawa}\\ \text{potential}}}{V}\!\!\!+\!\!\!\!\!\!\!\!\!\!\!\!\!\!\!\!\!\underbrace{VE}_{\substack{\quad\quad\quad\text{exchange Yukawa potential}}} &
\underset{\substack{$E$\text{ is the exchange operator}}}{E|\vec{r}_1,\vec{r}_2\rangle=|\vec{r}_2,\vec{r}_1\rangle}\\
V&|\vec{r}_1,\vec{r}_2\rangle \propto \frac{g^2e^{-\mu r}}{r}|\vec{r}_1,\vec{r}_2\rangle & r=|\vec{r}_1-\vec{r}_2|
\end{align*}

The exchange Yukawa potential is the source of a simple pole in the amplitude at $\Delta_c^2=-\mu^2$, also in a physically unobservable region. The NRQM amplitude is proportional to $\widetilde{V}(\vec{\Delta})+\widetilde{V}(\vec{\Delta}_{exch})$ ($\vec{\Delta}_{exch}=\vec{k}\,'+\vec{k}$). In a partial wave expansion
of the amplitude, the exchange potential gives a contribution $\left\{\begin{array}{c} \text{identical} \\ \text{opposite}\end{array}\right\}$ to the direct potential if $l$ is $\left\{\begin{array}{c} \text{even} \\ \text{odd}\end{array}\right\}$.

This is because in the COM an eigenstate of angular momentum is an eigenstate of the exchange operator $E$ with eigenvalue $(-1)^l$.
}{
 \sektion{12}{October 30}
\descriptiontwelve
\section*{$N + \overline{N} \rightarrow N + \overline{N}$ ``nucleon antinucleon" scattering}
\begin{center}
\includegraphics[width=15cm]{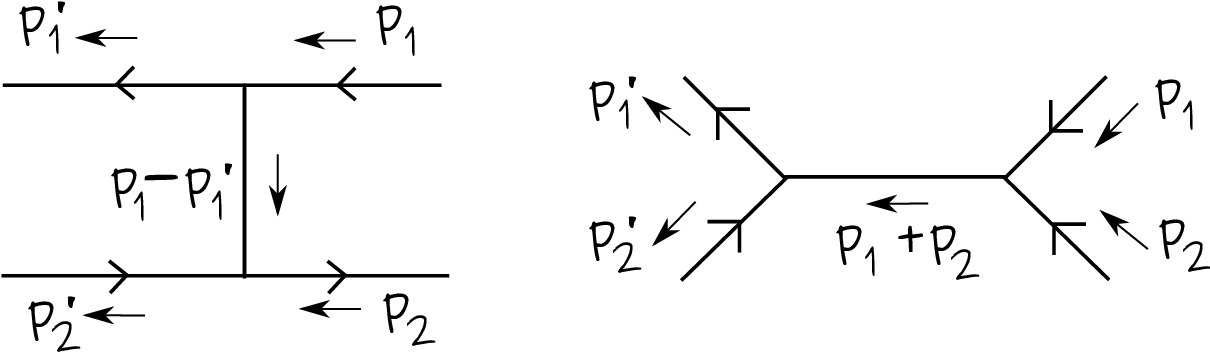}
\end{center}

Notice that my labeling of internal lines has been done so as to be consistent with energy-momentum conservation. This does away with the two steps of:
\begin{enumerate}
\item labelling the momenta arbitrarily and 
\item performing the trivial integration over the arbitrarily labelled momenta that are actually fixed by $\delta$ functions.
\end{enumerate}

From the diagram I write down
\[ ia = (-ig)^2 \left[ \frac{i}{(p_1 - p'_1)^2 - \mu^2} + \frac{i}{(p_1 + p_2)^2 - \mu^2} \right] \]

This has a less symmetric structure than than the amplitude for $N+N \rightarrow N + N$, but that is not unexpected. The symmetry of the amplitude for $N+N \rightarrow N + N$ was forced upon us because of identical particles in the incoming and outgoing states. Bose statistics does not apply to the the incoming and outgoing states of $N$ and $\overline{N}$.\\

We've already found what the first term is by going to the COM frame and taking the NR limit. It is a Yukawa potential. What about the second term? In the COM frame
\[ (p_1 + p_2)^2 = 4 \big( \sqrt{p^2 + m^2} \big)^2 = \big[2 \sqrt{p^2 + m^2}\big]^2 = E_T^2 \]

\noindent
where $E_T$ the total energy in the COM frame. Now 
\[ \frac{1}{(p_1 + p_2)^2 - \mu^2} = \frac{1}{E_T^2 - \mu^2} = \frac{1}{E_T - \mu} \frac{1}{E_T + \mu} \simeq \frac{1}{2m+\mu}\frac{1}{E_T - \mu} \]

\noindent
in the NR limit.\\

We have not replaced $E_T$ by $2m$ in the second term because $2m$ could be very near $\mu$. This can cause a rapid variation in this factor. We'll see this is because the intermediate state is spinless. Notice that this amplitude is independent of the scattering angle, $\theta$. A partial wave decomposition would show a contribution only to the $S$ wave.\\

What is the explanation of this in terms of NRQM?\\

Let us suppose there is an energy eigenstate just below threshold, i.e. an energy eigenstate with an energy slightly less than $2m$. Then even in perturbation theory, it may cause a significant contribution to the scattering amplitude in the \underline{second} term of the Born expansion
\[ a \propto \langle f | V | i \rangle + \sum_n \frac{\langle f | V | n \rangle \langle n |V | i \rangle}{E_T - E_n \pm i \epsilon} \]

\noindent
where $E_n$ could be a possible source of an energy eigenstate pole.
\begin{center}
\includegraphics[width=6cm]{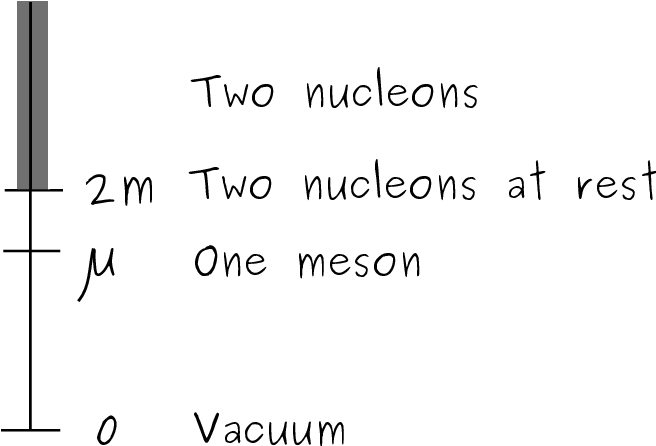}
\end{center}

In the COM frame the energy spectrum of \underline{possible} intermediate states looks like a state with $E_n = \mu$ and non-vanishing matrix elements could produce a pole in the amplitude. (Continuum states produce a branch cut in the amplitude.) A pole occurs in the partial wave that has the same angular momentum as the intermediate state.

\section*{$N + \phi \Rightarrow N + \phi$ meson-nucleon scattering}
\begin{center}
\includegraphics[width=15cm]{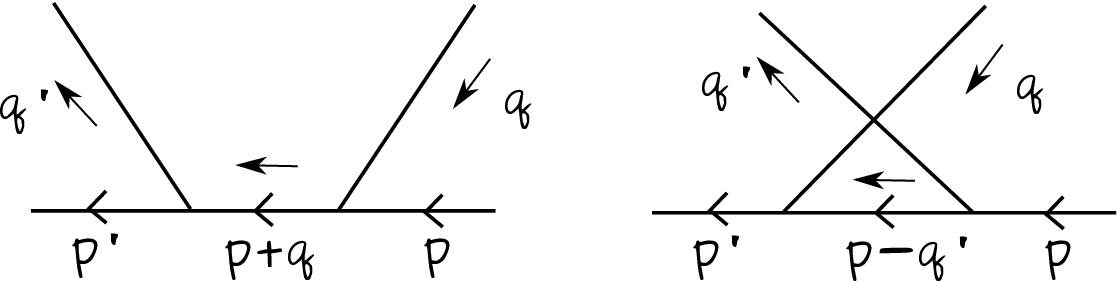}
\end{center}

The language that goes along with the second graph, the crossed graph, is that the outgoing meson is emitted first, then the incoming meson is absorbed. Of course this is no more kinematically possible for the intermediate nucleon than the first graph.
\[ i a = (-ig)^2 \left[ \frac{i}{(p+q)^2 -m^2 + i \epsilon} + \frac{i}{(p-q')^2 -m^2 + i \epsilon} \right] \]

This is an energy eigenstate pole and an exchange Yukawa potential. Perhaps it is clear to see that this is an exchange Yukawa potential if we redraw the second graph as 
\begin{center}
\includegraphics[width=6cm]{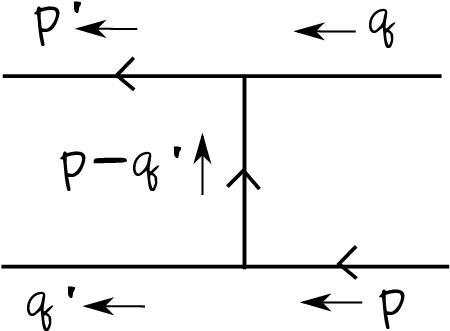}
\end{center}

Notice there is no direct Yukawa term. There is a resonance in the $p$ wave of pion nucleon scattering called the $N^*(12_{--})$. There is nothing in the $s$ wave. Usually one thinks of a resonance as caused by an attractive force that nearly creates a bound state. Usually the $s$ state is the most tightly bound. This is a classical expectation one has because the angular momentum barrier is the lowest. What kind of force could create a $p$ wave state but no $s$ wave state? A repulsive exchange force, which is attractive in odd partial waves. Because the particles have different masses this potential is different from the ones we've had before. The second term in the amplitude
\[ -g^2 \frac{1}{(p-q')^2 - m^2 + \underbrace{i\epsilon}_{\text{\tiny unnecessary}}} = g^2 \frac{1}{-(p-q')^2 + m^2} \]

\noindent
has a denominator in the COM frame of 
\[ -(p-q')^2 + m^2 = -(\sqrt{p^2 + m^2} - \sqrt{p^2 + \mu^2})^2 + \underbrace{p^2(1+2 \cos\theta)}_{\Delta_c^2} + m^2 \]

\noindent
since 
\[ p = (\sqrt{p^2 +m^2}, p \vec{e}\,) \]
\[ p' = (\sqrt{p^2 +m^2}, p \vec{e}\,') \]
\[ q = (\sqrt{p^2 +\mu^2}, -p \vec{e}\,) \]
\[ q' = (\sqrt{p^2 +\mu^2}, -p \vec{e}\,') \]

This has the usual $p^2(1+2\cos\theta)$ exchange Yukawa forward peak, but what we would call the range is p dependent, i.e energy dependent
\[ \frac{1}{\text{(range of pot)}^2} = -(\sqrt{p^2 +m^2} - \sqrt{p^2 + \mu^2})^2 + m^2 \]

The energy dependent part vanishes when $\mu^2 = m^2$. Note that as $p \rightarrow \infty$ this $\rightarrow m^2$ and as $p \rightarrow 0$ this $\rightarrow -(m-\mu)^2 + m^2 = 2m\mu - \mu^2 = \mu(2m-\mu)$.\\

It can have a long range at low energies if the mass $\mu$ is small. If one is bold, we can start applying these ideas to real pion nucleon interactions. However we still need to develop spin and isotopic spin to really get things right (the sign of the potential for one) and the pion-nucleon coupling is strong which means lowest order calculation can't be trusted (except at long range or high partial waves).

\section*{$N + \overline{N} \rightarrow \phi + \phi$ ``nucleon-anti-nucleon" annihilation.}
In 1930, this was sensational. 
\begin{center}
\includegraphics[width=15cm]{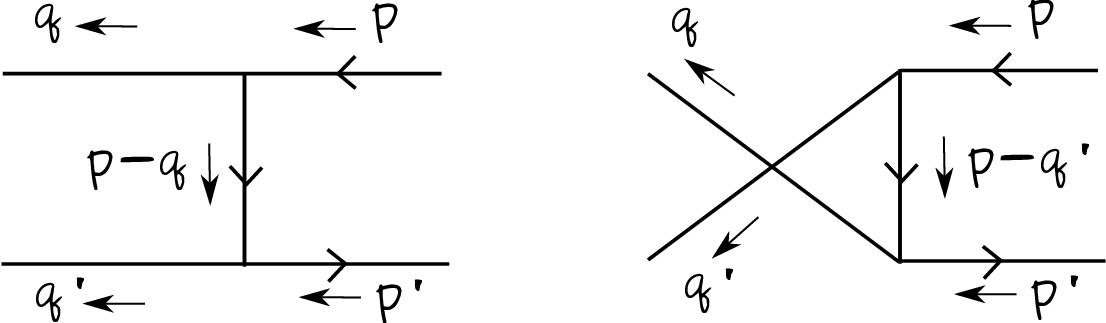}
\end{center}

The first graph is a Yukawa potential. The second is an exchange Yukawa potential.\\

Our next topic is a discussion of the connection between Yukawa potentials, exchange Yukawa potentials and energy eigenstate poles in relativistic scattering theory. 

In NRQM, there is absolutely no connection between these things. You can have any one (or two) without having all three. We'll develop some formalism which will be useful later to describe the connection.

\section*{Crossing [Symmetry]}

[Symmetry]: Brackets because this has nothing to do with symmetries and particles in the sense we have discussed them.\\

Imagine a general $2 \rightarrow 2$ scattering process $ 1+2 \rightarrow \overline{3} + \overline{4}$. We'll denote the amplitude (or some contribution to the amplitude) by 
\begin{center}
\includegraphics[width=8cm]{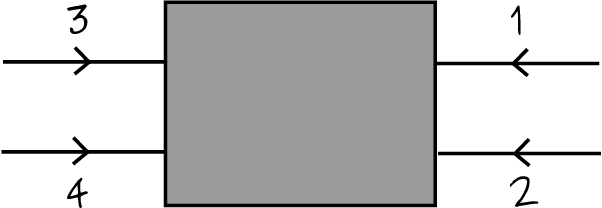}
\end{center}

\noindent
where the numbers on the lines tell you what type of particles propagate along these lines (with the arrows), and you aren't supposed to worry about what type of interactions are hidden from view when the lines go behind the shield. \\

The particle of type 1 is incoming, its momentum is $p_1$, the particle of type $2$ is also incoming, with momentum $p_2$, the particle of type $\overline{3}$ is outgoing, with momentum $p_3$, and the particle of type $\overline{4}$ is outgoing with momentum $p_4$. This you can tell because of our convention of putting incoming lines on the right and outgoing lines on the left, we read graphs as if time flowed from right to left, in analogy with the way we write down matrix elements.\\

Now we are going to abandon that convention. \\

Then who is to say this is not the amplitude for $3+4 \rightarrow \overline{1} + \overline{2}$, reading left to right, or $3+1 \rightarrow \overline{2} + \overline{4}$, reading top to bottom (or even $1 \rightarrow \overline{2} + \overline{3} + \overline{4}$??). \\

Well, we have another method for fixing a convention, which is useful (but not necessary) for discussing crossing. The honest to goodness physical momenta in the theory always are on the upper sheet of their mass hyperboloids, i.e
\[ p_3^2 = m_3^2 \] 
or 
\[ p_3^{0 \, 2} = m_3^2 + |\vec{p}_3|^2 \]
and 
\[ p_3^0 > 0 \]

There are no negative energy states in our theories.\footnote{Ignore everything written about QFT when it starts talking about negative energy particles.} Thus there will be no confusion if we flag a momentum by sending it to minus itself. If someone gives you a momentum $p$, with $p^0 <0$, you know what they are really giving you is a physical momentum, $-p$ and a wink, a flag, an extra bit of information. We'll use that extra bit of information to specify whether a particle is incoming or outgoing. \\

We will orient all momenta inward on our general $2\rightarrow 2$ graph:
\begin{center}
\includegraphics[width=8cm]{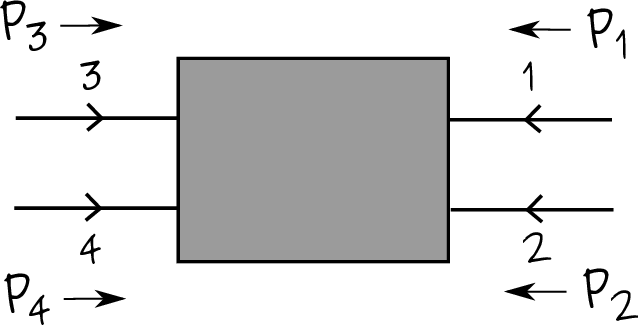}
\end{center}

The orientation on the page no longer matters. And if, say, $p_1^0$ and $p_3^0$ are less than zero and $p_2^0$ and $p_4^0$ are greater than zero, what this actually stands for is (what we used to mean by)
\begin{center}
\includegraphics[width=8cm]{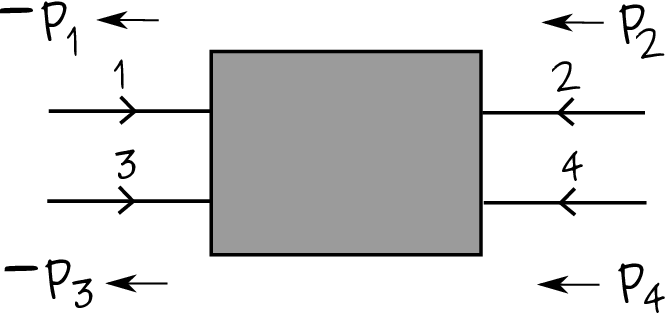}
\end{center}

\noindent
which is the amplitude for $2 + 4 \rightarrow \overline{1} + \overline{3}$. Note that the notation has been set up, so that in all cases, the energy momentum conserving delta function is $\delta^{(4)}(p_1 + p_2 + p_3 + p_4)$.\\

Mathematically, instead of graphically, what we have defined is a new function of three momenta. ($p_1 + p_2 + p_3 + p_4$ is restricted to be zero. If you like you could think of the function as a function of four momenta, which is zero whenever $p_1 + p_2 + p_3 + p_4$ is not equal to zero\footnote{[But don't.] $a_{fi}$ should be thought of as a function of parameters that parametrize the surface $p_1^2=p_2^2=p_3^2=p_4^2=m^2$. $p_1+p_2+p_3+p_4=0$. 3 independent momenta $\Rightarrow$ 6 Lorentz invariants, 4 constraints $\Rightarrow$ 2 parameters.}.) Just to keep an air of symmetry we'll display all four momenta in the function. 
\[ ia (p_1, p_2, p_3, p_4) \]

This function is the amplitude for a particle of type 2 with momentum $p_2$ and a particle of type 4 with momentum $p_4$, and scatters into a particle of type $\overline{1}$ with momentum $-p_1$, and a particle of type $\overline{3}$ with momentum $-p_3$, when $p_2^0$ and $p_4^0$ are $>0$ and $p_1^0$ and $p_3^0$ are $<0$. It is also the amplitude for a bunch of other processes when the time components of the three independent momenta take on their various possible signs. Another way of writing the amplitude for $2 + 4 \rightarrow \overline{1} + \overline{3}$ using this function is to take all the momenta, $p_1$, $p_2$, $p_3$ and $p_4$ to be their honest to goodness physical values ($p_1^0,p_2^0,p_3^0$ and $p_4^0>0$) and write 
\[ ia (-p_1, p_2, -p_3, p_4) \]

There is no reason we can't assemble the amplitude for all these different processes into a single process like this, but there is also no obvious reason it is any more useful that graphing the Dow Jones on the positive axis and the temperature in Miami on the same graph on the negative real axis. \\ 
\begin{center}
\includegraphics[width=10 cm]{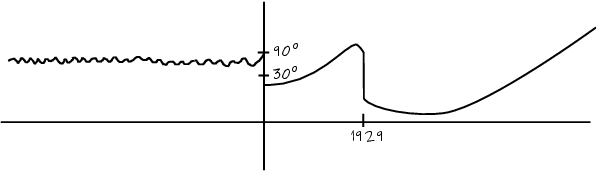}
\end{center}

\textbf{Food for thought ($ 3 \rightarrow 3 $ scattering)}
\begin{center}
\fcolorbox{white}{white}{
  \begin{picture}(217,164) (109,-34)
    \SetWidth{0.5}
    \SetColor{Black}
    \Line[arrow,arrowpos=0.5,arrowlength=5,arrowwidth=2,arrowinset=0.2](197,90)(117,119)
    \Line[arrow,arrowpos=0.5,arrowlength=5,arrowwidth=2,arrowinset=0.2](284,129)(196,90)
    \Line(196,90)(186,45)
    \Line[arrow,arrowpos=0.5,arrowlength=5,arrowwidth=2,arrowinset=0.2](186,45)(110,25)
    \Text(216,48)[lb]{{\Black{$\swarrow$ virtual nucleon}}}
    \Line[arrow,arrowpos=0.5,arrowlength=5,arrowwidth=2,arrowinset=0.2](252,25)(185,44)
    \Line[arrow,arrowpos=0.5,arrowlength=5,arrowwidth=2,arrowinset=0.2](320,44)(252,26)
    \Line(254,25)(260,-8)
    \Line[arrow,arrowpos=0.5,arrowlength=5,arrowwidth=2,arrowinset=0.2](260,-7)(182,-33)
    \Line[arrow,arrowpos=0.5,arrowlength=5,arrowwidth=2,arrowinset=0.2](325,-23)(261,-8)
  \end{picture}
}
\end{center}

{\small [Tools for analysis of this limit:
\begin{enumerate}
\item macro? causality?
\item Landau rules
\item Graphs with single poles]
\end{enumerate}}

Also think of this as $2 \rightarrow 2$ scattering followed 1 billion light years later by another $2 \rightarrow 2$ scattering. 
\begin{center}
\fcolorbox{white}{white}{
  \begin{picture}(378,134) (99,-34)
    \SetWidth{0.5}
    \SetColor{Black}
    \Line[arrow,arrowpos=0.5,arrowlength=5,arrowwidth=2,arrowinset=0.2](197,60)(117,89)
    \Line[arrow,arrowpos=0.5,arrowlength=5,arrowwidth=2,arrowinset=0.2](284,99)(196,60)
    \Line(196,60)(186,15)
    \Line[arrow,arrowpos=0.5,arrowlength=5,arrowwidth=2,arrowinset=0.2](186,15)(110,-5)
    \Line[arrow,arrowpos=0.5,arrowlength=5,arrowwidth=2,arrowinset=0.2](475,56)(407,38)
    \Line(407,39)(413,6)
    \Line[arrow,arrowpos=0.5,arrowlength=5,arrowwidth=2,arrowinset=0.2](413,7)(335,-19)
    \Line[arrow,arrowpos=0.5,arrowlength=5,arrowwidth=2,arrowinset=0.2](476,-9)(412,6)
    \Line[arrow,arrowpos=0.5,arrowlength=5,arrowwidth=2,arrowinset=0.2](409,37)(186,16)
    \Text(268,34)[lb]{{\Black{1 billion years}}}
    \Text(96,-38)[lb]{{\Black{A scattering in another galaxy}}}
    \Text(418,-39)[lb]{{\Black{One scattering}}}
  \end{picture}
}
\end{center}

There must be some appropriate limit where $2 \rightarrow 2$ followed by $2 \rightarrow 2$ is a limit of $3\rightarrow 3$. The virtual internal line must somehow become almost real. That's why you can get a vague description of virtual particles by thinking about them as real particles.\\ 

Let's define three relativistic invariants to describe $2\rightarrow 2$ scattering processes. 
\[ s \equiv (p_1 + p_2)^2 = (p_3 + p_4)^2 \]
\[ t \equiv (p_1 + p_3)^2 = (p_2 + p_4)^2 \]
\[ u \equiv (p_1 + p_4)^2 = (p_2 + p_3)^2 \]

If particle 3 is outgoing, $-p_3$ is its actual 4-momentum.

For the process $1+2 \rightarrow \overline{3} + \overline{4}$, $\sqrt{s}$ is the total COM energy, $-t$ is the momentum transfer squared, and $-u$ is the crossed momentum transfer squared. I have made these last two choices arbitrarily. If $1 \neq 2$ and $\overline{3} \neq \overline{4}$ and if $1 = \overline{3}$ or $2 = \overline{4}$, the choice is standard. If $1 \neq 2$ and $\overline{3} \neq \overline{4}$ and if $1 = \overline{4}$ or $2 = \overline{3}$, the choice is bassackwards; I ought to call $-u$ the momentum transfer$^2$ and $-t$ the crossed momentum transfer$^2$. In all other cases, anybody's designation is arbitrary.\\

Now, there are only two relativistic invariants describing a $2 \rightarrow 2$ scattering process of spinless particles. They are often taken as the COM total energy and scattering angle. $s$, $t$, and $u$ are three relativistic invariants. They must be redundant. Here is a (nice symmetric) derivation of their interdependence.
\begin{align*}
2 (s+t+u) &= (p_1 + p_2)^2 + ( p_3 + p_4)^2 + (p_1 +p_3)^2 + (p_2+p_4)^2 + (p_1 + p_4)^2 + (p_2 + p_3)^2 \\
&= 3 \sum_{a=1}^4 m_a^2 + 2 \sum_{a>b} p_a \cdot p_b
\end{align*}

Now use 
\[ 0 = (\sum_a p_a)^2 = \sum_a {m_a}^2 + 2 \sum_{a>b} p_a \cdot p_b \]
to see 
\[ 2 (s+t+u) = 2 \sum_{a=1}^4 m_a^2 \]
i.e 
\[ s+t+u = \sum_a {m_a}^2 \]

There is a symmetrical way of graphing three variables in the plane, when they are restricted like this. Look at the plane in $s-t-u$ space, $s+t+u = \sum_a m_a^2$
\begin{center}
\includegraphics[width=6 cm]{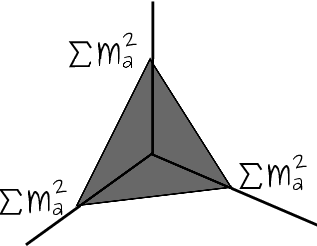}
\end{center}

Looking down perpendicular at this plane, you get the idea of representing $s$, $t$ and $u$ in the plane by 
\[ s = \vec{r}\cdot \widehat{e_s} + \frac{1}{3} \sum_a m_a^2 \]
\[ t = \vec{r}\cdot \widehat{e_t} + \frac{1}{3} \sum_a m_a^2 \]
\[ u = \vec{r}\cdot \widehat{e_u} + \frac{1}{3} \sum_a m_a^2 \]
\vspace{1cm}
\begin{center}
\includegraphics[width=6cm]{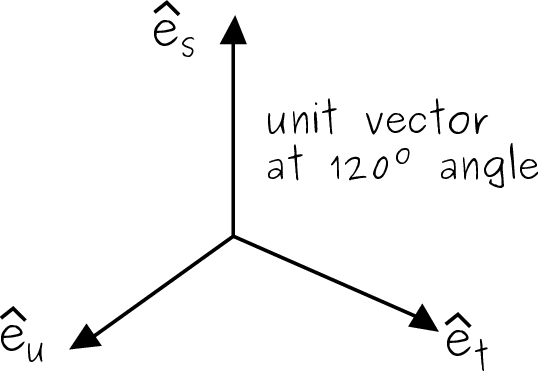}
\end{center}

Each vector $\vec{r}$ in the plane gives you a triple $s$, $t$, $u$, and since $\widehat{e_s} +\widehat{e_t} +\widehat{e_u}$ is obviously $\vec{0}$ (rotational invariance) the set satisfies $s+t+u = \sum_a {m_a}^2$. We have a ``Mandelstam-Kibble plot".
\begin{center}
\includegraphics[width=10cm]{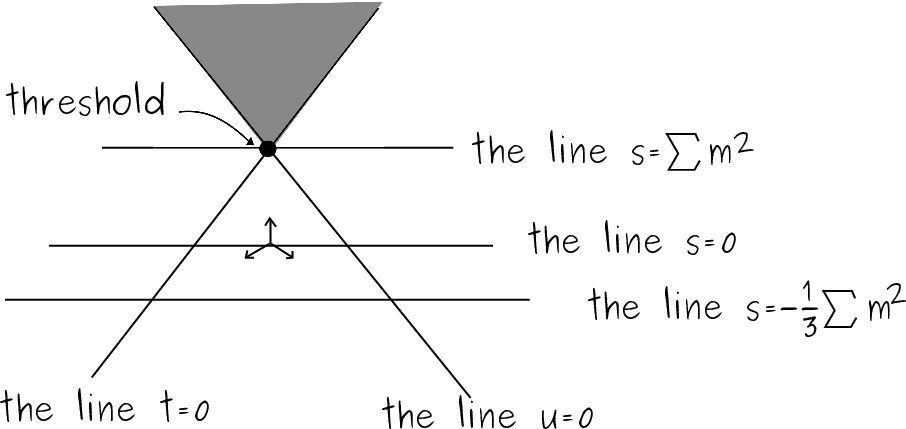}
\end{center}

When all four masses are equal, $m_a^2 = m^2 $, $a=1,2,3,4$, the shaded area, $s > 4m^2$, $u<0$, $t<0$ is the physically accesible region for the process $1 + 2 \rightarrow \overline{3} + \overline{4}$ and the process $3 + 4 \rightarrow \overline{1} + \overline{2}$.\\

[MORE FOOD: The article ``Uniqueness property of the Twofold Vacuum Expectation'' by Paul G.~Federbush and Kenneth A.~Johnson, Phys.~Rev.~120, 1926 (1960) was attached at this point.]\\

In an abuse of the scattering term ``channel'', the process $1 + 2 \rightarrow \overline{3} + \overline{4}$ is called the s-channel, and the crossed processes $1 + 3 \rightarrow \overline{2} + \overline{4}$ and $1 + 4 \rightarrow \overline{2} + \overline{3}$ are called the t-channel and u channel respectively because $\sqrt{t}$ and $\sqrt{u}$ are the total COM energy in these processes. In model 3, the lowest order scattering amplitude for the process $N+\phi \rightarrow N + \phi$ was (using our new wacky conventions, $p_3^0<0$, $p_4^0 < 0$).
\begin{center}
\includegraphics[width=12cm]{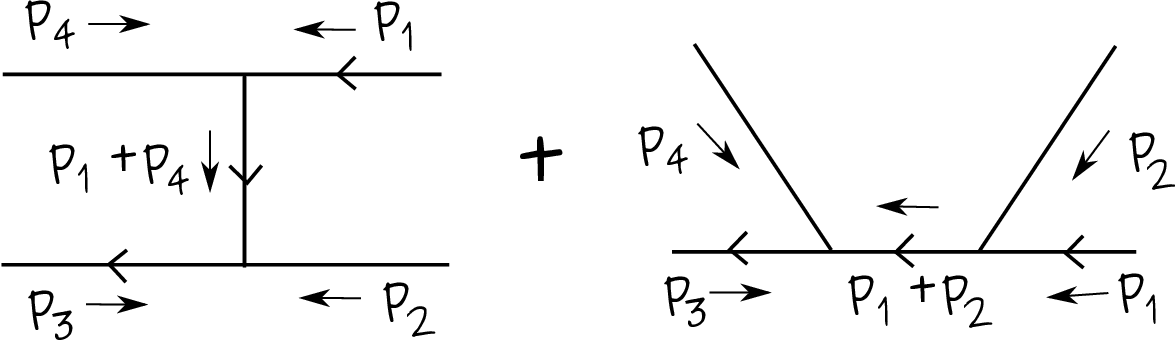}
\end{center}
\begin{align*} 
&= (-ig)^2 \left( \frac{i}{(p_1+p_4)^2 - m^2 + i\epsilon} + \frac{i}{(p_1+p_2)^2 - m^2 + i\epsilon} \right) \\
&= (-ig)^2\biggl(\underbrace{\frac{i}{u-m^2+i\epsilon}}_{\substack{\text{exchange Yukawa}\\\text{ interaction}}} + \underbrace{\frac{i}{s-m^2 + i \epsilon}}_{\substack{\text{energy eigenstate}\\\text{ pole}}} \biggr)
\end{align*}

The lowest order scattering amplitude for the corresponding $u$ channel process (I'm thinking of $1=N$, $2=\phi$, $3=N$, $4=\phi$) is $N + \phi \rightarrow N+ \phi$
\begin{center}
\includegraphics[width=12cm]{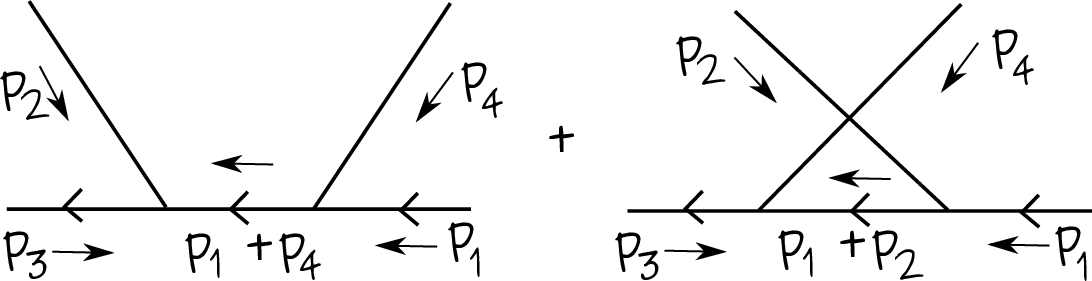}
\end{center}
\[ = (-ig)^2\biggl(\underbrace{\frac{i}{u-m^2+i\epsilon}}_{\substack{\text{energy eigenstate}\\\text{ pole}}} + \underbrace{\frac{i}{s-m^2 + i\epsilon}}_{\text{exchange Yukawa}}\biggr) \]

How about that: the amplitude are the same although the interpretation of the two terms are different. \\

What about the t channel process $N + \overline{N} \rightarrow 2 \phi$.
\begin{center}
\includegraphics[width=12cm]{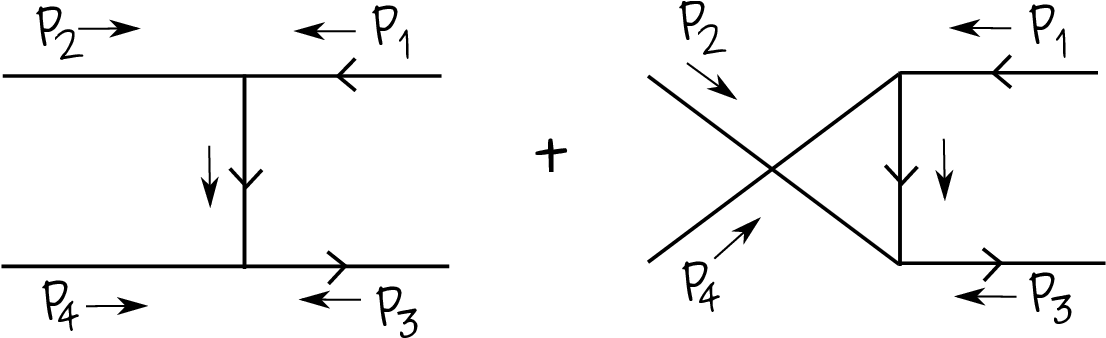}
\end{center}
\[ = (-ig)^2\left( \frac{i}{s-m^2 + i \epsilon} + \frac{i}{u-m^2+i\epsilon} \right) \]

\noindent
Again the amplitude is the same although the interpretation is different. (I'd rather not assign a NR interpretation to the two graphs because non-relativistically $N + \overline{N} \rightarrow 2 \phi$ can't occur.)

Every one of these amplitudes is the exact same function of $s$, $t$, and $u$. That is: the first amplitude is only defined in the shaded region 
\begin{center}
\includegraphics[width=10cm]{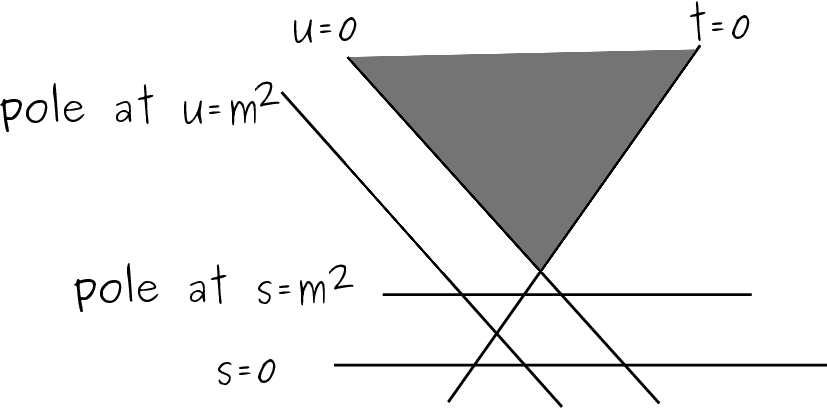}
\end{center}

\noindent
(for simplicity take $\mu = m$) where s, the COM energy squared, is greater than $4m^2$ and $t$ and $u$ are less than zero. \\

The second amplitude is only defined when $u > 4m^2 $ and $s$ and $t$ are less than zero. What we have observed is that if we analytically continue an amplitude for some process outside of its physical region to the physical region of some other process, we get the amplitude for that other process. \\

From our picture of the $s-t-u$ plane, it may look to you like the analytic continuation can't be performed even for the simple $\mathcal{O}(g^2)$ amplitudes we have discussed because the poles in $s$ and $u$ show up as lines which cut off one physical region from another. This is wrong because you can go around these poles by letting the variables become complex. Furthermore, they are avoidable singularities, that is, it doesn't matter how you go around them, you get the same analytic continuation. This brings up a tougher question: at this order in perturbation theory our amplitudes just have poles, but at higher orders they will have branch cuts, so can the analytic continuation from one physical region to another still be performed and if so, do you get the correct amplitude? The answer is yes and you do get the amplitude for one physical process by analytically continuing the amplitude for another, but you must follow specific prescriptions when going around the essential singularities.\\
\vspace{1cm}

Given this relation between amplitudes for different processes, we have related energy eigenstate poles, Yukawa interactions and exchange Yukawa interactions, three things which had no connection in nonrelativistic quantum mechanics. These effects are one and the same. A pole in $s$ in an $s$ channel process looks like an energy eigenstate pole. In the $u$ channel process that same pole looks like an exchange Yukawa potential.\\

They are two aspects of the same analytic function restricted to two disconnected regions of the plane. The next thing to ask is how do we lose the relationship when we take the nonrelativistic limit, $ c\rightarrow \infty$. As $c \rightarrow \infty$, the three physical regions on the Mandelstam plot which are separated by a distance of $\mathcal{O}(mc^2)$ get very far apart. They are only near each other for finite $c$. 
\begin{center}
\includegraphics[width=6cm]{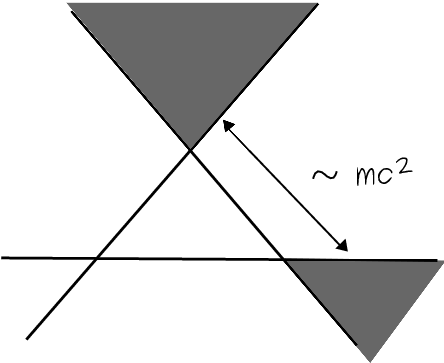}
\end{center}

The other thing we do in the NR limit is chuck terms of order $(v/c)^2$ which of course is an arbitrarily good approximation in this limit. The problem is that even if you have an excellent approximation to an analytic function if you analytically continue the approximation a long ways you may get something that doesn't remotely resemble the analytic function. An example will suffice. Consider $e^x$ on the real axis for $x< -1 \text{ million}$. In that region $0$ is a wonderful approximation to the function. But now analytically continue this wonderful approximation to $x = +1 \text{ million}$, and you discover you have completely missed the boat. It is in the way that the connection between different amplitudes is lost when you take the non-relativistic limit and chuck those teensie but important terms of $\mathcal{O}(\frac{v}{c})^2$ and higher. Although this has been illustrated only for $2 \rightarrow 2$ scattering it applies to any process.
\vspace{1cm}

\section*{CPT symmetry}

There are three scattering processes related to $1 + 2 \rightarrow \overline{3} + \overline{4}$ by various crossings. They are 
\begin{align*}
3 + 4 &\rightarrow \overline{1} + \overline{2} \\
2 + 4 &\rightarrow \overline{1} + \overline{3} \\
2 + 3 &\rightarrow \overline{1} + \overline{4} 
\end{align*}

Note that the physical region for $3 + 4 \rightarrow \overline{1} + \overline{2}$ is in the same region as that of $1 + 2 \rightarrow \overline{3} + \overline{4}$. There is no need to do any analytic continuation to show that the amplitudes for these two processes are the same. All you have to do is note that they are related by $p_a \rightarrow - p_a$, $a = 1, \ldots, 4$ and that all the Feynman rules are quadratic in the momenta so there is no way this operation can change the amplitude. Although we haven't discussed theories without parity invariance, if there is any grace in the world, parity violating interactions will involve an $\epsilon$ tensor contracted with four momenta and that too is an even power of momenta. This is an argument to all orders in perturbation theory that even in a parity violating theory the amplitudes for these two processes are equal. The argument applies to any process $n$ particles $\rightarrow$ $m$ particles by the same argument. The Feynman rules are invariant under $p_a \rightarrow -p_a$, $a= 1, \ldots, n, n + 1, \ldots, n+m$. This equality has nothing to do with analytic continuation. Graphically, if 
\begin{center}
\includegraphics[width=8cm]{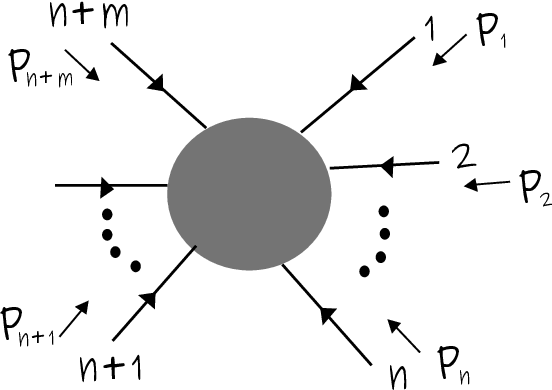}
\end{center}

\noindent
is calculated in a Lorentz invariant theory, it is invariant under $p_a \rightarrow -p_a $, $a = 1, \ldots, n+m$. This is the $CPT$ theorem. \\

Why is it called the $CPT$ theorem? I'll only explain why in the $2\rightarrow 2$ case, so I don't have to invent some notation. The most general $2\rightarrow 2$ process is 
\begin{align*} 
\begin{matrix}1\\ p_1\end{matrix} 
\begin{matrix} +\\ \\\end{matrix}
\begin{matrix}2\\ p_2\end{matrix} &
\begin{matrix} \rightarrow\\ \\ \end{matrix}
\underbrace{
\begin{matrix}\overline{3}\\ p_3\end{matrix}
\begin{matrix} +\\ \\\end{matrix}
\begin{matrix}\overline{4}\\p_4\end{matrix}}_{\substack{\text{physical}\\\text{4 momenta}}}\\
\text{amplitude}&=a(p_1, p_2, -p_3, -p_4)
\end{align*} 

\noindent 
If I charge conjugate the incoming and outgoing states, I get a related process 
\[ \overline{1} + \overline{2} \rightarrow 3+ 4. \]

\noindent 
In a charge conjugation invariant theory this process would have the same amplitude but in general it doesn't. Now let's consider the time reversed process. That would be 
\[ 3 + 4 \rightarrow \overline{1} + \overline{2}, \]

\noindent 
because if you run a movie backward the products of a reaction become the reagents and the reagents become the products. Furthermore, what once went north now goes south and what once went up now goes down, that is, the velocities are reversed. If we also apply parity we undo the reversal of velocities and the final process is
\begin{align*} 
\begin{matrix}3\\ p_3\end{matrix}
\begin{matrix} +\\ \\\end{matrix}
\begin{matrix}4\\ p_4\end{matrix} &
\begin{matrix} \rightarrow\\ \\ \end{matrix}
\underbrace{
\begin{matrix}\overline{1}\\ p_1\end{matrix}
\begin{matrix}+\\ \\\end{matrix}
\begin{matrix}\overline{2}\\p_2\end{matrix}}_{\substack{\text{physical}\\\text{4 momenta}}}\\
\text{amplitude}&=a(-p_1,-p_2, p_3, p_4)
\end{align*} 

\noindent 
Whether or not these three operations individually affect the amplitude, we have shown above that the combined effect of all three operations, $CPT$, can't change the amplitude on general grounds. If $CPT$ violated is ever observed, Lagrangian quantum field theory is cooked. Contrast: If $C$ violation is observed, we just write down $C$ non-invariant interactions.

\section*{Phase space and the $S$ matrix}
Our job is to make contact with the numbers experimenters measure. To do this we square our $S$ matrix elements and integrate over the possible final states a detector might register, and we get a probability that a counter will advance. Our $S$ matrix elements are proportional to $\delta^{(4)} (p_f - p_i)$. Squaring them is senseless. What went wrong?\\

What went wrong is that the states we are using are not normalizable. They extend throughout all of space. The scattering process occurs at every point in space, and since two plane wave states never get far apart no matter how long you wait, the scattering process goes on for all time.\\

A half-assed way to salvage the situation is to put the systems in a box, so that we can normalize the plane-wave states and to turn the interaction on for a finite amount of time $T$. A more satisfying way to salvage the situation is to build wave packets, which are normalizable, and do get far apart in the far past / future. We are in a hurry, so we'll put the world in a box of volume $V$. \\

The states in a box of volume $V$ with periodic boundary condition are $ |\vec{k_1}, \ldots, \vec{k_n} \rangle $ where $k_{i \, x,y,z} = \cfrac{2\pi n_{i \, x,y,z}}{L}$, $L^3 = V$. 
\begin{center}
\includegraphics[width=10 cm]{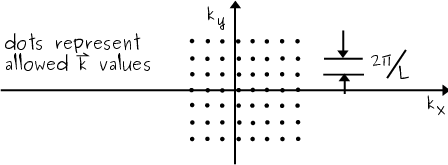}
\end{center}

\noindent 
The states $|\vec{k_1},\ldots, \vec{k_n} \rangle$ are built up from the vacuum by creation operators 
\[ |\vec{k_1}, \ldots, \vec{k_n} \rangle = a^\dagger_{\vec{k_1}} \cdots a^\dagger_{\vec{k_n}} | 0 \rangle \]
\[ a_{\vec{k}} |0 \rangle = 0 \]
\[ [a^\dagger_{\vec{k}}, a^\dagger_{\vec{k}\,'}] = 0 \]
\[ [a_{\vec{k}}, a_{\vec{k}\,'}] = 0 \]
\[ [a_{\vec{k}}, a^\dagger_{\vec{k}'}] =\underbrace{ \delta_{\vec{k}, \vec{k}\,'}}_{\substack{\text{\tiny Kronecker}\\ \text{\tiny delta}}} = \langle \vec{k} | \vec{k}\,' \rangle \]

\noindent 
The free field in the box has the expansion
\[ \phi(x) = \sum_{\vec{k}} \left( \frac{a_{\vec{k}} e^{-ik \cdot x}}{\sqrt{2E_{\vec{k}}} \sqrt{V}} + \frac{a^\dagger_{\vec{k}} e^{ik \cdot x}}{\sqrt{2E_{\vec{k}}} \sqrt{V}} \right) \]

\noindent
We want to know what is the probability of making a transition to some infinitesimal volume of phase space specified by $d^3k_1 \cdots d^3k_n$ ($n$ final particles). It is the probability of going to one of the final states in that infinitesimal region times the number of states in that infinitesimal region,
\[ \frac{d^3 k_1}{(2\pi)^3} V \cdots \frac{d^3 k_n}{(2\pi)^3} V \]

\noindent 
Let's look at the transition probability to go to one of the final states in that region 
\[ | \langle f | (S-1) | i \rangle | ^2. \]

\noindent 
We will restrict our attention to the two simplest and most important initial states: one particle and two particle initial states. We will normalize the two particle initial state unconventionally. We will consider
\[ |i \rangle = \left\{\begin{array}{l l} | \vec{k} \rangle & \quad \text{Decay}\\ | \vec{k_1}, \vec{k_2} \rangle \sqrt{V} & \quad \text{Scattering}\\ \end{array} \right.
\]

$| \vec{k} \rangle $ and $| \vec{k_1}, \vec{k_2} \rangle $ are box normalized. Why the factor $\sqrt{V}$? Without it, each particle has probability $1$ of being somewhere in the box. The probability that they are both near a given point and can scatter is $\propto \frac{1}{V^2}$. Of course, they could both be near any point in the box, so the probability that they will scatter from anywhere is $\propto \frac{1}{V}$. With the factor $\sqrt{V}$ you can think of one particle as having probability $1$ of being in any unit volume, and the other as having probability one of being somewhere in the box. With these conventions, we expect
\[ \langle f |S-1| i \rangle = i a_{fi}^{VT} (2 \pi)^4 \delta_{VT}^{(4)} (p_i - p_f) \left( \prod_{\text{final particles}} \frac{1}{\sqrt{2E_f}\sqrt{V}} \right) \left( \prod_{\text{initial particles}} \frac{1}{\sqrt{2E_i}} \right) \frac{1}{\sqrt{V}} \]

\noindent
where 
\[ (2 \pi)^4 \delta_{VT}^{(4)}(p) \equiv \int_V d^3 x \int dt f(t) e^{ip\cdot x} \]

The extra factors you have never seen before come from the expansion of the fields. The coefficient of the creation operator $a_{\vec{k}}^\dagger$, which annihilates a particle on the left and the coefficient of the annihilation operator, $a_{\vec{k}}$, which annihilates a particle on the right is 
\[ \cfrac{1}{\sqrt{V} \sqrt{2 \omega_k}}. \]

You get one of these factors for every particle that is annihilated on the left or right by the fields. We did not get these before because the coefficient of $a(k)$ and $a(k)^\dagger$ in the field is $\cfrac{1}{(2\pi)^3} \cfrac{1}{2 \omega_{\vec{k}}}$ which exactly cancels the factor we get when $a(k)$ hits the relativistically normalized state $|k' \rangle$ 
\[ a(k) | k' \rangle = (2 \pi)^3 2 \omega_{\vec{k}} |0 \rangle \]

The product over initial states in our formula has had the single factor of $\frac{1}{\sqrt{V}}$ pulled out in the decay case, and the two factors of $\frac{1}{\sqrt{V}}$ cancelled to just one by the $\sqrt{V}$ we put in by hand in the scattering case, leaving again a single factor of $\frac{1}{\sqrt{V}}$ which we have explicitly put in as the last factor of the formula.\\

To get the transition probability, we square the transition amplitude. We multiply this by the number of states in the infinitesimal region of final state phase space to get the differential transition probability. To get something that does not depend on the time we turn the interaction on for $T$, we divide by $T$. You should get 
\begin{multline*} 
\frac{\text{Differential Transition Probability}}{\text{Unit time}} = \\
\cfrac{1}{VT} |a_{fi}^{VT}|^2 \left( (2 \pi)^4 \delta_{VT}^{(4)}(p^i - p^f) \right)^2 \prod_{\text{final particles}} \frac{d^3 k_f}{(2\pi)^3 2 E_f} \prod_{\substack{\text{initial particles}\\\text{(1 or 2)}}} \frac{1}{2 E_i} 
\end{multline*}

\underline{Now} we take the limits $V$, $T \rightarrow \infty$. 
\[ a_{fi}^{VT} \longrightarrow a_{fi} \]
\[ |a_{fi}^{VT}|^2 \longrightarrow |a_{fi}|^2 \]
\[ \delta_{VT}^{(4)} \longrightarrow \delta^{(4)} \] 
\[ ((2 \pi)^4 \delta_{VT}^{(4)})^2 \rightarrow \text{Here's where we have to be careful} \]

Recall: $\delta^{(4)}_{VT}$ is a function concentrated near the origin. It becomes more and more so as $V$, $T \rightarrow \infty$. Also it is normalized to $1$ for all $V$, $T$. 
\[ \int d^4 p \, \delta_{VT}^{(4)}(p) = \frac{1}{(2 \pi)^4} \int_{-\infty}^{\infty} dt f(t) \int_V d^3x \underbrace{\int d^4 p e^{ip\cdot x}}_{(2\pi)^4 \delta^{(4)}(0)} = 1 \]

For these two reasons we say $\lim_{VT \rightarrow \infty} \delta_{VT}^{(4)} (p) = \delta^{(4)}(p)$. \\

What about $(\delta_{VT}^{(4)}(p))^2$ ?\\

In the limit $V,T \rightarrow \infty$, it is concentrated about $p =0$ just as surely as $\delta^{(4)}_{VT}(p)$ is. We can find its normalization 
\begin{align*}
\int d^4 p [ \delta_{VT}^{(4)}(p)]^2 &= \frac{1}{(2 \pi)^8} \int dt \int dt' f(t) f(t') \int_V d^3x \int_V d^3 x' \underbrace{\int d^4 p e^{ip\cdot x} e^{ip\cdot x'}}_{(2 \pi)^4 \delta^{(4)} (x+x')} \\
&= \frac{1}{(2\pi)^4} \int dt | f(t) |^2 \int_V d^3 x = \frac{1}{(2\pi)^4} VT 
\end{align*}

For these two reasons we say 
\[ \lim_{V,T \rightarrow \infty} \frac{1}{VT} (2 \pi)^4 (\delta_{VT}^{(4)} (p))^2 = \delta^{(4)}(p) \]

Thank God a factor of $\frac{1}{VT}$ appears in our formula for the differential transition probability per unit time so we can take the $V$, $T \rightarrow \infty$ limit to get 
\begin{multline*}
\frac{\text{differential transition probability}}{\text{unit time}} = \\ 
|a_{fi}|^2 \underbrace{(2\pi)^4 \delta^{(4)}(p_f - p_i) \prod_{\text{final particles}} \overbrace{\frac{d^3 k_f}{(2 \pi)^3 2E_f}}^{\substack{\text{\scriptsize L. I. measure}\\ \text{\scriptsize on the mass}\\ \text{\scriptsize hyperboloid}}}}_{\substack{\text{\scriptsize This factor which is manifestly Lorentz invariant}\\ \text{\scriptsize is called the ``invariant density of states", D,}\\ \text{\scriptsize or the "relativistic density of final states".}}} \prod_{\substack{\text{initial particles,}\\\text{1 or 2}}}\frac{1}{2E_i}
\end{multline*} 

Note that you have no excuse for not getting the $(2\pi)$'s right. Every $2\pi$ goes with a $\delta$ function and every $\frac{1}{2\pi}$ goes with a $k$ integration.}{
 \sektion{13}{November 4}
\descriptionthirteen
{\large Applications of }
\[ \frac{\text{Differential Transition Probability}}{\text{Unit time}} = |a_{fi}|^2 D\!\!\!\!\!\!\! \prod_{\substack{\\\text{initial particles}\\\text{(1 or 2)}} }\!\!\!\!\!\!\! \frac{1}{2E_i} \]
\begin{equation}\label{eq:13-page1-1} 
D = (2 \pi)^4 \delta^4(p_f - p_i)\!\!\!\!\!\! \prod_{\substack{\\\text{final particles}}}\!\!\!\!\!\! \frac{d^3 k_f}{(2\pi)^3 2 E_f} 
\end{equation}

\section*{Decay}
\[ d \Gamma = \frac{\text{Diff decay prob}}{\text{Unit time}} = \frac{1}{2E} |a_{fi}|^2 D \]

\noindent
The total decay probability per unit time is $d\Gamma$ summed and integrated over all possible final states 
\[ \frac{\text{Decay Probability}}{\text{Unit time}} = \frac{1}{2E}\!\!\!\underbrace{\thirteensuminta_{\substack{\\\text{final states}}} |a|^2 D}_{\substack{\text{It is obvious}\\ \text{that this part is L.I.}}} \]

\noindent
We'll evaluate the decay probability rate in the rest frame of the decaying particle. This is the ``decay width", $\Gamma$.
\begin{equation}\label{eq:13-page1-2}
\Gamma = \frac{\text{Rest decay probability}}{\text{Unit time}} = \frac{1}{2m} \thirteensumintb |a|^2 D 
\end{equation}

\noindent
Since the $\displaystyle \;\,\int\!\!\!\!\!\!\!\!\!\sum |a|^2 D $ is L.I.
\[ \frac{\text{Decay probability}}{\text{Unit time}} = \frac{m}{E} \Gamma = \frac{d \tau}{d t} \Gamma \]

\noindent
where $\tau$ is the particle's proper time. The shelf life of a moving particle is longer, its decay rate is slower exactly by a factor of elapsed proper time / elapsed observer time.

\section*{Cross sections}
\[ \quad\quad\quad \quad\quad\quad\quad d \sigma = \frac{\text{Diff Trans Prob}}{\text{Unit time} \times
\text{Unit flux}} = \underbrace{\frac{1}{4E_1 E_2}}_{} |a_{fi}|^2 D\!\!\!\!\!\!\!\!\!\!\!\!\!\!\!\!\!\! \!\!\!\!\!\!\!\!\! \!\!\!\!\!\!\!\!\! \!\!\!\!\!\!\!\!\! \!\!\!\!\!\!\!\!\! \!\!\!\!\! \underbrace{\frac{1}{|\vec{v_1} - \vec{v_2}|}}_{\text{These } \vec{v} \text{'s are \underline{intial} state particle velocities and energies}\quad\quad\quad\quad\quad\quad} \]

\noindent
The factor $\cfrac{1}{|\vec{v_1}-\vec{v_2}|}$ takes care of the per unit flux. Let's understand this factor with our conventions. 
$$\text{\underline{Our convention}}\quad\quad | i \rangle = \sqrt{V} |\vec{k_1}, \vec{k_2} \rangle $$

\noindent
The transition probability per unit time is some mess
\[ \frac{\text{t.p.}}{\text{u.t.}} = \text{(some mess)} \]

\noindent
Let's suppose the particle with momentum $\vec{k_2}$ presents some area, $A$ to the particle beam with momentum $\vec{k_1}$. Think of the particles with momentum $\vec{k_1}$ as having probability $1$ of being in any volume and the particle with momentum $\vec{k_2}$ as being located somewhere in the whole box with probability one.
\begin{center}
\includegraphics[scale=0.35]{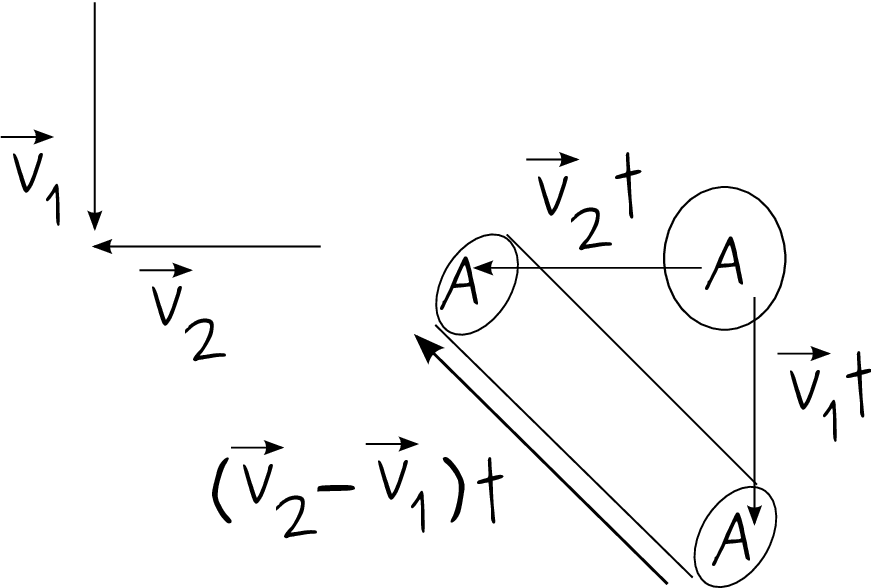}
\end{center}

\noindent
$\vec{v_2} t$ is the vector displacement of the particle with momentum $\vec{k_2}$ in a time $t$. $\vec{v_1} t$ is the motion of the beam in time $t$. The orientation of $A$ is so as to be $\bot$ to $\vec{v_2} - \vec{v_1}$, that is, so as to catch the most flux. The cylinder is the volume swept out in the beam in a time $t$. Its volume is 
\[ |\vec{v_2} - \vec{v_1}|tA \]

\noindent
The flux is thus $|\vec{v_2} - \vec{v_1}|$ and the 
\[ \frac{\text{t.p}}{\text{u.t.} \times {\text{unit flux}}} = \frac{\text{(some mess)}}{|\vec{v_2} - \vec{v_1}|} \]

\noindent
\underline{Another Convention} is to take $| i \rangle = | \vec{k_1}, \vec{k_2} \rangle $.\\

Then the transition probability per unit time would have come out as 
\[ \frac{\text{t.p}}{\text{u.t.}} = \frac{1}{V} (\!\!\!\!\!\!\!\!\!\!\!\!\!\!\!\!\!\!\!\!\!\!\!\!\!\!\!\!\!\!\!\!\!\!\!\!\!\underbrace{\text{some mess}}_{\substack{\text{\quad\quad\quad\quad\quad\quad\quad This is the same ``(some mess)'' as} \\ \text{\quad\quad\quad\quad\quad\quad\quad in the previous equation, whatever it is}}}\!\!\!\!\!\!\!\!\!\!\!\!\!\!\!\!\!\!\!\!\!\!\!\!\!\!\!\!\!\!\!\!\!\!\!\!\!) \]

\noindent
The flux for this normalization is however 
\[ \frac{1}{V} |\vec{v_1} - \vec{v_2}| \]

\noindent
So 
\[ \frac{\text{t.p.}}{\text{Unit time} \times \text{Unit flux}}=\frac{\cancel{\frac{1}{V}}\text{(some mess)}}{\cancel{\frac{1}{V}}|\vec{v_1}-\vec{v_2}|} \]

\noindent 
is the same. \\

\noindent
I want to emphasize that this is the nonrelativistic velocity and NR velocity addition formula that appears here. If two beams approach with speed $c$ head on the flux is $2c$.\\

\noindent
The \underline{total} cross section is 
\begin{equation}\label{eq:13-botpage3}
\sigma = \frac{1}{|\vec{v_1}- \vec{v_2}|} \frac{1}{4E_1 E_2} \thirteensuminta_{\text{final states}} |a_{fi}|^2 D
\end{equation}

\noindent
If $\vec{v_1}$ is parallel or antiparallel to $\vec{v_2}$ and the total cross section really has the interpretation of an area, then it should be unaffected by boosts along $\vec{v_1}$ and $\vec{v_2}$.
\begin{center}
\includegraphics[scale=0.3]{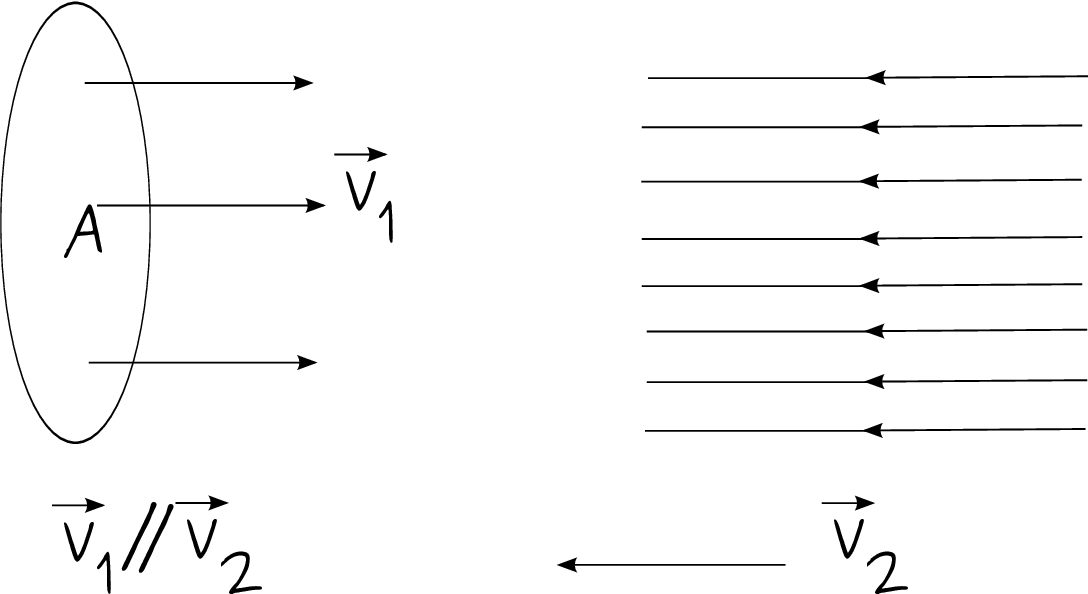}
\end{center}

The total number of particles that smash into the area $A$ will depend on the velocity of the observer. It will be proportional to the flux in that frame. However the idea of a perpendicular area should be Lorentz invariant for boosts along $\vec{v_1} - \vec{v_2}$ direction. Since 
\[ \sigma = \frac{1}{|\vec{v_1} - \vec{v_2}| } \, \frac{1}{4E_1 E_2} \underbrace{\thirteensumintb |a_{fi}|^2 D} \]

\noindent
is supposed to have the interpretation of an area, it should be unaffected by these boosts. The underbraced term is invariant under any Lorentz transformation. What about the factor in front? Take $\vec{v_1}$ and $\vec{v_2}$ to be along the $x$ direction. 
\[ P_1 = (E_1, p_{1x}, 0, 0) \]
\[ P_2 = (E_2, p_{2x}, 0,0) \]

\noindent
Then 
\[ E_1 E_2 |\vec{v_1} - \vec{v_2}| = E_1 E_2 \left| \frac{p_{1x}}{E_1} - \frac{p_{2x}}{E_2} \right| = |p_{1x} E_2 - p_{2x} E_1 | = |\epsilon^{23\mu \nu} p_{1\mu} p_{2\nu}| \]

\noindent
which is obviously invariant under rotations on the $0-1$ plane (boosts along $\vec{v_1}$). This justifies the interpretation of the cross section as an area.

\section*{$\boldsymbol{D}$ for a two body final state in the COM frame}
$D$ contains both $\delta$ functions and integrals that can be trivially performed by using those $\delta$ functions. We can do them once and for all. Of course this turns independent variables (in $|a_{fi}|^2$ and in where they arise kinematically) into dependent variables.\\

\noindent
In the center of mass frame $\vec{p^i} = 0$ and $E^i = E_T$ while 
\begin{align*}
D &= \frac{d^3 p_1 d^3 p_2}{(2\pi)^6 4 \!\!\!\!\!\!\!\!\!\!\!\! \underbrace{E_1 E_2}_{\text{final particle energies}}\!\!\!\!\!\!\!\!\!\!\!\!} (2 \pi)^4 \delta^{(3)}\!\!\!\!\!\!\!\!\!\!\underbrace{(\vec{p_1} + \vec{p_2})}_{\text{final particle momenta}}\!\!\!\!\!\!\!\!\!\delta (E_1 +E_2 - E_T) \\
&= \frac{d^3 p_1}{(2\pi)^3 4 E_1 E_2} 2 \pi \delta (E_1 + E_2 - E_T) 
\end{align*}
\[ \vec{p}_2 = - \vec{p}_1 \]

The $\delta^{(3)}(\vec{p}_1 + \vec{p}_2)$ is used to do the $\vec{p}_2$ integration. You must now remember that $\vec{p}_2$ depends on $\vec{p}_1$ whenever it appear (in $E_2$ or in $|a_{fi}|^2$).\\

Let's rewrite $d^3 p_1$ as $p_1^2 dp_1 d \Omega_1$ and use the energy delta function to do the $p_1$ integration. \\

Thought of as functions of $\vec{p_1}$, $E_1^2 = \vec{p_1}^2 +m_1^2$, and $E_2^2 = \vec{p_2}^2 +m^2 = \vec{p_1}^2 +m^2$, we have $E_1 dE_1 = p_1 dp_1$, $E_2 dE_2 = p_2 dp_2$. 
\[ \frac{\partial (E_1 + E_2)}{\partial p_1} = \frac{p_1}{E_1} + \frac{p_1}{E_2} = \frac{p_1 E_T}{E_1 E_2} \]

so 
\begin{equation}\label{eq:13-botpage5}
D = \frac{1}{16 \pi^2 E_1 E_2} d\Omega_1 \frac{p_1^2}{\left|\frac{\partial(E_1+E_2)}{\partial p_1}\right|} = \frac{1}{16 \pi^2} \frac{p_1 d \Omega_1}{E_T} 
\end{equation}

\subsection*{$2 \rightarrow 2$ scattering in the COM frame}
In the COM frame,
\begin{align}
4 E_{i1} E_{i2} | \vec{v_1} - \vec{v_2}| &= 4 | E_{i2} \vec{p}_{i1}-E_{i1} \vec{p}_{i2} | \nonumber\\
&= 4 | E_{i2} \vec{p}_{i1} + E_{i1} \vec{p}_{i1} | \nonumber\\
&= 4 E_T p_{i1} \nonumber\\
&= 4 E_T p_i\label{eq:13-toppage6}
\end{align}

(The $i$ subscript reminds you that these are initial particle momenta and energies.)
\[ d\sigma = \frac{1}{16\pi^2} \frac{p_f d \Omega_1}{E_T} \frac{1}{4 E_T p_i} |a_{fi}|^2 = \frac{1}{64 \pi^2 E_T^2} d\Omega_1 \frac{p_f}{p_1} |a_{fi}|^2 \]
\[ \frac{d\sigma}{d\Omega} = \frac{1} {64 \pi^2 E_T^2} \frac{p_f}{p_i} |a_{fi}|^2 \]

Note that for an exothermic reaction we can have $p_i = 0$ while $p_f\neq 0$. $\frac{d \sigma}{d \Omega}$ and hence $\sigma$ can be infinite even when the amplitude $a_{fi}$ is finite. This is why they slow down the neutrons in atomic piles. 

It is simple to understand this. As $p_i \rightarrow 0$, the amount of time the two particles spend in the danger zone near each other, which goes as $\frac{1}{p_i}$ becomes infinite. \\

We maximize the chance of neutron capture in the pile by making the neutron cruise out of the pile as slowly as possible.\\

\subsection*{Contact with elastic $2 \rightarrow 2$ scattering in NRQM}
Our formula for $\frac{d \sigma}{d \Omega}$ is 
\[ \frac{d \sigma}{d \Omega} = \frac{1}{64 \pi^2 E_T^2} \frac{p_f}{p_i} |a_{fi}|^2 = \frac{1}{64 \pi^2 E_T^2} |a_{fi}|^2 \]

\noindent
for elastic scattering. Compare this with 
\[ \frac{d \sigma}{d\Omega} = |f|^2 \] 

\noindent
from NRQM and see 
\[ |f| = \frac{1}{8 \pi E_T} |a_{fi}| \]

\noindent
We'll get the phase when we do the optical theorem.

\subsection*{Example, Model 3}
\[ \mathcal{L}' = - g \psi^* \psi \phi \]
\[ \Diagram{fdV \\ & f \\ fuA} = -ig \]
\[ \Diagram{\momentum[llft]{fdV}{p'\nwarrow} \\ & \momentum[bot]{f}{q} \\ \momentum[bot]{fuA}{\swarrow p}} = (-ig)(2\pi)^4\delta^{(4)}(p+p'-q) \]
\[ ia = -ig + \mathcal{O}(g^3), \quad \text{pretty simple, couldn't be simpler.} \] 

\noindent
From Eqs.~(\ref{eq:13-page1-2}) and (\ref{eq:13-botpage5})
\begin{align*}
\Gamma &= \frac{1}{2\mu} \;\int\!\!\!\!\!\!\!\!\!\sum |a|^2 D \\
&= \frac{1}{2\mu} g^2 \int \frac{p_1 d \Omega_1}{16 \pi^2 \underbrace{E_T}_{\mu}} \\
&= \frac{g^2 p_1}{8 \pi \mu^2} = \frac{g^2}{8\pi \mu^2} \sqrt{(\frac{1}{2} \mu)^2 - m^2} \\
&= \frac{g^2}{16\pi \mu^2} \sqrt{\mu^2 - 4m^2}
\end{align*}
 
\section*{Optical Theorem}
The optical theorem in NRQM is based on a simple idea. There is an incoming wave incident on a target, and an outgoing wave. The outgoing wave is the superposition of the incoming wave that passes right through the target and goes off in the forward direction, and the scattered wave which goes off in all directions. Since there is some probability that a particle in the beam is scattered off, and since probability is conserved, there must be a decrease in the intensity in the beam in the forward direction. The total probability for scattering in all directions but exactly forward, which mathematically is $\sigma$, the total cross section, must be equal to the decrease in probability of going exactly in the forward direction, which mathematically is an interference term between the wave that passes right through and the scattered wave in the forward direction.\\

There is nothing in this argument that is nonrelativistic, so we should be able to get a analog of the optical theorem in our relativistic scattering theory.\\

The mathematical statement of conservation of probability in the scattering process is 
\[ S S^\dagger = 1 \] 

We want to make a statement about our amplitudes, $a_{fi}$, which are proportional to matrix elements of $S-1$, so we'll rephrase the conservation of probability as 
\[ (S-1)(S-1)^\dagger = \underbrace{S S^\dagger}_{1} - S - S^\dagger + 1 = -(S-1) - (S-1)^\dagger \]

Now 
\[ \langle f | (S-1) | i \rangle = i a_{fi} (2 \pi)^4 \delta^4 (p_f - p_i) \]

and 
\[ \langle f| (S-1)^\dagger |i \rangle = -i a^*_{if} (2 \pi)^4 \delta^4 (p_f - p_i) \]

Our rephrased statement of the conservation of probability has matrix elements of 
\[ \langle f | (S-1) (S-1)^\dagger |i \rangle = - \langle f | (S-1) | i \rangle - \langle f | (S-1)^\dagger | i \rangle \] 

\noindent
By inserting a complete set of intermediate states, the left hand side (LHS) becomes 
\begin{align*}
\text{LHS} &= \;\;\int\!\!\!\!\!\!\!\!\!\!\!\!\!\!\!\! \sum_{\substack{\\ \\\text{intermediate} \\
\text{states }|m\rangle }}\!\!\!\!\!\langle f|(S-1)|m\rangle\langle m|(S-1)^\dagger|i\rangle \\
&= \!\!\!\!\!\!\!\sum_{\substack{\\ \text{intermediate}\\\text{states with }n_m\\ \text{particles}}}\!\!\!\!\!\!\!\!\!\!\!\!\!\!\!\!\!\!\!\!\!\!\!\!\!\!\!\!\!\!\!\!\!\!\!\!\!\!\!\!\overbrace{\frac{1}{n_m!}}^{\substack{\quad\quad\quad\quad\quad\text{an overcounting factor if the}\\ \quad\quad\quad\quad\quad n_m\text{ particles are identical}}}\!\!\!\!\!\!\!\!\!\!\!\!\!\!\!\!\!\!\!\!\!\!\!\!\!\!\!\!\!\!\!\!\!\!\!\! \int \frac{d^3 k_1}{(2\pi)^3 2 E_1} \cdots \frac{d^3 k_{n_m}}{(2\pi)^3 2 E_{n_m}} a_{fm} a_{im}^* (2\pi)^4 \delta^4(p_f - p_m)(2\pi)^4 \delta^4(p_m-p_i) 
\end{align*}

Because of the $\delta^4 (p_m - p_i)$, we can replace the $p_m$ in $\delta^4(p_f - p_m)$ by $p_i$, so that we explicitly have that LHS is proportional to $\delta^4 (p_f - p_i)$. 

The RHS of the rephrased statement of the conservation of probability is 
\[ \text{RHS} = -i a_{fi} (2 \pi)^4 \delta^4 (p_f - p_i) + i a_{if}^* (2\pi)^4 \delta^4(p_f - p_i) \]

Both the LHS and RHS are proportional to $(2\pi)^4 \delta^4 (p_f - p_i)$. 

Comparing the LHS with the RHS, we have
\begin{align*} 
\sum_{\substack{\\ \\\text{intermediate states} \\ \text{with } n_m \text{ particles}}}\!\!\!\!\!\!\!\!\!\frac{1}{n_m!}\int \underbrace{\frac{d^3 k_1}{(2 \pi)^3 2 E_1} \cdots \frac{d^3k_{n_m}}{(2\pi)^3 2E_{n_m}} (2\pi)^4 \delta^4 (p_m - p_i)} a_{fm} a_{im}^*& = -i a_{fi} + i a_{if}^* \\ 
&= 2 \, \text{Im} \, a_{fi} 
\end{align*} 

The underbraced factor is what we would call the invariant density of states for the process $i \rightarrow m$, $D_m$ (see Eq.~(\ref{eq:13-page1-1})). If we choose $f = i$, we get 
\[ \!\!\!\!\!\!\!\sum_{\substack{\\ \text{intermediate}\\\text{states with }n_m\\ \text{particles}}}\!\!\!\!\!\!\!\!\!\!\!\!\!\!\!\!\!\!\!\!\!\!\!\!\!\!\!\!\!\!\!\!\!\!\!\!\!\!\!\!\overbrace{\frac{1}{n_m!}}^{\substack{\quad\quad\quad\quad\quad\text{an overcounting factor if the}\\ \quad\quad\quad\quad\quad n_m\text{ particles are identical}}}\!\!\!\!\!\!\!\!\!\!\!\!\!\!\!\!\!\!\!\!\!\!\!\!\!\!\!\!\!\!\!\!\!\!\!\!\int D_m |a_{im}|^2 = 2 \, \text{Im} \, a_{ii} \]

This says the total transition probability (statement might be off by a factor of $E_T^2$) per unit time is equal to twice the imaginary part of the forward scattering amplitude.\\

If the process has a two particle initial state, we can rewrite this as a statement about cross sections. In the COM frame this says (see Eqs.~(\ref{eq:13-botpage3}) and (\ref{eq:13-toppage6}))
\[ \underset{2}{\cancel{4}} E_T p_i \sigma = \cancel{2} \, \text{Im} \, a_{ii} \]

\noindent
(since the LHS is zero till $\mathcal{O}(g^4)$ for $2N \rightarrow 2N$ scattering, we see that the RHS must be zero till $\mathcal{O}(g^4)$. This proves that the forward scattering amplitude for $2N \rightarrow 2N$ is real at $\mathcal{O}(g^2)$). \\

In NRQM the optical theorem for elastic scattering ($p_i=p_f=p$) is 
\[ \frac{p}{4\pi} \sigma = \text{Im} \, f |_{\theta =0} \]

Barring a different $\theta$ dependence in the phase conventions, we can finally state 
\[ f = \frac{1}{8\pi E_T} a \]

\subsection*{3 body final state phase space in the COM frame}
\begin{align*}
D &= \frac{d^3 p_1}{(2\pi)^3 2E_1} \frac{d^3 p_2}{(2\pi)^3 2E_2}\frac{d^3 p_3}{(2\pi)^3 2E_3} (2\pi)^4 \delta^{(3)}(\vec{p_1}+\vec{p_2}+\vec{p_3}) \cdot 
\delta(E_1+E_2+E_3-E_T) \\
&= \frac{1}{(2\pi)^5} d^3 p_1 d^3 p_2 \frac{1}{8E_1 E_2 E_3} \delta(E_1+E_2+E_3-E_T) 
\end{align*}

\noindent
The momentum conserving $\delta$ function has been used to eliminate $\vec{p_3}$. From now on $\vec{p_3}$ and $E_3$ are not independent variables.
\[ \vec{p_3} = -(\vec{p_1} + \vec{p_2}) \]
\[ E_3 = \sqrt{(\vec{p_1} + \vec{p_2})^2 + m_3^2} \]

\noindent
Now we'll rewrite $d^3 p_1$ as $p_1^2 dp_1 d \Omega_1$. Instead of writing $d^3 p_2$ as $p_2^2 dp_2 d\Omega_2$, let's rewrite it as $d^3 p_2 = p_2^2 dp_2 d\Omega_{12} = p_2^2 dp_2 d\phi_{12} d \cos \theta_{12}$, where $\phi_{12}$ is an azimuthal angle about $\vec{p_1}$ and $\theta_{12}$ is a polar angle measured from $\vec{p_1}$.
\begin{center}
\includegraphics[scale=0.7]{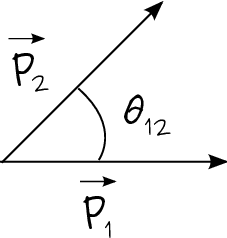}
\end{center}

We are going to use the energy conserving $\delta$ function to do the $\theta_{12}$ integration. $E_3$ depends on $\theta_{12}$,
\[ E_3^2 = p_1^2 + 2 \vec{p_1}\cdot \vec{p_2} + p_2^2 + m_3^2 = p_1^2 + p_2^2 + m_3^2 + 2 p_1 p_2 \cos \theta_{12} \]

Therefore $\frac{\partial E_3}{\partial \cos \theta_{12}} = \frac{p_1 p_2}{E_3}$ and thus $d \cos \theta_{12} \delta (E_1+E_2+E_3-E_T) = \frac{E_3}{p_1 p_2} $.

$\theta_{12}$ is now a dependent variable.
\begin{align*}
D &= \frac{1}{(2\pi)^5} p_1^2 d p_1 d \Omega_1 p_2^2 d p_2 d \phi_{12} \frac{E_3}{p_1 p_2} \frac{1}{8E_1 E_2 E_3} \\
&= \frac{1}{256 \pi^5} \underbrace{\frac{p_1dp_1}{E_1}}_{dE_1} \underbrace{\frac{p_2dp_2}{E_2}}_{dE_2} d \Omega_1 d \phi_{12} \quad\text{(Valid in COM frame)}
\end{align*}
 
(Amazing simple result if you use the right variable)\\

Suppose the amplitude $a$ is independent of $\Omega_1$ and $\phi_{12}$, as in the decay of a spinless meson (at rest), or for a particle with spin decaying, if you average over initial spin states, then we can do the angular integrations (which give $8\pi^2$) to get 
\[ \frac{1}{32\pi^3} |a|^2 d E_1 d E_2 \]

\noindent
as the differential transition probability per unit time into an energy range $E_1$ for particle $1$ and $E_2$ for particle $2$. \\

If I make a plot of experimental data points as a function of $E_1$ and $E_2$, they will be distributed according to $|a|^2$, because $d E_1 d E_2$ is the Euclidean measure on the plane 
\begin{center}
\includegraphics[scale=0.5]{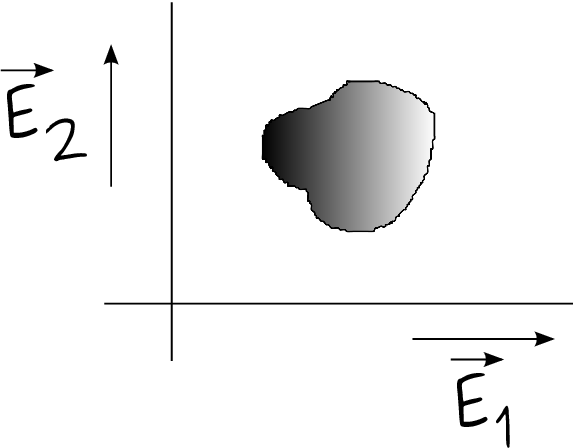}
\end{center}

\noindent
There will be a kinematically imposed boundary to the shaded region, but within those boundaries, the shaded region is directly proportional to $|a|^2$. \\

Our next topic is the beginning of a discussion of Green's functions, scattering with wave packets, and the LSZ reduction formula.

\section*{Feynman Diagrams With External Lines Off The Mass Shell}
We'll restrict ourselves (for notational simplicity only) to considering diagrams in which only one type of scalar meson appears on the external lines. (By scalar, I just mean uncharged, with no Lorentz indices on its field, that is, no spin, not a specification of its parity transformation properties. A parity need not even exist for the formalism we are about to develop to be applicable. (``charged scalar" means charged, but no spin. If I really wanted to specify parity's effect, I would say ``scalar under parity transformations", or whatever.)) We'll still let particles of all types run around on the internal lines.

Let a blob like this 
\[ \begin{array}{cc} \includegraphics[scale=0.5]{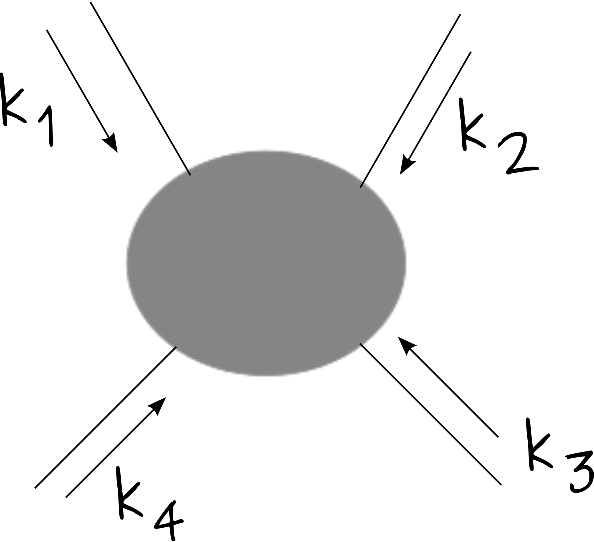}& \begin{array}{c}\equiv \widetilde{G}^{(4)} (k_1, \ldots,k_4 )\\ \\ \\ \\ \\ \\ \\ \\ \\ \\ \end{array} \end{array} \]
\vspace{-1in}

\noindent 
represent a sum of Feynman diagrams. It could be all Feynman diagrams to some order in perturbation theory, or in our imagination the sum of all diagrams to all orders in perturbation theory.\\

{\large \sc Can we assign any meaning to this blob if the momenta on the external lines are unrestricted, off the mass shell, maybe not even satisfying $k_1 + k_2 + k_3 + k_4 =0$?}\\

We are going to come up with three affirmative answers to this question. Something neat is that they all agree.

\subsection*{Answer 1}
The blob could be an internal part of a more complicated graph.
\begin{center}
\includegraphics[scale=0.5]{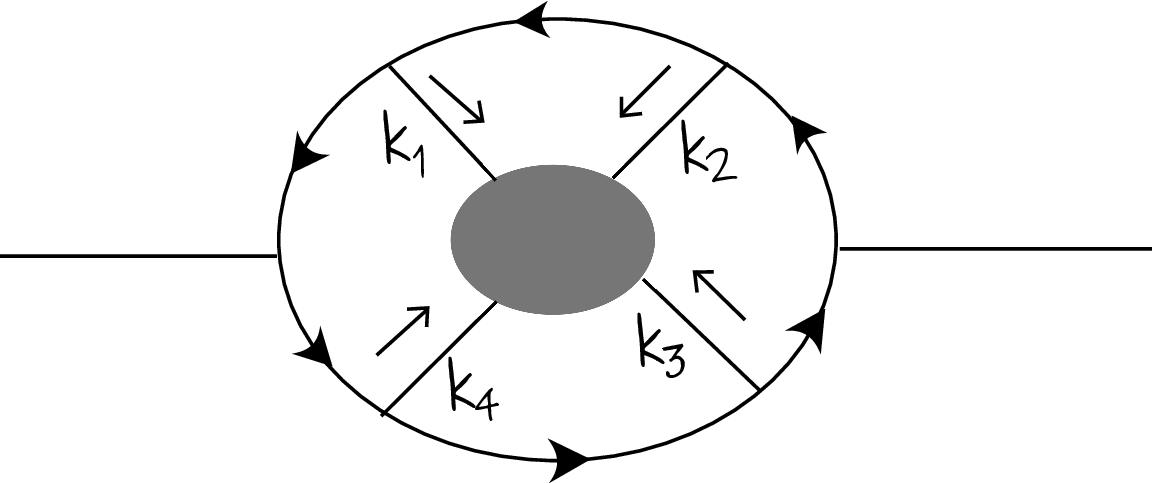}
\end{center}

\noindent
The Feynman rules instruct you to label all internal momenta arbitrarily and integrate over them. \\

\noindent
Suppose in our study of other graphs in the theory, for example
\begin{center}
\includegraphics[scale=0.45]{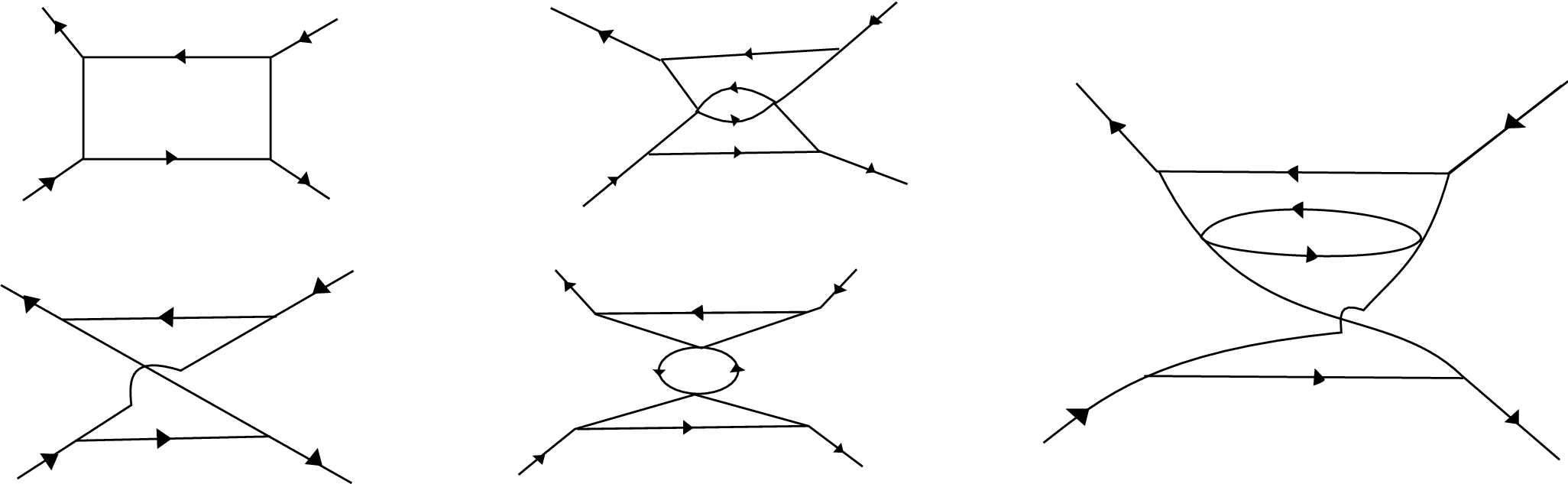}
\end{center}

all of which have the form
\begin{center}
\includegraphics[scale=0.5]{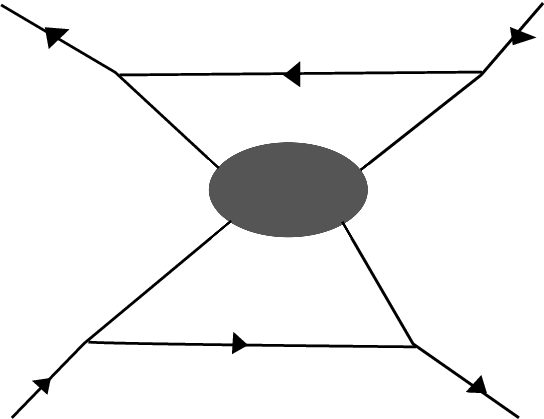},
\end{center}

\noindent
that we have already summed the blob, in our work calculating those graphs to some order. Then it would be nice not to repeat that work when calculating 
\includegraphics[width=5 cm]{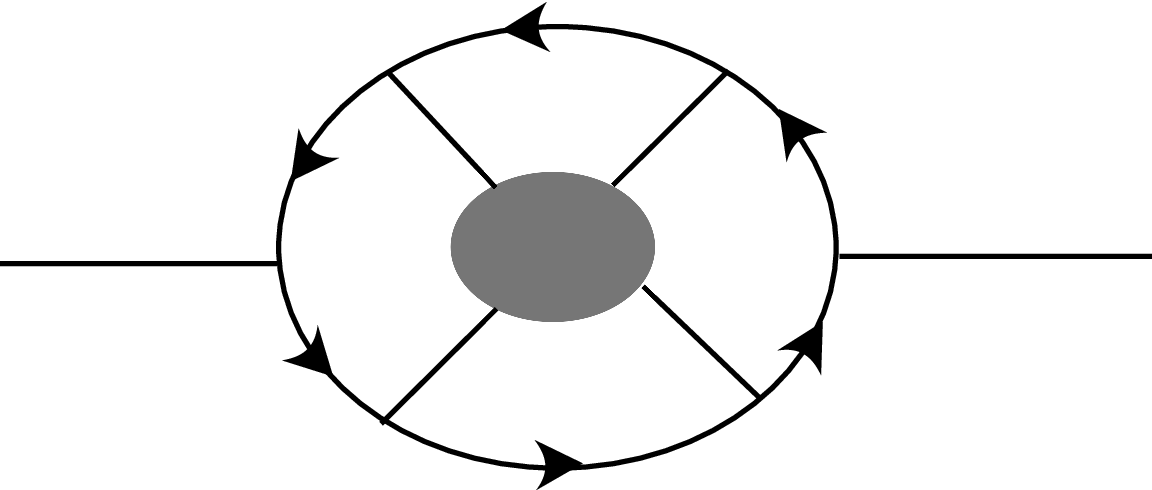}, it would be nice to just plug the result in from a table of blobs. \\

So we have one sensible, even useful, definition of the blob. We will define it to be what it would be if it were an internal part of a bigger graph (or a sum of internal parts in bigger graphs). Our Feynman rules for the bigger graph, which has \underline{its} external lines on the mass shell, then tell us exactly how to define the blob.\\

We still have a couple of conventional choices to make in defining a blob. We could include or not include the $n$ propagators that hang off $\widetilde{G}^{(n)}(k_1,\ldots, k_n)$ and we could include or not include the overall energy momentum conserving $\delta$ function. We'll include it all.\\

Here's a simple example. A big graph that contains $\widetilde{G}^{(2)}(k_1, k_2)$ is 
\begin{center}
\includegraphics[scale=0.4]{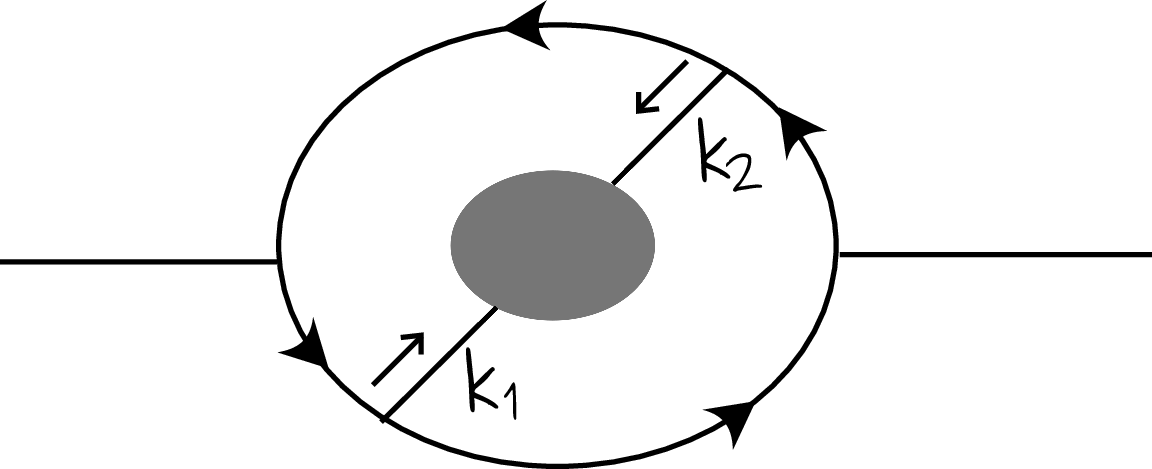}
\end{center}

More explicitly various contributions of this type are \\
\includegraphics[width=0.25\textwidth]{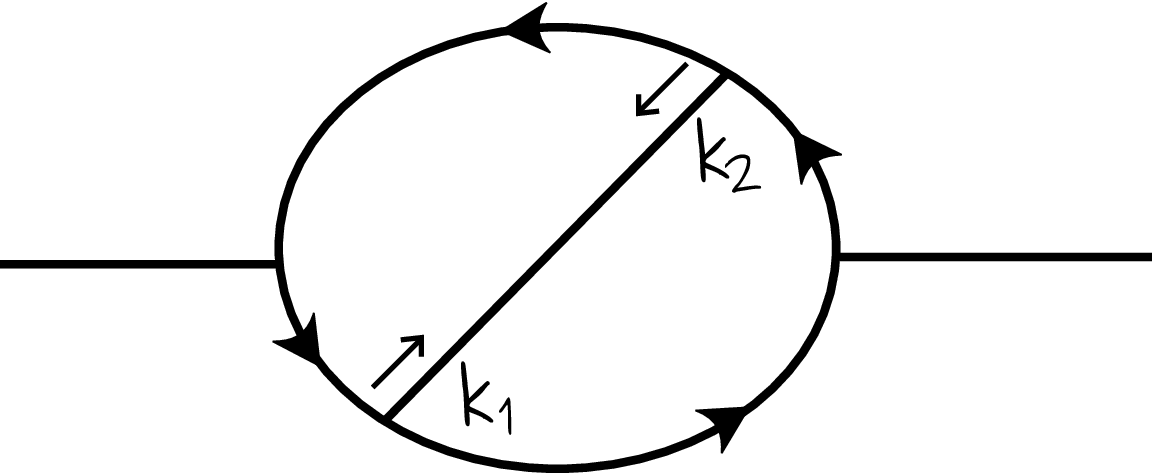}, \includegraphics[width=0.25\textwidth]{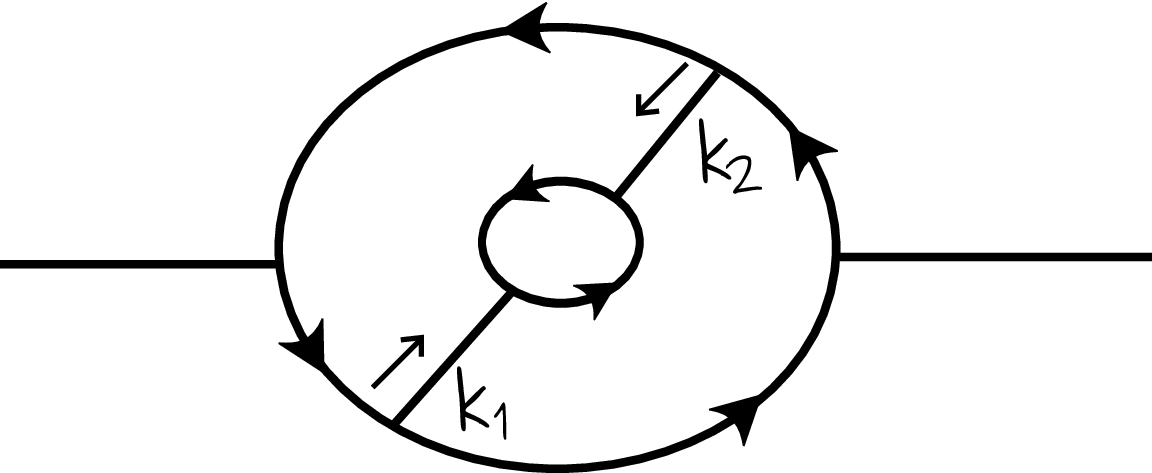} and \includegraphics[width=0.25\textwidth]{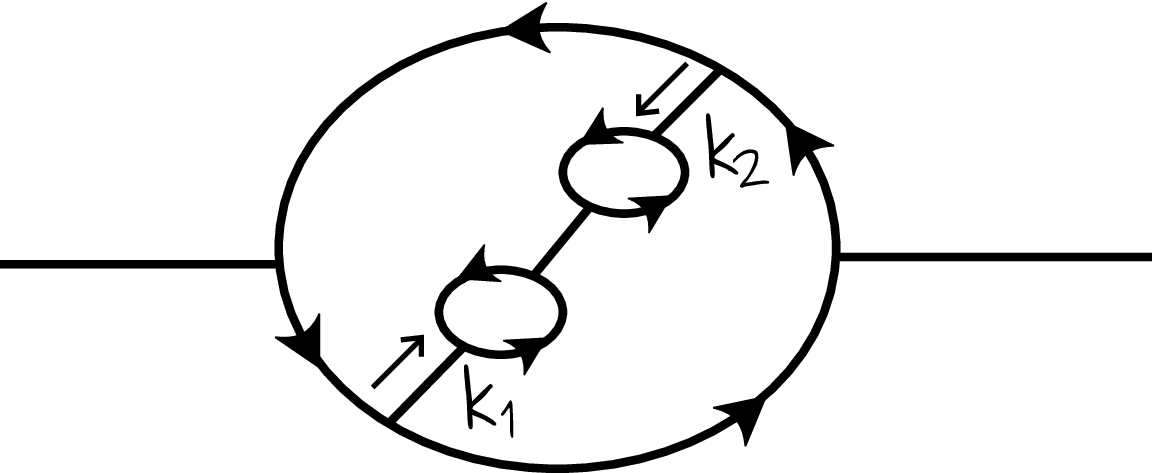}

These correspond to contributions to $\feyn{!{f}{k_1\rightarrow} p!{f}{\leftarrow k_2}}$ of $\feyn{!{f}{k_1\rightarrow}fs!{f}{\leftarrow k_2}}$, $\feyn{!{f}{k_1\rightarrow} fs0 !{flSV}{\leftarrow l} !{flSuA}{k_1+l\rightarrow} fs0 !{f}{\leftarrow k_2} }$ and $\feyn{!{f}{k_1\rightarrow}fs0 flSV flSuA fs0 fs fs0 flSV flSuA fs0 !{f}{\leftarrow k_2}}$.\\

To order $g^2$ we have
\begin{align*}
\feyn{!{f}{k_1\rightarrow} p!{f}{\leftarrow k_2}} &\equiv \widetilde{G}^{(2)}(k_1, k_2)\\
 &= (2 \pi)^4 \delta^{(4)}(k_1+k_2) \Bigg[ \frac{i}{k_1^2 - \mu^2 + i\epsilon} \\
&+ \int \frac{d^4 l}{(2\pi)^4} \frac{i}{(k_1 + l)^2 - m^2 + i\epsilon} \; \frac{i}{l^2 - m^2 + i\epsilon}\; \frac{i}{k_1^2 - \mu^2 + i\epsilon}\; \frac{i}{k_2^2 - \mu^2 + i\epsilon} \Bigg] 
\end{align*}

Because of the overall energy momentum conserving delta function, which enforces $k_1 = -k_2$, there is some ambiguity in the way to write down the contributions to $\widetilde{G}^{(2)}(k_1,k_2)$ to $\mathcal{O}(g^2)$. It could just as well have been written
\[ (2 \pi)^4 \delta^{(4)}(k_1+k_2) \Bigg[ \frac{i}{k_2^2 - \mu^2 + i\epsilon} + \left(\frac{i}{k_2^2 - \mu^2 + i\epsilon}\right)^2 \int \frac{d^4 l}{(2\pi)^4} \frac{i}{(-k_2 + l)^2 - m^2 + i\epsilon} \; \frac{i}{l^2 - m^2 + i\epsilon}\ \Bigg] \]

We can also write down a few contributions to 
\[ \begin{array}{ccc} \begin{array}{c} \widetilde{G}^{(4)} (k_1, k_2, k_3, k_4 )=\\ \\ \\ \\ \\ \\ \\ \end{array}
& \includegraphics[scale=0.3]{13-fig8.eps} & \begin{array}{c} = 
\begin{matrix}
\feyn{!{f}{k_1\rightarrow}fs!{f}{\leftarrow k_4}} \\
\feyn{![bot]{f}{k_2\rightarrow}fs![bot]{f}{\leftarrow k_3}}
\end{matrix} +
\begin{matrix}
\feyn{!{f}{k_1\rightarrow}fs!{f}{\leftarrow k_3}} \\
\feyn{![bot]{f}{k_2\rightarrow}fs![bot]{f}{\leftarrow k_4}}
\end{matrix} +
\begin{matrix}
\feyn{!{f}{k_1\rightarrow}fs!{f}{\leftarrow k_2}} \\
\feyn{![bot]{f}{k_3\rightarrow}fs![bot]{f}{\leftarrow k_4}}
\end{matrix} + 
\mathcal{O}(g^2) \\ \\ \\ \\ \\ \\ \\ \end{array} \end{array} \]
\vspace{-.8in}
\[ \quad\quad\quad\quad \quad= (2\pi)^4 \delta^{(4)}(k_1+k_4) \frac{i}{k_1^2 - \mu^2 + i\epsilon} (2\pi)^4 \delta^{(4)}(k_2 +k_3) \frac{i}{k_2^2 - \mu^2 + i \epsilon} + 2 \text{ permutations} + \mathcal{O}(g^2) \]

Since the second $\delta$ function enforces $k_2 = -k_3$ we can rewrite the first $\delta$ function as 
\[ \delta^{(4)}(k_1 + k_2 + k_3 +k_4) \]

\noindent
if you like, to display over all energy momentum conservation explicitly.\\

One thing we can do with these blobs is to recover $S$ matrix elements. We cancel off the external propagators and put the momenta back on their mass shells
\[ \langle k_1', k_2' | (S-1) | k_1,k_2 \rangle = \prod_{r =1,2,1', 2'} \!\!\!\!\!\!\!\!\!\!\!\!\!\!\!\!\!\!\!\!\!\!\!\!\!\!\overbrace{\frac{k_r^2 - \mu^2}{i}}^{\substack{ \text{to cancel the external} \\ \text{propagators we had included in } \widetilde{G}}} \!\!\!\!\!\!\!\!\!\!\!\!\!\!\!\!\!\!\!\! \widetilde{G}^{(4)}(-k_1',-k_2', k_1, k_2) \quad \text{(*) LSZ reduction formula} \]

\noindent
Because of the four factors of zero out front when the momenta are on mass shell, the graphs that we wrote out above do not contribute. Indeed, they should not contribute to $S-1$.
}{
  \sektion{14}{November 6}
\descriptionfourteen
Fourier transform (convention of Nov.~6):
\[ f(x) = \int \frac{d^4k}{(2\pi)^4} \widetilde{f}(k) e^{ik\cdot x} \]

\noindent
This is a little unfortunate because 
\[ e^{-iEt + i\vec{k}\cdot\vec{x}} \quad \quad (E>0) \] 

\noindent
is generally called a positive frequency plane wave (because $i \frac{\partial}{\partial t}$ acting on it give $E$ and $\frac{1}{i} \frac{\partial}{\partial \vec{x}}$ acting on it gives $\vec{k}$) and thus if $\widetilde{f}(k)$ has support for positive $k^0$, $f(x)$ negative frequency. \\

A source with positive frequencies creates particles while a source with negative frequencies absorbs particles. This is summed up in the Feynman rule
\[ \Diagram{\bullet\momentum[bot]{f}{\rightarrow k}} \Longleftrightarrow i \widetilde{\rho}(-k) \]

\noindent
(See Eqs.~(\ref{eq:14-frule1})-(\ref{eq:14-frule2}) if you don't know or remember how to get this Feynman rule)

\section*{Answer 1 (cont'd)}
We have found one meaning for our blob. We can use it to obtain another function, its Fourier transform. 

Using the Fourier transform convention 
\[ f(x) = \int \frac{d^4k}{(2\pi)^4} \widetilde{f}(k) e^{ik\cdot x} \]
\[ \widetilde{f}(k) = \int d^4x f(x) e^{-ik\cdot x} \]

\noindent
(which you'll notice has a different sign in the exponent from what we used on Oct.~21)\\

\noindent
The power theorem is $\displaystyle \int d^4 x f(x) g(x) = \int \frac{d^4 k}{(2\pi)^4} \widetilde{f}(k) \widetilde{g}(-k)$.\\

We have
\[ G^{(n)} (x_1, \ldots, x_n) = \int \frac{d^4 k_1}{(2\pi)^4} \cdots \frac{d^4 k_n}{(2\pi)^4} e^{ik_1\cdot x_1+ \cdots + ik_n \cdot x_n} \widetilde{G}^{(n)}(k_1, \ldots, k_n) \]

\noindent
where 
\[ \begin{matrix} \begin{matrix}\displaystyle \widetilde{G}^{(n)}(k_1, \ldots, k_n) =\\ \\ \\ \\ \\ \\ \\ \\ \\ \end{matrix}& \includegraphics[scale=0.35]{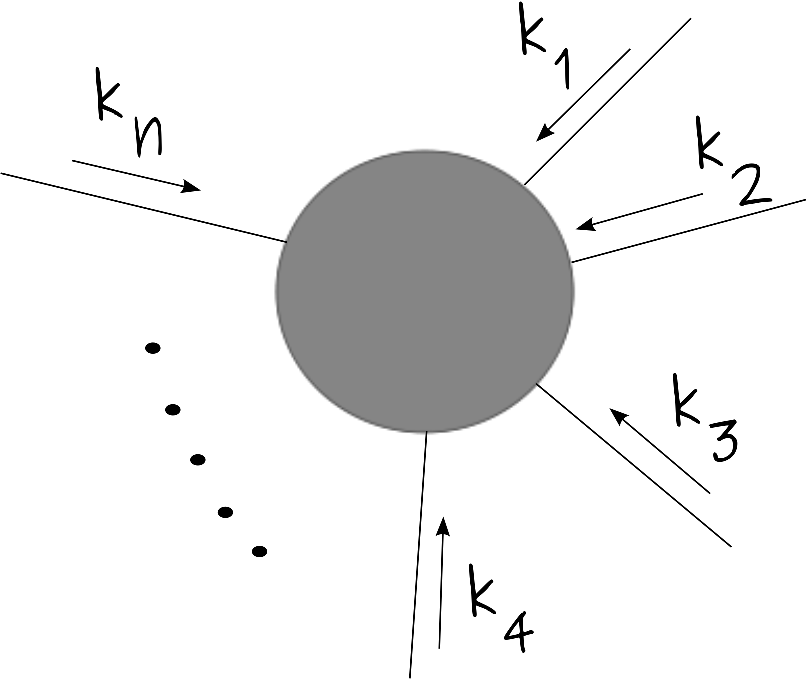} \end{matrix} \]
\vspace{-1in}

This is the sum of all (let's make this definite) Feynman diagrams to all orders (if the sum does not exist then the blob represents a formal power series in $g$) and the blob includes factors for the external propagators and the factor for overall energy momentum conservation $(2\pi)^4 \delta^{(4)}(k_1 + \cdots + k_n)$

\section*{Answer 2}
Consider modifying $\mathcal{H}$, say in model 3, 
\[ \mathcal{H} \rightarrow \mathcal{H} - \rho(x) \phi(x) \quad \quad (\mathcal{L} \rightarrow \mathcal{L} + \rho(x) \phi(x)) \]

\noindent
where $\rho(x)$ is a specified $c$ number source, not an operator. This adds a new vertex, a model 1 type vertex
\[ \Diagram{\bullet\momentum[bot]{f}{\rightarrow k}} = i \widetilde{\rho}(-k) \]

\noindent
The new Feynman rule was just quoted in class, let's see it arise in a simple example.\\

Let's suppose we have got the original Hamiltonian's vacuum counterterm all calculated out to some high order in perturbation theory so that there are no corrections to $\langle 0 | S | 0 \rangle $ to this high order. The modification of the Hamiltonian (density) $\mathcal{H} \rightarrow \mathcal{H} - \rho(x) \phi(x)$ spoils this. There are now contributions to $\langle 0 | S | 0 \rangle $ proportional to $\rho^n$ at low orders in $g$. At order $\rho$ and order $g$, we have $\Diagram{ \bullet f fs0 flSV flSuA fs0 } \quad $ At order $\rho$ and $\mathcal{O}(g^3)$ we have \includegraphics[width=1.5cm]{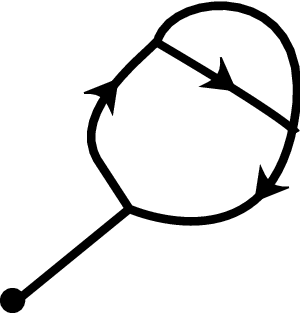}
\vspace{1cm}

Unfortunately, these are not interesting simple examples, because unless $\widetilde{\rho}(0)$ is nonzero they vanish because of energy-momentum conservation.\\

At order $\rho^2$ and order $g^0$ we have $\Diagram{\bullet \momentum[bot]{f}{} \bullet}$.

At order $\rho^2$ and order $g^2$ we have $\Diagram{\bullet f fs0 flSA flSuV fs0 f \bullet}$, as well as $\begin{matrix}\Diagram{ \bullet f fs0
flSA flSuV fs0 } \\ \\\Diagram{ \bullet f fs0 flSA flSuV fs0 }\end{matrix}\,\, $, $\quad\begin{matrix} \Diagram{ \bullet f \bullet} \\ \\ \Diagram{f fl fluA fV} \end{matrix}\,\,$, and $\quad \quad \Diagram{flSA flSuV fs0 f fs0 flSA flSuV } \quad \quad \Diagram{\bullet f \bullet}$ 
\vspace{1cm}

This is a nice simple example, let's look at it. It comes from the term second order in $\rho$ and second order in $g$ in 
\[ S = U_I(\infty,-\infty) = T e^{-i \int d^4 x ( g \psi^* \psi \phi - \rho \phi)} \]

\noindent
i.e.
\[ \frac{(-ig)^2}{2!} \frac{(i)^2}{2!} \int d^4x_1 d^4x_2 d^4x_3 d^4x_4 \rho (x_3) \rho (x_4) T (\psi^* \psi \phi(x_1) \psi^* \psi \phi(x_2) \phi(x_3) \phi(x_4)) \]

\noindent
The process is vacuum $\rightarrow$ vacuum, so we are looking for the completely contracted terms in the Wick expansion of the time ordered product. They are
\begin{equation*}
\begin{split}
\wick{1}{<1 \psi^* >1 \psi} \wick{21}{<2 \phi (x_1) <1 \psi^* >1 \psi >2\phi(x_2)} \wick{1}{<1 \phi(x_3) >1\phi(x_4)} \quad
&\longleftrightarrow \quad \quad \quad \Diagram{flSA flSuV fs0 f fs0 flSA flSuV } \quad\quad \quad \Diagram{ \bullet f\bullet}\\
\\
\left.
\begin{split}
\wick{1}{<1 \psi^* >1 \psi} \wick{213}{<2\phi (x_1) <1\psi^* >1\psi <3\phi(x_2) >3\phi(x_3) >2\phi(x_4)} \\ 
\wick{1}{<1 \psi^* >1 \psi} \wick{213}{<2\phi (x_1) <1\psi^* >1\psi <3\phi(x_2) >2\phi(x_3) >3\phi(x_4)} 
\end{split}
\right\} \quad &\longleftrightarrow \quad \quad \Diagram{ \bullet f fs0 flSA flSuV fs0 }\quad \quad \Diagram{ \bullet f fs0 flSA flSuV fs0 }
\\
\\
\wick{123}{<2 \psi^* <1 \psi <3\phi (x_1) >1\psi^* >2\psi >3\phi(x_2)} \wick{1}{<1\phi(x_3) >1\phi(x_4)} \quad &\longleftrightarrow \quad \quad \Diagram{ \bullet f \bullet} \quad \quad \quad \Diagram{f fl fluA fV} \\
\\
\left.
\begin{split}
\wick{1234}{<2 \psi^* <1 \psi <3\phi (x_1) >1\psi^* >2\psi <4\phi(x_2) >4\phi(x_3) >3\phi(x_4)} \\
\wick{1234}{<2 \psi^* <1 \psi <3\phi (x_1) >1\psi^* >2\psi <4\phi(x_2) >3\phi(x_3) >4\phi(x_4)} 
\end{split}\right\}\quad &\longleftrightarrow \quad \quad \Diagram{ \bullet f \bullet} \quad \quad \Diagram{\bullet f fs0 flSA flSuV fs0 f \bullet}
\end{split}
\end{equation*}

\vspace{1cm}
The last two are the ones I want to look at in detail. They differ by an exchange of $x_1 \leftrightarrow x_2$ only, and since these are dummy variables of integration they together make a contribution to $\langle 0 | S | 0 \rangle$ of 
\[ (-ig)^2 \frac{(i)^2}{2!} \int d^4x_1 d^4x_2 d^4 x_3 d^4 x_4 \rho(x_3) \rho(x_4) \underbrace{\wick{1}{<1\psi^*(x_1) >1\psi(x_2)}}_{(1)} \underbrace{\wick{1}{<1\psi(x_1) >1\psi^*(x_2)}}_{(2)} \underbrace{\wick{1}{<1\phi(x_1) >1\phi(x_3)}}_{(3)} \underbrace{\wick{1}{<1\phi(x_2) >1\phi(x_4)}}_{(4)} \]
\[ (1): \int \frac{d^4 p}{(2\pi)^4} e^{ip\cdot(x_1 - x_2)} \frac{i}{p^2 -m^2 + i\epsilon} \]
\[ (2): \int \frac{d^4 q}{(2\pi)^4} e^{iq\cdot(x_1 - x_2)} \frac{i}{q^2 -m^2 + i\epsilon} \]
\[ (3): \int \frac{d^4 k}{(2\pi)^4} e^{ik\cdot(x_1 - x_3)} \frac{i}{k^2 -\mu^2 + i\epsilon} \]
\[ (4): \int \frac{d^4 l}{(2\pi)^4} e^{il\cdot(x_2 - x_4)} \frac{i}{l^2 -\mu^2 + i\epsilon} \]

We have 
\begin{multline*}
\frac{1}{2!} \int \frac{d^4 p}{(2\pi)^4}\frac{d^4 q}{(2\pi)^4}\frac{d^4 k}{(2\pi)^4}\frac{d^4 l}{(2\pi)^4} \frac{i}{p^2 -m^2 + i\epsilon}\frac{i}{q^2 -m^2 + i\epsilon}\frac{i}{k^2 -\mu^2 + i\epsilon}\frac{i}{l^2 -\mu^2 + i\epsilon} \\ \times (-ig)^2 (i)^2 \int d^4 x_1 d^4 x_2 d^4 x_3 d^4 x_4 \rho(x_3) \rho(x_4)
e^{ix_1\cdot(p+q+k)}e^{ix_2\cdot (-p-q+l)}e^{-ix_3\cdot k}e^{-ix_4\cdot l}
\end{multline*}

Now if you go back to Eq.~(\ref{eq:11-botpage2}) to Eq.~(\ref{eq:11-page4}) and especially Eq.~(\ref{eq:11-botpage2}) in the lecture of Oct.~28, you'll see that when we have a factor $e^{ix_1\cdot (p+q+k)}$ that corresponds to a picture with $p$, $q$ and $k$ flowing out of $x_1$. Our picture for the integral at hand is 
\[ \Diagram{\bullet \momentum[bot]{f}{\leftarrow k} fs0 \momentum[bot]{flSV}{\rightarrow q} \momentum[bot]{flSuA}{\rightarrow p} fs0 \momentum[bot]{f}{\rightarrow l} \bullet} \]
\vspace{1cm}

The left vertex, corresponding to space-time point $x_1$ (but I hate to label it as such because it is just a dummy integration variable which is going to be integrated over, and in our combinatoric arguments for Feynman diagrams, we have kept the vertices unlabelled) is creating a nucleon with momentum $p$, creating an antinucleon with momentum $q$, and creating a meson with momentum $k$.\\

The $x$ integrals are easy to perform, we have 
\begin{multline*}
\frac{1}{2!} \int \frac{d^4 p}{(2\pi)^4}\frac{d^4 q}{(2\pi)^4}\frac{d^4 k}{(2\pi)^4}\frac{d^4 l}{(2\pi)^4} \frac{i}{p^2 -m^2 + i\epsilon}\frac{i}{q^2 -m^2 + i\epsilon}\frac{i}{k^2 -\mu^2 + i\epsilon}\frac{i}{l^2 -\mu^2 + i\epsilon} \\ \times (-ig)^2 (i)^2 \widetilde{\rho}(k) \widetilde{\rho}(l) (2\pi)^4 \delta^{(4)}(p+q+k) (2\pi)^4 \delta^{(4)}(-p-q+l) 
\end{multline*}

This is just what you would have directly written down using our old Feynman rules supplanted by 
\[ \Diagram{\bullet\momentum[bot]{f}{\leftarrow k}} = i \widetilde{\rho}(k) \]

\noindent
except for that $\cfrac{1}{2!}$ out front, which I'll explain in a moment.\\

This was a moderately interesting graph to show how the Feynman rule comes out. If you know how the Feynman rule comes out, but you want to check whether it's $\widetilde{\rho}(k)$ or $\widetilde{\rho}(-k)$ just look at the lowest order (in $\rho$, zeroth order in $g$) contribution to 
\begin{align}\label{eq:14-frule1}
\langle 0 | S | k \rangle &= \langle 0 | U_I(\infty, -\infty) |k \rangle \nonumber\\
&= \langle 0 | \left[ 1 + i \int d^4 x \rho(x) \phi(x) \right] |k \rangle + \cdots \nonumber\\
&= 0 + \Diagram{ \bullet\momentum[bot]{f}{\leftarrow k}} + \cdots
\end{align}

\noindent
($|k\rangle$ is relativistically normalized and I'll use the relativistically normalized creation and annihilation operators in the expansion for $\phi(x)$.)\\

\noindent
But, 
\begin{equation} 
\langle 0 | \phi(x) | k \rangle = \int \frac{d^3 k'}{(2\pi)^3 2 \omega_{k'}} e^{-ik\cdot x}\!\!\!\!\! \underbrace{\langle 0 | a(k') |k\rangle}_{(2\pi)^3 2 \omega_k \delta^{(3)}(\vec{k} -\vec{k'})}\!\!\!\!\! = e^{-ik\cdot x}
\end{equation}

\noindent
So
\begin{equation}\label{eq:14-frule2}
\Diagram{\bullet\momentum[bot]{f}{\leftarrow k}} = i \int d^4 x \rho(x) e^{-ik\cdot x} =i \widetilde{\rho}(k)
\end{equation}

Now for that $\frac{1}{2!}$ out front (see also the short argument after Eq.~(\ref{eq:14-shortarg})). Earlier (after Eq.~(\ref{eq:11-page5}) of lecture on Oct.~28), we sung and danced about how there were no symmetry factors in model 3. That argument still goes through, and it still only applies to diagrams where \underline{each connected part has at least one external line}. What we have here is a vacuum to vacuum diagram, no external lines, and we now have to worry about symmetry factors.\\

Suppose we have a graph in the Wick expansion with $n$ powers of $\rho \phi$ and $m$ powers of $g \psi^* \psi \phi$, which comes with a $\frac{1}{n!} \frac{1}{m!}$ from the exponential. 
\begin{center}
\includegraphics[scale=0.35]{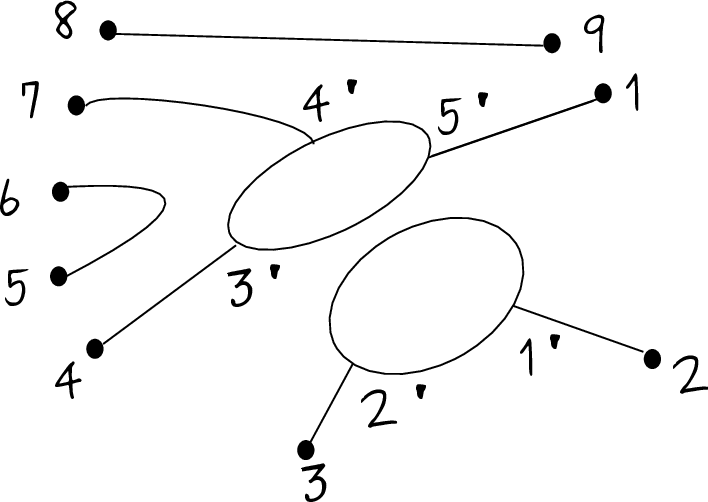}
\end{center}

\noindent
(In this example $n=9$ and $m=5$)\\

Each of the $m!$ permutations of the model 3 vertices, keeping the source vertices fixed, is a new term in the Wick expansion. They are all uniquely identified by the way they are attached to the source vertices. There are no model 3 vertices that are not somehow attached to a source vertex. That would be a disconnected bubble, which is a contribution to $\langle 0 | S | 0 \rangle$ with the source off, and by assumption, the vacuum energy counterterm has been adjusted so that there are no corrections to $\langle 0 | S | 0 \rangle$ with the source off.\\

Now what about the $n!$ permutations of source vertices? Some of them make no new contributions to the Wick expansion. For example $5\leftrightarrow 6$ and $8 \leftrightarrow 9$, but also $2 \leftrightarrow 3$, because that has already been counted as $1' \leftrightarrow 2'$.\\

Any of the $n!$ permutations that do make a new contribution to the Wick expansion, that is, that have not already been counted in the permutations of the $m$ model 3 vertices, are accounted for in another way.\\

\noindent
For example, $7 \leftrightarrow 8$ or $6 \leftrightarrow 8$ gives a new term in the Wick expansion.\\

To see the accounting work, look at the messy example in momentum space. It is 
\[ \begin{matrix} \begin{matrix} \displaystyle\frac{(i)^9}{9!} \int \frac{d^4 k_1}{(2\pi)^4} \cdots \frac{d^4 k_9}{(2\pi)^4} \widetilde{\rho}(-k_1) \cdots \widetilde{\rho}(-k_9) \times\\ \\ \\ \\ \\ \\ \\ \end{matrix} &\!\!\!\!\!\!\!\!\!\!\underbrace{\includegraphics[scale=0.35]{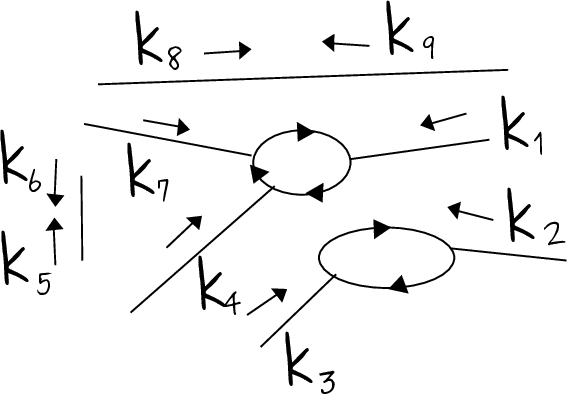}}_{\substack{\text{Feynman diagram with}\\ \text{external lines off the mass shell}}}\end{matrix} \]
\vspace{-.5in}

\noindent
Instead of using the permutation $7 \leftrightarrow 8$ to partially cancel off the $\frac{1}{9!}$ out front consider it as the same mess $m$ with a new diagram in the integrand.
\begin{center}
\includegraphics[scale=0.35]{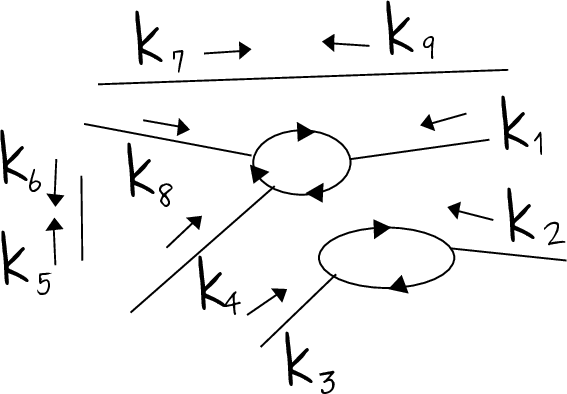}
\end{center}

\noindent
Of course when the momenta are integrated over, this is identical to the integral above.\\

Now both of these would be counted in $\widetilde{G} (k_1, \cdots, k_9)$ and in fact every permutation of the 9 source vertices that leads to a new term in the Wick expansion corresponds to a diagram in $\widetilde{G} (k_1, k_2, \cdots, k_9)$.\\

Of course there are diagrams with $n=9$ and $m=5$ that are not of the same pattern as the one drawn (differ by more than a permutation of vertices). For example 
\begin{center}
\includegraphics[scale=0.35]{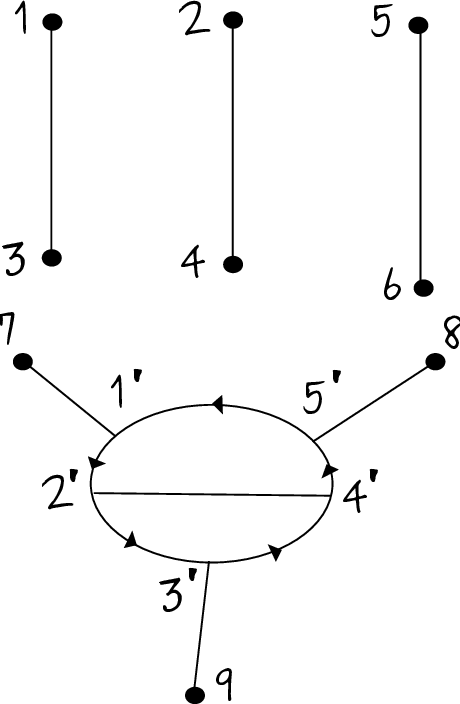}
\end{center}

\noindent
There is a Feynman diagram in $\widetilde{G}(k_1, \ldots, k_9)$ for this too
\begin{center}
\includegraphics[scale=0.35]{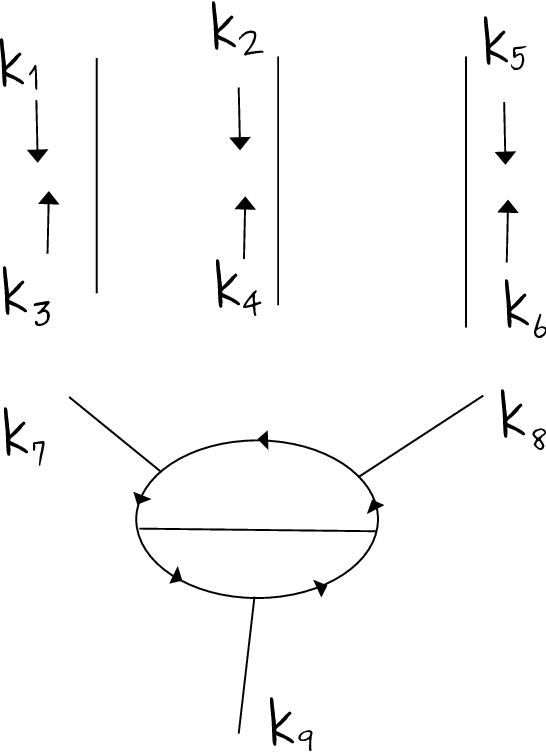}
\end{center}

Let
\begin{center}
\includegraphics[scale=0.35]{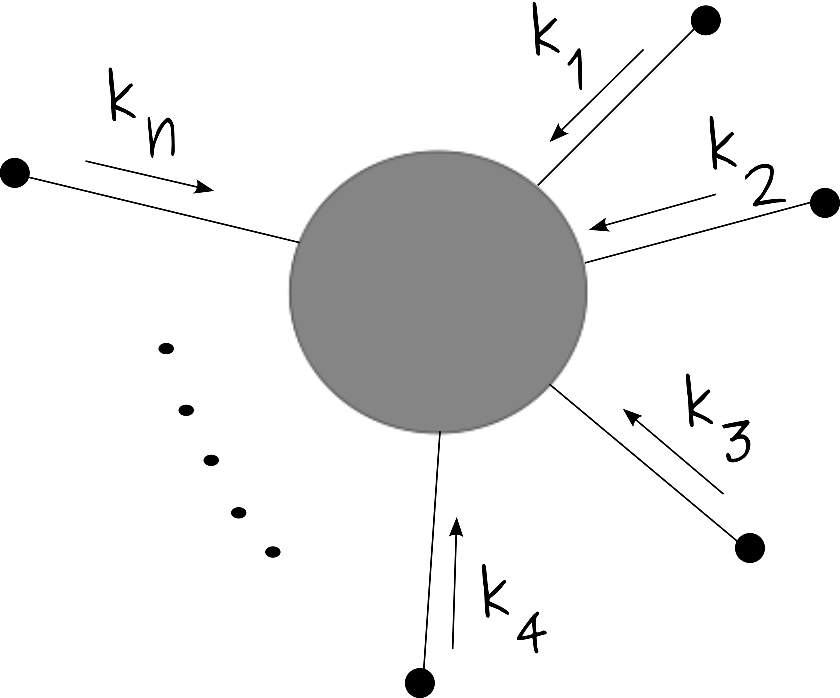}
\end{center}

\noindent
denote the sum of all diagrams to all orders in $g$ (i.e.~all $m$) and at $n$th order in $\rho$ that contribute to $\langle 0 | S | 0 \rangle$. \\

What these combinatoric arguments say is 
\begin{align}\label{eq:14-shortarg}
\includegraphics[scale=0.2]{14-fig7.eps} &\begin{matrix}\displaystyle = \frac{(i)^n}{n!} \int \frac{d^4 k_1}{(2\pi)^4} \cdots \frac{d^4 k_n}{(2\pi)^4} \widetilde{\rho}(-k_1) \cdots \widetilde{\rho}(-k_n) \times \\ \\ \\ \\ \\ \\\end{matrix}
\includegraphics[scale=0.2]{14-fig1.eps} \nonumber\\
\rule{0pt}{-1cm}&= \frac{(i)^n}{n!} \int \frac{d^4 k_1}{(2\pi)^4} \cdots \frac{d^4 k_n}{(2\pi)^4} \widetilde{\rho}(-k_1) \cdots \widetilde{\rho}(-k_n) \widetilde{G}(k_1,\ldots,k_n)
\end{align}

\noindent
Having gone all the way back to Wick expansion arguments to show the combinatorics are right for this, I'll try to make a shorter argument.\\

The source creates $n$ mesons, which are distinguishable by virtue of the fact that they all carry different momenta $k_1, \ldots, k_n$. They interact in all possible ways. That gives us \\
$(i)^n \widetilde{\rho}(-k_1)\cdots \widetilde{\rho}(-k_n) \widetilde{G}(k_1,\dots, k_n)$. Now we integrate over \underline{all} momenta $k_1, \dots, k_n$, and in doing so we make an overcounting by $n!$ .\\

\noindent
\textbf{BEST ARGUMENT}\\

One last way of arguing this. Instead of considering this as an $n$th order calculation in $\rho$, temporarily think of it as a first order calculation in each of $n$ different sources $\rho_1(x), \dots, \rho_n(x)$. Then the diagram where source 1 creates a particle with momentum $k_1$ and source 2 creates a particle with momentum $k_2$ really is distinguishable from a diagram where source 1 creates $k_2$ and source $2$ creates $k_1$. There is no overcounting when you integrate over all momenta. That contribution to $\langle 0 | S | 0 \rangle $ would be 
\[ (i)^n \int \frac{d^4 k_1}{(2\pi)^4} \cdots \frac{d^4 k_n}{(2\pi)^4} \widetilde{\rho}_1(-k_1) \cdots \widetilde{\rho}_n(-k_n) \widetilde{G}(k_1,\ldots,k_n) \]

How does this imagined calculation differ from ours? Well, in the exponential $\rho_1(x_1) \cdots \rho_1(x_n)$ comes with coefficient $1$, while $\rho(x_1) \cdots \rho(x_n)$ comes with coefficient $\frac{1}{n!}$.\\

To all orders
\begin{align*} 
\langle 0 | S | 0 \rangle &= 1 + \sum_{n=1}^\infty \frac{(i)^n}{n!} \int \frac{d^4 k_1}{(2\pi)^4} \cdots \frac{d^4 k_n}{(2\pi)^4} \widetilde{\rho}(-k_1) \cdots \widetilde{\rho}(-k_n) \widetilde{G}^{(n)}(k_1,\ldots,k_n) \\
&= 1 + \sum_{n=1}^\infty \frac{(i)^n}{n!} \int d^4 x_1 \cdots d^4 x_n\, \rho(x_1) \cdots \rho(x_n) G^{(n)}(x_1,\ldots,x_n)
\end{align*}

This is the second answer to our question. The Fourier transform of the sum of Feynman diagrams with $n$ external lines off the mass shell is a Green's function (that's what $G$ stands for). In a theory with linear response only $G^{(1)} \neq 0$. From conservation of probability alone, you can see that the response of a quantum mechanical system can't be linear. Green introduced the first Green's function in the early 19th century. From a prescribed charge distribution, $\rho(\vec{x})$, his Green's function gave you the electrostatic potential, $\phi (\vec{x})$.
\[ \phi(\vec{x}) = \int d^3 x' G(\vec{x},\vec{x}\,') \rho(\vec{x}\,') \]
\[ G(\vec{x}, \vec{x}\,') = \frac{1}{|\vec{x}-\vec{x}\,'|} \]

\noindent
satisfies 
\[ \nabla^2 \phi = - \vec{\nabla} \cdot \vec{E} = - 4 \pi \rho \quad \quad (\vec{E} = - \vec{\nabla} \phi) \]

Let's explicitly note that the vacuum to vacuum transition amplitude depends on $\rho$ by writing $\langle 0 | S | 0 \rangle_\rho $ (don't confuse the subscript $\rho$ with a $p$). 

$\langle 0 | S | 0 \rangle_\rho $ is a function\underline{al} of $\rho$. You give me a \underline{function} on spacetime, $\rho(x)$, and I give you back a number, $\langle 0 | S | 0 \rangle_\rho$. Actually, it is just a function of an infinite number of variables, the value of the source at each spacetime point, and the nomenclature ``functional" is redundant, we could just say ``function". Mathematicians don't call a vector in an infinite dimensional space ``vector\underline{al}". $\langle 0 | S | 0 \rangle_\rho $ comes up often enough it gets a name, $Z[\rho]$
\[ Z \!\!\!\!\!\!\!\!\!\!\!\!\!\!\!\!\! \underbrace{[\rho]}_{\substack{\text{The square brackets}\\ \text{remind you that}\\ \text{this is a function}\\ \text{of a function, } \rho}} \!\!\!\!\!\!\!\!\!\!\!\!\!\!\! \equiv \langle 0 | S | 0 \rangle_\rho \]

$Z[\rho]$ is called a generating functional for the Green's functions because in the infinite dimensional generalization of a Taylor series, we have 
\[ \frac{\delta^n Z[\rho]}{\delta \rho(x_1) \cdots \delta \rho(x_n)} \Bigg|_{\rho=0} = (i)^n G^{(n)} (x_1, \ldots, x_n) \]

\noindent
The $\delta$ instead of a $\partial$ reminds you that you are taking a partial derivative of $Z$ with respect to $\rho(x)$, holding a $(4-d)$ continuum of other variables fixed. These are called functional derivatives. \\

\underline{Ex nihil omnes}: All physical information (all Green's
functions, and hence all $S$ matrix elements) about the system is coded in the vacuum persistence amplitude in the presence of an external source $\rho$. \\

$Z[\rho]$ is called a generating functional in analogy with the functions of two variables which when you Taylor expand in one variable, the coefficients are a set of functions like the Legendre polynomials in the other. Sometimes it is very useful to put a whole set of functions in one neat package like that. An example will be the Ward identities, which are a statement about Green's functions resulting from a symmetry. It can be put very compactly in term of $Z$.\\

Because of our great theorem 
\[ \sum \text{all Wick diagrams} = \\\ :e ^{\sum \text{connected diagrams}} : \]

\noindent
the sums are sums of normal ordered terms. 

Apply this to a model which has had a source added. Take the vacuum expectation value of both sides. The LHS is just $\langle 0 | S | 0 \rangle_\rho $, i.e.~$Z[\rho]$.

We have
\[ Z[\rho] = \langle 0 | :e^{\sum \text{connected diagrams}} : | 0 \rangle= e^{\langle 0 | \sum \text{connected diagrams } |0 \rangle} \]

\noindent
This is true because the terms in the sum in the exponential are themselves normal ordered, convince yourself. Taking the natural logarithm, 
\begin{align*} 
\text{ln } Z [\rho] &= \langle 0 | \sum \text{connected diagrams } | 0 \rangle \\
&= \sum_{n=1}^\infty \frac{(i)^n}{n!} \int \frac{d^4 k_1}{(2\pi)^4} \cdots \frac{d^4 k_n}{(2\pi)^4} \widetilde{\rho}(-k_1) \cdots \widetilde{\rho}(-k_n) \widetilde{G}_c(k_1,\ldots,k_n)
\end{align*}

\noindent
$\widetilde{G}_c$ is the sum of all connected Feynman diagrams, with $k_1, \ldots, k_n$ possibly off shell, including the overall energy momentum conserving $\delta$ function, and the external propagators, which blow up on mass shell.

\section*{Answer 3}
One more way of interpreting $G^{(n)}(x_1,\ldots, x_n)$. By a cunning trick we will show that $G^{(n)}(x_1,\ldots, x_n)$ is a VEV (``Vacuum Expectation Value") of a time ordered string of Heisenberg fields. As we did in obtaining answer 2, let
\[ \mathcal{H} \rightarrow \mathcal{H}-\rho\phi(x) \]
\[ \mathcal{H}_0 + \mathcal{H}\,' \rightarrow \mathcal{H}_0 + \mathcal{H}\,'-\rho\phi(x) \]

\noindent
As far as Dyson's formula is concerned, you can break the Hamiltonian up into a ``free" and interacting part in any way you please. Let's take the ``free" part to be $\mathcal{H}_0 + \mathcal{H}\,'$ and the interaction to be $-\rho \phi(x)$. I put quotes around ``free", because in this new interaction picture, the fields evolve according to 
\[ \phi(\vec{x},t) = e^{iHt} \phi(\vec{x}, 0) e^{-iHt} \]
\[ H = \int d^3 x \mathcal{H} \quad \quad \mathcal{H} = \mathcal{H}_0 + \mathcal{H}\,' \]

These fields are not free. They do not obey the free field equations of motion. You can't define a contraction for these fields, and thus you can't do Wick's theorem. These fields are what we would have called Heisenberg fields if there was no source. For this reason we'll subscript them with an $H$.\\

Let's see what this tells us about $Z[\rho]$.
\begin{align*} 
Z [\rho] &= \langle 0 | S | 0 \rangle_\rho = \langle 0 | T e^{+i \int d^4 x \rho(x) \phi_H (x)} | 0 \rangle \\
\text{\small just expand }\quad &= 1 + \sum_{n=1}^\infty \frac{(i)^n}{n!} \int d^4 x_1 \cdots d^4 x_n\, \rho(x_1) \cdots \rho(x_n) \langle 0 | T(\phi_H(x_1) \cdots \phi_H(x_n) )|0 \rangle
\end{align*}

\noindent
and we read off 
\begin{equation}\label{eq:14-1}
G^{(n)}(x_1,\ldots,x_n) = \langle 0 | T(\phi_H(x_1) \cdots
\phi_H(x_n)) |0 \rangle 
\end{equation}

To summarize, we have found three meanings for the (sum of all) Feynman diagrams with ($n$) external lines off the mass shell.
\begin{enumerate}
\item It is a handy blob we can plaster into the interior of a larger diagram.
\item Its Fourier transform (times $\frac{(i)^n}{n!}$) is the coefficient of the $n$th order term in $\rho$ in the expansion of the vacuum to vacuum persistence amplitude in the presence of a source, $\rho$.
\item Its Fourier transform is the VEV of a time ordered string of Heisenberg fiels. 
\end{enumerate}

\noindent
This can all be taken as motivation, because we are going to start from scratch and do a 

\begin{center}
REFORMULATION OF SCATTERING THEORY \\
No more turning on and off function 
\end{center}

Imagine you have a well-defined theory, with a time independent Hamiltonian, $H=\int d^3x\mathcal{H}$ (the turning on and off function is gone for good), whose spectrum is bounded below, whose lowest lying state is not part of a continuum, and the Hamiltonian has actually been adjusted so that this state, $\!\!\!\!\!\!\!\!\!\!\!\!\!\!\!\!\!\underbrace{|0\rangle_p}_{\text{don't confuse } p \text{ with } \rho}\!\!\!\!\!\!\!\!\!\!\!\!\!\!\!$, the physical vacuum, satisfies
\[ H |0\rangle_p = 0 \]

\noindent
The vacuum is translationally invariant and normalized to one 
\[ \vec{P} |0 \rangle_p = 0 \quad \text{and} \quad _p\langle 0 | 0 \rangle_p = 1 \]

Now let $\mathcal{H} \rightarrow \mathcal{H} - \rho(x) \phi(x) $ and define 
\begin{align*}
Z [\rho] &\equiv \ _p\langle 0 | S | 0 \rangle_p \, \Big|_{\text{in the presence of the source }\rho}\\
&=\ _p\langle 0 | \!\!\!\!\!\!\!\!\!\!\!\!\!\!\!\!\!\!\!\!\! \underbrace{U(\infty,-\infty)}_{\substack{\text{Schr\"odinger picture evolution operator}\\\text{for the Hamiltonian} \int d^3x (\mathcal{H} - \rho \phi)}}\!\!\!\!\!\!\!\!\!\!\!\!\!\!\!\!\!\!\!\!\! |0 \rangle_p
\end{align*}

\noindent 
and define 
\[ G^{(n)} (x_1, \ldots, x_n) = \frac{1}{i^n} \frac{\delta^{n} Z[\rho]}{\delta \rho(x_1) \cdots \delta\rho(x_n)} \bigg|_{\rho=0} \]

\textbf{Two Questions:}
\begin{enumerate}
\item Is $G^{(n)}$ defined this way (the F.T.) of the sum of all Feynman graphs? Let's call the $G^{(n)}$ defined as the sum of all Feynman graphs $G_F^{(n)}$ and the $Z$ which generated those $Z_F$. The question is: Is $G^{(n)} = G^{(n)}_F$ or equivalently, is $Z = Z_F$?
\begin{center} 
Answer will be ``yes."
\end{center}
\item Are $S-1$ matrix elements obtained from Green's functions in the same way as before? For example, is 
\[ \langle k'_1,k'_2 | (S-1) |k_1,k_2 \rangle = \prod_{a=1,2,1',2'} \frac{k_a^2 - \mu^2}{i} \widetilde{G}(-k_1',-k_2',k_1,k_2) \quad \text{?} \]
\begin{center}
Answer will be ``almost."
\end{center}
\end{enumerate}

\textbf{Answer to question 1:} Is $G^{(n)} = G_F^{(n)}$? 

Using Dyson's formula, in the exact same way as we did in Eq.~(\ref{eq:14-1}), gives 
\begin{equation}\label{eq:14-2} G^{(n)} 
(x_1,\ldots, x_n) =\ _p \langle 0 | T (\phi_H(x_1) \cdots \phi_H(x_n)) |0 \rangle_p
\end{equation}

\noindent
Does $Z_F[\rho]$, the generating functional you get by blindly summing graphs generate the same Green's functions? \\

$\mathcal{H}$ splits up into $\mathcal{H}_0 + \mathcal{H}\,'$. Let $\mathcal{H}_I = \mathcal{H}\,'(\phi_I)$. The thing which after Wick's theorem and a combinatoric argument or two had a graphical expansion is 
\[ Z_F [\rho] = \lim_{t_\pm \rightarrow \pm \infty} \!\!\!\!\!\!\!\!\! \underbrace{\langle 0 |}_{\substack{\text{Base eigenstate}\\\text{of } H_0\\ H_0 | 0 \rangle = 0}} \!\!\!\!\!\!\!\!\!\! T e^{-i \int^{t_+}_{t_-} d^4 x [ \mathcal{H}_I - \rho \phi_I ]} |0 \rangle \]

We used to adjust the constant part of $\mathcal{H}_I$ to eliminate vacuum bubbles in our old scattering theory when $\rho = 0$. That is, we adjusted the vacuum energy counterterm, so that the vacuum to vacuum graphs (with no source vertices) summed to zero. There is an equivalent way of throwing away the vacuum bubbles. You divide out of $Z_F[\rho]$ the sum of all vacuum to vacuum graphs with no source vertices explicitly, and then you don't have to worry about a vacuum energy c.t., i.e.~you divide by the same thing with $\rho = 0$.
\[ Z_F [\rho] = \lim_{t_\pm \rightarrow \pm \infty} \frac{\langle 0 | T e^{-i \int^{t_+}_{t_-} d^4 x [ \mathcal{H}_I - \rho \phi_I ]} |0 \rangle}{\langle 0 | T e^{-i \int^{t_+}_{t_-} d^4 x \mathcal{H}_I} |0 \rangle} \]

\noindent
To get $G_F^{(n)} (x_1, \ldots, x_n)$ we do $n$ functional derivatives w.r.t.~$\rho$ and then set $\rho = 0$ (and divide by $i^n$)
\[ G_F^{(n)} (x_1, \ldots, x_n) = \lim_{t_\pm \rightarrow \pm \infty} \frac{\langle 0 | T \big[ \phi_I(x_1) \cdots \phi_I(x_n) e^{-i \int^{t_+}_{t_-} d^4 x \, \mathcal{H}_I } \big] |0 \rangle}{\langle 0 | T e^{-i \int^{t_+}_{t_-} d^4 x \, \mathcal{H}_I} |0 \rangle} \]

We have got a little work to do to show this is the same as $G^{(n)}$ in Eq.~(\ref{eq:14-2}). Fortunately, both these expressions are manifestly symmetric under the $n!$ permutations of the $x_1, \ldots, x_n$, so it suffices to prove they are equal for one ordering which for convenience we choose so that 
\[ x_1^0 > x_2^0 > \cdots > x_n^0 \quad \quad \text{or for short } t_1 > t_2 > \cdots > t_n \]

The time ordering in the expression for $G^{(n)}$ is just lexicographic ordering.
\begin{align*} 
G^{(n)}(x_1, \ldots, x_n) &= \ _p \langle 0 | T (\phi_H(x_1) \cdots \phi_H(x_n)) | 0 \rangle_p \\
&= \ _p \langle 0 | \phi_H(x_1) \cdots \phi_H(x_n) | 0 \rangle_p 
\end{align*}

\noindent
Using the standard shorthand for $\displaystyle e^{-i \int^{t_b}_{t_a} d^4x \mathcal{H}_I} = U_I(t_b, t_a) $, the time ordering in the expression for $G_F^{(n)}$ is 
\[ G^{(n)}_F(x_1,\ldots,x_n) = \lim_{t_\pm \rightarrow \pm \infty} \frac{ \langle 0 | U_I(t_+,t_1) \phi_I(x_1) U_I(t_1,t_2) \phi_I(x_2) \cdots \phi_I(x_n)U_I(t_n,t_-) |0 \rangle}{\langle 0 | U(t_+,t_-) |0 \rangle} \]

\noindent
at least in the $\lim_{t_\pm \rightarrow \pm \infty}$ when $t_+ > t_1 > \cdots > t_n > t_- $. Convince yourself.\footnote{I usually put ``convince
yourself" when I haven't written enough to make something clear, but if I wrote more it would take just as long to figure out what I was saying as it would take to convince yourself.}\\

Everywhere $U_I(t_a,t_b)$ appears, rewrite it as $U_I(t_a,0) U_I(0,t_b)$ and then use 
\[ \phi_H(x_i) = U_I(t_i, 0 )^\dagger \phi_I(x_i) U_I(t_i,0) = U_I(0, t_i ) \phi_I(x_i) U_I(t_i,0) \]

to get
\[ G^{(n)}_F(x_1,\ldots,x_n) = \lim_{t_\pm \rightarrow \pm \infty} \frac{ \langle 0 | U_I(t_+, 0) \phi_H(x_1) \phi_H(x_2) \cdots \phi_H(x_n)U_I(0, t_-) |0 \rangle }{ \langle 0 | U_I(t_+, 0) U_I(0, t_-)|0 \rangle } \]

Considering the whole mess sandwiched with $U_I(0,t_-) |0 \rangle$ in the numerator or denominator as some fixed state $\langle \phi |$, let's work on 
\begin{align*} 
\lim_{t_- \rightarrow -\infty} \langle \phi | U_I(0,t_-) |0 \rangle &= \lim_{t_- \rightarrow -\infty} \langle \phi | U_I(0,t_-) \!\!\!\! \underbrace{e^{i H_0 t_-}}_{\substack{\text{fancy way of}\\\text{inserting 1}}} \!\!\!\! |0 \rangle\\ &\!\!\!\overbrace{=}^{\substack{*\text{see}\\\text{below}}}
\lim_{t_- \rightarrow -\infty} \langle \phi | \!\!\!\!\!\!\! \underbrace{U(0,t_-)}_{\text{Schr\"odinger picture}} \!\!\!\!\!\!\!\! |0 \rangle \quad \quad \substack{\text{Using an easily derivable}\\\text{relationship between the}\\\text{evolution operator in the}\\\text{various pictures, Oct.~16, Eq.~
(\ref{eq:08-interactionpicture})}}\\ \substack{\text{insert a}\\\text{complete}\\\text{set}} &= \lim_{t_-\rightarrow -\infty} \langle \phi | U(0,t_-) \underbrace{\left[ |0 \rangle_p \, _p\langle 0 | + \fourteensumintb |n \rangle \langle n| \right]}_{\substack{\text{all other eigenstates of}\\\text{the full Hamiltonian, } H\\ H |0 \rangle_p = 0, \, H|n\rangle = E_n | n \rangle }} |0 \rangle \\ &= \langle \phi | 0 \rangle_p \ _p\langle 0 | 0 \rangle + \lim_{t_- \rightarrow - \infty} \fourteensuminta_{\substack{\text{all other}\\\text{eigenstates}}} e^{i E_n t_-} \langle \phi | n \rangle \langle n | 0 \rangle 
\end{align*}

\noindent
*: You see it doesn't really matter from this point on that it is a bare vacuum that $U(0,t_-)$ is acting on. What we are showing is that for two arbitrary fixed states $\lim_{t_- \rightarrow - \infty} \langle \phi | U(0,t_-) | \psi \rangle = \langle \phi | 0 \rangle_p \ _p \langle 0| \psi \rangle $ .\\

Now every state but the vacuum is part of a continuum. As long as $\langle \phi | n \rangle \langle n | 0 \rangle $ is a continuous function, the limit vanishes. The integral is a continuous function that oscillates more and more wildly as $t_- \rightarrow - \infty$. In the limit it integrates to zero. A similar argument shows 
\[ \lim_{t_+ \rightarrow \infty} \langle 0 | U_I(t_+,0) |\psi \rangle = \langle 0 | 0 \rangle_p \ _p\langle 0 | \psi \rangle \]

Physically what this theorem about oscillating integrands (the Riemann-Lebesgue lemma) says is that if you look at a state in some fixed region (take its inner product with some fixed state $\langle \phi |$) and you wait long enough, the only trace of it that will remain is its (true) vacuum component. All the one and multi-particle states will have run away.
\begin{align*} 
G_F^{(n)}(x_1, \ldots, x_n) &= \frac{\cancel{\langle 0| 0 \rangle_p} \ _p\langle 0 |\phi_H(x_1)\cdots \phi_H(x_n) |0 \rangle_p \cancel{\ _p\langle 0 | 0 \rangle}}{\cancel{\langle 0 | 0 \rangle_p} \ _p\langle 0 | 0 \rangle_p \cancel{\ _p\langle 0 | 0 \rangle}} \\
&= G^{(n)}(x_1, \ldots, x_n)
\end{align*}

\noindent
and there is no longer any reason to distinguish between them. \\

\textbf{Question 2}: Are $S-1$ matrix elements obtained from Green's functions in the same way as before?\\

By introducing a turning on and off function, we were able to show that 
\[ \langle l_1, \ldots, l_s | (S-1) | k_1, \ldots, k_r \rangle = \prod_{a=1,\ldots,s} \frac{l_a^2 - \mu^2}{i} \prod_{b=1,\ldots,r} \frac{k_b^2 - \mu^2}{i} \widetilde{G}^{(r+s)}(-l_1, \ldots, -l_s, k_1, \ldots, k_r) \]

The real world does not have a turning on and off function. Is this formula right? \\

The answer is ``almost."\\

We will show how to obtain $S-1$ matrix elements from Green's function without resorting to perturbation theory. We will make no reference to free Hamiltonia, bare vacua, interaction picture fields, etc. Accordingly, take 
\[ \phi(x) \equiv \phi_H(x) \]
\[ | 0 \rangle \equiv | 0 \rangle_p \]

\noindent
$|0 \rangle$ is the ground state of the full Hamiltonian which as usual we assume to be translationally invariant and not part of a continuum, i.e.~normalizable. 
\[ P^\mu | 0 \rangle = 0 \]
$$ \langle 0 | 0 \rangle = 1$$

We will assume there are physical one meson states, $|k \rangle$, relativistically normalized,
\[ H | k \rangle = \sqrt{\vec{k}^2 + \mu^2 } |k \rangle \]
\[ \vec{P} | k \rangle = \vec{k} | k \rangle \]
\[ \langle k' | k \rangle = (2 \pi)^3 2 \omega_{\vec{k}} \delta^{(3)}(\vec{k} - \vec{k}\,') \]

The reason that the answer to question 2 is ``almost" is because the field, $\phi$, which enters the formula for $S-1$ matrix elements through $G^{(n)}$ ($G^{(n)}(x_1,\ldots,x_n) = \langle 0 | T(\phi(x_1) \cdots \phi(x_n)) |0 \rangle$), does not have quite the right properties. First, it may have a VEV, and second, in general it is not normalized so as to create a one particle state from the vacuum with a standard amplitude. It is normalized to obey the canonical commutation relations. We correct for these things by defining a renormalized field, $\phi'$ in terms of $\phi$.\\

More precisely, $\langle 0 | \phi(x) | 0 \rangle $ may not be zero. However this VEV is independent of $x$ by translational invariance. 
\[ \langle 0 | \phi(x) | 0 \rangle = \langle 0 | e^{iP\cdot x} \phi(0) e^{-iP\cdot x} |0 \rangle = \langle 0 | \phi(0) | 0 \rangle \]

\noindent
We also have by translational invariance
\[ \langle k | \phi(x) |0 \rangle = \langle k | e^{iP\cdot x} \phi(0) e^{-iP\cdot x} |0 \rangle = e^{ik\cdot x} \langle k | \phi(0) |0 \rangle \]

By Lorentz invariance, you can easily see that $\langle k | \phi(0) |0 \rangle$ is independent of $k$. It is some number, $Z_3^{\frac{1}{2}}$ (traditionally called the ``wave function renormalization"), in general $\neq 1$, 
\[ Z_3^{\frac{1}{2}} \equiv \langle k | \phi(0) |0 \rangle \]

\noindent
which we hope is nonzero. \\

We define a new field $\phi'$ which has zero VEV and is normalized to have a standard amplitude to create one meson 
\[ \phi'(x) = Z_3^{-\frac{1}{2}} (\phi(x) - \langle 0 | \phi(0) | 0 \rangle) \]
\[ \langle 0 | \phi'(x) | 0 \rangle = 0 \]
\[ \langle k | \phi'(x) | 0 \rangle = e^{ik\cdot x} \]

\section*{LSZ formula stated}
Define the renormalized Green's functions, $G'^{(n)}$, 
\[ G'^{(n)} (x_1, \ldots, x_n) \equiv \langle 0 | T (\phi'(x_1) \cdots \phi'(x_n)) |0 \rangle \]

\noindent
and $\widetilde{G}\,'^{(n)}$, their Fourier transforms, then $S-1$ matrix elements are given by 
\[ \langle l_1, \ldots, l_s | (S-1) | k_1, \ldots, k_r \rangle = \prod_{a=1,\ldots,s} \frac{l_a^2 - \mu^2}{i} \prod_{b=1,\ldots,r} \frac{k_b^2 - \mu^2}{i} \widetilde{G}\,'^{(r+s)}(-l_1, \ldots, -l_s, k_1, \ldots, k_r) \]

This is the Lehmann-Symanzik-Zimmermann reduction formula.\\

The \underline{only} assumptions needed about the local scalar field $\phi'$ is that it satisfy 
\[ \langle 0 | \phi'(x) | 0\rangle = 0 \quad \text{and} \quad \langle k | \phi'(x) |0 \rangle = e^{ik\cdot x} \]

In particular, its relationship to $\phi(x)$, the field that appears in the Lagrangian with a standard kinetic term, is not used. Any field that satisfies these properties, whose Green's functions you have, gives you the $S$ matrix.\\

The proof of the LSZ reduction formula is as follows. First we'll find a way of making one meson wave packets. The method will be inspired by the way a limiting process gave us the physical vacuum when we started with the bare vacuum. Once we know how to make one meson states we'll wave our arms some and describe how to get many meson in and out states. Then we'll be set to get $S$ matrix elements in terms of the Green's function of the renormalized fields that were used to create the in and out states.

\subsection*{LSZ reduction formula proof}

A notation for normalizable wave packet states
\[ |f \rangle \equiv \int \frac{d^3 k}{(2\pi)^3 2 \omega_{\vec{k}}} F(\vec{k}) |k \rangle \]

\noindent
(You can recover $F(\vec{k})$ from $|f \rangle$: $F(\vec{k}) = \langle k | f \rangle $.)\\

\noindent
Associated with each of these wave packets, we have a negative frequency solution of the K.-G.~equation.
\[ f(x) \equiv \int \frac{d^3k}{(2\pi)^3 2\omega_{\vec{k}}} F(\vec{k}) e^{-ik\cdot x} \]
\[ (\Box + \mu^2) f(x) = 0 \]

\noindent
Note that as $|f\rangle \rightarrow | k \rangle $ (i.e.~$F(\vec{k'}) \rightarrow (2 \pi)^3 2 \omega_{\vec{k}'} \delta^{(3)} (\vec{k} - \vec{k'})$)
\[ f(x) \rightarrow e^{-ik\cdot x} \]

Define an operator which is a function of time only out of any operator which is a function of $\vec{x}$ and $t$ (here taken to be $\phi'$):
\[ \phi'^f(t) = i \int d^3x (\phi' \partial_0 f - f \partial_0 \phi') \]

\noindent
Those of you who are familiar with the initial value theory of the K.-G.~equation will not find this a strange combination. It will turn out to create a particle in state $|f \rangle$, in the limit $t \rightarrow \pm \infty $.

From the properties of $\phi'(x)$ and $f(x)$
\[ \langle 0 | \phi'^f (t) | 0 \rangle = 0 \quad \quad \text{and} \]
\begin{align*} 
\langle k | \phi'^f(t) | 0 \rangle &= i \int d^3 x \int \frac{d^3 k'}{(2 \pi )^3 2 \omega_{\vec{k'}}} F(\vec{k'}) \langle k | \Big[ \phi' \partial_0 e^{-ik'\cdot x} - e^{-ik'\cdot x} \partial_0 \phi'(x,t) \Big] | 0 \rangle \\
&= i \int d^3 x \int \frac{d^3 k'}{(2 \pi )^3 2 \omega_{\vec{k'}}} F(\vec{k'}) \Big[-i \omega_{\vec{k'}} e^{-ik'\cdot x} - e^{-ik'\cdot x} \partial_0 \Big] \underbrace{\langle k | \phi'(x,t) | 0 \rangle}_{e^{ik\cdot x}} \\
&= i \int d^3 x \int \frac{d^3 k'}{(2 \pi )^3 2 \omega_{\vec{k'}}} F(\vec{k'}) ( -i \omega_{\vec{k'}} -i \omega_{\vec{k}} ) e^{-ik'\cdot x + ik\cdot x} \\
&= F(\vec{k}) \quad \quad \text{independent of time}
\end{align*}

A similar derivation except for one crucial minus sign shows 
\[ \langle 0 | \phi'^f(t) |k\rangle =0 \]

In these few matrix elements, $\phi'^f(t)$ is acting like a creation operator for a physical meson wave packet. We will now take the limit $t\rightarrow \pm \infty$, and in this limit we'll see that many more matrix elements of $\phi'^{f}(t)$ look like the matrix elements of a creation operator.\\

Consider any state with two or more particles satisfying $P^\mu |n\rangle = P^{\mu}_n |n\rangle $. 
\begin{align*} 
\langle n | \phi'^f(t) | 0 \rangle &= \langle n | i \int d^3 x \Big[ \phi'(\vec{x},t) \partial_0 f - f \partial_0 \phi'(\vec{x},t) \Big] | 0 \rangle \\
&= i \int d^3 x \Big[ \partial_0 f - f \partial_0 \Big] \underbrace{\langle n | \phi'(\vec{x},t) | 0 \rangle}_{e^{iP_n\cdot x} \langle n | \phi'(0) | 0 \rangle} \\
&= i \int d^3 x \Big[ \partial_0 f - f i P_n^0 \Big] e^{iP_n\cdot x} \langle n | \phi'(0) | 0 \rangle \\
&= i \int d^3 x \Big[ \partial_0 - i P_n^0 \Big] \int \frac{d^3k}{(2\pi)^3 2\omega_{\vec{k}}} F(\vec{k}) e^{-ik\cdot x} e^{iP_n\cdot x} \langle n | \phi'(0) | 0 \rangle \\
&= i \int \frac{d^3k}{(2\pi)^3 2\omega_{\vec{k}}} F(\vec{k}) (-i\omega_{\vec{k}} - i P^0_n) \int \!\!\!\!\!\!\!\!\!\! \underbrace{d^3 x e^{-ik\cdot x} e^{iP_n\cdot x}}_{(2\pi)^3 \delta^{(3)} (\vec{k} - \vec{P}_n) \cdot e^{-i \omega_{\vec{k}}t + i P^0_n t}} \!\!\!\! \langle n | \phi'(0) | 0 \rangle \\
&= \frac{\omega_{\vec{P}_n} + P_n^0}{2 \omega_{\vec{P}_n}} F(\vec{P}_n) \langle n | \phi'(0) | 0 \rangle e^{-i(\omega_{\vec{P}_n} - P_n^0)t}
\end{align*}

The important thing to notice is that this matrix element contains $e^{-i(\omega_{\vec{P}_n} - P_n^0)t}$ and that $\omega_{\vec{P}_n} < P_n^0$ for any multiparticle state. A multiparticle state with momentum $\vec{P}_n$ always has more energy than a single particle state with momentum $\vec{P}_n$. A two particle state with $\vec{P} = 0$ can have any energy from $2\mu$ to $\infty$. The one particle state with $\vec{P} = 0$ has energy $\omega_{\vec{P}} = \mu$. \\

Now consider $\langle \psi | \phi'^{f}(t) | 0 \rangle $ in the limit $t \rightarrow \pm \infty$ where $|\psi \rangle$ is a definite (not varying with $t$) normalizable state. Insert a complete set.
\begin{align*} 
\lim_{t \rightarrow \pm \infty} \langle \psi | \phi'^f(t) |0 \rangle &= \lim_{t \rightarrow \pm \infty} \langle \psi | \left( |0 \rangle \langle 0 | +\int \frac{d^3 k}{(2\pi)^3 2 \omega_{\vec{k}}} |k \rangle \langle k | + \fourteensuminta_{\substack{\text{multiparticle}\\\text{states} |n\rangle }} |n\rangle \langle n | \right) \phi'^f(t) |0 \rangle \\
&= 0 + \int \frac{d^3 k}{(2\pi)^3 2 \omega_{\vec{k}}} \langle \psi | k \rangle F(\vec{k}) \\ 
& \;\;\;\; + \lim_{t \rightarrow \pm \infty} \fourteensumintb_{|n \rangle} \langle \psi | n \rangle \frac{F(\vec{P}_n) (\omega_{\vec{P}_n} + P_n^0)} {2 \omega_{\vec{P}_n}} \langle n | \phi'(0) |0\rangle e^{-i(\omega_{\vec{P}_n} - P_n^0)t} \\
&= \langle \psi | f \rangle + 0
\end{align*}

The integrals over the various \underline{continua} of multiparticle states have integrands which oscillate more and more wildly as $t\rightarrow \pm
\infty$. They integrate to zero in the limit by the Riemann-Lebesgue lemma.\\

The phases were arranged to cancel in the one particle state matrix elements only. \\

The analogous derivation showing $\lim_{t \rightarrow \pm \infty} \langle 0 | \phi'^{f} (t) | \psi \rangle = 0$ goes through because the
phases add (and thus obviously never cancel) for every momentum eigenstate, one or multiparticle in the inserted complete set. \\

Physically what we have shown is this: We have created a state which in part looks like a one meson wave packet, plus a little multiparticle garbage. If we look at this state in some definite region in space time (take its inner product with some definite state $|\psi \rangle$, and send the time of reaction to $-\infty$), all the multiparticle states will run away. Of course the one particle state may run away too. We prevent this by modifying the state we create in such a way that the one meson packet always has the same relationship to the observer that's the funny combination $\phi'^f(t)$. No multiparticle state has the right dispersion relation to keep the same relationship to the observer under this modification.
 }{
 \sektion{15}{November 13}
\descriptionfifteen
What we have done so far has been rigorous, at least it can be made rigorous, with only a little effort. For the creation of multiparticle in and out states we are going to have to do some vigorous handwaving.\\

Consider two normalizable wave packets described by $F_1(\vec{k})$ and $F_2(\vec{k})$ which have no common support in momentum space. This excludes scattering at threshold. \\

What is $\displaystyle \lim_{t \rightarrow +(-) \infty} \langle \psi | \phi'^{f_2} (t) | f_1 \rangle$ ?\\ 

The handwaving requires we picture this in position space. Because $F_1$ and $F_2$ describe packets headed in different directions, if you wait long enough (go back far enough) the two wave packets will be widely separated in position. Then the application of $\phi'^{f_2}(t)$ to $|f_1 \rangle$ is, for all purposes to an observer near the $f_2$ packet, like an application of $\phi'^{f_2}(t)$ to the vacuum.\\

You may be bothered that a position space picture doesn't really exist, there is no $\vec{x}$ operator. However there is still some concept of localization up to a few Compton wavelengths. If a teensy exponential tail with $\frac{1}{e}$ distance $\frac{1}{m}$ is too big for you for some purpose, wait another thousand years.\\

This physical argument says
\[ \lim_{t \rightarrow +(-) \infty} \langle \psi | \phi'^{f_2} (t) |f_1 \rangle = \langle \psi | f_1, f_2 \rangle^{\text{out (in)}} \]

By definition, the $S$ matrix is what tells you the probability amplitude that a state looks like a given state in the far past will look like another given state in the far future.
\begin{align*}
\langle f_3, f_4 | S | f_1, f_2 \rangle &\equiv \ ^\text{out} \langle f_3, f_4 | f_1, f_2 \rangle^{\text{in}} \\
&= \lim_{t_4 \rightarrow \infty} \lim_{t_3 \rightarrow \infty} \lim_{t_2 \rightarrow - \infty} \lim_{t_1 \rightarrow - \infty} \langle 0 | \phi'^{f_3 \dagger}(t_3) \phi'^{f_4 \dagger}(t_4) \phi'^{f_2}(t_2) \phi'^{f_1}(t_1) | 0 \rangle 
\end{align*}
 
Note that in the limits, this is time ordered and thus we have succeeded in writing $S$ matrix elements in terms of the renormalized Green's functions. So we could quit now, but we are going to massage this expression. In doing so, we'll extend the idea of an $S$ matrix element. Physically there is no way to create plane wave states. Thus there is no way to measure or define $S$ matrix elements of plane wave states. However, after we get done massaging the RHS of the above equation, we will get an expression that you can put plane wave states into without getting nonsense. We'll make this the definition of $S$
matrix elements of plane wave states, $\langle k_3, k_4 | (S-1) |k_1, k_2 \rangle$. The utility of this object is that you can integrate it, smear it a little, to recover physically measurable $S$ matrix elements. Now I'll tell you the answer, that is, what we will soon show is a sensible definition for 
\begin{multline*} 
\langle k_3, k_4 | (S -1 )| k_1, k_2 \rangle = \int d^4 x_1 \cdots d^4 x_4 e^{ik_3 \cdot x_3 + ik_4 \cdot x_4 - ik_1 \cdot x_1 - ik_2 \cdot x_2}\times \\
(i)^4 \prod_r (\Box_r + \mu^2) \langle 0 | T (\phi'(x_1) \cdots \phi'(x_4) ) |0 \rangle 
\end{multline*}

That looks unfamiliar and messy, but it actually isn't. Recall that 
\begin{align*} 
\langle 0 | T (\phi'(x_1) \cdots \phi'(x_4)) |0 \rangle &\equiv G'^{(4)}(x_1, \ldots, x_4) \\
&= \int \frac{d^4 l_1 }{(2\pi)^4} \cdots \frac{d^4 l_4 }{(2\pi)^4} e^{i l_1 \cdot x_1 + \cdots + i l_4 \cdot x_4} \widetilde{G}'^{(4)}(l_1, \ldots, l_4) 
\end{align*}

If you substitute this in the expression for $\langle k_3, k_4 | (S -1 )| k_1, k_2 \rangle$, it collapses to 
\[ \langle k_3,k_4 | (S -1 )| k_1,k_2 \rangle = \prod_r \frac{k_r^2 - \mu^2}{i} \widetilde{G}'^{(4)}(k_1,k_2,-k_3,-k_4) \]

This says that an $S-1$ matrix element is equal to a Green's function with the external propagators removed. This is almost exactly the result that came out of our low budget scattering theory, with the only difference being that the Green's function is of renormalized fields. The result we will first obtain won't be an expression for $\langle k_3, k_4 | (S -1 )| k_1, k_2 \rangle $. We get that by abstracting the expression for $\langle f_3, f_4 | (S -1 )| f_1, f_2 \rangle$ which looks just like the expression for $\langle k_3, k_4| (S -1 )| k_1, k_2 \rangle $ stated above except $e^{-ik_1\cdot x_1}$
is replaced by $f_1(x_1)$, $e^{-ik_2\cdot x_2}$ by $f_2(x_2)$, $e^{ik_3\cdot x_3}$ by $f^*_3(x_3)$ and $e^{ik_4\cdot x_4}$ by $f^*_4(x_4)$. That is, what we will show is 
\begin{multline*}
\langle f_3, f_4 | (S -1 )| f_1, f_2 \rangle = \int d^4 x_1 \cdots d^4 x_4 f_3^*(x_3) f_4^*(x_4) f_1(x_1) f_2(x_2)\times \\
(i)^4 \prod_r (\Box_r + \mu^2) \langle 0 | T (\phi'(x_1) \cdots \phi'(x_4)) |0 \rangle 
\end{multline*}

Let's get on with the proof, beginning with a lemma. \\

Given any function, $f(x)$, satisfying $(\Box + \mu^2) f(x) =0 $, and $f \rightarrow 0$ as $|x| \rightarrow \infty $ and a general field $A$, then 
\begin{align*}
i \int d^4 x f(\Box + \mu^2) A &= i \int d^4 x \Big[ f \partial^2_0 A + A(-\nabla^2 + \mu^2) f \Big] \\
&= i \int d^4 x ( f \partial_0^2 A - A \partial_0^2 f ) \\
&= \int dt \; \partial_0 \underbrace{\int d^3 x \; i (f \partial_0 A -A \partial_0 f )}_{\substack{\text{this is something that appears}\\\text{often enough that it is worth}\\\text{giving it a name. It is a function}\\\text{of time only, call it } -A^f(t)}}\\
&= - \int dt \; \partial_0 A^f(t) \\
&= \left(\lim_{t\rightarrow - \infty} - \lim_{t\rightarrow \infty}\right) A^f(t)
\end{align*}

Also, if $A$ is hermitian,
\[ i \int d^4 x f^*(x) (\Box + \mu^2) A = \!\!\!\!\!\!\!\!\!\!\!\!\!\!\!\underbrace{\left(\lim_{t\rightarrow \infty} - \lim_{t\rightarrow -\infty}\right)}_{\substack{\text{note difference in sign from conjugating}\\\text{the } i \text{ in the def'n of } A^f(t)}} \!\!\!\!\!\!\!\!\!\!\!\!\!\!\! A^{f\dagger}(t) \]

Now we'll apply this equation to the RHS of the equation we want to prove. First do the $x_1$ integration, you get 
\begin{multline*}
\left(\lim_{t_1\rightarrow - \infty} - \lim_{t_1\rightarrow \infty}\right) \int d^4 x_2 \, d^4 x_3 \, d^4 x_4 f_3^*(x_3) f_4^*(x_4) f_2(x_2) \times\\
(i)^3 \prod_{r=2,3,4} (\Box_r + \mu^2) \langle 0 | T (\phi'^{f_1}(t_1) \phi'(x_2)\phi'(x_3) \phi'(x_4)) |0 \rangle 
\end{multline*}

\noindent
We push a time derivative through a time-ordered product in these steps. It is OK in the limit we have. Then the $x_2$ integration
\begin{multline*}
\left(\lim_{t_1\rightarrow - \infty} - \lim_{t_1\rightarrow \infty}\right)\left(\lim_{t_2\rightarrow - \infty} - \lim_{t_2\rightarrow \infty}\right) \int d^4 x_3 \, d^4 x_4 f_3^*(x_3) f_4^*(x_4) \times\\
(i)^2 (\Box_3 + \mu^2) (\Box_4 + \mu^2)\langle 0 | T (\phi'^{f_1}(t_1) \phi'^{f_2}(t_2) \phi'(x_3) \phi'(x_4)) |0 \rangle
\end{multline*}

\noindent
Etcetera.
\begin{align*}
\left(\lim_{t_1\rightarrow - \infty} - \lim_{t_1\rightarrow \infty}\right) \underbrace{\left(\lim_{t_2\rightarrow - \infty} - \lim_{t_2\rightarrow \infty}\right) \left(\lim_{t_3\rightarrow \infty} - \lim_{t_3\rightarrow -\infty}\right)}_{\text{note the difference in sign}}&
\left(\lim_{t_4\rightarrow \infty} - \lim_{t_4\rightarrow - \infty}\right) \\
& \langle 0 | T (\phi'^{f_1}(t_1) \phi'^{f_2}(t_2) \phi'^{f_3\dagger}(t_3) \phi'^{f_4\dagger}(t_4)) |0 \rangle
\end{align*}

If we had reduced the integrals in some other order we would have a different order of limits here. All $4!$ orderings lead to the same result however, and we'll just do the order we have arrived at. \\

When $t_4 \rightarrow - \infty$, it is the earliest time and thus the time ordering puts it on the right. However $\phi'^{f_4\dagger}(t_4)$ with the vacuum on the right and any other state on the left vanishes in the limit. When $t_4 \rightarrow + \infty$, it is the latest time and the time ordering puts it on the left, and we get the matrix element of $\langle f_4 |$ with the rest of the mess. We have 
\[ \left(\lim_{t_1\rightarrow - \infty} - \lim_{t_1\rightarrow \infty}\right) \underbrace{\left(\lim_{t_2\rightarrow - \infty} - \lim_{t_2\rightarrow \infty}\right) \left(\lim_{t_3\rightarrow \infty} - \lim_{t_3\rightarrow -\infty}\right)}_{\text{note the difference in sign}} \langle f_4 | T (\phi'^{f_1}(t_1) \phi'^{f_2}(t_2) \phi'^{f_3\dagger}(t_3)) |0 \rangle \]

The exact same considerations apply to the $t_3$ limits except we get the matrix elements of $^{\text{out}}\langle f_3,f_4 |$ with the remaining mess `out' because both fields are applied to the vacuum in the far future. Doing the $t_2$ limits does not result in such a simplification. We get 
\[ \left(\lim_{t_1\rightarrow - \infty} - \lim_{t_1\rightarrow \infty}\right) \Big( \ ^{\text{out}}\langle f_3,f_4 | \phi'^{f_1}(t_1) |f_2 \rangle - \lim_{t_2 \rightarrow \infty} \ ^{\text{out}}\langle f_3,f_4 | \phi'^{f_2}(t_2) \phi'^{f_1}(t_1) |0 \rangle \Big) \]

The first term was expected. The second term looks real bad. Let's compartmentalize our ignorance by just giving a name to this state we have created 
\[ \langle \psi | \equiv \lim_{t_2 \rightarrow \infty} \ ^{\text{out}}\langle f_3,f_4 | \phi'^{f_2}(t_2) \]

On to the evaluation of the $t_1$ limit. We get 
\[ ^{\text{out}} \langle f_3, f_4 | f_1, f_2 \rangle^{\text{in}} - \ ^{\text{out}} \langle f_3, f_4 | f_1, f_2 \rangle^{\text{out}} - \cancel{\langle \psi | f_1 \rangle} + \cancel{\langle \psi | f_1 \rangle} \]

The last two terms cancel, because there is no difference between a matrix element of $\lim_{t_1 \rightarrow + \infty} \phi'^{f_2}(t_2) | 0 \rangle$ and $\lim_{t_1 \rightarrow - \infty} \phi'^{f_2}(t_2) | 0 \rangle$ $\ddot{\smile}$ . The two terms remaining are exactly what we wanted to get. We have obtained 
\[ \langle f_3, f_4 | (S-1) | f_1, f_2 \rangle \]

\noindent
\textbf{REMARKS: }
\begin{enumerate}
\item The mathematical expression we started with makes sense even for $f_1$, $f_2$, $f_3$, $f_4$ plane waves. We'll make that expression the definition of an $S-1$ matrix element of plane waves. Of course you only get something physically measurable when you integrate, smear, the expression. The situation is very analogous to $V(\vec{x} - \vec{y})$ in the expression $U = \int d^3x d^3y V(\vec{x} - \vec{y}) \rho(\vec{x}) \rho(\vec{y})$. No one can build a point charge, and thus no one can make a charge distribution that directly measures $V(\vec{x} - \vec{y})$, that is, one for which the interaction energy is $V(\vec{x} - \vec{y})$. All you can do is measure $U$ for various charge distributions. Then you can abstract to the notion of $V(\vec{x} - \vec{y})$, ``the potential energy of between two point charges." You only recover something physically measurable when you integrate, smear, the expression for $S-1$ matrix elements of plane waves. The formula analogous to $U = \int d^3 x d^3 y \; V(\vec{x} - \vec{y}) \rho(\vec{x}) \rho(\vec{y}) $ is 
\begin{multline*}
\langle f_3, f_4 | (S-1) | f_1,f_2 \rangle = \int \frac{d^3 k_1}{(2\pi)^3 2 \omega_{\vec{k_1}}} \cdots \frac{d^3 k_4}{(2\pi)^3 2 \omega_{\vec{k_4}}} F^*_3(\vec{k_3}) F^*_4(\vec{k_4}) F_1(\vec{k_1}) F_2(\vec{k_2}) \times\\ 
\langle k_3,k_4 | (S-1) |k_1,k_2 \rangle 
\end{multline*}

\item The proof only required that the field you begin with have a nonzero vacuum to one particle matrix element. Then you shift that field by some constant, and multiply it by another constant to get the renormalized field, whose Green's functions are what actually entered the proof. There is thus a many to one correspondence between fields and particles. From the point of view of the reduction formula, $\widetilde{\phi} = \phi + \frac{1}{2} g \phi^2$ is just as good a field (at least except for one exceptional value of $g$ that makes the vacuum to one particle matrix element of $\widetilde{\phi}$ vanish). You do not have to begin with one of the fields that seemed to be fundamental in the Lagrangian.

\item There is no problem in principle of obtaining scattering matrix elements of composite particles and bound states. In the QCD theory of the strong interactions, the mesons are bound states of a quark and an antiquark. If $q(x)$ is a quark field, you would expect $\overline{q} q(x)$ to have a nonvanishing vacuum to one meson matrix elements. ``All" you need to calculate $2\rightarrow 2$ meson scattering then would be 
\[ G^{\prime(4)} (x_1,x_2,x_3,x_4) \equiv \langle 0 | T ( \overline{q'} q' (x_1) \overline{q'} q'(x_2) \overline{q'} q'(x_3) \overline{q'} q'(x_4) ) | 0 \rangle \]

where $\overline{q} q(x)$ is the renormalized field. Of course no one has gotten $G'^{(4)}$.

\item If we have some exact knowledge of the position space properties of a field, it may be possible to use these properties in the LSZ formula to get some exact knowledge about $S-1$ matrix elements.

\item Using methods of the same type as those used in the derivation of the LSZ formula, other formulas can be derived. For example, one can ``stop half way" in the reduction formula and obtain
\begin{multline*}
\langle k_3, k_4 | (S-1) |k_1 k_2 \rangle = \int d^4 x_3 \, d^4 x_4 e^{ik_3\cdot x_3} e^{ik_4\cdot x_4}\times \\ 
(i)^2 ( \Box_3 + \mu^2 ) ( \Box_4 + \mu^2 ) \langle 0 | T (\phi'(x_3) \phi'(x_4) )|k_1, k_2 \rangle^{\text{in}} 
\end{multline*}

This is used to derive theorems about the production of ``soft" (low energy) photons.\\

We can also use LSZ methods to derive expressions for the matrix elements of fields between in and out states. For example, I can show 
\begin{multline*}
^\text{out}\langle k_1, \ldots, k_n | A(x) |0 \rangle = \int d^4 x_1 \cdots d^4x_n e^{ik_1\cdot x_1 + \cdots + i k_n\cdot x_n}\times \\
(i)^n \prod_r (\Box_r + \mu^2) \langle 0 | T (\phi'(x_1) \cdots \phi'(x_n) A(x) |0 \rangle 
\end{multline*}

\noindent 
where $\phi'(x)$ is a correctly normalized field that can create the outgoing mesons and $A(x)$ is an arbitrary field. Of course this is really an abstraction of 
\begin{align*}
^\text{out}\langle f_1, \ldots, f_n | A(x) |0 \rangle = \int d^4 x_1 \cdots d^4x_n & f^*(x_1) \cdots f^*(x_n) \\
& (i)^n \prod_r (\Box_r + \mu^2) \langle 0 | T (\phi'(x_1) \cdots \phi'(x_n) A(x) |0 \rangle 
\end{align*}

Applying the methods used above, after the proof of the lemma, we get 
\[ \left(\lim_{t_1 \rightarrow \infty} -\lim_{t_1 \rightarrow - \infty}\right) \cdots \left(\lim_{t_n \rightarrow \infty} -\lim_{t_n \rightarrow - \infty}\right)\langle 0 | T (\phi'^{f_1\dagger}(t_1) \cdots \phi'^{f_n \dagger}(t_n) A(x) |0 \rangle \]

Just as easily as we evaluated the $t_3$ and $t_4$ limits, these limits can be evaluated to get 
\[ ^{\text{out}}\langle f_1, \ldots, f_n | A(x) |0 \rangle \]
\end{enumerate}

\textbf{A second look at model 3 and its renormalization: }
\[ \mathcal{L} = \frac{1}{2} (\partial_\mu \phi)^2 - \frac{\mu_0^2}{2} \phi^2 + \partial_\mu \psi^* \partial^\mu \psi -m_0^2 \psi^* \psi - g_0 \psi^* \psi \phi \]

The upshot of what we have done so far is that some $0$ subscripts have been added to the Lagrangian. The coefficient of $-\frac{1}{2} \phi^2$ in $\mathcal{L}$ may not be the meson mass squared, $m_0^2$ may not be the charged muon (nucleon) mass squared. Furthermore $g_0$ may not be what we want to call the coupling constant. In real electrodynamics there is a parameter $e$ defined by some experiment. It would be lucky, extremely lucky, if that were the coefficient of some term in the electrodynamics Lagrangian. In general it isn't. We'll subscript the coupling constant, compute the conventionally
defined coupling constant from it, and then invert the equation to eliminate $g_0$, which is not directly measured, from our expressions for all other quantities of interest. Also, when calculating our scattering matrix elements, we need Green's functions. What we have a perturbative expansion for if we treat $-g_0 \psi^* \psi \phi$ as our interaction Lagrangian, is the Green's functions of $\phi$. Those Green's functions aren't exactly what we are interested in. We want the Green's function of $\phi'$, the field satisfying 
\[ \langle 0 | \phi' |0 \rangle = 0 \]
\[ \langle p | \phi'(0) |0 \rangle = 1 \]
\[ \phi' = Z_3^{-\frac{1}{2}} (\phi - \langle \phi \rangle_0) \]

So along the way in calculating quantities of interest, we'll have to calculate the Green's function of $\phi'$ from the Green's functions of $\phi$. This determination of $\phi'$, $m$, $\mu$, and $g$ from $m_0$, $\mu_0$, $g_0$ and the above conditions, and then the pluggin in of the inverse of these equations into other quantities of interest sounds like a mess. It can be avoided.\\

We rewrite $\mathcal{L}$ with six new parameters, $A,\ldots, F$.
\[ \mathcal{L} = \frac{1}{2} (\partial_{\mu} \phi')^2 - \frac{\mu^2}{2} \phi^{\prime 2} + \partial_{\mu} \psi^{*\prime} \partial^\mu \psi' - m^2 \psi^{*\prime} \psi' - g \psi^{*\prime} \psi' \phi' + \mathcal{L}_{CT} \]
\[ \mathcal{L}_{CT} = A \phi' + \frac{B}{2}(\partial_\mu \phi')^2 - \frac{C}{2}\phi^{\prime 2} + D \partial_\mu \psi^{*\prime} \partial^\mu \psi' - E \psi^{*\prime}\psi' - F \psi^{*\prime} \psi' \phi' + \text{const} \]

The six new parameters are going to be determined order by order in perturbation theory by six renormalization conditions.
\begin{enumerate}
\item $\langle 0 | \phi' | 0 \rangle =0$
\item $\underbrace{\langle q |}_{\text{one meson}} \!\!\!\!\!\! \phi'(0) | 0 \rangle =1$
\item $\underbrace{\langle p |}_{\text{one anti-nucleon}} \!\!\!\!\!\!\!\!\!\!\! \psi'(0) | 0 \rangle =1$ 
\item The meson mass is $\mu$
\item The nucleon mass is $m$
\item $g$ agrees with the conventionally defined $g$.
\end{enumerate}

Six unknowns, six conditions.\\

Of course, if you actually wanted to know the relationship of $\phi'$ to $\phi$, the field whose kinetic term has coefficient 1 in the Lagrangian (and thus obeys the canonical commutation relation)\footnote{This parenthetical remark should be emphasized more. It is $\phi$ that satisfies $[\phi, \dot{\phi}] = i \delta^{(3)}$ with coefficient 1, not $\phi'$.}, and has no linear term, you can read it off. You can also read off the bare meson mass, the bare nucleon mass, $g_0$, and the relationship of $\psi'$ to $\psi$. 
\[ \frac{1}{2}(1+B)(\partial_\mu \phi')^2 + \frac{1}{2}(\mu^2 +C)\phi^{\prime 2} + A \phi' + \text{const} = \frac{1}{2} (\partial_\mu \phi)^2 - \frac{\mu_0^2}{2} \phi^2 \]
\[ Z_3 = 1+B \]
\[ \mu_0^2 = \frac{1}{Z_3}(\mu^2 +C) \text{,     etc.} \]

\noindent
(See the discussion from Eq.~(\ref{eq:10-page13}) to the end of the lecture of Oct.~23 for these same ideas expressed before we knew / worried about $Z_3$ and $\langle 0 | \phi | 0 \rangle$).\\

What you have now is a perturbation theory for the quantities you are really interested in, in terms of the conditions on $\phi'$ and $\psi'$, and experimentally input parameters. \\

The differences in the two kinds of perturbation theory is what you call the interaction Lagrangian. We'll be taking $-g \psi^{*\prime}\psi'\phi' + \mathcal{L}_{CT}$ as the interaction. This is called renormalized perturbation theory.\\

Instead of computing scattering matrix elements in term of $\mu^2$, $m^2$ and $g$ from $\mu_0$, $m_0$ and $g_0$, the wrong parameters to hold fixed, and then inverting to get $S$ matrix elements in terms of $\mu^2$, $m^2$ and $g$, we compute everything in terms of the right quantities, the experimentally input parameters, $\mu^2$, $m^2$ and $g$.\\

This procedure has a bonus. As long as you stick to observable quantities expressed in terms of physical parameters, you avoid the infinities which plague quantum field theory. \\

There are three technical obstacles we will have to overcome to implement this program.\label{15-page9}
\begin{enumerate}
\item There are derivative interactions in $\mathcal{L}_{CT}$.
\item Renormalization conditions (4), (5), (6) are not expressed in terms of Green's functions, the things we usually compute.
\item We have to make contact with the committee definition of $g$. Then we may still have to worry about defining it in terms of Green's functions [(2)].
\end{enumerate}
}{
 \sektion{16}{November 18}
\descriptionsixteen
Let's go into more detail on how $\langle 0 | \phi' | 0 \rangle = 0$ determines $A$. \\

$A$ is going to be some power series in $g$.\\
\[ A = \sum_r A_r \]
\[ A_r \propto g^r \]

Diagrammatically,
\begin{align*}
\Diagram{x\momentum[bot]{f}{k\leftarrow}}\text{ corresponds to }&i A (2\pi)^4 \delta^{(4)}(k) \\
\Diagram{\momentum[bot]{x}{\substack{\\
\\(r)}}\momentum[bot]{f}{k\leftarrow}}\text{ corresponds to }&i A_r (2\pi)^4 \delta^{(4)}(k) \\
\Diagram{x f} = \sum_r \quad \Diagram{\momentum[bot]{x}{\substack{\\ \\(r)}}\momentum[bot]{f}{}}&
\end{align*}

I'll now explain how to determine $A$ order by order in perturbation theory.\\

Suppose that we know all Feynman graphs and have determined all counterterms to order $g^n$.\\

To determine $A$ to order $g^{n+1}$, that is to get $A_{n+1}$ we apply the renormalization condition $\langle 0 | \phi' | 0 \rangle = 0$. Graphically, 
\[ \feyn{ p!{f}{}} = 0\text{ at order }g^{n+1}. \]

We demand this for all values of $g$, so the coefficient of $g^{n+1}$ in its power series must vanish.\\

We can break $\feyn{ p!{f}{}}$ at $\mathcal{O}(g^{n+1})$ into two parts
\[ \underbrace{\feyn{ p!{f}{}}}_{\text{at order } g^{n+1}} = \sum \!\!\!\!\!\!\!\!\!\! \underbrace{\feyn{ p!{f}{}}}_{\substack{\text{graphs of order } g^{n+1}\\\text{ with more than one vertex} \\\text{behind the shield}}} \!\!\!\!\!\!\! + \sum \!\!\!\!\!\!\! \underbrace{\feyn{ p!{f}{}}}_{\substack{\text{graphs of order } g^{n+1} \\\text{ with only one vertex}}} \]

The graphs with more than one vertex behind the shield have a special property. If they are going to be of order $g^{n+1}$ every vertex has to be of order $g^n$ or less. Thus these graphs only contain known stuff, by hypothesis.\\

The graph with only one vertex at order $g^{n+1}$ with one external line also have a special property. There is only one of them.
\[ \Diagram{\momentum[bot]{x}{\substack{\\ \\(n+1)}}\momentum[bot]{f}{\quad k\leftarrow}} \]
\vspace{1cm}

Setting $\quad \quad \Diagram{\momentum[bot]{x}{\substack{\\ \\(n+1)}}\momentum[bot]{f}{\quad k\leftarrow}} = - \sum \!\!\!\!\!\!\!\!\!\! \underbrace{\feyn{ p!{f}{\leftarrow k}}}_{\substack{\text{graphs of order } g^{n+1}\\\text{ with more than one vertex} \\\text{behind the shield}}}$
\vspace{1cm}

\noindent
determines $A_{n+1}$. We can cancel completely a potentially momentum dependent sum of graphs by adjusting a single number because it is always $\propto \delta^{(4)}(k)$. $\displaystyle \feyn{ p!{f}{\leftarrow k}} = 0$ for all $k$. Now there is a nice simplification in all graphs because of this cancellation. Suppose we have
\begin{center}
\includegraphics[scale=0.3]{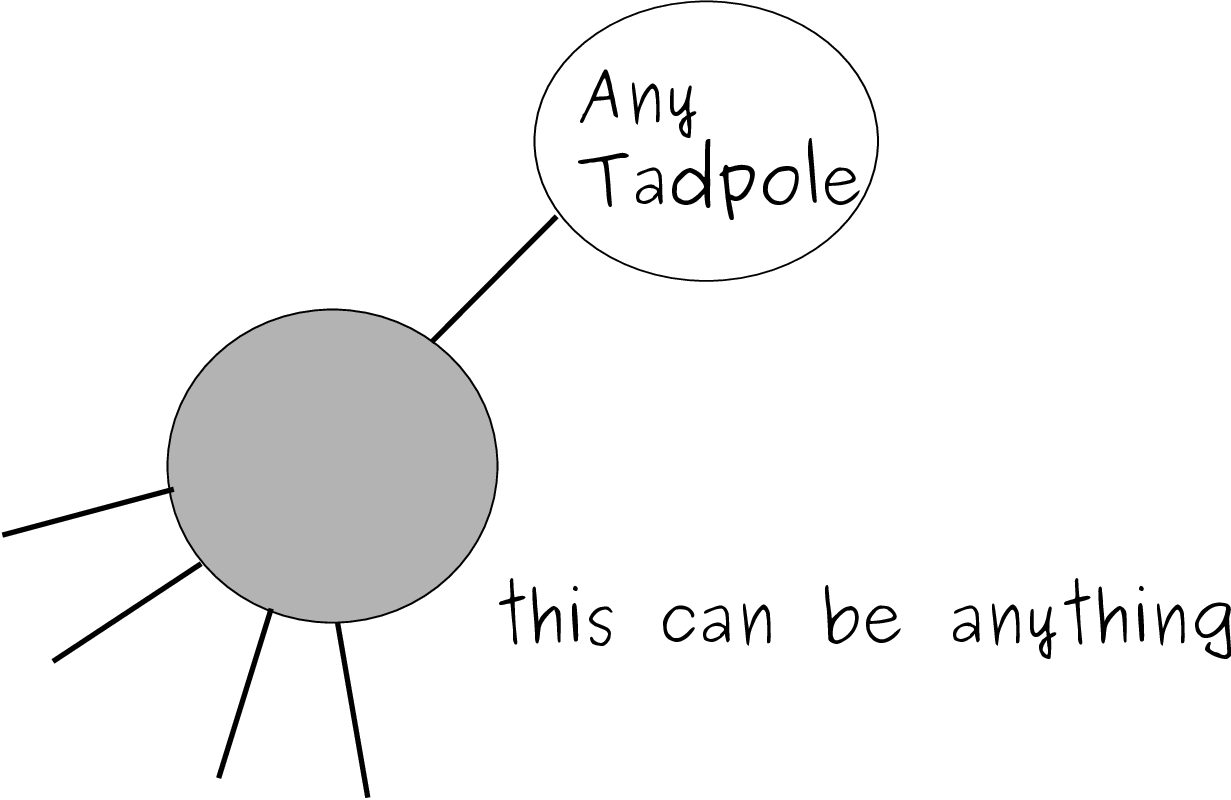}
\end{center}

A part of graph that is connected to the rest of the graph by one and only one line is called a tadpole. Physical Review's editors rejected spermion.\\

Consider \underline{the same} anything but summed over all possible tadpoles that can be attached to that same line.
\[ \begin{matrix}\displaystyle \sum_{\substack{\text{all tadpoles}\\\text{including}\\\text{counterterms}}}  \\ \\ \\ \\ \\ \\ \end{matrix}
\includegraphics[scale=0.25]{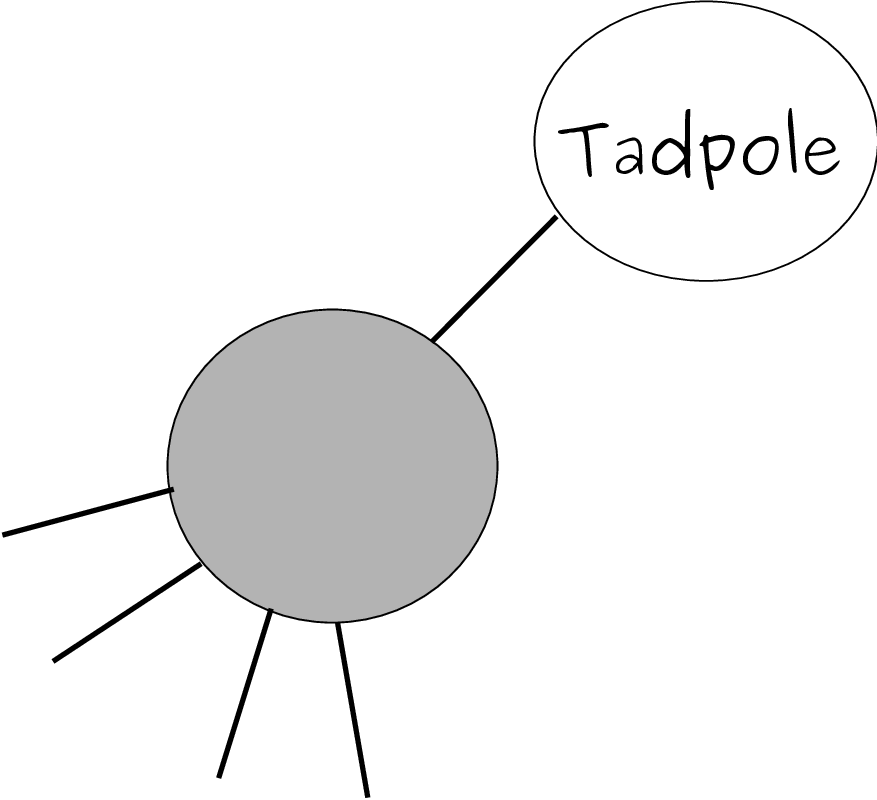} 
\!\!\!\!\!\!\!\!\!\! \begin{matrix} = \\ \\ \\ \\ \\ \\ \end{matrix} \quad
\includegraphics[scale=0.25]{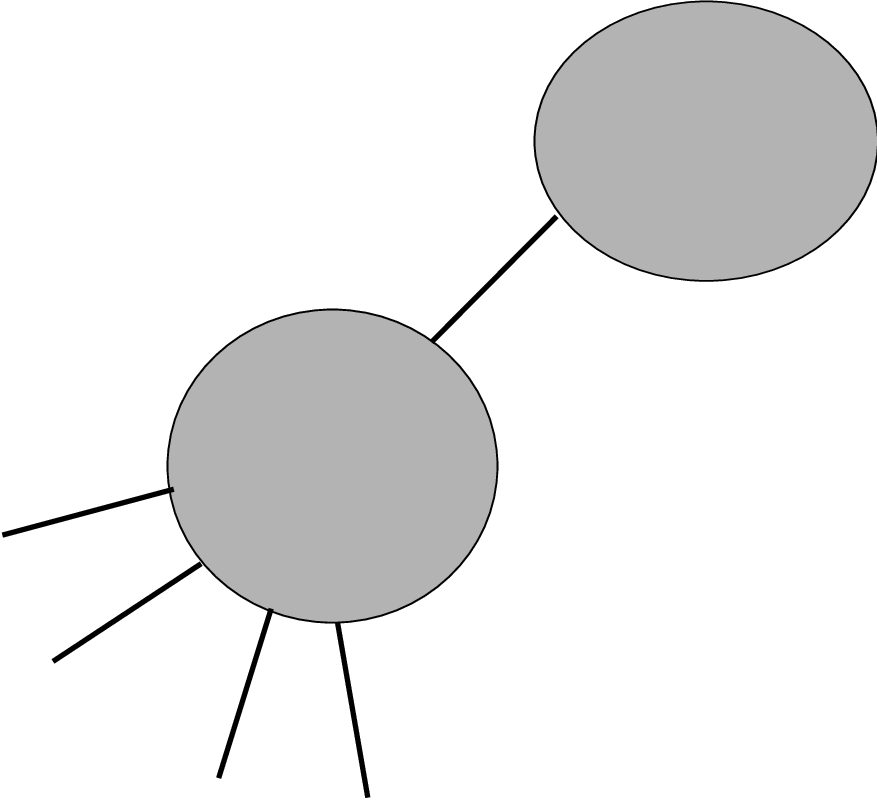} 
\begin{matrix} = 0 \\ \\ \\ \\ \\ \\ \end{matrix} \]
\vspace{-1.6cm}

The total result is that you can just \underline{ignore all tadpoles}. (Unless you cared about $A$).\\

The program for determining $B,\ldots, F$ successively will be similar, but first we have to surmount three obstacles (see page \pageref{15-page9} in the lecture of Nov.~13) before we can do anything.\\
\vspace{1cm}

\noindent \textbf{(1) Problems with derivative couplings and why they don't arise here} \\

In the presence of a derivative interaction, 
\[ \pi^\mu \neq \partial^\mu \phi,\text{ in general.} \]

This means that the interaction Hamiltonian is not just $-\mathcal{L}_I$, the interaction Lagrangian.\\

A second problem is that to get from Dyson's formula to Feynman diagrams, we had to employ the Wick expansion which turns time ordered products of free fields into normal ordered products. The Wick expansion does not apply to derivatives of fields, and we can't pull the derivatives out of the time ordered product
\[ T(\partial_\mu \phi(x) \cdots) \neq \partial_\mu T(\phi(x) \cdots) \]

\noindent
and then apply the Wick expansions. \\

Later we will develop a new method to deal with these problems. \\

For now, we'll just note that these two problems frequently cancel out, and that in a few simple examples, we can explicitly show this.\\

(Act as if $\mathcal{H}_I = - \mathcal{L}_I$ and as if $\partial_\mu \rightarrow i k_\mu$) \\

What do I mean by ``cancel out"?\\

If you are naive, and you act as if $\mathcal{H}_I = - \mathcal{L}_I$ and as if $T(\partial_\mu \phi \cdots )= \partial_\mu T (\phi \cdots)$ you get the right answer.\\
\vspace{1cm}

\noindent \textbf{A simple example}\\

Take the simplest field theory
\[ \mathcal{L} = \frac{1}{2} (\partial_\mu \phi)^2 - \frac{\mu^2}{2} \phi^2 \]

\noindent
and introduce a new field $\phi' = Z_3^{-\frac{1}{2}} \phi $, take $Z_3$ to be arbitrary. In terms of $\phi'$
\begin{align*}
\mathcal{L} &= Z_3 \left[ \frac{1}{2} (\partial_\mu \phi')^2 - \frac{\mu^2}{2} \phi^{\prime 2} \right] \\
&= \frac{1}{2} (\partial_\mu \phi')^2 - \frac{\mu^2}{2} \phi^{\prime 2} + (Z_3 - 1)\left[ \frac{1}{2} (\partial_\mu \phi')^2 - \frac{1}{2}\mu^2 \phi^{\prime 2} \right]
\end{align*}

The Green's functions of $\phi'$ are simply related to the Green's functions of $\phi$, because $\phi' = Z_3^{-1/2}\phi$.
\[ \langle 0 | T(\phi(x_1) \cdots \phi(x_n)) | 0 \rangle = Z_3^{n/2} \langle 0 | T (\phi'(x_1) \cdots \phi'(x_n) )| 0 \rangle \]

We'll show that this holds perturbatively using the naive method above. Actually, first we will only show it for one Green's function but we'll be more general in a moment.\\

Define a \underline{connected} Green's function $\widetilde{G}_c^{(n)}(k_1,\dots,k_n)$, to be the sum of all \underline{connected} graphs with $n$ external line that contribute to $\widetilde{G}^{(n)}(k_1,\ldots, k_n)$. The only nonzero connected Green's function for one scalar field with no interactions is 
\[ \widetilde{G}_c^{(2)}(k_1, k_2) = (2\pi)^4 \delta^4(k_1 + k_2) \frac{i}{k_1^2 - \mu^2 + i \epsilon} \]

\noindent
The only contribution is $\Diagram{ \momentum[bot]{f}{k_2 \rightarrow} \momentum[bot]{f}{\leftarrow k_1}}$
\vspace{1cm}

That's the right answer in this exactly soluble theory. What does naive perturbation theory give for $\widetilde{G}_c^{(2)\prime}(k_1,k_2)$, the sum of all connected graphs that contribute to $\widetilde{G}^{(2)\prime}(k_1,k_2)$? \\

The interaction is $\displaystyle (Z_3-1) \left[ \frac{1}{2}(\partial_\mu \phi')^2 - \frac{1}{2} \mu^2 \phi^{\prime 2}\right]$\\

It has as Feynman rule
\begin{multline*} 
\Diagram{ \momentum[bot]{f}{k_2 \rightarrow} x \momentum[bot]{f}{\leftarrow k_1} } \longleftrightarrow i(Z_3 -1) \frac{1}{2} 2! ((-i k_{1\mu})(-i k_2^\mu) - \mu^2 ) (2\pi)^4 \delta^{(4)} (k_1 + k_2) \\
= - i (Z_3 -1) (-k_1^2 + \mu^2) (2\pi)^4 \delta^{(4)} (k_1 + k_2)
\end{multline*}

The connected graphs contributing to $\widetilde{G}_c^{(2)\prime}(k_1, k_2)$ are 
\[ \Diagram{ \momentum[bot]{f}{k_1 \rightarrow} \momentum[bot]{f}{\leftarrow k_2} } + \Diagram{ \momentum[bot]{f}{k_1 \rightarrow} x \momentum[bot]{f}{\leftarrow k_2} } + \Diagram{\momentum[bot]{f}{k_1 \rightarrow} x \momentum[bot]{f}{k_1 \rightarrow} x \momentum[bot]{f}{\leftarrow k_2} } + \Diagram{ \momentum[bot]{f}{k_1 \rightarrow} x \momentum[bot]{f}{k_1 \rightarrow} x \momentum[bot]{f}{k_1 \rightarrow} x \momentum[bot]{f}{\leftarrow k_2} } + \cdots \text{ birds on a rail} \]
\begin{align*}
&= (2\pi)^4 \delta^{(4)}(k_1+k_2) \frac{i}{k_1^2 - \mu^2 + i \epsilon} \Bigg[ 1 + \frac{-i(Z_3 - 1)(-k_1^2 + \mu^2) i}{k_1^2 - \mu^2 + i \epsilon} \\ 
& \quad \quad \quad \quad \quad \quad \quad \quad \quad \quad \quad \quad \quad \quad \quad + \left( \frac{-i(Z_3 - 1)(-k_1^2 + \mu^2) i}{k_1^2 - \mu^2 + i \epsilon} \right)^2 + \cdots \Bigg] \\
&= (2\pi)^4 \delta^{(4)}(k_1+k_2) \frac{i}{k_1^2 - \mu^2 + i \epsilon} \left[ 1 - (Z_3 - 1 ) + (Z_3 - 1)^2 + \cdots \right] \\
&= (2\pi)^4 \delta^{(4)}(k_1+k_2) \frac{i}{k_1^2 - \mu^2 + i \epsilon} \; \frac{1}{1+(Z_3 -1)} \\
&= Z_3^{-1} (2\pi)^4 \delta^{(4)}(k_1+k_2) \frac{i}{k_1^2 - \mu^2 + i \epsilon}
\end{align*}

We have shown 
\[ \widetilde{G}_c^{(2)\prime}(k_1,k_2) = Z_3^{-1} \widetilde{G}_c^{(2)} (k_1,k_2) \]

\noindent
using a naive method, but this agrees with the right result.\\
\vspace{1cm}

\noindent \textbf{A slightly less simple example.}\\

Consider a theory of one scalar meson with arbitrary nonderivative self interactions
\[ \mathcal{L} = \frac{1}{2} (\partial_\mu \phi)^2 - \frac{1}{2} \mu^2 \phi^2 + \sum_{r=3}^N g_r \phi^r \]

Again let $\phi' = Z_3^{-1/2} \phi$
\[ \mathcal{L} = \frac{1}{2} (\partial_\mu \phi')^2 - \frac{1}{2} \mu^2 \phi^{\prime 2} + (Z_3 - 1) \left[ \frac{1}{2} (\partial_\mu \phi')^2 - \frac{1}{2} \mu^2 \phi^{\prime 2} \right] + \sum_{r=3}^N g_r \phi^{\prime r} Z_3^{r/2} \]

We are going to compute a general Green's function in this theory in terms of the Green's function of the theory without $Z_3$ by making a graphical relation. \\

Consider any graph in the Green's function of $\phi$.\\
\begin{center}
\includegraphics[scale=0.35]{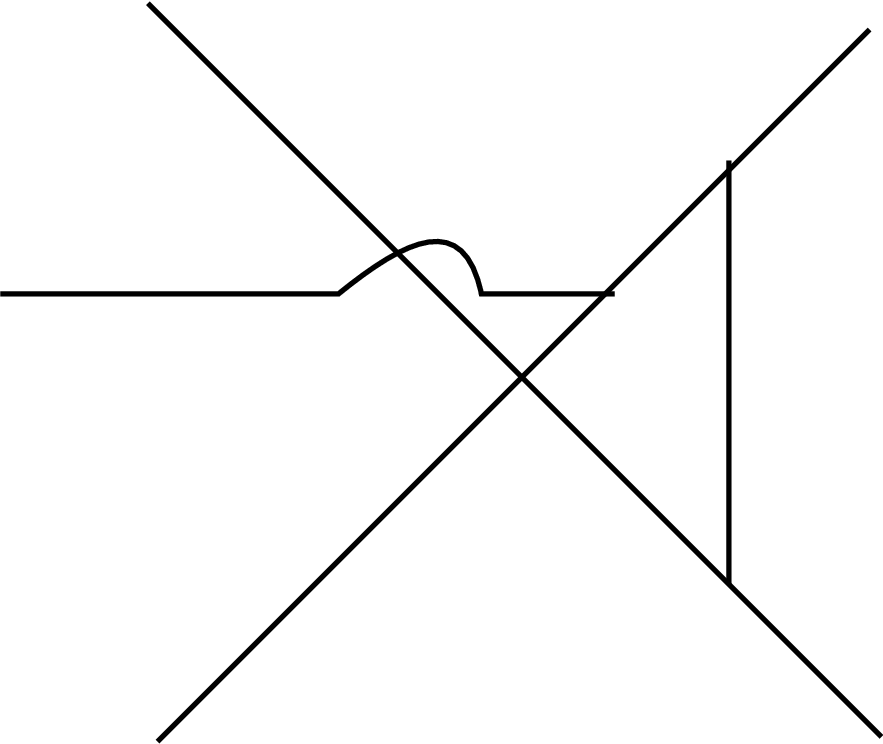}
\end{center}

Corresponding to this graph, the progenitor, there are a whole bunch of graphs in the Green's function of $\phi'$ which look just like this graph except there are an arbitrary number of birds on each line.
\[ \includegraphics[scale=0.35]{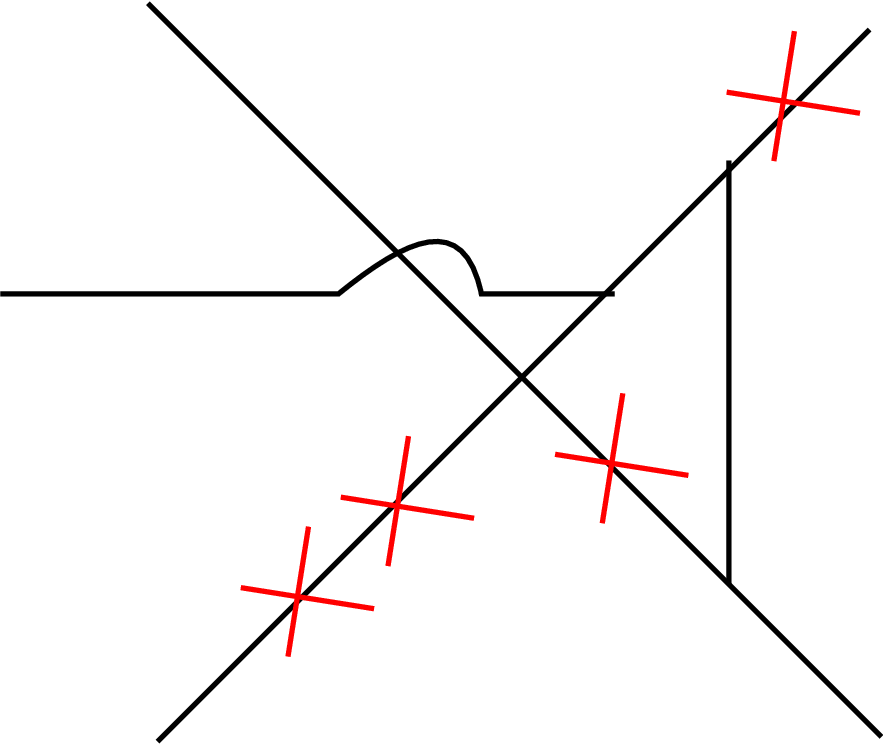} \begin{matrix}\text{ and}\\ \\ \\ \\ \\ \\ \\ \\ \\ \end{matrix} \includegraphics[scale=0.35]{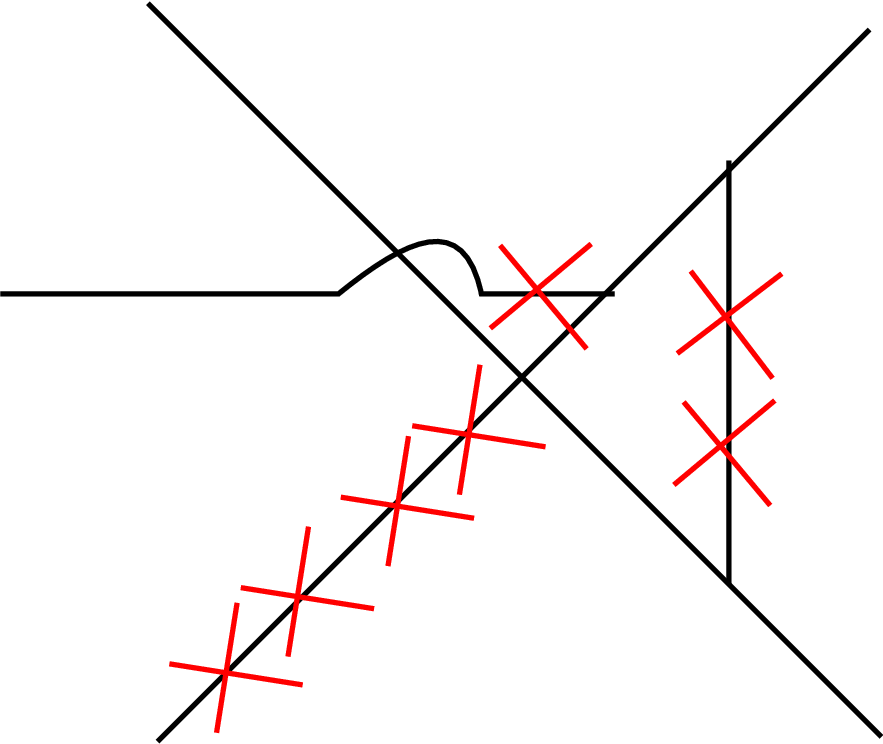} \begin{matrix}\text{ etc.}\\ \\ \\ \\ \\ \\ \\ \\ \\ \end{matrix} \]
\vspace{-2cm}

We can sum up this bunch of graphs with our naive Feynman rule. The only effect of all these birds is to replace each internal and external propagator by $Z_3^{-1}$ times the free propagator. There is also an effect on the value of the graph coming from all those $Z_3^{r/2}$ factors at the vertices. Suppose there are $n$ external lines, $I$ internal lines and $V_r$ vertices with $r$ legs. The graphs we have summed give 
\[ \underbrace{Z_3^{-n} Z_3^{-I}}_{\substack{\text{product of } n+I \\\text{ independent geometric series}}} \!\!\!\!\!\!\!\!\!\!\!\prod_r Z_3^{r V_r/2} = Z_3^{-n-I + \sum_r r V_r/2} \]

\noindent
times the progenitor's contributions to $\widetilde{G}^{(n)}$. It looks like the contributions to $\widetilde{G}^{(n)\prime}$ depend on $Z_3$ in a graph dependent way, but we aren't done yet.\\

There is a conservation law, conservation of ends.\\

Every external line ends on a vertex. Every internal line has both ends on a vertex. Every $r$ legged vertex connects to $r$ of these ends. Therefore
\[ n+2I = \sum r V_r \]
or 
\[ -n-I + \sum_r \frac{r V_r}{2} = -\frac{n}{2} \]

The graphs we have summed give $Z_3^{-n/2}$. Since all the contributions to $\widetilde{G}^{\prime (n)}$ have this factor 
\[ \widetilde{G}^{\prime (n)} = Z_3^{-n/2} \widetilde{G}^{(n)} \]

\noindent
as expected.\\

This is the right result. It justifies the naive treatment we will apply to $\mathcal{L}_{CT}$ on Model 3.

\vspace{1cm}

\noindent \textbf{Overcoming the second obstacle}, that renormalization conditions (2), (3), (4) and (5) aren't expressed in terms of Green's function (we will worry about (6) later). This will require a study of $\widetilde{G}^{(2)\prime}$.
\begin{align*}
\widetilde{G}^{(2)\prime}(k_1,k_2) &= \feyn{!{f}{k_1 \rightarrow} p!{f} {\leftarrow k_2}}\\ 
& = \int d^4 x \, d^4y e^{-ik_1\cdot x - ik_2\cdot y} \langle 0 | T (\phi'(x) \phi'(y)) |0 \rangle 
\end{align*}

Now
\[ T(\phi'(x) \phi'(y)) = \theta(x^0 - y^0) \phi'(x) \phi'(y) + \theta(y^0 - x^0) \phi'(y) \phi'(x) \]

\noindent
so it is sufficient to study $\langle 0 | \phi'(x) \phi'(y) |0 \rangle$ and then take this combination at the end. $ \langle 0 | \phi'(x) \phi'(y) |0 \rangle$ is called a Wightman function.
\[ \langle 0 | \phi'(x) \phi'(y) |0 \rangle = \sixteensumintaa_{\substack{\text{complete set of}\\\text{intermediate}\\\text{momentum}\\\text{eigenstates } |n\rangle \\ P_\mu |n\rangle = P_{n\mu} |n\rangle}} \!\!\!\!\langle 0 | \phi'(x) |n \rangle \langle n | \phi'(y) |0 \rangle \]

Now,
\[ \langle 0|\phi'(x)|n\rangle = \langle 0| e^{iP \cdot x} \phi'(0) e^{-iP\cdot x} |n\rangle = e^{-iP_n \cdot x}\langle 0 | \phi'(0) | n \rangle \] 
 
so
\begin{align}
\langle 0 | \phi'(x) \phi'(y) |0 \rangle &= \sixteensumintb_{|n\rangle} e^{-iP_n\cdot (x-y)} | \langle 0 | \phi'(0) |n \rangle|^2 \nonumber\\
\label{eq:16-page8} &= | \underbrace{\cancel{\langle 0 | \phi'(0) |0 \rangle}}_{=0}|^2 + \int \frac{d^3 p}{(2\pi)^3 2 \omega_{\vec{p}}} e^{-ip\cdot (x-y)}|
\underbrace{\langle 0 | \phi'(0) \!\!\!\!\!\!\! \overbrace{|p \rangle}^{\text{one meson}}}_{=1} \!\!\!\!\!\!\! |^2 \\
& \quad \quad \quad + \sixteensumintaaa_{\substack{\text{all other momentum}\\\text{eigenstates } |n \rangle \text{besides}\\\text{vacuum and one meson}}}\!\!\!\!\!\!\!\!\!\!\!\!\!\!\! e^{-iP_n\cdot (x-y)} |\langle 0 | \phi'(0) |n \rangle |^2 \nonumber
\end{align}

We have broken up the sum into vacuum, one meson and all other intermediate states and applied renormalization conditions (1) and (2). We have an name for 
\[ \int \frac{d^3 p}{(2\pi)^3 2 \omega_{\vec{p}}} e^{-ip\cdot(x-y)} \]

\noindent
it is $\Delta_+ (x-y, \mu^2) $.\\

$\mu^2$, the physical meson mass squared is what comes out here. It is in $\omega_{\vec{p}} $ ( $\omega_{\vec{p}} = \sqrt{\vec{p}^2 + \mu^2}$ ) and it comes from inserting physical one meson momentum eigenstates.\\

Let's massage the sum over all other momentum eigenstates
\[ \sixteensuminta_{\text{all other }|n\rangle} \!\!\!\!\!\! e^{-iP_n\cdot(x-y)} |\langle 0 | \phi'(0) |n \rangle |^2 = \sixteensuminta_{\text{all other }|n\rangle} \!\!\!\!\! e^{-iP_n\cdot(x-y)} \int d^4p \,\,\, \delta^4 (p- P_n) |\langle 0 | \phi'(0) |n \rangle |^2 \]

The integral over $p$ is just a fancy way of writing 1, but now we can do something tricky with it. Take $e^{-i P_n \cdot (x-y)}$ inside the
$p$ integration and rewrite it as $e^{-ip\cdot(x-y)}$. We have 
\[ \int d^4p \, e^{-ip\cdot(x-y)} \underbrace{\sixteensuminta_{\text{all other }|n\rangle} \delta^4 (p- P_n) |\langle 0 | \phi'(0) |n \rangle
|^2}_{\substack{\text{This is a manifestly Lorentz}\\\text{invariant function of } p \text{, that}\\\text{vanishes when } p_0 < 0 \text{. It} \\\text{is conventionally called}\\ \frac{1}{(2\pi)^3} \sigma(p^2) \theta(p^0)}} = \int \frac{d^4p}{(2\pi)^3} \, e^{-ip\cdot(x-y)} \sigma(p^2) \theta(p^0) \]

(To agree with unfortunate but longstanding conventions, we are abandoning our `every $p$ integration gets a $\frac{1}{2\pi}$, every $\delta$ function gets a $2\pi$' rule.)

The density $\sigma(p^2)$ has some definite properties. It is always $\geq 0$. In perturbation theory, it equals zero if $p^2 < \min (4m^2, 4 \mu^2)$, because there are no bound states in P.T. Outside of P.T., it still is zero for $p^2 < \text{mass}^2$ of the lightest neutral bound state, call it $\mu^2 + \epsilon$, $\epsilon > 0$. (If the lightest neutral bound state has a mass less than the meson mass, then that is what we would be calling the meson.) \\

So what we have found so far is 
\begin{align*}
\langle 0 | \phi'(x) \phi'(y) |0 \rangle &= \Delta_+ (x-y, \mu^2) + \int \frac{d^4p}{(2 \pi)^3} e^{-ip\cdot(x-y)} \sigma(p^2) \theta(p^0) \\
&= \Delta_+ (x-y, \mu^2) + \int \frac{d^4p}{(2 \pi)^3} e^{-ip\cdot(x-y)} \int_0^\infty da^2 \, \delta (a^2 -p^2) \sigma(a^2) \theta(p^0) \\
&= \Delta_+ (x-y, \mu^2) + \int_0^\infty da^2 \sigma(a^2) \Delta_+ (x-y, a^2)
\end{align*}

(Sometimes $\rho (a^2) \equiv \delta (a^2 - \mu^2) + \sigma(a^2) $ is used)
\[ \sigma(a^2) \geq 0 \]
\[ \sigma(a^2) = 0 \quad \text{for } a^2 < \mu^2 + \epsilon \]

This is the Lehmann-K\"all\'en (``Chalain") spectral decomposition. \\

We can use this to make a statement about $Z_3$ using $\phi' = Z_3^{1/2} \phi$ and the fact that $\phi$ obeys the canonical commutation relation.
\[ \langle 0 | [ \phi'(\vec{x},t), \dot{\phi}'(\vec{y},t) ] | 0\rangle = Z_3^{-1} i \delta^{(3)}(\vec{x} - \vec{y}) \quad \text{by c.c.r} \]
\[ \langle 0 | [ \phi'(\vec{x},t), \dot{\phi}'(\vec{y},t) ] | 0\rangle = i \delta^{(3)}(\vec{x} - \vec{y}) + \int_0^\infty da^2 \, \sigma(a^2) i \delta^{(3)}(\vec{x} - \vec{y}) \]

by using
\[ \frac{\partial}{\partial y^0} \Delta_+ (\vec{x} - \vec{y}) = \frac{i}{2} \delta^{(3)}(\vec{x} - \vec{y}) \]
\[ \text{hence } Z_3^{-1} = 1 + \int_0^{\infty} da^2 \, \sigma(a^2) \geq 1 \]

\noindent
$ \Rightarrow $ In general $Z_3 < 1 $. We will show that if $Z_3 = 1$ you have free field theory, later in the course. \\

\noindent
(*: $[ \phi_{\text{in}}, \dot{\phi}_{\text{in}} ] = i \delta^{(3)} (\vec{x} - \vec{y})$ because $\phi$ and $\pi$ can be changed to $\phi_{\text{in}}$, $\pi_{\text{in}}$ by a canonical transformation.)\\

We set out to study $\widetilde{G}^{(2)\prime}(k, k')$. What we have shown implies that 
\begin{align*} 
\widetilde{G}^{(2)\prime}(k, k') &= (2\pi)^4 \delta^{(4)}(k+k') \left( \frac{i}{k^2 - \mu^2 + i \epsilon} + \int^\infty_0 da^2 \sigma(a^2) \frac{i}{k^2 - a^2 + i \epsilon} \right)\\
&= (2\pi)^4 \delta^{(4)}(k+k') D' (k^2) 
\end{align*}
\begin{align}\label{eq:16-Dpage11}
D' (k^2) &= \text{``renormalized propagator"} \nonumber\\
&= \frac{i}{k^2 - \mu^2 + i\epsilon} + \int^\infty_0 da^2 \sigma(a^2) \frac{i}{k^2 - a^2 + i\epsilon}
\end{align}

\noindent
(Note that $[-iD'(p^2)]^* = -iD'(p^{2*}) $ Schwarz reflection property.)\\

This is a highly nontrivial expression. It defines a function everywhere in the complex $k^2$ plane (even though the propagator was not originally defined there). The function is analytic except at $k^2 = \mu^2$ where it has a pole with residue $i$ and along the positive real axis beginning at $k^2 = \mu^2 + \epsilon$, where it has a branch cut. The value on the positive real axis is given by the $i \epsilon$ prescription, which says you take the value just above the cut.
\begin{center}
\includegraphics[scale=0.2]{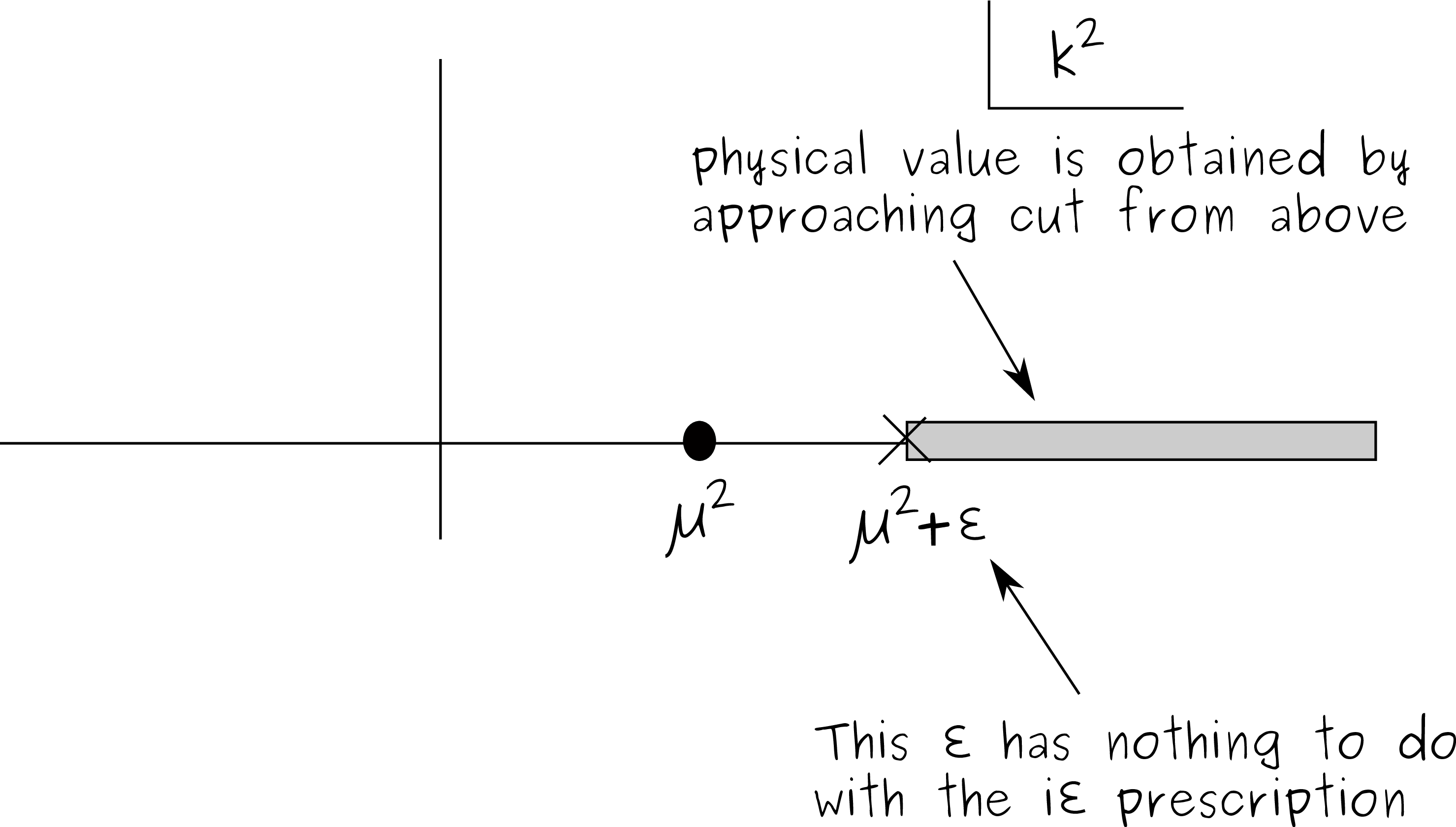}
\end{center}

Our renormalization conditions, (2) and (4) are encoded in the function.\\

\noindent
(4) The meson mass is $\mu \Leftrightarrow D'$ has a pole at $\mu^2$. \\
(2) $\langle 0 | \phi'(0) | q \rangle = 1 \Leftrightarrow$ The residue at this pole is $+i$.\\
\noindent
(look back and see where (2) was used in the derivation of the expression for $D'$) \\

We are going to keep massaging $\widetilde{G}^{(2)\prime}$ to find a slicker statement of our renormalization conditions.\\

Define another new kind of Green's function, the one particle irreducible (1PI) Green's function. Again it will be defined graphically.\\
\[ \includegraphics[scale=0.15]{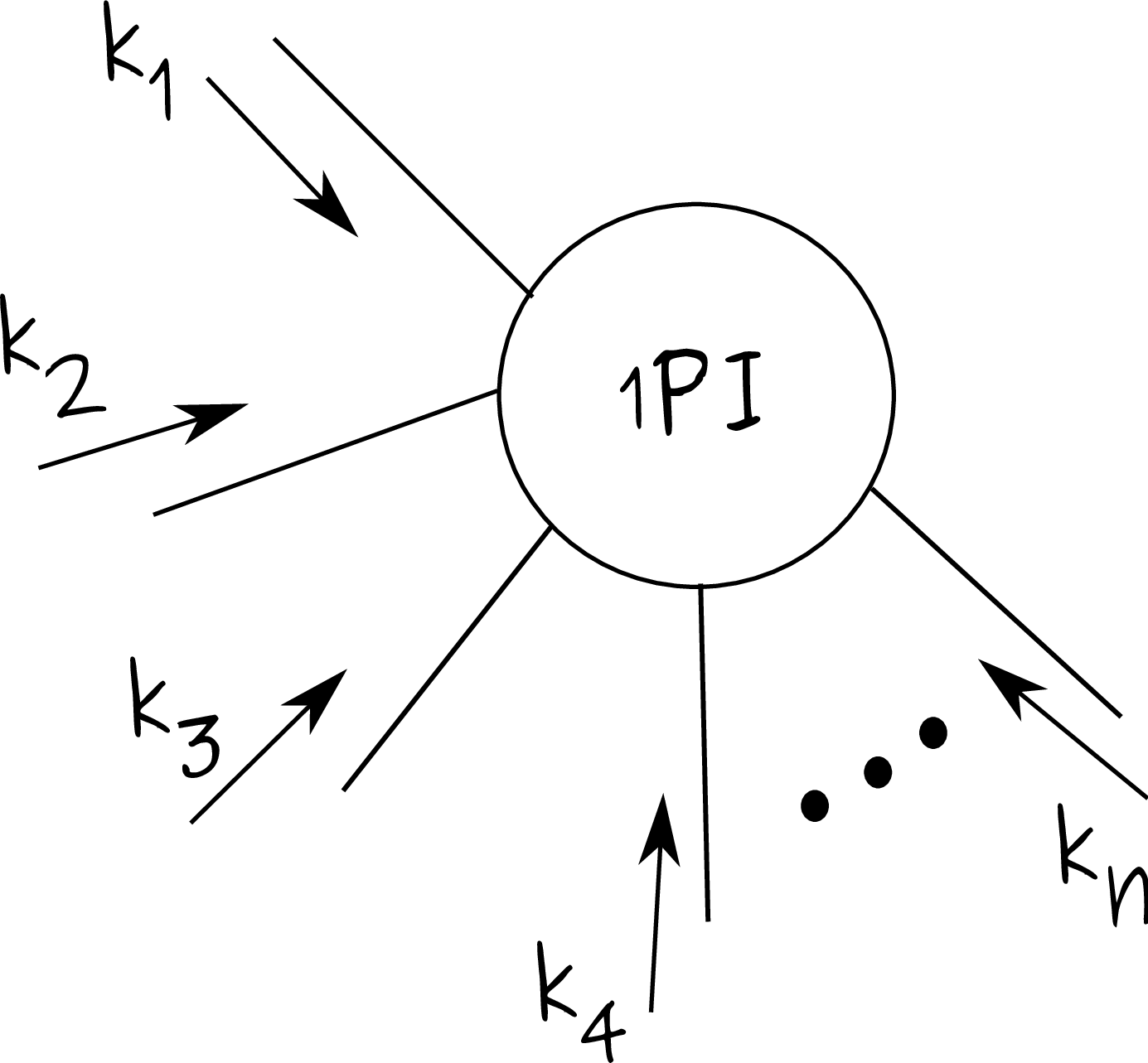}
\begin{matrix}\displaystyle\equiv \substack{\text{the sum of all connected graphs}\\\text{that cannot be disconnected by}\\\text{cutting a single internal line}}\\ \\ \\ \\ \\ \\ \\ \\ \end{matrix} \]
\vspace{-2cm}

Our convention will be that this does \underline{not} include the overall energy momentum conserving $\delta$ function or the external propagators.\\

The cute thing about this Green's function when $n=2$ is the following expression for $\widetilde{G}^{(2)\prime}$
\[ \feyn{!{f}{k \rightarrow} p!{f}{\leftarrow k'}} = \feyn{f f} + \feyn{!{f}{} !{p}{\text{1PI}}!{f}{}} + \feyn{!{f}{} !{p}{\text{1PI}}!{f}{} !{p}{\text{1PI}}!{f}{}} + \cdots \]

That is, nothing can happen, or we can have an interaction, but before we get to the other external line there is never a point where we get just one line, or there is only one line like that or ... \\

By definition, the LHS is $(2\pi)^4 \delta^{(4)}(k+k') D'(k^2)$. If we define
\[ \feyn{!{f}{k \rightarrow} !{p}{\text{1PI}}!{f}{k \rightarrow}} = - i \pi'(k^2) \]

\noindent
$\pi' (k^2) = \text{``self-energy,"}$ we can sum the series on the RHS. It is 
\begin{multline*} 
\frac{i}{k^2 -\mu^2 + i \epsilon} \left[ 1 + \frac{\pi' (k^2)}{k^2 - \mu^2 + i \epsilon} + \Big(\frac{\pi' (k^2)}{k^2 - \mu^2 + i \epsilon}\Big)^2 + \cdots \right] (2\pi)^4 \delta^{(4)}(k+k')\\
= \frac{i}{k^2 -\mu^2 + i \epsilon} \frac{1}{1 - \frac{\pi' (k^2)}{k^2 - \mu^2 + i \epsilon}} (2\pi)^4 \delta^{(4)}(k+k') = \frac{i}{k^2 -\mu^2 - \pi' (k^2) + i \epsilon}  (2\pi)^4 \delta^{(4)}(k+k')
\end{multline*}

Now you can see why $\pi'(k^2)$ is called the ``self-energy". It is like a momentum dependent mass. Identifying the coefficient of $(2\pi)^4 \delta^{(4)}(k+k')$ on the LHS and RHS, 
\[ D'(k^2) = \frac{i}{k^2 - \mu^2 - \pi'(k^2) + i\epsilon} \]
\vspace{1cm}

Now for the slick rephrasing of the renormalization conditions:\\

$D'$ has a pole at $\mu^2 \Leftrightarrow \pi'(\mu^2) = 0$

The residue of this pole is $i \Leftrightarrow \frac{d\pi'}{dk^2} \big|_{k^2 = \mu^2} = 0$ 

Perhaps this is easier to see if you think of expanding $\pi'(k^2)$ around $k^2 = \mu^2$ in a power series.
\[ \pi'(k^2) = \pi'(\mu^2) + \frac{d\pi'}{dk^2} \Big|_{\mu^2} (k^2 - \mu^2) + \cdots \]

\noindent
These two terms must vanish or it screws up the location and residue of the pole.
}{
 \sektion{17}{November 20}
\descriptionseventeen
Having succeeded in expressing renormalization conditions (2) and (4) as statements about the 1PI two-point function, I'll now explain how to determine $B$ and $C$ order by order in perturbation theory (those of you that have taken quantum field theory once or twice before probably recognize that this is going to be a rerun of the argument for determining $A$).
\[ \mathcal{L}_{CT} = \cdots + \frac{1}{2} B (\partial_\mu \phi')^2 - \frac{1}{2} C \phi^{\prime 2} +\cdots \]
\vspace{.2cm}

\[ \feyn{!f{} !{p}{1PI} !f{}} = - i\pi'(k^2) \]
\[ \pi'(\mu^2) = 0 \]
\[ \left.\frac{d\pi'}{dk^2}\right|_{\mu^2} = 0 \]

We can express the Feynman rule for the $B$ and $C$ counterterms together as 
\begin{multline*} 
\feyn{!f{ k \rightarrow} x !f{\leftarrow k'}} \text{corresponds to } i(2\pi)^4\delta^{(4)}(k+k')(-Bk\cdot k' - C) = i(2\pi)^4\delta^{(4)}(k+k')(Bk^2 - C) 
\end{multline*}

Writing $B$ and $C$ as power series expansions
\begin{align*}
B &= \sum_r B_r \quad B_r \propto g^r\\
C& = \sum_r C_r \quad C_r \propto g^r
\end{align*}

We can also write 
\[ \feyn{f x f } = \sum_r \feyn{f !x{(r)} f} \]
\[ \feyn{!f{k \rightarrow \quad} !x{(r)} !f{\quad \leftarrow k'}} \text{ corresponds to } i(2\pi)^4\delta^{(4)}(k+k')(B_r k^2 - C_r) \]

Assume everything is known to $\mathcal{O}(g^n)$, including all the counterterms, and we'll show that $B_{n+1}$ and $C_{n+1}$ can be determined.
\[ \underbrace{\feyn{!f{k \rightarrow \; \;} !p{1PI} !f{\; \; \rightarrow k}}}_{\text{at order } g^{n+1}} = \underbrace{\text{known stuff}}_{\substack{\text{sum of all 1PI graphs}\\\text{with more than one}\\\text{vertex at order } g^{n+1}}} + \underbrace{\feyn{!f{k \rightarrow \quad \quad} !x{(n+1)} !f{\quad \quad \rightarrow k}}}_{\substack{\text{the only }\mathcal{O}(g^{n+1})\\\text{1PI graph with}\\\text{only one vertex}}} \]
\[ i B_{n+1}\mu^2 - i C_{n+1} = -\left.(\text{known stuff})\right|_{\mu^2} \]
\[ i B_{n+1} = - \left.\frac{d(\text{known stuff})}{dk^2} \right|_{k^2=\mu^2} \]

Similar arguments apply to the nucleon self energy.
\vspace{0.2cm}
\[ \feyn{!{fV}{p \leftarrow \; \;} !p{1PI} !{fV}{\; \; \leftarrow p}} = -i\Sigma'(p^2) \]

\noindent 
which can be used to express renormalization conditions (3) and (5) as 
\[ \Sigma'(m^2) = 0 \]
\[ \left.\frac{d\Sigma'}{dp^2}\right|_{p^2 =m^2} = 0 \]

Of course these subtractions are not going to allow you to ignore corrections to the 1PI two point function.\\

$\feyn{!{f}{k \rightarrow \; \;} !p{1PI}!{f}{\; \; \rightarrow k}}$ has a complicated momentum dependence in general, which is not eliminated by just subtracting a constant and a term linear in $k^2$. However, this does allow you to ignore corrections \underline{to lines on the mass shell}, that is external lines in the computations of $S$ matrix elements. That is because in the computation of an $S$ matrix element, the only thing that matters about an external line is the location and residue of the pole.
\begin{multline*}
\lim_{k^2 \rightarrow \mu^2} \frac{k^2-\mu^2}{i} \times \includegraphics[scale=0.2]{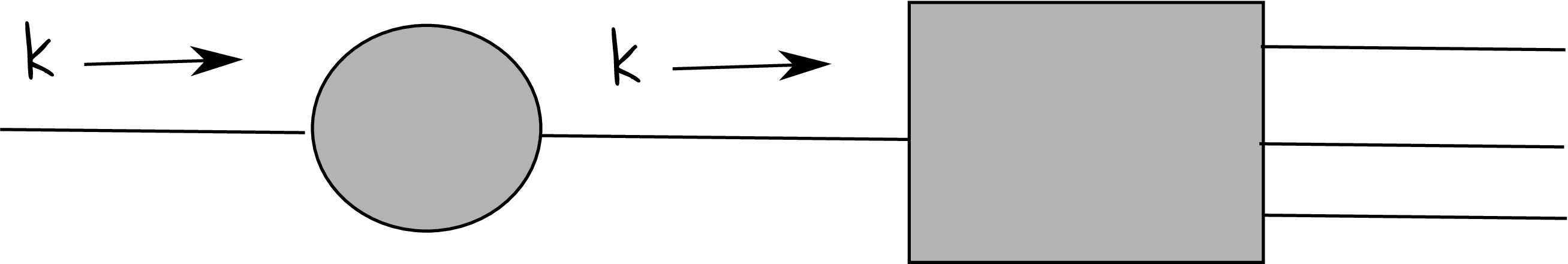} \\
= \lim_{k^2 \rightarrow \mu^2} \frac{k^2-\mu^2}{i} \times \includegraphics[scale=0.2]{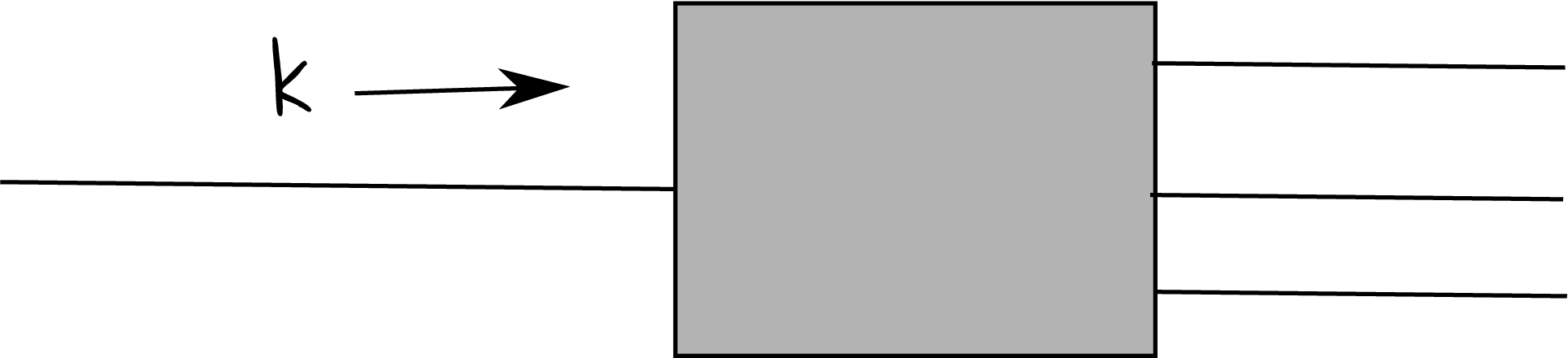} 
\end{multline*}

The location and residue of the pole in the full propagator are, in renormalized perturbation theory, the exact same as that of the free propagator.\\

We can do some examples before worrying about obstacle (3), that is renormalization condition (6).
\begin{center}
\textbf{Calculation of $\pi'(k^2)$ to order $g^2$}
\end{center}
\begin{align*}
- i \pi'(k^2) &= \feyn{!{f}{} !p{1PI} !{f}{}} \\
&\\
&= \feyn{!{f}{} fs0 !{flSV}{} !{flSuA}{} fs0 !{f}{} } + \feyn { f !x{(2)} f } \\
&\\
&= -i \pi_f(k^2) + i B_2 k^2 - iC_2
\end{align*}

\noindent 
where 
\[ - i \pi_f(k^2) \equiv \feyn{!{f}{} fs0 !{flSV}{} !{flSuA}{} fs0 !{f}{} } \]

The renormalization conditions are 
\[ \pi_f(\mu^2) - B_2 \mu^2 + C_2 = 0 \]
\[ \left.\frac{d \pi_f}{dk^2}\right|_{\mu^2} - B_2 = 0 \]

If you don't care what $B_2$ and $C_2$ are, these can be rephrased as 
\begin{equation}\label{eq:17-pipage3}
\pi'(k^2) = \pi_f(k^2) - \pi_f(\mu^2) - (k^2 - \mu^2) \left.\frac{d \pi_f}{dk^2}\right|_{\mu^2}
\end{equation}

\noindent 
we should check that $B_2$ and $C_2$ are real however.
\[ -i\pi_f(k^2) = \feyn{!{f}{\leftarrow k} fs0 !{flSV}{\leftarrow k+q} !{flSuA}{\rightarrow q} fs0 !{f}{\leftarrow k} } = (-ig)^2 \int \frac{d^4 q}{(2\pi)^4} \frac{i}{q^2 - m^2 + i \epsilon} \frac{i}{(q+k)^2 - m^2 + i \epsilon} \]
\vspace{.5cm}

There are three problems in doing this integral.
\begin{enumerate}
\item Not spherically symmetric. I suppose we could parametrize the integral with a polar angle measured from $k$ but,
\item We are in Minkowski space, and it isn't even spherical symmetry we have.
\item The integral is divergent; at high $q$ it looks like $\int \frac{d^4 q}{(2\pi)^4} \frac{1}{q^4}$ which if the integral was spherically symmetric would be $\sim \int \frac{q^3 dq}{q^4}$, and if it was cut off at some large radius in momentum space $\Lambda$, would be $\sim \text{ln } \Lambda$. (This is called log divergent.)
\end{enumerate}

This last problem is the easiest to take care of: $\pi_f(k^2) - \pi_f(\mu^2)$ is not divergent. \\

Renormalized perturbation theory, which was implemented to make expansions in the right parameters has saved us from this unexpected infinity.\\

To make this thing manifestly spherically symmetric (actually L.I.) we use Feynman's trick for combining two denominators.
\begin{equation}\label{eq:17-page5}
\int_0^1 dx \frac{1}{[ax+b(1-x)]^2} = \frac{1}{b-a} \; \; \left.\frac{1}{ax + b(1-x)} \right|^1_0 = \frac{1}{b-a} \left(\frac{1}{a} - \frac{1}{b}\right) = \frac{1}{ab}
\end{equation} 

Apply this to the two denominators in $\pi_f$, with 
\[ a = (q+k)^2 - m^2 + i \epsilon \]
\[ b = q^2 -m^2 +i\epsilon \]
\begin{align}\label{eq:17-pipage5}
- i \pi_f(k^2) &= g^2 \int \frac{d^4q}{(2\pi)^4} \int^1_0 dx \frac{1}{[((q+k)^2 - m^2 + i \epsilon)x + (q^2 - m^2 + i \epsilon) (1-x)]^2}\nonumber\\
&= g^2 \int \frac{d^4q}{(2\pi)^4} \int^1_0 dx \frac{1}{[q^2 + xk^2 + 2k\cdot qx - m^2 + i \epsilon]^2}\nonumber\\
[q' = q + kx] \quad \quad &=g^2 \int^1_0 dx \int \frac{d^4q'}{(2\pi)^4} \frac{1}{[q'^2 + k^2 x - k^2 x^2 - m^2 + i \epsilon]^2}
\end{align}
 
We could do this integral in a moment if we were living in Euclidean space. It is not spherically symmetric though.\\

So now we'll study integrals of the form
\[ I_n(a) = \int \frac{d^4 q}{(2\pi)^4} \frac{1}{(q^2 + a)^n} = \int\frac{d^3 q \, dq^0}{(2\pi)^4} \frac{1}{(q^{0\,2} - \vec{q}\,^2 + a)^n} \]

\noindent
where $a$ has a positive imaginary part. (The case of interest has $n=2$, $a=k^2x(1-x)-m^2 + i\epsilon$.)

The location of the poles in the $q^0$ integration splits into two cases\\
\begin{center}
\begin{tabular}{cc}
Case: Re$(\vec{q}\,^2 -a ) > 0$ & Case: Re$(\vec{q}\,^2 -a ) < 0$\\
\includegraphics[scale=0.2]{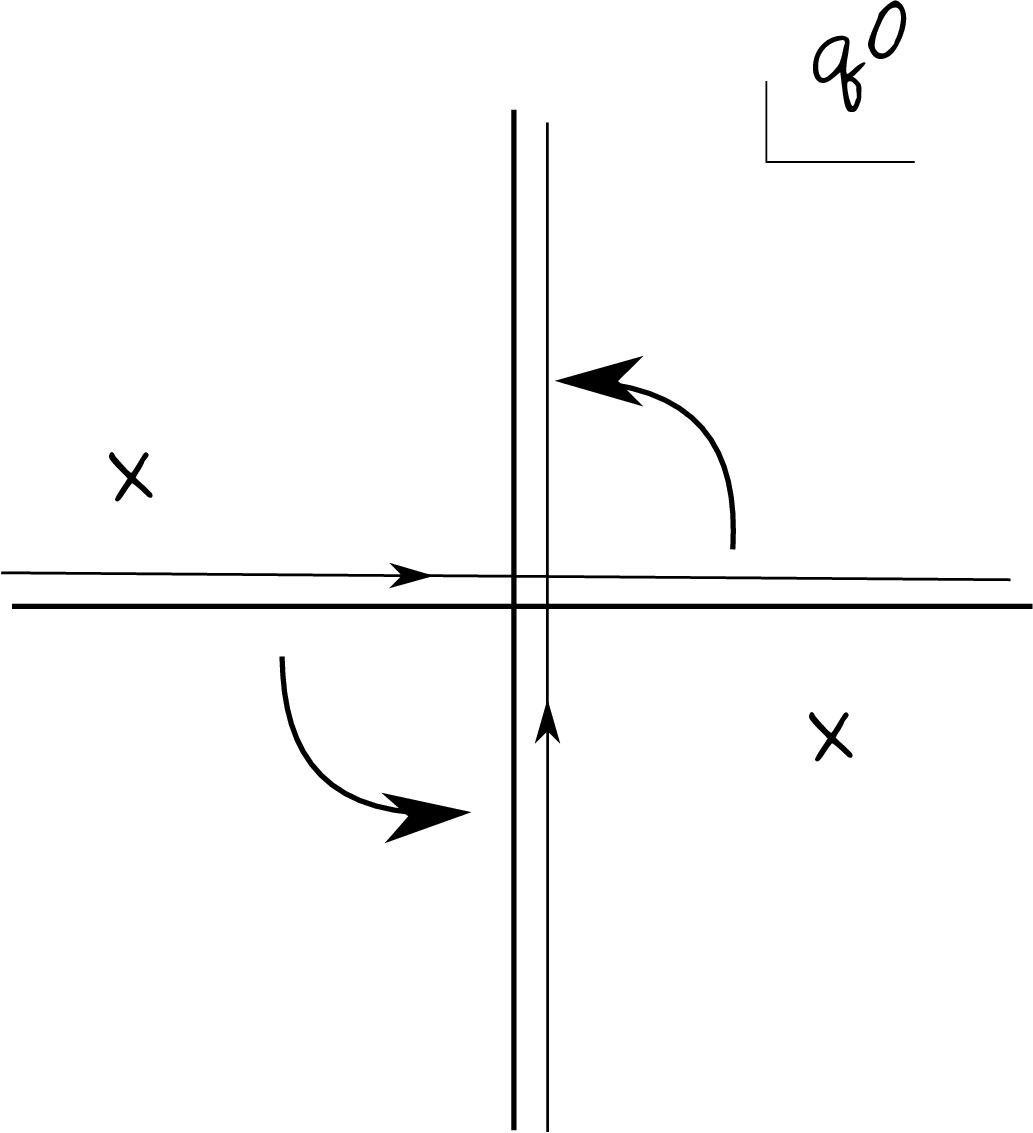} & \includegraphics[scale=0.2]{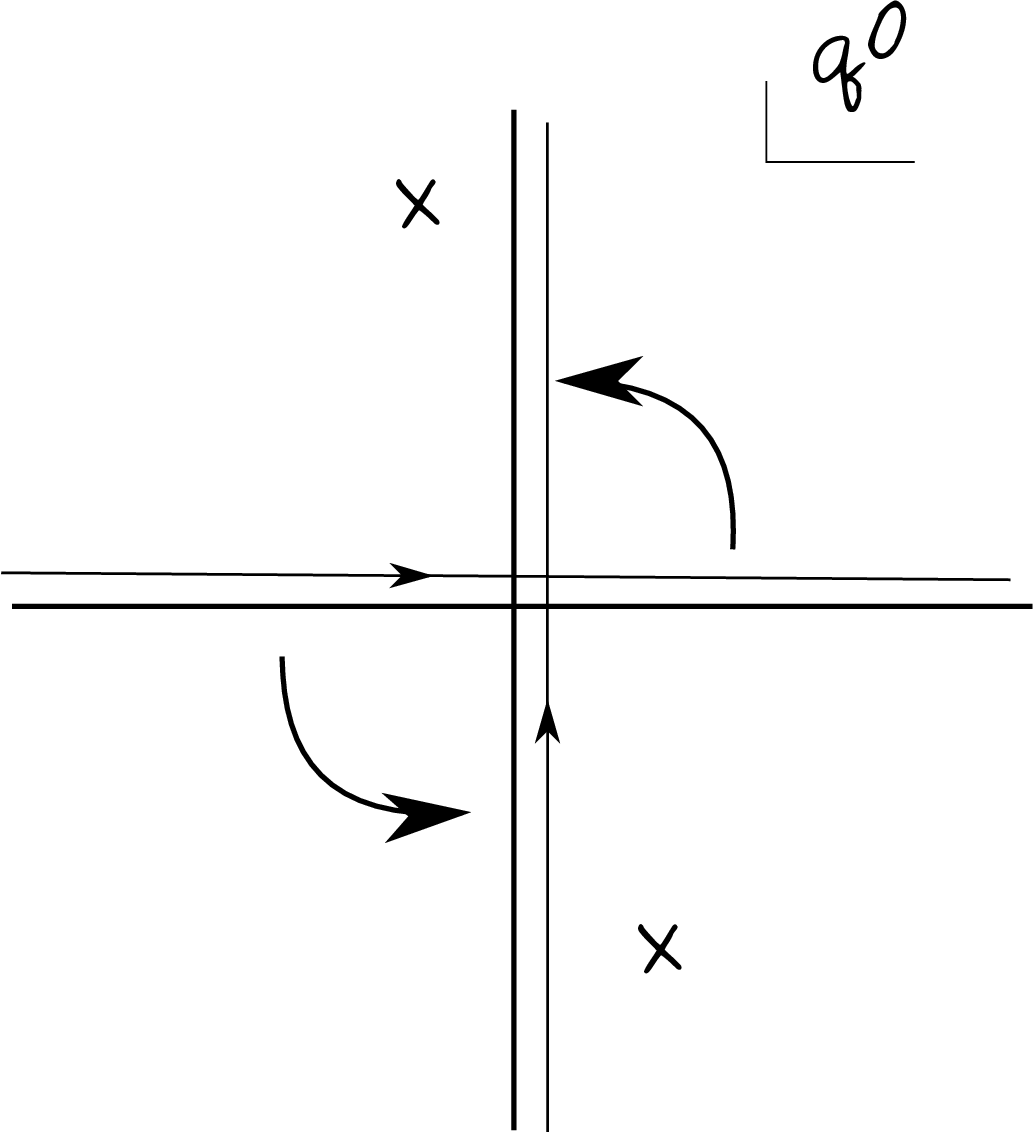}
\end{tabular}
\end{center}

In either case, the contour can be rotated as shown (called the ``Wick rotation''), so that it runs up the imaginary $q^0$ axis. Because this rotation does not cross any poles the value of the integral is unchanged. Now that $q^0$ runs from $-i\infty$ to $+i\infty$, define a new variable $q_4$ that runs from $-\infty$ to $\infty$.
\[ q_4 = - i q_0 \]
\[ dq^0 = i dq_4 \]
\[ d^4q = i d^4 q_E = idq_4 \;d^3 q \]
\[ I_n(a) = i \int \frac{d^4 q_E} {(2\pi)^4 } \frac{1}{[-q_4^2 - \vec{q}\,^2 + a]^n} = i \int \frac{d^4q_E}{(2\pi)^4}\frac{1}{(-q_E^2 + a)^n} \]

This is now a spherically symmetric integral in 4-d Euclidean space. Using $V(S^3)=2\pi^2$ and setting $z = q_E^2$, $q_E^3 dq_E = \frac{1}{2} z dz$, we have 
\begin{align*}
I_n(a) &= i \frac{\pi^2}{(2\pi)^4} \int_0^\infty z dz \frac{1}{(-z+a)^n}\\
&= \frac{i}{16\pi^2}\frac{(-1)^{n-1}}{(n-1)!} \frac{d^{n-1}}{da^{n-1}} \int^\infty_0 z dz \frac{1}{-z+a} \\
&= \frac{(-1)^{n-1}}{(n-1)!} \frac{d^{n-1} I_1(a)}{da^{n-1}}
\end{align*}

This is only a formal expression because $I_1(a) = \int_0^\infty z dz \frac{1}{-z+a}$ has a divergent part. If we cut the integral off at some large value $\Lambda^2$ (in a bit we'll send $\Lambda\rightarrow \infty$) we have
\[ I_1(a) = \int^{\Lambda^2}_0 dz \frac{z-a+a}{-z+a} = \int^{\Lambda^2}_0 dz \left(-1 + \frac{a}{-z+a}\right) = \int^{\Lambda^2}_0 dz \left( -1 + \frac{a}{-z+a} \right) \]
 
For large $z$, the integrand is $-1- \frac{a}{z} + \mathcal{O}\big(\frac{a^2}{z^2}\big)$. 

I'll evaluate $I_1(a)$ in a way which is only valid when the integral is part of a convergent combination
\[ \int^\infty_0 dz\; z \sum_n \frac{C_n}{-z+a_n} \]

\noindent
where $\sum_n C_n = 0$ and $\sum_n a_n C_n = 0$. This will guarantee that those first two terms in the integrand of order $1$ and $\frac{1}{z}$ have coefficient zero.
\begin{align*}
I_1 (a) &= \frac{-i}{16\pi^2} \lim_{\Lambda\rightarrow \infty} \int_0^{\Lambda^2} dz \left( 1 + \frac{a}{z-a} \right) \\
&= \frac{-i}{16\pi^2} \lim_{\Lambda\rightarrow \infty} \left.\left[ z + a \text{ln }(z-a) \right] \right|^{\Lambda^2}_0 \\
&= \frac{-i}{16\pi^2} \lim_{\Lambda\rightarrow \infty} \bigg[\!\!\! \underbrace{\cancel{\Lambda^2} + \cancel{a \text{ln }\Lambda^2}}_{\substack{\text{vanishes in convergent}\\\text{combinations}}}\!\!\!\!\bigg(1+\mathcal{O}\bigg(\!\!\!\!\!\!\!\!\!\!\!\underbrace{\cancel{\frac{a}{\Lambda^2}}}_{=0, \text{in } \Lambda\rightarrow \infty \text{ limit}}\!\!\!\!\!\!\!\!\!\!\!\!\bigg)\bigg) - a \text{ln }(-a) \bigg] \\
\text{(for our purposes)} \quad \quad &= \frac{i}{16 \pi^2} a \text{ln }(-a) 
\end{align*} 

What about $I_2 (a)$ ?
\begin{align*} 
-16 \pi^2 i I_2(a) &= \int^\infty_0 z dz \frac{1}{(z-a)^2} \\
&= \int_0^\infty dz \frac{z-a+a}{(z-a)^2}\\ 
&= \int^\infty_0 dz \left(\frac{1}{z-a} + \frac{a}{(z-a)^2} \right)
\end{align*}

For large $z$ the integrand is $\frac{1}{z} + \mathcal{O}(\frac{1}{z^2})$.

What follows is only valid in either of two cases.\\

I. The integral is part of a convergent combination 
\[ \int_0^\infty z dz \sum_n \frac{C_n}{(z-a_n)^2} \quad \text{where } \sum_n C_n =0 \]

II. You plan to differentiate $I_2$ with respect to $a$ to get $I_3$, $I_4$, etc.\\

\noindent
As before we make sense of $I_2(a)$ by itself by cutting the integral off at some large value $\Lambda^2$. The limit $\Lambda\rightarrow
\infty$ will be taken at the end.
\begin{align*}
I_2(a) &= \frac{i}{16\pi^2} \int^{\Lambda^2}_0 dz \left(\frac{1}{z-a} + \frac{a}{(z-a)^2} \right)\\
&= \frac{i}{16\pi^2} \left[ \text{ln }(z-a) - a\frac{1}{z-a} \right]^{\Lambda^2}_0 \\
&= \frac{i}{16\pi^2} \bigg[\underbrace{\cancel{\text{ln }(\Lambda^2)}}_{\text{vanishes}} \Big( 1 +\underbrace{\cancel{\mathcal{O}(\frac{a}{\Lambda^2})}\Big) - \cancel{\frac{a}{\Lambda^2}(1+\mathcal{O}(\frac{a}{\Lambda^2}))}}_{=0 \text{ in } \Lambda\rightarrow \infty \text{ limit}} - \text{ln }(-a) -
\!\!\!\!\! \underbrace{\cancel{1}}_{\text{vanishes}} \bigg] 
\end{align*}
 
Let's see why the terms I claim vanish, vanish in either case.\\
 
I. The condition $\sum_n C_n = 0$ which was put in to make the coefficient of $\frac{1}{z}$ vanish makes the infinite terms as $\Lambda \rightarrow \infty$ vanish. It also gets rid of the $-1$, since that is independent of $a$.\\
 
II. $\text{ln }(\Lambda^2)$ and $1$ are both constants independent of $a$. Taking a derivative w.r.t.~$a$ eliminates these terms.\\
 
So $I_2(a) = \frac{-i}{16\pi^2} \text{ln }(-a)$ for our purposes.\\
 
Note that if you take $-\frac{dI_1(a)}{da}$ to get $I_2(a)$ you get $\frac{-i}{16\pi^2}(\text{ln }(-a)+1)$, and the $1$ that vanishes in convergent combinations or when differentiated to get $I_3$, $I_4$, etc., can be chucked. \\
 
Let's get $I_3$, $I_4$, etc. For $n\geq 3$, 
\begin{align*}
I_n(a) &= \frac{(-1)^{n-1}}{(n-1)!} \frac{d^{n-1}I_1(a)}{da^{n-1}} \\
&= \frac{(-1)^{n-1}}{(n-1)!} \frac{d^{n-1}}{da^{n-1}} \left( \frac{i}{16\pi^2} a \text{ln }(-a) \right) \\ 
&= \frac{i}{16\pi^2}\frac{(-1)^{n-1}}{(n-1)!} \frac{d^{n-2}}{da^{n-2}}(\text{ln }(-a) +1) \\
&= \frac{i}{16\pi^2}\frac{(-1)^{n-1}}{(n-1)!} \frac{d^{n-3}}{da^{n-3}}\left(\frac{1}{a}\right) \\
&= \frac{i}{16\pi^2}(-1)^{n-1}(-1)^{n-3} \frac{(n-3)!}{(n-1)!}\frac{1}{a^{n-2}} \\
&= \frac{i}{16\pi^2} \frac{1}{(n-1)(n-2)a^{n-2}}
\end{align*}

These facts are summarized on the following table of integrals.\\

The Minkowski-space integral, 
\[ I_n(a) = \int \frac{d^4q}{(2\pi)^4} \frac{1}{(q^2 + a)^n}, \]
with $n$ integer and $\text{Im } a > 0 $, is given by 
\[ I_n(a) = i [ 16 \pi^2 (n-1)(n-2) a^{n-2} ]^{-1}, \]
for $n \geq 3$. For $n = 1,2$, 
\[ I_1 =\frac{i}{16 \pi^2} a\text{ln }(-a) + \cdots, \]
and
\[ I_2 =\frac{-i}{16\pi^2} \text{ln }(-a) + \cdots, \]

\noindent
where the triple dots indicate terms that cancel in a sum of such terms such that the total integrand vanishes for high $q$ more rapidly than $q^{-4}$.\\

Eq.~(\ref{eq:17-pipage5}) is an expression for $\pi_f$ to which we can apply our expression for $I_2$, with $a= k^2 x - k^2 x^2 - m^2 + i\epsilon$.
\[ \pi_f(k^2) = \frac{g^2}{16\pi^2} \int^1_0 dx \; \text{ln } (-k^2 x (1-x) + m^2 - i\epsilon ) + \text{terms that vanish in convergent combinations} \]

Eq.~(\ref{eq:17-pipage3}) is an expression for $\pi'$ in terms of $\pi_f$. 
\begin{align*} 
\pi'(k^2) &= \pi_f(k^2) - \pi_f(\mu^2) - (k^2 - \mu^2) \left.\frac{d \pi_f}{dk^2}\right|_{\mu^2} \\
&= \frac{g^2}{16\pi^2} \int^1_0 dx \left[ \text{ln } \frac{-k^2x(1-x)+m^2 - i \epsilon}{-\mu^2 x (1-x) + m^2} + \frac{(k^2-\mu^2)(+x(1-x))}{-\mu^2 x (1-x) + m^2} \right] 
\end{align*}

This thing, $\pi_f(\mu^2)$, which was subtracted off of $\pi_f(k^2)$, corresponds to the mass counterterm in $\mathcal{L}$, $-\frac{1}{2} C\phi'^2$. It is infinite, $C$ is infinite, the bare mass of the meson is infinites.\\

However, that is unimportant. The bare mass of the meson does not enter into any expression relating physical quantities. We should worry whether this infinite term we have stuck into $\mathcal{L}$ is real.\\

The only way the expression for $\pi_f(\mu^2)$ (and $\displaystyle \left.\frac{d \pi_f}{dk^2}\right|_{\mu^2}$) gets an imaginary part is when the argument of the logarithm in the integral becomes negative, which can happen for ranges of $x$ within $[0,1]$ if $\mu^2 > 4m^2$. This can be seen by graphing $x(1-x)$.
\begin{center}
\includegraphics[scale=0.3]{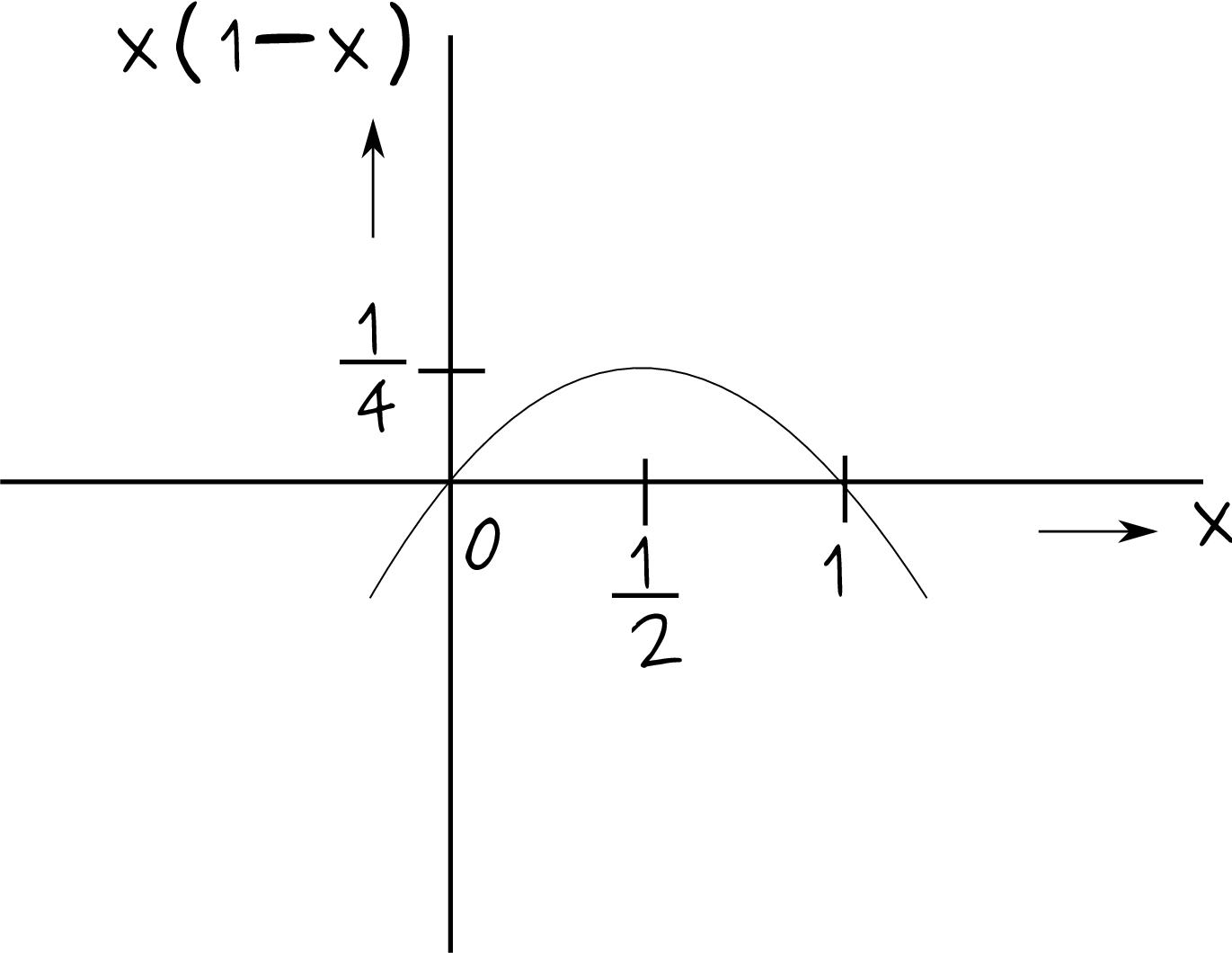}
\end{center}

Of course in this case we have no business treating the meson as a stable particle anyway.\\

Recall that (Nov.~18, after Eq.~(\ref{eq:16-Dpage11})) we have already found the analytic structure of $D'(k^2)$, and 
\[ D'(k^2) = \frac{i}{k^2 - \mu^2 - \pi'(k^2) + i \epsilon} \]

\noindent
(Note that $[ i D'(k^2)]^* = i D'(k^{2*}) \Rightarrow \pi'(k^2)^* = \pi'(k^{2*}) $)\\

Let's look at the analytic structure of $\pi'(k^2)$ to second order in perturbation theory, the function we have just obtained an expression for, for real $k^2$, but which can be defined by this expression for complex $k^2$. 
\[ \pi'(k^2) = \frac{g^2}{16\pi^2} \int^1_0 dx \left\{ \text{ln } \frac{-k^2x(1-x)+m^2 - i\epsilon}{-\mu^2 x (1-x) + m^2} + \frac{(k^2-\mu^2)x(1-x)}{-\mu^2 x (1-x) + m^2} \right\} \]

This expression is not only well defined, it is analytic for $\text{Im} k^2 \neq 0$. It is also analytic for $\text{Im } k^2 =0$, $-\infty < k^2 < 4m^2$, but starting at $4m^2$, because the branch cut in the logarithm needs to be defined for ranges of $x\in [0,1]$, there is a branch cut in $\pi'$.
\begin{center}
\includegraphics[scale=0.2]{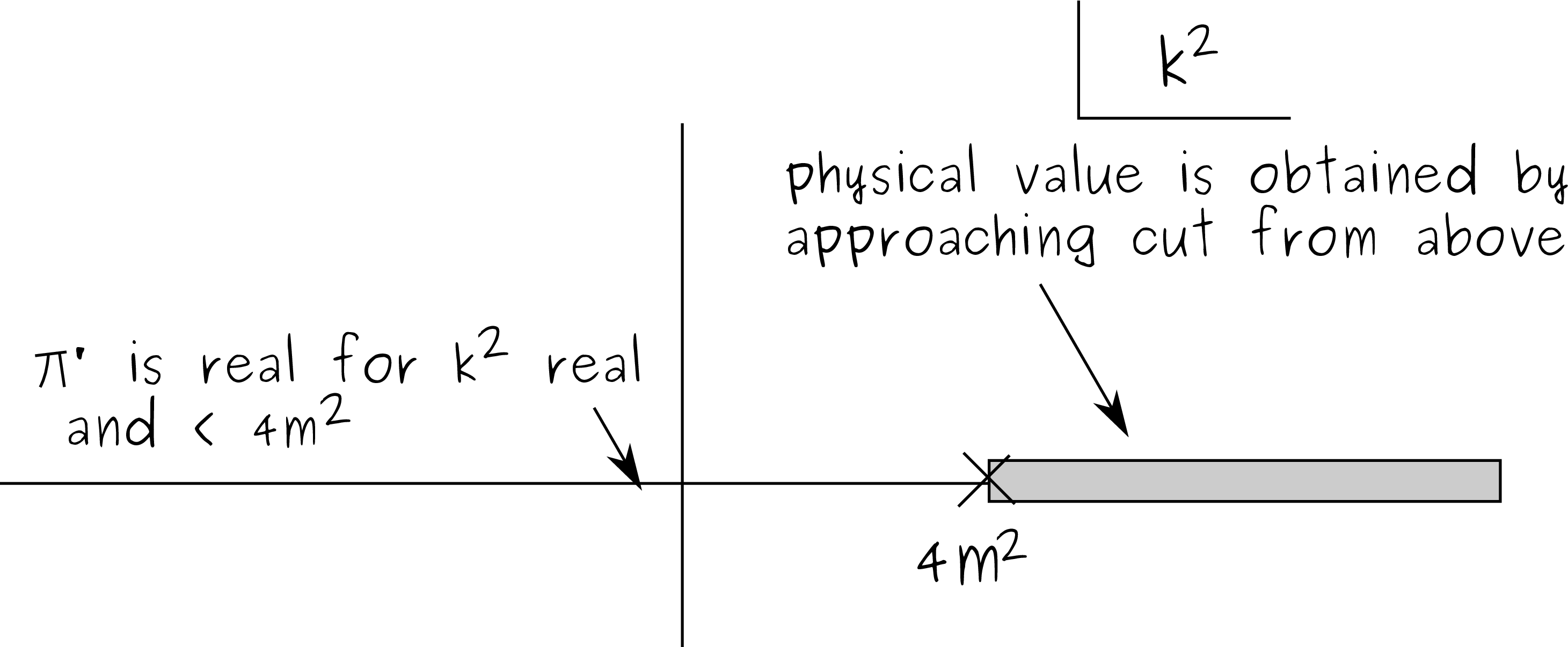}
\end{center}

The $i \epsilon$ prescription when $k^2$ is real and greater than $4m^2$ says to define the logarithm and hence $\pi'$ by approaching the cut from above. Compare this with the analytic structure of $D'$ and you'll see that perturbation theory is satisfying formulas obtained outside of perturbation theory. ($D'$ had a pole at $\mu^2$. This is in agreement. When you invert $D'$ to get $\pi'$ you get a zero.) 

This and the fact that our counterterm was real when $\mu^2 < 4m^2$ (a requirement necessary for the existence of a physical meson), are satisfying consistency checks. All right theories are internally consistent. (However, all internally consistent theories are not right.)

\section*{Loop Lore}
How do you generalize the wonderful tricks done here to graphs with more propagators and more loops?
\begin{center}
\includegraphics[scale=0.2]{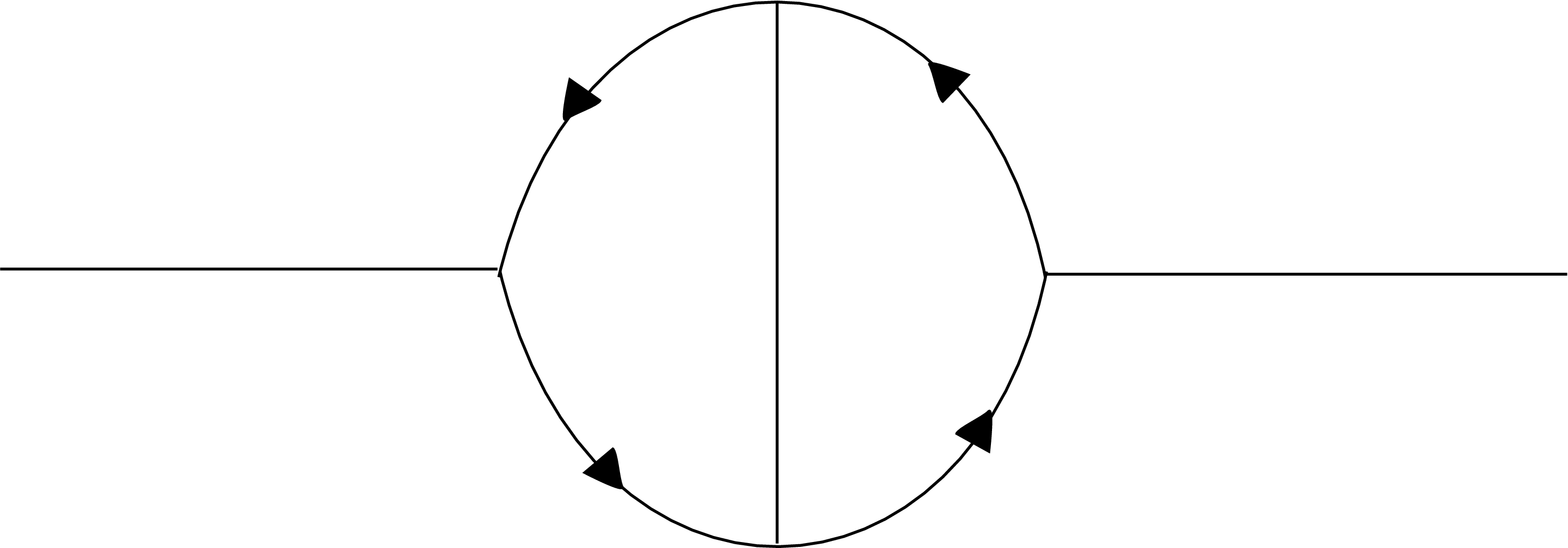}
\end{center}

We'll introduce Feynman parameters, $n-1$ of them if there are $n$ propagators, that combine all the propagators into one denominator. Then we'll be able to do a shift and a Wick rotation and then a spherically symmetric integral. What remains and is very difficult is the integration over the Feynman parameters. In difficult but important applications those integrations are done accurately by computer.
\vspace{1cm}

\textbf{Combining denominators}
\begin{align*}
\prod_{r=1}^n \frac{1}{a_r + i \epsilon} &= \prod_r \left[ - i \int^\infty_0 d\beta_r e^{i\beta_r(a_r+ i\epsilon)}\right]\\
&= (-i)^n \int^\infty_0 d\beta_1 \cdots d\beta_n e^{i \sum_r \beta_r (a_r + i \epsilon) } \underbrace{\int_0^\infty d\lambda \; \delta\left(\lambda - \sum_s \beta_s\right)}_{\substack{\text{Fancy way of}\\\text{inserting 1}\\\text{into the integrand}}}\\
&= (-i)^n \int^\infty_0 d\lambda \int_0^\infty d\beta_1 \cdots d\beta_n \; e^{i \sum_r \beta_r (a_r i \epsilon)} \delta\left(\lambda - \sum_s \beta_s\right) 
\end{align*}

Now for some rescalings. First rewrite
\[ \delta \left(\lambda - \sum_s \beta_s \right) \quad \text{as} \quad \frac{1}{\lambda} \delta \left(1- \frac{\sum \beta_s}{\lambda}\right) \]

Then introduce new integration variables $\alpha_i= \frac{\beta_i}{\lambda}$
\begin{align*}
\prod_{r=1}^n \frac{1}{a_r + i\epsilon} &= (-i)^n \int^\infty_0 d\lambda \int^\infty_0 d\alpha_1 \cdots d\alpha_n \lambda^{n-1} e^{i\lambda \sum \alpha_r (a_r + i\epsilon)} \delta\left(1 - \sum_i \alpha_i\right) \\
&= \int^\infty_0 d\alpha_1 \cdots d\alpha_n \delta \left(1 - \sum_i \alpha_i\right) (-i)^n \int^\infty_0 \lambda^{n-1} d \lambda e^{i\lambda \sum \alpha_r (a_r + i\epsilon)} \\
&= \int^\infty_0 d\alpha_1 \cdots d\alpha_n \delta \left(1 - \sum_i \alpha_i\right) \frac{(-i)^n}{[-i \sum_r \alpha_r (a_r + i\epsilon) ]^n} (n-1)! 
\end{align*}

So,
\[ \prod_{r=1}^n \frac{1}{a_r + i\epsilon} = (n-1)! \int^1_0 d\alpha_1 \cdots d\alpha_n \delta \left(1 - \sum_i \alpha_i\right) \frac{1}{[\sum_r \alpha_r (a_r + i\epsilon) ]^n} \]

\noindent
where I have noticed that the $\delta$ function vanishes whenever any of the $\alpha_i$ are greater than one, and used that to stop the $\alpha_i$ integration at $1$. This is a nice symmetric form; easy to remember. However, we can use the $\delta$ function to do the integral over one of the Feynman parameters, leaving $n-1$ of them, as advertised.
\begin{align*} \prod_{r=1}^n \frac{1}{a_r + i\epsilon} =(n-1)! \int^1_0 d\alpha_1 \int^{1-\alpha_1}_0 d\alpha_2 \int^{1-\alpha_1 - \alpha_2}_0 d\alpha_3 &\cdots \int^{1-\alpha_1-\alpha_2-\cdots-\alpha_{n-2}}_0 d\alpha_{n-1} \\
&\frac{1}{[\sum_{r=1}^{n-1} \alpha_r a_r + (1-\sum_{r=1}^{n-1} \alpha_r) a_n + i \epsilon]^n}
\end{align*}

This generalizes the result of Eq.~(\ref{eq:17-page5}). Take $n=2$, $\alpha_1 =x$, and you have
\[ \frac{1}{A_1 + i\epsilon} \frac{1}{A_2 + i\epsilon} = \int^1_0 dx \frac{1}{[A_1 x + A_2(1-x) + i\epsilon]^2} \]

\section*{A shorter derivation using the $\Gamma$ function}
Uses: $\displaystyle \Gamma(x) \equiv \int^\infty_0 dt \; t^{x-1} e^{-t} $, $\Gamma(n+1) = n!$
\vspace{0.5cm}
\noindent
\textbf{Feynman parameters}           $A_j$ real, $\alpha_j > 0$
\[ I = \int^\infty_0 dt \; t^{\alpha -1} e^{-At} = \frac{1}{A^\alpha} \Gamma(\alpha) \]
\begin{align*}
\frac{1}{\prod_j (A_j^{\alpha_j})} &= \prod_j \frac{\int^\infty_0 dt_j \; t_j^{\alpha_j -1} e^{-A_jt_j}}{\Gamma(\alpha_j)} \underbrace{\int^\infty_0 ds \; \delta(s - \sum_j t_j)}_{\substack{\text{Fancy way of}\\\text{writing 1}}} \\
(t_j = s x_j) \quad \quad&= \int^\infty_0 ds \prod_j \frac{\int^\infty_0 s^{\alpha_j} dx_j \; x_j^{\alpha_j -1} e^{-sA_j x_j}}{\Gamma(\alpha_j)} \frac{\delta(1 - \sum_j x_j) }{s} \\
&= \prod_j \frac{1}{\Gamma(\alpha_j)} \int^1_0 dx_j \; x_j^{\alpha_j -1} \delta \big(1 - \sum_j x_j \big) \int^\infty_0 ds\; s^{(\sum_j \alpha_j) -1} e^{-s\sum_j x_j A_j} \\
&= \prod_j \frac{1}{\Gamma(\alpha_j)} \int^1_0 dx_j \; x_j^{\alpha_j -1} \delta \big(1 - \sum_j x_j \big) \frac{\Gamma(\sum_j \alpha_j)}{(\sum_j x_j A_j)^{\sum_j\alpha_j}}
\end{align*}
\[ \frac{1}{\prod_j (A_j^{\alpha_j})} = \frac{\Gamma(\sum_j \alpha_j)}{\prod_j \Gamma(\alpha_j)} \int dx_1 \cdots dx_j \delta\big(1-\sum_j x_j\big) \frac{\prod_j x^{\alpha_j -1}}{(\sum_j x_j A_j)^{\sum_j \alpha_j}} \]

\noindent \textbf{Examples}
\begin{enumerate}
\item $\displaystyle \frac{1}{AB} = \int^1_0 dx \frac{1}{(xA + (1-x)B)^2}$
\item Take $A_j =1$, $\displaystyle \frac{\prod_j \Gamma(\alpha_j)}{\Gamma(\sum_j \alpha_j)} = \int \delta\left(1-\sum_j x_j\right) \prod_j dx_j \; x_j^{\alpha_j-1}$, generalized binomial expansion.
\item Beta function $\displaystyle \frac{\Gamma(\alpha_1) \Gamma(\alpha_2)}{\Gamma(\alpha_1 + \alpha_2)} = \int_0^1 dx\, x^{\alpha_1-1}(1-x)^{\alpha_2-1}$
\end{enumerate}

Now that we have introduced the Feynman parameters into the integral, how do we make the loop integration trivial?\\

Suppose we have a graph with $I$ internal lines, and $L$ loops, that is $L$ momentum integrals still left to be done after using the energy-momentum conserving $\delta$ functions.
\begin{center}
\includegraphics[scale=0.2]{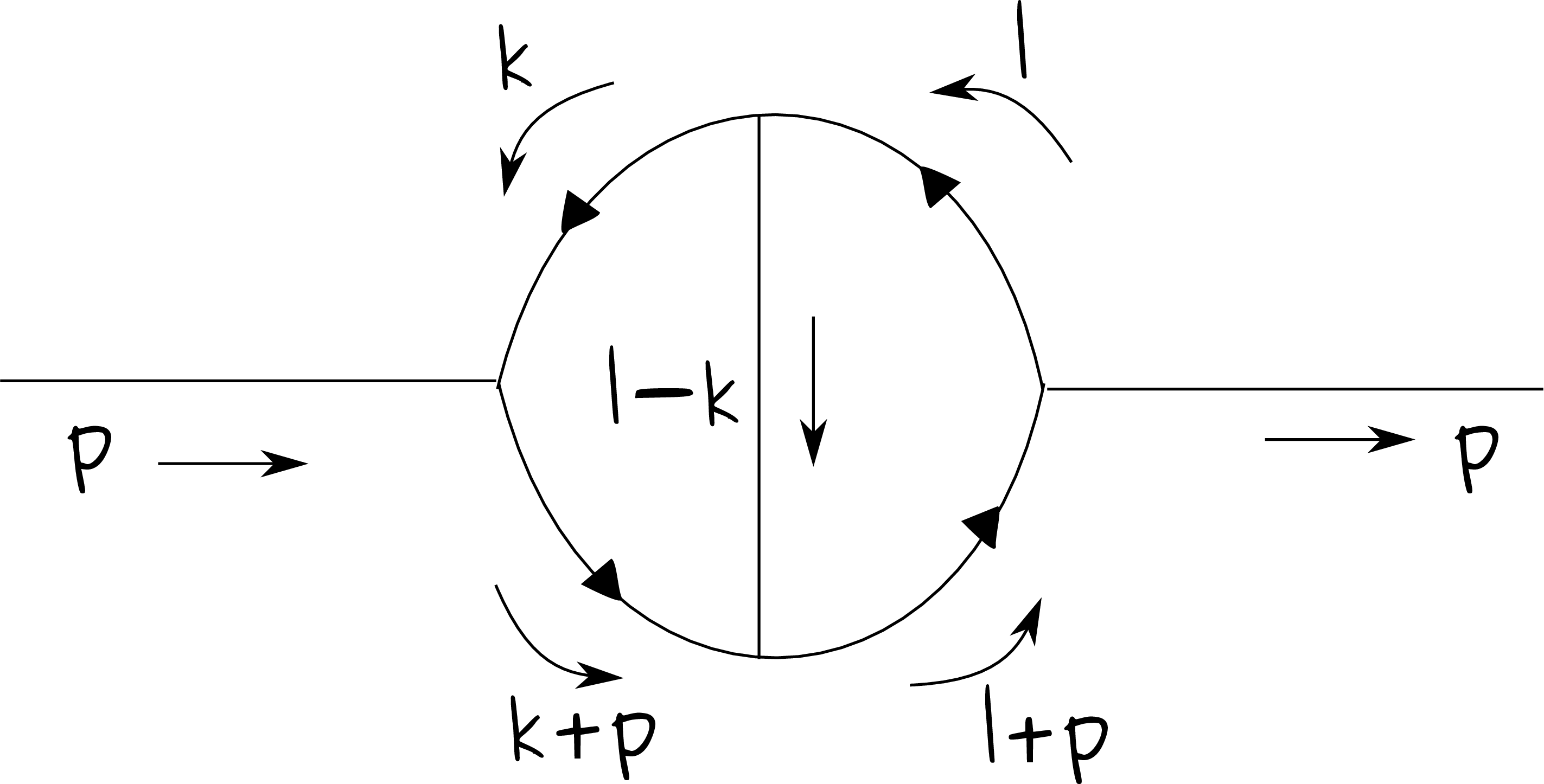}
\end{center}

\noindent
has $L=2$, $I=5$. The integral to be done looks like 
\[ \int \frac{d^4k\; d^4l}{(k^2 -m^2)(l^2-m^2)((k-l)^2 - \mu^2)((k+p)^2 - m^2) ((l+p)^2 - m^2)} \]

To this we would apply our denominator combining identity. Let's write down the general case. Call the independent loop momenta $k_i$, $i=1,\ldots,L$, and the external momenta, $q_j$. All momenta on the $I$ internal lines are linear combinations of the $k_i$ and $q_j$. After introducing the Feynman parameters, the integral to be done is of the form 
\[ \int^1_0 d\alpha_1 \cdots d\alpha_I \delta\left(1- \sum \alpha\right) \int \frac{d^4k_1 \cdots d^4 k_L}{D^I} \]

\noindent
where $\displaystyle D = \sum_{i,j=1}^L A_{ij} k_i \cdot k_j + \sum_{i=1}^L B_i \cdot k_i + C$

$A$ is an $L \times L$ matrix that is linearly dependent on the Feynman parameters. It is positive definite except at the endpoints of the Feynman parameter integrations. $B$ is a vector with $L$ four vector components. It is linear in the Feynman parameters and linear in the external momenta. $C$ is a number, depending linearly on the Feynman parameters, and the external momenta squared and the masses squared that appear in the propagators. It has a small positive imaginary part.\\

Now shift the $k$ integration to eliminate the terms linear in $k$.
\[ k_i' = k_i +\frac{1}{2} \sum_j (A^{-1})_{ij} B_j \]
\[ d^4 k_i = d^4 k_i' \]
\[ D = \sum_{ij=1}^L A_{ij} k'_i\cdot k'_j + C' \]

\noindent
where $\displaystyle C' = C - \frac{1}{4} \sum_{ij} B_i A^{-1}_{ij} B_j$\\

$C$ is still linear in external momenta squared and the masses squared, but now it has some awful dependence on the Feynman parameter because of $A_{ij}^{-1}$. It still has a small positive imaginary part.\\

Now diagonalizing $A_{ij}$ with an orthogonal transformation on the set of four vectors $k_i'$
\[ k_i' = O_{ij} k_j^{\prime\prime} \quad \quad \det O = 1 \]
\[ \prod^L_{i=1} d^4 k_i' = \prod^L_{i=1} d^4 k_i^{\prime\prime} \]
\begin{align*}
D &= \sum_{i,j=1}^L A_{ij} O_{ik} k_k^{\prime\prime} \cdot O_{jl} k_l^{\prime\prime} + C' \\
&= \sum_{i,j=1}^L (O^T A O)_{ij} k_i^{\prime\prime} \cdot k_j^{\prime\prime} + C' \\
&= \sum_{i=1}^L a_i k_i^{\prime\prime} \cdot k_i^{\prime\prime} + C'
\end{align*}

\noindent
where $\displaystyle (O^T A O)_{kl} = \delta_{kl} a_l$.\\

Finally we'll make a transformation to eliminate the $a_i$
\[ k_i^{\prime\prime} = \frac{1}{\sqrt{a_i}} k_i^{\prime\prime\prime} \]
\begin{align*} 
\prod_{i=1}^L d^4 k_i^{\prime\prime} &= \prod_{i=1}^L
\left(\frac{1}{\sqrt{a_i}}\right)^4 d^4 k_i^{\prime\prime\prime} \\
&= (\det A)^{-2} \prod_{i=1}^L d^4 k_i^{\prime\prime\prime} 
\end{align*}

The integral to be done has been reduced to 
\[ \int^1_0 d\alpha_1 \cdots d\alpha_I \; \delta\left(1 - \sum \alpha\right) (\det A)^{-2} \int \frac{d^4 k_1^{\prime\prime\prime} \cdots d^4 k_L^{\prime\prime\prime}}{D^I} \]

\noindent
where $\displaystyle D = \sum_{i=1}^L k_i^{\prime\prime\prime} \cdot k_i^{\prime\prime\prime} + C'$ \\

Now we can perform Wick rotations on each of the $k_i^{\prime\prime\prime \; 0}$ variables independently to get 
\[ \int^1_0 d\alpha_1 \cdots d\alpha_I \; \delta\left(1 - \sum \alpha\right) (\det A)^{-2} i^L \int \frac{d^4 k_{1\,E} \cdots d^4 k_{L\,E}}{D^I} \]
\[ d^4 k_i^{\prime\prime\prime} = i d^4k_{i\, E} \]
\[ k_{i\,4} = i k_i^{\prime\prime\prime\,0} \]
\[ D = - \sum_{i=1}^L k_E^2 + C' \]

This is one big spherically symmetric integral in $4L$ dimensions$!$ Easily done with only a slight generalization of our integral table. We have reduced a general graph to an awful integral over Feynman parameters; this is progress. Note that you don't actually have to diagonalize $A$ when applying this formula. All you need in the end is $\det A$.
}{
 \sektion{18}{November 25}
\descriptioneighteen
\section*{The definition of $g$ in Model 3}
Renormalization condition (6), the committee definition of $g$, has not been stated or turned into an equation among Green's functions. The statement is needed to fix the counterterm in $\mathcal{L}$
\[ \mathcal{L} = \cdots - F \psi^{*\prime}\psi' \phi' + \cdots \]

\noindent
which has Feynman rule 
\[ \Diagram{\momentum[llft]{fdV}{p'\searrow } & x \momentum[bot]{f}{\leftarrow q} \\ \momentum[bot]{fuA}{\nearrow p}} = -iF(2\pi)^4 \delta^{(4)}(p+p' + q) \]

\noindent
Model 3 does not exist in the real world, so no committee has actually gotten together to define $g$. We'll play committee.\\

Define 
\[ \Diagram{\momentum[bot]{fdV}{p'\searrow } & !p{1PI} \momentum[bot]{f}{\leftarrow q} \\ \momentum[bot]{fuA}{\nearrow p}} = \!\!\!\!\!\!\!\!\!\!\!\!\!\!\!\!\! \underbrace{-i}_{\substack{\text{The } -i \text{ is a sensible}\\\text{convention, put there so}\\\text{that at lowest}\\\text{order } \Gamma = g}} \!\!\!\!\!\!\!\!\!\!\!\!\!\!\!\Gamma'(p^2, p^{\prime 2}, q^2) \]

\noindent
Why can we consider $\Gamma'$ to be a function of $p^2$, $p^{\prime 2}$ and $q^2$? \\

\noindent
$\Gamma'$ is a Lorentz invariant, so it must be a function of Lorentz invariants only. There are only two independent momenta, $q = - p -p'$, so the only Lorentz invariants are $p^2$, $p^{\prime 2}$ and $p\cdot p'$. However $ p \cdot p'$ can be traded in for $q^2$.\\

So here is our committee definition of $g$:
\[ g \equiv \Gamma'(\underline{p}^2,\underline{p}^{\prime 2}, \underline{q}^2) \]

\noindent
The bars mean some specific point in momentum space.\\

This is a reasonable if not obvious generalization of the types of conditions we used to determine $A,B,C,D$ and $E$. The proof of the iterative determination of $F$ is identical.\\

While all points $\underline{p}^2$, $\underline{p}^{\prime 2}$, $\underline{q}^2$ are equally good as far as determining $F$ is concerned, there is one that is more equal than others. It might well be the one the committee picks, because as we will show, it has some experimental significance. The point is 
\[ \underline{p}^2 = \underline{p}^{\prime 2} = m^2 \quad \quad \quad \underline{q}^2 = \mu^2 \]

To actually find a trio of four-vectors satisfying these conditions, as well as $p+p'+q=0$, you have to make some of their components complex. This point is not kinematically accessible. One can show in general however, that the domain of analyticity of $\Gamma'$, considered as a function of three complex variables, is sufficiently large to define the analytic continuation of $\Gamma'$ from any of its physically accessible regions to this point.\\

What is this point's experimental significance?\\

Look at the process $\phi + N \rightarrow \phi +N $. Diagrammatically,
\begin{center}
\includegraphics[scale=0.5]{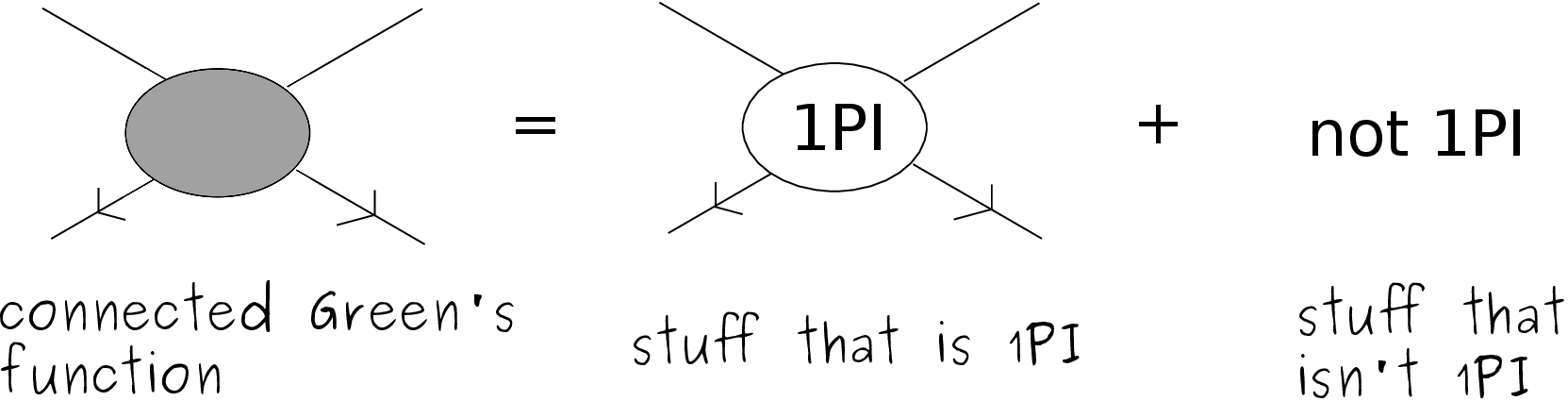}
\end{center}

We can say more about the stuff that isn't 1PI. By definition, there is some line in the graph which can be cut and the graph falls into two pieces. If I ignore interactions on the external legs, the cutting of the internal line separates the graph into two pieces each having two external lines. The $\displaystyle \frac{1}{2} \binom{4}{2} = 3$ possibilities look like $s$, $t$ and $u$ channel graphs. If cutting the internal line separates the incoming meson and nucleon from the outgoing ones the graph must be a contribution to (I hope you can convince yourself)
\begin{center}
\includegraphics[scale=0.4]{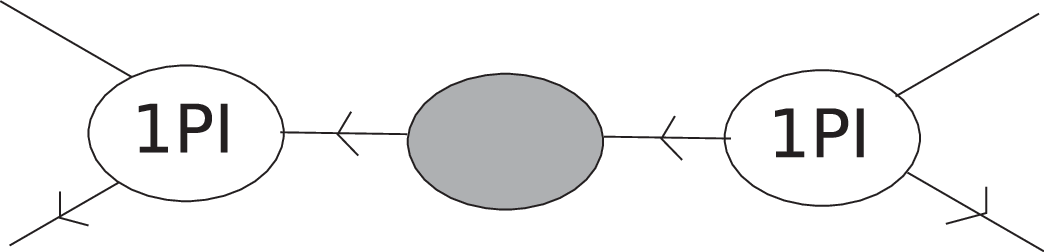}
\end{center}

\noindent 
the other two possibilities are
\begin{center}
\includegraphics[scale=0.4]{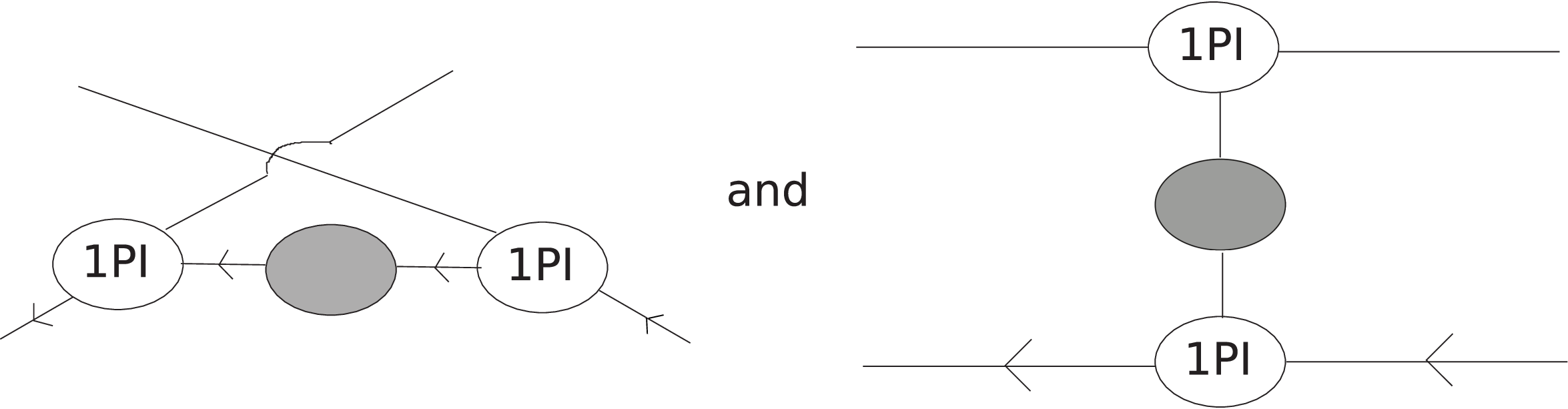}
\end{center}

Using our definitions, the first graph (on mass shell) is 
\[ -i\Gamma'(s,m^2,\mu^2) D'(s)(-i\Gamma'(m^2,s,\mu^2)) \]

This has a pole at $s=m^2$ because the full nucleon propagator has a pole there, and because of our renormalization conditions, we can write down what the residue of that pole is. $s=m^2$, we have
\[ -ig\frac{i}{s-m^2}(-ig) + \text{ analytic stuff at } s = m^2 \]

Because of our definitions, the residue of the pole of these graphs is $-ig^2$. \\

Now what about the other two graphs? They look like they have poles at $u = m^2$ and $t=\mu^2$, but we don't expect them to have a pole at $s=m^2$. Furthermore, the graph $\Diagram{\momentum[bot]{fd}{} !p{1PI} & \momentum[bot]{fu}{} \\ \momentum[bot]{fuV}{} & \momentum[bot]{fdV}{}} $
probably has all sorts of cuts, but it is unlikely that it has a pole at $s=m^2$ because there is no propagator on the inside of the graph that carries the whole incoming momentum.\\

To summarize
\[ \Diagram{\momentum[bot]{fd}{} !p{} & \momentum[bot]{fu}{} \\ \momentum[bot]{fuV}{} & \momentum[bot]{fdV}{}} = \frac{-ig^2}{s-m^2} + \text{ plus analytic or at least no pole, near } s = m^2 \]

Experimentally, the residue of this pole can be measured by looking at $\phi + N$ scattering in the physical region, and extrapolating down to $s=m^2$. You just measure the $s$ wave scattering. When this was done, they found pole-like behavior with $g\approx 13.5$. Actually, they weren't very good at making pion beams back when they did this, so they measured the pole in $\gamma + p \rightarrow p + \pi$, which measures $eg$. When $g=13.5$ was determined this way, it put the last nail in the coffin for the attempts to consider the strong interactions perturbatively with the pion and nucleons as fundamental particles.\\

Consider the process $N+N \rightarrow N+N$. By similar arguments, we can split up 
\begin{center}
\includegraphics[scale=0.4]{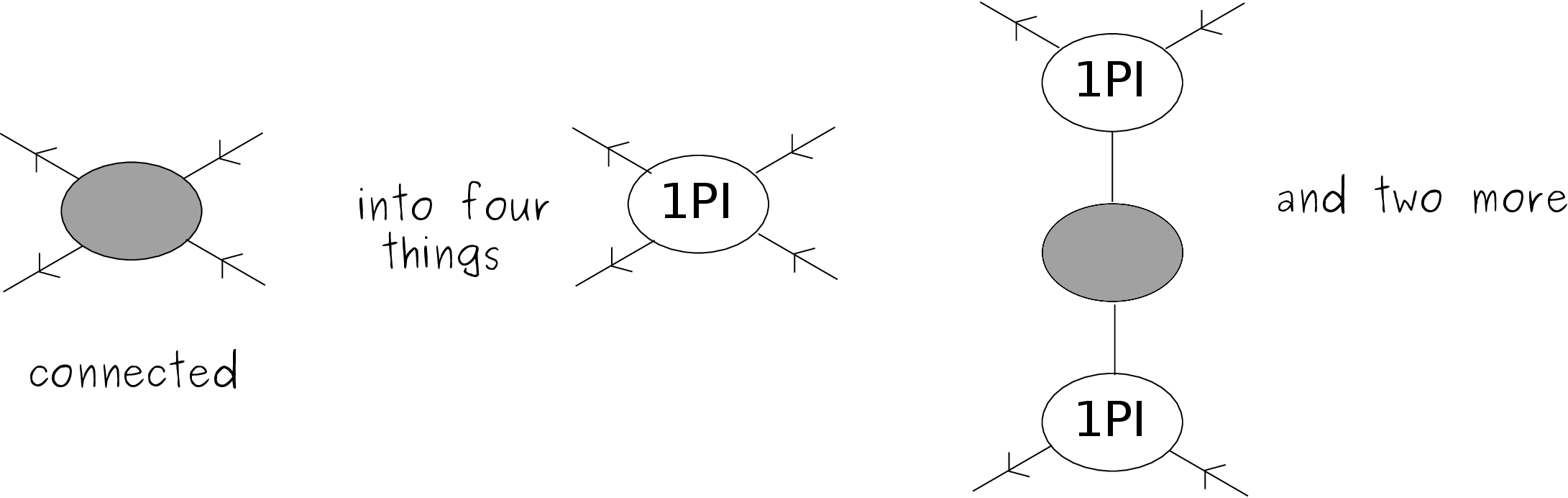}
\end{center}

The thing to note is that this decomposition leads us to expect
\[ \Diagram{\momentum[bot]{fdV}{} !p{} & \momentum[bot]{fuV}{} \\ \momentum[bot]{fuV}{} & \momentum[bot]{fdV}{}} = (-ig)^2 \frac{i}{t-\mu^2} + \text{stuff with no pole at } t =\mu^2 \]

\noindent
Since $t \approx \mu^2$ is unphysical, to measure the effect of this pole, you again have to extrapolate.\\

Our simple model states that the residue of this pole is \underline{the same}, $-ig^2$. When they did this experiment with $p+p \rightarrow p+p$, after doing some work to eliminate electromagnetic effects, they got agreement (to within $10\%$). Futhermore, the fit showed that the location of the pole was at $t= m^2_\pi$ (They only fit the high partial waves, where they felt justified calculating with P.T.)

\section*{Renormalization vs. infinities}
The $\mathcal{O}(g^3)$ correction to $\Gamma'$ in model 3
\includegraphics[scale=0.2]{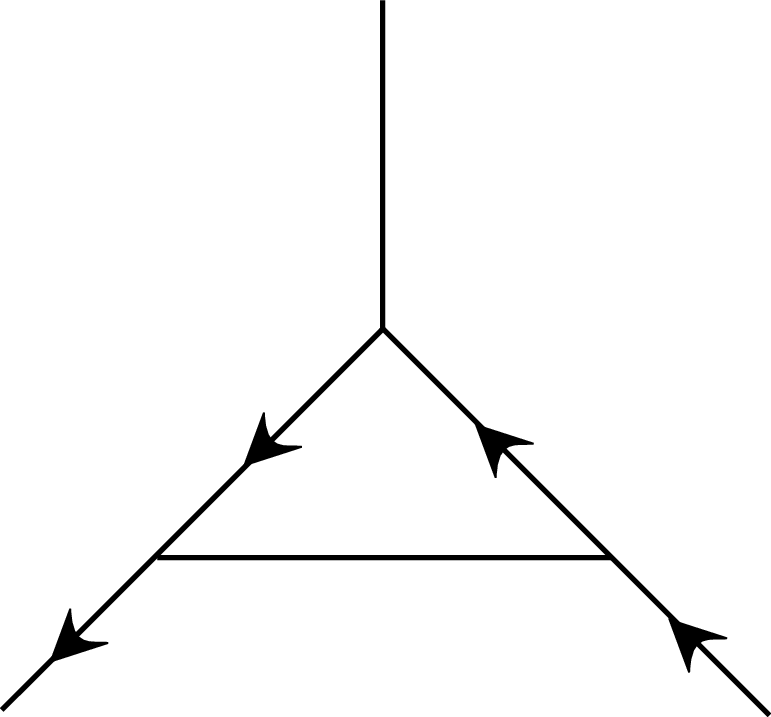}
is finite. The counterterm 
\includegraphics[scale=0.2]{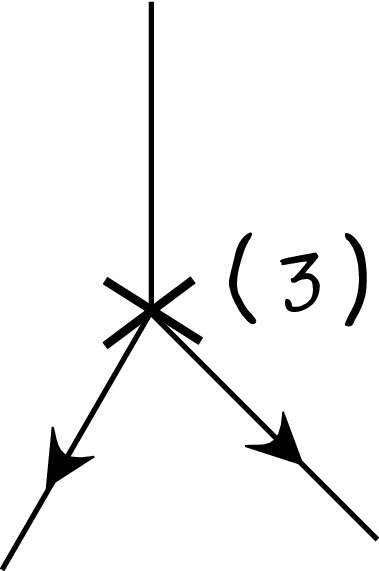}
is needed only to make the theory agree with the committee definition of $g$. To see that the graph is finite, look at its high momentum behavior. Without even combining denominators you can see that at high $q$ the integral looks like $\int \frac{d^4q}{q^6}$. This extreme convergence is peculiar to model 3 (and other models where all the couplings have positive mass dimension as we will later see). \\

Consider a model with a four scalar field interaction.
\[ \mathcal{L} = \cdots + g\phi^4 + \cdots \]

\noindent
$\phi^4$ could be $ABCD$ or $(\psi^* \psi)^2$. \\

Look at the lowest order correction to a propagator:
\begin{center}
\includegraphics[scale=0.3]{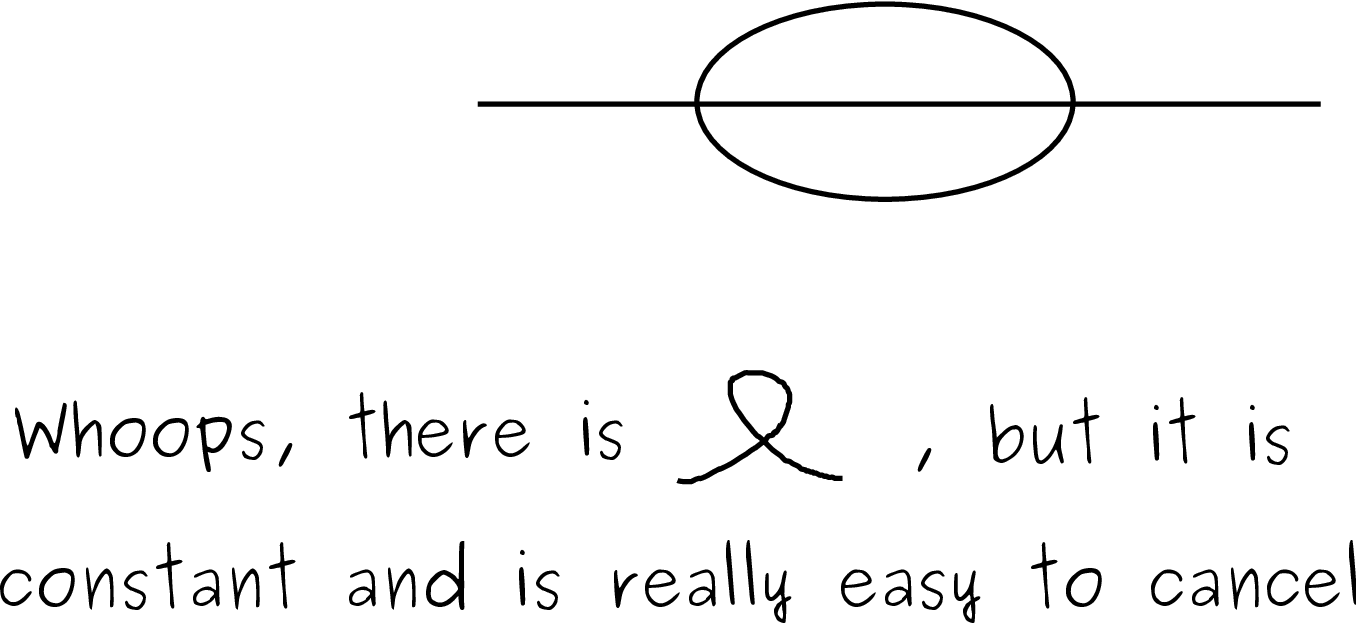}
\end{center}

After combining denominators you see that the integral looks like at high $q$ 
\[ \int \frac{\overbrace{d^8 q}^{\text{2 loops}}}{\underbrace{q^6}_{\text{three propagators}}} \quad \quad \text{quadratic divergence} \]

Fortunately there are renormalization counterterms $\Diagram{f x f}$ to cancel this infinity.\\

What about other graphs in this theory? There is 
\[ \includegraphics[scale=0.3]{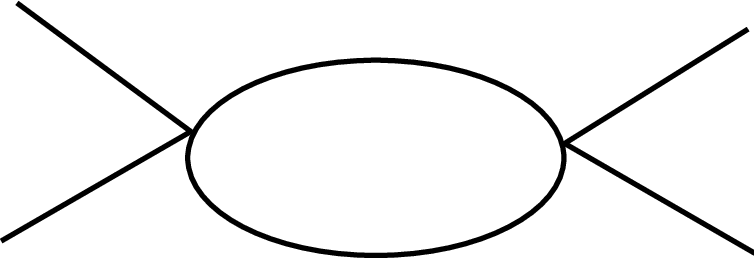} \sim \int \frac{d^4q}{q^4} \quad \text{log divergent} \]

There is a committee definition of $g$, and a $\phi^4$ counterterm which can cancel off the log divergence
\begin{center}
\includegraphics[scale=0.3]{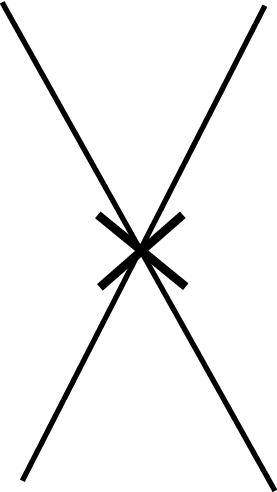}
\end{center}

What about
\begin{center}
\includegraphics[scale=0.3]{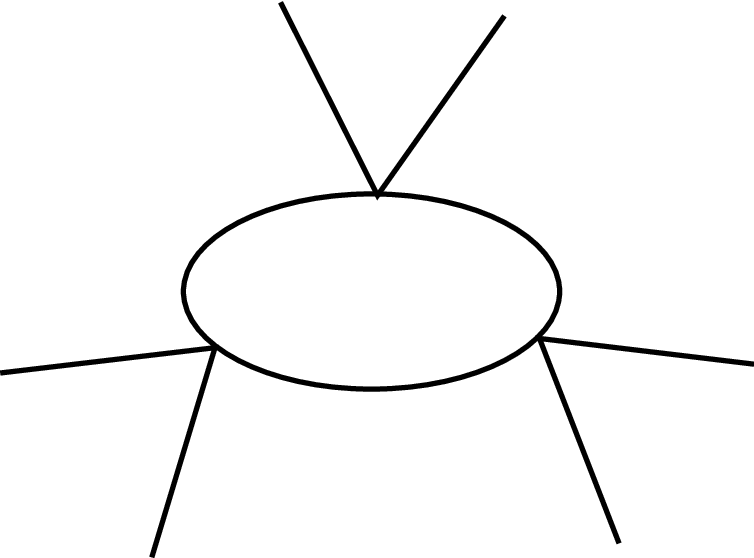}
\end{center}

This is finite, which is good. We would need a $\phi^6$ counterterm to cancel this graph's divergence if it weren't convergent. That would require another committee definition, say for $3 \rightarrow 3$ scattering at some momentum.\\

\underline{Definition} (This is a more stringent definition than is often used.) A Lagrangian is renormalizable only if all the counterterms required to remove infinites from Green's functions are terms of the same type as those present in the original Lagrangian.\\

Suppose a theory has a $\phi^5$ ($ABCDE$) interaction. Then

$\includegraphics[scale=0.3]{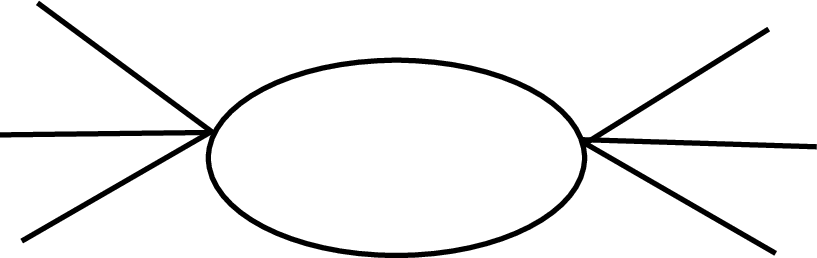}$ is log divergent, and you would need a $\phi^6$ counterterm to cancel it. Since the theory did not originally contain a $\phi^6$ interaction, we say $\phi^5$ theory is not renormalizable.\\

It seems fairly clear that to correct this defect, you just add a $\phi^6$ term to your Lagrangian, then your $\phi^6$ counterterm will be of the same type as the interaction term in the original Lagrangian. But then, there is 
\begin{center}
\includegraphics[scale=0.3]{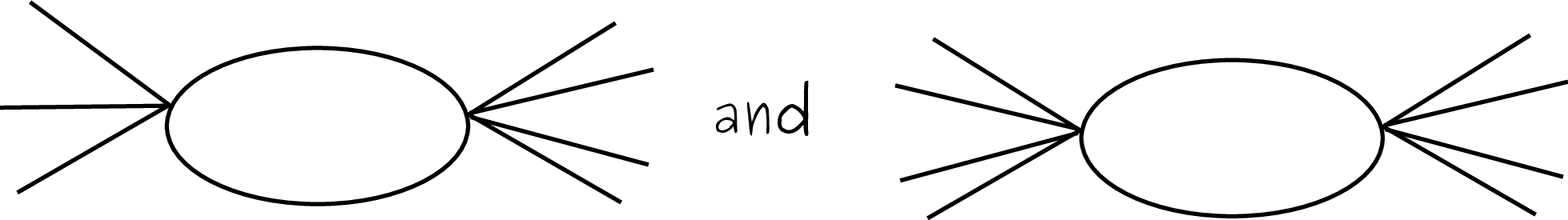}
\end{center}

\noindent
to worry about. These are also log divergent and they require $\phi^7$ and $\phi^8$ counterterms to cancel them. So $g\phi^5 + h\phi^6$ is not a renormalizable interaction either. You can see that adding $j\phi^7$ and $k\phi^8$ to the Lagrangian is not going to help.\\

This shows that any polynomial interaction of degree higher then 4 is not renormalizable. We have not shown that polynomials of degree 4 or less are renormalizable, but what we have found so far suggests it.\\

Now these theories with an infinite series of interactions are disgusting because they contain an infinite number of independently adjustable parameters. Unless you make some additional statement, you cannot make any predictions. One possibility is to hunt for some relationship among the terms in the infinite series. Perhaps
$\displaystyle \mathcal{L} = \frac{1}{2} \partial_\mu \phi^2 - \frac{\mu^2}{2} \phi^2 - \lambda \cos \alpha \phi$ is renormalizable.\\

Perhaps $S = \int d^4 x \sqrt{-g} R $, the Einstein-Hilbert action is renormalizable. Suffice it to say that no one has ever been able to construct a renormalizable non-polynomial interaction that is not equivalent to free field theory.

\section*{Unstable particles}
Let's look at model 3 in the regime $\mu > 2m$. In that case, $\pi_f(k^2)$ is not real at the subtraction point. You can see this by looking at $\pi_f$ or you can look at the nonperturbative formula
\[ \text{Im } \pi' (k^2) = - \pi \frac{\sigma(k^2)}{|D'(k^2)|^2} \]

\noindent
$\sigma (k^2) \neq 0$ when $k^2 = \mu^2 > 4m^2$ so $\text{Im }\pi' \neq 0$. Our subtraction, which says 
\[ \pi'(\mu^2) = 0 \quad \quad \text{and} \quad \quad \frac{d\pi'}{dk^2}\Big|_{k^2 =\mu^2} = 0 \]

\noindent
would be causing us to subtract imaginary terms from the Lagrangian. This is unacceptable because a non-Hermitian Hamiltonian is unacceptable. One road is to just say, for $\mu > 2m$, the meson is unstable, I have no business calculating meson-meson scattering or nucleon-meson scattering, or anything else involving an external meson, so just drop all renormalization conditions and subtractions related to the meson.\\

This road is not ideal for two reasons. The definition of the theory does not change in any smooth way as $\mu$ increases beyond $2m$, and we lose the bonus of renormalization, the elimination of infinities. We will modify our subtraction procedure for $\mu > 2m$ so that it still removes $\infty$'s, and is continuously related to the subtractions made for $\mu < 2m$, but so that we do not make imaginary subtractions. Our modified procedure is to quite a degree ad hoc, but we will see that it is useful. For $\mu > 2m$, demand 
\begin{align*} 
\text{Re }\pi'(\mu^2) &= 0\\
\text{Re }\frac{d\pi'}{dk^2}\Big|_{k^2=\mu^2} &= 0 
\end{align*}

We will see that with these renormalization conditions for $\mu > 2m$, and the usual ones for $\mu < 2m$, that as you increase $\mu$, the pole
in $D'(k^2)$ moves up the real axis until it touches the branch cut, and then it moves onto the second sheet (The poles can run, but they can't hide.)
\begin{center}
\includegraphics[scale=0.4]{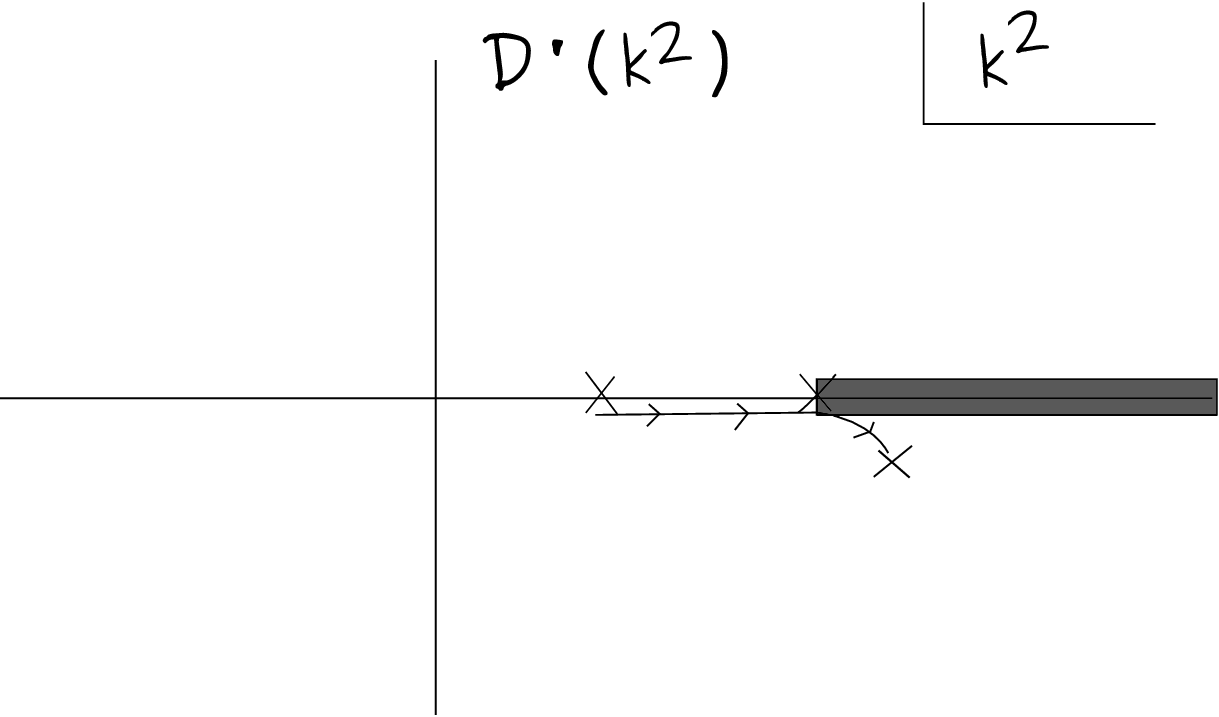}
\end{center}

What do I mean by second sheet? The value of $D'(k^2)$ for $k^2 > 4m^2$ (and real) is obtained by taking the limit from positive imaginary $k^2$ down onto the real axis, according to the $i\epsilon$ prescription. The value of $D'(k^2)$ for $\text{Im }k^2 < 0$ is defined by that integral expression for $D'(k^2)$ in terms of $\sigma(a^2)$. $D'(k^2)$ has a discontinuity across the cut. What we get when we \underline{analytically continue} $D'(k^2)$ to $\text{Im } k^2 < 0$ from its value for $\text{Im } k^2 >0$, to get a function that is continuous along the old cut, is called $D'(k^2)$ on the second sheet.\\

A couple other ways of saying this:\\

The branch point is fixed, the value along the real axis is physical and can't be changed, but within those restrictions, we can move the cut leading from the branch point to $\infty$ around any way we like.\\

The second sheet is what you get from peering down from above the cut. In some sense this is much closer to the physical region, because it is not separated by a discontinuity.\\

In model 3, for $\mu > 2m$, we will now compute $(-iD')^{-1}$ to $\mathcal{O}(g^2)$ when $k^2-\mu^2$ is order $g^2$.
\[ [-iD'(k^2)]^{-1} = k^2 - \mu^2 - \pi'(k^2) \]

\noindent
A formula we will use is
\begin{align*}
\text{Im } \pi'(k^2) &= |D'(k^2)|^{-2} (-\!\!\!\underbrace{\pi}_{3.14\ldots} \!\!\!) \sigma(k^2) \\
&= - \frac{1}{2} |D'(k^2)|^{-2} \eighteensumintaa_{\substack{|n\rangle \neq | 0 \rangle, | p\rangle \\ | p \rangle \text{: one meson}}}|\langle n | \phi'(0) | 0\rangle|^2 (2\pi)^4 \delta^{(4)}(k-P_n)
\end{align*} 

\noindent
using the definition of $\sigma$.\\

Now let's work on $[-iD'(k^2)]^{-1}$
\begin{align*} 
[-iD'(k^2)]^{-1} &= k^2 - \mu^2 - \pi'(\mu^2) - \underbrace{(k^2 - \mu^2)}_{\text{this is }\mathcal{O}(g^2)} \underbrace{\frac{d\pi'}{dk^2}}_{\text{this is } \mathcal{O}(g^2)}\Big|_{k^2=\mu^2} + \mathcal{O}(g^4)\\
&= k^2 - \mu^2 - \!\!\!\! \underbrace{\text{Re }\pi'(\mu^2)}_{\substack{\text{O by our convenient}\\\text{renormalization}\\\text{condition}}}\!\!\!\! - i
\text{ Im } \pi' (\mu^2) + \mathcal{O}(g^4) 
\end{align*}

Now using that formula for $\text{Im }\pi'(k^2)$,
\begin{align*}
[-iD'(k^2)]^{-1} &= k^2 - \mu^2 + \frac{i}{2} \frac{\;\; \eighteensumintc_{n \neq |0\rangle, |k\rangle} |\langle n | \phi'(0) | 0 \rangle |^2 (2\pi)^4 \delta^{(4)}(k-P_n)}{[D'(k^2)]^2} \Bigg|_{k^2=\mu^2} + \mathcal{O}(g^4) \\
&= k^2 - \mu^2 + i \mu\Gamma + \mathcal{O}(g^4) \\
&= k^2 -\left(\mu - \frac{i\Gamma}{2}\right)^2 + \mathcal{O}(g^4)
\end{align*}

The nice thing I have noticed in the next to last step is that what multiplies $\frac{i}{2}$ in the first expression is $\mu \Gamma$ when $k^2 = \mu^2$. Compare with Eq.~\ref{eq:13-page1-2} from Nov.~4. The $|D'|^{-2}$ serves to exactly eliminate the external propagators you would get in relating $\langle n | \phi'(0) | 0 \rangle$ to $\langle n | (S-1) |k\rangle \propto ia$.\\

This does not prove that $\Gamma$ is a $\text{lifetime}^{-1}$. That $\Gamma$ was an inverse lifetime in the theory with a turning on and off function does not suffice to show that it is a lifetime in our full-blown scattering theory.\\

To summarize what we have found so far, we have found that in model 3, with $\mu > 2m$, and some ad hoc renormalization conditions, in the small $g$ limit, there is a pole in $D'(k^2)$ at $k^2 = (\mu - \frac{i\Gamma}{2})^2$ on the second sheet. $D'(k^2)$ is still analytic on the cut complex plane. In a sense, this pole is close to the physical region. Our perturbative analysis shows that as $g\rightarrow 0$, $\Gamma\rightarrow 0$, and the actual value of $D'(k^2)$ along the real axis should be more and more dominated by the presence of this pole when $k^2 \approx \mu^2$.\\

What we have done so far has depended on perturbation theory in model 3, although the way $\Gamma$ appeared, it is clear how a perturbative calculation would go in other models. What we will do next does not depend on perturbation theory in the coupling constant, or on any model.\\

Our only assumption now will be that 
\[ D'(k^2) = \frac{i}{k^2-\mu^2+\mu i \Gamma} + \text{ small terms} \]

\noindent
for some range of the real axis near $k^2 = \mu^2$. That is the pole on the second sheet dominates the behavior of $D'(k^2)$ near $k^2 = \mu^2$.\\

We will now do two thought experiments and show that this behavior is what experimentalists are talking about when they say they have discovered an unstable particle.\\

Our first thought experiment is to blast the vacuum at $\vec{x} = t = 0$. A theorist blasts the vacuum by turning on a source 
\[ \mathcal{L} \rightarrow \mathcal{L} + \rho(\vec{x},t) \phi'(\vec{x},t) \]
\[ \rho(\vec{x},t) = \lambda \delta^{(4)}(x) \]

\noindent
An experimentalist blasts the vacuum at $\vec{x}=t=0$ by crashing two protons together at the origin of coordinates.\\

The amplitude that you'll get any momentum eigenstate $|n\rangle$ is proportional to 
\[ \lambda \langle n | \phi'(0) | 0 \rangle + \mathcal{O}(\lambda^2) \]

The probability of having momentum $k$ in the final state is proportional to 
\[ \lambda^2 \eighteensumintb_{|n\rangle} |\langle n | \phi'(0) | 0 \rangle |^2 (2\pi)^4 \delta^{(4)} (P_n - k) + \mathcal{O}(\lambda^3) = 2 \pi \lambda^2 \sigma(k^2) \theta(k^0) + \mathcal{O}(\lambda^3) \]

In the sum, I don't have to specify $|n\rangle \neq |0 \rangle, |k\rangle$ (one meson) (as long as I stay away from $k=0$ so $\delta^{(4)}(P_n - k) = 0$ when $|n\rangle = |0 \rangle$) since there are no physical one meson states to emerge from blasting the vacuum when ``the meson is unstable".\\

$\sigma(k^2)$ is in turn proportional to $-\text{Im }\pi'$ 
\[ \sigma(k^2)= - (\text{Im } \pi'(k^2)) | D'(k^2)|^2 \]

So the probability of finding momentum $k$ not equal to zero, $k^0 > 0$ is proportional to 
\[ -\lambda^2 \text{Im } \pi'(k^2) |D'(k^2)|^2 + \mathcal{O}(\lambda^3) \]

Finally using the form of $D'$ which is assumed to dominate near $k^2 =\mu^2$ ($\pi'(k^2) = - \mu i \Gamma$) we have 
\[ \frac{\lambda^2 \mu \Gamma}{(k^2 - \mu^2)^2 + \mu^2 \Gamma^2} + \mathcal{O}(\lambda^3) \]
\begin{center}
\includegraphics[scale=0.4]{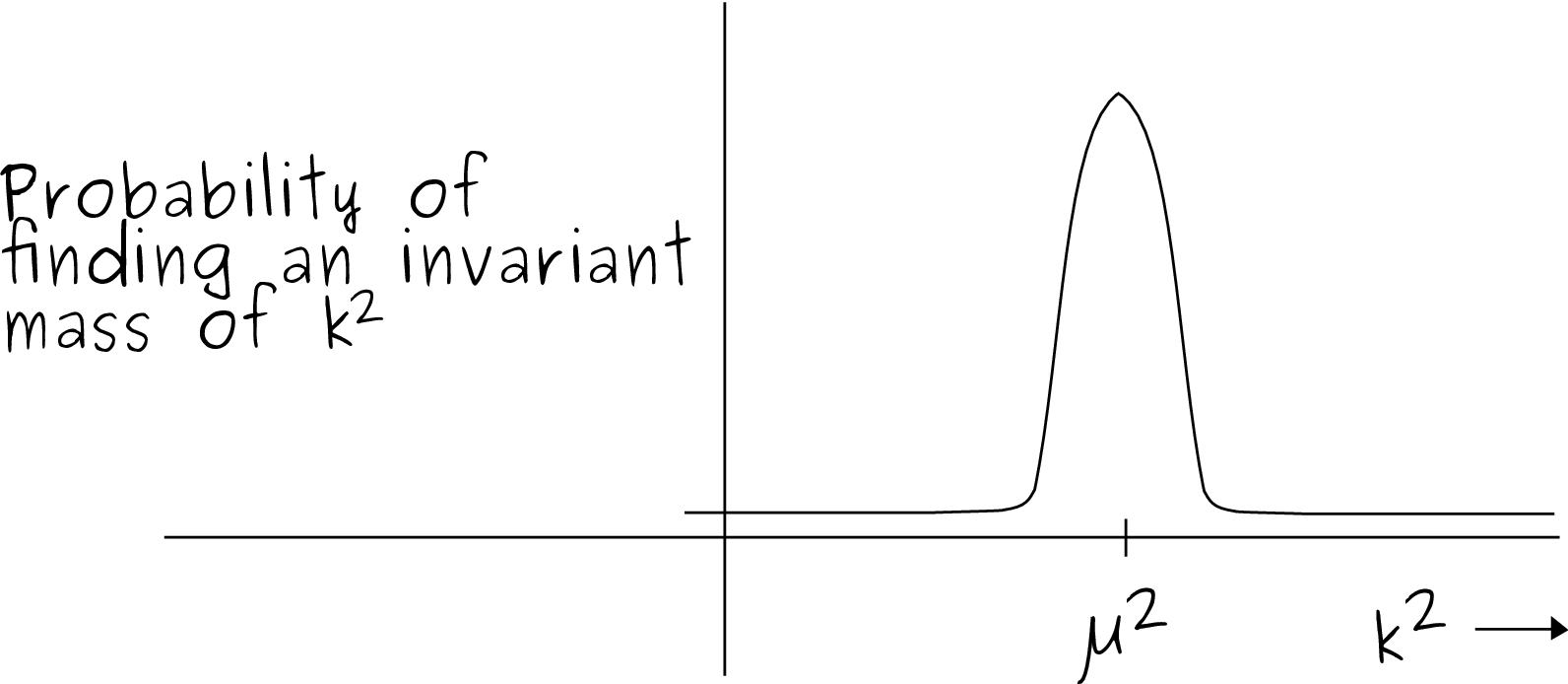}
\end{center}

We can look at the center of mass energy of the decay products. That is, we can think of this probability distribution as a function of $E$, the decay products' COM energy, instead of as a function of $k^2$. The probability of finding a COM energy $E$ in the decay products is proportional to (drop the $\mathcal{O}(\lambda^3)$)
\begin{align*}
\frac{\mu\Gamma}{(E^2-\mu^2)^2 + \mu^2 \Gamma^2} &= \frac{\mu \Gamma}{(E-\mu)^2 (E+\mu)^2 + \mu^2 \Gamma^2} \\
&\approx \frac{\mu \Gamma}{(E-\mu)^2(2\mu)^2 + \mu^2 \Gamma^2} \\
&= \frac{\mu \Gamma}{(4\mu^2)\left[(E-\mu)^2 +\frac{\Gamma^2}{4}\right]} 
\end{align*}
 
\noindent
(Approximation preserves the character of the function if $\Gamma \ll \mu$)\\
\begin{center}
\includegraphics[scale=0.4]{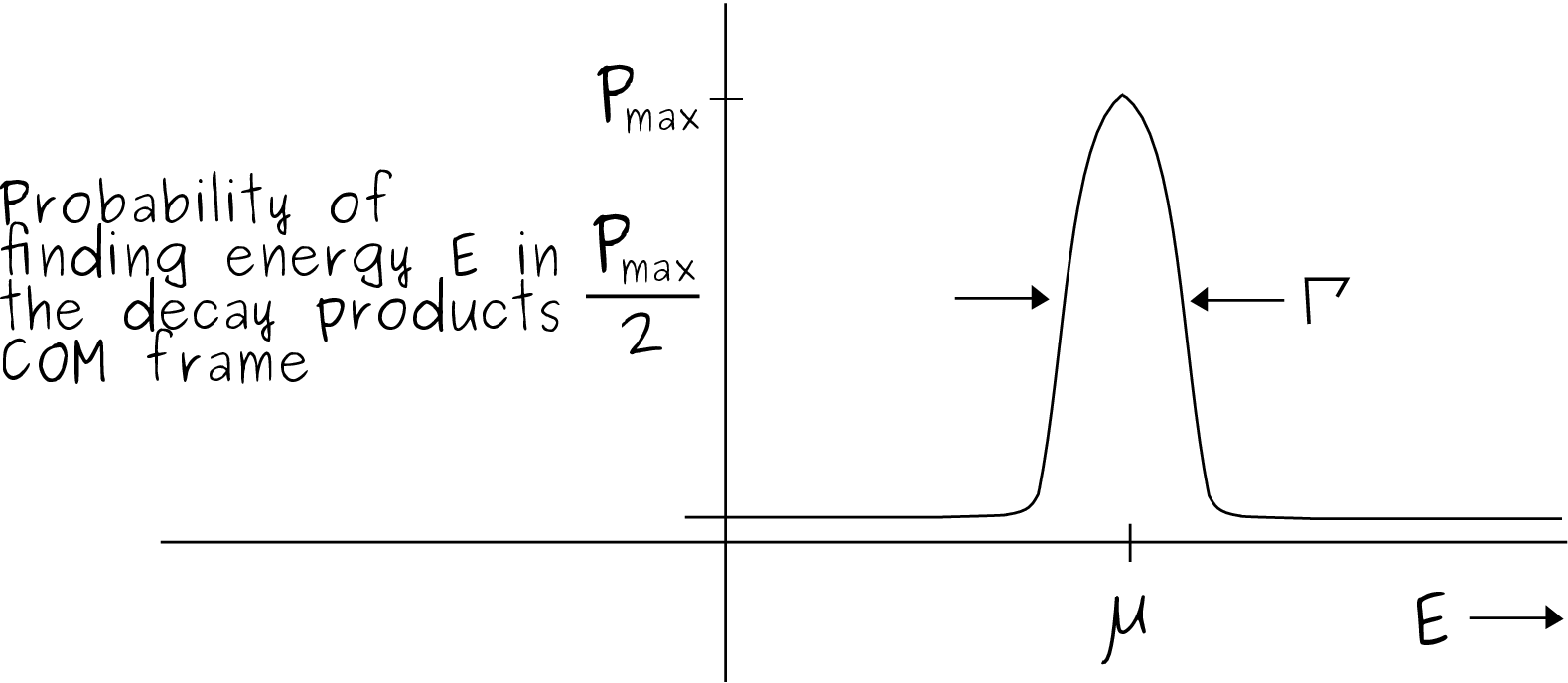}
\end{center}

This is called a Breit-Wigner or Lorentzian line shape, and it is familiar from QM. $\Gamma$ is the full width at half maximum, or decay width. As $\Gamma$ gets smaller, the peak gets narrower and higher.\\

So we have shown that $\mu$ and $\Gamma$, which locate the pole on the second sheet of $D'(k^2)$ are the mass and decay width respectively that an experimenter reports when she says she has found an unstable particle.\\

Experimenters have another way of measuring $\Gamma$, which is purported to be equivalent. They use a clock, and the average lifetime is $\Gamma^{-1}$. We will now do a second thought experiment to show that this second way of determining $\Gamma$ is equivalent.

}{
 \sektion{19}{December 2}
\descriptionnineteen
We have explained ``width" in the phrase ``decay width". With a second thought experiment we'll explain ``decay". In thought experiment 2, we'll produce an unstable particle near the origin and detect it a long ways away at some region near $y$. The region of production cannot be too sharply localized as we are going to make states only with $k^2 \approx \mu^2$.\\
\begin{center}
\includegraphics[scale=0.4]{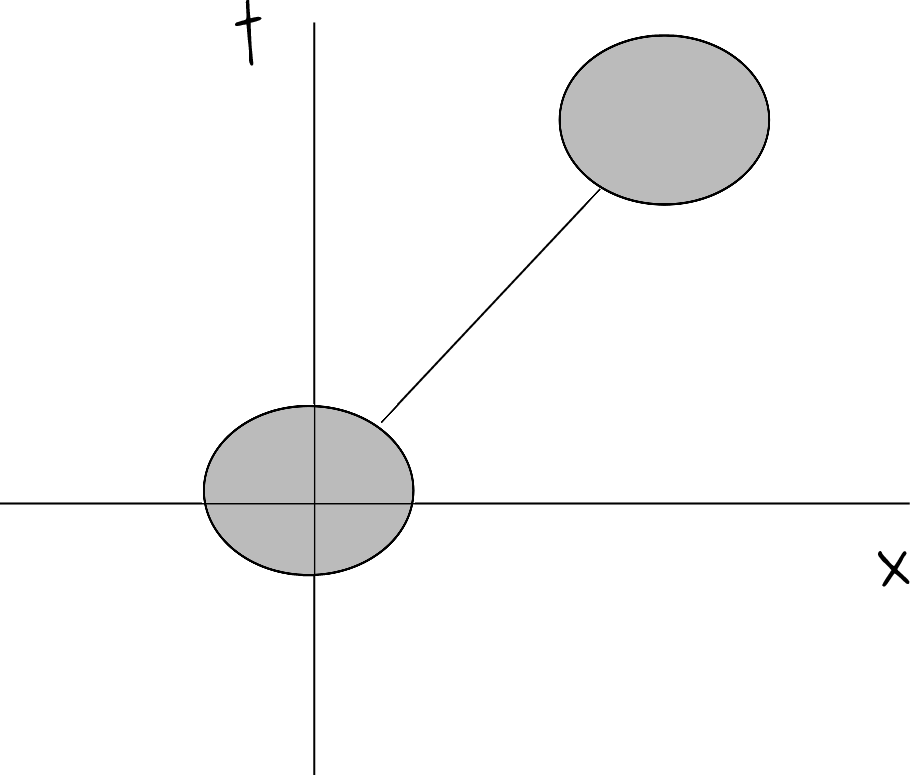}
\end{center}

We'll make the initial state by hitting the vacuum with $\int d^4x \; f(x) \phi'(x)$, $f(x)$ is fairly well localized in position space and its Fourier transform ($\widetilde{f}(k)=\int d^4x\;e^{ik\cdot x}f(x)$) is fairly well localized in momentum space about a momentum \underline{$k$}. Initial state is 
\[ \int d^4 x \; f(x) \phi'(x) |0 \rangle \]
\begin{center}
\includegraphics[scale=0.4]{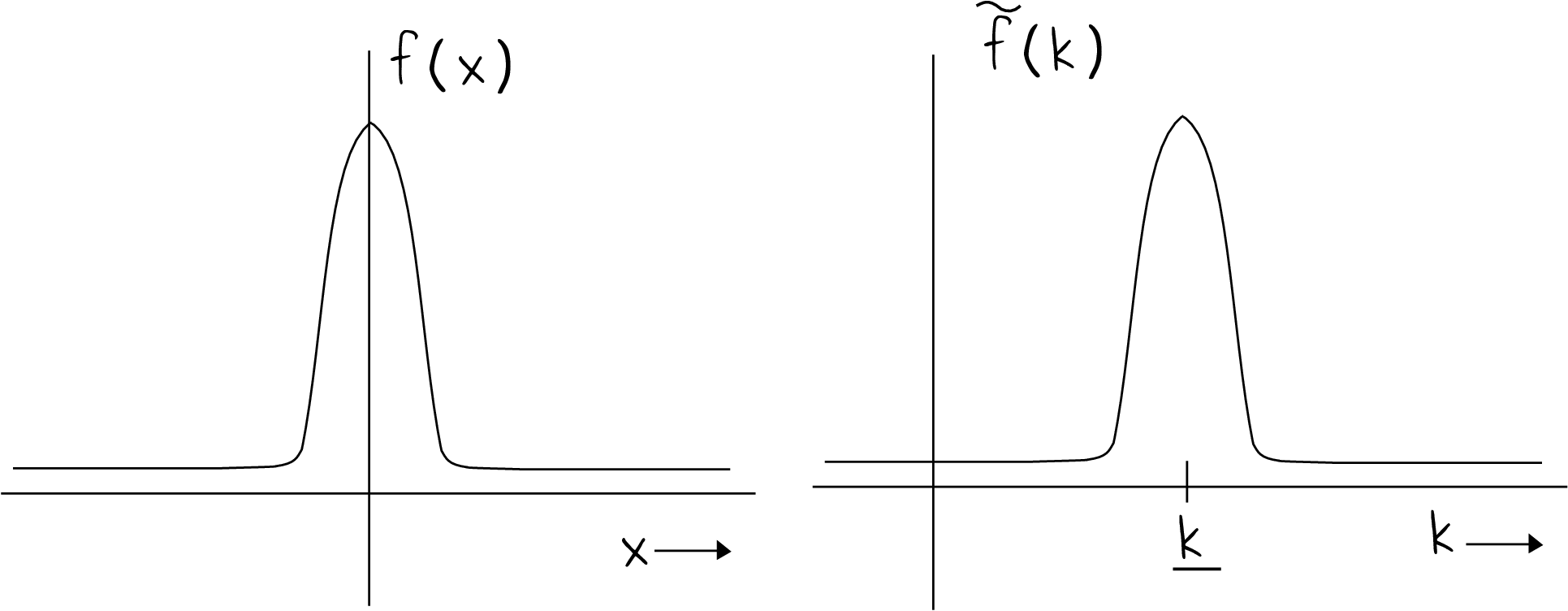}
\end{center}

I'll detect the particle by finding the amplitude that this state becomes the state 
\[ \int d^4x \; g(x-y) \phi'(x) |0 \rangle \]
\begin{center}
\includegraphics[scale=0.45]{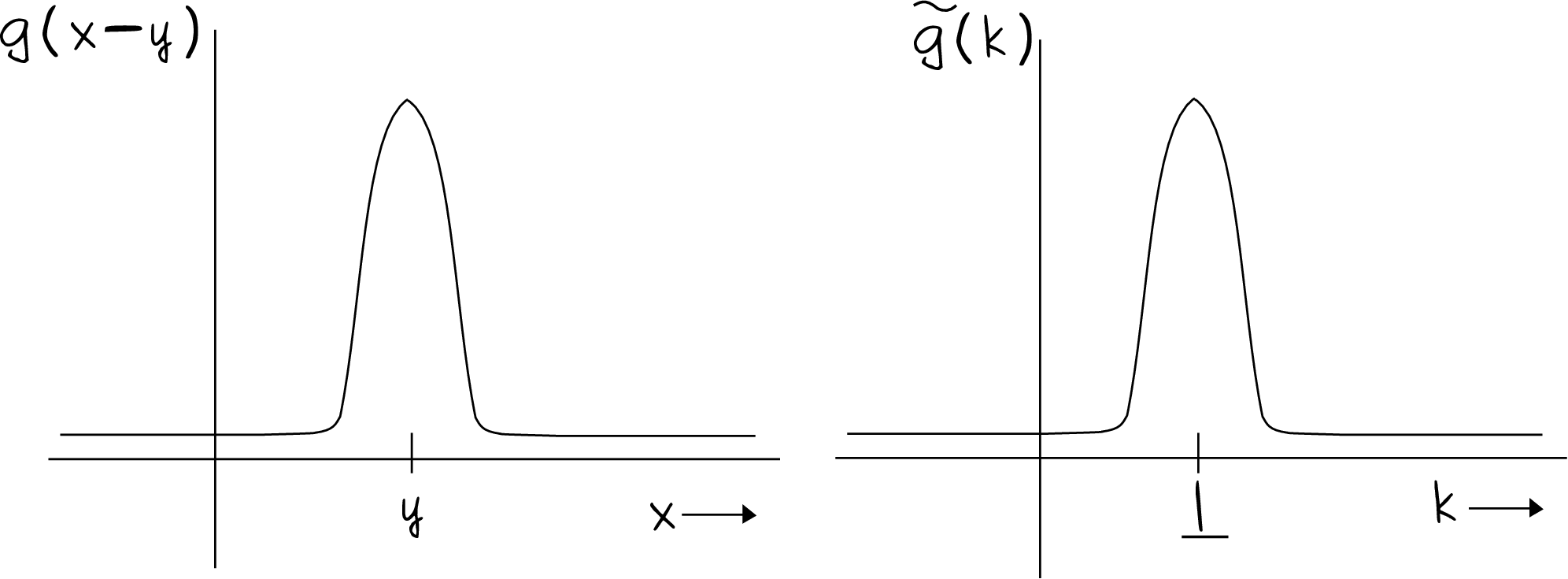}
\end{center}

$g(x-y)$ is concentrated around $y$. $\widetilde{g}(k)$ is concentrated around \underline{$l$}, The amplitude, parameterized by $y$ is 
\[ A(y) = \langle 0 | \int d^4x' g^*(x'-y) \phi'(x') \int d^4 x f(x) \phi'(x) | 0 \rangle \]

If the point $y$ is far later in time than the origin, we can make this a time ordered vacuum expectation value with negligible error.
\begin{align*}
A(y) &= \int d^4 x d^4 x' \; g^* (x'-y)f(x) \langle 0 | T(\phi'(x') \phi'(x) |0 \rangle \\
&= \int d^4 x d^4 x' \int \frac{d^4 k}{(2\pi)^4} \frac{d^4 k'}{(2\pi)^4} e^{ik\cdot (x'-y)} \widetilde{g}^*(k) e^{-ik'\cdot x} \widetilde{f}(k') \langle 0| T(\phi(x') \phi'(x) )|0\rangle \\
&= \int \frac{d^4k}{(2\pi)^4} \frac{d^4 k'}{(2\pi)^4} e^{-ik\cdot y} \widetilde{g}^*(k) \widetilde{f}(k') \!\!\!\!\!\!\!\!\!\!\!\!\!\!\!\!\!\!\!\!\!\!\!\!\!\!\!\! \underbrace{\int d^4 x d^4 x' \; e^{ik\cdot x'} e^{-ik'\cdot x} G'(x',x)}_{\substack{\widetilde{G}\,'(-k,k') \equiv (2\pi)^4 \delta^{(4)} (k -k') D'(k^2)\\ \text{Compare this F.T. convention with the one for $f$ on the previous page}}}\\
&= \int \frac{d^4 k}{(2\pi)^4} e^{-ik\cdot y} \widetilde{g}^*(k) \widetilde{f}(k) D'(k^2) \qquad \text{This integral gives zero, unless \underline{$k$} $\approx$ \underline{$l$}}
\end{align*} 

Recall that $\widetilde{g}(k)$ is concentrated around \underline{$l$} and $\widetilde{f}(k)$ is concentrated around \underline{$k$}, with \underline{$k^2$} $\approx \mu^2$. Assume that $D'(k^2)$ is dominated by a stable or unstable particle pole at $k^2 \approx \mu^2$ and that $\widetilde{f}(k)$ is sufficiently tightly concentrated around \underline{$k$} that we can make the approximation
\[ A(y) = \int \frac{d^4k}{(2\pi)^4} e^{-ik\cdot y} \widetilde{g}^*(k) \widetilde{f}(k) \frac{i}{k^2 - \mu^2 + i \mu \Gamma} \]

(The stable case is handled by taking the limit $\Gamma \rightarrow 0^+$).

We want to analyze this for large $y$, which is difficult because the phase of the exponential is varying rapidly as $k$ changes. Furthermore as $k^2$ increase through $\mu^2$, the phase of the propagator changes rapidly. There is a method for handling these kinds of integrals.\\

\uline{\textbf{Method of stationary phase}}

Given $\displaystyle I = \int dt e^{i\phi(t)} f(t)$, where $\phi$ is a rapidly varying function of $t$ except for a few points where $\frac{d\phi}{dt} =0$, call them $t_i$. (See Whittaker and Watson \uline{Modern Analysis}).
\begin{center}
\includegraphics[scale=0.43]{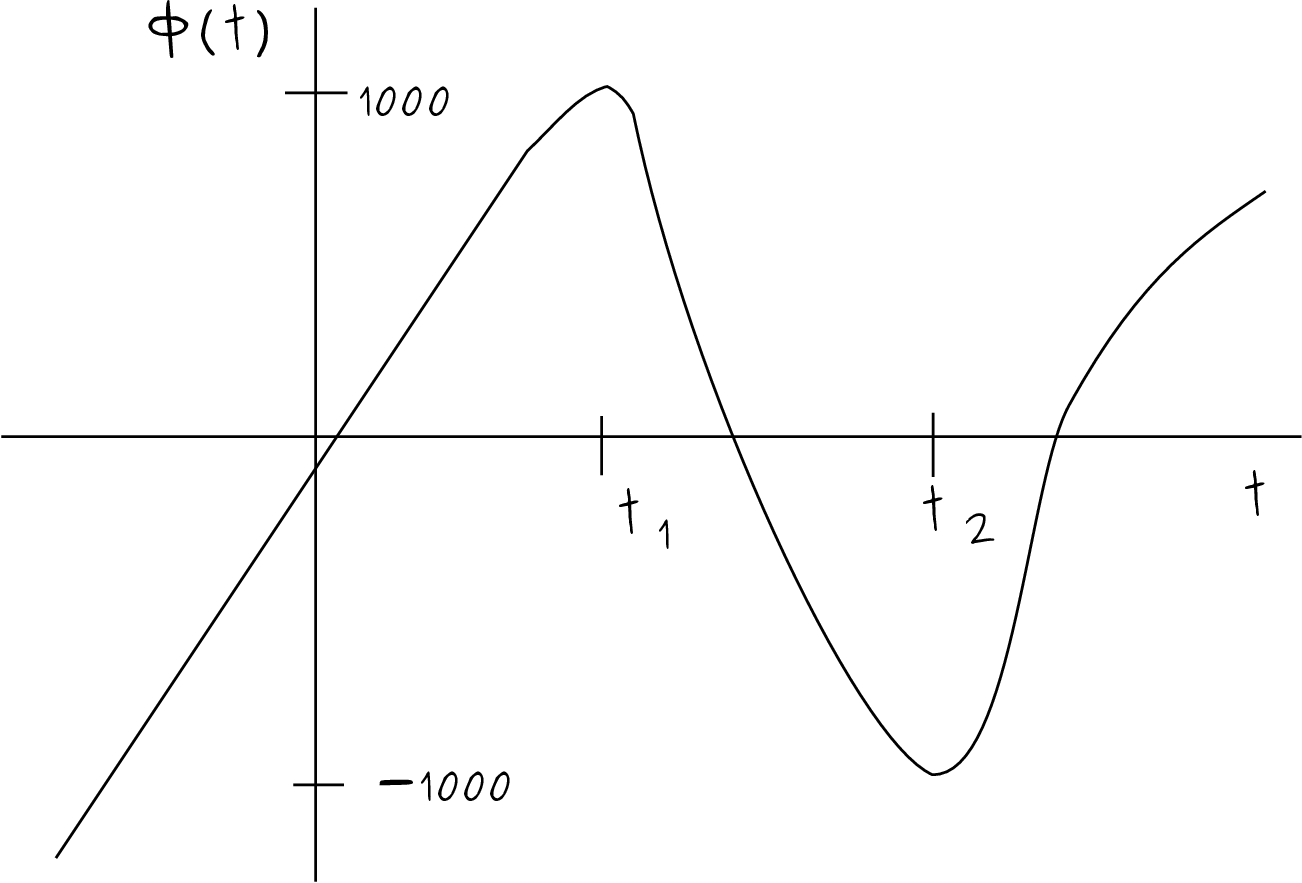}
\end{center}

Then the integral gives nothing almost everywhere, except at those points where the phase stops wildly varying for a moment. At those points $t_i$, the contribution can be approximated by 
\[ I = \sum_i e^{i \phi(t_i)} f(t_i) \underbrace{\int dt \; e^{\frac{i}{2} \phi''(t_i) (t-t_i)^2}}_{\sqrt{\frac{2\pi}{\left|d^2 \phi / dt^2\right| |_{t_i}}} e^{i\pi/4}} \]

We will rewrite $\displaystyle \frac{i}{k^2-\mu^2 + i \mu \Gamma}$ as an exponential so we can use this method. Unfortunately we introduce another integral, but it can also be done by the method of stationary phase.
\[ \frac{i}{k^2-\mu^2 + i \mu \Gamma} = \int ^\infty_0 \frac{ds}{2\mu} e^{i\frac{s}{2\mu} (k^2 -\mu^2 + i \mu \Gamma)} \]

Assume the variation in phase of $\widetilde{g}^*(k)\widetilde{f}(k)$ is slow compared to the variation in the exponential, or take $g=f$.
\[ A(y) = \int^\infty_0 d \Big( \frac{s}{2\mu} \Big) \int \frac{d^4k}{(2\pi)^4} e^{-ik\cdot y + i \frac{s}{2\mu} (k^2 - \mu^2 + i\mu \Gamma)} \widetilde{g}^*(k) \widetilde{f}(k) \]

To do the $k$ integration by stationary phase, set 
\[ 0 = \frac{\partial}{\partial k_\mu} \Big( -k\cdot y + \frac{s}{2\mu} (k^2 -\mu^2) \Big) = -y^\mu + \frac{s}{\mu} k^\mu \]

Thus the only stationary phase point in the $k$ integrations is at $\displaystyle k_0 = \frac{\mu}{s}y$, $\displaystyle \left| \frac{d^2 \phi}{d k^{\substack{\mu \;\;\;\;\;\; 2 \\ \text{\tiny No sum}}}} \right| \Bigg|_{k_0} = \frac{s}{\mu}$
\[ A(y) \approx - \int^\infty_0 \frac{ds}{2\mu} \frac{1}{(2\pi)^2} \Big(\frac{\mu}{s}\Big)^2 \widetilde{g}^* \Big(\frac{\mu}{s}y \Big) \widetilde{f} \Big( \frac{\mu}{s} y \Big) e^{i\frac{s}{2} (-\mu + i \Gamma)} e^{-i \frac{\mu}{2s} y^2} \]

To do the $s$ integration by stationary phase, set 
\[ 0 = \frac{\partial}{\partial s} \Big( - \frac{s}{2} \mu - \frac{\mu}{2s}y^2 \Big) \]
\[ s = \sqrt{y^2} \]

Note that there is no stationary phase point if $y^2$ is spacelike. As $y \rightarrow \infty$, $y^2 < 0$, there is no probability that a particle will be detected. We have recovered causality. \\

Call $ \sqrt{y^2}$ ``$s_0$", then $\displaystyle \left| \frac{d^2 \phi}{ds^2} \right|_{s_0} = \frac{\mu}{s_0}$, and 
\[ A(y) = - e^{i\pi/4} \sqrt{\frac{2\pi s_0}{\mu}} \frac{1}{2\mu} \frac{1}{(2\pi)^2} \left( \frac{\mu}{s_0} \right)^2 \widetilde{g}^* \left( \frac{\mu}{s_0}y \right) \widetilde{f}\left( \frac{\mu}{s_0}y \right) e^{-i\mu s_0} e^{-\frac{\Gamma s_0}{2}} \]

These factors can be understood. Suppose you \uline{classically} propagate a stable particle with velocity $\displaystyle v_\mu = \frac{k_\mu}{\mu}$. In a proper time $s$, it will arrive at a point $y_\mu = v_\mu s$, where $v_\mu^2 =1 \Rightarrow s =\sqrt{y^2}$. This is just classical kinematics, but you see we have recovered it in the limit of large $y$ from quantum field theory. The conditions of stationary phase are the equations of classical kinematics.\\

The factor $e^{-i\mu s_0}$ is just $e^{-iEt}$ of quantum mechanics that has come out in a Lorentz invariant generalization. There is a factor $s_0^{-3/2}$. That is there because if you wait long enough, every packet because of an initial uncertainty in velocity is spreading out in all directions linearly with time. In 3-D this means that the probability density at the center of the packet goes down like $\displaystyle \frac{1}{t^3}$. So the amplitude falls like $\displaystyle \frac{1}{t^{3/2}}$. The Lorentz invariant generalization of this is that the amplitude at the center of the packet falls like $\displaystyle \frac{1}{\text{(proper time)}^{3/2}}$. \\

Finally there is the unstable case. We have $\displaystyle e^{-\Gamma s_0/2}$ in the amplitude, which means that the probability has a factor $e^{-\Gamma s_0}$. They are indeed decaying and again we have gotten the Lorentz invariant generalization of $e^{-\Gamma t}$; $\Gamma$ is the decay rate per unit proper time.\\

>> This talk of ``correct generalization" must be made more precise. It should be possible to do the computation in the frame where $y = (\sqrt{y^2}, \vec{0})$. However our stationary phase computations are not justified in that case ??

\begin{center}
\uline{\textbf{``WHERE IT BEGINS AGAIN"}}\\
\end{center}

If we had been proceeding logically, starting from first principles, making the most general statements about relativistic quantum field theory we could and, only after exhausting those, made simplifying assumptions and approximations, we would have begun the course by listing all possible field transformation laws, then we would have constructed all possible quadratic Lagrangians, that is all possible combinations that are at most quadratic in the fields and transform as scalars under the Lorentz group. Then we would do canonical quantization and in the process discard many of the Lagrangians because of one or another inconsistency, like the Hamiltonian not being bounded below. At this point we would have all possible free particle theories, and we could start adding interactions, higher order polynomials, to the Lagrangian. The actual order we have been doing this course, is to spend a lot of time studying relativistic invariants made up of the simplest kind of fields, scalar fields. Under Lorentz transformations a set of scalar fields transform like 
\[ \phi^a (x) \quad \quad \quad a = 1,\ldots,n \quad \quad \quad \quad \Lambda \in \text{SO}(3,1) \]
\[ \Lambda:\phi^a (x) \rightarrow \phi^a(\Lambda^{-1}x) \]

The only Lorentz scalars you can construct have derivatives, $0,2,4,\ldots$ of them, which act on the scalars and are completely contracted with $g\mn$ or $\epsilon\mnls$.
\[ \Box \phi_1 \, \pmu \phi_2 \pmuu \phi_3 \]
\[ \epsilon\mnls \pmu \phi_1 \pnu \phi_2 \plu \phi_3 \psu \phi_4 \]

The list of possible quadratic Lagrangians is pretty short, and we have gone a long ways toward exploring them. In fact we have even gone a long ways toward studying the total list of interacting scalar fields since the renormalization vs infinities arguments pretty well rule out Lagrangian that are more than quartic in the fields. We haven't exhausted the study of scalars, but we are now going to go on to\\

\textbf{DISCOVERING ALL POSSIBLE LORENTZ TRANSFORMATION LAWS OF FIELDS} \\

We'll phrase the analysis in a quantum language, but all that we are about to do can be carried through classically. Assume we have a \uline{finite} number of fields,
\[ \phi^a(x) \quad \quad a= 1,\ldots, N \]

Let $\Lambda$ denote an abstract element of $\text{SO}(3,1)$, the part of the Lorentz group connected to the identity. More concretely, $\Lambda$ can also be thought of as a $4\times 4$ matrix, that preserves the metric, $\Lambda^\mu_\alpha \Lambda^\nu_\beta g\ab = g\mn$, is proper, $\text{det}(\Lambda)=1$, and is orthochronous, $\Lambda^0_0 > 0$. For each $\Lambda$ there is a unitary transformation
\[ U(\Lambda)^\dagger \phi^a(x) U(\Lambda) = D^a{}_b (\Lambda) \phi^b (\Lambda^{-1} x) \quad \quad (\Sigma_b \text{ implied}) \]

For each $\Lambda$ there is some $N\times N$ matrix $D^a{}_{b}(\Lambda)$ that gives a \uline{linear} relationship between the complete set of commuting observables at $\Lambda^{-1} x$ and those at $x$, for all $x$. If we think of the $D$'s as matrices and the $\phi$'s as column vectors, we can write 
\[ U(\Lambda)^\dagger \phi(x) U(\Lambda) = D(\Lambda) \phi(\Lambda^{-1} x) \]

($D$'s are $N \times N$ matrices, ``The dimension of $D$ is $N$")\\

A property of the $U$'s reflects itself in the $D$'s (It only takes a couple of lines to prove this)
\[ U(\Lambda_1) U(\Lambda_2) = U(\Lambda_1 \Lambda_2) \Longrightarrow D(\Lambda_1 \Lambda_2) = D(\Lambda_1) D(\Lambda_2) \]

Also $U(1) =1 \Longrightarrow D(1) =1 $ and from these two properties of the $D$'s, $D(\Lambda^{-1}) = D(\Lambda )^{-1}$. It seems that the $D$ matrices obey all the properties of the group, and you might think from any set of $D$'s you could reconstruct the group. You can't though. Many elements of the group can map into a single $D$ matrix, that is it is possible that $D(\Lambda) = D(\Lambda ')$ while $\Lambda \neq \Lambda '$. The trivial prototypical example is (all fields are scalar) $D(\Lambda) =1$ for all $\Lambda$. A set of $D$'s that obey the group laws is called a representation. If $D(\Lambda) = D(\Lambda') \Longrightarrow \Lambda = \Lambda'$, the representation is ``faithful". \\

\begin{center} 
\textbf{A person who is tired of group theory is tired of life.}
\end{center} 

An additional complication that we are only going to consider in a very cavalier way is the possibility which is impossible to rule out in quantum mechanics that 
\[ U(\Lambda_1) U(\Lambda_2) \neq U(\Lambda_1 \Lambda_2), \quad U(\Lambda_1) U(\Lambda_2) = U(\Lambda_1 \Lambda_2) e^{i\phi(\Lambda_1,\Lambda_2)} \]

The product law is not exactly true in general. It only need be true up to a phase. It turns out that for $\SO (3)$ and $\SO (3,1)$, the phases can be removed except in representations called spinor representations where a rotation by $\pi$ about any axis $\vec{e}$ followed by another rotation by $\pi$ about $\vec{e}$ gives 
\[ U(\!\!\!\!\!\!\!\!\!\!\!\!\!\!\!\!\!\!\!\!\!\!\!\!\!\!\!\!\underbrace{\vec{e} \pi}_{\substack{\text{notation for the unitary}\\\text{operator that
rotates around } \vec{e} \text{ by } \pi}}\!\!\!\!\!\!\!\!\!\!\!\!\!\!\!\!\!\!\!\!\!\!\!\!\!\!\!\!\!) U(\vec{e} \pi) = -\mathds{1} \]

Two rotations by $\pi$ are physically equivalent to no rotation at all, so you would expect to have gotten $1$. (If you want to study this the only good reference I know of is Bargmann, V., ``On Unitary Ray Representations of Continuous Groups," Annals of Mathematics, Vol. 59,(1954) p.1, and it is in the basement of Cabot. I am not recommending this however. We will get all the right results with much less effort by being cavalier and lucky.) The possibility that the product of the unitary operators is only the unitary operator of the product up to a phase reflects itself identically in the composition law for the $D$'s. (Another thing to note about the $D$'s is that they are not in general unitary. Try as you may, you cannot use the unitary of the $U$'s to prove unitary of the $D$'s).\\

Our task of finding all possible Lorentz transformation laws of fields has been reduced to the task of making a catalog of all finite dimensional representations of $\SO(3,1)$.\\

\begin{center}
\textbf{Shortening the catalog of finite dimensional representations of $\SO (3,1)$}
\end{center}

Suppose I have a representation $D(\Lambda)$. I can make a new representation of $\SO(3,1)$, that obeys all three conditions, by defining 
\[ D'(\Lambda) = S D(\Lambda) S^{-1} \]

where $S$ does not vary with $\Lambda$, it is some definite invertible matrix. Equivalent to this though is just a redefinition of the basis fields. If the fields transform as $\Lambda:\phi(x) \rightarrow D(\Lambda) \phi(\Lambda^{-1} x)$, the new basis $\phi'(x) = S \phi(x)$ transforms as $\Lambda: \phi'(x) \rightarrow D'(\Lambda) \phi'(\Lambda^{-1} x)$. This is not worth listing as a new kind of field theory. Define two representations $D$ and $D'$ to be equivalent if there exists an invertible $S$ such that 
\[ D'(\Lambda ) = S D(\Lambda ) S^{-1} \quad \quad \text{for all } \Lambda \]

and write $ D' \sim D$. If there is no such $S$, $D$ and $D'$ are inequivalent $D' \not \sim D$. Our catalog will only include inequivalent representations of $\SO(3,1)$, one representation from each ``equivalence class."\\

I'll give a useful example. The Lorentz transformation of parity is not in $\SO (3,1)$, because it is not connected to the identity, it has determinant $-1$. (You can think of parity, $P$, abstractly or as a $4 \times 4$ matrix.) For every $\Lambda \in \SO (3,1)$, I can obtain another element of $\SO (3,1)$, $\Lambda_p \equiv P \Lambda P$ ($P=P^{-1}$). The association is one to one, and it preserves the group multiplication law. $\Lambda_p \Lambda_p' = (\Lambda \Lambda')_p$. These properties make it an ``automorphism". With this automorphism of $\SO(3,1)$, I can construct a new representation of the group from any given representation. Starting with a representation $D$, define
\[ D_p(\Lambda) = D(\Lambda_p) \]

This new rep obeys all three conditions. 

It may or may not be true that $D_p \sim D$. 

To make the example more concrete, let's look at what the parity automorphism does to one of the representations of $\SO (3,1)$ we all know and love, say the two index tensor. 
\[ \Lambda: T\mn \rightarrow \Lambda^\mu{}_{\alpha} \Lambda^\nu{}_{\beta} T\ab \qquad \text{(Transformation law of a tensor)} \]

We read off $D\mn{}_{\alpha \beta} (\Lambda) = \Lambda^\mu{}_{\alpha} \Lambda^\nu{}_{\beta}$. The $D$'s in the representation induced by the parity automorphisms are 
\[ D_{p}{}^{\mu\nu}{}_{\alpha\beta} (\Lambda) = D\mn{}_{\alpha \beta} (\Lambda_p) = (P \Lambda P)^\mu{}_{\alpha} (P\Lambda P)^\nu{}_{\beta} \]

Now we expect that we have not constructed an inequivalent representation of $\SO (3,1)$ this way. After all what parity is really doing is just turning $T^{00} \rightarrow T^{00}$, $T^{ij} \rightarrow T^{ij}$ and $T^{i0} \rightarrow -T^{i0}$, $T^{0j} \rightarrow -T^{0j}$. If it is interpretable as a change of basis, we must be able to find the similarity transformation relating the two representations. Let's massage the expression for $D_p$ until we find it.
\begin{align*}
D_{p}{}^{\mu\nu}{}_{\alpha\beta} &= (P \Lambda P)^\mu{}_{\alpha} (P \Lambda P)^\nu{}_{\beta} \\
&= P^\mu{}_{\sigma} \Lambda^\sigma{}_{\tau} P^\tau{}_{\alpha} P^\nu{}_{\phi} \Lambda^\phi{}_{\psi} P^\psi{}_{\beta} \\
&= P^\mu{}_{\sigma} P^\nu{}_{\phi} \Lambda^\sigma{}_{\tau} \Lambda^\phi{}_{\psi} P^\tau{}_{\alpha} P^\psi{}_{\beta} \quad\quad S\mn{}_{\sigma\phi}\equiv P^\mu{}_\sigma P^\nu{}_\phi\\
&= S\mn{}_{\sigma\phi} D(\Lambda)^{\sigma\phi}{}_{\tau \psi} S^{-1}{}^{\tau \psi}{}_{\alpha\beta}\quad\quad\text{ i.e.}D_p(\Lambda)=SD(\Lambda)S^{-1}
\end{align*} 

(you can do the vector case. I did the two index tensor because it is a little less trivial)\\

\begin{center}
\textbf{Shortening the catalog of finite dimensional inequivalent representations of $\SO(3,1)$}
\end{center}

Suppose someone has two theories, one with a set of fields
\[ \phi_{1a} \quad \quad \quad a = 1, \ldots, N_1 \]

transforming as 
\[ \Lambda: \phi_1 (x) \rightarrow D^{(1)}(\Lambda) \phi_1 (\Lambda^{-1}x) \]

and the other with a set of fields 
\[ \phi_{2a} \quad \quad \quad a = 1, \ldots, N_2 \]

transforming as 
\[ \Lambda: \phi_2(x) \rightarrow D^{(2)}(\Lambda) \phi_2 (\Lambda^{-1} x) \]

Now this person comes to you and says, I have a new theory with $N_1+N_2$ fields, which he has assembled into a vector $\phi$, and they transform like 
\[ \Lambda: \phi (x) \rightarrow D(\Lambda) \phi (\Lambda^{-1}x) \]
\[ D(\Lambda) = \begin{pmatrix} D^{(1)} (\Lambda) & 0 \\ 0 & D^{(2)}(\Lambda) \end{pmatrix} \]

For example, theory 1 could contain a vector and theory 2 could contain a scalar. The new theory would have to be five dimensional. This is hardly a big breakthrough. It is such a simple extension of what was previously known that it is not worth including in our catalog. Define this representation $D$ to be the ``direct sum" of $D^{(1)}$ and $D^{(2)}$, $D= D^{(1)} \oplus D^{(2)}$. It has dimension $N_1 + N_2$. The $D$'s are in block diagonal form. Call a representation ``reducible" if it is equivalent to a direct sum, otherwise, call it ``irreducible". \\

Our task is to build the remarkably shorter catalog, the catalog of finite dimensional inequivalent irreducible representations of $\SO(3,1)$. \\

By a wonderful fluke, peculiar to living in $3+1$ dimensions, the representations of $\SO(3,1)$ can be rapidly obtained from the representations of $\SO(3)$. You know all about the representations of $\SO(3)$ from undergraduate QM, so we will be able to wrap the catalog up by the end of next lecture. If we lived in $9+1$ dimensions, there would be no quick reduction of the problem of finding the representations of $\SO(9,1)$ to the problem of finding
the representations of $\SO(9)$, which $9+1$ dimensional students solve as undergraduates. We will review the representations of $\SO(3)$ just enough to refresh your memory. They are carefully constructed in a few pages \uline{in a way that generalizes to other groups} beginning on page 16 of Howard Georgi's \uline{Lie Algebras in Particle Physics}. Actually what is constructed there are the representations of the Lie algebra of $\SO(3)$ rather than the representations of the Lie Group but you'll see that is what we want.\\

\begin{center}
\textbf{THE FINITE DIMENSIONAL INEQUIVALENT IRREDUCIBLE REPRESENTATIONS OF $\SO(3)$}
\end{center}

An element $R$ of $\SO (3)$ can be thought of abstractly or as a $3 \times 3$ matrix. It is specified by giving an axis of rotation, and an angle of rotation about the axis, $\vec{e}$ and $\theta$. Let's standardize the vector that defines the axis by taking it to be a unit vector. The product $\vec{e} \, \theta$ defines the rotation completely, $R(\vec{e} \, \theta)$. Its length gives the amount of rotation in the counterclockwise direction when looking down toward the tail of the vector $\vec{e}$.
\begin{center}
\includegraphics[scale=0.4]{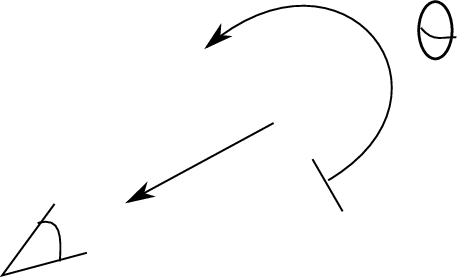}
\end{center}

If we let $\theta$ take on any value from $0$ to $2 \pi$ we have included twice every element of $\SO (3)$. The reason is that a rotation about $\vec{e}$ by an angle $\theta$ is exactly the same as a rotation about $-\vec{e}$ by an angle $2\pi - \theta$. So we'll restrict $\theta \in [0,\pi]$. This still includes twice the rotations by $\pi$ because
\[ R(\vec{e} \pi) = R (-\vec{e} \pi) \]

The group $\SO (3)$ is topologically like the \uline{ball} (not the sphere $S^2$, the ball $B^3$) in three space of radius $\pi$, except antipodal points on the surface of the ball are identified.\\

Just so you have some impressive jargon at your disposal, the ball just described, with the identification, is topologically like the projective $3$ sphere.\footnote{[BGC note: The standard name for these spaces are real projective 3-\uline{space} $\mathbb{RP}^3$ and real projective 2-space or the real projective plane $\mathbb{RP}^2$.]} I can explain that in one lower dimension where I can visualize it. The projective $2$ sphere is $S^2$ with each pair of antipodal points identified. Each point on the whole bottom half of the sphere below the equator has a point it is identified with in the half of the sphere above the equator. Chuck the whole bottom half of the sphere leaving the top half and the equator. Each point on the equator is still identified with one other point on the equator. But it is clear that
\begin{center}
\includegraphics[scale=0.4]{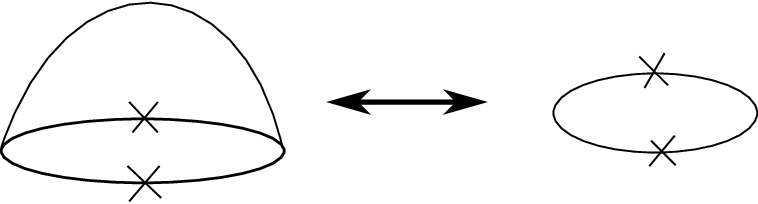}
\end{center}

(x's represents a pair of identified points) Just flatten the sphere out into the disk and this disk is the ``ball" in two space, with antipodal points on the ball identified.\\

\begin{center}
\textbf{A method of infinitesimal analysis}
\end{center}
\[ \frac{d}{d\theta} R(\vec{e} \, \theta) |_{\theta =0} \equiv -i \vec{e} \cdot \vec{J} \]

This expression defines $\vec{J}$, a set of three $3 \times 3$ matrices if you think of $R$ as a $3 \times 3$ matrix, and something more abstract if you think of $R$ more abstractly. The three matrices are called the Lie algebra of $\SO (3)$. There are three because $\SO(3)$ is a three parameter group. ($\SO (n)$ is an $\frac{n(n-1)}{2}$ parameter group). How do you see that this derivative is linear in $\vec{e}$? Recall the picture of $\SO (3)$ as a ball. Assume all the differentiability you desire. What this derivative is is a directional derivative in the direction $\vec{e}$ at the center of the ball and $-i \vec{J}$ is the gradient.\\

Someplace in this analysis we have to put in the properties of $\SO(3)$. Rather than making mathematical statements about $3 \times 3$ matrices, we'll put in two properties physically that are enough to specify the group.
\begin{itemize}
\item[(1).] $R(\vec{e} \, \theta') R(\vec{e} \, \theta) = R(\vec{e} \, [\theta + \theta']) $
\item[(2).] $R'^{-1}R(\vec{e} \, \theta) R' = R (R'^{-1} \vec{e} \, \theta ) $
\end{itemize}

These are physically motivated statements about properties of rotations.

Property (1) is obvious. How to see property (2)? A way of characterizing the axis of rotation is to say it is the axis such that any vector parallel to this axis is unchanged by the rotation. $R'^{-1} \vec{e}$ is unchanged by the RHS of (2). It is also unchanged by the LHS because $R'$ turns it into $\vec{e}$, which is unchanged by $R(\vec{e} \, \theta)$ and $R'^{-1}$ turns it back into $R'^{-1}\vec{e}$. Thus the LHS is a rotation by $\theta$ about $R'^{-1} \vec{e}$. We won't actually use all the information contained in (1) and (2). We will only use them in infinitesimal form, that is, we'll take derivatives with respect to $\theta$ and $\theta'$. This loss of information decreases the restrictions on the form of the representations, and is the reason we pick up representation up to a phase, spinors, even though our formalism hasn't explicitly included them.\\

Applying the infinitesimal analysis to a representation. Take 
\[ \frac{d}{d\theta} D(R(\vec{e} \, \theta)) |_{\theta=0} \equiv - i \vec{e} \cdot \vec{J} \]

(The $\vec{J}$'s are the ``generators" of this representation)\\

This is a very concrete equation. The $D$'s are some representation of dimension $N$ and the three $\vec{J}$'s are some $N \times N$ matrices. (1) can be made into a statement about representations.
\[ D(R(\vec{e} \, \theta')) D(R(\vec{e} \, \theta)) = D(R(\vec{e} \, [ \theta + \theta' ] )) \]

Take $\cfrac{d}{d\theta'}$, and set $\theta' = 0$ to get (on the RHS $\cfrac{d}{d\theta} = \cfrac{d}{d\theta'}$)
\[ -i\vec{e} \cdot \vec{J} \; D(R(\vec{e} \, \theta)) = \cfrac{d}{d\theta} D(R(\vec{e} \, \theta)) \]

This is a simple differential equation. The solution with the boundary condition $D(R(\vec{e}0)) = \mathds{1}$ is 
\[ D(R(\vec{e} \, \theta)) = e^{-i\vec{e}\cdot \vec{J} \theta} \]

(The $\vec{J}$'s ``generate the representation")\\

Now we can transfer our definitions about inequivalence and irreducibility to the generators. If the rep $D$ is generated by $\vec{J}$ and the representation $D'$ by $\vec{J}\,'$ and $D \sim D'$ that is $D(R) = S D'(R) S^{-1}$ for some $S$ and all $R$ then 
\[ S e^{-i \vec{e}\cdot \vec{J}\,' \theta } S^{-1} = e^{-i \vec{e} \cdot \vec{J} \theta } \Longleftrightarrow S \vec{J}\,' S^{-1} = \vec{J} \quad \text{i.e.} \vec{J} \sim \vec{J}\,' \]

Equivalence of two reps is the same as equivalence of their generators. What about irreducibility? If a representation is reducible, then it is equivalent to a representation that has block diagonal form for all rotations. So $\vec{e}\cdot \vec{J} = i \frac{d}{d\theta} D(R(\vec{e} \, \theta)) |_{\theta=0}$ has block diagonal form. Reducibility of a rep is the same as reducibility of its generators. I can even phrase this a little more strongly.
\[ \text{If }D(R) = \begin{pmatrix} D^{(1)}(R) & 0 \\ 0 & D^{(2)}(R) \end{pmatrix} \text{ then } \vec{J} = \begin{pmatrix} \vec{J}^{(1)} & 0 \\ 0 & \vec{J}^{(2)} \end{pmatrix} \]

and if $ D \sim D^{(1)} \oplus D^{(2)}$ then $\vec{J} \sim \vec{J}^{(1)} \oplus \vec{J}^{(2)}$\\

The task of finding inequivalent irreducible finite dimensional representations $D$ of $\SO (3)$ has been reduced to the task of finding inequivalent irreducible sets of $3$ matrices $\vec{J}$, whose properties we are elucidating.\\

As a statement about representations, (2) is 
\[ D(R'^{-1}) D (R(\vec{e} \, \theta)) D(R') = D(R(R'^{-1} \vec{e} \, \theta)) \]

Take $i\frac{d}{d\theta}$ at $\theta = 0$ to get 
\[ D (R'^{-1}) \; \vec{e} \cdot \vec{J} \; D(R') = (R'^{-1} \vec{e})\cdot \vec{J} = \vec{e} \cdot R'\vec{J} \]

(the last equality using the fact rotation matrices preserve scalar products)\\

Dropping the primes, and using the fact that $\vec{e}$ is an arbitrary unit vector, this says 
\[ D(R^{-1}) \vec{J} D(R) = R\vec{J} \]

which is the statement that the generator of the rotations, the 3 $\vec{J}$'s ($N\times N$ matrices) transform like a vector.\\

You can go further by writing $R$, parametrizing $R$ by $\vec{e}$ and $\theta$ (it shouldn't cause confusion to use these variables again). What we have found is 
\[ e^{i\vec{J}\cdot \vec{e} \, \theta} \vec{J} e^{- i \vec{e}\cdot \vec{J} \theta} = \vec{J} + \theta \vec{e} \times \vec{J} + \mathcal{O}(\theta^2) \]

where on the RHS I have used a physical property of the rotation group, that for small $\theta$, a rotation matrix acting on a vector changes it by 
\[ \theta \vec{e} \times \vec{V} + \mathcal{O}(\theta^2) \]

(use the RH rule to make sure this agrees with the convention and picture before). Take $-i \frac{d}{d\theta}$ of this equation to get 
\[ [\vec{J} \cdot \vec{e}, \vec{J}] = - i\vec{e} \times \vec{J} \]

Take $\vec{e} = \widehat{e}_x$ and look at $y$ component to get 
\[ [J_x, J_y] = iJ_z \]

also can get
\[ [J_i, J_j] = i\epsilon_{ijk} J_k \]

The generators $\vec{J}$ form a representation of the Lie algebra of the group. They satisfy the same commutation relations.\\

\begin{center}
\textbf{Facts about finite dimensional inequivalent irreducible representation of the Lie algebra of the rotation group}
\end{center}

A complete set of them is the set
\[ \vec{J}^{(s)} \qquad s=0,\frac{1}{2},1,\frac{3}{2}, \cdots \qquad \text{``spin" $s$} \]
\[ \vec{J}^{(0)} = \vec{0} \quad \vec{J}^{(\frac{1}{2})} = \frac{\vec{\sigma}}{2} \qquad (J_i^{(1)})^j{}_{k} = - i\epsilon_{ijk} \]
\[ J_z^{(s)}|m\rangle = m|\!\!\!\!\underbrace{m}_{\text{no sum}} \!\!\!\!\rangle \qquad m =-s,-s+1,-s+2,\ldots,s-2,s-1,s \]

in a usual basis for the $2s+1$ dimensional vector space the $\vec{J}^{(s)}$ act on. \\

The $\vec{J}^{(s)}$ are hermitian. Every representation is equivalent to a hermitian representation. \\

\begin{center}
\textbf{Facts about finite dimensional inequivalent irreducible representation up to a phase of the rotation group, $D^{(s)}(R(\vec{e} \, \theta)) = e^{-i\vec{e}\cdot \vec{J}^{(s)}\theta}$}
\end{center}

\begin{itemize}
\item[(1).] The reps of the Lie algebra just listed not only generate the reps of the rotation group, they generate the reps up to a phase. The integer $s$ are representations. The half integers $s$ are reps up to a phase. More specifically, they are double valued 
\[ D^{(s)}(R(2\pi \vec{e})) = (-1)^{2s} \mathds{1} \]

\item[(2).] $\text{dim} \; D^{(s)} = 2s+1 $

\item[(3).] The hermiticity of the $\vec{J}^{(s)}$ implies the $D^{(s)}$ are unitary ($D^{(s)}(R^{-1}) = [D^{(s)}(R)]^{-1} = [D^{(s)}(R)]^\dagger $). Every representation of the rotation group is equivalent to a unitary representation of the rotation group. Of course, in dumb bases, like $\vec{i}, \vec{j}, 7\vec{k}$ for the space $D^{(1)}$ acts on, the $D$'s preserve $x^2+y^2+\frac{1}{49} z^2$, and they are not unitary.

\item[(4).] If I have any representation of any group $G$, $g\in G$, $D^a{}_{b}(g)$, I can define a new representation, $g\in G$, $D^{*a}{}_{b} (g) = (D^a{}_{b} (g) )^* $ (no matrix transpose). \\

This new representation has the same dimension as the original representation. It may or may not be equivalent to the representation you obtained it from. If $D$ is irreducible, the new rep $D^*$ is irreducible. In $\SO (3)$ since there is only one inequivalent irreducible representation of a given dimension, a rep must be equivalent to its associated complex conjugate rep. Furthermore
\begin{equation}\label{eq:19-page17}
D^{(s)} \sim D^{(s)*} \quad \text{ and } \quad D^{(s)}(R(\vec{e} \, \theta)) = e^{-i \vec{e}\cdot \vec{J} \theta}
\end{equation}

implies $\vec{J}^{(s)} \sim - \vec{J}^{(s)*} $. The $-$ sign is present because of the $i$ in the exponential.

\item[(5).] Direct product of representations. A new notation, the notation of kets and linear operators, ($J_z^{(s)}|m\rangle = m |m\rangle $, no sum implied) appeared out of the blue. It is intuitively clear that a matrix of numbers acting on a column vector, can be reinterpreted as a linear operator acting on a vector space, but I would like to make the connection precise.\\
\end{itemize}

Suppose I have an $n$ dimensional representation, that is, an $n \times n$ matrix for every element of a group, $D^i{}_{j}(g)$. Now let me define a linear operator, $D(g)$, (I am sorry this notation also been used for the matrix although we shall see why the ambiguity is small), which will act on an $n$ dimensional vector space, with an orthonormal basis $|i\rangle, \;\; i =1,\ldots, n$. $D(g)$ is defined by 
\[ D(g) |i \rangle = \sum_j |j\rangle \langle j | D(g) |i \rangle = \sum_j |j\rangle D^j{}_{i}(g) \]

The intermediate step was motivational, the definition is 
\[ D(g) |i\rangle \equiv \sum_j |j \rangle D^j{}_{i}(g) \]

From the definition, and the fact that the basis is orthonormal, you can recover the matrix $D^j{}_{i}(g)$ from the abstract linear operator.
\[ \langle j | D(g) | i \rangle = \langle j | (\sum_k |k \rangle D^k{}_{i}(g)) = \sum_k \delta_{jk} D^k{}_{i}(g) = D^j{}_{i}(g) \]

Let's see what the composition law
\[ \sum_j D^i{}_{j}(g) D^j{}_{k}(g') =D^i{}_{k}(gg') \]

which must be satisfied by a representation, implies about our new abstract linear operator
\begin{align*}
D(g) D(g') | i \rangle &= D(g) \Big( \sum_j |j \rangle D^j{}_{i}(g') \Big)\\
&= \sum_{j,k} |k \rangle D^k{}_{j}(g) D^j{}_{i}(g') \\
&= \sum_k |k \rangle D^k{}_{i}(g g') \\
&= D(gg') |i \rangle 
\end{align*}

Therefore $D(g)D(g')=D(gg')$. \\

It is nice to see this work out, because we also write the composition law of the matrices as 
\[ D(g) D(g') =D(gg') \]

Now we see that is true for the matrices and the abstract linear operators. \uline{There was some danger} this was not going to work out. We might have gotten $D(g)D(g')=D(g'g)$ for the operator composition law.\\

In this spiffy notation, I'll define the tensor product of two representations. First to define tensor product space. If I have two vector spaces \\

$V_1$ with a basis $|i\rangle_1, \quad i=1,\ldots,d_1$ and $V_2$ with a basis $|j \rangle_2, \quad j=1,\ldots,d_2$. I define a vector space $V_1 \times V_2$ with a basis 
\[ |i,j \rangle \equiv |i\rangle_1 \otimes |j\rangle_2 \]

The dimension of $V_1 \times V_2$ is $d_1 \cdot d_2$. \\

Given a linear operator on $V_1$, $A$, and a linear operator on $V_2$, $B$, I can define a linear operator on $V_1 \times V_2$ called $A \otimes B$ by 
\[ (A\otimes B) |i,j\rangle = (A|i\rangle_1) \otimes (B|j\rangle_2) \]

or 
\[ \langle i',j' | A \otimes B | i,j \rangle = \, _1\langle i' | A| i \rangle_1 \, _2 \langle j' | B | j \rangle_2 \]

Now if we have a representation $D^{(1)}$ acting on $V_1$ and $D^{(2)}$ acting on $V_2$, we can define a new representation denoted $D^{(1)} \otimes D^{(2)}$ acting on $V_1 \times V_2$:
\[ D^{(1)} \otimes D^{(2)}(g) = D^{(1)}(g) \otimes D^{(2)}(g) \]

The new representation may or may not be reducible. \\

If the two representations are the same, that is $d =d_1 = d_2$, $V= V_1 = V_2$ and $D= D^{(1)} = D^{(2)}$, we can show that $D\otimes D$ is reducible (except in the case $d=1$). The trick is that in this case it makes sense to talk about 
\[ \frac{1}{\sqrt{2}}\Big(|i\rangle \otimes |j \rangle + |j\rangle \otimes |i \rangle \Big) \quad \text{ and } \frac{1}{\sqrt{2}}\Big( |i \rangle \otimes |j \rangle - | j \rangle \otimes |i \rangle \Big) \]

(or in another notation $\frac{1}{\sqrt{2}}\Big( |i,j\rangle + | j, i \rangle \Big) $ and $ \frac{1}{\sqrt{2}}\Big( |i,j\rangle - |j,i \rangle \Big)$).\\

You can think of this as a new basis. The number of basis elements of the first type is $\frac{d(d+1)}{2}$, the number of the second is $\frac{d(d-1)}{2}$. You can get fancy by defining projection operators 
\[ P_+ ( | i \rangle \otimes | j \rangle ) \equiv \frac{1}{2} ( | i \rangle \otimes | j\rangle + | j\rangle \otimes | i \rangle) \]

and
\[ P_- ( | i \rangle \otimes | j \rangle ) \equiv \frac{1}{2} ( | i \rangle \otimes | j\rangle - | j\rangle \otimes | i \rangle) \]

(The $\frac{d(d+1)}{2}$ elements of the symmetric part of the basis satisfy $P_+|\text{sym}\rangle=|\text{sym}\rangle $, $P_-|\text{sym}\rangle=0$).\\

The $\frac{1}{2}$ is put in so that $P_+^2 = P_+$ and $P_-^2 =P_-$, also $P_+ + P_- = \mathds{1} $, $ P_+ P_- =0$. \\

With these projection operators, I'll show that $D\otimes D$ is reducible. What I need to show is that 
\[ (D \otimes D)(g) P_{\pm} = P_\pm (D\otimes D)(g) \]

It's just a matter of using definitions (somehow math proofs are always just a matter of using the definitions although it is usually beyond me to do it)
\begin{align*}
(D \otimes D)(g) P_\pm |i \rangle \otimes |j \rangle &= \frac{1}{2}(D\otimes D)(g) \Big( |i \rangle \otimes | j \rangle \pm | j \rangle \otimes | i\rangle \Big)\\
&= \frac{1}{2} \Big( D(g) | i \rangle \otimes D(g) |j\rangle \pm D(g) |j \rangle \otimes D(g) |i \rangle \Big)\\
&= P_\pm D(g) |i\rangle \otimes D(g) |j\rangle = P_\pm (D\otimes D) (g) |i\rangle \otimes |j \rangle 
\end{align*}
 
\uline{That does it.} It may not be obvious to you that this shows that the representation is reducible, since our definition of reducibility was in terms of matrices, so I'll make the connection precise and I'll try to phrase the connection so that you can see a representation is reducible whenever you have a set of projection operator like $P_+$ and $P_-$, commuting with it.\\

If you have a set of projection operators 
\[ P_i, \;\; i=1,\ldots,m \qquad P_i^2 = P_i \qquad P_iP_j =0, i \neq j \]
\[ \sum_i P_i = \mathds{1} \]

then I can choose the basis of the vector space they act on so that it breaks up into bases for various subspace that are either annihilated or unaffected by the $P_i$.\\

(In the example of importance, there are $\frac{n(n+1)}{2}$ basis vectors unaffected by $P_+$ and annihilated by $P_-$, while the other $\frac{n(n-1)}{2}$ basis vector unaffected by $P_-$ and annihilated by $P_+$)\\

I'll write the basis vector as $|i,\alpha \rangle$, ($i$ has nothing to do with direct products, just a way of labelling the basis; and $\alpha = 1,\ldots,d_i$, where $d_i$ is the dimension of the $i$th subspace), where 
\[ P_i |i,\alpha \rangle = | i, \alpha \rangle \]
\[ P_i | j,\alpha \rangle = 0 \qquad j \neq i \]

The big assumption about these projection operators is that they commute with the representation operators. Let's call the rep $D$. 
\[ P_i D(g) = D(g) P_i \qquad \text{for all $i$ and $g$} \]

Let's look at the matrix associated with $D(g)$ in this basis and see what we can show about it.
\begin{align*}
D^{i\alpha}{}_{j \beta} &= \langle i,\alpha | D(g) | j, \beta \rangle \\
&= \langle i,\alpha | D(g) \sum_k P_k |j,\beta \rangle \\
&= \langle i, \alpha | D(g) P_j | j, \beta \rangle \\
&= \langle i, \alpha | P_j D(g) | j,\beta \rangle \propto \delta_{ji} 
\end{align*}

(The proportionality constant depends $g$, $i$, $j$, $\alpha$ and $\beta$ but that \uline{doesn't} matter.)\\

This is the statement that in this basis, the matrix $D(g)$ is block diagonal. 
\[ \begin{pmatrix} d_1 \times d_1 & & & & \\ & d_2 \times d_2 & & & \\ & & \ddots & & \\ & & & \ddots & \\ & & & & d_m \times d_m \end{pmatrix} \]

In the example of importance, by finding $P_+$ and $P_-$ that commute with $D \otimes D$, we have shown that $D\otimes D$ is equivalent to 
\[ [(D\otimes D)^{i \alpha}{}_{j \beta} (g)] = \begin{pmatrix} \text{Some } \frac{n(n+1)}{2} \times \frac{n(n+1)}{2} \text{ matrix} & 0 \\ 0 & \frac{n(n-1)}{2} \times \frac{n(n-1)}{2} \end{pmatrix} \]

These two blocks may or may not be further reducible. If the $\frac{n(n+1)}{2}$ dim block is reducible into $m$ irreducible components, $D^{(1)}, \ldots, D^{(m)}$, each of these representations is said to be in the symmetric part of the tensor product. If the $\frac{n(n-1)}{2}$ dimension block is reducible into $m'$ irreducible components, $D'^{(1)},\ldots, D'^{(m')}$, each of these is said to be in the antisymmetric part of the tensor product. \\

So after a multipage rambling explanation of tensor product, I'll finally state the fifth fact about the rotation group. Tensoring two irreducible reps together 
\begin{align*}
D^{(s_1)}\otimes D^{(s_2)} &\sim D^{(s_1+s_2)} \oplus D^{(s_1 +s_2 -1)} \oplus D^{(s_1 +s_2 -2)} \oplus \cdots \oplus D^{(|s_1 -s_2|)}\\
&= \oplus \sum_{s=|s_1-s_2|}^{s_1+s_2} D^{(s)} 
\end{align*}

Tensoring two identical reps together
\[ D^{(s)} \otimes D^{(s)} = D^{(2s)} \oplus D^{(2s-1)} \oplus D^{(2s-2)} \oplus \cdots \oplus D^{(0)} \]

where $D^{(2s)}$, $D^{(2s-2)},\cdots$ are in the symmetric part. $D^{(2s-1)}, \cdots$ are in the antisymmetric part. $D^{(0)}$ is symmetric if $s$ is an integer, antisymmetric if $s$ is a $\frac{1}{2}$ integer.\\

They just alternate.
}{
 \sektion{20}{December 4}
\descriptiontwenty
\begin{center}
\textbf{Parametrizing the connected homogeneous Lorentz group}
\end{center}

The rotation group was parametrized by a direction $\vec{e}$ and an angle $\theta$. We'll show that $\SO (3,1)$ can be parametrized by $\vec{e}$, $\theta$, and another direction and ``angle", by showing that any Lorentz transformation can be decomposed into a rotation and a boost. A Lorentz transformation is called a rotation if it takes 
\[ t \rightarrow t \qquad \text{ and } \qquad \vec{x} \rightarrow R\vec{x} \]

We'll denote such a Lorentz transformation by $R$ and you'll have to understand from context when $R \in \SO (3,1)$ and when $R$ is a $3 \times 3$ orthogonal matrix.\\

A boost in the $x$ direction by an ``angle", velocity parameter, $\phi$, takes
\begin{align*}
t &\rightarrow t \cosh \phi + x \sinh \phi \\
x &\rightarrow t \sinh \phi + x \cosh \phi \\
y &\rightarrow y \\
z &\rightarrow z 
\end{align*}

We'll denote such a L.T.~$A(e_x \phi)$, restricting $0 \leq \phi < \infty $ to avoid parametrizing each boost more than one way. In general 
\begin{align*}
A(\ve \phi ) \quad: \quad & t \rightarrow t \cosh \phi + \vec{e} \cdot \vec{x} \sinh \phi \\
&\vec{x} \rightarrow \vec{e}\;t \sinh \phi + \vec{x} + \vec{e}\; (\cosh \phi -1)\; \vec{e} \cdot \vec{x} 
\end{align*}

(This generalization is forced upon you by Eq.~(\ref{eq:20-page3-1}).)\\

To go along with the formula 
\begin{equation}\label{eq:20-page1-1}
R(\vec{e} \, \theta') R(\ve \, \theta) = R( \ve \, [ \theta + \theta' ] )
\end{equation}

We also have the formula (which can be verified with a little algebra) (the algebra involves using formulas for $\sinh (\phi_1 + \phi_2)$ and
$\cosh (\phi_1 + \phi_2)$).
\begin{equation}\label{eq:20-page1-2}
A(\ve \phi') A(\ve \phi) = A (\ve \,[ \phi + \phi' ] ) 
\end{equation}

This is why the velocity parameter is such a useful parameter for boosts, it just adds.\\

Now to prove that any Lorentz transformation can be uniquely decomposed into a rotation followed by a boost. The proof is by construction and the construction is unambiguous, it has no freedom, which implies uniqueness.\\

Starting with a general Lorentz transformation $\Lambda$ consider its action on the vector $e_0 \equiv (1,\vec{0})$. Since this vector has time component $>0$ and since all L.T.~connected to the identity preserve this when the vector is timelike, we must have 
\[ \Lambda:e_0 \rightarrow (\gamma, \alpha \ve) \]

where $\alpha, \gamma$ real and greater then zero but otherwise unknown and $\ve$ is some unit vector. The key thing about Lorentz transformations is that they leave the length of a vector unchanged, so we know there is a restriction on $\gamma$ and $\alpha$
\[ \gamma^2 - \alpha^2 = 1 \quad \text{and} \quad \gamma > 0 \]
\[ \Longrightarrow \gamma = \sqrt{1+ \alpha^2} \]

Let's rename $\alpha = \sinh \phi$, $\phi >0$, then $\gamma = \cosh \phi$ and the most general thing that $e_0$ can transform into under a Lorentz transformation is 
\[ \Lambda:e_0 \rightarrow (\cosh \phi, \vec{e} \sinh \phi) \]

This determination of an ``angle" $\phi$ and a direction $\vec{e}$ allows me to (uniquely) read off the boost that will bring $e_0$ back to rest, it is 
\[ A\one (\ve \phi) \]
\[ A\one (\ve \phi) \; \Lambda:e_0 \rightarrow e_0 \]

This means this product is some rotation, call it $R$.
\[ A\one (\ve \phi) \Lambda = R \]
\[ \Lambda = A (\ve \phi) R   \]

Just like when we were working with $\SO(3)$, formulas like Eqs.~(\ref{eq:20-page1-1}) and (\ref{eq:20-page1-2}) show that rotations and boosts can be written as exponentials. In a rep $D$ it implies 
\[ \frac{d}{d\theta} D(R(\ve \, \theta)) |_{\theta = 0 } \equiv -i \vec{L}\cdot \ve, \qquad D(R(\ve \, \theta)) = e^{-i \vec{L}\cdot \ve \, \theta} \]
\[ \frac{d}{d\phi} D(A(\ve \phi)) |_{\phi = 0 } \equiv -i \vec{M}\cdot \ve, \qquad D(A(\ve \phi)) = e^{-i \vec{M} \cdot \ve \phi} \]

Just as for $\SO(3)$, if you know the inequivalent irreducible reps of $\vec{L}$ and $\vec{M}$, the generators of $\SO(3,1)$, you know the inequivalent irreducible reps up to a phase of $\SO(3,1)$. $\vec{L}$ is playing the exact same role as $\vec{J}$ did in our discussion of $\SO(3)$, in fact we would have reused $\vec{J}$ if it weren't that it is conventionally used for something else. So from the properties of rotations, we have 
\[ [L_i, L_j] = i \epsilon_{ijk} L_k \qquad (\sum_k \text{ implied}) \]

(that came from $ R'^{-1} R(\ve \, \theta ) R' = R(R'^{-1} \ve \, \theta) $)\\

A property of the Lorentz group is that 
\begin{equation}\label{eq:20-page3-1}
R\one A(\ve \phi) R = A( R\one \ve \phi)
\end{equation}

(can be used to get the general boost from a boost in the $\vec{x}$ direction)\\

(you can convince yourself that both sides are boosts by $\phi$ along $R\one \ve$) Out of this (take $i \frac{d}{d\phi}$) comes the statement (applied to rep $D$). 
\[ D(R\one ) \vec{e} \cdot \vec{M}  D(R) = R\one \vec{e} \cdot \vec{M} \]

which implies
\[ D(R\one)\vec{M} D(R) = R\vec{M} \] 

(These are just like the $\SO (3)$ arguments, so I haven't written them in detail)

which implies after a little more work
\[ [L_i, M_j] = i\epsilon_{ijk} M_k \]

Using 
\[ D(R)\one \; M_i \; D(R) = \!\!\!\!\!\!\!\!\!\! \underbrace{R_{ij}}_{\substack{ 3 \times 3 \text{ orthogonal}\\\text{rotation matrix}}} \!\!\!\!\!\!\!\!\!\! M_j \qquad (\sum_j \text{ implied}) \]

It is easy to see that 
\[ D(R)\one M_i M_j D(R) = R_{ik} R_{jl} M_k M_l \]

that is $M_i M_j$ transforms like a two index tensor under rotations. Therefore $[M_i,M_j] $ is a two index antisymmetric tensor under rotation. If the Lie algebra of $\SO (3,1)$ is going to close, the commutator of two boost generators must be a linear combination of a boost generator and a rotation generator. The most general thing I can make that is a two index antisymmetric tensor that is linear in the boost and rotation generators, which transform like vectors, is
\[ \alpha \epsilon_{ijk} M_k + \beta \eijk L_k \]

Therefore it must be that 
\[ [M_i,M_j] = i \eijk [\alpha M_k+\beta L_k] \]

Still more evasive reasoning shows that $\alpha =0$. Using the Parity automorphism introduced on December 2, we get a new representation from the one we were working with by defining $D_P(\Lambda)= D(\Lambda_P)$. Now for a rotation, $PRP = R$, so $D_P(R) = D(R)$. The generator of rotations in $D_p$ are the same as those in $D$. However for $P A(\ve \phi) P = A (- \ve \phi)$ so the generators of boosts are minus the generators of boosts in $D$. This can be summarized, $L_i \rightarrow L_i$, $M_i \rightarrow -M_i$. The commutation relations have to still work under this transformation. $[L_i, L_j] = i\eijk L_k$ is OK. $[L_i, -M_j] = i \eijk (-M_k)$ is OK but $[-M_i, -M_j] = i\eijk [ \alpha (- M_k) + \beta L_k ]$ is OK only if $\alpha = 0$.\\

In fact with a fair amount of work, you can check from the definitions, $[M_i,M_j] = - i \eijk L_k$. (The $-$ sign would not be present if this were $\SO (4)$ instead of $\SO (3,1)$). From the commutation relations of the Lie algebra of any group, there is a general method called the method of highest weight. Fortunately, a miracle occurs, and we will not have to go through that method.\\ 

Define 
\[ \vec{J}^{(\pm )} = \frac{1}{2} (\vec{L} \pm \!\!\!\!\!\!\!\!\!\!\!\!\!\!\! \underbrace{i}_{\substack{\text{no $i$ would be here}\\\text{if it were $SO(4)$}\\\text{we were studying}}} \!\!\!\!\!\!\!\!\!\!\!\!\!\!\! \vec{M}) \]
\[ \vec{L} = \vec{J}^{(+)} + \vec{J}^{(-)} \]
\[ -i\vec{M} = \!\!\!\!\!\!\!\!\!\!\!\!\!\!\!\!\!\! \underbrace{-}_{\substack{\text{This $-$ sign is obvious }}} \!\!\!\!\!\!\!\!\!\!\!\!\!\!\!\!\! \Big(\vec{J}^{(+)} - \vec{J}^{(-)}\Big) \]

You can verify from the commutators of the $\vec{L}$'s and $\vec{M}$'s that 
\[ [J_i^{(\pm)}, J_j^{(\pm)}] = i \eijk J^{(\pm)}_k \]

and
\[ [J_i^{(\pm)}, J_j^{(\mp)}] = 0 \]

The $\vec{J}^{(+)}$'s and $\vec{J}^{(-)}$'s form two independent, commuting $SO(3)$ algebras. What we want to find is matrices (my brain is better equipped to think about matrices than abstract linear operator, but you can think about it either way) that have these commutation relations -- we want a complete set of inequivalent irreducible ones.\\

Here are some: For any $s_1$, $s_2$, $\quad s_1 = 0,\frac{1}{2}, 1, \dots$, $\quad s_2 = 0, \frac{1}{2}, 1,\cdots$, take 
\begin{equation}\label{eq:20-page5-1}
J^{(+)}_{i} = J^{(s_1)}_{i} \otimes \!\!\!\!\!\!\!\!\! \underbrace{\id_{2s_2 + 1}}_{\substack{(2s_2+1)\times(2s_2+1)\\\text{dimensional identity}}}
\end{equation}

and 
\begin{equation}\label{eq:20-page5-2}
J^{(-)}_{i} = \!\!\!\!\!\!\!\!\underbrace{\id_{2s_1+1}}_{\substack{(2s_1+1)\times(2s_1+1)\\\text{dimensional identity}}} \!\!\!\!\!\!\!\!\! \otimes \;
J^{(s_2)}_{i} 
\end{equation}

The $J^{(s)}_{i}$ are our friends from the last lecture. Let me show that this satisfies one of the commutation relations. (You can check the others).
\begin{align*}
[J_i^{(+)},J_j^{(+)}] &= [J_i^{(s_1)} \otimes \id_{2s_2 +1}, J_j^{(s_1)} \otimes \id_{2s_2+1}] \\
&= J_i^{(s_1)} \otimes \id_{2s_2+1} \;\; J_j^{(s_1)} \otimes \id_{2s_2+1} \quad - i \leftrightarrow j \\
&= J_i^{(s_1)}J_j^{(s_1)} \otimes \id_{2s_2 +1} - i\leftrightarrow j\\
&= [J_i^{(s_1)}, J_j^{(s_1)} ] \otimes \id_{2s_2+1} \\
&= i \eijk J^{(s_1)}_k \otimes \id_{2s_2+1} \\
&= i \eijk J^{(+)}_k 
\end{align*}

Now you can solve for the $\vec{L}$'s and $\vec{M}$'s in term of the $\vec{J}^{(\pm)}$. 
\[ \vec{L} = \vec{J}^{(+)} + \vec{J}^{(-)}= \vec{J}^{(s_1)} \otimes \id_{2s_2 +1} + \id_{2s_1+ 1}\otimes \vec{J}^{(s_2)} \]

Because the $\vec{J}^{(s)}$ are hermitian\footnote{If you can't prove this from $(A\otimes B)^\dagger=A^\dagger\otimes B^\dagger$ please feel free to come ask me to elaborate.}, $\vec{L}$ is hermitian.
\[ \vec{M} = \frac{1}{i} \Big(\vec{J}^{(+)} - \vec{J}^{(-)}\Big) \quad \text{is antihermitian (because of the $\frac{1}{i}$}) \]

The commutation relations among the $\vec{J}^{(+)}$ and $\vec{J}^{(-)}$ (which were derived by using the definitions in term of $\vec{L}$ and $\vec{M}$), give us back the correct commutation relations for $\vec{L}$ and $\vec{M}$.\\

Exponentiating $\vec{L}$ and $\vec{M}$ for any choice of $s_1$ and $s_2$ gives a representation called $D^{(s_1,s_2)}$. I'll define 
\[ \Lambda(\ve \, \theta, \vf \phi) \]
\begin{align*}
D^{(s_1,s_2)} (\Lambda(\ve \, \theta, \vf \phi)) &= e^{-i\vl\cdot \ve\,\theta} e^{-i \vm \cdot \vf \phi} \\
&= e^{-i(\vj^{(+)} + \vj^{(-)}) \cdot \ve \, \theta} e^{- (\vj^{(+)} - \vj^{(-)}) \cdot \vf \phi} \\
\text{(only for $\ve \parallel \vf$)} \qquad &= e^{-i \vj^{(+)} \cdot ( \ve \, \theta - i \vf \phi)} e^{-i\vj^{(-)} \cdot (\ve \, \theta + i \vf \phi)}
\end{align*}

This has a simpler form for a pure boost or a pure rotation.
\[ D^{(s_1,s_2)} (A(\ve \phi)) = e^{-(\vj^{(+)} - \vj^{(-)}) \cdot \ve \phi} \]
\[ D^{(s_1,s_2)} (R(\ve \, \theta)) = e^{- i(\vj^{(+)} + \vj^{(-)}) \cdot \ve \, \theta} \]

It turns out that these are only representations up to a phase of $\SO (3,1)$ if $s_1+ s_2$ is a half integer. The claim is that these are a complete set of inequivalent irreducible reps up to a phase of the Lorentz group.\\

I would like to explain the distinction between $D^{(s_1,s_2)}$, a representation of $\SO (3,1)$, and $D^{(s_1)}\otimes D^{(s_2)}$, the direct product of two representations of $\SO(3)$, which is in general reducible. 
\[ D^{(s_1)} \otimes D^{(s_2)} (R (\ve \, \theta) ) = e^{-i \vj^{(s_1)} \cdot \ve \, \theta} \otimes e^{-i\vj^{(s_2)} \cdot \ve \, \theta} \]

This is a representation of $\SO (3)$, it has three generators, and they are given by taking 
\[ i \frac{d}{d \theta} (D^{(s_1)} \otimes D^{(s_2)})\big(R(\ve \, \theta)\big) \Big|_{\theta=0} = (\vj^{(s_1)} \otimes \id_{2s_2+1} + \id_{2s_1+1} \otimes \vj^{(s_2)}) \cdot \ve \]

These three generators (one for each direction $\ve$ can point), bear some resemblance to the \uline{six} generators of $\SO (3,1)$:
\[ \vj^{(+)} = \vj^{(s_1)} \otimes \id_{2s_2+1} \qquad \text{and} \qquad \vj^{(-)} = \id_{2s_1+1} \otimes \vj^{(s_2)} \]

and if you restrict yourself to elements of $\SO (3,1)$ such that the coefficient of $J_i^{(+)}$ is the same as the coefficient of $J_i^{(-)}$, that is, the rotations, you really have got the same thing. But, the coefficient of $J^{(+)}$ is independent of that of $J^{(-)}$. $D^{(s_1)}\otimes D^{(s_2)}$ is a reducible rep of $\SO (3)$, $D^{(s_1,s_2)}$ is an irreducible rep of $\SO (3,1)$.\\

A standard basis for the representation $D^{(s_1,s_2)}$ is the basis $|m_+, m_- \rangle$, 
\[ m_+ = - s_1, -s_1+1,\ldots, s_1 -1, s_1 \]
\[ m_- = - s_2, -s_2+1,\ldots, s_2 -1, s_2 \]

which simultaneously diagonalizing the two commuting hermitian operators $J_z^{(+)}$ and $J_z^{(-)}$: 
\[ J_z^{(\pm)} |m_+, m_- \rangle = m_\pm | m_+, m_- \rangle \]

It would have been possible to carry out the analysis in a more Lorentz invariant fashion by defining (according to Ramond, p.10) 
\[ J^{ij} = \eijk L_k \]
\[ J^{0i} = - J^{i0} = -M_i \]
\[ J^{00} = J^{ii} = 0 \qquad \text{no sum} \]

The commutation relation would have been unified into 
\[ [J\mn, J\ab] = i g^{\nu\alpha} J^{\mu\beta} - i g^{\mu\alpha} J^{\nu \beta} - ig^{\nu\beta} J^{\mu\alpha} + ig^{\mu\beta} J^{\nu \alpha} \]

A Lorentz transformation would be parametrized by an antisymmetric matrix $\epsilon_{\mu\nu}$
\[ D(\Lambda(\epsilon_{\mu\nu})) = e^{-i \epsilon_{\mu\nu} J^{\mu\nu}/2} \]

This approach buys you nothing but elegance, at the expense of clarity.\\

Another thing worth noting is that a parallel analysis can be made of $\SO(4)$. It is kind of nice to look through the calculations and see where $i$'s are changed to $-i$'s and where $g\mn$ becomes $\delta\mn$.\\

Carefully spelling out and proving any of the statements before would be instructive. I feel I have put enough of the outlines here that I could pursue the proofs to my satisfaction. I would enjoy making some of these statement more concrete. If people come ask me about them it will force me to do so.\\

Facts about $D^{(s_1,s_2)}$ summarized
\begin{itemize}
\item[(1).] The dimension of $D^{(s_1,s_2)}$ is $(2s_1 +1)(2s_2 +1)$.
\item[(2).] \uline{When} $s_1+s_2$ is a half integer, we have representation up to a phase
\[ D\big(R(\ve \pi) \big) D\big(R(\ve \pi) \big) = -1 \]
\item[(3).] For $R\in \SO (3,1)$, a rotation $D^{(s_1,s_2)}(R)$ is unitary, however for $A \in \SO (3,1)$, a boost, $D^{(s_1, s_2)}(A)$ is not. In fact it is hermitian. \\

This can be seen from the hermiticity of $\vj\pp$ and $\vj\mm$. \\

This agrees with some general theorem of group theory, worth noting.\\

The finite dimensional representations of a compact group are always equivalent to unitary representations; the generators can always be chosen to be hermitian in some basis.\\

$\SO (3,1)$ is not a compact group. The range of boosts is infinite: $0 \leq \phi < \infty$. \\

The unitary representations of a \uline{non}-compact group are always infinite dimensional. The finite dimensional ones are never unitary.\\

Although the $\vec{J}^{\pm}$ we have found are hermitian, their coefficients, when $\Lambda$ is not a rotation, in the exponential are not purely imaginary. \\

One can consider $\infty$ dimensional unitary reps. These would presumably describe an $\infty$ number of particle types. To agree with reality, infinitely many particles would somehow have to be hidden...\\

\item[(4).] What rep do we get by taking the complex conjugate of $D^{(s_1,s_2)}$. Since $\vj^{(\pm)} \sim - \vj^{(\pm)*}$ (follows from their expressions in terms of $\vj^{(s_1)}$ and $\vj^{(s_2)}$ and the properties of $\vj^{(s)}$. (Dec.~2, Eq.~(\ref{eq:19-page17}))). The representation of a rotation is equivalent\footnote{The sketch of fact $(4)$ can be made more concrete by giving a name $S$ to the matrix that satisfies $S\one \vj^{(s)*} S = -\vj^{(s)}$ and using it to display the similarity transformation between $D^{(s_1,s_2)*}$ and $D^{(s_1,s_2)}$.}
\[ D^{(s_1,s_2)*} \Big(R (\ve \, \theta) \Big) = e^{+i (\vj^{(+)} + \vj^{(-)})^*\cdot \ve \, \theta} \]
\[ \sim e^{-i (\vj\pp + \vj\mm) \cdot \ve \, \theta} = D^{(s_1,s_2)}\Big(R(\ve \, \theta) \Big) \]

The representation of a boost however is screwed up in a way that cannot be undone by some equivalence.
\[ D^{(s_1,s_2)*} (A (\ve \phi)) = e^{-(\vj^{(+)} - \vj^{(-)})^*} \cdot \ve \phi \]
\[ \sim e^{-(\vj\mm - \vj\pp) \cdot \ve \phi} \]

The roles of $\vj\mm$ and $\vj\pp$ have been exchanged. This is true for both the rotations, $\vj\pp \leftrightarrow \vj\mm$ does nothing to them, and the boosts. We can identify the new rep we have made by noting that the exchange of $\vj\pp \leftrightarrow \vj\mm$ is just like the exchange of $s_1$ and $s_2$. That is 
\[ D^{(s_1,s_2)*} (\Lambda) \sim D^{(s_2,s_1)} (\Lambda) \]

If you believe that $D^{(s_1,s_2)}$ is not equivalent to $D^{(s_2,s_1)}$ (unless $s_2 =s_1$), this shows that $D^{(s_1,s_2)*}$ is not equivalent to $D^{(s_1,s_2)}$ (unless $s_1 = s_2$).\\

\item[(5).] The effect of parity on a rep was already discussed $L_i \rightarrow L_i$, $M_i \rightarrow - M_i$, is also interpretable as 
\[ J\pp \leftrightarrow J\mm \]

so
\[ D^{(s_1,s_2)}_p \sim D^{(s_2,s_1)} \]

\item[(6).] $\displaystyle D^{(s'_1, s'_2)} \otimes D^{(s''_1, s''_2)} = \oplus \sum_{s_1=|s'_1-s''_1|}^{s'_1+s''_1} \sum_{s_2=|s'_2-s''_2|}^{s'_2+s''_2} D^{(s_1,s_2)} $. If $s_1= s'_1 =s''_1$ and $s_2 = s'_2 = s''_2$, then in the tensor product 
\begin{itemize}
\item $D^{(2s,2s)}$ is symmetric, $D^{(2s,2s-1)}$ is antisymmetric. 
\item $D^{(2s,2s-2)}$ is symmetric, $D^{(2s,2s-3)}$ is antisymmetric, etc.
\item $D^{(2s-1,2s-1)}$ is symmetric, $D^{(2s-1,2s-2)}$ is antisymmetric.
\item $D^{(2s-2,2s-2)}$ is symmetric, etc.
\end{itemize}
The proof of these statements would probably be instructive to construct. I expect with some thought they can be derived from the analogous statements about $\SO(3)$ in short order.\\

\item[(7).] If you understood the distinction between $D^{(s_1,s_2)}$ and $D^{(s_1)}\otimes D^{(s_2)}$, this fact about $D^{(s_1,s_2)}$ will be easy to understand.\\

Any time you have a representation of a group (up to a phase), you have a representation of any of its subgroups (up to a phase). \\

Even if the representation of the group is irreducible, the representation of the subgroup may be reducible.\\

By restricting ourselves to the rotations, a subgroup of $\SO(3,1)$, we get a representation of $\SO(3)$ which is in general reducible. In fact we are down to the case where the coefficients of $\vj\pp$ and $\vj\mm$ are identical, and for those coefficients, we have exactly the same possible representation matrices as if we had taken $D^{(s_1)} \otimes D^{(s_2)}$. So $D^{(s_1,s_2)}$ ``induces" a rep of $\SO(3)$ which is 
\[ D^{(s_1)} \otimes D^{(s_2)} \sim \oplus \sum_{s=|s_1 - s_2|}^{s_1 + s_2} D^{(s)} \]

$(s,0)$ and $(0,s)$ are irreducible reps of $\SO(3,1)$ and $\SO(3)$. 
\end{itemize} 

\vspace{1cm}

\begin{center}
\textbf{Examples}\\
\textbf{Where is the vector?}
\end{center}

If the vector representation which we all know and love, is irreducible (I've never been able to reduce it), it must be equivalent to one of the four dimensional representations we have constructed. The only irreducible four dimensional reps on our list are 
\[ \begin{matrix} D^{(3/2,0)} & D^{(0,3/2)} & \text{and} & D^{(1/2,1/2)} \\ (2\cdot 3/2 +1)(0+1) & (2\cdot 0 + 1 ) (2 \cdot 3/2 + 1) & & (2\cdot 1/2 +1) (2 \cdot 1/2 + 1) \\ 4 & 4 & & 4 \end{matrix} \]

(there are only three ways to factor $4$).\\

$D^{(3/2,0)}$ cannot be the vector for any of several reasons. 
\begin{enumerate}
\item The complex conjugate of $D^{(3/2,0)}$ is $D^{(0,3/2)}$ which is inequivalent to $D^{(3/2,0)}$. However the vector is equivalent to its complex conjugate. Manifestly, because in the usual basis, the matrices that transform a vector are purely real. 
\item A similar argument applies by considering the effect of parity.
\item If you restrict yourself to the rotation subgroup of $\SO (3,1)$, the four vector representation is reducible into a three vector and a rotational scalar. $D^{(3/2,0)}$ however remains irreducible under this restriction. It is a spinor with spin $3/2$ under rotation.
\end{enumerate}

Identical arguments rule out $D^{(0,3/2)}$. \\

$D^{(1/2,1/2)}$ must be the vector. It looks funny, but it must be right. It certainly can't be ruled out along the lines of above three arguments. Later we'll make a vector out of $D^{(1/2,0)}\otimes D^{(0,1/2)}$, which is equivalent to $D^{(1/2,1/2)}$.

\begin{center}
\textbf{What about rank 2 tensors}
\end{center}

Usually we get a rank $2$ tensor by taking the tensor product of two vectors. If $A^\mu$ and $B^\nu$ transform like vectors, $A^\mu B^\nu$
transforms like a tensor. If $D^{(1/2,1/2)}$ is a vector then $D^{(1/2,1/2)} \otimes D^{(1/2,1/2)}$ must be a tensor. Now 
\[ \begin{matrix} D^{(1/2,1/2)} \otimes D^{(1/2,1/2)} \sim & D^{(1,1)} & \oplus & D^{(0,1)} & \oplus & D^{(1,0)} & \oplus & D^{(0,0)} \\ & S & & A & & A & & S \\ & \text{dim} \; 9 & & \text{dim} \; 3 && \text{dim} \; 3 && \text{dim} \; 1 \end{matrix} \]

There must be a way to reduce the tensor $T\mn$ into $9,3,3,$ and $1$ dimensional subspaces that transform independently under $\SO(3,1)$. We can rewrite 
\begin{align*}
T\mn &= \frac{1}{2} (T\mn + T\nm) + \frac{1}{2} (T\mn - T\nm) \\
&\equiv \underbrace{S\mn}_{10\text{ dim}} + \underbrace{A\mn}_{6\text{ dim}}
\end{align*}

and from our general arguments about symmetric and antisymmetric tensor products, these two subspaces must transform independently. The 10 dim symmetric subspace must contain a 9 dim and 1 dim subspace which transform independently. Indeed, the linear combination 
\[ g\mnd S\mn \quad \text{is a Lorentz invariant,} \]

so that $1$ dimensional subspace transforms independently of the other $9$ components of $S\mn$ which are in 
\[ S\mn - \frac{1}{4} g_{\tau \sigma} S^{\tau\sigma} \mbox{Id}\mn \]

the ``traceless" part of $S$. This is $D^{(1,1)}$, the symmetric traceless tensor. In general, $D^{(n/2,n/2)}$ is the symmetric traceless tensor of rank $n$.\\

Just for completeness, let's find the two irreducible parts of $A\mn$. Given any antisymmetric two index tensor, you can define a new antisymmetric tensor by 
\[ A^D{}_{\mu\nu} = \frac{1}{2} \epsilon_{\mu\nu\alpha\beta} A\ab \qquad \epsilon_{0123} = +1 \]
\[ A^D{}_{01} = A^{23} \qquad \text{the $\frac{1}{2}$ was inserted to avoid $A^D{}_{01} = 2 A^{23}$} \]
\[ A^{D01} = - A^{23}, \quad A^{D23} = A^{01} \]

The cute thing about this operation is that 
\[ A^{DD\, \mu\nu} = - A\mn \]

The square of the dualing operation is $-1$, and the eigenvalues must be $\pm i$. The dualing operation also commutes with Lorentz transformations (that have det $(\Lambda ) = + 1$). Thus, the Lorentz transformations transform the subspaces of each eigenvalue of the dualing operation independently.
\begin{align*}
A\mn &= \frac{1}{2}(A\mn + i A^{D\mu\nu}) + \frac{1}{2} (A\mn - i A^{D\mu\nu}) \\
&\equiv A^{(+) \mu\nu} + A^{(-) \mu\nu}
\end{align*}
\begin{align*}
A^{(\pm)D\mu\nu} &= \frac{1}{2}\Big[ A^{D\mu\nu} \pm i A^{DD \mu\nu} \Big] \\
&= \frac{1}{2}\Big[ A^{D\mu\nu} \mp i A^{\mu\nu} \Big] \\
&= \mp i A^{(\pm) \mu\nu} 
\end{align*}

The six dimensional antisymmetric part of the tensor product has been broken into two $3$ dimensional parts, which must be $D^{(1,0)}$ and $D^{(0,1)}$. It is no surprise that this splitting required taking complex combinations because $D^{(1,0)}$ and $D^{(0,1)}$ are not equivalent to their complex conjugates.
\vspace{1cm}

\begin{center}
\textbf{Spinors}
\end{center}

This is the example we are really interested in. We'll be using, for the rest of the course, fields transforming like $D^{(1/2,0)}$ and $D^{(0,1/2)}$. In fact, this is really the only application of the last couple lectures that is going to be used. (The scalar and vector reps you (presumably) already understood.) So if you get a good handle on how to manipulate these two representations, you don't really have to understand all the representation theory that has gone before.\\

We'll study $D^{(0,1/2)}$. Using the formulas in Eqs.~(\ref{eq:20-page5-1}) and (\ref{eq:20-page5-2}), with $s_1=0$, $s_2 = \frac{1}{2}$, and using 
\[ \vec{J}^{(0)} = \vec{0}, \qquad \vj^{(1/2)} = \frac{\vec{\sigma}}{2} \]

we have
\[ \vec{J}^{(+)} = \vec{0}, \qquad \vj^{(-)} = \frac{\vec{\sigma}}{2} \]

In this representation, a rotation is given by 
\begin{align*} 
D(R(\ve \, \theta)) &= e^{-i (\vj\pp + \vj\mm)\cdot \ve \, \theta} \\
&= e^{-i \vec{\sigma}\cdot \ve \, \theta /2 }
\end{align*}
 
A boost is represented by 
\[ D(A(\ve \phi)) = e^{-(\vj\pp - \vj\mm) \cdot \ve \phi }= e^{+ \vec{\sigma}\cdot \ve \phi /2} \]

Because of the plus sign in $e^{+ \vec{\sigma}\cdot \ve \phi /2 }$, let's call a field that transforms under $D^{(0,1/2)}$, $u_+$.\\

In the $D^{(1/2,0)}$ rep, a rotation is represented the same way \footnote{It better be. Up to equivalences there is only one way for an irreducible rep of a given dimension to transform under rotations.}, but a boost has a minus sign in the exponential, so the field transforming as $D^{(1/2,0)}$ will be called $u_-$.\\

Because $D^{(0,1/2)*} \sim D^{(1/2,0)}$, $u_+^*$ must transform like $D^{(1/2,0)}$. If we want to think of $u_+^*$ as a new vector, we'll write $u_+^\dagger$. Since 
\[ D^{(1/2,0)} \otimes D^{(0,1/2)} \sim D^{(1/2,1/2)} \]

the product of $u$ and $u^\dagger$ must be a four vector, and if we can find the right combination, we can make that explicit. An easy way to find the right combinations is to recall the transformations of spinors under the subgroup of rotations. Both $u_+$ and $u_+^\dagger$ transform as rotational spinors, and a rotational scalar can be made by taking $u_+^\dagger u_+$ and a rotational vector is $u_+^\dagger \vec{\sigma}u_+$.

The four vector must be $(u_+^\dagger u_+, \alpha u_+^\dagger \vec{\sigma} u_+ )$. $\alpha$ is unknown from this argument, but we can find it by looking at a boost along $e_z$ with rapidity $\phi$.
\begin{align*}
A(\vec{e}_z \phi): \quad &u_+ \rightarrow e^{+\sigma_z \phi /2 } u_+\\
\text{and}\quad & u_+^\dagger \rightarrow u_+^\dagger (e^{\sigma_z \phi /2})^\dagger = u_+^\dagger e^{\sigma_z \phi /2} 
\end{align*}

so 
\begin{align*}
u_+^\dagger u_+ &\rightarrow u_+^\dagger e^{\sigma_z \phi} u_+\\
&= u_+^\dagger (\cosh \phi + \sigma_z \sinh \phi ) u_+ \\
&= \cosh \phi \, u_+^\dagger u_+ + \sinh \phi \, u^\dagger_+ \, \sigma_z \, u_+ 
\end{align*}

If we take $\alpha =1$, i.e.~$v^0 = u_+^\dagger u_+$, $\vec{v} = u_+^\dagger \vec{\sigma} u_+$, this says 
\[ v^0 \longrightarrow \cosh \phi v^0 + \sinh \phi v_z \]

Let's see what $v_z$ ($v^3$ if you want to keep your indices up) and $v_x$ (or $v_y$) transform into 
\begin{align*}
v_z = u_+^\dagger \sigma_z u_+ &\longrightarrow u_+^\dagger e^{\sigma_z \phi /2} \sigma_z e^{\sigma_z \phi /2} u_+ \\
&= u_+^\dagger (\sigma_z \cosh \phi + \sinh \phi ) u_+ \\
&= \cosh \phi \, v_z + \sinh \phi \, v^0 
\end{align*}
\[ v_x = u_+^\dagger \sigma_x u_+ \longrightarrow u_+^\dagger e^{\sigma_z \phi /2} \sigma_x e^{\sigma_z \phi /2} u_+ \]

But using $\displaystyle e^{\sigma_z \phi /2} = \cosh \frac{\phi}{2} + \sinh \frac{\phi}{2} \sigma_z$ and $\sigma_z \sigma_x = - \sigma_x \sigma_z$, we see that we have found 
\[ v_x \longrightarrow u_+^\dagger e^{+\sigma_z \phi /2} e^{-\sigma_z \phi /2} \sigma_x u_+ = v_x \]

Similarly $ v_y \longrightarrow v_y$.\\

If you had gone through this procedure for $u_-$, you would have found that the vector is 
\[ w^\mu = (u_-^\dagger u_-, - u_-^\dagger \vec{\sigma} u_-) \]

It looks like the two component fields we have found have a fighting chance of describing spin $1/2$ particles.
}{
 \sektion{21}{December 9}
\descriptiontwentyone
Promote the two component objects $u_\pm$ into two component functions of space time, spinor fields.\\

\begin{center}
\textbf{Criteria for a free theory made up of a $u_+$ field.} 
\end{center}
\[ \mathcal{L}(\up, \upd, \pmu \up, \pmu \upd) \]
\begin{itemize}
\item[(i).] $S = \int d^4 x \; \mathcal{L} $ had better be real, $S= S^*$ .
\item[(ii).] $\ml$ must be a Lorentz scalar, but not necessarily parity invariant.
\item[(iii).] $\ml$ bilinear in the fields so we get a linear equation of motion and a free field theory
\item[(iv).] No more than two derivatives in $\ml$. If we can't construct anything of this type we'll go to three, four or more derivatives
\item[(v).] Want the theory to have a conserved charge 
\[ \up \longrightarrow e^{i\theta} \up \qquad \upd \longrightarrow e^{-i\theta} \upd \]
(Rule out Majorana neutrinos) because all known spin $1/2$ particles in the world do carry some conserved quantum number, like baryon number.
\end{itemize}

Property (v) with property (iii) force us to have one $\up$ and one $\upd$ factor in each term. Now the product $\up \upd$ is a four vector, which is neither a scalar itself, nor can a scalar be built from it with an even number of derivatives. However, a scalar can be built from it and one derivative. So (ii) and (iv) imply 
\[ \ml \propto (\upd \partial_0 \up + \upd \vec{\sigma}\cdot \vec{\nabla} \up) \]

The coefficient of proportionality must be purely imaginary to satisfy (i) (proof involves a parts integration). By rescaling $\up$ and $\upd$ we have
\[ \ml = \pm i \Big[ \upd \partial_0 \up + \upd \vec{\sigma}\cdot \vec{\nabla} \up \Big] \]

(For $u_-$ we would have arrived at $\ml = \pm i \Big[ \umd \partial_0 \um - \umd \vec{\sigma}\cdot \vec{\nabla} \um \Big]$) \\

This is called the Weyl Lagrangian. Let's derive the equation of motion by varying w.r.t. $\upd$
\[ \partial_0 \up + \vec{\sigma}\cdot \vec{\nabla} \up = 0 \]

We can see that any solution of this equation is a solution of the Klein-Gordon equation by acting on it with $\partial_0 - \vec{\sigma} \cdot \vec{\nabla}$ to get 
\[ (\partial_0^2 -\nabla^2 ) \up = 0. \]

(An identity matrix has been suppressed). To derive this you need to use equality of mixed particles and $\sigma_i \sigma_j = i \eijk \sigma_k + \delta_{ij}$, where a $2 \times 2$ identity is again suppressed in $\delta_{ij}$.\\

So all the solutions of the Weyl equation satisfy the wave equation, and they must be of the form 
\[ \up(x) = \up e^{-ik\cdot x} \qquad k^2 =0. \quad \text{(no need yet for $k^0 >0$}) \]

where $u_+$ is some constant two component column vector.\\

(For $u_-$ we would have gotten $\partial_0 u_- - \sigma \cdot \vec{\nabla} u_- = 0$, which leads to $u_-(x) = \um e^{-ik\cdot x}$)\\

Now we plug these potential solutions back into the equation to get the condition on the constant spinors 
\[ (\!\!\!\!\!\!\!\!\!\underbrace{k^0}_{\substack{2\times2 \text{ identity}\\\text{is understood}}} \!\!\!\!\!\!\!\!\!-\vec{\sigma}\cdot\vec{k}) u_+ = 0 \]

Let's take $\vec{k} = k^0 \widehat{e_z} $. This then says 
\[ (1-\sigma_z) \up = 0 \qquad u_+ = \begin{pmatrix} 1 \\ 0 \end{pmatrix} \]

(At this point we would have gotten $u_- = \begin{pmatrix} 0 \\ 1 \end{pmatrix} $.)\\

In general $u_+$ is the eigenstate of 
\[ \frac{\vec{\sigma} \cdot \vec{k}}{k^0} \quad \text{with eigenvalue 1} \quad \text{which since} \quad |\vec{k}| = k^0 \]

has the interpretation of being the spin along the direction of motion (at least for $k^0 > 0$). \\

Unlike the normal theory of spinors, for a given $k^\mu$, we only have one solution, one direction of spin. This would not be possible if these particles weren't massless. If they were massive, you could always boost to their rest frame, turn their spin around and then boost back, and you'll have a particle with spin pointed the opposite direction. The spin of a massless particle is usually referred to as helicity, the component of angular momentum along the direction of motion. Spin is usually reserved for massive particles. It is the angular momentum in the rest frame.\\

Because $D_P^{(0,1/2)} = D^{(1/2,0)}$ which is inequivalent to $D^{(0,1/2)}$ there is no parity transformation in this theory. More physically, we have found that a theory with a single $u_+$ spinor has only one helicity. Since parity reverses linear momentum, but leaves angular momentum unaffected, a theory with particles of only one helicity can't be parity invariant.\\

\begin{center}
\textbf{Some guesses about the quantum field $\up$ and the particles it will annihilate and create}
\end{center}

The solution of the Weyl equation going like $e^{-ik\cdot x}$ with $k^0 > 0$ will probably multiply an annihilation operator in the expansion of the quantum field $u_+$. By the known transformation properties of the solution of the field equation, we can obtain the transformation properties of the states it annihilates. We expect
\[ \langle 0 | \up (x) | k \rangle \propto e^{-ik\cdot x} \begin{pmatrix} 1\\ 0 \end{pmatrix} \]
\[ k^2 =0,\quad k^0 >0, \quad k_x = k_y = 0, \quad k_z = k^0 \]

or 
\[ \langle 0 | \up (0) |k \rangle \propto \begin{pmatrix} 1 \\ 0 \end{pmatrix} \]

We are going to find the $J_z$ value of $|k \rangle$:
\[ J_z | k \rangle = \lambda | k\rangle, \qquad \text{ find $\lambda$} \]

In the quantum theory
\[ U(R (\vec{e}_z \, \theta) ) = e^{-i J_z \theta } \]

and thus 
\[ U(R (\vec{e}_z \, \theta) ) | k \rangle = e^{-i\lambda \theta} | k \rangle \]
\[ U(R (\vec{e}_z \, \theta) ) | 0 \rangle = | 0 \rangle \]

On the other hand $U^\dagger \big( R ( \vec{e}_z \theta ) \big) \up(0) U\big( R (\vec{e}_z \, \theta) \big) = D^{(0,1/2)} \big( R (\vec{e}_z \, \theta) \big) \; \up(0) $ and this equation can be used by taking the $ \langle 0 | -|k\rangle $ matrix elements. We get 
\[ e^{-i\lambda \theta } \begin{pmatrix} 1 \\ 0 \end{pmatrix} \propto e^{-i \sigma_z \theta /2 } \begin{pmatrix} 1 \\ 0 \end{pmatrix} = e^{-i\theta/2} \begin{pmatrix}1 \\ 0 \end{pmatrix} \]

i.e. $\displaystyle \lambda = \frac{1}{2} $. The annihilation operator multiplying $\up e^{-ik\cdot x}$, $k_0 > 0$, will annihilate particles with helicity $\frac{1}{2}$ along the direction of motion.\\

(When Weyl came up with this theory, which describes neutrinos, the world thought the world was parity invariant, and his theory was dismissed quickly. People thought he was just playing with irrelevant mathematics).\\

Creation operators in the expansion of $u_+(x)$ will create $\lambda = -\frac{1}{2}$ particles. The field always changes the helicity by the same amount. The field $\upd$ will annihilate particles with helicity $\lambda = - \frac{1}{2}$ and create particles with $\lambda = + \frac{1}{2}$.\\

The annihilation operators in the $u_-$ field annihilate particles with helicity $-\frac{1}{2}$, like neutrinos are observed to have, and the creation operators create particles with helicity $+ \frac{1}{2}$, the antineutrinos. Conventionally, such a field is called ``left handed". By the right hand rule, a ``right handed" field annihilates ``right handed" particles, that is, ones whose angular momentum is along the direction of motion (agrees with I+Z, p.88).\\

Instead of canonically quantizing this theory now, we are going to move on and find a Lagrangian for massive particles that can include parity. The representation the fields are in must be equivalent to the representation obtained by parity. We could look at the irreducible reps of this type, $D^{(n/2,n/2)}$, but when you restrict to rotations, these reps do not contain spinors, in fact as already stated, they turn out to be rank $n$ symmetric traceless tensors under $\SO(3,1)$. The simplest parity invariant reducible rep with spinors is $D^{(1/2,0)} \oplus D^{(0,1/2)}$, a set of two complex doublets, $\up$ and $\um$. We'll restrict the possible Lagrangian by a set of conditions like those at the start of this lecture. \\

$\ml (u_+, u_-, \upd, \umd, \pmu \up, \ldots )$ must be 
\begin{itemize}
\item[(i).] Bilinear
\item[(ii).] Real, at least $S=S^*$ 
\item[(iii).] No more than one derivative.
\item[(iv).] $\ml$ is a Lorentz scalar.
\item[(v).] The Lagrangian has a $U(1)$ symmetry under which $\up$ and $\um$ transform the same way:
\[ u_\pm \longrightarrow e^{i\theta} u_\pm \qquad u_\pm^\dagger \longrightarrow e^{-i\theta} u_\pm^\dagger \]
\item[(vi).] The theory has a parity operation
\[ u_+ (\vec{x},t) \longrightarrow a \; u_-(-\vec{x},t) \]
\[ u_- (\vec{x},t) \longrightarrow b \; u_+(-\vec{x},t) \]
under which $\ml$ is invariant.
\end{itemize}

($D^{(1/2,0)}(\Lambda) \otimes D^{(1/2,0)}(\Lambda) \sim D^{(1,0)}(\Lambda) \oplus D^{(0,0)} (\Lambda) \propto \upd \um$)\\

After rescaling $\up$ and $\um$, the only Lagrangian satisfying the first five conditions is 
\[ \ml = \pm \Big[ i \upd (\partial_0 + \vec{\sigma} \cdot \vec{\nabla}) \up + i \epsilon \umd (\partial_0 - \vec{\sigma} \cdot \vec{\nabla}) \um - m \upd \um - m^* \umd \up \Big] \]

where $\epsilon = \pm 1$.\\

By adjusting the relative phase of $\um$ and $\up$, we can always take $m$ to be real and nonnegative.\\

Let's find out if we can define a parity of the type in condition (vi).
\begin{align*}
\upd (\partial_0+\vec{\sigma}\cdot\vec{\nabla})\up &\longrightarrow \umd (-\vec{x},t)\; a^* \; (\partial_0 + \vs \cdot \vn ) \; a \; \um (-\vec{x},t) \\
&= |a|^2 \umd (-\vec{x},t) \Big(\partial_0 - \vs \cdot \frac{\partial}{\partial(-\vec{x})} \Big) \um (-\vec{x},t) 
\end{align*}

If this term in the parity transformed Lagrangian is going to equal the term in the original Lagrangian
\[ i \epsilon \umd (\partial_0 - \vs \cdot \vn ) \um \]

we must have $\epsilon = + 1$ and $|a|^2 = 1$. By considering the effect of parity on the other terms, you also get 
\[ |b|^2 = 1 \qquad \text{and} \qquad ab^* = 1 \]

The conditions on $a$ and $b$ can be summarized by saying 
\[ a = b = e^{i \lambda} \]
\[ P:\quad u_\pm (\vec{x},t) \longrightarrow e^{i\lambda} u_\mp (-\vec{x},t) \]

Now if a theory has an internal symmetry (and this one has an internal $U(1)$ symmetry), then I can redefine parity to be the old parity composed with any element of the internal symmetry group, and I'll have just as good a definition of parity. In this case, compose parity with the symmetry
\[ u_\pm (x) \longrightarrow e^{-i\lambda} u_\pm (x) \]

Then the new parity has the effect 
\[ P:\quad u_\pm (\vec{x},t ) \longrightarrow u_\mp (-\vec{x},t) \]

The Lagrangian we have found is the Dirac Lagrangian, although it doesn't look like it yet. Let's derive the equations of motion. By varying $\upd$, you get 
\[ i (\po + \vs \cdot \vn ) \up = m \um \]

and by varying $\umd$
\[ i (\po - \vs \cdot \vn ) \um = m \up \]

This is Dirac's equation although it doesn't look like it yet. The solutions of the Dirac equation are solutions of the Klein-Gordon equation. To see this, multiply the first equation by $-i(\po - \vs \cdot \vn)$ and use the second equation to get 
\[ (\po^2 - \vn^2) \up = - m^2 \up \]

or 
\[ (\Box +m^2) \up = 0 \]

This verifies that the thing we have been calling $m$ actually is a mass, and not $m/2$ or $m^{3/5}$ or whatever. \\

We are now going to modify the equations in order to make them more obscure and sophisticated looking. Actually we'll be building a machinery which will speed up calculations. Assemble the two two-component fields into a single four component one.
\[ \psi \equiv \begin{pmatrix} \up \\ \um \end{pmatrix} \]

(This is not the only way to do this, see below).\\

Then 
\[ \ml = \pm \Big[ i \upd (\!\!\!\!\!\!\!\!\!\!\!\!\!\!\!\!\!\!\!\!\!\overbrace{\po}^{\substack{\text{Remember a $2\times2$ identity}\\\text{matrix is supressed here.}}} \!\!\!\!\!\!\!\!\!\!\!\!\!\!\!\!\!\!\!\!\!+ \vs \cdot \vn) \up + i \umd (\po - \vs \cdot \vn) \um - m \upd \um - m \umd \up \Big] \]

can be rewritten as
\[ \ml = \pm \Big[ i \big(\psid \po \psi + \psid \va \cdot \vn \psi \big) - m \psid \beta \psi \Big] \]

where $\va$'s and $\beta$ are \uline{$4 \times 4$} hermitian matrices
\[ \va = \begin{pmatrix} \vs & 0 \\ 0 & - \vs \end{pmatrix} \qquad \beta = \begin{pmatrix} 0 & 1 \\ 1 & 0 \end{pmatrix} \]

in which each entry represents a $2\times 2$ block.\\

In this spiffy notation, we can also write the effect of parity as 
\[ P:\quad \psi (\vx,t) \longrightarrow \beta \psi (-\vx,t) \]

The equation of motion obtained by varying $\psid$ is 
\[ i(\po + \va \cdot \vn) \psi = \beta m \psi \]

This is the Dirac equation (1929).\\

A pleasant surprise is that in this notation we can also give the effect of a Lorentz boost without defining a whole bunch more matrices. The equations
\begin{align*}
\Lambda:\quad &\up \longrightarrow e^{\vs \cdot \ve \phi /2} \up \\
& \um \longrightarrow e^{-\vs \cdot \ve \phi /2} \um
\end{align*}

can be assembled into 
\[ \Lambda:\psi \longrightarrow e^{\va \cdot \ve \phi /2} \psi \]

The generator of boosts (what is dotted into $-i\ve \phi$ in the exponential) is 
\[ \vec{M} = \frac{i\va}{2} \]

We can get the generators of rotations quickly by using 
\[ [M_i, M_j] = - i\eijk L_k \]
\[ \vec{L} = \frac{1}{2} \begin{pmatrix} \vs & 0 \\ 0 & \vs \end{pmatrix} \]

I said $ \psi = \begin{pmatrix} \up \\ \um \end{pmatrix}$ is not the only way of making a four component spinor out of two two-component spinor. Another way is 
\[ \psi = \bpm \up + \um \\ \up - \um \epm \frac{1}{\sqrt{2}} \]

If we had done that, the $\va$'s and $\beta$ would have come out differently. In fact this second way is the way Dirac originally did it.\\

To summarize, in the first basis, called the \uline{WEYL BASIS}\footnote{usually called the Weyl representation, but the terminology is incorrect}
\[ \va = \bpm \vs & 0 \\ 0 & - \vs \epm \qquad \beta = \bpm 0 & 1 \\ 1 & 0 \epm \qquad \psi = \bpm \up \\ \um \epm \]
\[ \vm = \frac{i\va}{2} \qquad [M_i,M_j] = - i\eijk L_k \Longrightarrow \vl = \frac{1}{2} \bpm \vs & 0 \\ 0 & \vs \epm \]

\uline{DIRAC (OR STANDARD) BASIS}
\[ \va = \bpm 0 & \vs \\ \vs & 0 \epm \qquad \beta = \bpm 1 & 0 \\ 0 & -1 \epm \qquad \psi = \frac{1}{\sqrt{2}} \bpm \up + \um \\ \up - \um \epm \]
\[ \vm =\frac{i \va}{2} \qquad \vl = \frac{1}{2} \bpm \vs & 0 \\ 0 & \vs \epm \]

In either basis, the Dirac Lagrangian is 
\[ \ml = \pm \Big[ i \psid (\!\!\!\!\!\!\!\!\!\!\!\!\underbrace{\po}_{\substack{4\times 4 \text{ identity}\\\text{matrix suppressed}}}\!\!\!\!\!\!\!\!\!\!\!\!\!\! + \va \cdot \vn) \psi - m \psid \beta \psi \Big] \]

The Dirac basis is the standard basis because the solutions of the Dirac equation become especially simple in the nonrelativistic limit in this basis.\\

\begin{center}
\textbf{PLANE WAVE SOLUTIONS OF THE DIRAC EQUATION}
\end{center}

We are going to look for solutions\footnote{There are extensive discussions of the solutions of the Dirac equation in the literature. All you have to do is ignore all references to holes and negative solutions. Just because something was understood in a poor way 50 years ago, doesn't mean you have to learn it that way today.} of the form 
\[ \psi = \uvp \, e^{-ip\cdot x} \qquad \text{or} \qquad \psi = \vvp \, e^{i p\cdot x} \]

where $\uvp$ and $\vvp$ are space-time independent four component spinors. Of course, to have a chance of satisfying the Dirac equation a proposed solution must satisfy the Klein-Gordon equation, so $p^2=m^2$, or 
\[ p^0 \equiv \sqrt{\vp^2 + m^2} \]

$p$ is a forward pointing vector on the mass shell.\\

(The other solution, $p^0 = - \sqrt{\vp^2 + m^2}$, is taken care of in the way we have busted up the problem into two cases: one with $e^{-ip\cdot x}$ dependence and one with $e^{+ip\cdot x}$ dependence.)\\

If you plug the first type of solution,
\[ \psi = \uvp \, e^{-ip\cdot x} \]

(called ``positive frequency", a convention that goes all the way back to $H\psi = i\hbar \frac{\partial \psi}{dt} \longrightarrow \psi_n = e^{-i \omega_n t}$) into the Dirac equation, 
\[ i(\po + \va \cdot \vn ) \psi = \beta m \psi \]

you get $(p^0 - \va \cdot \vp) \uvp = \beta m \uvp $.\\

Let's look at a special case: $\vp = 0$, $p^0 =m$. \\

Then the equation says
\[ u_{\vec{0}} = \beta u_{\vec{0}} \]

In the standard basis, $\beta = \bpm 1 & 0 \\ 0 & -1 \epm$ (each entry represents a $2\times2$ block), and there are two linearly independent solutions to this equation, they are of the form 
\[ u_{\vec{0}} = \bpm a \\ b \\ 0 \\ 0 \epm \]
}{
 \sektion{22}{December 11}
\descriptiontwentytwo
I can choose the direction my two standard linearly independent solutions point in the subspace they are allowed in, as well as their total normalization, any way I like. A convenient choice is 
\[ u_{\vec{0}}^{(1)} = \sqrt{2m} \bpm 1 \\ 0 \\ 0 \\ 0 \epm \]
\[ u_{\vec{0}}^{(2)} = \sqrt{2m} \bpm 0 \\ 1 \\ 0 \\ 0 \epm \]

Whatever basis you choose, make it satisfy
\[ \uvo^{(r)\dagger} \uvo^{(s)} = 2m \delta_{rs} \]

and 
\[ \uvo^{(r)\dagger} \va \uvo^{(s)} = 0 \]

The second condition follows fairly easily from the form of $\va$. In the standard basis it connects the upper two components with the lower two.\\

By arguments like those used for the solutions of the Weyl equation, we expect that in the expansion of the quantum field $\psi$, an annihilation operator that annihilates $J_z = +\frac{1}{2}$ electrons with momentum $p$ will multiply 
\[ \uvp^{(1)} \eipxm \]

and an annihilation operator that annihilates electrons with $J_z =-\frac{1}{2}$ will multiply 
\[ \uvp^{(2)} \eipxm \]

The $2m$ in the normalization of the $u$'s is there to agree with conventional normalization for relativistic states:
\[ \langle p' | p \rangle = (2\pi )^3 2 \!\!\!\!\!\!\!\!\!\underbrace{\omega_{\vp}}_{\substack{\text{reduces to } 2m \\ \text{when }\vec{p} =0 }}\!\!\!\!\!\!\!\!\delta^{(3)} (\vp - \vp\,') \]

In many equations it will allow us to take a smooth $m \rightarrow 0$ limit. (Good for neutrinos or extremely high energy physics).\\

The Dirac Lagrangian was constructed to be a Lorentz scalar, so you expect that the equations of motion derived from it are Lorentz invariant in the sense that given one solution of the Dirac equation, I ought to get another solution by Lorentz transforming it. So while we could just go ahead and solve the Dirac equation for arbitrary $\vp$ (it's just the problem of finding the eigenvalues and eigenspinors of some $\vp$ dependent $4 \times 4$ matrix), we will flaunt Lorentz invariance by getting solutions with momentum $\vp$ by boosting those with momentum $\vec{0}$.\\
\[ \uvp^{(r)} = e^{\va \cdot \ve \phi /2} \uvo^{(r)} \]

where $\ve = \frac{\vp}{|\vp|}$ and $\sinh \phi = \frac{| \vp|}{m} $ ($\cosh \phi = \frac{E}{m}$).\\

You can mechanically verify that the conditions
\[ \uvp^{(r)\dagger} \uvp^{(s)} = 2p^0 \delta_{rs} \qquad \uvprd \va \uvps = 2 \vp \delta_{rs} \]

are satisfied, but it is actually not necessary.\\

$(\uvprd \uvps, \uvprd \va \uvps)$ is a four vector, and you know it is $(2m \delta_{rs}, \vz )$ when $\vp =0$: that determines it completely for arbitrary $\vp$. (This proof is sweet because it is basis independent.)\\

Using $(\va \cdot \ve)^2 = 1$, you can rewrite 
\[ \uvpr = \Big[ \!\!\!\!\!\!\!\!\!\!\!\!\!\underbrace{\cosh \frac{\phi}{2}}_{4 \times 4 \text{ identity suppressed}} \!\!\!\!\!\!\!\!\!\!\!\!\!\!\!+ \va \cdot \ve \sinh \frac{\phi}{2} \Big] \uvo^{(r)} \]

Using $\cosh \frac{\phi}{2} = \sqrt{\frac{1 + \cosh \phi}{2}}$ and $\sinh \frac{\phi}{2} = \sqrt{\frac{\cosh \phi - 1}{2}}$ and $\cosh \phi = \frac{E}{m}$, you can rewrite
\[ \uvpr = \Big[ \sqrt{\frac{E+m}{2m}} +\sqrt{\frac{E-m}{2m}} \va \cdot \ve \Big] \uvo^{(r)} \]

In the standard basis, with $\vp$ (hence $\ve$) pointing in the $z$ direction, this is 
\[ \uvp^{(1)} = \bpm \sqrt{E+m} \\ 0 \\ \sqrt{E-m} \\ 0 \epm \qquad \uvp^{(2)} = \bpm 0\\\sqrt{E+m} \\ 0 \\ - \sqrt{E-m} \epm \]

(The normalization is already proving useful. Thanks to that factor of $\sqrt{2m}$, this doesn't blow up when $m \rightarrow 0$.)\\

A similar set of relations holds for the $\vvp^{(r)}$
\[ \vvo^{(1)} = \sqrt{2m} \bpm 0 \\ 0 \\ 1 \\0 \epm \qquad \vvo^{(2)} = \sqrt{2m} \bpm 0 \\ 0 \\ 0 \\ 1 \epm \]
\[ \vvpr = e^{\va \cdot \ve \phi /2 } \vvo^{(r)} \]
\[ (\vvprd \vvps, \vvprd \va \vvps ) = ( 2p^0 \delta_{rs}, 2 \vp \delta_{rs}) \]

In the standard basis, with $\vp$ in the $z$ direction.
\[ \vvp^{(1)} = \bpm \sqrt{E-m} \\0 \\ \sqrt{E+m} \\ 0 \epm \qquad \vvp^{(2)} = \bpm 0 \\ - \sqrt{E-m} \\0 \\ \sqrt{E+m} \epm \]

The $\va$'s and $\beta$ satisfy a simple algebra.
\[ \alpha_i^2 = 1 \qquad \Big\{ \alpha_i, \alpha_j \Big\} = 0 \qquad i \neq j \]
\[ \beta^2 = 1 \qquad \Big\{ \beta, \alpha_i \Big\} = 0 \]

Every $\va$ and $\beta$ squares to $1$ and anticommutes with all three others.\\
\vspace{1cm}

\begin{center}
\textbf{A famous theorem due to Pauli}
\end{center}

Any set of 4 $4 \times 4$ matrices obeying these equations is equivalent to any other set.\\

The theorem says ``everything is in here." Anything we get by manipulating a set of $4 \times 4$ matrices satisfying this algebra can be obtained by manipulating the algebra.\\

``Anything" means any equation that is unaffected by a similarity transformation, or any result that is basis independent, like a cross section summed over final spin states and averaged over initial ones.\\

An example of a statement that is not basis independent is 
\[ \beta^\dagger = \beta \]

This is true in the Weyl or standard basis but in general, just because $\beta$ is hermitian, it does not follow that
\[ S^{-1} \beta S \qquad \text{is hermitian} \]

It is true though if $S$ is unitary. Sometimes we'll restrict ourselves to bases that are related by a unitary $S$. They are called ``unitarily equivalent". \\

Coleman's proof uses something we already believe (but in fact takes some effort to prove): That up to equivalences we have found all the finite dimensional reps of the Lorentz group.\\

Start by constructing a representation of the Lorentz group from the $\va$'s and $\beta$.\\

Let $M_i \equiv i \frac{\alpha_i}{2}$. Define $L_k$ by 
\[ \Big[M_i, M_j\Big] = - i \eijk L_k \quad (L_k = \frac{i}{2} \eijk \Big[M_i, M_j\Big]) \] 

Using the algebra the $\alpha$'s are supposed to obey, it is easy to show that
\[ \Big[ L_i,M_j \Big] = i \eijk M_k \quad \text{and} \quad \Big[ L_i, L_j \Big] = i \eijk L_k \]

Thus we have defined a representation of the Lorentz group. Furthermore, it is a four dimensional representation, and the rotation generators square to $\frac{1}{4}$, $L_i^2 = \frac{1}{4}$. Thus it must be made of just spin $\frac{1}{2}$ reps when you restrict this four dimensional rep of the Lorentz group to the rotation subgroup. What we have made must be equivalent to 
\[ D^{(1/2,0)} \oplus D^{(0,1/2)}, \quad D^{(0,1/2)} \oplus D^{(0,1/2)} \quad \text{ or} \quad D^{(1/2,0)} \oplus D^{(1/2,0)} \]

But now the existence of $\beta$ can be used to rule out the second two possibilities. Note that $\beta^2 = 1 \Longleftrightarrow \beta = \beta^{-1}$ so using the algebra 
\[ \beta^{-1} \va \beta = \beta \va \beta = -\va \]

For our generators this implies 
\begin{equation}\label{eq:22-page5-1}
\beta^{-1} \vm \beta = - \vm \qquad \text{and} \qquad \beta^{-1} \vl \beta = \vl 
\end{equation}

Recall the stuff about parity: Given a representation of the Lorentz group, $D$, I can define a new rep $D_P$ by
\[ D_P (\Lambda) = D(\Lambda_P) \]

The generators of the rotations in this new rep were the same. The generators of boosts in $D_P$ were minus the generators of boosts in $D$. In general the representation obtained from
\[ D^{(n/2, m/2)} \qquad \text{was} \qquad D^{(m/2,n/2)} \]

There are equivalent only if $m=n$. The parity transform of 
\[ D^{(1/2,0)} \oplus D^{(1/2,0)} \qquad \text{is} \qquad D^{(0,1/2)} \oplus D^{(0,1/2)} \]

and vice versa. These (reducible) reps are not equivalent to their parity transformed reps. But $\beta$, by Eq.~(\ref{eq:22-page5-1}) is such an equivalence. These reps cannot be candidates for what we have constructed.\\

We must have constructed the rep $D^{(1/2,0)} \oplus D^{(0,1/2)}$ (it is equivalent to its parity transform), and there is only one such rep up to equivalence transformation.\\

A jargony way of saying what we have found is:\\

There is only one rep of $\SO (3,1)$ plus parity that is four-dimensional and only contains spin $\frac{1}{2}$ particles.

A little bit of the proof remains to be done. We have shown the $\va$'s are always equivalent, but it remains to be shown that $\beta$ is equivalent by the same transformation.\\

So suppose I have found a similarity transformation that puts the $\va$'s into standard form
\[ \va = \bpm \vs & 0 \\ 0 & - \vs \epm \]

Can I find a further similarity transformation that leaves the $\va$'s unchanged, but brings $\beta$ into standard form? If I could, this would show that any set of $\va$'s and $\beta$ is equivalent to any other, since they are all equivalent to a standard form. With $\va = \bpm \vs & 0 \\ 0 & - \vs \epm $, the algebra
\[ \Big\{ \va, \beta \Big\} = 0 \Longrightarrow \beta = \bpm 0 & \lambda_2 \\ \lambda_1 & 0 \epm \]

where the $\lambda_1$, $\lambda_2$ are blocks proportional to a $2 \times 2$ identity.\\

(This is fairly easy to show. Write $\beta = \bpm A & B \\ C & D \epm$ and find out what the conditions are on each of the $2 \times 2$ matrices $A,B,C$ and $D$. No matrix anticommutes with all three Pauli matrices, and the only matrix that commutes with all three of them is the identity.)\\

The other condition on the $\beta$ from the algebra is $\beta^2 =1$, which implies 
\[ \lambda_1 \lambda_2 = 1 \qquad \text{or} \qquad \lambda = \lambda_1 = \lambda_2^{-1} \]

So $ \va = \bpm \vs & 0 \\ 0 & - \vs \epm $ and $\beta = \bpm 0 & \lambda^{-1} \\ \lambda & 0 \epm $ and we need to find a similarity transformation that leaves the $\va$'s unaffected but puts $\beta$ into the standard form
\[ \beta = \bpm 0 & 1 \\ 1 & 0 \epm \] 

The similarity transformation is $ S = \bpm \lambda^{-1} & 0 \\ 0 & 1 \epm $
\begin{align*}
S^{-1}\beta S &= \bpm \lambda & 0 \\ 0 & 1 \epm \bpm 0 & \lambda^{-1} \\
\lambda & 0 \epm \bpm \lambda^{-1} & 0 \\ 0 & 1 \epm \\
&= \bpm \lambda & 0 \\ 0 & 1 \epm \bpm 0 & \lambda^{-1} \\ 1 & 0 \epm = \bpm 0 & 1 \\ 1 & 0 \epm 
\end{align*}

A lot of people including some Nobel laureates did lengthy calculations using explicit representations of the Dirac algebra in the 1930's. This must have been unnecessary since the whole thing is tied up in the commutation relations.\\

From now on, we are going to assume we are in a basis where
\[ \va = \va^\dagger \qquad \text{and} \qquad \beta = \beta^\dagger \]

A basis which is obtained from this basis by a unitary transformation will also satisfy these relations. All popular representations satisfy these relations.
\vspace{1cm}

\begin{center}
\textbf{DIRAC\footnote{due to Pauli} ADJOINT, PAULI-FEYNMAN NOTATION}
\end{center}

Since $\psid \beta \psi$ is a Lorentz invariant ($-m \psid \beta \psi$ appears in the Lagrangian), that is since under 
\[ \Lambda:\quad \psi \longrightarrow D(\Lambda) \psi \qquad \psid \beta \psi \longrightarrow \psid \beta \psi \]

We are going to define a new adjoint 
\[ \psib = \psid \beta \] 

This new adjoint has every property you'd like an adjoint to have except that $\psib \psi$ is not always greater than zero.\\

Then we can write, oh so slickly
\[ \Lambda: \psib \psi \longrightarrow \psib \psi \]

(or $\Lambda:\overline{\chi} \psi \longrightarrow \overline{\chi} \psi$ for two Dirac spinors )\\

[ The situation is a lot like $\SO(3,1)$. The usual inner product between two four vectors 
\[ y^{T} x = \sum_\mu y^\mu x^\mu \]

is not a Lorentz invariant. The combination that is is
\[ y^T g x \qquad g=\bpm 1 & 0 & 0 & 0 \\ 0 & -1 & 0 & 0 \\ 0 & 0 & -1 & 0 \\ 0 & 0 & 0 & -1 \epm \]

So we define a ``new transpose" $y^T g$ and call it a covariant vector, and put its indices down. Then we can write, oh so slickly, 
\[ \Lambda:\sum_\mu y_\mu x^\mu \longrightarrow \sum_\mu y_\mu x^\mu. \] 

As a statement about the $ 4 \times 4$ matrices $\Lambda$ this is 
\[ \Lambda^T g \Lambda =g \qquad \text{or} \qquad g \Lambda^T g \Lambda = 1] \]

The definition of the adjoint of an operator is obtained from the definition of the adjoint of a vector by 
\[ \phi^\dagger A^\dagger \psi \equiv (\psi^\dagger A \phi)^* \]

That tells you an arbitrary matrix element of $A^\dagger$. Similarly, the Dirac adjoint of an operator is obtained by 
\[ \overline{\phi} \, \overline{A} \psi \equiv ( \psib A \phi)^* \]

It is the work of a moment to show that 
\[ \overline{A} = \beta A^\dagger \beta \]

All your favorite equations for the adjoint of an operator follow for the Dirac adjoint, and the proofs are the same since it all comes from the definition of the adjoint.
\[ \overline{ \alpha A + \beta B} = \alpha^* \overline{A} + \beta^* \overline{B} \]
\[ \overline{AB} = \overline{B} \, \overline{A} \]
\[ \overline{A \psi} = \overline{\psi} \, \overline{A} \]

This last relation implies that
\[ \Lambda:\psib \psi \longrightarrow \psib \,\overline{D(\Lambda)} D(\Lambda) \psi \]

but this equals $\psib \psi$ so $\overline{D(\Lambda)} D(\Lambda) = 1 $ is the statement about the $ 4 \times 4$ matrices $D(\Lambda)$. They aren't unitary, but they are ``Dirac unitary". 

Remember that 
\[ V^\mu = (\chi^\dagger \psi, \chi^\dagger \va \psi) \]

transforms like a four-vector. Here comes some more notation to make this look slick too. We can rewrite $V^\mu$ as 
\[ V^\mu = ( \overline{\chi} \beta \psi, \overline{\chi} \beta \va \psi ) = \overline{\chi} \gamma^\mu \psi \]

where
\[ \gamma^\mu = (\beta, \beta \va ) \]

These are the famous Dirac $\gamma$ matrices.\\

With a slight abuse of language, we can say that the $\gamma$ matrices transform like a vector. Of course, they don't transform at all. What is meant by this, no more, no less, is that 
\begin{equation}\label{eq:22-page11-1}
\overline{D(\Lambda)} \gamma^\mu D(\Lambda) = \Lambda^\mu{}_{\nu} \gamma^\nu 
\end{equation}

We can also say that the product $\gamma^\mu \gamma^\nu$ transforms like a tensor. The proof is:
\begin{equation}\label{eq:22-page11-2}
\begin{split}
\overline{D(\Lambda)}\gamma^\mu \gamma^\nu D(\Lambda) &= \overline{D(\Lambda)} \gamma^\mu \!\!\!\!\!\!\!\underbrace{D(\Lambda) \overline{D(\Lambda)}}_{\text{fancy way of inserting 1}} \!\!\!\!\!\!\!\! \gamma^\nu D(\Lambda) \\
&= \Lambda^\mu{}_{\sigma} \gamma^\sigma \Lambda^\nu{}_{\tau} \gamma^\tau \\
&= \Lambda^\mu{}_{\sigma} \Lambda^\nu{}_{\tau} \gamma^\sigma \gamma^\tau 
\end{split}
\end{equation}

This is the transformation law for a two index tensor. \\

The anti-commutation relations for the $\gamma$ matrices follow from those for the $\va$'s and $\beta$. They can all be summed up in 
\[ \Big\{ \gamma^\mu, \gamma^\nu \Big\} = 2g^{\mu\nu} \]

which you should check.\footnote{If in some math book you start reading about Clifford algebras, it is a special case of them we are studying. More generally, $\mu,\nu = 1,\ldots, N$ and any $\#$ of diagonal components of $g^{\mu\nu}$ can be $-1$.}

Also, in the special class of bases we have restricted ourselves to 
\[ \gamma^{0 \dagger} = \gamma^0 \qquad \gamma^{i \dagger} = - \gamma^i \]

An elegant way of summarizing these four relations is 
\[ \overline{\gamma^\mu} = \gamma^\mu \]

which you could check, but here is a high-powered proof instead:\\

In the same sense as in Eqs.~(\ref{eq:22-page11-1}) and (\ref{eq:22-page11-2}), both sides of this equation transform like a four-vector. (The RHS we have already accepted this for;
\[ \overline{D(\Lambda)} \gamma^\mu D(\Lambda) = \Lambda^\mu{}_{\nu} \gamma^\nu. \]

The LHS transforms the same way as you can see by taking the bar of this equation to get
\[ \overline{D(\Lambda)} \, \overline{\gamma^\mu} D(\Lambda) = \Lambda^\mu{}_\nu \overline{\gamma^\nu}) \]

($ \Lambda^\mu{}_{\nu} $: This matrix is unaffected. It is real, and it is not transposed because we are transposing only in the spinor indices. Maybe I should first say, for any given $\mu$, this is just a set of $4$ real coefficients.) \\

So $\overline{\gamma^\mu} = \gamma^\mu $ is a Lorentz covariant equation. To see if it is correct we only have to check one of its components, say $\mu = 0$. For $\mu =0 $ it reduces to $\gamma^0 \gamma^{0\dagger} \gamma^0 = \gamma^0 $, $\checkmark\checkmark$.\\

Now that we have these Dirac $\gamma$ matrices, we can rewrite the Dirac Lagrangian as 
\[ \ml = \pm \Big[ i \psib \gamma^\mu \pmu \psi - m \psib \psi \Big] \]

The equation of motion is $i \gamma^\mu \pmu \psi - m \psi = 0 $ (from varying $\psib$).\\

We can make this look even more sophisticated and obscure by introducing a super compact notation due to Feynman. Let 
\[ \cancel{a} \equiv a_\mu \gamma^\mu \quad ( = a^\mu \gamma_\mu = a\cdot \gamma) \]

The algebra of the $\gamma$ matrices can be summarized in this notation as 
\[ \Big\{\cancel{a}, \cancel{b} \Big\} = 2 a \cdot b \qquad (\cancel{a}^2 = a^2) \]

The Dirac Lagrangian and equation of motion are 
\[ \ml = \pm \psib ( i \notd - m ) \psi \qquad (i \notd - m) \psi = 0 \]

The proof that each component of every solution of the Dirac equation satisfies the Klein-Gordon equation is 
\[ (i\notd - m) \psi = 0 \Longrightarrow (-i \notd - m ) (i \notd - m) \psi = 0 \]
\[ \Longrightarrow (\Box + m^2 ) \psi = 0 \]
\vspace{1cm}

\begin{center}
\textbf{Parity and $\gamma_5$}
\end{center}
\[ P:\quad \psi(\vx,t) \longrightarrow \!\!\!\!\!\!\!\!\underbrace{\beta}_{\substack{\text{could also be}\\\text{written } \gamma^0}} \!\!\!\!\!\!\!\! \psi (-\vx,t) \] 
\[ \psib (\vx,t) \longrightarrow \overline{\beta \psi (-\vx,t)} = \psib(-\vx,t) \overline{\beta} = \psib (-\vx,t)\beta \]

So
\[ P: \psib \psi(\vx,t) \longrightarrow \psib \psi(-\vx,t) \]

Not only is $\psib \psi$ a scalar under the Lorentz transformations connected to the identity, it is a scalar under parity.
\begin{align*} 
P: \quad \psib \gamma^\mu \psi &\longrightarrow \psib \beta \gamma^\mu \beta \psi \\
&= \begin{cases} \psib \gamma^0 \psi & \mu = 0 \\ -\psib \gamma^i \psi & \mu = i \end{cases}
\end{align*}

This is how you expect a vector to transform under parity.\\

What about $\psib \gamma^\mu \gamma^\nu \psi $? It's clear that the $00$ component will be unaffected by parity, as will the $ii$ components, while the $0i$ components will go into minus themselves. This is a tensor under L.T.~and parity. Actually we have obtained nothing new from the $00$ and $ii$ components because $\gamma^{\mu \, 2} =1$. The only new quantities we have are the antisymmetric parts. Define
\[ \sigma^{\mu\nu} = \frac{1}{2i} \Big[ \gamma^\mu, \gamma^\nu \Big] \qquad \overline{\sigma^{\mu\nu}} =\sigma\mn \]

$ \psib \sigma\mn \psi $ is an antisymmetric tensor under L.T.~and parity.\\

We can proceed on building tensors of higher and higher rank. In the product of two four component objects there are only $16$ possible bilinears. So far we have found
\[ \underbrace{1}_{\text{scalar}} + \underbrace{4}_{\text{vector}} + \underbrace{6}_{\text{antisymmetric tensor}} = 11 \qquad \text{of them} \]

Let's jump up to tensors of the fourth rank and see what we can make
\[ \psib \gamma^\mu \gamma^\nu \gamma^\alpha \gamma^\beta \psi \]

is a fourth rank tensor, but if any two of the indices are the same, this reduces to something we have already found. There is only one possibility if all four indices have to be different, it is
\[ \gamma^0 \gamma^1 \gamma^2 \gamma^3 \]

It is conventional to define 
\begin{align*}
\gamma_5 &\equiv i \gamma^0 \gamma^1 \gamma^2 \gamma^3 \qquad (\equiv \gamma^5) \\
&= \frac{i}{4!} \epsilon_{\mu \nu \alpha \beta} \gamma^\mu \gamma^\nu \gamma^\alpha \gamma^\beta \qquad (\epsilon_{0123} = + 1)
\end{align*}

Unfortunately, because of the $i$, $\overline{\gamma_5} = - \gamma_5 $, but $\gamma_5^\dagger = \gamma_5$ and $\gamma_5^2 = \gamma_5 \gamma_5^\dagger = +1 $.\\

Except for $\overline{\gamma_5} = -\gamma_5$, $\gamma_5$ is a lot like a fifth $\gamma$ matrix, that is 
\[ \Big\{ \gamma_5, \gamma^\mu \Big\} = 0 \]

We'll write $i\gamma_5$, because $\overline{i \gamma_5} = i \gamma_5 $. 

Now $\epsilon_{\mu\nu\alpha\beta} \gamma^\mu \gamma^\nu \gamma^\alpha \gamma^\beta $ transforms like a scalar under L.T.~but goes into minus itself under parity, that is
\[ P:\psib i \gamma_5 \psi \longrightarrow - \psib i \gamma_5 \psi \qquad \text{ is a pseudoscalar} \]

$\psib i \gamma_5 \psi $ is hermitian, it can appear in a Lagrangian with a real coefficient.\\

$ \psib \gamma^\mu \gamma_5 \psi $ is also hermitian and 
\[ P:\quad \psib \gamma^\mu \gamma_5 \psi \longrightarrow \begin{cases} - \psib \gamma^0 \gamma_5 \psi & \mu = 0 \\ \psib \gamma^i \gamma_5 \psi & \mu = i \end{cases} \qquad \text{an axial vector} \]

So now we have found a total of $16$ bilinears transforming in distinct ways under parity and L.T.
\begin{align*}
&S &1 \qquad &\text{scalar} \\
&P &1 \qquad &\text{pseudoscalar} \\ 
&V &4 \qquad &\text{vector} \\
&A &4 \qquad &\text{axial vector}\\
&T &6 \qquad &\text{antisymmetric tensor}
\end{align*}

(Any other bilinear we might construct must be expressible in terms of these.)\\

We could start building Lagrangians with interactions like (but we are going to proceed with canonical quantization)\\
\begin{itemize}
\item $g \phi \psib i \gamma_5 \psi$ (to conserve parity $\phi$ must be a pseudoscalar)\\

\item or $g \phi \psib \psi$ or $g \pmu \phi \psib \gamma^\mu \psi$ (under parity $\phi$ would be a scalar)\\

\item or $g \phi \psib i \gamma_5 \psi + h \phi \psib \psi$ (parity violating, no choice of parity is possible).\\
\end{itemize}

Some things that are very useful when deriving any basis independent result are \uline{orthogonality and completeness} conditions for the $u^{(r)}$ and $v^{(r)}$ \\

The $\uvpr$ and $\vvpr$ satisfy
\[ (\notp - m) \uvpr = 0 \qquad (\notp + m)\vvpr = 0 \]

Taking the bar of these equations we also have 
\[ \overline{\uvp}^{(r)} (\notp - m ) = 0 \qquad \overline{\vvp}^{(r)}(\notp + m) = 0 \]
\vspace{1cm}

\begin{center}
\textbf{ORTHOGONALITY CONDITIONS}
\end{center}

We have already derived
\[ \overline{\uvp}^{(r)} \gamma^\mu \uvps = 2 p^\mu \delta_{rs} \qquad \text{and} \qquad \overline{\vvp}^{(r)} \gamma^\mu \vvps = 2 p^\mu \delta_{rs} \]

Now $\overline{\vvp}^{(r)} \gamma^\mu \uvps$ is also a four vector, and we can find what it is by looking at 
\[ \overline{v_{\vec{0}}}^{(r)} \gamma^0 u_{\vec{0}}^{(s)} = v_{\vec{0}}^{(r)\dagger} u_{\vec{0}}^{(s)} = 0 \]

so
\[ \overline{\vvp}^{(r)} \gamma^\mu \uvps = \overline{\uvp}^{(r)} \gamma^\mu \vvps = 0 \]

There are also the scalars
\[ \overline{\uvp}^{(r)} \uvps \qquad \overline{\vvp}^{(r)} \uvps \qquad \text{and} \qquad \overline{\vvp}^{(r)} \vvps \]

Evaluating them for $\vp =0 $ is easy, and since they are scalar, that gives their value for general $\vp \qquad$ ($u$'s are $\beta = + 1$ eigenstates, $v$'s are $\beta = - 1$ eigenstates).
\[ \overline{\uvp}^{(r)} \uvps = 2 m \delta_{rs} \qquad \overline{\vvp}^{(r)} \vvps = - 2m \delta_{rs} \]
\[ \overline{\uvp}^{(r)} \vvps = \overline{\vvp}^{(r)} \uvps = 0 \]
\vspace{1cm}

\begin{center}
\textbf{COMPLETENESS RELATIONS}
\end{center}

Suppose I have an orthogonal normalized basis for $\mathbb{R}^n$, 
\[ \ve\,^{(r)} \qquad r=1,\cdots,n \]

Then $\sum_r \ve\,^{(r)} \ve\,^{(r)T}$ is the identity matrix. We are going to get the analog of this for our $4$ solutions of the Dirac equation for any $p$. Define
\[ A = \sum_r \uvpr \overline{\uvp}^{(r)} \]

Let's see what $A$ is by seeing what it does to a basis for our $4$-dim spinor space.
\[ A \uvps = \sum_r \uvpr 2m \delta_{rs} = 2m \uvps \]
\[ A \vvps = \sum_r \uvpr \cdot 0 = 0 \]

But we already know a matrix that has this effect on the basis, $\notp + m$, so 
\[ \sum_r \uvpr \overline{\uvp}^{(r)} = \notp + m \]

Similarly,
\[ \sum_r \vvpr \overline{\vvp}^{(r)} = \notp - m \]
\vspace{1cm}

\begin{center}
\textbf{What Every 253a Student Needs to Know about The Dirac Equation}
\end{center}

I have heard that some of you have had trouble keeping the Dirac equation in view through a cloud of $\SO (3,1)$ representation theory. This sheet has been prepared to help you. It contains \uline{all} results we have derived to date that we will need in the remainder of the course, \uline{without proofs}.
\begin{itemize}
\item[(1).] \uline{Dirac Lagrangian, Dirac Equation, Dirac Matrices} \\

The theory is defined by the Lagrange density,
\[ \ml = \psi^\dagger [ i \partial_0 + i \va \cdot \vec{\nabla} - \beta m ] \psi. \]

Here $\psi$ is a set of four complex fields, arranged in a column vector (\uline{a Dirac bispinor}), and the $\alpha$'s and $\beta$ are a set of four $4\times 4$ hermitian matrices (\uline{the Dirac matrices}). The equation of motion (\uline{the Dirac equation}) is 
\[ (i \partial_0 + i \va \cdot \vec{\nabla} - \beta m) \psi = 0. \]

The Dirac matrices obey \uline{the Dirac algebra},
\[ \Big\{ \alpha_i, \alpha_j \Big\} = 2 \delta_{ij}, \quad \Big\{\alpha_i, \beta \Big\} = 0, \qquad \beta^2 = 1. \]

Any set of $ 4 \times 4 $ matrices obeying this algebra is equivalent to any other set. Two representations of the Dirac algebra that will be useful to us are the \uline{Weyl representation},
\[ \va = \bpm \vs & 0 \\ 0 & -\vs \epm, \qquad \beta = \bpm 0 & 1 \\ 1 & 0 \epm, \]

and the \uline{standard representation},
\[ \va = \bpm 0 & \vs \\ \vs & 0 \epm, \qquad \beta = \bpm 1 & 0 \\ 0 & -1 \epm. \]
\vspace{1cm}

\item[(2.)] \uline{Space-Time Symmetries}\\

The Dirac equation is invariant under both Lorentz transformation and parity.\\

Under a Lorentz transformation characterized by a $4 \times 4$ Lorentz matrix, $\Lambda$,
\[ \Lambda:\quad \psi(x) \longrightarrow D(\Lambda) \psi(\Lambda^{-1} x), \]

where the matrix $D$ is defined from the $\alpha$'s by the following rules:\\

For an acceleration by rapidity $\phi$ in a direction $\ve$,\footnote{$\alpha^i = \gamma^0 \gamma^i$, follows from $\gamma^0 = \beta, \gamma^i = \beta \alpha^i$ and $\beta^2 = 1$}
\[ D\big(A(\ve \phi)\big) = e^{\va \cdot \ve \phi /2}. \]

For a rotation by angle $\theta$ about an axis $\ve$ 
\[ D\big(R(\ve \, \theta)\big) = e^{-i \vl \cdot \ve \, \theta}, \]

where $\vl$ is defined by\footnote{$L_k = \frac{i}{4} \epsilon_{kij} \gamma^i \gamma^j $}
\[ \Big[ \alpha_i, \alpha_j \Big] = 4 i \eijk L_k. \]

In both the Weyl and standard representations 
\[ \vl = \frac{1}{2} \bpm \vs & 0 \\ 0 & \vs \epm. \]

Under parity, 
\[ P:\quad \psi (\vx, t) \longrightarrow \beta \psi (-\vx,t). \]
\vspace{1cm}

\item[(3.)] \uline{Dirac Adjoint, $\gamma$ Matrices} \\

The \uline{Dirac adjoint} of a Dirac bispinor is defined by 
\[ \psib = \psi^\dagger \beta, \]

of a $4 \times 4$ matrix by 
\[ \overline{A} = \beta A^\dagger \beta. \]

These obey the usual rules for adjoints, e.g., 
\[ (\psib A \phi)^* = \overline{\phi} \, \overline{A} \psi, \]

The \uline{$\gamma$ matrices} are defined by 
\[ \gamma^0 = \beta, \qquad \gamma^i = \beta \alpha^i, \]

These are not all hermitian,
\[ \gamma^{\mu\dagger} = \gamma_\mu \equiv g_{\mu\nu} \gamma^\nu, \]

but they are self-Dirac adjoint (``\uline{self-bar}"),
\[ \overline{\gamma}^\mu = \gamma^\mu. \]

The $\gamma$ matrices obey the \uline{$\gamma$ algebra},
\[ \Big\{ \gamma^\mu, \gamma^\nu \Big\} = 2g^{\mu\nu}. \]

They also obey
\[ \overline{D(\Lambda)} \gamma^\mu D(\Lambda) = \Lambda^\mu_{\;\;\nu} \gamma^\nu. \]

For any vector, $a$, we define
\[ \cancel{a} = a_\mu \gamma^\mu. \]

It follows from the $\gamma$ algebra that
\[ \cancel{a} \cancel{b} + \cancel{b} \cancel{a} = 2 a \cdot b. \]

In this notation, the Dirac Lagrange density is 
\[ \psib (i\notd -m)\psi, \]

and the Dirac equation is 
\[ (i\notd -m) \psi = 0. \]
\vspace{1cm}

\item[(4).] \uline{Bilinear Forms}\\

There are sixteen linearly independent bilinear forms we can make from a Dirac bispinor and its adjoint. We can choose these sixteen to form the components of objects that transform in simple ways under the Lorentz group and parity.\\

The \uline{scalar} is 
\[ S = \psib \psi. \]

The \uline{vector} is 
\[ V^\mu = \psib \gamma^\mu \psi. \]

The \uline{tensor} is 
\[ T^{\mu\nu} = \psib \sigma^{\mu\nu} \psi, \]

where
\[ \sigma\mn = \frac{1}{2i} \Big[ \gamma^\mu, \gamma^\nu \Big]. \]

The \uline{pseudoscalar} is 
\[ P = \psib i \gamma_5 \psi, \]

where
\[ \gamma_5 = i \gamma^0 \gamma^1 \gamma^2 \gamma^3 \equiv \gamma^5. \]

The \uline{axial vector} is 
\[ A^\mu = \psib \gamma^\mu \gamma_5 \psi. \]

$\gamma_5$ is in many ways ``the fifth $\gamma$ matrix". It obeys
\[ (\gamma_5)^2 =1,\quad \gamma_5 = \gamma_5^\dagger = - \overline{\gamma_5}, \quad \Big\{ \gamma_5, \gamma^\mu \Big\} = 0. \]
\vspace{1cm}

\item[(5).] \uline{Plane-wave Solutions} 

The positive-frequency solutions of the Dirac equation are of the form
\[ \psi = u e^{-ip\cdot x}, \]

where $p^2 = m^2$ and $p^0$ is positive. The negative-frequency solutions are of the form 
\[ \psi = v e^{ip\cdot x}. \]

There are two positive-frequency and two negative-frequency solutions for each $p$. The Dirac equation implies that 
\[ (\notp - m) u = 0 = (\notp + m) v. \]

For a particle at rest, $p = (m, \vec{0})$, we can choose the two independent $u$'s in the standard representation to be, 
\[ u_{\vec{0}}^{(1)} = \bpm \sqrt{2m} \\ 0 \\ 0 \\ 0 \epm, \qquad u_{\vec{0}}^{(2)} = \bpm 0 \\ \sqrt{2m} \\ 0 \\ 0 \epm, \]

and the two independent $v$'s to be
\[ v_{\vec{0}}^{(1)} = \bpm 0 \\ 0 \\ \sqrt{2m} \\ 0 \epm, \qquad v_{\vec{0}}^{(2)} = \bpm 0 \\ 0\\ 0\\ \sqrt{2m} \epm. \]

We can construct the solutions for a moving particle, $\uvpr$ and $\vvpr$, by applying a Lorentz acceleration (see (2)).\\

These solutions are normalized such that 
\[ \overline{\uvp}^{(r)} \uvps = 2m \delta^{rs} = - \overline{\vvp}^{(r)} \vvps, \qquad \overline{\uvp}^{(r)} \vvps = 0. \]

They obey the completeness relations
\[ \sum_{r=1}^2 \uvpr \overline{\uvp}^{(r)} = \notp + m, \qquad \sum_{r=1}^2 \vvpr \overline{\vvp}^{(r)} = \notp -m. \]

Another way of expressing the normalization condition is 
\[ \overline{\uvp}^{(r)} \gamma^\mu \uvps = 2 \delta^{rs} p^\mu = \overline{\vvp}^{(r)} \gamma^\mu \vvps. \]

This form has a smooth limit as $m$ goes to zero. 
\end{itemize}
}{
 \sektion{23}{December 16}
\descriptiontwentythree
\begin{center}
\textbf{CANONICAL QUANTIZATION OF DIRAC LAGRANGIAN}
\end{center}
\[ \ml = \pm \Big[ \psi^\dagger ( i \po + i\va \cdot \vn - \beta m) \psi \Big] \]
\[ \pi_\psi \equiv \frac{\partial \ml}{\partial ( \po \psi ) } = \pm i \psid \qquad \text{$\psi, \psid$ completely characterize system} \]

\begin{center}
(more generally $\displaystyle \pi_A = \frac{\partial \ml}{\partial \po \psi_A} = \pm i \psi_A^\dagger$) 
\end{center}
\[ \mh = \pm i \psid \po \psi - \ml = \pm \Big[ \psid ( - i\va \cdot \vn + \beta m ) \psi \Big] = \pm i \psid \po \psi \quad \text{using E-L Eq} \]
\[ \pm i \; \Big[ \psi_\alpha (\vx,t), \psid_\beta (\vy,t) \Big] = i \delta^{(3)}(\vx - \vy) \delta_{\alpha \beta} \qquad \alpha,\beta = 1,2,3,4 \]
\[ \text{suppress $\alpha,\beta$ indices} \qquad \Big[ \psi (\vx,t),\psid (\vy,t) \Big] = \pm 1 \delta^{(3)} (\vx - \vy) \]
\[ \Big[ \psi (\vx,t),\psi (\vy,t) \Big] = 0 = \Big[ \psid (\vx,t),\psid (\vy,t) \Big] \]
\[ \Big[ \psi (\vx,t), \psib (\vy,t) \Big] = \pm \gamma^0 \deltat (\vx -\vy) \qquad \text{easily show by reinserting indices} \]
\[ \psi(x) = \sum_{r=1}^2 \int \dtp \frac{1}{(2\pi)^{3/2}} \; \frac{1}{\sqrt{2E_{\vec{p}}}} \Big[\bvp^{(r)} \uvp^{(r)} \eipxm + \cvp^{(r)\dagger} \vvp^{(r)} \eipx \Big] \]
\[ \psid(x) = \sum_{r=1}^2 \int \dtp \tptep \Big[ \bvp^{(r)\dagger} \uvp^{(r)\dagger} \eipx + \cvp^{(r)} \vvp^{(r)\dagger} \eipxm \Big] \]
\[ \psib(x) = \sum_{r=1}^2 \int \dtp \tptep \Big[ \bvp^{(r)\dagger} \overline{\uvp}^{(r)} \eipx + \cvp^{(r)} \overline{\vvp}^{(r)} \eipxm \Big] \]

\uline{Ansatz}: (To avoid doing Fourier inversion): 
\[ \Big[ \bvp^{(r)}, \bvpp^{(s)\dagger} \Big] = \delta^{rs} \deltat (\vp - \vp\,') \; B \]
\[ \Big[ \cvp^{(r)\dagger}, \cvpp^{(s)} \Big] = \delta^{rs} \deltat (\vp - \vp\,')\; C \]
\[ \Big[b,b \Big] = \Big[ c,c \Big] = \Big[ b,c \Big] =0 \Longrightarrow \Big[ \psi,\psi \Big] = \Big[ \psid, \psid \Big] = \Big[ \psib, \psib \Big] = 0 \]

\[ \underbrace{\Big[ \psi (\vx,t),\psid (\vy,t) \Big]}_{\substack{\text{Formula also true}\\\text{with $\psid$ replaced by $\psib$}\\\text{and $u^\dagger$, $v^\dagger$ replaced by $\overline{u}$, $\overline{v}$}}} = \sum_r \int \dtp \frac{1}{(2\pi)^3 2E_{\vec{p}}} \Big[ B e^{i\vec{p}\cdot(\vx-\vy)} \uvp^{(r)} \uvp^{(r)\dagger} + C e^{-i\vp \cdot(\vx-\vy)} \vvp^{(r)} \vvp^{(r)\dagger} \Big] \]
\begin{align*}
\sum_r \uvp^{(r)} \uvp^{(r)\dagger} &= \sum_r \uvp^{(r)} \overline{\uvp}^{(r)} \beta = (\notp + m )\beta \\
&= (E_{\vec{p}} \beta - \vp \cdot \beta \va + m ) \beta = E_{\vec{p}} + \vp \cdot \va + \beta m 
\end{align*}

Similarly,
\[ \sum_r \vvp^{(r)} \vvp^{(r)\dagger} = (\notp - m) \beta = \evp + \vp \cdot \va - \beta m \]

If $B=C$, terms $\propto$ $ \va, \beta$ vanish, so that we may have integral proportional to $1$, choose $B=C=\pm1$, and we have canonical quantization relations:
\[ \Big[ \psi (\vec{x},t), \psid (\vec{y},t) \Big] = \pm \int \frac{\dtp}{(2\pi)^3} e^{i\vec{p}\cdot (\vec{x}-\vec{y})} \]

However, our expression for $\psi$ has two annihilation or creation operators leading to problems with $H$:
\begin{align*}
H &= \int d^3 x \; \mh = \pm \sum_{rs} \int \frac{\dtp}{2 \evp} \Big( \bvp^{(r)\dagger} \bvp^{(s)} \overbrace{\uvp^{(r)\dagger} \uvp^{(s)}}^{\delta_{rs} 2 \evp} \evp - \cvp^{(r)} \cvp^{(s)\dagger} \overbrace{\vvp^{(r)\dagger} \vvp^{(s)}}^{\delta_{rs} 2\evp} \evp \Big) \\
H &= \pm \sum_r \int \dtp \evp \Big( \bvp^{(r)\dagger} \bvp^{(r)} - \cvp^{(r)} \cvp^{(r)\dagger} \Big) 
\end{align*}

which is an unbounded below energy. For $(+)$ sign c-type quanta carry negative energy.\\
\vspace{1cm}

5 TOPICS FOR THE REST OF THIS LECTURE
\begin{enumerate}
\item Canonical Anticommutation
\item Solves energy crisis
\item Fermi-Dirac statistics
\item Fields as observables
\item Classical Limit ...
\end{enumerate}
\vspace{1cm}

\begin{itemize}
\item[(1)] $\left\{ \begin{array}{l} \text{Fermi} \\ \text{Bose} \end{array} \right\}$ $p$'s and $q$'s have $\left\{ \begin{array}{l} \frac{1}{2} \text{odd int.} \\ \text{int.} \end{array} \right\}$ spin.\\

At equal times 
\begin{itemize}
\item Bose-Bose $[p^a, p^b ] = [ q^a, q^b ] = 0$, $[q^a, p^b ] = i \delta^{ab}$. 
\item B-F everything commutes.
\item Fermi-Fermi $\{ p^a, p^b \} = \{ q^a, q^b \} = 0 $, $\{ q^a, p^b \} = i \delta^{ab}$.
\end{itemize}

Hence we make the changes $\displaystyle \Big\{ \bvp^{(r)}, \bvp^{(s)\dagger} \Big\} = \delta^{rs} \deltat (\vp-\vp\,') B$ etc.\\
$\displaystyle \Big\{ \psi ( \vx,t ), \psid (\vy,t) \Big\} = \pm \sum_r \int \dtp \frac{1}{(2\pi)^3} \cdots $

\vspace{1cm}
IF WE ARE NOT CAREFUL WE WILL LOSE POSITIVE DEFINITENESS IN THE HILBERT SPACE
\[ A \equiv \int \dtp \sum_r f_r(\vec{p}) \; b_{\vec{p}}^{(r)} \{ A, A^\dagger \} = \pm \int \sum | f_r (\vp) |^2 \]

LOOK AT QUANTITY OF THE FORM
\[ \langle \phi | \Big\{ A, A^\dagger \Big\} | \phi \rangle = \langle \phi | A A^\dagger | \phi \rangle + \langle \phi | A^\dagger A | \phi \rangle \underset{!}{\geq} 0 \]

Hence we must choose $+$ sign
\vspace{1cm}

\item[(2)] $ \ml = \psib (i\notd - m) \psi $
\[ \Big\{ \bvp^{(r)}, \bvpp^{(s)\dagger} \Big\} = \Big\{ \cvp^{(r)}, \cvpp^{(s) \dagger} \Big\} = \delta^{rs} \deltat (\vp-\vp\,') \]

\begin{center}
all others zero
\end{center}
\[ H = \sum_r \int \dtp \evp \Big[ \bvp^{(r)\dagger} \bvp^{(r)} + \cvp^{(r)\dagger} \cvp^{(r)} \Big] - \!\!\!\!\!\!\!\!\!\!\!\!\! \underbrace{\cancel{\deltat (0)}}_{\substack{\text{from anticommutation}\\\text{of $C$ and $C^\dagger$}}} \]

which is bounded below.
\vspace{1cm}

\item[(3)] Consider pedagogical simplification of Hamiltonian for a moment.
\[ H = \sum_{\vp} \evp \; \bvp^\dagger \; \bvp \qquad \Big\{ \bvp^\dagger, \bvpp \Big\} = \delta_{\vp, \vp\,'}, \qquad \Big\{ b, b \Big\} = \Big\{ b^\dagger, b^\dagger \Big\} = 0 \]
\[ \Big[ AB,C \Big] = A \Big\{ B, C \Big\} - \Big\{A, C \Big\} B \]
\[ \Big[H, \bvp \Big] = \underbrace{- \evp \; \bvp}_{\text{energy lowering}} \]
\[ \Big[H, \bvp^\dagger \Big] = \underbrace{+ \evp \; \bvp^\dagger}_{\text{energy raising}} \]

\vspace{1cm}
Define $\qquad \bvp \sta = 0$ all $\vp$, $\qquad \langle 0 | 0 \rangle = 1 \qquad H |0 \rangle = 0$
\[ \bvp^\dagger \sta = | \vp \rangle \qquad H | \vp \rangle = \evp | \vp \rangle \]
\[ \langle p' | p \rangle = \langle 0 | \bvpp \bvp^\dagger | 0 \rangle = - \stai \bvp^\dagger \bvpp \sta + \delta_{\vp,\vp\,'} = \delta_{\vp,\vp\,'} \]
\[ | \vp_1, \vp_2 \rangle = b_{\vec{p}_1}^\dagger b_{\vec{p}_2}^\dagger \sta = - | \vec{p}_2, \vec{p}_1 \rangle \]
\[ H | \vp_1, \vp_2 \rangle = (E_{\vp_1} + E_{\vp_2}) | \vp_1, \vp_2 \rangle \]

Pauli Exclusion principle
\[ \bvp^\dagger | \vp \rangle = (\bvp^\dagger)^2 | 0 \rangle = 0 \]
\vspace{1cm}

\item[(4)] Observable made out of Fermi fields.\\

Recall Bose Fields
\[ \Big[ \phi (x), \phi (y) \Big]_{\text{E.T.}} = 0 \]

and by Lorentz invariance this is true for all spacelike separated $x$ and $y$. With Fermi fields
\[ \Big\{ \psi_\alpha (x), \psi_\beta (y) \Big\} = 0 \]

If $\psi(x)$ were an observable, we would have observables that did not commute at spacelike separation. Observables can only be made of products with an even number of Fermi Fields. \\

This is also necessary for just rotational properties. Under a rotation by $2\pi$, 
\[ \psi(x) \longrightarrow - \psi(x) \]

No meter on any experimental apparatus ever gives a different reading when the experiment is rotated by $2 \pi$. \\

All observables are in single valued representations of the Lorentz group.

\item[(4)] Classical Limit $(\hbar \longrightarrow 0)$ 
\begin{itemize}
\item[(a)] Two classical limits physically. Take some physical situation. $N$ particles in a box all in the same energy and momentum eigenstate. 
\[ E = N \hbar \omega \]

\begin{itemize}
\item [(i)] $\hbar \longrightarrow 0 \qquad N, E \text{ and } \vp \text{ fixed}$
\[ \omega, \vec{k} \longrightarrow \infty \qquad \text{wavelength} \longrightarrow 0 \]
No diffraction. This is the classical particle limit.\\

\item [(ii)] $\hbar \longrightarrow 0 \qquad E, \omega \text{ and } \vp, \vec{k} \text{ fixed.}$
$N \longrightarrow \infty$ Lots of wavy behavior, but lose quantum granularity.
\end{itemize}

\vspace{0.5cm}
For fermions we can only do the first limit because of the Pauli exclusion principle. There is no analog of the wave limit. There will never be a competing theory as there was for light with the corpuscular and wave theories. \\

\item[(b)] Classical limits formally. 
\[ \hbar \longrightarrow 0 \qquad \text{ in canonical algebras} \]
\begin{align*}
\text{Bose fields} &\longrightarrow \text{commuting quantities (numbers)} \\
\text{Fermi fields} &\longrightarrow \text{anticommuting quantities (Grassmann variables)} 
\end{align*}
\end{itemize}
\end{itemize} 
\vspace{1cm}

\begin{center}
\textbf{Working with classical Fermi Fields or Grassmann variables}
\end{center}

Never exchange the order of two terms in the classical field equations without a compensating minus sign or you would have no hope of the classical limit of the quantum theory agreeing with the classical theory even at order $\hbar^0$.\\

Derivation of the Euler-Lagrange equations:
\[ dL = \underbrace{\frac{\partial L}{\partial \dot{q}^a }}_{\equiv p_a} d\dot{q}^a + \frac{\partial L}{\partial q^a} dq^a \]

If both the derivatives are kept to the same side of the differentials then I can integrate by parts in the action and get the usual E-L equations.
\[ \dot{p}_a = \frac{\partial L}{\partial q^q} \]

rather than something else. Define 
\[ H = p_a \dot{q}^a - L \]

Not $H = \dot{q}^a p_a - L$, for example, which would give something different.
\[ dH = dp_a \dot{q}^a - \dot{p}_a dq_a - d L \]
\[ \frac{\partial H}{\partial p_a} = \dot{q}^a \qquad \frac{\partial H}{\partial q^a} = - \dot{p}_a \]
\vspace{1cm}

Try using the quantum relations $\displaystyle \dot{p}_a = - i \Big[ p_a,H \Big] $, $\displaystyle \dot{q}^a = - i \Big[ q^a, H \Big] $ with the canonical anticommutation relation, and see if you can reproduce the Heisenberg equation of motion in the Dirac theory. 
}{
 \sektion{24}{December 18}
\descriptiontwentyfour
\section*{Perturbation theory for spinors}
Because scalar fields commute at spacelike separations, the idea of time ordering is Lorentz invariant. That is
\begin{center}
\includegraphics[scale=0.3]{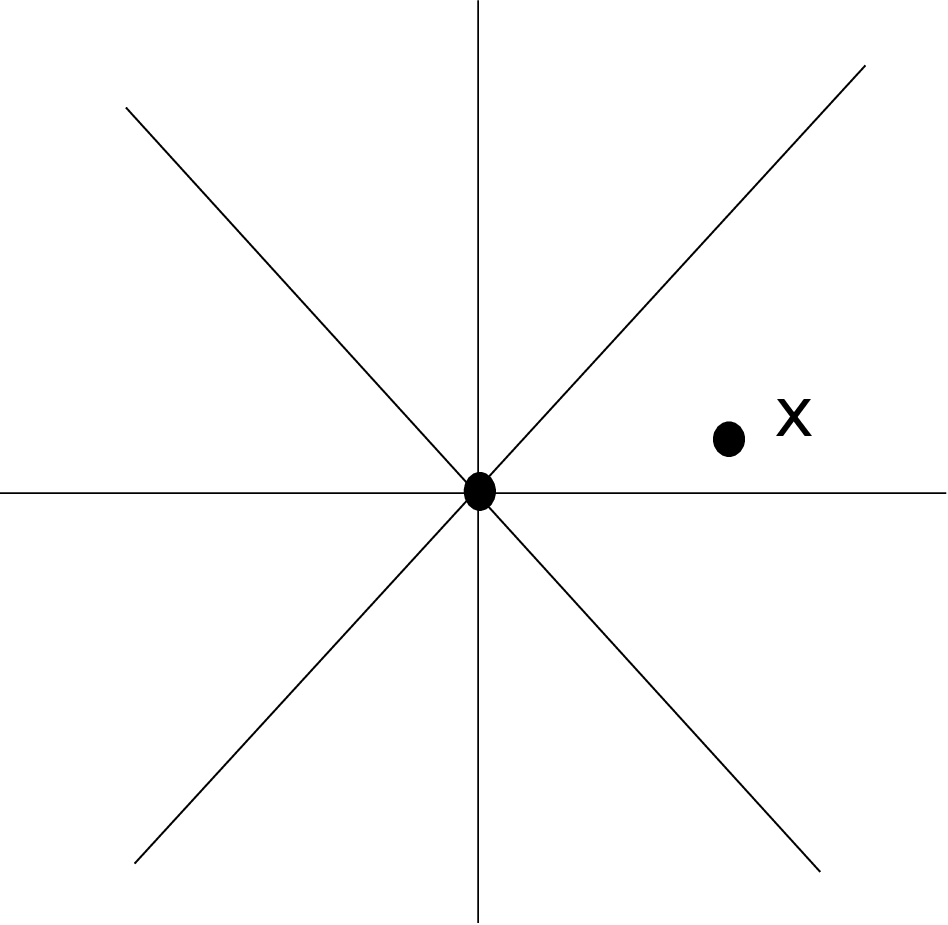}
\end{center}

If $x^2 < 0 $,
\[ T(\phi(0)\phi(x)) = \phi(x) \phi(0) \]

In the situation pictured, but in another frame, whose axes are represented by dotted lines $x^0$ is less than $0$ and 
\begin{center}
\includegraphics[scale=0.3]{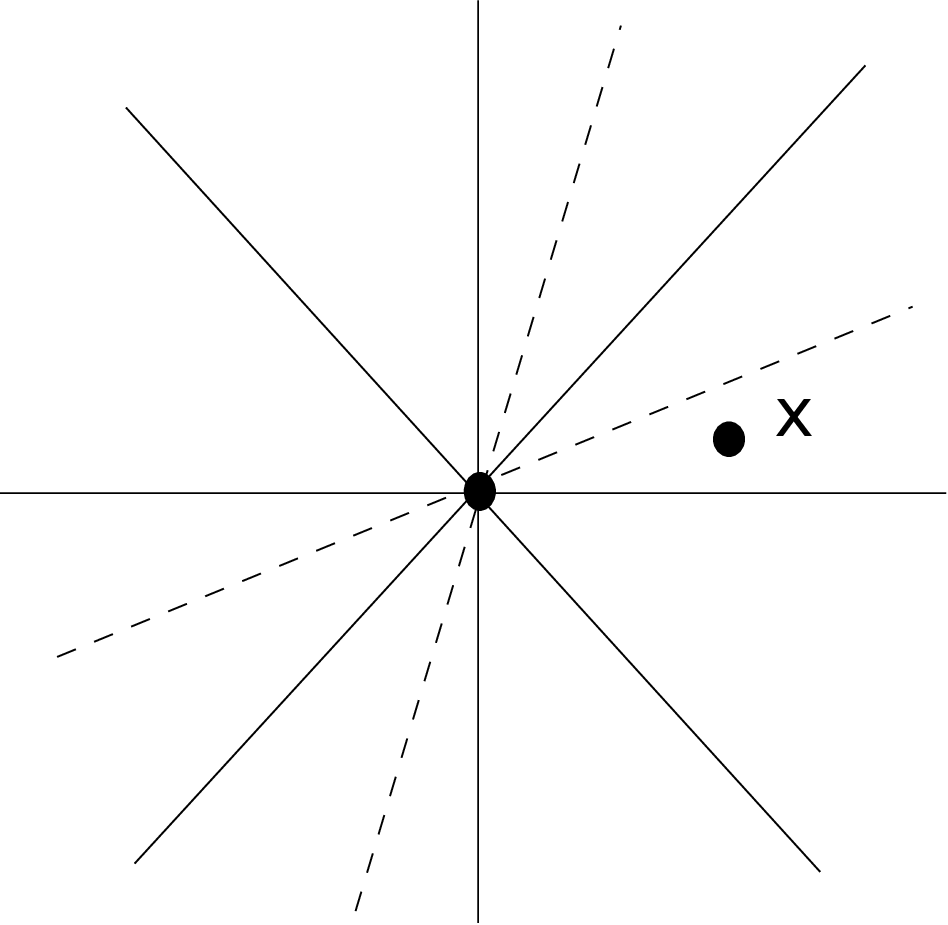}
\end{center}
\[ T(\phi(0)\phi(x)) = \phi(0) \phi(x) \]

This ambiguity is not a problem for scalar fields since $[\phi(0),\phi(x)] = 0$ when $x^2 < 0$. When $x^2 >0$, there is no ambiguity.\\

For spinor fields, this definition of time ordering is a failure. In one frame 
\[ T(\psi_\alpha(x) \overline{\psi}_\beta(0)) = \psi_\alpha(x) \overline{\psi}_\beta (0) \]

and if $x^2<0$, in another frame it may be that 
\[ T(\psi_\alpha(x) \overline{\psi}_\beta(0)) = \overline{\psi}_\beta (0) \psi_\alpha(x) = -\psi_\alpha(x) \overline{\psi}_\beta (0) \qquad \text{OH OH$!$} \]

The way to patch this up is to put an extra minus sign into the definition of the time ordered product whenever the number of permutations of Fermi fields required to turn a product into a time ordered product is odd.\\

\uline{\textbf{Assertion}:} Most of the derivations we did, expressing $S$ matrix elements in terms of physical vacuum expectation values of time ordered products of renormalized Heisenberg picture fields, and showing that 
\[ _p \stai T(\phih '(x_1) \cdots \phih '(x_n)) \sta_p =\frac{\stai T\Big[ \phii'(x_1)\cdots \phii'(x_n) e^{-i \int_{-\infty}^\infty d^4x \mh_I} \Big] \sta}{\stai T e^{-i \int_{-\infty}^\infty d^4x \mh_I} \underbrace{\sta}_{\text{bare vacuum}}} \]

are unaffected by the fact that the fields may now be spinors and the time ordered product now has a $(-1)^p$ in it.\\

A way of seeing that this is probably true is to think about how we obtained $S= U_I(\infty,-\infty)$ in the formalism with the turning on and off function. About the only place we could have problem is in obtaining the expression
\[ U_I(\infty,-\infty) = T e^{- i \int_{-\infty}^\infty d^4x \, \mh_I} \]

But you expect no problem there because the Hamiltonian is quadratic in spinor fields, so you always move spinor fields around in pairs, and the permutation is thus always even. The new minus sign in the time ordered product doesn't matter.\\

Once we have that big messy expression on the RHS above, we used Wick's theorem to turn the time ordered products into normal ordered products, and then wrote down Wick diagrams to represent operations in the Wick expansion, and Feynman diagrams to represent $S$ matrix elements.\\

\uline{\textbf{Assertion}:} Wick's theorem can be proven for spinor fields and with the extra minus sign in the time ordered product \uline{provided} you also put an extra minus sign in the normal ordered product.\\

For example if $A_1$ and $A_2$ are Fermi fields
\[ : A_1 A_2 : = \aonep \atwop + \aonem \atwop + \aonem \atwom - \atwom \aonep \]

Note that 
\[ : \aone \atwo = - : \atwo \aone : \]

also 
\[ T(\aone \atwo) = - T(\atwo \aone) \]

(FERMI FIELDS ANTICOMMUTE INSIDE THE TIME ORDERED AND NORMAL ORDERED PRODUCTS).\\

The contraction $\wick{1}{<1\aone >1\atwo}$ is defined as usual to be the time ordered product minus the normal ordered product
\[ \wick{1}{<1\aone >1\atwo} = T(\aone \atwo) -:\aone \atwo : \]

The contraction is a $c$-number. Here is a proof by cases. Take $x_1^0 > x_2^0$, then
\[ T(\aone \atwo) = \aone \atwo = \aonep \atwop + \aonep \atwom + \aonem \atwop + \aonem \atwom \]

and 
\begin{align*}
\wick{1}{<1\aone >1\atwo} &= \aonep \atwom + \atwom \aonep \\
&= \{ \aonep, \atwom \} 
\end{align*}

NOTATION: 
\[ :\wick{1}{<1 \aone \atwo >1 A_3} A_4: \equiv - \wick{1}{<1 \aone >1 A_3} :A_2 A_4: \]

$-$ sign for odd permutations is needed to prove Wick's theorem.\\

If we hadn't stuck that extra $-$ sign into the definition of the ordered product we would have gotten
\[ \wick{1}{<1 \aone >1 \atwo} = [ \aonep, \atwom ] \]

For $x_1^0 > x_2^0$, which is not a $c^\#$. Things are looking good though, $\{ A_1^{(+)},A_2^{(-)}\}$ is a $c^\#$. The case $x_2^0 > x_1^0$ clearly goes through too.
\vspace{1cm}

\begin{center}
\textbf{Calculation of the contraction (propagator)}
\end{center}

Since the contraction of two Fermi fields is a $c$-number, we can use the same trick for evaluating it as we did when we calculated the contraction of two scalar fields, i.e.~take vacuum expectation value.
\begin{align*}
\psi(x) &= \sum_r \int \frac{d^3p}{(2\pi)^{3/2} \sqrt{2E_{\vec{p}}}} \Big[ \bvp^{(r)} \uvp^{(r)} e^{-ip\cdot x} + \cvp^{(r)\dagger} \vvp^{(r)} e^{ip\cdot x} \Big] \\
\psib (y) &= \sum_{r'} \int \intpp \Big[ \bvpp^{(r')\dagger} \overline{\uvpp}^{(r')} e^{ip' \cdot y} + \cvpp^{(r')} \overline{\vvpp}^{(r')} e^{-ip' \cdot y} \Big] 
\end{align*}

For $x^0 > y^0$, 
\begin{align*}
\wick{1}{<1\psi(x) >1{\overline{\psi}}(y)} &= \stai \wick{1}{<1\psi(x) >1{\overline{\psi}}(y)} \sta \\
&= \stai \Big[ T \Big(\psi (x) \overline{\psi}(y)\Big) - \underbrace{\cancel{: \psi(x) \overline{\psi}(y):}}_{0} \Big] \sta \\
&= \stai \psi(x) \psib(y) \sta \\
&= \sum_{r r'} \int \intp \intpp \underbrace{\stai \bvp^{(r)} \bvpp^{(r')\dagger} \sta }_{\delta_{r r'} \delta^{(3)} (\vp - \vp\,') } \eipxm e^{ip'\cdot y} \uvp^{(r)} \overline{\uvpp}^{(r')}\\
&= \int \frac{d^3 p}{(2\pi)^3 2E_{\vec{p}}} e^{-ip\cdot (x-y)} \underbrace{\sum_r \uvp^{(r)} \overline{\uvp}^{(r)}}_{\cancel p + m} \\
&= (i \cancel{\partial}_x + m ) \int \frac{d^3 p}{(2\pi)^3 2E_{\vec{p}}} e^{- i p \cdot (x-y)}
\end{align*}

For $y^0 > x^0$, you get 
\begin{align*}
\wick{1}{<1\psi(x) >1{\overline{\psi}}(y)} &= \overbrace{-}^{(*)} \int \intps \eipxy (\cancel{p} - m) \\
&= (i \cancel{\partial}_x + m ) \int \intps \eipxy
\end{align*}

(*): CRITICAL MINUS SIGN OUT FRONT IS THE ONE WE PUT INTO OUR TIME ORDERED PRODUCT FOR FERMI FIELDS\\

The nice thing about this is that it is the same for $x^0$ less than or greater than $y^0$ (the sign of $p$ in the exponential no longer matters once $\cancel{p}$ is turned into a derivative and pulled out). In either case
\[ \wick{1}{<1\psi(x) >1{\overline{\psi}}(y)} = (i \cancel{\partial}_x + m) \underbrace{\wick{1}{<1\phi (x) >1\phi(y)}}_{\substack{\text{contraction of a}\\\text{scalar field}\\\text{of mass } m}} \]

We've already massaged $\wick{1}{<1\phi (x) >1\phi(y)}$:
\[ \wick{1}{<1\phi (x) >1\phi(y)} = \int \frac{d^4p}{(2\pi)^4} \eipxym \frac{i}{p^2-m^2 + i \epsilon} \]

So without further ado we can rewrite 
\begin{equation}\label{eq:24-page5}
\begin{split}
\wick{1}{<1\psi(x) >1{\overline{\psi}}(y)} &= (i \cancel{\partial}_x + m) \int \frac{d^4p}{(2\pi)^4} \eipxym \frac{i}{p^2-m^2 + i \epsilon} \\
&= \int \frac{d^4 p}{(2\pi)^4} \frac{i (\cancel{p}+m)}{p^2 -m^2 + i \epsilon} \eipxym
\end{split}
\end{equation}

Both sides of this equation are $4 \times 4$ matrices. If you prefer, 
\[ \wick{1}{<1\psi_\alpha (x) >1{\overline{\psi}}_\beta(y)} = \int \frac{d^4 p}{(2\pi)^4} \frac{i (\cancel{p}_{\alpha\beta} + m \id_{\alpha\beta})}{p^2 -m^2 + i\epsilon} \eipxym \]

\begin{center}
\textbf{3 Comments on the Propagator}
\end{center}
\begin{itemize}
\item[(1).] The propagator $\displaystyle \ipmpme$ is going to play the same role in the perturbation theory for Dirac fields as $\displaystyle \ipme$ played in the perturbation theory for scalar fields. Recall though that when you wrote down\\
\[ \overset{\text{charged scalar}}{\Diagram{\momentum[bot]{fV}{\leftarrow p}}} \text{ which stands for } \displaystyle \ipme \]

it did not matter whether $p$ was routed in the same direction or the opposite direction as charge flow, simply because
\[ \frac{i}{(-p)^2 - m^2 + i \epsilon} = \frac{i}{p^2 - m^2 + i\epsilon} \]

Now however the propagator is not even in $p$.\\
\[ \overset{\text{charged fermion}}{\Diagram{\momentum[bot]{fV}{\leftarrow p}}} \text{ will stand for }\displaystyle \ipmpme \]

while 
\[\overset{\text{charged fermion}}{\Diagram{\momentum[bot]{fV}{\rightarrow p}}} \text{ will stand for }\frac{i(-\cancel{p} + m)}{p^2 -m^2 + i\epsilon} \]

The propagator is a kind of projection operator. 

\item[(2).] There is more common way of writing the fermion propagator, more common because it is shorter. Rewrite
\[ p^2 - m^2 + i\epsilon = (\cancel{p} - m + i\epsilon) (\cancel{p} + m - i\epsilon) \]

(this still gives the right prescription for the location of poles as $\epsilon \rightarrow 0$, $m>0$). Then
\[ \ipmpme = \frac{i (\cancel{p} + m - \!\!\!\!\!\!\!\!\!\!\!\!\!\!\!\!\!\!\!\!\!\!\overbrace{i\epsilon}^{\substack{\text{adding this here does}\\\text{nothing to location of pole}}}\!\!\!\!\!\!\!\!\!\!\!\!\!\!\!\!\!\!\!\!\!\!)}{(\cancel{p} - m + i\epsilon)(\cancel{p} + m - i\epsilon)} = \frac{i}{\cancel{p} - m + i\epsilon} \]

There is no danger that these hokey looking matrix manipulations are wrong because $\cancel{p}-m+i\epsilon$ commutes with $\cancel{p}+m-i\epsilon$.\\

\item[(3).] For bosons the action is 
\begin{align*}
S &= \int d^4 x \; ( \pmu \psi^* \pmuu \psi - m^2 \psi^* \psi ) \\
\text{parts integration} &= \int d^4 x \;\psi^* (-\Box -m^2) \psi
\end{align*}

In momentum space, i.e.~when acting on $\eipxm$, $-\Box - m^2$ becomes $p^2-m^2$.\\

For fermions, the action is 
\[ S = \int d^4 x \; \psib (i\cancel{\partial} -m) \psi \]

In momentum space $i \notd - m$ becomes $\cancel{p} - m$. The fact that the charged boson propagator came out to be $\frac{i}{p^2-m^2}$ while the charged fermion propagator came out to be $\frac{i}{\notp -m}$ is at least an interesting coincidence.
\end{itemize}
\vspace{1cm}

\begin{center}
\textbf{Feynman diagrams in $\ml = \psib (i\notd - m) \psi + \frac{1}{2}(\pmu \phi)^2 - \frac{\mu^2}{2}\phi^2 - g \psib \Gamma \psi \phi$}
\end{center}

($\Gamma =1$ (meson is a scalar) or $i\gamma_5$ (meson is a pseudoscalar).)\\

We'll ``derive" the Feynman rules by looking at a couple of processes and hoping that the general process at general orders has an obvious generalization. Let's look at the order $g^2$ term in $T e^{-i \int d^4 x \mh_I}$, i.e. 
\[ \frac{(-ig)^2}{2!} \int d^4x_1 d^4x_2 \; T \Big( \psib \Gamma \psi \phi(x_1) \psib \Gamma \psi \phi(x_2) \Big) \]

This can contribute to many processes, let's look at how it contributes to $N+ \phi \longrightarrow N+\phi$. The relevant terms in the Wick expansion of the time ordered product are
\[ \frac{(-ig)^2}{2!} \int d^4x_1 d^4 x_2 \; \Big(:\psib \Gamma \wick{1}{<1\psi \phi (x_1) >1{\overline{\psi}}} \Gamma \psi \phi(x_2): + :\wick{1}{<1{\overline{\psi}} \Gamma \psi \phi(x_1) {\overline{\psi}} \Gamma >1\psi}\phi(x_2) : \Big) \]

The Feynman diagrams for these two terms are 
\[ \includegraphics[scale=0.3]{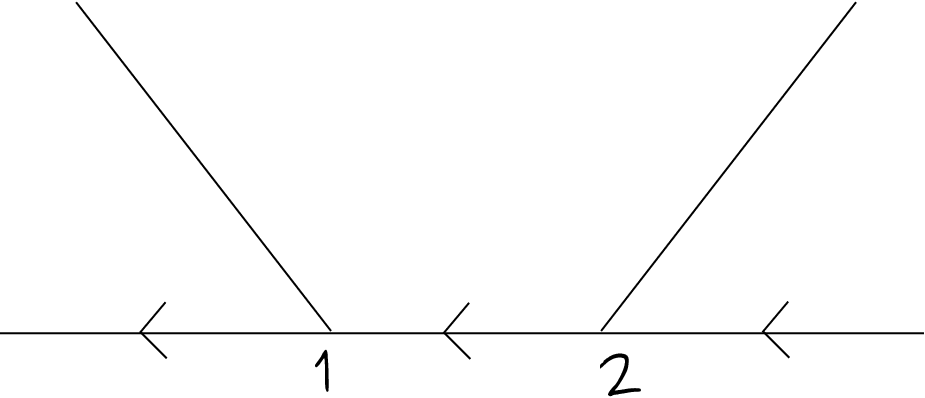} \qquad \text{and} \qquad \includegraphics[scale=0.3]{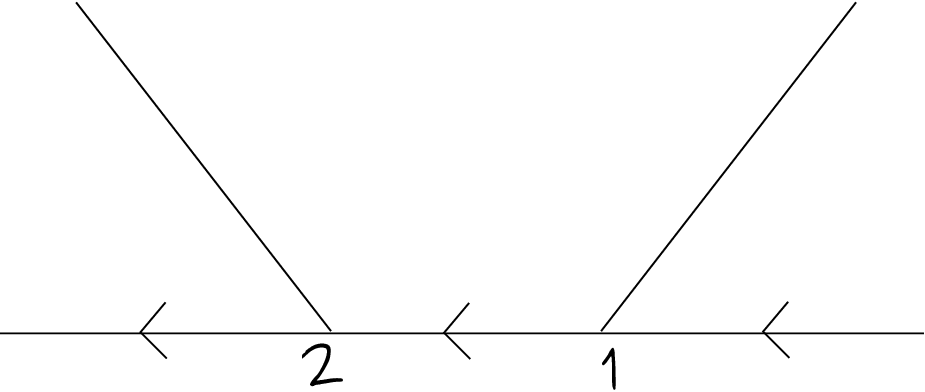} \] \\
respectively.\\

The picture for the second term looks identical to the first picture (they are of the same pattern) except for an exchange of the dummy variable $1 \longleftrightarrow 2$. Is the second operator the same as the first? Well,
\[ :\wick{1}{<1 \psib \Gamma \psi \phi(x_1) \psib \Gamma >1\psi} \phi(x_2) : = : \psib \Gamma \wick{1}{<1 \psi \phi(x_2) >1\psib} \Gamma \psi \phi(x_1): \]

because interaction Hamiltonians are made out of fermion bilinears which \uline{commute} inside the normal ordered product, and this expression clearly differs from the one in the first term by an exchange of dummy indices $1 \longleftrightarrow 2$.\\

I'll just look at how the first term contributes to $N+\phi \longrightarrow N+\phi$ ($N$: an electron or ``nucleon") since the second one is identical and serves only to cancel the $2!$. We'll look at the matrix element of the operator between relativistically normalized states.
\[ \langle p',r';q'| (-ig)^2 \int d^4x_1 d^4 x_2 \;:\psib \Gamma \wick{1}{<1\psi \phi(x_1) >1\psib} \Gamma \psi \phi(x_2) : |p,r;q \rangle \]

Notice that the incoming and outgoing electrons have to have their spin specified as well as their momentum; that is what the $r$ does. [Alternatively, you could just give an arbitrary spinor for the incoming and outgoing electrons, say $u$ and $u'$.] There is only one field in the normal ordered product that has an annihilation operator that can annihilate the incoming electron, $\psi(x_2)$.
\begin{align*}
\langle 0 | \psi(x_2) | p,r\rangle &= \stai \sum_s \int \frac{d^3 l}{(2\pi)^3 2 \omega_{\vec{l}}} b^{(s)}(l) u^{(s)}_{\vec{l}} e^{-il\cdot x_2} | p, r \rangle \\
&= \sum_s \int \frac{d^3 l}{(2\pi)^3 2 \omega_{\vec{l}}} u^{(s)}_{\vec{l}} e^{-i l\cdot x_2} \underbrace{\stai b^{(s)}(l) | p, r \rangle}_{(2\pi)^3 2 \omega_{\vec{p}} \delta^{(3)}(\vec{p}-\vec{l}) \delta_{sr}} \\
&= u_{\vec{p}}^{(r)} e^{-i p \cdot x_2}
\end{align*}

Similarly when $\psib(x_1)$ is used to create the outgoing electron it becomes 
\[ \overline{\uvpp}^{(r')} e^{+ip'\cdot x_1} \]

because that is the coefficient of $b^{(r')}(p')^\dagger$ in the expansion of $\psib(x_1)$. We also have derived an expression for $\psi (x_1) \psib (x_2)$, which I'll use and our matrix element simplifies to
\[ (-ig)^2 \int d^4 x_1 d^4 x_2 \int \frac{d^4 l }{(2\pi)^4} e^{ip'\cdot x_1} e^{-ip \cdot x_2} e^{-il\cdot (x_1-x_2)} \overline{\uvpp}^{(r')} \Gamma \frac{i}{\cancel{l} - m + i\epsilon} \Gamma \uvp^{(r)} \langle q' |:\phi(x_1) \phi(x_2) : | q \rangle \]
 
There are two terms in $\langle q' |:\phi (x_1) \phi (x_2) : | q\rangle $,
\[ e^{iq' \cdot x_1} e^{-i q\cdot x_2} + x_1 \leftrightarrow x_2 \]

Let's just consider the first one for a second. The integral is 
\[ (-ig)^2 \int d^4 x_1 d^4 x_2 \int \frac{d^4 l}{(2\pi)^4} e^{i(p'-l+q') \cdot x_1} e^{i(-p+l-q) \cdot x_2} \overline{\uvpp}^{(r')} \Gamma \frac{i}{\cancel{l}-m+i\epsilon} \Gamma \uvp^{(r)} \]

The exponentials go along with an interpretation. At $x_1$ an electron with momentum $p'$ and a meson with momentum $q'$ are created and a virtual electron with momentum $l$ is absorbed ($+$ signs go with creation, $-$ with absorption). At $x_2$ an electron with momentum $p$ is absorbed and a meson with momentum $q$ is absorbed, while a virtual electron with momentum $l$ is created. This can happen at any space-time points $x_1$ and $x_2$, so they are integrated over, and in fact the integrals are trivial to do.
\[ (-ig)^2 \int \frac{d^4 l}{(2\pi)^4} (2 \pi)^4 \delta^{(4)}(p'-l+q') (2\pi)^4 \delta^{(4)}(-p+l-q) \overline{\uvpp}^{(r')} \Gamma \frac{i}{\cancel{l}-m+i\epsilon} \Gamma \uvp^{(r)} \]

The $l$ integration can be done because the energy momentum of the internal electron is fixed by the $\delta$ function. We get
\[ (-ig)^2 (2 \pi)^4 \delta^{(4)} (p'+q'-(p+q)) \overline{\uvpp}^{(r')} \Gamma \frac{i}{\cancel{p} + \cancel{q} -m+i\epsilon} \Gamma \uvp^{(r)} \]

The Feynman diagram for this contribution to $N+ \phi \longrightarrow N + \phi$ is 
\begin{center}
\includegraphics[scale=0.3]{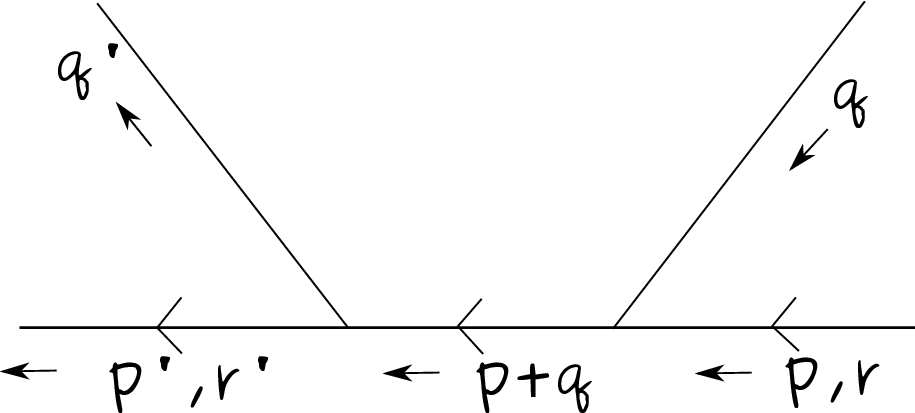}
\end{center}

What about the second term in $\langle q'|:\phi(x_1) \phi(x_2) : | q \rangle$? It contributes to the same process, but the diagram is different:
\begin{center}
\includegraphics[scale=0.3]{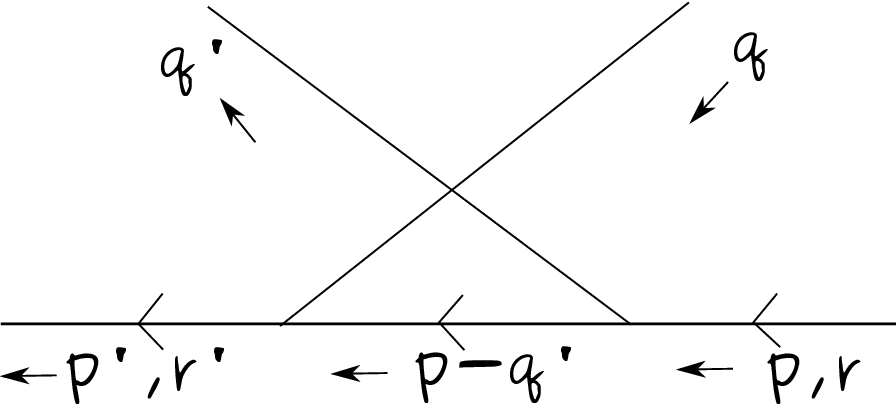}
\end{center}

The contribution is 
\[ (-ig)^2 (2 \pi)^4 \delta^{(4)}(p'+q'-(p+q)) \overline{\uvpp}^{(r')} \Gamma \frac{i}{\cancel{p}-\cancel{q}'-m+i\epsilon} \Gamma \uvp^{(r)} \]
\vspace{1cm}

Let's look at another process $\overline{N} + \phi \longrightarrow \overline{N} + \phi $ that the exact same operator in the Wick expansion can contribute to.
\[ \underbrace{\langle p',r';q'|}_{\text{outgoing positron and meson}} \!\!\!\!\!\!\!\!\!\!\!\!(-ig)^2 \int d^4 x_1 d^4 x_2:\psib \Gamma \wick{1}{<1 \psi \phi(x_1) >1\psib} \Gamma \psi \phi(x_2) : \!\!\!\!\!\!\!\!\!\!\!\!\!\!\! \underbrace{|p,r;q \rangle}_{\text{incoming positron and meson}} \]

The field $\psi(x_2)$ has to create the outgoing positron. \uline{But}, it has to be anticommuted past three (actually 1: two have been contracted) fermionic fields to do it. The coefficient of $\cvpp^{(r')\dagger}$ in $\psi(x_2)$ is $e^{ip' \cdot x_2} \vvpp^{(r')}$, so we get 
\[ -\langle q'| (-ig)^2 \int d^4 x_1 d^4 x_2 \;:\psib \Gamma \wick{1}{<1\psi \phi(x_1) >1 \psib }\Gamma \vvpp^{(r')} e^{ip' \cdot x_2} \phi(x_2) :|p,r;q \rangle \]

The $\psib$ field then only has to get by two (0) Fermi fields to annihilate the positron on the right. The coefficient of $\cvp^{(r)}$ in $\psib (x_1)$ is $e^{-ip\cdot x_1} \vvp^{(r)}$ so I get 
\[ -(-ig)^2 \int d^4 x_1 d^4 x_2 \; \int \frac{d^4l}{(2\pi)^4} e^{-ip \cdot x_1} e^{ip' \cdot x_2} e^{-il \cdot (x_1 - x_2)} \overline{\vvp}^{(r)} \Gamma \frac{i}{\cancel{l} -m + i\epsilon} \Gamma \vvpp^{(r')} \langle q' |:\phi(x_1) \phi(x_2) :|q \rangle \]

The main difference to notice for this process is the overall minus sign, and the change from $u$'s to $v$'s. There are still going to be two Feynman diagrams from the two terms in $ \langle q' |:\phi(x_1) \phi(x_2):| q \rangle $. I'll just write down the result of doing the $x_1$, $x_2$ and $l$ integrals.
\begin{center}
\includegraphics[scale=0.3]{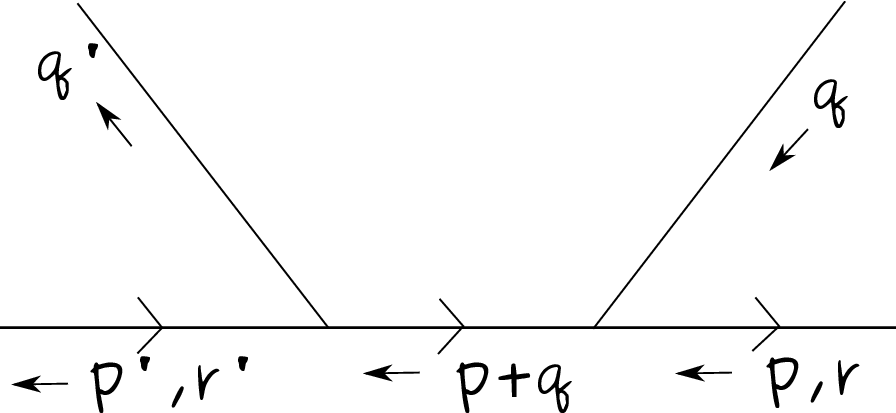}
$\quad\quad$\includegraphics[scale=0.3]{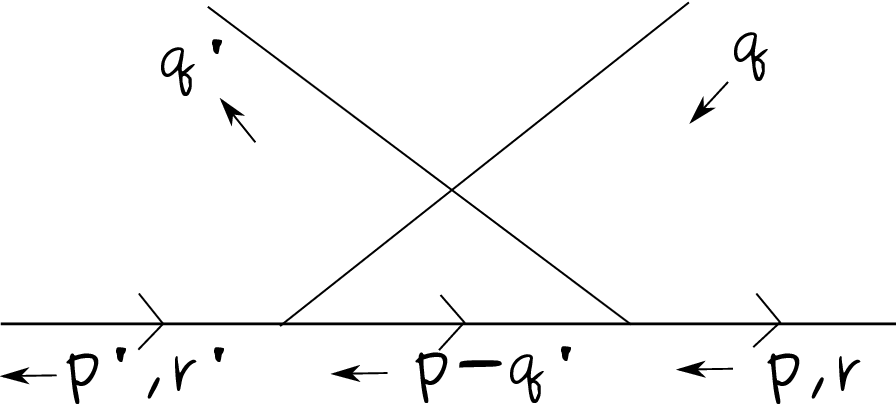}
\end{center}
\[ -(-ig)^2 (2\pi)^4 \delta^{(4)} (p+q - (p' + q')) \overline{\vvp}^{(r)} \Gamma \Big( \frac{i}{- \notp - \notq - m+ \ie} + \frac{i}{- \notp + \notq ' - m + \ie} \Big) \Gamma \vvpp^{(r')} \]

The initial antinucleon state gave us a $\overline{\vvp}^{(r)}$. If the initial state had been specified simply by some spinor $v$ (rather than a type $r$, one of our basis spinors), the amplitude is antilinear in $v$, that is linear in $\overline{v}$. (Why do we expect this?)\\

\begin{center}
\textbf{\uline{Feynman Rules} - Factors}
\end{center}

$\Diagram{\momentum[bot]{f}{q}}$ internal meson $\displaystyle \frac{i}{q^2 - \mu^2 + \ie} $\\

$\Diagram{\momentum[bot]{fA}{\rightarrow p}}$ internal nucleon
$\frac{i}{\cancel{p}-m+\ie}$ oriented along arrow (along charged flow for positively charged particles).\\

$\Diagram{\momentum[llft]{fdV}{p'\nwarrow } \\ & \momentum[bot]{f}{\leftarrow q} \\  \momentum[bot]{fuA}{\nearrow p}} \qquad -ig \Gamma (2\pi)^4 \delta^{(4)}(p+q-p')$\\

Integrate over internal momenta\\

\begin{itemize}
\item For every incoming nucleon (annihilated by a $ \psi$) get a $u$
\item For every incoming antinucleon (annihilated by a $ \psib$) get a $\overline{v}$
\item For every outgoing nucleon (created by a $\overline{\psi}$) get a $\overline{u}$
\item For every outgoing antinucleon (created by a $\psi$) get a $v$
\end{itemize}

Because fermions always appear bilinearly (we'll soon see quadrilinears are ruled out) in a Lagrangian, a fermion line either goes all the way through a graph or in a loop. Since we haven't done any examples with a fermion loop yet, we'll just do the matrix multiplication rules for a line going all the way through a graph first.
\vspace{1cm}

\begin{center}
\textbf{\uline{Feynman Rules - Matrix multiplication}}
\end{center}

Go to the head of any fermion line that goes all the way through a graph. You will either be at an incoming antinucleon, in which case you write down a $\overline{v}$, or at an outgoing nucleon in which case you write down a $\overline{u}$. (I start at the head of the line because I habitually write from left to right, so I start with the row vector then go through the matrices and finish with a column vector.)\\

Working against the arrows, the next thing you get is a vertex. Write down the matrix for the vertex. Then you'll get an internal line followed by another vertex some number of times. Write down the propagator matrix, then the interaction matrix each time.\\

When you get to the tail end of the line, you'll either be at an incoming nucleon, in which case write down a $u$, or at an outgoing antinucleon, in which case write down a $v$.\\

\uline{EXAMPLE}
\begin{center}
\includegraphics[scale=0.3]{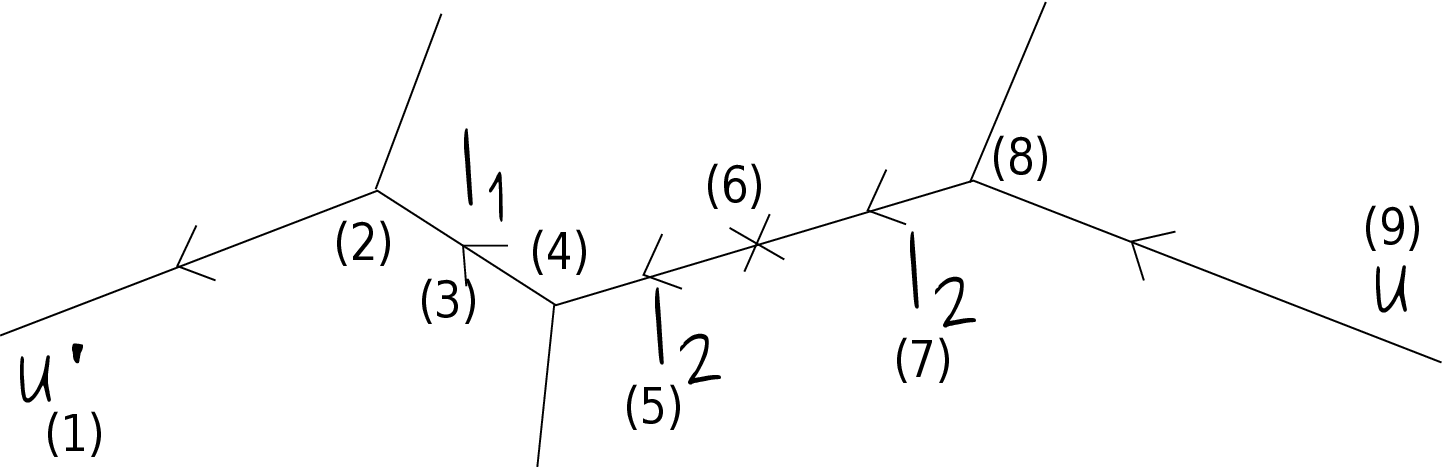}
\end{center}

(6): mass renormalization; comes from $-B\overline{\psi} \psi$ term in $\mathcal{L}_{CT}$.
\[ \underset{(1)}{\overline{u'}} \quad \underset{(2)}{\Gamma} \quad \underset{(3)}{\frac{i}{\cancel{l}_1 -m+\ie}} \quad \underset{(4)}{\Gamma} \quad \underset{(5)}{\frac{i}{\cancel{2}_1 -m+\ie}} \quad \underset{(6)}{\id} \quad \underset{(7)}{\frac{i}{\cancel{2}_1 -m +\ie}} \quad \underset{(8)}{\Gamma} \quad \underset{(9)}{u} \]

\vspace{1cm} 
What you finally obtain is a number. You get another product like this for each Fermi line that goes through a graph.\\

What about Fermi lines that go in loops? Here is a graph: 
\[ \Diagram{f fs0 flSA flSuV fs0 f} \] 

Another is 
\[ \Diagram{ fd & & & fu \\ & fV \\ & fvV & fvA \\ & fA \\ fu & & & fd } \]

This is an order $g^4$ diagram for $2$ meson $\longrightarrow$ $2$ meson scattering. The factor of interest in this contribution to the process is
\[ \wick{2111}{<4\psib \Gamma <1\psi(x_1) >1\psib \Gamma <2\psi(x_2) >2\psib \Gamma <3\psi(x_3) >3\psib \Gamma >4\psi (x_4)} \]

I'll rewrite this as 
\[ -\text{Tr } \wick{1111}{<1 \psi(x_4) >1\psib \Gamma <2 \psi(x_1) >2 \psib \Gamma <3 \psi(x_2) >3 \psib \Gamma <4\psi (x_3) >4 \psib(x_4) \Gamma} \]

Notice the minus sign. To put this in standard form for replacing $\wick{1}{<1 \psi >1\psib}$ by our integral over $\cfrac{i}{\notp-m}$, I not only had to move $\psi(x_4)$ all the way to the left and write the sum on $\alpha$ in $\Gamma_{\beta\alpha} \psi_\alpha (x_4)$ as a trace, I had to anticommute $\psi(x_4)$ by an odd number of Fermi fields to get it there. First $\overline{\psi} (x_4)$, then a bunch of bilinears. This is why there is a minus sign out front.\\

To conclude the matrix multiplication rules the rule is thus: If a Fermi line goes in a loop, start anywhere in the loop, and working against the arrows write down vertex and propagator matrices until you get back to where you started. Then take the trace to get a number. You get a factor like this for each Fermi loop. There are no $u$'s or $v$'s in the factor coming from a Fermi loop.
\vspace{1cm}

\begin{center}
\textbf{\uline{Feynman Rules + Fermi minus signs}}
\end{center}

One thing is for sure, each Fermi loop gives you a minus sign. \\

To get the rest of the minus signs, I'll do an example. Back on Oct.~28, we did ``nucleon"-``nucleon" scattering at $\mathcal{O}(g^2)$. We'll do \uline{the analogous calculation} for nucleon-nucleon scattering (no quotes). Eq.~(\ref{eq:11-page2}) is the expression whose spinor analog is going to give us some troublesome minus signs. We need to simplify:
\[ \frac{1}{2!} \langle p',r';q',s'|:\psib \Gamma \psi(x_1) \psib \Gamma \psi(x_2): |p,r;q,s \rangle \]

It is slightly ambiguous to write $|p,r;q,s\rangle$ and $\langle p',r';q',s'|$. The two possibilities for the ket are 
\begin{align*}
& |p,r;q,s \rangle = \!\!\!\!\!\!\!\!\!\!\!\!\!\!\!\!\!\! \overbrace{b^r(p)^\dagger}^{\text{relativistically normalized}} \!\!\!\!\!\!\!\!\!\!\!\!\!\!\!\!\! b^s(q)^\dagger |0 \rangle \qquad \text{and} \qquad |p,r;q,s \rangle = b^s(q)^\dagger b^r (p)^\dagger |0 \rangle = - b^r(p)^\dagger b^s(q)^\dagger |0 \rangle \\
& \{ b^r(p), b^s(p')^\dagger \} = (2\pi)^3 \delta^{rs} \delta^{(3)} (\vp - \vpp) 2E_{\vec{p}}
\end{align*}

It doesn't matter which choice you take - I'll take the second - as long as you choose $\langle p',r';q',s'| $ to be the corresponding bra.
\[ \langle p',r';q',s'| = ( |p ',r' ; q',s' \rangle) ^\dagger = \stai b^{r'}(p') b^{s'}(q') \]

That way the forward scattering amplitude when there is no interaction (or at zeroth order when there is) is positive, not negative. So what we have to simplify is 
\[ \frac{1}{2!} \stai b^{r'}(p') b^{s'}(q'):\psib \Gamma \psi(x_1) \psib \Gamma \psi (x_2) : b^s(q)^\dagger b^r(p)^\dagger \sta \]

$\psi(x_1)$ and $\psi(x_2)$ both contain operators which could annihilate either of the incoming nucleons. Let's say $\psi(x_2)$ annihilates the nucleon with momentum $q$. If $\psi(x_1)$ annihilates the nucleon with momentum $q$, we can rewrite what follows, and the only difference will be $x_1 \longleftrightarrow x_2$. Since these are dummy variables in an otherwise symmetric integration, I'll ignore this second case and cancel the $\frac{1}{2!}$. The coefficient of $b^s(q)$ in $\psi (x_2)$ is $\displaystyle u^s_q \frac{e^{-iq\cdot x_2}}{(2\pi)^3 2 E_{\vec{q}}}$, so what I have is 
\[ \stai b^{r'}(p') b^{s'}(q'):\psib \Gamma \psi (x_1) \psib (x_2) \Gamma u^s_q : b^r(p)^\dagger \sta e^{-iq\cdot x_2} \]

Move the bilinear past $\psib (x_2)$ and let $\psi (x_1)$ annihilate the remaining incoming nucleon:
\[ \stai b^{r'}(p') b^{s'}(q'):\psib(x_2) \Gamma u^s_q \psib (x_1) \Gamma u^r_p : \sta e^{-iq\cdot x_2} e^{-i p \cdot x_1} \]

Now either $\psib (x_2)$ or $\psib (x_1)$ can take care of the outgoing nucleon with momentum $q'$, but this does not just give us another factor of $2$. These two possibilities are different because a distinction between $x_1$ and $x_2$ has now been made by the way we annihilate the incoming nucleon.\\

Suppose $\psib (x_2)$ takes care of the outgoing nucleon with momentum $q'$. The coefficient of $b^{s'}(q')^\dagger$ in $\psib (x_2)$ is $\displaystyle \frac{\overline{u_{q'}}^{s'} e^{+i q' \cdot x_2}}{(2\pi)^3 2 E_{\vec{q}}}$, so what I have is
\[ \stai \, b^{r'}(p') \, \overline{u_{q'}}^{s'} \, \Gamma u^s_q \, \psib (x_1)\, \Gamma u_p^r \, \sta e^{-iq\cdot x_2}e^{-ip\cdot x_1} e^{iq'\cdot x_2} \]

The final simplification gives 
\[ \overline{u_{q'}}^{s'} \, \Gamma \, u^s_q \, \overline{u_{p'}}^{r'} \, \Gamma \, u^r_p \, e^{-iq\cdot x_2} e^{-ip\cdot x_1} e^{iq' \cdot x_2} e^{ip' \cdot x_1} \]

There are other factors, but the graph these factors are from is 
\begin{center}
\includegraphics[scale=0.3]{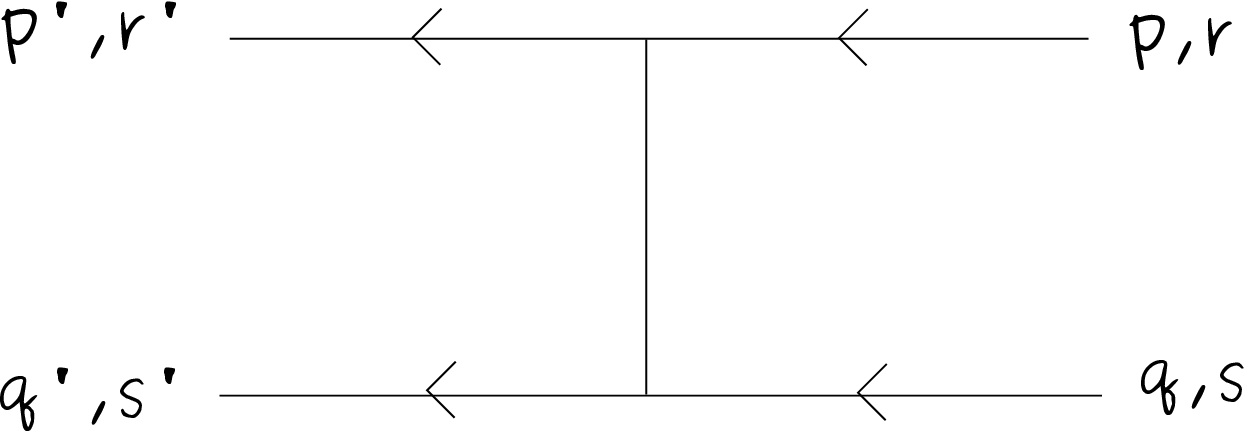}
\end{center}

$p$ is absorbed at the same spacetime point as $p'$ is created; $q$ is absorbed at the same spacetime point as $q'$ is created. \\

What about the contribution where $\overline{\psi}(x_1)$ takes care of the outgoing nucleon with momentum $q'$? First I'll anticommute the two $\psib$ fields to get: 
\[ -\stai | b^{r'}(p') b^{s'}(q'):\psib (x_1) \Gamma u^r_p \psib (x_2) \Gamma u^s_q \sta e^{-iq\cdot x_2} e^{-i p\cdot x_1} \]

Now the coefficient of $b^{s'}(q')^\dagger$ in $\psib (x_1)$ is $\displaystyle \frac{\overline{u_{q'}}^{s'} e^{iq' \cdot x_1}}{(2\pi)^3 2 E_{\vec{q}\,'}}$. So I get

The final simplification gives 
\[ -\overline{u_{q'}}^{s'} \Gamma u^r_p \overline{u_{p'}}^{r'} \Gamma u^s_q e^{-iq\cdot x_2 - ip\cdot x_1 + i q' \cdot x_1 + i p' \cdot x_2} \]

The differences with the previous expression worth noting are the minus sign and the different spinor structure. The exponential factors are different in the expected way. The graph is
\begin{center}
\includegraphics[scale=0.3]{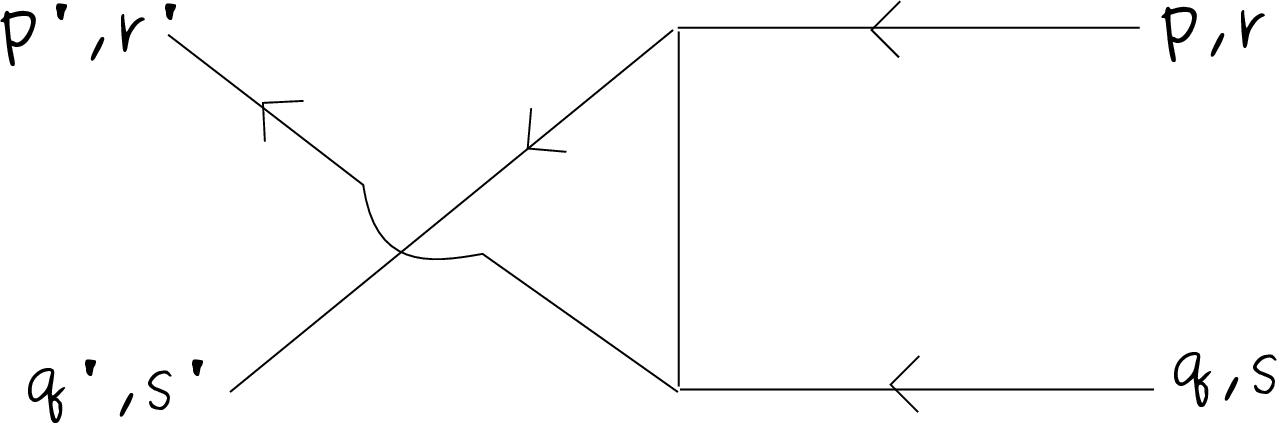}
\end{center}

Having obtained the spinor factor and the minus sign, you can continue on by doing the $x$ integrations in the fashion leading up to Eq.~(\ref{eq:11-page5})
for the ``nucleon"-``nucleon" scattering.\\

This was how reordering Fermi operators gives minus signs. I don't have a tidy little rule to summarize the sign of the general case, but of course you won't have to be so detailed when you are just checking how many reorderings of Fermi fields it takes to get a given contribution to a matrix element.
\vspace{1cm}

\textbf{\uline{EXAMPLE: COMPLETE EXPRESSION FOR} $N + N \longrightarrow N + N \qquad \Gamma = i \gamma^5$}
\begin{center}
\includegraphics[scale=0.3]{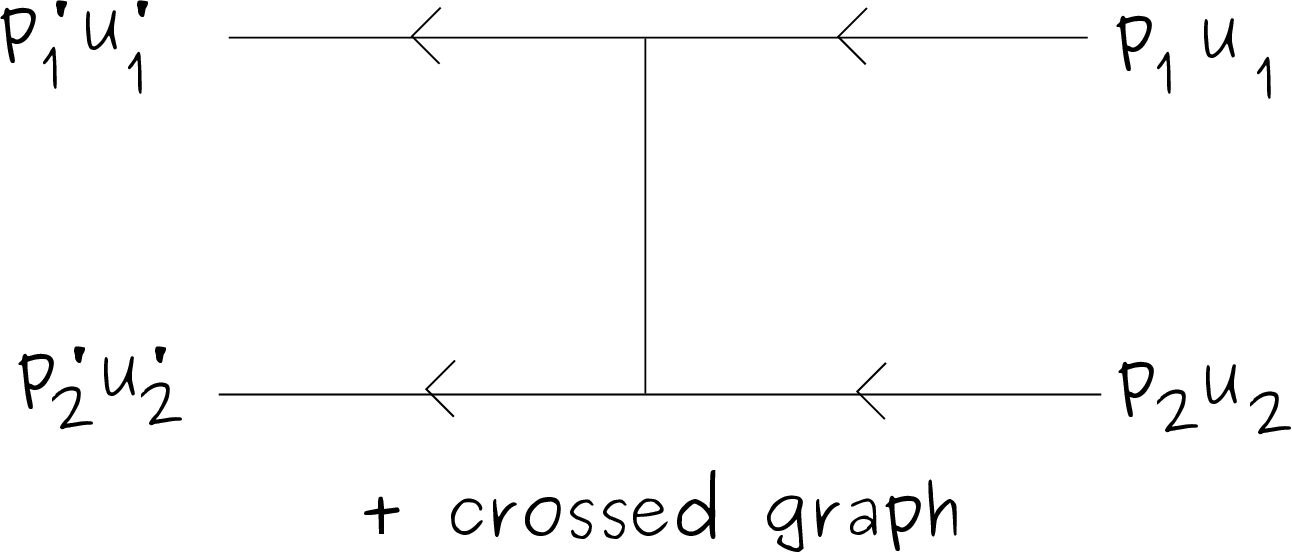}
\end{center}
\[ iA = (-ig)^2 \Bigg[ \overline{u'_1} i \gamma^5 u_1 \overline{u'_2} i \gamma^5 u_2 \frac{i}{(p_1-p'_1)^2 - \mu^2} (+1) + \overline{u'_1} i \gamma^5 u_2 \overline{u'_2} i \gamma^5 u_1 \frac{i}{(p_1-p'_2)^2 - \mu^2} (-1) \Bigg] \]
\vspace{1cm}

\textbf{\uline{EXAMPLE: COMPLETE CALCULATION OF} $N + \phi \longrightarrow N + \phi \qquad \Gamma = i \gamma^5$}
\begin{center}
\includegraphics[scale=0.3]{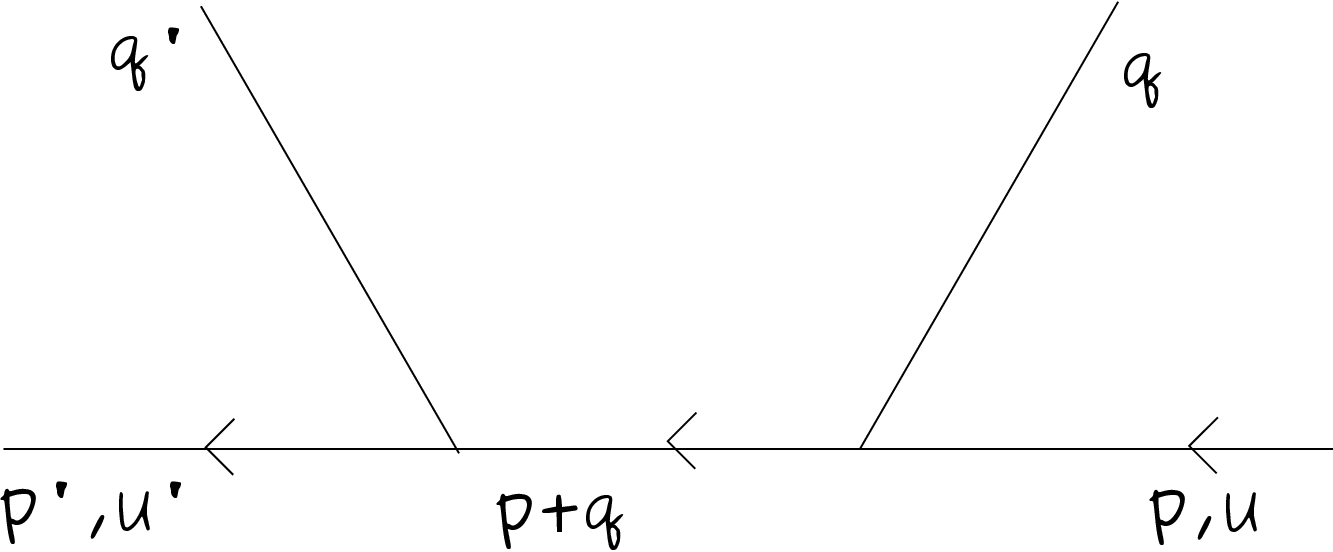}
\end{center}
\begin{center}
\includegraphics[scale=0.3]{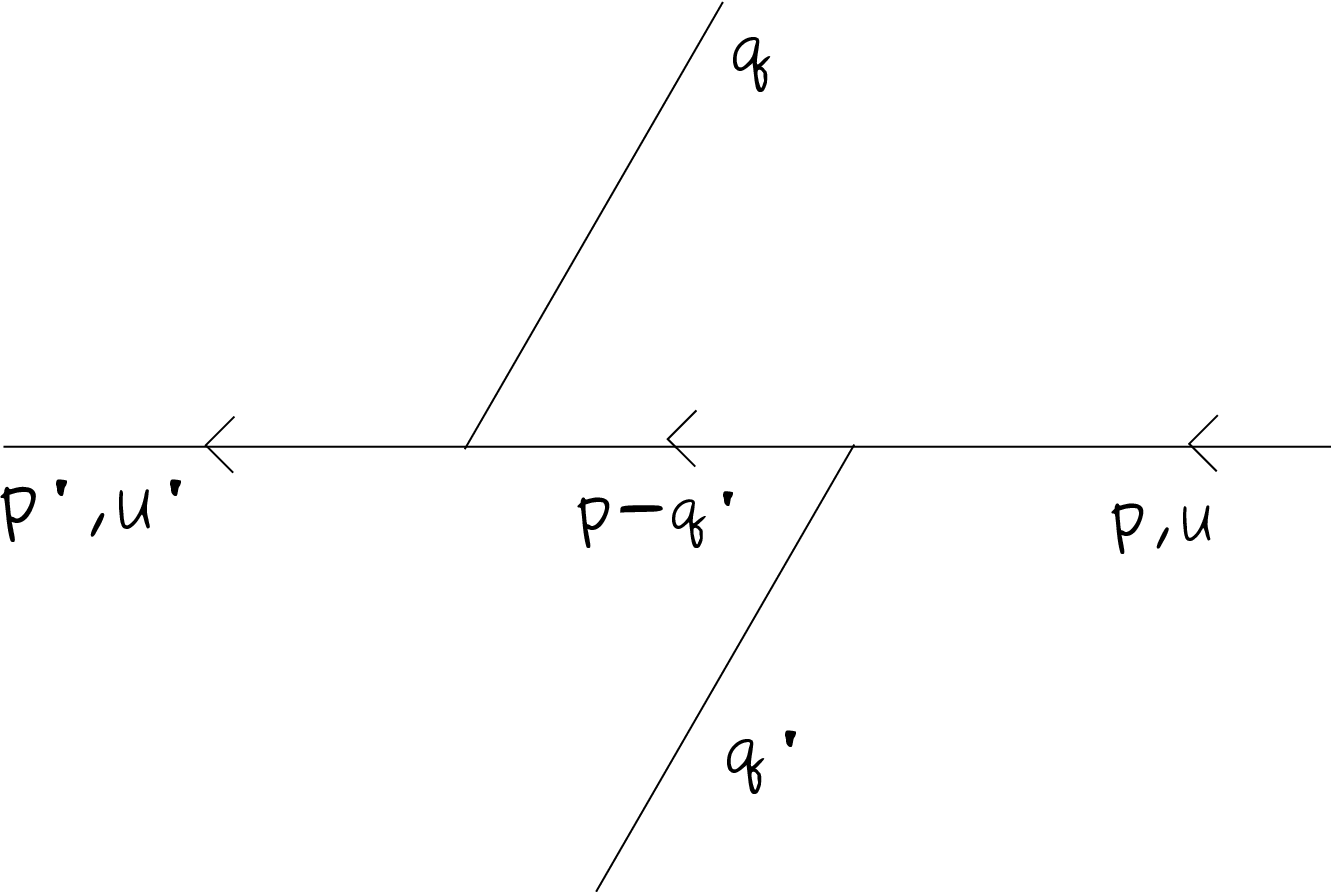}
\end{center}
\begin{align*}
iA &= (-ig)^2 \Bigg[ \overline{u'} i \gamma^5 \frac{i}{\notp + \notq -m+\ie} i\gamma^5 u + \overline{u'} i \gamma^5 \frac{i}{\notp' - \notq -m+\ie } i\gamma^5 u \Bigg] \\
A &= g^2 \overline{u'} \gamma^5 \Bigg[ \frac{\notp + \notq + m}{(p+q)^2-m^2} + \frac{\notp' - \notq +m}{(p'-q)^2-m^2} \Bigg] \gamma^5 u \qquad \text{USING $(\gamma^5)^2=1$ AND $\{ \gamma^5, \notp \} =0$} \\
A &= g^2 \overline{u'} \Bigg[ \frac{-\notp - \notq + m}{(p+q)^2-m^2} + \frac{-\notp' + \notq + m}{(p'-q)^2-m^2} \Bigg] u \qquad \text{USING $\notp u = mu$ AND $\overline{u'} \notp' = \overline{u'} m$} \\
&= g^2 \overline{u'} \notq u \Bigg[ \frac{1}{(p'-q)^2-m^2} - \frac{1}{(p+q)^2-m^2} \Bigg] 
\end{align*}

\textbf{\uline{SPIN AVERAGING AND SPIN SUMMING}}

INITIAL (RESP.~FINAL) SPINS IN EXPT.~UNKNOWN, AVERAGE (RESP.~SUM) TRANS.~PROB OVER THEM\footnote{Do not sum or average the Feynman amplitude. Average or sum \uline{probability}.}
\begin{align*}
M &= \overline{u_1} \overbrace{O^I}^{\text{MATRIX}} u_2 \overbrace{( \quad )_I }^{\substack{\text{KINEMATIC}\\\text{FACTORS}}} \qquad \text{i.e. } A= \overline{u_1} \gamma^\mu u_2 q_\mu, \quad I:\text{ collective indices} \\
|M|^2 &= \overline{u_{2 \, \gamma}} \Big[ \gamma^0 O^\dagger \gamma^0 u_1 \overline{u_1} O \Big]_{\gamma \delta} u_{2 \, \delta} \; \Big[\qquad \Big] \qquad \text{using $\sum_{\text{spin 1}} u_{1\, \alpha} \overline{u_{1 \, \beta}} = (\notp_1 + m)_{\alpha \beta}$} \\
&= \Big[ \gamma^0 O^\dagger \gamma^0 \frac{\notp_1 + m }{2m} O \Big]_{\gamma\delta} u_{2\,\delta}\overline{u_{2\,\delta}} \left[\qquad\right]\qquad\text{using $\sum_{\text{spin 2}}u_{2\alpha}\overline{u_{2\beta}}=(\notp_2+m)_{\alpha\beta}$}\\
&=\text{Tr}\left[\gamma_0O^\dagger\gamma^0\frac{\notp_1+m}{2m}O\frac{\notp_2+m}{2m}\right]
\end{align*}

In other calculations you will need
\[ \sum_{\text{spins}}v_3\overline{v_3}=\notp_3-m \]
\begin{align*}
\underbrace{\frac{1}{2}\sum_r}_{\text{initial av.}}\underbrace{\sum_s}_{\text{final sum}}|a_{rs}|^2&=\frac{1}{2}|F|^2\sum_{r,s}\text{Tr}\left[\overline{u_{\vec{p}\,'}}\notq u_{\vec{p}}^{(s)}\overline{u_{\vec{p}}}^{(s)}\notq u_{\vec{p}\,'}^{(r)}\right]\\
&=\frac{1}{2}|F|^2\text{Tr }\notq(\notp+m)\notq(\notp'+m)\\
&=\frac{1}{2}|F|^2(4m^2\mu^2+8p\cdot qp'\cdot q-4p\cdot p'\mu^2)
\end{align*}
}{
 \sektion{25}{January 6}
\descriptiontwentyfive
\begin{center}
\textbf{$P$, $C$, and $PT$ for spinor fields}
\end{center}

We know what $P$ does to a Dirac field
\[ P:\quad \psi (\vx,t ) \longrightarrow \beta \psi(-\vx,t) \]

We expect (general QM theorem coming from uniqueness of canonical commutation relations) that in the quantum theory there is a unitary operator effecting this change
\[ P:\quad \psi (\vx,t )\longrightarrow U_P^\dagger \psi ( \vx,t ) U_P = \beta \psi(-\vx,t) \]

$U_P$ is a unitary operator in Hilbert space. It has no spinor indices. (A Dirac field has a spinor index; each entry in the Dirac field is itself an operator in Hilbert space.)\\

From the effect of $U_P$ on $\psi$ and the expansion of $\psi$ in terms of creation and annihilation operators, we get the effect of $U_P$ on the creation and annihilation operators.
\[ \psi(\vx,t) = \sum_r \int \frac{d^3 p}{(2\pi)^{3/ 2 } \sqrt{2 E_{\vec{p}}}} \Big[ \bvp^{(r)} \!\!\!\!\!\! \underbrace{\uvp^{(r)} e^{-ip\cdot x}}_{\substack{\text{positive frequency}\\\text{solution of}\\\text{Dirac equation}}} \!\!\!\!\!\!+ \;\cvp^{(r) \dagger} \!\!\!\!\!\!\!\!\! \underbrace{\vvp^{(r)} e^{ip\cdot x}}_{\text{negative frequency}} \!\!\!\!\!\!\!\!\Big] \]

We need an expression for $\beta \psi(\vx,t)$. The first step is to evaluate $\beta \uvp^{(r)}$ and the first step to evaluating that is to find $\beta u_{\vec{0}}^{(r)}$. But that's easy, $\beta u_{\vec{0}}^{(r)} = u_{\vec{0}}^{(r)}$ comes right from $(\notp -m) = 0$ for $p = (m,\vec{0})$. To get the effect of $\beta$ on $\uvp^{(r)}$ we now use $\uvp^{(r)} = e^{\va \cdot \ve \, \phi/2} u_{\vec{0}}^{(r)} $.
\begin{align*}
\beta \uvp^{(r)} &= \beta e^{\va \cdot \ve \, \phi/2} u_{\vec{0}}^{(r)} \qquad \qquad \ve = \frac{\vp}{|\vp |} \quad \sinh \phi = \frac{| \vp |}{m} \\
&= e^{-\va \cdot \ve \, \phi/2} \beta u_{\vec{0}}^{(r)} \qquad (\text{because } \{ \beta,\alpha_i \} = 0) \\
&= e^{- \va \cdot \ve \, \phi /2} u_{\vec{0}}^{(r)} \\
&= u_{-\vp}^{(r)} 
\end{align*}

We have shown that parity does to positive frequency solutions of the Dirac equation what you would expect it to; it reverses the direction of motion but doesn't do anything to the spin.\\

So the effect of $U_P$ on $\bvp^{(r)}$ must be 
\[ U_P^\dagger \bvpr U_P = b_{-\vec{p}}^{(r)} \]

(and the h.c.~equation $U_P^\dagger \bvprd U_P = b_{-\vec{p}}^{(r)\dagger} $) \\

A very similar argument goes through for the $\cvprd$ which are multiplied by $\vvpr$ except for one thing 
\[ \beta v_{\vec{0}}^{(r)} = -v_{\vec{0}}^{(r)} \]

Because of that minus sign the effect of $U_P$ on $\cvprd$ is $U_P^\dagger \cvprd U_P= -c_{-\vp}^{(r) \dagger}$ (and the h.c.~eqn), $U_P^\dagger \cvpr U_P= - c_{-\vp}^{(r)}$.\\

The $b$'s have positive intrinsic parity, but the $c$'s have negative intrinsic parity. $U_P$ acting on a state with $n$ elections does nothing. $U_P$ acting on a state with $n$ positrons gives $(-1)^n$. Fermion and antifermion have opposite intrinsic parity. This is unlike the charged scalar, where both particle and antiparticle were scalar or pseudoscalar.\\

Is this just an artifact of some unfortunate convention\footnote{i.e.~could I redefine $P$ to be $P \; \times$ some internal symmetry which would not have this relative minus sign.}, or does this counterintuitive result have observable consequences.\\

Consider the process $N + \overline{N} \longrightarrow 2 \pi$. The nucleon and antinucleon are taken to be at rest, because this implies (no momentum $\longrightarrow$ no angular momentum) that they are in an $l=0$ state. There are two possible $l=0$ states, either $s=0$ or $s=1$.
\[ l=0 \qquad s=0 \qquad J=0 \qquad P = -1 \]
\[ l=0 \qquad s=1 \qquad J=1 \qquad P = -1 \]

The total angular momentum $J$ is absolutely conserved, and the parity is $-1$ because of what we have just found. \\

The two pseudoscalar pions either have 
\[ l=0 \qquad J=0 \qquad P=+1 \]
\[ l=1 \qquad J=1 \qquad P=-1 \]

($l=0$ is an even function of relative momentum so it is positive parity, and $l=1$ is an odd function of relative momentum so it has negative parity; in general you get $(-1)^l$.) The fact that the pion is pseudoscalar doesn't affect the outcome because there are two (an even number) of them.\\

Looking at the possibilities, in particular at the $J$ and $P$, you see that this process is forbidden. (Except by the $P$ violating weak interaction, but the strong interactions which are $P$ conserving and which would make this process occur very quickly compared to the weak interaction aren't allowed to do it).\\

This is a convention independent consequence of the opposite intrinsic parity of $N$ and $\overline{N}$.
\vspace{1cm}

\begin{center}
\textbf{$C$ Charge conjugation}
\end{center}

Recall how charge conjugation acted on a charged scalar:
\[ C:\quad \psi(x) \longrightarrow \psi^* (x) \qquad (\psi^*(x) \longrightarrow \psi(x)) \]

was a symmetry of the action. Alternatively, you could see that $\psi^*(x)$ was a solution of the equation of motion whenever $\psi$ was. In the free case:
\[ (\Box+ m^2) \psi = 0 \Longleftrightarrow (\Box+ m^2) \psi^* = 0 \]

This follows because $\Box + m^2$ is real. \\

(The free case is worth looking at because in general, when you add interactions you break symmetries, not create them.)\\

In the Dirac theory
\[ (i\notd -m) \psi = 0 \nLeftrightarrow (i\notd -m)\psi^* = 0 \]

Unless we are in a representation in which all of the $\gamma^\mu$ are purely imaginary. [This doesn't mean that a symmetry worth calling charge conjugation does not exist in a general basis; it just means it takes a more complicated form.] \\

I'll show that a representation in which all of the $\gamma^\mu$ are purely imaginary exists by constructing one. Any real change of coordinates preserves this property, so there are lots of possibilities. You already know four matrices with square one that anticommute with each other:
\[ \bpm 1 & 0 \\ 0 & -1 \epm \quad \text{and} \quad \bpm 0 & \vec{\sigma} \\ \vec{\sigma} & 0 \epm \]

Of these four, one is pure imaginary $\bpm 0 & \sigma_y \\ \sigma_y & 0 \epm $, and the others are pure real. So
\[ \gamma^0 = \bpm 0 & \sigma_y \\ \sigma_y & 0 \epm \quad \gamma^1 = i \bpm 1 & 0 \\ 0 & -1 \epm \quad \gamma^2 = i \bpm 0 & \sigma_x \\ \sigma_x & 0 \epm \quad \gamma^3 = i \bpm 0 & \sigma_z \\ \sigma_z & 0 \epm \]

satisfies $\Big\{ \gamma^\mu, \gamma^\nu \Big\} = 2 g\mn $ and $ \gamma^{\mu \, *} = - \gamma^\mu$, the ``Majorana" condition.\\

In this basis $ C:\; \psi(x) \longrightarrow \psi^*(x)$ is a symmetry of the free equation of motion and the free Dirac action (provided the classical field is thought of as anticommuting). \\

Thinking quantum mechanically, we expect a unitary operator, $U_C$, will exist such that 
\[ U_C^\dagger \psi(x) U_C = \psi^* (x) \]

Why write $\psi^*(x)$ and not $\psi^\dagger(x)$? Because we have been thinking of $\psi^\dagger$ as a row vector, not only have we been using $\dagger$ to mean complex conjugation of numbers and hermitian conjugation of operators, but transpose in spinor indices. We don't always want to do a transpose when we hermitian conjugate a spinor, so we'll use $\psi^* (x)$ to mean hermitian conjugation without transposition in spinor indices. If you like, $\psi^*(x) = \psi^{\dagger \, T} (x) $.\\

[There is always a similarity transformation between any two sets of $\gamma^\mu$ (that's Pauli's theorem), and so as not to completely jettison the approach that most books take to charge conjugation, I'll show what charge conjugation looks like in a general basis. Let 
\[ \psi_s = S \psi_m \qquad C:\; \psi_m \longrightarrow \psi_m^* \]

$\psi_m:$ Dirac spinor in Majorana basis, $\psi_s:$ some other basis like the standard one.\\

What does $C$ do to $\psi_s$?
\begin{align*}
C: \quad \psi_s \longrightarrow S \psi_m^* &= S S^{* -1} S^* \psi_m^* \\
&= S S^{* -1} (S \psi_m)^* \\
&= S S^{*-1} \psi_s^* 
\end{align*}

The matrix $SS^{*-1}$ is usually denoted $C$. The representation dependent computations that we'll do are so much simpler, that this is all we'll have to say about $C$.]

\vspace{1cm}

\begin{center}
LORENTZ TRANSFORMATION IN A MAJORANA BASIS
\end{center}
\[ M_i = \frac{i \alpha_i}{2} = \frac{i \gamma^0 \gamma^i}{2} \qquad \text{ so } M_i = - M_i^* \]
\[ L_k = i \epsilon_{ijk} M_i M_j \qquad \text{ so } L_k = - L_k^* \]
\[ D(A(\ve \phi)) = e^{-i\vm \cdot \ve \phi} = D(A)^* \]
\[ D(R(\ve \, \theta)) = e^{-i \vl \cdot \ve \, \theta} = D(R)^* \]
\[ \text{i.e.~} D(\Lambda) = D(\Lambda)^* \text{ real} \]

Charge conjugation commutes with Lorentz transformations. (The utility of this simple little result will become clear in the following.)\\

We want to find the effect of $U_C$ on creation and annihilation operators. You get that from the expansion of $\psi$ in terms of creation and annihilation operators and the assumed effect of $U_C$ on $\psi$; $U_C^\dagger \psi (x) U_C = \psi^* (x) $. We need to get the expansion of $\psi^*(x)$ in terms of creation and annihilation operators explicitly. Thus we need to know what $\uvp^{(r)*}$ and $\vvp^{(r)*}$ are. From the eqn. 
\[ (\notp -m)\uvpr = 0, \]

complex conjugated in a Majorana basis, we have 
\[ (-\notp - m) \uvp^{(r)*} = 0 \quad \text{i.e. } (\notp +m) \uvp^{(r)*} = 0. \]

This says that the complex conjugate of a solution of the Dirac equation that has positive frequency is a solution of the Dirac equation with negative frequency.\\

We are free to choose the $\uvpr$ and $\vvpr$ any way we want. So to make the action of charge conjugation simple, choose 
\[ \vvpr = \uvp^{(r)*} \]

This is consistent with
\begin{align*} 
\uvpr &= D(A(\ve \phi)) u_{\vec{0}}^{(r)} \\
\vvpr &= D(A(\ve \phi)) v_{\vec{0}}^{(r)} 
\end{align*}

because charge conjugation commutes with Lorentz transformation (in a Majorana basis). \\

Let's see what this implies about the effect of charge conjugation on spins. From the complex conjugate of 
\[ L_z u_{\vec{0}}^{(1)} = \frac{1}{2} u_{\vec{0}}^{(1)} \]

we have $-L_z u_{\vec{0}}^{(1)*} = \frac{1}{2} u_{\vec{0}}^{(1)*}$, i.e. 
\[ L_z v_{\vec{0}}^{(1)} = - \frac{1}{2} v_{\vec{0}}^{(1)} \]

Since $u$ multiplies an annihilation operator and $v$ multiples a creation operator, this is exactly what we expect for charge conjugation. It shouldn't have any effect on spin.
\[ \psi(x) = \sum_r \int \frac{d^3 p}{(2\pi)^{3/2} \sqrt{ 2 E_{\vec{p}}}} \Big[ \bvp^{(r)} \uvp^{(r)} \eipxm + \cvprd \vvpr \eipx \Big] \]

So, 
\begin{align*} 
\psi^*(x) &= \sum_r \int \frac{d^3 p}{(2\pi)^{3/2} \sqrt{ 2 E_{\vec{p}}}} \Big[ \bvp^{(r)\dagger} \uvp^{(r)*} \eipx + \cvpr \vvp^{(r)*} \eipxm \Big] \\
&= \sum_r \int \frac{d^3 p}{(2\pi)^{3/2} \sqrt{ 2 E_{\vec{p}}}} \Big[ \bvp^{(r)\dagger} \vvp^{(r)} \eipx + \cvpr \uvp^{(r)} \eipxm \Big]
\end{align*}

We equate this with 
\[ U_C^\dagger \psi(x) U_C = \sum_r \int \frac{d^3 p}{(2\pi)^{3/2} \sqrt{ 2 E_{\vec{p}}}} [U_C^\dagger \bvpr U_C \uvpr \eipxm + U_C^\dagger \cvprd U_C \vvpr \eipx \Big] \]

Matching coefficients gives ($U_C^* \psi(x) U_C^\dagger = \psi^*(x)$)
\[ U_C^\dagger \bvpr U_C = \cvpr \quad \text{and} \quad U_C^\dagger \cvprd U_C = \bvprd \]

The h.c.~equations are 
\[ U_C^\dagger \bvprd U_C = \cvprd \quad \text{and} \quad U_C^\dagger \cvpr U_C = \bvpr \]

This couldn't be simpler. Thanks to the way we set up the correspondence, complex conjugation does not mix up the spin ups and spin downs. Spin ups transform into spin ups and spin downs transform into spin downs, exactly as if the spin up electron was a boson whose antiparticle is a spin up positron.
\vspace{1cm}

\begin{center}
\textbf{Construction of nucleon-antinucleon state}
\end{center}

\uline{Scalar case} (``nucleon" -``antinucleon" state for warm up and comparison) 
\[ \!\!\!\!\!\!\!\!\!\!\!\!\!\!\!\!\!\underbrace{\cvpp^\dagger}_{\text{adds an ``anti-nucleon"}} \!\!\!\!\!\!\!\!\!\!\!\!\!\!\!\!\!\!\!\!\!\!\!\!\!\overbrace{\bvp^\dagger}^{\text{creates ``nucleon"}}\!\!\!\!\!\!\!\!\!\!\sta \]
\[ U_C \cvppd \bvpd \sta = \bvppd \cvpd \sta \]
\[ | \psi \rangle \equiv \int d^3 p \; d^3 p' \; F(\vp, \vp\,') \cvppd \bvpd \sta \]

Then $U_C | \psi \rangle = \pm | \psi \rangle \quad \text{if} \quad F(\vp\,' \vp) = \pm F(\vp, \vp\,')$\\
\vspace{1cm}

\uline{Fermionic case}
\[ |\psi \rangle = \sum_{rs} \int d^3p \; d^3 p' \; F_{rs} (\vp, \vp\,') \cvpp^{(s)\dagger} \bvprd \sta \]
\[ U_C | \psi \rangle = \mp | \psi \rangle \quad \text{ if } F_{sr} (\vp, \vp\,') = \pm F_{rs} (\vp\,',\vp) \]

An \uline{antisymmetric} state of fermion and antifermion is charge conjugation \uline{even}$!$ Came from (anti-)commutation relations. 
\vspace{1cm}

\begin{center}
\textbf{Charge conjugation properties of fermion bilinears}
\end{center}

Consider $\overline{A}MB$. So we can anticommute $\overline{A}$ and $B$ without worrying about the anticommutator, either consider the Fermi fields $A$ and $B$ to be classical Fermi fields, or consider the normal ordered product, $:\overline{A}MB:$. $M$ is just some matrix in spinor space, like $i\gamma_5$. \\

Under charge conjugation
\[ U_C^\dagger A U_C = A^* \qquad U_C^\dagger B U_C = B^* \]

From this 
\[ U_C^\dagger A^* U_C = A \qquad U_C^\dagger B^* U_C = B \]

(or 
\[ U_C^\dagger A^\dagger U_C = A^T \qquad U_C^\dagger B^\dagger U_C = B^T \]

same statement as a row vector.)\\

Now $\overline{A} = A^\dagger \gamma^0$ so 
\begin{align*}
U_C^\dagger \overline{A} U_C &= U_C^\dagger A^\dagger U_C \gamma^0 = A^T \gamma^0 \\
&= A^{\dagger *} ( - \gamma^{0 *}) = - (A^\dagger \gamma^0 )^* \\
&= -\overline{A}^* 
\end{align*} 

also,
\[ U_C^\dagger \overline{B} U_C = - \overline{B}^* \]
\vspace{1cm} 

$\overline{B} \; \overline{M}^* A$:
\[ \psib \gamma^\mu \psi \longrightarrow -\psib \gamma^\mu \psi \qquad \psib i \gamma_5 \psi \longrightarrow \psib i \gamma_5 \psi \]
\[ \psib \sigma\mn \psi \longrightarrow - \psib \sigma\mn \psi \qquad \psib \gamma_\mu \gamma_5 \psi \longrightarrow \psib \gamma_\mu \gamma_5 \psi \]
\vspace{1cm}

\uline{Exercise}
In quantum mechanics you frequently use $(\mathcal{O}_1 \mathcal{O}_2 )^\dagger = \mathcal{O}_2^\dagger \mathcal{O}_1^\dagger $. Prove this from the definition of the adjoint $(\phi, \mathcal{O}^\dagger \psi) = (\psi, \mathcal{O} \phi )^* $. Is this formula changed if $\mathcal{O}_1$ and $\mathcal{O}_2$ are Fermi? No.
\begin{align*}
U_C^\dagger:\overline{A} M B : U_C &= : U_C^\dagger \overline{A} M B U_C: \qquad \substack{\text{This step is allowed}\\\text{ because $U_C$ does not}\\\text{ mix up creation and}\\\text{ annihilation operators}} \\
&=:U_C^\dagger \overline{A} U_C M U_C^\dagger B U_C : \\
&= -:\overline{A}^* M B^*:
\end{align*}

To do the next step, I am going to explicitly display the spinor matrix multiplications so I don't have to worry about keeping matrices and spinors in a given order. The idea of the next step is to write this as the complex conjugate of something. We have 
\begin{align*}
-:\overline{A}^* M B^* : &= - : \overline{A}_\alpha^* M_{\alpha \beta} B^*_\beta :\\
&= -:B_\beta M_{\alpha \beta}^* \overline{A}_\alpha : ^* \qquad \qquad \substack{\text{remember $*$ means}\\\text{adjoint without}\\\text{transverse and}\\\text{adjoint reverses order}} \\
&= +:\overline{A}_\alpha M_{\alpha\beta}^* B_\beta :^* \qquad \substack{\text{Fermi fields anticommute}\\\text{inside normal ordered product}}
\end{align*}

This last anticommutation puts things back in the right order to use the conventions of spinor matrix multiplication. What I have shown is 
\begin{align*}
U_C^\dagger:\overline{A} M B : U_C &= : \overline{A} M^* B : ^* \\
&=:\overline{B} \, \overline{M^*} A: \qquad \substack{\text{look back}\\\text{when we introduced}\\\text{the bar of a matrix}}
\end{align*}

So all you have to do to calculate the effect of $C$ on our $16$ bilinears is to calculate things like 
\begin{align*}
\overline{1}^* &= 1 &\\
\overline{\gamma^\mu}^* &= - \gamma^\mu &\text{in a Majorana basis} \\
\overline{i\gamma_5}^* &= i \gamma_5 &\text{in a Majorana basis} \\
\overline{\gamma_5 \gamma^\mu}^* &= (\gamma_5 \gamma^\mu)^* = \gamma_5 \gamma_\mu & \\
\overline{\sigma\mn}^* &= - \sigma\mn
\end{align*}

(You can be sloppy and not distinguish between $\overline{M}^*$ and $\overline{M^*}$ in a Majorana basis.)\\

So $\psib \psi$ is charge conjugation invariant. That's good; it would be bad to find out that our mass term breaks $C$.\\

$g_1 \phi \psib \psi + g_2 \phi \psib \gamma_5 \psi$ is charge conjugation invariant\\

$-e \!\!\!\!\!\!\!\!\!\underbrace{A_\mu}_{\text{vector mesons}} \!\!\!\!\!\!\!\!\! \psib \gamma^\mu \psi $ is charge conjugation invariant only if $A_\mu$ is charge conjugation odd, \\$U_C^\dagger A_\mu U_C = - A_\mu $.\\

$ a \!\!\!\!\!\!\!\! \underbrace{W_\mu}_{\text{vector mesons}}\!\!\!\!\!\!\! \psib \gamma^\mu \gamma^5 \psi + v W_\mu \psib \gamma^\mu \psi $ is parity violating and $C$ violating, but it preserves $CP$.
\[ : \psib \sigma\mn \psi:\longrightarrow - :\psib \sigma\mn \psi : \]

$10$ odd fermion bilinears from symmetric combinations and $6$ even ones from antisymmetric combination 
\[ D^{(1/2,1/2)} \otimes D^{(1/2,1/2)} = D^{(1,1)} \oplus D^{(0,0)} + D^{(0,1)} \oplus D^{(1,0)} \]
\vspace{1.5cm}

\[ | \psi \rangle = \int d^3 p \; d^3 p' \; f(\vp, \vp\,') \bvpd \, \cvppd \sta \]

Bosons:
\[ U_C |\psi \rangle = \pm | \psi \rangle \qquad \text{ if } f(\vp,\vp\,') = \pm f (\vp\,',\vp) \]

Fermions:
\[ U_C |\psi \rangle = \pm | \psi \rangle \qquad \text{ if } f(\vp,\vp\,') = \mp f(\vp\,', \vp) \]
\vspace{1.5cm}

\begin{center}
\textbf{Example of Charge Conjugation in QED}
\end{center}

The QED Lagrangian contains $A_\mu J^\mu$ where 
\[ J^\mu = e \psib \gamma^\mu \psi \]

and $A_\mu$ is the photon field. If $J^\mu$ is charge conjugation odd and if the Lagrangian is to preserve charge conjugation, then $A_\mu$ must be charge conjugation odd and a state with $N$ photons satisfies:
\[ U_C | N \gamma \rangle = (-1)^N | N \gamma \rangle \]

We'll use this to evaluate the relative decay rates of ortho and para positronium, the two lowest nearly degenerate hydrogen like bound states of an electron and a positron
\[ l=0 \quad \left\{\begin{array}{l} \text{ortho}: \quad s=1 \quad J =1 \quad C = -1 \\ \text{para }: \quad s=0 \quad J=0 \quad C=+1 \end{array} \right. \]

(``para" means opposite to)\\

The charge conjugation properties are deduced: a state of one electron and one positron stands a chance of being a charge conjugation eigenstate. The orbital wave function when $l$ is even is symmetric. When two spin $\frac{1}{2}$ are put together in a symmetric combination, you get a spin 1 state. When they are put together antisymmetrically, you get a spin 0 state. These facts, and the crucial Fermi minus sign from anticommuting particle and antiparticle
creation operators gives the charge conjugation eigenvalue. Now what are the possible decay products? Electron and positron are the lightest charged particle antiparticle pairs so the decay must be into $n$ photons. By kinematics alone, one photon is not allowed. Two or more photons are kinematically allowed, but each additional photon comes with a factor of $e$ in the amplitude, or $e^2$ in the probability. Assuming there are no numerical surprises (without doing some calculations there is no way to rule out factors like $(2\pi)^4$ in relative amplitudes), the partial decay rate into $n+1$ photons
should be down by a factor of $e^2 \approx \frac{1}{137}$ compared to the partial decay rate into $n$ photons, assuming they are both allowed. The decay that goes fastest will be the one that goes into the lowest number of photons. The lowest possibilities are 
\[ 2 \text{ photons} \qquad C=+1 \]
\[ 3 \text{ photons} \qquad C=-1 \]
\begin{align*}
\text{para} \longrightarrow 2 \gamma \qquad &\text{allowed}\\
\text{ortho} \longrightarrow 2 \gamma \qquad &\text{not allowed}\\
\text{ortho} \longrightarrow 3 \gamma \qquad &\text{allowed}
\end{align*}

For another explanation along these lines, see I+Z p.154, where the experimental values are also given.
\[ U_CU_P=U_PU_C(-1)^{N_F}=U_PU_CU(R(2\pi\ve)) \]
\vspace{1cm}

\uline{Show: $U_C U_P = U_P U_C U(R(\ve \, 2 \pi))$}

$U(R (\ve\,2\pi))$ is a fancy way of writing the operator $(-1)^{N_F}$ ($N_F = \#$ of fermions)\\

Write $(-1)^{N_F}$ this way because it reminds you that it is a symmetry of the theory.\\

One way to show this is to show it is true when acting on an arbitrary state, or at least a basis. Consider 
\[ b_{\vec{p}_1}^\dagger \cdots b_{\vec{p}_n}^\dagger c_{\vec{p}_1}^\dagger \cdots c_{\vec{p}_n}^\dagger \sta \] 

States of this form are a basis and states of this form it is easy to convince yourself the identity is true.
\vspace{1cm}

\begin{center}
\textbf{PT}
\end{center}

Just as in a scalar theory, Lorentz invariance makes it easier to consider $PT$ than $T$. Recall $PT$ in the classical scalar theory
\[ (\Box + m^2) \phi(x) = 0 \Longrightarrow (\Box +m^2) \phi(-\vx,-t) = 0 \]

so we expect there to be an antiunitary operator having the effect 
\[ \Omega_{PT}^{-1} \phi(x) \Omega_{PT} = \phi(-x) \]

In the Dirac theory
\[ (i\notd -m)\psi(x) = 0 \nRightarrow (i\notd - m) \psi(-x) = 0 \]

Now an operation worth calling $PT$ can have a more general form.
\[ PT: \psi(x) \longrightarrow M \psi(-x) \]

where $M$ is some four-by-four matrix in spinor space. What we need is
\[ (i\notd - m) \psi(x) = 0 \Longrightarrow (i\notd - m) M \psi(-x) = 0 \]

i.e.~$M$ must anticommute with $\gamma^\mu$. We'll take $M= i\gamma_5$. Up to a factor this choice is unique.\\

[ Proof: Suppose there is a second $M$ anticommuting with the $\gamma^\mu$, call it $M'$. Then $MM'$ commutes with every one of the $16 \;\; \Gamma$ matrices, $MM'$ must be proportional to the identity and $M'$ must be proportional to $\gamma_5^{-1} = \gamma_5$ ] 
\[ PT: \quad \psi (x) \longrightarrow i \gamma_5 \psi(-x) = \Omega_{PT}^{-1} \psi(x) \Omega_{PT} \]

Consider applying $PT$ twice
\begin{align*}
\Omega_{PT}^{-2} \psi(x) \Omega_{PT}^2 &= \Omega_{PT}^{-1} i \gamma_5 \psi (-x) \Omega_{PT}\\
&= i\gamma_5 \Omega_{PT}^{-1} \psi(-x) \Omega_{PT} \qquad \substack{i \gamma_5 \text{ is real so}\\\text{it goes through}\\ \Omega_{PT}}\\
&= i \gamma_5 i \gamma_5 \psi(x) = - \psi(x) \qquad \gamma_5^2 =1 \\
&= U(R (\ve\,2 \pi) ) \psi(x) 
\end{align*}

$PT$ is a rotation half way around the rotation group. This proof is unaffected by giving $M$ an arbitrary phase\\
\[ (\Omega_{PT}^2 = U(R(\ve\,2\pi)) \]
\vspace{1cm}

Using $(\Omega^{-1} A \Omega )^\dagger = \Omega^{-1} A^\dagger \Omega$ \\

[ Proof: 
\begin{align*}
(b, (\Omega^{-1} A \Omega)^\dagger a ) &= ( a,\Omega^{-1} A \Omega b )^* &\text{definition of adjoint of } \Omega^{-1} A \Omega \\
&= (\Omega a, \Omega \Omega^{-1} A \Omega b )& \text{antiunitary of } \Omega \\
&= (\Omega a, A \Omega b) &\\
&= ( \Omega b, A^\dagger \Omega a )^* & \text{definition of adjoint of } A \\
&= (\Omega^{-1} \Omega b, \Omega^{-1} A^\dagger \Omega a ) & \text{antiunitarity of } \Omega^{-1} \\
&= ( b, \Omega^{-1} A^\dagger \Omega a )\quad]
\end{align*}
\begin{align*} 
\Omega_{PT}^{-1} \psid(x) \Omega_{PT} &= ( i \gamma_5 \psi(-x))^\dagger \\
& = - \psid(-x) i \gamma_5 
\end{align*}

This tells how $\psib = \psi^\dagger \beta $ transforms
\begin{align*}
\Omega_{PT}^{-1} \psib \Omega_{PT} &= \Omega_{PT}^{-1} \psid \Omega_{PT} \overbrace{\beta^*}^{-\beta}\\
&= - \psid (-x) i \gamma_5 ( - \beta) \\
&= - \psib(-x) i\gamma_5
\end{align*}

So
\[ PT:\quad \psib(x) \psi(x) \longrightarrow \psib(-x) \psi(-x) \]
}{
 \sektion{26}{January 8}
\descriptiontwentysix
\begin{center}
\textbf{$PT$ (cont'd)}
\end{center}
($PT$ commutes with Lorentz transformations)\\

$\Omega_{PT}$ on states.\\

We expect $PT$ to do nothing to momenta 
\[ \vec{k} \underset{P}{\longrightarrow} - \vec{k} \underset{T}{\longrightarrow} \vec{k} \]

Reflection turns the momenta around but running the movie backward does it again.\\

The spin of a particle is affected though.
\[ \vec{S} \underset{P}{\longrightarrow} \vec{S} \underset{T}{\longrightarrow} - \vec{S} \]

Thus we can't expect to find a basis where single particle states are left unchanged. I.e.
\[ \Omega_{PT}^{-1} \left\{ \begin{array}{l} \bvpr \\ \cvpr \end{array} \right\} \Omega_{PT} = \left\{ \begin{array}{l} \bvp^{(r)'} \\ \cvp^{(r)'} \end{array} \right\} \]

where
\begin{align*}
\bvp^{(1)'} &= \text{some phase } \times \bvp^{(2)} \\
\bvp^{(2)'} &= \text{some other phase } \times \bvp^{(1)}
\end{align*}

($ \text{some other phase } = \text{ - some phase}^* $)\\

So that $\Omega_{PT}$ applied twice gives $- \left\{ \begin{array}{l} \bvpr \\ \cvpr \end{array} \right\} $. 
\vspace{1cm}

\begin{center}
\textbf{Tricky choice of spinor basis}
\end{center}

If $\bvpr$ is associated with the solution of the Dirac equation $\uvpr$, let $\bvp^{(r)'}$ be associated with $ \uvp^{(r)'} \equiv - i \gamma_5 \uvp^{(r)*} $.

If $\cvpr$ is associated with $\vvpr$, let $\cvp^{(r)'}$ be associated with $ \vvp^{(r)'} \equiv - i \gamma_5 \vvp^{(r)*} $.\\

What's good about this choice? It makes the action of $\Omega_{PT}$ on creation and annihilation operator simple. Furthermore it agrees with the expectations of the previous page.\\

If $(\notp -m) u = 0 $, then (taking c.c.), $ (-\notp - m ) u^* = 0 \text{ and } (\notp -m)(-i \gamma_5 u^*) = 0 $.\\

Given $L_z u_{\vec{0}}^{(1)} = + \frac{1}{2} u_{\vec{0}}^{(1)} $, then (taking c.c.), $ L_z u_{\vec{0}}^{(1)*} = - \frac{1}{2} u_{\vec{0}}^{(1)*}$ and \\
$L_z \big(-i\gamma_5 u_{\vec{0}}^{(1)*}\big) = - \frac{1}{2} \big(- i \gamma_5 u_{\vec{0}}^{(1)*} \big)$\\

Because Lorentz transformations are real in a Majorana basis, this generalizes to moving states.\\

The action of $\Omega_{PT}$ on creation and annihilation operators is derived from the action of $\Omega_{PT}$ on the field, and the expansion of the field $ \vvp^{(r)'} \equiv - i \gamma_5 \vvp^{(r)*} $.\\

So now we'll see that the definitions of $\bvp^{(r)'}$ and $\cvp^{(r)'}$ and 
\[ \Omega_{PT}^{-1} \left\{ \begin{array}{l} \bvpr \\ \cvpr \end{array} \right\} \Omega_{PT} = \left\{ \begin{array}{l} \bvp^{(r)'} \\ \cvp^{(r)'} \end{array} \right\} \]

are consistent with
\[ \Omega_{PT}^{-1} \psi(x) \Omega_{PT} = i \gamma_5 \psi(-x) \]

which is equivalent to
\[ -i\gamma_5 \Omega_{PT}^{-1} \psi(-x) \Omega_{PT} = \psi(x) \]

We can write the expansion of $\psi(x)$ two ways 
\[ \psi(x) = \sum \int \!\!\!\!\!\!\!\!\!\!\!\!\!\!\!\!\!\!\!\!\!\!\! \underbrace{(\cdots)}_{\substack{\text{kinematic factors}\\\text{unimportant to the argument}}} \!\!\!\!\!\!\!\!\!\!\!\!\!\!\!\!\!\!\!\!\!\! \Big[ \bvpr \, \uvpr \, e^{-ip \cdot x} + \cvprd \, \vvpr \, \eipx \Big] \]

or 
\[ \psi(x) = \sum \int (\cdots) \Big[ \bvp^{(r)'} \uvp^{(r)'} \eipxm + \cvp^{(r)'\dagger} \vvp^{(r)'} \eipx \Big] \]

Use the first way in the LHS and the second way in the right. Writing out the LHS, we have 
\begin{align*}
\text{LHS} &= - i \gamma_5 \Omega_{PT}^{-1} \int (\cdots) \Big[ \bvpr \uvpr \eipx + \cvprd \vvpr \eipxm \Big] \Omega_{PT} \\
&= -i\gamma_5 \int (\cdots) \Big[ \Omega_{PT}^{-1} \bvpr \Omega_{PT} \uvp^{(r)*} \eipxm + \Omega_{PT}^{-1} \cvprd \Omega_{PT} \vvp^{(r)*} \eipx \Big] \\
&= \int (\cdots) \Big[ \bvp^{(r)'} \uvp^{(r)'} \eipxm + \cvp^{(r)'\dagger} \vvp^{(r)'} \eipx \Big] \\
&= \text{RHS}
\end{align*}

\begin{center}
\textbf{Proof of $PCT$ within perturbation theory} 
\end{center}

For scalars $TCP$ invariance of the $S$ matrix was equivalent to
\[ a(p_1,\cdots,p_n) = a (-p_1,\cdots,-p_n) \]

This says that the amplitude with all incoming particles turned into outgoing antiparticles with the same 3-momentum is the same. What is the corresponding statement when there are Dirac particles in the theory? We'll simplify by looking only at
\[ \text{1 fermion + any number of mesons} \longrightarrow \text{1 fermion + any other number of mesons} \]
\begin{center}
\includegraphics[scale=0.6]{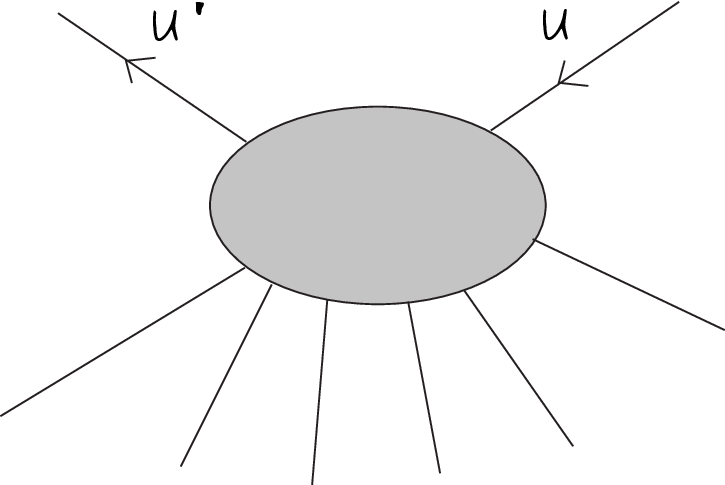}
\end{center}
\[ a \text{ is of the form} \]
\[ a = \overline{u'} M(p_1, \cdots, p_n) u \]

Instead of having an outgoing fermion characterized by $\overline{u'}$, the CPT reversed process has an incoming antifermion with the opposite spin characterized by $- i \overline{u'} \gamma_5$. Instead of an incoming fermion characterized by $u$ the CPT reversed process has an outgoing antifermion characterized by $-i \gamma_5 u$. If you want to understand this in two steps 
\[ \underbrace{u}_{\substack{\text{describes}\\\text{an incoming}\\\text{fermion with}\\\text{some spin}}} \underset{PT}{\longrightarrow} \underbrace{-i\gamma_5 u^*}_{\substack{\text{describes}\\\text{an outgoing}\\\text{fermion of the}\\\text{opposite spin}}} \underset{C}{\longrightarrow} \underbrace{-i \gamma_5 u}_{\substack{\text{describes}\\\text{an outgoing}\\\text{antifermion}\\\text{of the opposite spin}}} \]

(This is a definite choice for the transformed spin, a choice of another phase would screw up the CPT theorem.)\\

There is also an additional minus sign in the amplitude. Because an operator has to have an odd number of reordering of Fermi fields to contribute to this CPT reversed process.\\

Equality of the amplitude for this process and the CPT transformed process is thus 
\[ \overline{u'} M(p_1,\cdots, p_n) u = \!\!\!\!\!\!\!\!\!\!\!\!\!\!\!\!\!\! \underbrace{-}_{\text{switching the operators}} \!\!\!\!\!\!\!\!\!\!\!\!\!\!\!\!\!\! (-i)^2 \overline{u'} \gamma_5 M(-p_1, \cdots, -p_n) \gamma_5 u \]

The proof that these two are equal only uses L.I.~of the Feynman rules. Whatever the Feynman rules are, L.I.~tells us that 
\[ \overline{u'} M(p_1, \cdots,p_n) u = \overline{u'} \overline{D(\Lambda)} M(\Lambda p_1, \cdots, \Lambda p_n) D(\Lambda) u \]

Consider the case when $\Lambda$ is a boost in any direction by an angle $\phi$. Is the RHS an analytic function of $\phi$? $\overline{D(\Lambda)}$ contains complex conjugation, so we're off to a bad start. However, $\overline{D(\Lambda)} = D(\Lambda)^{-1} $. In this form, and using that 
\[ D(A(\ve \phi) ) = e^{\va \cdot \ve \phi / 2} \]

We see that both $D(\Lambda)$ and $\overline{D(\Lambda)}$ are analytic; they are just exponentials. \\

What about $M(\Lambda p_1, \cdots, \Lambda p_n) $. $M$ is of the form
\[ M = \int d^4 k_1 \cdots d^4 k_m \; \frac{N(p_1, \cdots, p_n; k_1, \cdots, k_m)}{D(p_1,\cdots, p_n; k_1, \cdots, k_m)} \]

The denominator is Lorentz invariant. The numerator may be an unbelievably complex matrix, but at any finite order in perturbation theory, it is still a polynomial. So the whole RHS is an analytic function of $\phi$, and we can use the equation for complex $\phi$.\\

[ If LHS$(\phi)$ $=$ RHS$(\phi)$ for real $\phi$, and if both sides are analytic functions of $\phi$ in some domain of the complex plane containing a segment of the real line, then both sides are equal in that domain.] \\

Consider the Lorentz transformation
\[ \Lambda = R(\ve_z \pi) A (\ve_z i \pi) \]
\begin{align*}
A(\ve_z \phi):\quad p^0 &\longrightarrow p^0 \cosh \phi + p^3 \sinh \phi \\
p^1 &\longrightarrow p^1 \\
p^2 &\longrightarrow p^2 \\
p^3 &\longrightarrow p^3 \cosh \phi + p^0 \sinh \phi
\end{align*}

so 
\begin{align*}
A(\ve_z i \pi):\quad p^0 &\longrightarrow - p^0 \\
p^1 &\longrightarrow p^1 \\ 
p^2 &\longrightarrow p^2 \\
p^3 &\longrightarrow - p^3
\end{align*}

while 
\begin{align*}
R(\ve_z \pi): \quad -p^0 &\longrightarrow - p^0 \\
p^1 &\longrightarrow - p^1 \\
p^2 &\longrightarrow -p^2 \\
-p^3 &\longrightarrow - p^3
\end{align*}

so $\Lambda:\quad p^\mu \longrightarrow - p^\mu $.\\

A rotation by $\pi$ in the $z$, $it$ plane and a rotation by $\pi$ in the $x,y$ plane. \\
\vspace{1cm}

What is $D(\Lambda)$? 
\[ L_z = \frac{i}{4} \epsilon_{3ij} \gamma^i \gamma^j = \frac{i}{4} (\gamma^1 \gamma^2 - \gamma^2 \gamma^1) = \frac{i}{2} \gamma^1 \gamma^2 \]
\begin{align*}
D(R(\ve_z \pi)) &= e^{-i L_z \pi} \\
&= e^{\frac{\pi}{2} \gamma^1 \gamma^2} \\
&= \cos \frac{\pi}{2} + \gamma^1 \gamma^2 \sin \frac{\pi}{2}\\
&= \gamma^1 \gamma^2 
\end{align*}
\begin{align*}
D(A (\ve_z i \pi)) &= e^{\alpha_z i \pi /2 } \\
&= e^{i\frac{\pi}{2} \gamma^0 \gamma^3 }\\
&= \cos \frac{\pi}{2} + i \gamma^0 \gamma^3 \sin \frac{\pi}{2} \\
&= i \gamma^0 \gamma^3 
\end{align*}

So
\[ D(\Lambda) = D(R(\vec{e}_z \pi)) D(A(\vec{e}_z i \pi)) = \gamma^1 \gamma^2 i \gamma^0 \gamma^3 = \gamma^5 \]

($PCT$ is the analytic continuation of the Lorentz-transformation)\\

$\text{Since } \quad \gamma_5^2 = 1 \qquad D(\Lambda)^{-1} = \gamma_5 \qquad \text{also} $\\

Lorentz invariance says
\begin{align*}
\overline{u'} M (p_1,\cdots,p_n) u &= \overline{u'} D(\Lambda)^{-1} M(\Lambda p_1, \cdots, \Lambda p_n) D(\Lambda) u \\
&= \overline{u'} \gamma_5 M(-p_1,\cdots, - p_n) \gamma_5 u 
\end{align*}
 
and this exactly the statement of equality between an amplitude and the CPT transformed amplitude.\\

Only analyticity of Feynman amplitude was used in the proof of this theorem. This suggests that the theorem has very little to do with perturbation theory.\\

When we talked about parity invariance we had to hunt for the correct transformation of the field. Depending on the interactions that transformation may have to be chosen in various ways. A scalar meson may be forced to be scalar or pseudoscalar. For CPT invariance, you don't have to hunt for the right transformation. You just compute $D(\Lambda)$ for the funny Lorentz transformation with complex rapidity. It will be a symmetry of the Lagrangian as long as the Lagrangian is Lorentz invariant and hermitian.\\

This proof easily generalizes to higher spin: You just compute $D(\Lambda)$ for the higher spin field.\\

(The restriction to one incoming and one outgoing fermion was totally inessential.) \\

NEXT: renormalization of spinor fields.\\
\vspace{1cm}

\begin{center}
\textbf{Renormalization of spinor theories}
\end{center}

To have a simple example in mind 
\[ \ml = \frac{1}{2} (\pmu \phi)^2 - \frac{1}{2} \mu_0^2 \phi^2 + \psib ( i \notd - m_0 ) \psi - \lambda_0 \phi^4 - g_0 \psib i \gamma_5 \psi \phi \]

The meson is pseudoscalar. This ensures $\stai \phi (x) \sta = 0$. Not so if the meson nucleon interaction is $g_0 \psib \psi \phi $. \\

$m_0$ and $\mu_0$ have no necessary connection with physical masses. $g_0$ and $\lambda_0$ have no necessary connection with the couplings measured in the standard scattering process. $\phi$ and $\psi$ are not necessarily good fields from the standpoint of the LSZ reduction formula.\\

Define $Z_3$ by
\[ \stai \phi(0) \!\!\!\!\!\!\! \underbrace{| k \rangle}_{\text{one meson}} \!\!\!\!\!\!\!\! \equiv Z_3^{1/2} \]
\[ \phi' \equiv Z_3^{-1/2} \phi \]
\[ \stai \phi'(0) |k \rangle = 1 \]

$\phi'$ is a good field from the standpoint of the LSZ reduction formula (but of course it does not have conventionally normalized equal time commutation relations).\\

$\stai \psi(x) \sta = 0$ by Lorentz invariance. So it also only needs rescaling to get a good field for LSZ. However, the various components of this field may need different rescalings.\\

Let $|r,p \rangle$ (relativistically normalized so as to make L.T.~properties simple) be a one fermion state with momentum $p$ and spin labelled by $r$. We'll just study $\stai \psi(0) |r,p \rangle $ in the rest frame of $p$. \uline{Anything else can be obtained} by a Lorentz transform \\
\[ \overbrace{\stai}^{\text{physical vacuum}}\!\!\!\!\!\!\!\!\!\!\! \psi(x)\!\!\!\!\!\!\!\! \underbrace{| p,s \rangle}_{\text{physical nucleon}} \]

(Matrix elements of $\psib$ are related to matrix elements of $\psi$ by CPT or just by $C$ if the theory had that invariance.)\\

For definiteness, label the $J_z = +\frac{1}{2}$ state by $r=1$ and $J_z = - \frac{1}{2}$ by $r=2$. Let 
\[ u_0 \equiv \stai \psi(0) | 1,p \rangle \]

We can obtain some restrictions on the form of $u_0$ by using $L_z$ conservation
\begin{align*}
u_0 &\equiv \stai \psi(0) | 1,p \rangle \\
&= \underbrace{\stai e^{-i J_z \theta}}_{\stai} \underbrace{e^{iJ_z \theta} \psi(0) e^{-iJ_z \theta}}_{e^{-i L_z \theta} \psi(0)} \underbrace{e^{iJ_z \theta} | 1, p \rangle}_{e^{i\theta / 2} |1,p \rangle}\\
&= e^{-i L_z \theta} e^{i\theta /2} \stai \psi(0) | 1, p \rangle \\
&= e^{-iL_z \theta} e^{i\theta/2} u_0 
\end{align*}
 
In the standard basis $L_z = \frac{1}{2} \bpm \sigma_z & 0 \\ 0 & \sigma_z \epm $ so in the standard basis this restricts $u_0$ to be of the form 
\[ u_0 = \bpm a \\ 0 \\ b \\ 0 \epm \]

A less formal way of getting what we have just shown is to say that of the four components of $\psi$, in the standard basis these two (marked with $x$) $\bpm x \\ . \\ x \\ . \epm$ lower $J_z$ by $\frac{1}{2}$ and the other two raise $J_z$ by $\frac{1}{2}$ so only the first two can have a nonzero $J_z = \frac{1}{2}$ to zero matrix element.\\

To simplify life, let's also assume the theory has parity invariance
\[ U_p^\dagger \psi(0) U_p = \beta \psi(0) \]
\[ U_p |1,p \rangle = |1,p \rangle \qquad \substack{\text{remember we are in}\\\text{the rest frame of $p$}} \]

Now this is an assumption\footnote{ASIDE: Strong coupling scenarios could violate the assumption that the parity transformation property of the physical nucleon are the same as that of the bare nucleon. Starting with weak coupling, as you turn up the coupling a nucleon meson bound state may form. Turn up the coupling and it may become lighter than the nucleon. What you had called the nucleon is now unstable. If the meson is a pseudoscalar, the $s$ wave bound state will not have the same parity the perturbation theory nucleon did.} about the parity transformation properties of a physical nucleon, but in perturbation theory, the transformation properties of the physical nucleon should be the same as those of the bare nucleon for whatever symmetries are not broken by the interaction.\\

This assumption simplifies the possible form of $u_0$ 
\begin{align*}
u_0 &= \stai \psi(0) | 1, p \rangle \\
&= \underbrace{\stai U_p}_{\stai} \underbrace{U_p^\dagger \psi(0) U_p }_{\beta \psi(0)} \underbrace{U_p^\dagger | 1, p \rangle}_{| 1, p \rangle} \\
&= \beta \stai \psi (0) | 1,p \rangle \\
&= \beta u_0 
\end{align*}

In the standard basis $\beta = \bpm 1 & 0 \\ 0 & -1 \epm $, so in the standard basis is now restricted to be 
\[ u_0 = \bpm a \\ 0 \\ 0 \\0 \epm \]

Define $Z_2^{1/2}$ by $a = Z_2^{1/2} \sqrt{2m} $ and $\psi'$ by $\psi' = Z_2^{1/2}\psi$ then 
\[ \stai \psi'(0) |1,p \rangle = \bpm \sqrt{2m} \\ 0 \\ 0 \\0 \epm \]

For general $p$, $r=1,2$, and any $x$ we then have
\[ \stai \psi'(x) | r,p \rangle = \eipxm \uvpr \]

which has been arranged to be exactly like the free theory. The LSZ reduction formula goes through as before. The Lagrangian you proceed from to do renormalized perturbation theory is 
\begin{align*} 
\ml = & \frac{1}{2} (\pmu \phi') ^2 - \frac{\mu^2}{2} \phi'^2 + \psipb (i\notd - m) \psi' \\
& - \lambda \phi'^4 - g \psipb i \gamma_5 \psi' \phi' \\
& +\frac{1}{2} A(\pmu \phi')^2 - \frac{1}{2} B \phi'^2 + C \psipb i \notd \psi' - D \psipb \psi \\
& - E \psipb i \gamma_5 \psi' \phi' - F \phi'^4 
\end{align*}
\vspace{1cm}

\begin{center}
\textbf{Digression on spinor renormalization in parity nonconserving theories.}
\end{center}

$\gamma_5$ commutes with Lorentz transformation, so $\gamma_5 \psi(x)$ transforms in the same way as $\psi(x)$ under Lorentz transformations. In the standard rep $\gamma_5 = \bpm 0 & 1 \\ 1 & 0 \epm$ so 
\[ \stai \gamma_5 \psi(0) | 1, \!\!\!\!\!\underbrace{p}_{\text{at rest}} \!\!\!\!\! \rangle = \bpm b \\ 0 \\ a \\ 0 \epm \]

($\psi$ and $\gamma_5 \psi$ have opposite parity transformation properties.)\\

The field $\displaystyle \psi'(x) = \frac{a \psi (x) - b \gamma_5 \psi(x) }{a^2 - b^2}$ is the one satisfying
\[ \stai \psi'(x) | r, p \rangle = \eipxm \uvpr \]
\vspace{1cm}

In parallel with the method for determining $A$ and $B$ order by order in perturbation theory done on November 18 and 20, we'll show how $C$ and $D$ are determined order by order in perturbation theory.\\

Define 
\begin{align*}
\Diagram{\momentum[bot]{fV}{\leftarrow p'} p \momentum[bot]{fV}{\leftarrow p}} &= \int d^4x d^4 y \; e^{ip' \cdot x} e^{- i p \cdot y} \stai T(\psi'(x) \psipb (y) \sta \\
&\equiv (2\pi)^4 \delta^{(4)} (p' - p) \mathcal{S}'(p) 
\end{align*}

where $\mathcal{S}' (p)$ is some $4 \times 4$ matrix function of $p$ (this form is dictated by translational invariance; $\stai T(\psi'(x) \psipb (y) | 0 \rangle $ is a function of $x-y$ alone)\\

Let's check that the conventions are right by comparing with the free field theory result. On December 18, we calculated (Eq.~(\ref{eq:24-page5}))
\begin{align*}
\wick{1}{<1 \psi(x) >1\psib(y) } &= \stai T(\psi(x) \psib(y)) | 0 \rangle \\
&= (i \notd_x +m) \int \frac{d^4 q}{(2\pi)^4} e^{-iq \cdot(x-y)} \frac{i}{q^2 -m^2 + i\epsilon} \\
&= \int \frac{d^4 q}{(2\pi)^4} e^{-iq\cdot(x-y)} \frac{i (\cancel{q} + m)}{q^2 -m^2 + i \epsilon} \\
&= \int \frac{d^4 q}{(2\pi)^4} e^{-iq \cdot( x-y)} \frac{i}{\cancel{q} -m + i\epsilon}
\end{align*}

in free field theory. Putting this in above we have
\begin{align*}
&=\int \frac{d^4 q}{(2\pi)^4} \frac{i}{\cancel{q} - m + \ie} \int d^4x d^4y \; e^{-iq \cdot (x-y)} e^{i p' \cdot x} e^{-i p\cdot y} \\
&= \int \frac{d^4q }{(2 \pi)^4} \frac{i}{\cancel{q} - m + \ie} (2\pi)^4 \delta^{(4)} (p' -q ) (2\pi)^4 \delta^{(4)} (p-q) \\
&= (2\pi)^4 \delta^{(4)} (p ' - p) \frac{i}{\notp - m + \ie} 
\end{align*}

The conventions are right; this is what we write down upon seeing $\qquad \Diagram{\momentum[bot]{fV}{\leftarrow p' \quad \leftarrow p}}$ \\

Lorentz invariance restricts the form of $\mathcal{S}'(p)$.
\begin{align*}
\mathcal{S}'(p) &= \int d^4 x \; \eipx \stai T\big(\psi'(x) \psipb (0) \big)\sta \\ 
&= \int d^4 x \; \eipx \underbrace{\stai U(\Lambda)}_{\stai} U(\Lambda)^\dagger T \big(\psi'(x) \psipb (0) \big) U(\Lambda) \underbrace{U(\Lambda)^\dagger |0 \rangle}_{\sta} \\
&= \int d^4 x \; \eipx \stai T \big(U(\Lambda)^\dagger \psi'(x) U(\Lambda) U(\Lambda)^\dagger \psipb (0) U(\Lambda)\big) \sta \\
&= \int d^4 x \; \eipx D(\Lambda) \stai T \big(\psi'(\Lambda^{-1} x) \psipb (0) \big) | 0 \rangle \overline{D(\Lambda)} \\
&= D(\Lambda) \int d^4 x \; e^{ip \cdot \Lambda x} \stai T \big(\psi'(x) \psipb (0) \big) \sta \overline{D(\Lambda)} \\
&= D(\Lambda) \int d^4 x \; e^{i\Lambda^{-1} p \cdot x } \stai T\big( \psi'(x) \psipb (0) \big) \sta \overline{D(\Lambda)} \\
&= D(\Lambda) \mathcal{S}'(\Lambda^{-1} p) \overline{D(\Lambda)} 
\end{align*}

You can use this and the Lorentz transformation properties of the $16$ $\Gamma$ matrices (which are a complete set of $4 \times 4$ matrices) to get 
\begin{align*} 
\mathcal{S}'(p) &= a(p^2) \quad + \overbrace{\cancel{b(p^2) \gamma_5}}^{\substack{\text{Ruled out if we}\\\text{assume parity}\\\text{invariance}}} + \quad c(p^2) \gamma^\mu p_\mu \quad + \quad \overbrace{\cancel{d(p^2) \gamma_5 \gamma^\mu p_\mu}}^{\substack{\text{Ruled out if we}\\\text{assume
parity}\\\text{invariance}}} \quad + \quad \underbrace{\cancel{e(p^2) \sigma_{\mu\nu} p^\mu p^\nu}}_{0 \text{ by antisymmetry}} \\
&= a(p^2) + c(p^2) \notp 
\end{align*}

Define a new function $S'(z) = a(z^2) + z c(z^2) $, a function of a single complex variable. Then because $\notp^2 = p^2$ 
\[ \mathcal{S}'(p) = S'(\notp) \]

The propagator is characterized by a single function of $\notp$, a function of one variable$!$ (There is a one-to-one correspondence between functions of one variable and functions of $1$ matrix. A function of two matrices is far more complicated than a function of two numbers unless the two matrices commute)
}{
 \sektion{27}{January 13}
\descriptiontwentyseven
So far we have found 
\[ \Diagram{\momentum[bot]{fV}{\leftarrow p'} p \momentum[bot]{fV}{\leftarrow p}} = (2\pi)^4 \delta^{(4)} (p'-p) S'(\notp) \]

Now we define a one particle irreducible Green's function (defined to not include $(2\pi)^4 \delta^{(4)} (p' -p ) $ or external propagators) 
\[ \Diagram{\momentum[bot]{fV}{} !p{1PI} \momentum[bot]{fV}{\leftarrow p}} = - i \Sigma' (\notp) \]

For example, if a term in the Lagrangian is 
\[ -\delta m \psib \psi \]

There is a contribution to $-i \Sigma'(\notp)$ of 
\[ \Diagram{\momentum[bot]{fV}{} x \momentum[bot]{fV}{\leftarrow p}} \quad - i \delta m \qquad \text{i.e. to } \Sigma' (\notp) \text{ of } \delta m \]

The nice thing about the $1PI$ function is that it gives us an expression for the full Green's function (without the $(2\pi)^4 \delta^{(4)} (p' -p)$), i.e.~it gives us $S'(\notp)$.
\begin{align*}
\Diagram{\momentum[bot]{fV}{} p \momentum[bot]{fV}{}} &= \Diagram{\momentum[bot]{fV}{}} + \Diagram{\momentum[bot]{fV}{} !p{1PI} \momentum[bot]{fV}{}} + \Diagram{\momentum[bot]{fV}{} !p{1PI} \momentum[bot]{fV}{} !p{1PI} \momentum[bot]{fV}{}} + \cdots \\
&= \Diagram{\momentum[bot]{fV}{}} \left( \frac{1}{1-\Diagram{\momentum[bot]{fV}{} !p{1PI} \momentum[bot]{fV}{}}} \right)
\end{align*}

Mathematically,
\begin{align*}
S'(\notp) = \frac{i}{\notp - m + \ie} &+ \frac{i}{\notp - m + \ie} (-i \Sigma'(\notp)) \frac{i}{\notp - m + \ie} \\
&+ \frac{i}{\notp - m + \ie}(- i\Sigma'(\notp)) \frac{i}{\notp - m + \ie} (- i \Sigma' (\notp) )\frac{i}{\notp - m + \ie} + \cdots 
\end{align*}

which sums to 
\begin{align*}
S'(\notp) &= \frac{i}{\notp - m + \ie} \left[ 1 + \frac{\Sigma'(\notp)}{\notp - m +\ie} + \left(\frac{\Sigma'(\notp)}{\notp - m + \ie} \right)^2 + \cdots \right] \\
&= \frac{i}{\notp - m + \ie} \; \frac{1}{1 - \frac{\Sigma'(\notp)}{\notp -m + \ie}} \\
&= \frac{i}{\notp - m - \Sigma'(\notp) + \ie } 
\end{align*}

(Remember $\notp$ is the only matrix in the problem so it commutes with ``every other matrix" and manipulations in which $\notp$ is treated like a number are correct. This simplification does not persist in the spin $\frac{3}{2}$ problem.)\\

To get a spectral representation for $S'(\notp)$ we insert a complete set into $\stai \psi' (x) \psipb (y) \sta $. 
\[ \quad\;\int\!\!\!\!\!\!\!\!\!\!\!\!\!\!\!\!\!\!\!\sum_{\substack{\text{complete set $|n \rangle$}\\\text{of momentum}\\\text{eigenstates}}} \stai \psi'(x) | n \rangle \langle n | \psipb (y) \sta = \sumintb_{| n\rangle } e^{-i P_n \cdot (x-y)} \stai \psi'(0) |n \rangle \langle n | \psipb (0) \sta \]

Now we break the $| n \rangle $ up into physical vacuum, physical one electron, one positron and all other states. One positron does not contribute because $\displaystyle \langle \text{one positron} | \psipb \sta = 0$ (fermion $\#$ conservation). As on November 18, (Eq.~(\ref{eq:16-page8})), we use the renormalization conditions to eliminate the physical vacuum contribution and to simplify the one electron contribution.
\begin{align*}
\stai \psi'(x) \psipb (y) \sta = & \sum_r \int \frac{d^3 q}{(2\pi)^3 2 \omega_{\vec{q}}} e^{- i q\cdot(x-y) } \stai \psi'(0) \!\!\!\!\! \overbrace{| q, r\rangle}^{\text{one electron}} \!\!\!\!\langle q,r | \psipb(0) | 0 \rangle \\
&+ \quad\;\int\!\!\!\!\!\!\!\!\!\!\!\!\!\sum_{\substack{\text{all other}\\\text{states } |n \rangle}} e^{-i P_n \cdot (x-y)} \stai \psi'(0) | n \rangle \langle n | \psipb (0) | 0 \rangle \\
=& \sum_r \int \frac{d^3 q}{(2\pi)^3 2 \omega_{\vec{q}} }e^{- i q \cdot (x-y) } u_{\vec{q}}^{(r)} \overline{u_{\vec{q}}}^{(r)} \\
&+ \int d^4 p \; e^{- i p \cdot (x-y)} \!\!\!\!\!\! \suminta_{\text{all other } |n \rangle} \!\!\!\!\!\! \delta^{(4)} (p - P_n) \langle 0 | \psi'(0) | n \rangle \langle n | \psipb (0) \sta 
\end{align*}

The sum on $|n \rangle$ only contributes when the state $|n \rangle$ has spin $\frac{1}{2}$ in its rest frame. In a parity invariant theory, we can split these states into $J^P = \frac{1}{2}^+$, like a nucleon and meson in a $p$ wave, and $J^P = \frac{1}{2}^-$, like a nucleon and meson in an $s$ wave.\\

The parity $+$ states only give nonzero contributions to the upper two components of $\stai \psi'(0) | n \rangle $ (in a standard basis which is easiest for states at rest to work with). The parity $-$ state only give nonzero contributions to the lower two components of $\psi$. Furthermore the contribution of a $J_z = + \frac{1}{2}$ state to the top component is the same as a contribution to the second component of the same state hit with the lowering operator $J_x - i J_y$. A matrix which reduces to $2 \sqrt{p^2} \bpm 1 & 0 \\ \underbrace{0}_{2\times 2 \text{ blocks}} & 0 \epm$ (in the standard basis) when $p$ is at rest and is covariant in $\notp + \sqrt{p^2}$, so 
\[ \quad\;\int\!\!\!\!\!\!\!\!\!\!\!\!\!\!\!\!\!\sum_{\substack{\text{all other } |n \rangle \\\text{with $J^P = \frac{1}{2}^+$}}} \!\!\!\!\!\!\!\! \delta^{(4)} (p-P_n) \stai \psi'(0) |n \rangle \langle n | \psipb (0) | 0 \rangle = \frac{\theta (p^0)}{(2\pi)^3} \sigma_+ (\sqrt{p^2}) ( \notp + m) \]

Similarly \footnote{Can shortcut some work by noticing that $\widetilde{\psi'} = \gamma_5 \psi'$ has the same matrix elements with a $\frac{1}{2}^+$ state as $\psi'$ has with a $\frac{1}{2}^-$ state. ``As the $\frac{1}{2}^+$ states are to $\psi$, the $\frac{1}{2}^-$ states are to $\gamma_5 \psi$." }
\[ \quad\;\int\!\!\!\!\!\!\!\!\!\!\!\!\!\!\!\!\sum_{\substack{\text{all other } |n \rangle \\\text{with } J^P = \frac{1}{2}^-}} \delta^{(4)} (p-P_n) \stai \psi'(0) | n\rangle \langle n | \psipb (0) | 0 \rangle = \frac{\theta(p^0)}{(2\pi)^3} \sigma_- (\sqrt{p^2}) (\notp-m) \]

$ \sigma_+ $ and $\sigma_-$, which are defined by these equations are both positive semidefinite by the positivity of the norm on Hilbert space. You might worry that $\notp - m $ is negative in the rest frame of $p$, but it should be because $ \stai \psi'(0) | n\rangle \langle n | \psipb (0) \sta $ differs from $ \stai \psi'(0) | n\rangle \langle n | \psi'^\dagger(0) \sta $ by the matrix $\gamma^{0}$ which is negative in its lower two components. In perturbation theory, $\sigma_+ = \sigma_- = 0 $, when $p^2 < (m + \mu)^2 $.\\

Putting this together, we have 
\begin{align*}
\stai \psi'(x) \psipb (y) \sta &= \int \frac{d^3 q}{(2\pi)^3 2 \omega_{\vec{q}}} e^{-i q \cdot (x-y)} (\cancel{q} + m ) \\
& \qquad + \int \frac{d^4 p}{(2\pi)^3} e^{-i p \cdot (x-y)} \left[\theta(p^0) \sigma_+ ( \sqrt{p^2}) (\notp + \sqrt{p^2}) + \theta (p^0) \sigma_- (\sqrt{p^2}) (\notp - \sqrt{p^2}) \right] \\
&= ( i \notd_x + m ) \Delta_+ (x-y) \\
&\qquad + \int^\infty_0 da \; \sigma_+ (a) \;\; \int \frac{d^4 p}{(2\pi)^3} \theta (p^0) \delta(p^2 -a^2) (\notp + a) e^{-ip\cdot (x-y)} \\
& \qquad + \int^\infty_0 da \; \sigma_- (a) \;\; \int \frac{d^4 p}{(2\pi)^3} \theta (p^0) \delta (p^2 - a^2) (\notp - a) e^{-i p\cdot (x-y)} \\
&= ( i \notd_x + m ) \Delta_+ (x-y) + \int^\infty_0 da \; \sigma_+ (a) ( i \notd_x + a) \Delta_+(x-y; a^2) \\
& \qquad + \int^\infty_0 da \; \sigma_-(a) ( i \notd_x - a) \Delta_+ (x-y; a^2) \\
&= \int^\infty_0 da \; \left[ \rho_+ (a) ( i \notd_x + a ) \Delta_+ (x-y) + \rho_- (a) (i\notd_x - a)\Delta_+(x-y) \right] 
\end{align*}

For compactness in the last step I have introduced 
\[ \rho_+ (a) = \sigma_+ (a) + \delta(a-m) \]
\[ \rho_- (a) = \sigma_- (a) \qquad \text{and dropped the $;a^2$ in $\Delta_+$} \]

Rather than redoing a lot of steps, I can get
\[ \stai \psipb_\beta (y) \psi'_\alpha (x) \sta \]

from what we've just calculated using $\Omega_{CPT} (=\Omega)$. (In a theory with $C$ invariance, it would be easier to just use $U_c$, but I'll be more general.) Then we'll be set to write down the time ordered product
\begin{align*} 
\stai T (\psi_\alpha'(x) \psipb_\beta (y) ) \sta = &\theta (x^0 - y^0) \stai \psi'_\alpha (x) \psipb_\beta (y) \sta \\
&-\theta (y^0 -x^0) \stai \psipb_\beta (y) \psi'_\alpha (x) \sta 
\end{align*}

I'll do the calculation in a Majorana basis
\begin{align*}
\stai \psipb_\beta (y) \psi'_\alpha (x) \sta &= \stai \Omega \Omega^{-1} \psipb_\beta(y) \Omega \Omega^{-1} \psi'_\alpha(x) \Omega \Omega^{-1} \sta \\
&= ( \stai \underbrace{\Omega^{-1} \psipb_\beta (y) \Omega}_{-i(\gamma_5 \gamma^0 \psi'(-y))_\beta } \underbrace{\Omega^{-1} \psi'_\alpha(x) \Omega}_{-i (\psipb (-x) \gamma^0 \gamma_5)_\alpha} \sta)^* \qquad \substack{\text{In this step when $\langle 0 | \Omega$ is simplified to $\stai$}\\ \text{ the resulting matrix element must be}\\ \text{complex conjugation because $\Omega$ is antiunitary}}\\ 
&= \left(- i (\gamma_5 \gamma^0)_{\beta \sigma}\stai \psi'_\sigma (-y) \psipb_\tau (-x) \sta (\gamma^0 \gamma_5)_{\tau \alpha} \right)^* 
\end{align*}

Notice the indices $\beta, \alpha$ come out in the wrong order, so to think of $\stai T (\psi'(x)\psipb (y)) \sta$ as a matrix, we actually need this thing transposed. 
\begin{multline*}
\stai T(\psi'(x) \psipb (y)) \sta = \theta (x^0 - y^0) \int^\infty_0 da \; \left[ \rho_+(a) (i \notd_x + a) \Delta_+ (x-y) + \rho_- (a) (i \notd_x - a) \Delta_+ (x-y) \right] \\
+ \theta(y^0 -x^0) \int^\infty_0 da \;\left[ \rho_+(a) \gamma_5 \gamma^0 (i \notd_x +a ) \Delta_+(x-y) \gamma^0 \gamma_5 + \rho_-(a) \gamma_5 \gamma^0 (i \notd_x - a) \Delta_+(x-y) \gamma^0 \gamma_5 \right]^{*T}
\end{multline*}

Now
\[ (i \gamma_5 \gamma^0 \gamma^\mu \gamma^0 \gamma_5 )^{*T} = i \gamma^\mu \]

and 
\[ (\gamma_5 \gamma^0 1 \gamma^0 \gamma_5)^{*T} = 1 \]

so (in this case we can pull the time derivative through the time ordered product)
\begin{align*}
\stai T(\psi'(x) \psipb (y) \sta = \int^\infty_0 da \;& \left( \rho_+ (a) (i \notd_x +a ) + \rho_- (a) (i\notd_x -a) \right) \qquad \text{acting on}\\
& \left[ \theta(x^0 -y^0) \Delta_+(x-y) + \theta(y^0 -x^0) \Delta_+(y-x) \right]
\end{align*}
(Using that $\rho_+ (a)$ and $\rho_-(a)$ are real and that $\Delta_+(x-y)^* = \Delta_+(y-x) $)\\

The object in brackets is 
\[ \int \frac{d^4 p}{(2\pi)^4} \frac{i}{p^2 -a^2 + \ie} e^{- i p\cdot (x-y)} \]

$i \notd_x $ hitting this gives $\notp$. So the result is
\[ \stai T(\psi'(x) \psipb (y)) \sta = \int \frac{d^4 p}{(2\pi)^4} e^{-i p\cdot(x-y)} \int^\infty_0 da \; \left( \rho_+(a) \frac{i(\notp + a)}{p^2-a^2 + \ie} + \rho_-(a) \frac{i(\notp - a)}{p^2 -a^2 + \ie} \right) \]

Or in a more suggestive form
\begin{align*}
\stai T(\psi'(x) \psipb (y) ) \sta &= \int \frac{d^4 p}{(2\pi)^4} e^{-ip\cdot (x-y)} \int^\infty_0 da \; \left(\rho_+(a) \frac{i}{\notp - a + \ie} + \rho_- (a) \frac{i}{\notp + a + \ie} \right) \\
&= \int \frac{d^4 p}{(2\pi)^4} e^{-i p \cdot (x-y) } S'(\notp) 
\end{align*}

where
\[ S'(z) = \int^\infty_0 da \; \left( \rho_+ (a) \frac{i}{z-a+\ie} + \rho_-(a) \frac{i}{z+a+\ie} \right) \]

This result for $S'(z)$ has the renormalization conditions built in. They say $S'$ has a pole at $z=m$ with residue $i$.
\begin{center}
\includegraphics[scale=0.7]{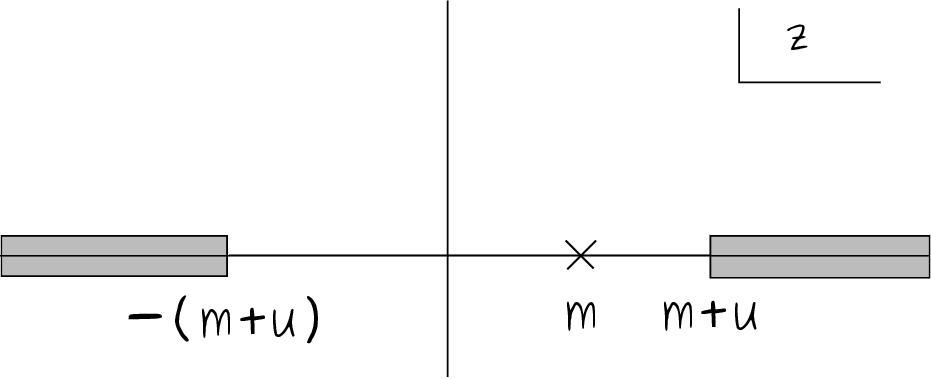}\\
Analytic structure of $S'$ in perturbation theory
\end{center}

Compare this with our other expression for $S'$ 
\[ S'(z) =\frac{i}{z-m- \Sigma'(z) +\ie} \]

In terms of $\Sigma'$, we see the renormalization conditions are
\begin{align*}
\Sigma' (m) &= 0 \qquad \text{pole is at $m$} \\
\left.\frac{d \Sigma'}{dz}\right|_{z=m} &= 0 \qquad \text{residue is $i$} \qquad \text{(often written $\displaystyle \left.\frac{d\Sigma'}{d\notp}\right|_{\notp = m} = 0 $)}
\end{align*}

\vspace{1cm}
In the model with $C$ and 
\[ \ml' = - g\psib' i \gamma_5 \psi' \phi' + C \psib' i \notd \psi' - D\psib' \psi' \qquad \text{$C,D$: $\infty$ power series in g: $C^{(n)} \propto g^n$} \]

\vspace{0.5cm}
Get $\Sigma'(\notp)$ to order $g^2$ 
\[ \feyn{!{fV}{} fs0 !{flS}{} !{flSuV}{} fs0 !{fV}{} } \qquad + \qquad \Diagram{!{fV}{} !{x}{(2)} !{fV}{}} \]
\begin{align*}
-i\Sigma'(\notp) &= - i \Sigma^f (\notp) + i C^{(2)} \notp - iD^{(2)} \\
\Sigma'(\notp) &= \Sigma^f (\notp) - \Sigma^f (m) -\left.\frac{d\Sigma^f}{d\notp} \right|_m (\notp -m) \qquad \text{Only knocks off 1 power of $p$}
\end{align*}

\[ \ml' = - g \psib i \gamma_5 \psi \phi \]
$\feyn{!{fV}{p} fs0 !{flS}{k \rightarrow} !{flSuV}{p+k} fs0 !{fV}{p} }$ $\longleftarrow$ a $4 \times 4$ matrix, this is a propagator without external\\

\vspace{0.3cm}
propagators included.
\begin{align*}
-i\Sigma^f &= (-ig)^2 \int \frac{d^4 k}{(2\pi)^4} \; \frac{i}{k^2 -\mu^2 + \ie} i\gamma_5 \frac{i(\notp + \cancel{k}-m)}{(p+k)^2-m^2+\ie} i\gamma_5 \\
&= -\frac{g^2}{(2\pi)^4} \int d^4 k \; \frac{1}{k^2 - \mu^2 + \ie} \; \frac{-\notp - \cancel{k} + m}{(p+k)^2 - m^2 + \ie} \\
&= -\frac{g^2}{(2\pi)^4} \int d^4 k \int^1_0 dx \; \frac{-\notp - \cancel{k} + m }{[k^2 + 2kpx + p^2 x -m^2 x - \mu^2 (1-x) + \ie ]^2} \\
k'=k+px \qquad &= - \frac{g^2}{(2\pi)^4} \int d^4 k' \int^1_0 dx \frac{- \notp (1-x) + m -\overbrace{\cancel{k}'}^{\text{ODD}}}{[k'^2 + p^2 x (1-x) - m^2 x - \mu^2 (1-x) + \ie]^2} 
\end{align*}
\begin{align*}
\Sigma' &= \frac{-ig^2}{(2\pi)^4} \int d^4 k' \int^1_0 dx \Bigg\{ \frac{-\notp (1-x) +m }{[k'^2 +p^2 x(1-x) - m^2x - \mu^2 (1-x) + \ie]^2} \\
&\qquad \qquad \qquad - \frac{-m (1-x) +m}{[k'^2 + m^2 x(1-x) - m^2x - \mu^2 (1-x) + \ie ]^2} \\
&\qquad \qquad \qquad - (\cancel{p}-m) \bigg[ \frac{-(1-x)}{[k'^2 + m^2 x (1-x) - m^2 x - \mu^2 (1-x) + i\epsilon]^2} \\
&\qquad \qquad \qquad \qquad - \frac{ 4m^2x^2(1-x)}{[k'^2+m^2x(1-x)-m^2x - \mu^2 (1-x) + i\epsilon]^3} \bigg] \Bigg\} \\
\substack{\text{DIVERGENT}\\\text{PART}} \qquad &\propto \int \frac{d^4k'}{k'^4} \left[ -\notp (1-x) + m + m (1-x) - m + (\notp-m)(1-x) \right] =0
\end{align*}

A quicker way of seeing if the result is finite is to compute $\displaystyle \frac{d^2 \Sigma'}{d\notp^2} $ . $\Sigma'$ is completely determined by this second derivative. \\

Note that one derivative is not enough to give a finite integral. Two does the job $\Longrightarrow$ Need two subtractions to remove $\infty$'s, unlike scalar case where diagram was only log divergent. \\
\vspace{1cm}

\begin{center}
\textbf{Coupling constant renormalization in spinor theory (parallels scalar case)} 
\end{center}

\[ \Diagram{\momentum[llft]{fdA}{p \searrow } \\ & !{p}{1PI} \momentum[bot]{fV}{q=p'-p} \\ \momentum[bot]{fuV}{\swarrow p'}} = -i \!\!\!\!\! \underbrace{\Gamma}_{\substack{\text{some awful}\\ 4 \times 4 \text{ matrix}}} \!\!\!\!\! (p',p) \] 

Contributions up to order $g^3$ are 
\[ \Diagram{\momentum[llft]{fdA}{} \\ & \momentum[bot]{f}{} \\ \momentum[bot]{fuV}{}} \qquad + \qquad \includegraphics[scale=0.2]{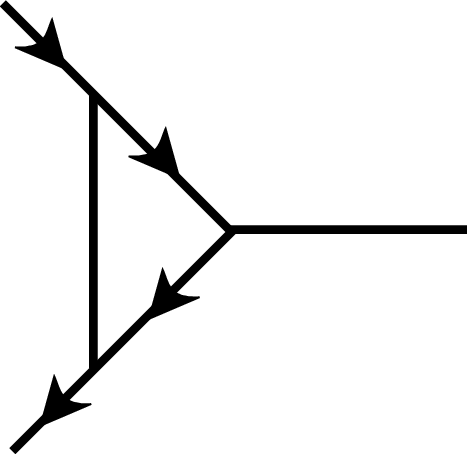} \qquad + \qquad \Diagram{\momentum[llft]{fdA}{} \\ & !{x}{(3)} \momentum[bot]{f}{} \\ \momentum[bot]{fuV}{}} \]

You might give your renormalization condition as 
\[ \Gamma = ig\gamma_5 \]

at some astutely chosen value of $p'$ and $p$. However, $\Gamma$ may not be proportional to $\gamma_5$. It may have $\notp \gamma_5$ or $p_\mu p_\nu \sigma\mn \gamma_5$. We can remedy this by cutting down the $4\times 4$ matrix by sandwiching it between projectors. We'll show
\[ \left.(\cancel{p}\,' + m) \Gamma (p',p) (\notp +m) \right|_{p^2 = p'^2 = m^2} \]

must be $\propto$ to $(\notp' + m ) \gamma_5 (\notp +m )$. $\frac{\notp +m}{2m}$ projects onto incoming nucleons or outgoing antinucleon.\\

\vspace{1cm}
Consider this graph as contributing to 
\[ \phi (\text{off shell}) \longrightarrow N + \overline{N} \]

and look at the process in the COM frame where 
\[ q = (q^0, \vec{0}) \]

The initial state is $J^P = 0^-$ .\\

The two spin $\frac{1}{2}$'s in the final state can make $S=1$ or $S=0$. To get $J=0$ the only possible final states are
\begin{align*}
l &= 0 \qquad S= 0 \qquad \text{which has} \qquad P=-1 \\
l &= 1 \qquad S=1 \qquad \text{which has} \qquad P = +1 
\end{align*}

Only the first final state is allowed. There is only one amplitude. (It may vary with $q^0$)
\[ (\notp' +m) \gamma_5 (\notp +m) \]

is nonzero when sandwiched between a $\overline{u}$ and a $u$ (remember $p^0<0$ for this process with the momentum conserving conditions $\Diagram{\momentum[llft]{fdV}{p' \nwarrow } \\ & !{p}{} \momentum[bot]{f}{} \\ \momentum[bot]{fuA}{\nearrow p}}$ )\\

So 
\[ \left. (\notp' +m) \Gamma(p, p') (\notp +m) \right|_{p^2 = p'^2 = m^2} = (\notp' +m) i \gamma_5 (\notp +m) G(q^2) \]

We'll take 
\[ G(q^2 = \mu^2 ) \equiv g \]

as our renormalization condition.\\
\vspace{1cm}

Utility of this choice
\begin{align*}
\includegraphics[scale=0.3]{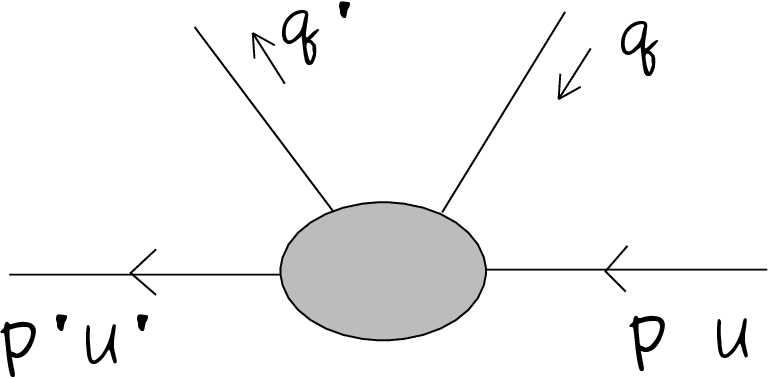} \qquad = \qquad &\includegraphics[scale=0.3]{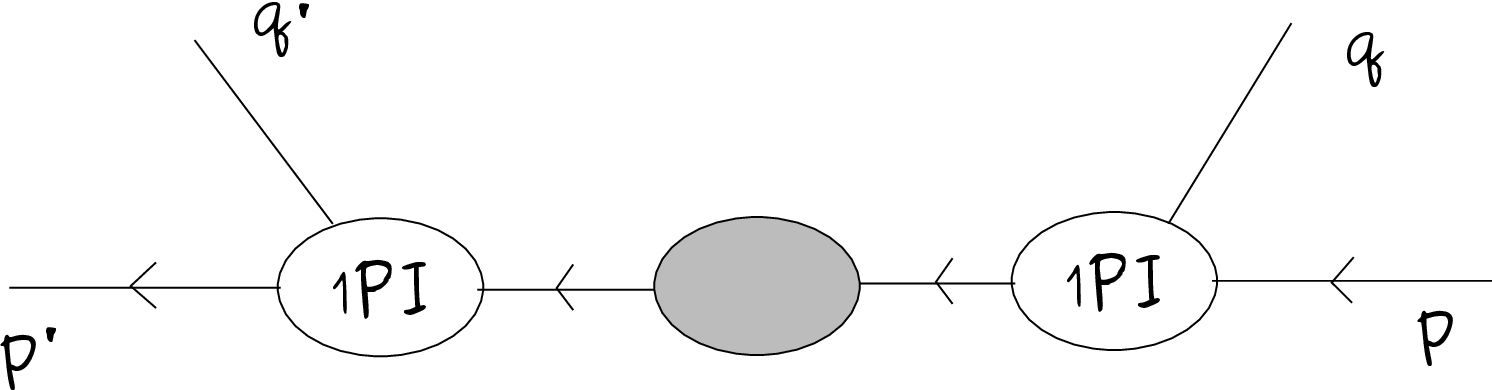} \\
&+\text{ graphs with no pole at } (p+q)^2 = s = \mu^2 
\end{align*}

Pole piece in $ia$
\begin{multline*} 
\qquad- \overline{u'} \Gamma (p', p+q) S'(p+q) \Gamma (p+q, p ) u \\
= - \overline{u'} \underbrace{\frac{\notp'+m}{2m}}_{\substack{\text{insert it is the}\\\text{identity on $\overline{u'}$}}} \Gamma(p', p +q) \underbrace{S'(p+q)}_{\substack{\text{near $s=\mu^2$ this is}\\ \frac{i(\notp + \cancel{q} + m )}{(p+q)^2 - m^2 + \ie}\\\ + \text{analytic}}} \Gamma (p+q,p) \underbrace{\frac{\notp + m}{2m}}_{\text{insert}} u 
\end{multline*}

So near $s=\mu^2$, we have $\Gamma$ sandwiched between projection operators and we can use the renormalization condition to get that the pole piece in $ia$ is 
\[ -\overline{u'} i \gamma_5 g \frac{i(\notp + \cancel{q} + m ) }{(p+q)^2 - m^2} i \gamma_5 g u \]

This simplification allows for unambiguous comparison with experiment to set $g$.\\
\vspace{1cm}

Is renormalization necessary and sufficient to get rid of $\infty$'s ?\\

Let's look at the contributions to $\Gamma$ at $\mathcal{O}(g^3)$.\\
\[ \includegraphics[scale=0.2]{27-fig2.eps} \qquad + \qquad \Diagram{\momentum[llft]{fdA}{} \\ & !{x}{(3)} \momentum[bot]{f}{} \\ \momentum[bot]{fuV}{}} \]

At high $k$, the integral for the first Feynman graph goes like
\[ \int d^4 k \frac{1}{k^2} \gamma_5 \frac{1}{\cancel{k}} \gamma_5 \frac{1}{\cancel{k}} \gamma_5 \sim \gamma_5 \int d^4 k \frac{1}{k^4} \]

This is divergent, but only logarithmically divergent, and it multiplies $\gamma_5$. So the second graph cancels the divergent part.\\

So far our slovenliness has been good enough. \\

\begin{center}
\textbf{Regularization and renormalization}
\end{center}

Throwing around ill-defined quantities, and discovering they always end up in convergent combinations isn't good enough.\\

The infinities came because the theory has an infinite $\#$ of degrees of freedom, both from the $\infty$ extent of spacetime (which gives IR $\infty$'s) and from the fact that in any given volume there is an $\infty \;\; \#$ of degrees of freedom (which gives UV $\infty$'s). Next lecture we'll talk about ways to cut down the $\#$ degree of freedom in a given volume.
}{
 \sektion{28}{January 15}
\descriptiontwentyeight
This lecture 
\begin{itemize}
\item[I.] Regularization
\begin{itemize}
\item[A.] Regulator fields (Feynman)
\item[B.] Dimensional Regularization ('t Hooft-Veltman)
\end{itemize}
\item[II.] BPHZ renormalization
\end{itemize}
\vspace{1cm}

\begin{center}
\textbf{I. Regularization}
\end{center}

No one knows of a quantum field theory that is nontrivial and finite.\\

In all theories worth studying, so as not to be making ad hoc cancellations of infinities with infinities, you have to hack up the theory in some way to make it finite. For example, you could throw away all Fourier components in the Feynman integrals with momentum greater than some cutoff value $\Lambda$. Then you would renormalize as usual. Instead of making subtractions of $\infty$'s from $\infty$'s to satisfy the renormalization conditions, you will be subtracting finite (but big; proportional to $\Lambda$, $\Lambda^2$ or $\text{ln } \Lambda$) things from other finite things to satisfy the renormalization conditions. Then you try to undo your hatchet job by sending $\Lambda$ to $\infty$. The big job is to prove that the properties you expect of the theory (Lorentz invariance, gauge invariance, positivity of the Hilbert space inner product) are recovered as $\Lambda \longrightarrow \infty$, and that nothing depends on $\Lambda$ in this limit. A scattering amplitude should not depend on the method some theorist used to make an infinity large but finite.\\
\vspace{1cm}

\begin{center}
\textbf{Regulator fields or Propagator Modification}
\end{center}

A good regularization method should 
\begin{itemize}
\item[(1)] be analytically tractable 
\item[(2)] ruin as few properties of the theory as possible. (The less you ruin the less you have to laboriously prove you recover in the $\Lambda \longrightarrow \infty $ limit.)
\end{itemize}

Regulator fields (or at least a variant we'll discuss March 3 called Pauli-Villars) only wreck positivity of the Hilbert space metric in QED with massive charged particles and the only kinds of integrals that have to be evaluated are of the same type we have already studied. The idea is to replace propagators in the Feynman integrals by propagators that fall off faster at high momentum so that loop integrals will be finite. To do this we'll let
\[ \frac{i}{k^2 - m^2} \qquad \text{ become a combination of propagator. For example } \]
\begin{equation}\label{eq:28-page2}
\frac{i}{k^2-m^2} \longrightarrow \frac{i}{k^2 - m^2} - \frac{i}{k^2-M^2} 
\end{equation}

$M$ plays the role of the cutoff. For $k^2 \gg M^2$ this combination falls off like 
\[ \frac{1}{k^4} \qquad \text{instead of } \qquad \frac{1}{k^2} \]

Similarly
\[ \frac{i}{\notp -m} \longrightarrow \frac{i}{\notp -m} - \frac{i}{\notp - M} \propto \frac{1}{p^2} \qquad \text{at high } p \]

After modifying the propagators enough to make the diagrams convergent, you adjust the counterterms to satisfy the renormalization conditions, and then send $M \longrightarrow \infty $.\\

Making a propagator go like $\displaystyle \frac{1}{k^4}$ may not be enough to make diagrams convergent. Here's how to make them go like $\displaystyle \frac{1}{k^{2n}}$ for $n$ as big as you need. Let
\[ \frac{i}{k^2 - m^2} \longrightarrow \frac{i}{k^2-m^2} + \sum_{r=1}^n \frac{i C_r^2}{k^2 - M_r^2} \]

(I write the coefficient as $C_r^2$, but don't let me mislead you into thinking $C_r^2 > 0$)\\

We can look at the behavior of this for high $k^2$ by expanding
\begin{align*}
\frac{1}{k^2-m^2} &= \frac{1}{k^2} \Big( \frac{1}{1- \frac{m^2}{k^2}} \Big) \\
&= \frac{1}{k^2} \Big( 1 + \frac{m^2}{k^2} + \Big( \frac{m^2}{k^2} \Big)^2 + \cdots \Big) 
\end{align*}

By choosing 
\begin{align*}
&& 1 + \sum_{r=1}^n C_r^2 = 0 & \qquad \text{makes propagator} & \sim \frac{1}{k^4} \\
&\text{and} & m^2 + \sum_{r=1}^n M_r^2 C_r^2 = 0 & \qquad \text{makes propagator} & \sim \frac{1}{k^6} \\
&\text{and} & m^4 + \sum_{r=1}^n M_r^4 C_r^2 = 0 & \qquad \text{makes propagator} & \sim \frac{1}{k^8} \\
&\text{and} & m^6 + \sum_{r=1}^n M_r^6 C_r^2 = 0 & \qquad \text{makes propagator} & \sim \frac{1}{k^{10}}
\end{align*}

etc...\\

($n=1$, $C_1=1$, $M_1 =M$ creates the simplest example in Eq.~(\ref{eq:28-page2}).)\\

By making $n$ large enough, you can clearly make the propagator fall off as fast as you like, and still have freedom to send all the $M_r \longrightarrow \infty$. (The $C_r$ must remain finite.) Of course some of the $C_r^2$ are going to have to be less than zero, or you are just going to have $i$ times a sum of things with the same sign at large $k^2$. There is no way this can happen in any realistic theory of the world. $C_r^2 > 0$ is a consequence of the Lehmann-K\"all\'en spectral representation.\\

We can construct an operator theory that is unrealistic that has these sicko propagators though.\\

Suppose the original theory had 
\[ \ml = \frac{1}{2} (\pmu \phi)^2 - \frac{\mu^2}{2} \phi^2 + \ml'(\phi) \]

The unrealistic theory that has these propagators is 
\begin{align*}
\ml = & \frac{1}{2} (\pmu \phi)^2 - \frac{m^2}{2} \phi^2 \\
&+ \sum_{r=1}^n \bigg[ \frac{1}{2} (\pmu \phi_r )^2 - \frac{M_r^2}{2} \phi_r^2 \bigg]\\
&+ \ml'(\Phi) \qquad \Phi = \sum_{r=1}^n C_r \phi_r 
\end{align*}

I talk about why this gives the right propagator combination on March 3.\\

About this point you may be wondering why we are trying to construct a Lagrangian that reproduces our hatchet job. Answer: It helps you ascertain what properties of the theory you have or haven't ruined.\\

Because some of the $C_r^2$ are less than zero, some of the $C_r$ are imaginary, and the Hamiltonian is not Hermitian.\\

We can gain some insight into what is going on by defining a new inner product. \\

In the theory that embodies the simplest propagator modification,
\[ \ml = \frac{1}{2}(\pmu \phi)^2 - \frac{m^2}{2} \phi^2 + \frac{1}{2}(\pmu \phi_1)^2 - \frac{M^2}{2} \phi_1^2 + \ml'(\Phi) \qquad \Phi = \phi + i\phi_1 \]

define a new inner product
\[ \langle a| b \rangle_{\text{new}} = \langle a | (-1)^{N_1} | b\rangle \]

$N_1$ counts the number of mesons of the sicko type. This metric is not positive definite.
\[ \langle a | a \rangle_{\text{new}} < 0 \qquad \text{if } |a\rangle \text{ has an odd number of } \phi_1 \text{ mesons in it} \]

The great thing about this metric is that in it $\Phi$ is hermitian.
\[ (\phi_1)^\dagger_{\text{new}} = -\phi_1 \]

because $\phi_1$ anticommutes with $(-1)^{N_1}$ . So 
\[ (\Phi)^\dagger_{\text{new}} = (\phi + i \phi_1)^\dagger_{\text{new}} = \Phi \]

To summarize. In the old metric, which was positive definite, the Hamiltonian wasn't hermitian and thus didn't conserve probability. In the new metric, we have a new definition of probability, and although it is not always greater than zero, the Hamiltonian is hermitian, and the new probability is conserved.\\

Here is why you might hope that a sensible theory will be recovered when the $M\longrightarrow \infty$ limit is taken. We won't be interested in amplitudes that have those phony particles in the initial states, and when $M\longrightarrow \infty$, it will be impossible to produce them in the final state, just for lack of energy.\\

The only initial and final states possible will thus be the ones with sensible particles in them, and for them, the inner product is normal.\\

The good things about regulator fields are that they preserve Lorentz invariance, internal symmetries in theories with massive particles (they spoil symmetries that depend on masslessness), conserve probability at energies low compared to the cutoff, with some modification, will be seen to preserve gauge invariance in QED, and they are computationally easy to introduce.\\

\underline{A note on computation}: In practice you don't try to combine the various propagators to make the integrals manifestly convergent. You just work along with each propagator separately, and use the integral tables that are valid when you work with a convergent combination.\\
\vspace{1cm}

\begin{center}
\textbf{Dimensional Regularization}
\end{center}

Begin with an example. Let's evaluate 
\[ I = \int \frac{d^d k }{(k^2 + a^2)^n} \qquad \text{which is convergent if } n > \frac{d}{2} \]

in an arbitrary number of Euclidean space dimensions $d$.\\

Here is a trick to turn a denominator into an exponential. Start with 
\[ \Gamma (n) = \int^\infty_0 t^{n-1} e^{-t} \; dt \]

(which is sometimes taken as the definition of the $\Gamma$ function).\\

Change variables in the integrand to $\lambda$ given by $\alpha\lambda = t$; $\alpha$ real, $>0$.
\[ \Gamma (n) = \int^\infty_0 (\alpha \lambda)^{n-1} e^{-\alpha \lambda} \; d(\alpha \lambda) \]

or 
\[ \frac{1}{\alpha^n} = \frac{1}{\Gamma(n)} \int^\infty_0 \; \lambda^{n-1} e^{-\alpha \lambda} \; d\lambda \]

Our (Euclidean space) integral becomes
\begin{align*}
I &= \frac{1}{\Gamma (n)} \int^\infty_0 \lambda^{n-1} \; d\lambda \underbrace{\int d^d k \; e^{-\lambda (k^2 + a^2)} }_{e^{-\lambda a^2} ( \frac{\pi}{\lambda} )^{d/2}} \\
&= \frac{\pi^{d/2}}{\Gamma (n)} \int^\infty_0 \lambda^{n- \frac{d}{2} - 1} e^{-\lambda a^2} \; d\lambda \\
&= \frac{\pi^{d/2}}{\Gamma (n)} \frac{\Gamma (n - \frac{d}{2})}{a^{2n-d}} 
\end{align*}

Here is 't Hooft and Veltman's whammy: adopt this formula for arbitrary complex $d$. If you stay away from even integers $ d \geq 2n$, this expression is well defined. As you head toward $d=4$, you approach poles in the $\Gamma$ function. You do your renormalization in arbitrary $d$ and only after you have your expressions for the graphs plus counterterms in convergent combinations (that is with poles in $d-4$ cancelling) you send $d\longrightarrow 4$.\\

You have to be careful formulating a theory in an arbitrary $\#$ of dimensions. \\

You can't just maintain $\displaystyle \frac{e^2}{4\pi} = \frac{1}{137}$ in an arbitrary number of dimensions because only in four dimensions is $e$ dimensionless. There are simpler examples than QED to demonstrate the effect of this. Take 
\[ \ml = \frac{1}{2} (\pmu \phi')^2 - \frac{m^2}{2} \phi'^2 - \frac{\lambda}{4!} \phi'^4 + \ml_{c.t.} \]

$[\phi'] = \frac{d-2}{2}$, so $m$ is a mass as it appears. \\

But to keep the Lagrangian having dimension $d$, we must have 
\[ d = [\lambda ] + 4 [\phi' ] \qquad [ \lambda ] = d -2(d-2) = -d+4 \]

Only in four dimensions is $\lambda$ dimensionless. It cannot remain constant as we change $d$. It has to acquire dimension. So we rewrite the interaction:
\[ \underbrace{\lambda}_{\substack{\text{Now dimensionless}\\\text{for any $d$}}} \!\!\!\!\!\!\!\!\!\! \mu^{4-d} \frac{\phi'^4}{4!} \]

where $\mu$ is a parameter that has appeared uninvited into the theory.\\

You might think that after renormalizing, when we set $d=4$, all $\mu$ dependence will go away.\\

We'll look at a contribution to the four point function. An $\mathcal{O}(\lambda^2)$ diagram\\
\begin{center}
\includegraphics[scale=0.3]{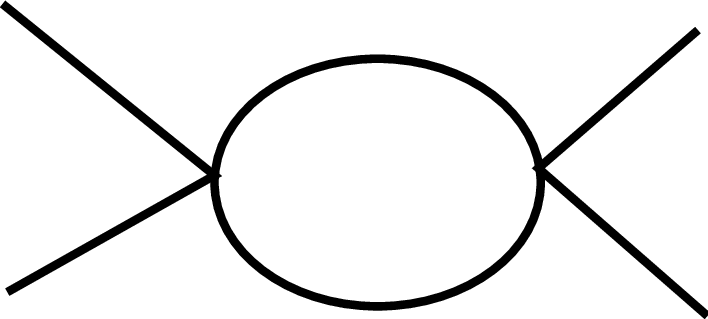}
\end{center}

It leads to an integral like 
\[ (\lambda\mu^{4-d})^2 \int \frac{d^dk}{(2\pi)^d} \frac{1}{(k^2+a^2)^2} \]

where $a$ contains masses, external momenta and Feynman parameters, and I have suppressed the Feynman parameter integral and a lot of factors.\\

Our result for this integral is 
\[ \frac{1}{(2\pi)^d} (\lambda \mu^{4-d})^2 \frac{\pi^{d/2}}{\Gamma(2)} \Gamma \Big(2 - \frac{d}{2}\Big) a^{d-4} \]

$\Gamma$ has a pole piece near $d=4$. For integer $n$ 
\[ \Gamma (-n + \epsilon) = \frac{(-1)^n}{n!} \Big[ \frac{1}{\epsilon} + \!\!\!\!\!\!\!\!\! \underbrace{\psi (n+1)}_{\substack{\text{some number like } \pi\\\text{except more complicated}}} \!\!\!\!\!\!\!\!\! + \mathcal{O}(\epsilon) \Big] \]

see Ramond, [Field Theory: A Modern Primer (1st edition, 1981)] p.~152.\\

You can't just set $d=4$ in the rest of the expression. That would give you the right coefficient of the pole, but the wrong finite part.\\

Let's see what the finite part is. In our example $n=0$ and $\displaystyle \epsilon = 2 - \frac{d}{2}$. \\

We'll pull out the a factor of $\mu^{4 -d}$ since that is the dimension of this Green's function (the lowest order contribution is proportional to $\lambda \mu^{4-d}$). We expand the dimensionless thing that is left
\begin{align*}
\lambda^2 \mu^{4-d} \frac{\pi^{d/2}}{1!} \Gamma \Big(2-\frac{d}{2}\Big) a^{d-4}
&= \lambda^2 \pi^2 \Gamma \Big( 2- \frac{d}{2} \Big) \Big(\frac{\mu}{a} \Big)^{4-d} \pi^{\frac{d}{2} -2} \\
&= \lambda^2 \pi^2 \Gamma \Big(2 - \frac{d}{2}\Big) \Big(\frac{\mu^2}{\pi a^2}\Big)^{\frac{4-d}{2}} \\
&= \lambda^2 \pi^2 \Bigg[ \frac{(-1)^0}{0!} \Big( \frac{1}{2-\frac{d}{2}} + \psi(1) + \mathcal{O}(d-4) \Big) \Bigg] \underbrace{e^{\frac{4-d}{2}\ln \frac{\mu^2}{\pi a^2}}}_{1 + \frac{4-d}{2} \text{ln } \frac{\mu^2}{\pi a^2} } \\
&= \lambda^2 \pi^2 \Bigg[ \frac{1}{2-\frac{d}{2}} + \psi(1) + \text{ln } \frac{\mu^2}{\pi a^2} + \mathcal{O}(d-4) \Bigg] 
\end{align*}

You would have lost the $\displaystyle \text{ln } \frac{\mu^2}{\pi a^2}$ piece if you prematurely set $d=4$.\\

Now you can renormalize as usual although you need an extension of the renormalization conditions for arbitrary dimension.\\
\vspace{1cm}

\begin{center}
\textbf{Minimal Subtraction (or MS)}
\end{center}

MS is another renormalization prescription, that is, a way of determining counterterms. It makes no reference to the physical mass and coupling so it is not good for comparison with experiment. It is a companion to dimensional regularization. Theorists like it because they no longer make comparison with experiment and the minimal subtraction renormalization prescription is easy. It amounts to just chucking the pole terms in the dimensionally regularized integrals. I'll do it in our example.\\

Again suppressing the Feynman parameter integral and whatever else, we have found
\[ \includegraphics[scale=0.3]{28-fig1.eps} = \mu^{4-d} \lambda^2 \pi^2 \Bigg[ \frac{1}{2 - \frac{d}{2} } + \text{finite as } d\rightarrow 4 \Bigg] \]

The coefficient of the pole is unambiguous. Minimal subtraction says introduce a counterterm to exactly cancel it. In this example we need a term in $\ml_{c.t.}$
\[ \mu^{4-d} \lambda^2 \pi^2 \frac{1}{2- \frac{d}{2}} \frac{\phi^4}{4!} \qquad \text{(up to $i$s and minus signs)} \]

In what follows, another renormalization prescription is heavily used. It also makes no reference to physical masses and coupling either. It's called BPH. It is useful for proving that renormalization removes the $\infty$'s .\\
\vspace{1cm}

\begin{center}
\textbf{Renormalization and symmetry: \\ a review for non-specialists (1971)}\footnote{[BGC: Annotated photocopies of Coleman's lecture from \underline{Aspects of Symmetry} were attached here]}
\end{center}
\vspace{1cm}

\begin{center}
\textbf{Discussion \\ Chairman: Prof. S. Coleman \\ Scientific Secretary: B.W. Keck}\footnote{[BGC: included in the notes was a photocopy of the discussion from the original publication of the lecture ``Renormalization and symmetry'' in {\em Properties of the Fundamental Interactions} (Editrice Compositori, Bologna, 1973)]}
\end{center}
\[ \hbar = c = 1 \]
\[ [ML] =1 \]
\[ [S] = [\hbar] = 1 \]
\[ [\ml] = M^4 \]
\[ [\phi] = M \qquad [\psi] = M^{3/2} \qquad [\partial_\mu] = M \]

dim $\mathcal{L}_i$ (in mass units) $= b_i + \frac{3}{2} f_i + d_i = \delta_i + 4$\\

Given a set of bosons + fermions the most general interactions of renormalizable type defines a renormalizable theory.\\

The same is true if we restrict the theory to be invariant under parity and internal symmetry.\\ 

The same is true if we allow symmetry breaking interactions if we allow all sym-breaking interactions with dim $\leq n$ ($n =3,2,1$)
\vspace{2cm}

As an example of a matrix element of a composite operator, let's calculate the matrix element of $\frac{1}{2} \phi^2$ between single nucleon states in our meson nucleon theory. As equation (9) in ``Renormalization and Symmetry'' suggests we add 
\[ \frac{1}{2} \mathcal{J}(x) \phi^2(x) \]

to $\ml$. This gives us a new Feynman rule \\
\[ \includegraphics[scale=0.5]{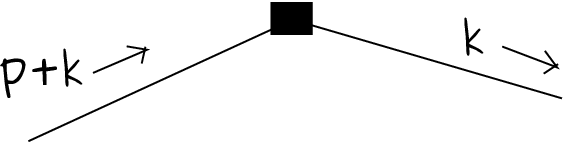} \Longleftrightarrow i \widetilde{\mathcal{J}}(p) \]

To $\mathcal{O}(g^2) $, $\langle l', u' | \frac{1}{2} \phi^2 (x) | l, u \rangle $ is calculated by evaluating\\
\begin{center}
\includegraphics[scale=0.5]{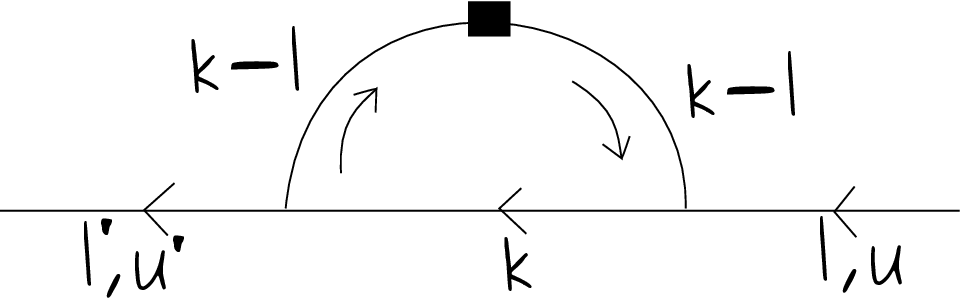}
\end{center}
\begin{align*}
&= i \widetilde{\mathcal{J}} (l -l') (-ig)^2 \int \frac{d^4 k}{(2\pi)^4} \overline{u'} \frac{i}{\cancel{k} -m} u \frac{i}{(k-l')^2 - \mu^2} \frac{i}{(k-l)^2 - \mu^2} \\
&= - \widetilde{\mathcal{J}}(l-l') g^2 \int \frac{d^4k}{(2\pi)^4} \frac{\overline{u'} (\cancel{k} +m ) u }{k^2 - m^2} \frac{1}{(k-l')^2-\mu^2} \frac{1}{(k-l)^2- \mu^2} \\
&= -2g^2 \widetilde{\mathcal{J}} (l - l') \\
& \quad \int^1_0 dx \int^{1-x}_0 dy \int \frac{d^4 k}{(2\pi)^4} \frac{\overline{u'} (\cancel{k} +m) u }{ \{ (1-x-y)(k^2- m^2) + x [(k-l')^2 - \mu^2 ] + y [(k-l)^2 - \mu^2 ] \}^3} \\
&= -2g^2 \widetilde{\mathcal{J}} (l - l') \\
& \quad \int^1_0 dx \int^{1-x}_0 dy \int \frac{d^4 k}{(2\pi)^4} \frac{\overline{u'} (\cancel{k} +m) u }{ [ k^2 - (1-x-y)m^2 - 2xk\cdot l' - 2y k \cdot l - (x+y)\mu^2 + xl'^2 + yl^2 ]^3} \\
&\qquad \qquad \qquad (k' = k - xl' -yl) \\
&= - 2g^2 \widetilde{\mathcal{J}}( l-l') \\
&\quad\int^1_0 dx \int^{1-x}_0 dy \int \frac{d^4 k'}{(2\pi)^4} \frac{\overline{u'} (\cancel{k}' + x \cancel{l}' + y \cancel{l} + m ) u }{[ k'^2 + x(1-x)l'^2 + y(1-y)l^2 - 2xyl \cdot l' - (1-x-y) m^2 - (x+y)\mu^2 ]^3 } 
\end{align*}

The $\cancel{k}'$ term in the numerator is seen to be odd. Also $l^2 =l'^2 = m^2$, $\cancel{l}u = mu $ and $\overline{u'} \cancel{l}' = m \overline{u'}$ are simplifications. We have (dropping prime on $k$)
\[ -2g^2 \widetilde{\mathcal{J}} (l - l') \int^1_0 dx \int^{1-x}_0 dy \int \frac{d^4 k}{(2\pi)^4} \frac{(x+y+1)m \overline{u'} u}{[k^2 - M^2 (x,y)]^3 } \]

where
\[ M^2(x,y) = [-x(1-x)-y(1-y)+1-x-y]m^2 + (x+y) \mu^2 + 2xy l \cdot l' \]

The $k$ integration is in our tables.
\[ 2ig^2 \widetilde{\mathcal{J}}(l-l') m \overline{u'} u \frac{1}{32\pi^2} \int^1_0 dx \int^{1-x}_0 dy \frac{x+y+1}{M^2(x,y)} \]

Let's just call the result of the Feynman parameter integrations
\[ F(m^2, \mu^2, (l-l')^2) \]

so what we have is 
\[ \frac{g^2}{16\pi^2} i \widetilde{\mathcal{J}}(l-l') m \overline{u'} u F (m^2, \mu^2, (l -l\,')^2) \]

To get $\langle l',u' | \frac{1}{2} \phi^2(x) | l, u \rangle $ from this we have to write $\widetilde{\mathcal{J}}(l-l')$ in terms of $\mathcal{J}(x)$, divide by $i$ and then take $\displaystyle \frac{\delta}{\delta \mathcal{J}(x)}$
\begin{align*}
\langle l', u' | \frac{1}{2} \phi^2(x) | l, u \rangle &= \frac{\delta}{\delta \mathcal{J} (x) } \Big[ \frac{g^2}{16 \pi^2} \int d^4 x \; e^{-i (l - l') \cdot x } \mathcal{J}(x) m \overline{u'} u F \Big] \\
&= \frac{g^2}{16 \pi^2} e^{-i (l-l') \cdot x} m \overline{u'} u F (m^2, \mu^2, (l - l\,')^2) 
\end{align*}
\vspace{1cm}

\begin{center}
\textbf{Renormalization of composite operators}
\end{center}

Unfortunately even in a theory that was finite to some order in perturbation theory, the matrix elements of composite operators will not necessarily be finite to that order.\\

Redefinitions of the composite operator are necessary and additional renormalization conditions to make these redefinitions definite are needed.\\

In the method of getting Feynman rules for composite operators by adding a source coupled to the operator to $\ml$ the redefinitions come as further additions multiplied by the same source. For example we will see that at order $\lambda$ in a theory with a $\phi^4$ interaction it is necessary to add to $\ml$ in addition to 
\[ \mathcal{J} (x) \frac{1}{2} \phi^2(x) \]

further terms
\[ \mathcal{J} (x) (\frac{A}{2} \phi^2(x) + B) \]

The total coefficient of $\mathcal{J}(x)$ is
\[ \frac{1}{2} \phi^2 (x) (1 +A) + B \equiv \frac{1}{2} \phi_R^2 \]

We'll have cutoff independent matrix elements in the limit of large cutoff. The finite parts of $A$ and $B$ will be determined by renormalization conditions.\footnote{\[ \ml \longrightarrow \ml + J(x) \Big(\Theta(x) - \!\!\!\!\!\!\!\!\!\!\!\!\underbrace{\sum}_{\substack{\text{all ops with}\\\text{right sym prop}\\\text{of lower dim}}}\!\!\!\!\!\!\!\ \Theta \Big) \]}

Rather than calculate a matrix element of $\frac{1}{2} \phi_R^2$, let's calculate
\[ \stai T (\phi^2_R(x) \phi'(y_1) \phi'(y_2) \cdots \phi'(y_n) ) \sta \]

at least for $n=0$ and $n=2$, to order $\lambda$. We'll just calculate the Fourier transform
\begin{multline}\label{eq:28-page27}
\widetilde{G}(p;q_1,\cdots, q_n) = \includegraphics[scale=0.3]{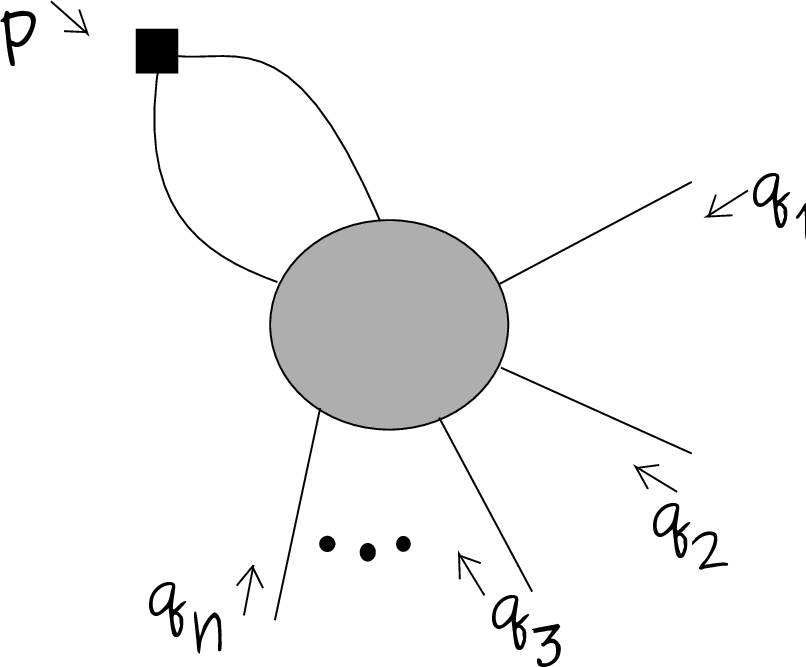} \\
=\int d^4x\; d^4y_1 \cdots d^4 y_n \; e^{-ip\cdot x} e^{-i (q_1 \cdot y_1 + \cdots + q_n \cdot y_n)} \stai T(\phi^2_R(x) \phi'(y_1) \cdots \phi'(y_n)) \sta 
\end{multline}

For $n=2$ the contributions to $\mathcal{O}(\lambda)$ are 
\begin{center}
\includegraphics[scale=0.4]{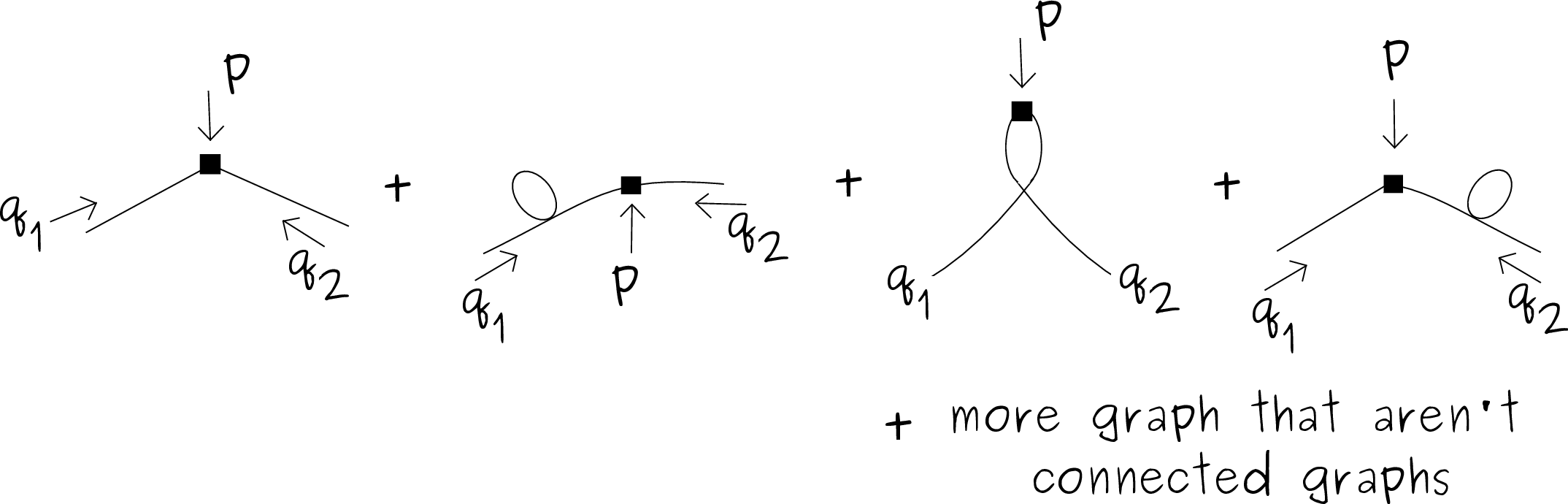}
\end{center}

I'll discuss these contributions in reverse order. The disconnected graphs will be disposed of by only computing the connected part of $\widetilde{G}$, denoted 
\[ \widetilde{G}_c (p; q_1, \cdots, q_n) \]

The second and fourth graphs will be exactly cancelled by the $\mathcal{O}(\lambda)$ mass renormalization counterterm graphs
\begin{center}
\includegraphics[scale=0.4]{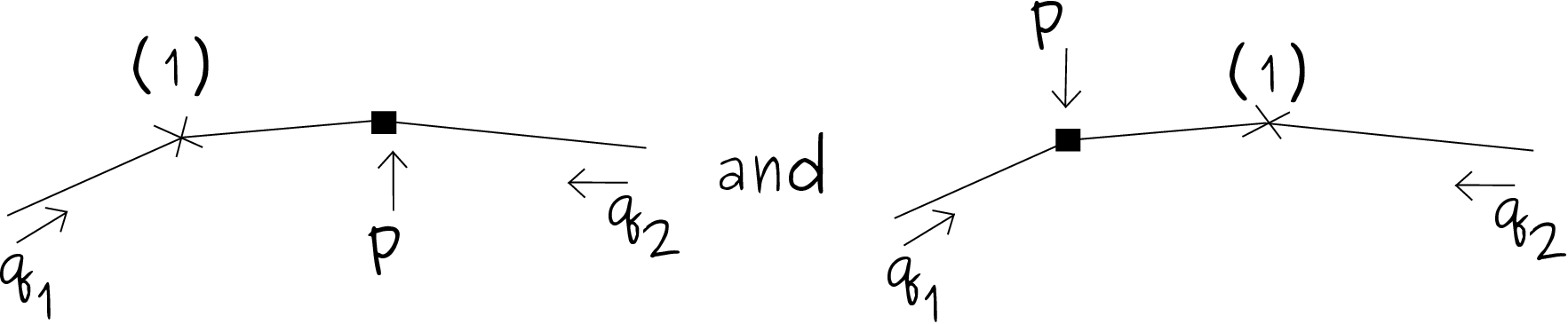}
\end{center}

The third graph is moderately interesting. It is
\[ \frac{i}{q_1^2-\mu^2} \frac{i}{q_2^2-\mu^2} (2\pi)^4 \delta^{(4)}(q_1+q_2-p) \cdot \frac{(-i\lambda)}{2} \int \frac{d^4k}{(2\pi)^4} \frac{i}{(k-\frac{p}{2})^2-\mu^2} \frac{i}{(k+\frac{p}{2})^2-\mu^2} \]

This integral is logarithmically divergent which is why we need the $\mathcal{O}(\lambda )$ graph coming from the $\mathcal{O}(\lambda )$ part of 
\[ A \phi^2 \]

in $\phi_R^2$. I'll denote that $A^{(1)} \phi^2$ and we get one more graph 
\begin{center}
\includegraphics[scale=0.4]{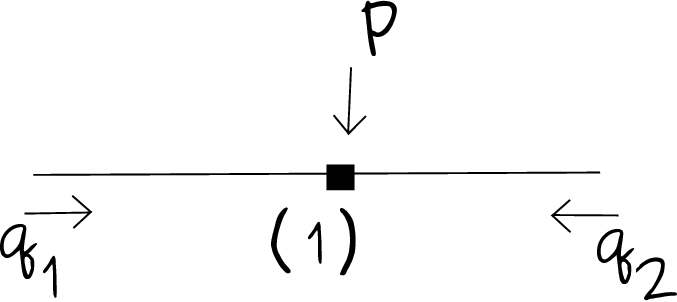}
\end{center}

This graph and the first graph give 
\[ (1 + A^{(1)}) \frac{i}{q_1^2 - \mu^2} \frac{i}{q_2^2 -\mu^2} (2\pi)^4 \delta^{(4)} (q_1 + q_2 - p) \]

$A^{(1)}$ is a divergent constant chosen to cancel the logarithmically divergent part of the second graph. It is sufficient to do that. A superficially log divergent graph only needs one subtraction (in any order of perturbation theory) according to BPHZ.\\

We need a renormalization condition to determine the finite part of $A$. A logical one is that $\widetilde{G}_c$ at zero momentum be given exactly by its lowest order contribution. This means
\[ A^{(1)} + \frac{(-i\lambda)}{2} \int \frac{d^4k}{(2\pi)^4} \frac{i^2}{(k^2- \mu^2)^2} = 0 \]

So finally to $\mathcal{O}(\lambda)$ 
\begin{align*}
\widetilde{G}_c (p; q_1, q_2) = &(2\pi)^4 \delta^{(4)} ( q_1 + q_2 - p) \frac{i}{q_1^2 - \mu^2} \frac{i}{q_2^2 - \mu^2} \\
&\cdot \Bigg\{ 1 - \frac{i \lambda}{2} \int \frac{d^4k}{(2\pi)^4} \Bigg[ \frac{i}{(k-\frac{p}{2})^2 - \mu^2 } \frac{i}{(k+ \frac{p}{2})^2 - \mu^2 } - \frac{i^2}{(k^2-\mu^2)^2} \Bigg] \Bigg\} 
\end{align*}

$B^{(1)}$ could be chosen so that 
\[ \stai \frac{1}{2} \phi_R^2 (x) \sta = 0 \]

It is surprising to me that there is so much arbitrariness in the definition of Green's function with a composite operator that has to be fixed by renormalization conditions.\\

If someone hands you $T\mn = \partial^\mu \phi \partial^\nu \phi - g\mn \ml $ (the energy-momentum tensor obtained through Noether's theorem), you'd think something like
\[ \stai T\mn \sta \]

being such a physical thing would not be susceptible to redefinition. Apparently the counterterms for a conserved current can usually be pinned down by calling upon cherished properties such as
\[ T^{\mu\nu} = - T^{\nu\mu} \qquad \text{and} \qquad \partial_\mu T^{\mu\nu} = 0 \]

\vspace{1cm}
A stupid example to make sure I have my F.T.~conventions right. Let's compute 
\[ \stai T(\psi(x) \psib(y) ) \sta \qquad \text{in free Dirac theory} \]

According to Eq.~(\ref{eq:28-page27}) this should be
\begin{align*}
&\qquad \int \frac{d^4 q_1}{(2\pi)^4} \frac{d^4 q_2}{(2\pi)^4} e^{iq_1 \cdot x + i q_2 \cdot y} \includegraphics[scale=0.3]{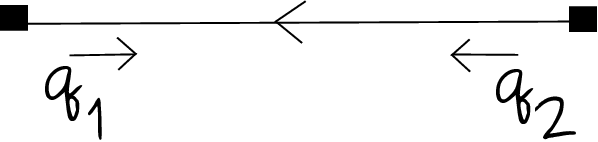} \\
&=\int \frac{d^4 q_1}{(2\pi)^4} \frac{d^4 q_2}{(2\pi)^4} e^{iq_1 \cdot x + iq_2 \cdot y } (2\pi)^4 \delta^{(4)} (q_1 + q_2) \cdot
\frac{i}{\cancel{q}_2 - m + \ie} \\
&= \int \frac{d^4 q}{(2\pi)^4} e^{-iq \cdot (x-y)} \cdot \frac{i}{\cancel{q} - m + \ie} 
\end{align*}

in agreement with Dec.~18, Eq.~(\ref{eq:24-page5}) and following ``3 comments''.
}{

}\end{document}